\documentclass[aps,rmp,twocolumn,amsfonts,amssymb,longbibliography]{revtex4-1}
\newcommand{\setbasefolder}[1]{\newcommand{\basefolder}{#1}}
\setbasefolder{./}
\usepackage[toc,page]{appendix}
\usepackage[T1]{fontenc}
\usepackage{url}
\usepackage{upgreek}
\usepackage{epsfig}
\usepackage{newfloat}
\usepackage{mdframed}
\usepackage{amssymb,amsmath}
\usepackage{cancel}
\usepackage{graphicx}
\usepackage{epstopdf}
\usepackage[mathscr]{eucal}
\usepackage[usenames,dvipsnames]{pstricks}
\usepackage{pst-grad} 
\usepackage{pst-plot} 
\usepackage{verbatim}
\usepackage{physics}
\usepackage{textgreek}
\usepackage{siunitx}
\usepackage{amsthm}
\usepackage{accents}
\usepackage{appendix}
\usepackage{cancel}
\usepackage{bm}
\usepackage{xifthen}
\usepackage{amsthm}
\usepackage{amsmath}
\usepackage{upgreek}
\usepackage{esvect}
\usepackage{tensor}
\usepackage{natbib}

\let\realcitep\citep
\setcitestyle{square}
\renewcommand*{\citep}[1]{{\fontfamily{qcs}\selectfont\footnotesize \textit{\realcitep{#1}}}}
\let\realcitet\citet
\renewcommand*{\citet}[1]{{\fontfamily{qcs}\selectfont \small \textit{\realcitet{#1}}}}
\newcommand{\REF}{}
\newcommand{\REFS}{}
\newcommand{\linecite}[1]{\onlinecite{#1}}

\let\reallinecite\linecite
\renewcommand*{\linecite}[1]{{\fontfamily{qcs}\selectfont \small \textit{\reallinecite{#1}}}}
\usepackage{etoolbox}

\newcounter{tempsection}

\newcommand{\Walpha}{W_\alpha^{\scriptscriptstyle(\Nelec-1)}}
\newcommand{\fourierconst}{\mathfrak{o}}
\newcommand{\energy}{\varepsilon}

\newcommand{\hamtwo}{\hat{w}}
\newcommand{\pdf}{\mathfrak{p}}
\newcommand{\spin}{\upsilon}

\newcommand{\bbraket}[2]{(#1|#2)}

\newcommand{\Hamint}{\Hamtwo_{\mathrm{int}}}
\newcommand{\CPenergy}[1]{\mathcal{E}^{\scriptscriptstyle (#1)}}

\newcommand{\sxe}{\mathcal{I}_{\scriptscriptstyle -}}
\newcommand{\sxn}{\mathcal{I}_{\scriptscriptstyle +}}
\newcommand{\sye}{\mathcal{J}_{\scriptscriptstyle -}}
\newcommand{\syn}{\mathcal{J}_{\scriptscriptstyle +}}

\newcommand{\lcontract}[2]{\ket{#1\rfloor #2}}
\newcommand{\rcontract}[2]{\bra{#1\lfloor #2}}
\newcommand{\lcontractN}[2]{\ket{#1\rfloor #2}_1}
\newcommand{\rcontractN}[2]{\bra{#1\lfloor #2}_1}

\newcommand{\hamone}{\hamsmall}

\newcommand{\Energy}{W}

\newcommand{\spinup}{\uparrow}
\newcommand{\spindown}{\downarrow}
\newcommand{\spinupdown}{\uparrow\downarrow}

\newcommand{\pden}{\rho}

\newcommand{\antisymmetrizer}{\hat{\mathcal{A}}}
\newcommand{\PsiHF}{\Psi_{\scriptscriptstyle\mathrm{HF}}}

\newcommand{\Esingle}{E_s}
\newcommand{\Esingleapprox}{\Esingle^{\scriptscriptstyle \approx}}
\newcommand{\Emf}{E_{\scriptscriptstyle \mathrm{MF}}}
\newcommand{\Estatic}{E_{\scriptstyle \mathrm{MF}}}
\newcommand{\Estatarg}[1]{E_{\scriptstyle \mathrm{MF}}^{\scriptscriptstyle #1}}
\newcommand{\Edyn}{E_{\scriptscriptstyle \mathrm{Cor}}}
\newcommand{\Edynarg}[1]{E_{\scriptscriptstyle \mathrm{Cor}}^{\scriptscriptstyle #1}}
\newcommand{\edynarg}[1]{\epsilon_{\scriptscriptstyle \mathrm{Cor}}^{\scriptscriptstyle #1}}
\newcommand{\Erestapprox}{\Delta^{\scriptscriptstyle \approx}}
\newcommand{\Erest}{\Delta}

\DeclareMathOperator{\Hamiltonian}{H}
\DeclareMathOperator{\Wamiltonian}{W}
\newcommand{\Ham}{\hat{\Hamiltonian}}
\newcommand{\Hamone}{\Ham_{\scriptstyle \mathrm{ind}}}
\newcommand{\Hamtwo}{\hat{\Wamiltonian}}

\newcommand{\hamsmalltwo}{\hat{w}}

\newcommand{\identity}{\hat{\mathcal{I}}}

\newcommand{\occ}{\mathfrak{f}}

\newcommand{\dmeasure}[1]{\dd{\scriptscriptstyle [#1]}}
\newcommand{\dmeasureA}[1]{\mathrm{d}^{\scriptscriptstyle \mathsf{CP1}}_{\scriptscriptstyle [#1]}}
\newcommand{\dmeasureB}[1]{\mathrm{d}^{\scriptscriptstyle \mathsf{CP2}}_{\scriptscriptstyle [#1]}}

\newcommand{\vext}{v^{\text{ext}}}
\newcommand{\densityn}{n^{\scriptscriptstyle (N-1)}}

\newcommand{\dmatrixarg}[1]{\Gamma^{\scriptscriptstyle (#1)}}

\newcommand{\Wn}{W^{\scriptscriptstyle (N-1)}}

\newcommand{\dbarvp}[1]{\delta\bar{v}_{\scriptscriptstyle +}^{\scriptscriptstyle (#1)}}
\newcommand{\dbarvm}[1]{\delta\bar{v}_{\scriptscriptstyle -}^{\scriptscriptstyle (#1)}}

\newcommand{\pdensuper}[1]{\pden^{\scriptscriptstyle (#1)}}
\newcommand{\pdensuperp}[1]{\pden^{\scriptscriptstyle (#1)}_{\scriptscriptstyle +}}
\newcommand{\pdensuperm}[1]{\pden^{\scriptscriptstyle (#1)}_{\scriptscriptstyle -}}

\newcommand{\numpx}{N^{\scriptscriptstyle (1)}_{\scriptscriptstyle +}}

\newcommand{\nummy}{N^{\scriptscriptstyle (2)}_{\scriptscriptstyle -}}

\newcommand{\pdenup}[1]{\pden^{\scriptscriptstyle (#1)}}
\newcommand{\pdenarg}[1]{\rho^{\scriptscriptstyle (#1)}}
\newcommand{\pdfarg}[1]{\pdf^{\scriptscriptstyle (#1)}}
\newcommand{\rhoarg}[1]{\rho^{\scriptscriptstyle (#1)}}

\newcommand{\entropy}{\mathcal{S}}
\newcommand{\tentropy}{\tilde{\mathcal{S}}}
\newcommand{\den}{n}

\newcommand{\denmm}{\den_{\scriptscriptstyle --}}
\newcommand{\denpp}{\den_{\scriptscriptstyle ++}}
\newcommand{\denpm}{\den_{\scriptscriptstyle +-}}
\newcommand{\denmp}{\den_{\scriptscriptstyle -+}}
\newcommand{\dencmm}[1]{\den_{\scriptscriptstyle -|-}^{\scriptscriptstyle (#1)}}
\newcommand{\dencpp}[1]{\den_{\scriptscriptstyle +|+}^{\scriptscriptstyle (#1)}}
\newcommand{\dencpm}[1]{\den_{\scriptscriptstyle +|-}^{\scriptscriptstyle (#1)}}
\newcommand{\dencmp}[1]{\den_{\scriptscriptstyle -|+}^{\scriptscriptstyle (#1)}}
\newcommand{\ddencmm}[1]{\delta\den_{\scriptscriptstyle -|-}^{\scriptscriptstyle (#1)}}
\newcommand{\ddencpp}[1]{\delta\den_{\scriptscriptstyle +|+}^{\scriptscriptstyle (#1)}}
\newcommand{\ddencpm}[1]{\delta\den_{\scriptscriptstyle +|-}^{\scriptscriptstyle (#1)}}
\newcommand{\ddencmp}[1]{\delta\den_{\scriptscriptstyle -|+}^{\scriptscriptstyle (#1)}}

\newcommand{\denargp}[1]{\den^{\scriptscriptstyle (#1)}_{\scriptscriptstyle +}}
\newcommand{\denargm}[1]{\den^{\scriptscriptstyle (#1)}_{\scriptscriptstyle -}}

\newcommand{\interact}[2]{w(#1,#2)}

\newcommand{\p}{{\scriptscriptstyle +}}
\newcommand{\subminus}{{\scriptscriptstyle -}}
\newcommand{\Eint}{E_{\mathrm{int}}}

\newcommand{\what}{\hat{w}}

\newcommand{\Eintmf}{E_{\scriptscriptstyle 12}^{\scriptscriptstyle \mathrm{MF}}}
\newcommand{\barv}{\bar{v}}

\newcommand{\Eintcp}[1]{E_{\mathrm{int}}^{\scriptscriptstyle #1}}

\newcommand{\PRLsep}{\noindent\makebox[\linewidth]{\resizebox{0.3333\linewidth}{1pt}{$\bullet$}}\bigskip}

\newcommand{\bloch}{b}
\newcommand{\ftsbloch}{\fts{\bloch}}
\newcommand{\ftspbloch}{\fts{\pbloch}}

\newcommand{\pbloch}{u}

\newcommand{\tpbloch}{\tilde{u}}

\newcommand{\reciplatt}{\mathfrak{L}_{G}}

\newcommand{\reciplattg}{\mathfrak{L}_{g}}

\newcommand{\bandgap}{\Delta_\text{gap}}
\newcommand{\dimension}{\mathfrak{d}}
\newcommand{\Ndof}{{N_d}}
\newcommand{\Nparticle}{N_p}
\newcommand{\Nident}{N}
\newcommand{\precmicro}{h_{0}}
\newcommand{\Nelec}{\Nident}
\newcommand{\Nelecbulk}{\Nelec_{\scriptscriptstyle \bulk}} 

\newcommand{\Neleccell}{\Nelec_{\unitcell}}

\newcommand{\braketT}[2]{\braket{#1}{#2}_{\scriptscriptstyle \onetorus}}
\newcommand{\braketR}[2]{\braket{#1}{#2}_{\scriptscriptstyle \realone}}
\newcommand{\braketD}[2]{\braket{#1}{#2}_{\scriptscriptstyle \domain}}
\newcommand{\braketcell}[2]{\braket{#1}{#2}_{\scriptscriptstyle \unitcell}}

\newcommand{\densityset}{\boldsymbol{\Psi}}
\newcommand{\phaseset}{\boldsymbol{\Theta}}
\newcommand{\densityfunction}{\varphi^{\scriptscriptstyle (n)}}
\newcommand{\indexsetN}{\indexset_{\Nident}}

\newcommand{\varphiset}{\{\varphi_i(\zeta)\}}

\newcommand{\varphisetcell}{\{\varphi_i(\zeta)\}_\unitcell}
\newcommand{\tvarphi}{\tilde{\varphi}}

\newcommand{\closedopen}[1]{[#1)}

\newcommand{\tG}{\tilde{G}}

\newcommand{\barF}{\bar{F}}

\newcommand{\smallhamop}{\hat{\mathfrak{h}}}
\newcommand{\smallham}{\mathfrak{h}}

\newcommand{\momspace}{\mathbb{P}}

\newcommand{\discreal}{\mathscr{X}}
\newcommand{\indexset}{\mathscr{I}}

\newcommand{\gradx}{\partial_x}

\newcommand{\partialt}{\partial_t}

\newcommand{\gradt}{\partial_t}
\newcommand{\gradG}{\partial_{\scriptscriptstyle \Gamma}}
\newcommand{\gradp}{\partial_\pi}

\newcommand{\hilbertv}{\hilbert_{\hilbv}}

\newcommand{\sqhilbv}{\sqrt{\hilbv}}

\newcommand{\phasespace}{\mathbb{G}}

\newcommand{\configspace}{\mathbb{Q}}

\newcommand{\intconfig}{\int_{\configspace}}

\newcommand{\alattx}{a_x}

\newcommand{\psdist}{\pdf_{\scriptscriptstyle \phasespace}}

\newcommand{\cfgdist}{\pdf_{\scriptscriptstyle \configspace}}
\newcommand{\momdist}{\pdf_{\scriptscriptstyle \momspace}}
\newcommand{\momcfgdist}{\pdf_{\scriptscriptstyle \momspace|\configspace}}
\newcommand{\cfgmomdist}{\pdf_{\scriptscriptstyle \configspace|\momspace}}
\newcommand{\intervalmeas}{\interval_{\text{meas}}}
\newcommand{\pdfO}{\mathfrak{p}_{\Obs}}

\newcommand{\compress}{\Lambda}
\newcommand{\compressx}{\compress_x}
\newcommand{\compressp}{\compress_\pi}

\newcommand{\prs}{\pdf^{\scriptscriptstyle (\subject)}}
\newcommand{\prp}{\pdf^{\scriptscriptstyle (\probe)}}
\newcommand{\prcm}{\pdf^{\scriptscriptstyle (\combined)}}
\newcommand{\prps}{\pdf^{\scriptscriptstyle (\probe|\subject)}}
\newcommand{\prsp}{\pdf^{\scriptscriptstyle (\subject|\probe)}}
\newcommand{\psicm}{\psi^{\scriptscriptstyle (\combined)}}

\newcommand{\psis}{\psi^{\scriptscriptstyle (\subject)}}

\newcommand{\uppsis}{\uppsi^{\scriptscriptstyle (\subject)}}

\newcommand{\Gammasub}[1]{\Gamma_{#1}}

\newcommand{\Gammat}{\Gammasub{\!t}}

\newcommand{\intmom}{\int_{\momspace}}

\newcommand{\subject}{\mathscr{S}}
\newcommand{\probe}{\mathscr{P}}
\newcommand{\combined}{\mathscr{C}}

\newcommand{\configS}{\configspace}
\newcommand{\configP}{\ydomain}
\newcommand{\glide}{\hat{\mathfrak{g}}}
\DeclareMathOperator{\hU}{\hat{U}}

\DeclareMathOperator{\hB}{\hat{B}}
\DeclareMathOperator{\hone}{\hat{\mathbb{I}}}
\DeclareMathOperator{\hO}{\hat{O}}

\newcommand{\hOx}{\hO_x}

\DeclareMathOperator{\hH}{\hat{H}}

\newcommand{\tPsi}{\psi}

\newcommand{\tPsiv}{\tPsi_{\hilbv}}
\newcommand{\phix}{\phi_{x}}
\newcommand{\phik}{\phi_{k}}

\newcommand\thickbar[1]{\accentset{\rule{.6em}{0.5pt}}{#1}}

\newcommand{\barbsigmaNu}{\thickbar{\boldsymbol{\sigma}}_\mathcal{V}}

\newcommand{\bsigmaNu}{\boldsymbol{\sigma}_\mathcal{V}}
\newcommand{\bsigmadNu}{\boldsymbol{\sigma}_{\mathcal{V}^{(1)}}}

\newcommand{\wannierset}{\mathcal{W}}
\newcommand{\wannierxset}{\mathscr{X}}
\newcommand{\barx}{\bar{x}}

\newcommand{\htx}{\hat{x}}

\newcommand{\qfig}[1]{\mathbf{q}^{\scriptscriptstyle (#1)}}

\newcommand{\denop}{\hat{n}}

\newcommand{\vextop}{\hat{v}^{\text{ext}}}
\newcommand{\vmfop}{\hat{v}^{\scriptscriptstyle \text{MF}}}
\newcommand{\vmf}{v^{\scriptscriptstyle \text{MF}}}
\newcommand{\Dop}{\hat{\mathcal{D}}}
\newcommand{\functionalarg}[1]{\mathscr{F}_{\scriptscriptstyle #1}}
\newcommand{\lagrangearg}[1]{\uplambda_{\scriptscriptstyle #1}}
\newcommand{\Df}{\mathcal{D}}

\newcommand{\A}{\vec{A}}
\newcommand{\avec}{\vec{a}}

\newcommand{\D}{\vec{D}}

\newcommand{\G}{\mathcal{G}}

\newcommand{\M}{\vec{M}}
\newcommand{\manifoldn}[1]{\mathscr{M}^{\expval{#1}}}
\newcommand{\manifold}{\mathscr{M}}
\newcommand{\volumeform}{\mathit{m}}
\newcommand{\B}{\vec{B}}

\newcommand{\hh}{\vec{H}}
\newcommand{\Rho}{{\boldsymbol{\varrho}}}
\newcommand{\DRho}{\boldsymbol{\Delta\varrho}}
\newcommand{\Rhofree}{{\boldsymbol{\varrho}^{\bm{\text{free}}}}}
\newcommand{\Rhobound}{{\boldsymbol{\varrho}^{\bm{\text{bound}}}}}
\newcommand{\Jbound}{{\vec{J}^{\bm{\text{bound}}}}}
\newcommand{\Jfree}{{\vec{J}^{\bm{\text{free}}}}}

\newcommand{\bphi}{\boldsymbol{\Phi}}
\newcommand{\dbphi}[1]{\bm{\Phi^{(#1)}}}

\newcommand{\dphi}{\dot{\phi}}

\newcommand{\bphibethe}{\boldsymbol{\Phi}^\textrm{Bethe}}
\newcommand{\phir}{{\phi_r}}

\newcommand{\bpsi}{\boldsymbol{\Psi}}
\newcommand{\Psigs}{\Psi_0}

\newcommand{\micro}{\mathfrak{m}}
\newcommand{\dmicro}{\dot{\mathfrak{m}}}
\newcommand{\dwidth}{\upsilon}
\newcommand{\dmacro}{\dot{\mathfrak{M}}}
\newcommand{\macro}{\mathfrak{M}}

\newcommand{\volume}{\abs{\Omega}}

\newcommand{\phibar}{\bar{\Phi}}
\newcommand{\rhobar}{\bar{\rho}}
\newcommand{\Dbphi}{\boldsymbol{\Delta\Phi}}
\newcommand{\bsigmadot}{{\mathbf{\dot{\boldsymbol{\sigma}}}}}
\newcommand{\bqdot}{\mathbf{\dot{\mathbfcal{Q}}}}
\newcommand{\linecharge}{\mathbfcal{L}}

\newcommand{\linechargeNuind}[1]{\mathbfcal{L}_{\mathcal{V},#1}}
\newcommand{\bsigmaNuind}[1]{\boldsymbol{\sigma}_{\mathcal{V},#1}}
\newcommand{\pointcharge}{\mathbf{q}}

\newcommand{\pointchargeNuind}[1]{\mathbf{q}_{\mathcal{V},#1}}
\newcommand{\dbsigma}{\dot{\boldsymbol{\sigma}}}
\newcommand{\bsigma}{{\boldsymbol{\sigma}}}
\newcommand{\Dbsigma}{{\boldsymbol{\Delta\sigma}}}

\newcommand{\bsigmafree}{\boldsymbol{\sigma}^{\bm{\text{free}}}}
\newcommand{\bsigmabound}{\boldsymbol{\sigma}^{\bm{\text{bound}}}}
\newcommand{\notiff}{%
  \mathrel{{\ooalign{\hidewidth$\not\phantom{"}$\hidewidth\cr$\iff$}}}}

\newcommand{\Eext}{{\vec{E}_\mathrm{ext}}}
\newcommand{\EAfar}{{\vec{E}_\mathrm{A}^\mathrm{far}}}
\newcommand{\EAnear}{{\me_\mathrm{A}^{\scriptsize \mathrm{near}}}}
\newcommand{\EAnearmacro}{{\vec{E}_\mathrm{A}^\mathrm{near}}}
\newcommand{\EAnearmicro}{{\me_\mathrm{A}^\mathrm{near}}}

\newcommand{\bh}{\bm{h}}

\newcommand{\bnu}{\bar{\nu}}

\newcommand{\brho}{\bar{\rho}}

\newcommand{\ftsnu}{\fts{\nu}}
\newcommand{\yznu}{\bnu_{yz}}
\newcommand{\yzNu}{\bbNu_{yz}}
\newcommand{\valnu}{\upsilon}

\newcommand{\coincidence}[1]{\left[#1\right]_{\Lequiv}}

\newcommand{\Nu}{\mathbfcal{V}}

\newcommand{\DNu}{\boldsymbol{\Delta}\mathbfcal{V}}
\newcommand{\bDelta}{\boldsymbol{\Delta}}
\newcommand{\bDx}{\boldsymbol{\Delta}\vec{x}}

\newcommand{\bNu}{{\overline{\mathcal{V}}}}
\newcommand{\bbNu}{{\bm{\overline{\mathcal{V}}}}}

\newcommand{\dDNu}[1]{\bm{\Delta\mathcal{V}^{(#1)}}}
\newcommand{\dNu}[1]{\bm{{\mathcal{V}^{(#1)}}}}

\newcommand{\depth}{{\xi_b}}

\newcommand{\pp}{\vec{P}}

\newcommand{\mbp}{\mathbfcal{P}}
\newcommand{\mbd}{\mathbfcal{D}}

\newcommand{\mpp}{\mathcal{P}}
\newcommand{\mbpdot}{{\mathbf{\dot{\mathbfcal{P}}}}}
\newcommand{\mbddot}{{\mathbf{\dot{\mathbfcal{D}}}}}

\newcommand{\mss}{\mathcal{S}^{[\Delta\nu]}}
\newcommand{\mbs}{\mathbfcal{S}^{[\DNu]}}

\newcommand{\bsms}{\bar{\mathcal{S}}^{[\Delta\nu]}}
\newcommand{\mssx}[1]{\mathcal{S}^{[#1]}}
\newcommand{\mbsx}[1]{\mathbfcal{S}^{\expval{#1}}}

\newcommand{\bsmsx}[1]{\bar{\mathcal{S}}^{\expval{#1}}}
\newcommand{\Dnu}{{\Delta\nu}}
\newcommand{\bDnu}{\overline{\Delta\nu}}

\newcommand{\Vregion}{V}
\newcommand{\Volume}{\abs{\Vregion}}

\newcommand{\hilbert}{{\mathcal{H}}}
\newcommand{\tpsi}{\tilde{\psi}}

\newcommand{\hilbv}{\mathrm{v}}
\newcommand{\ftshilbv}{\fts{\hilbv}}

\newcommand{\myket}[1]{\big|#1\big\rangle}

\newcommand{\mybraket}[2]{\big\langle #1 \big| #2\big\rangle}
\newcommand{\myexpval}[2]{\big\langle #2 \big|#1\big| #2\big\rangle}

\newcommand{\Navogadro}{N_{\scriptscriptstyle \mathrm{A}}}

\newcommand{\bchi}{\boldsymbol{\chi}}

\newcommand{\E}{\vec{E}}
\newcommand{\DE}{\boldsymbol{\Delta}\E}

\newcommand{\wavekmin}{k_\mathrm{min}}

\renewcommand{\H}{\vec{H}}
\newcommand{\me}{\mathcal{E}}
\newcommand{\melo}{\mathcal{E}_{\scriptscriptstyle \mathrm{LO}}}

\newcommand{\J}{\vec{J}}

\renewcommand{\j}{{j}}

\newcommand{\rhopm}{\rho^{\rhoscriptsize \pm}}

\newcommand{\rhoscriptsize}{\scriptscriptstyle}
\newcommand{\rhop}{\rho^{\rhoscriptsize +}}

\newcommand{\drho}{\dot{\rho}}

\newcommand{\rhom}{\rho^{\rhoscriptsize -}}

\newcommand{\Jconv}{\J^{(p)}}
\newcommand{\kinetic}{\hat{t}}
\newcommand{\dipcell}{d_{\unitcell}}
\newcommand{\Drho}{\Delta\rho}
\newcommand{\Drhobar}{\overline{\Delta\rho}}

\newcommand{\dipcelldot}{\dot{d}_{\scriptscriptstyle \unitcell}}

\newcommand{\Nunitcell}{M_{\unitcell}}
\newcommand{\NUnitcell}{M_{\unitcell}}
\newcommand{\nunitcell}{n_{\unitcell}}

\newcommand{\rhocell}{\rho_{ \unitcell}}

\newcommand{\rhomcell}{\rho^{ -}_{\unitcell}}

\newcommand{\unitcell}{\Omega}
\newcommand{\unitcellscr}{{\scriptscriptstyle \Omega}}

\newcommand{\Jconvp}{{\J^{(p)}_+}}

\newcommand{\Jconvm}{{\J^{(p)}_-}}

\newcommand{\Jcond}{\J^{(c)}}

\newcommand{\jconv}{\j^{(p)}}
\newcommand{\jconvp}{{\j^{(p)}_+}}
\newcommand{\jconvm}{{\j^{(p)}_-}}

\newcommand{\im}{\operatorname{Im}}

\newcommand{\hamsmall}{\hat{h}}

\newcommand{\hamsmallx}{\hamsmall^{\scriptscriptstyle(x)}}

\newcommand{\ksuper}{{\scriptscriptstyle(k)}}
\newcommand{\hamfrak}{\hat{\mathfrak{h}}}
\newcommand{\hamfrakx}{\hat{\mathfrak{h}}^{\scriptscriptstyle (x)}}
\newcommand{\hamfrakr}{\hat{\mathfrak{h}}^{\scriptscriptstyle (r)}}
\newcommand{\Span}{\mathit{span}}
\newcommand{\SPAN}{\Span}

\renewcommand{\P}{\hat{\mathrm{P}}}

\newcommand{\trans}{\mathrm{\hat{T}}}

\newcommand{\BZ}{\fts{\Omega}}

\newcommand{\ftsf}{\fts{f}}
\newcommand{\ftsg}{\fts{g}}
\newcommand{\intbz}{\int_{\BZ}}
\newcommand{\intthree}{\int_{\mathbb{R}^3}}

\newcommand{\intone}{\int_{\mathbb{R}}}
\newcommand{\intfull}{\int_{-\infty}^\infty}

\renewcommand{\vec}[1]{\ensuremath{\mathbf{#1}}}

\newcommand{\phinuc}{\phi^{\text{nuc}}}

\newcommand{\trho}{\tilde{\varrho}}
\newcommand*{\bdot}[1]{%
  \accentset{\mbox{\normalsize\bfseries .}}{#1}}

\newcommand{\xl}{x_{\scriptscriptstyle \! L}}
\newcommand{\xr}{x_{\scriptscriptstyle \! R}}

\newcommand{\iset}{\mathcal{I}}
\newcommand{\xppr}[1][m]{x_{#1}^{+}}
\newcommand{\xmpr}[1][m]{x_{#1}^{-}}
\newcommand{\dxppr}[1][m]{x_{#1}^{+\prime}}
\newcommand{\dxmpr}[1][m]{x_{#1}^{-\prime}}

\newcommand{\mm}{{\scriptscriptstyle -M}}
\newcommand{\m}{{\scriptscriptstyle M}}
\newcommand{\pmm}{{\scriptscriptstyle \pm M}}
\newcommand{\mmnew}{{\scriptscriptstyle -M^\text{new}}}
\newcommand{\mnew}{{\scriptscriptstyle M^\text{new}}}

\newcommand{\mzero}{{\mathcal{M}^{\expval{0}}_{\Delta\nu}}}
\newcommand{\mone}{{\mathcal{M}^{\expval{1}}_{\Delta\nu}}}
\newcommand{\many}{{\mathcal{M}^{\expval{n}}_{\Delta\nu}}}
\newcommand{\mtwo}{{\mathcal{M}^{\expval{2}}_{\Delta\nu}}}

\newcommand{\monerho}{{\mathcal{M}^{\expval{1}}_{\rho}}}
\newcommand{\manyrho}{{\mathcal{M}^{\expval{n}}_{\rho}}}

\newcommand{\boldmone}{{\boldsymbol{\mathcal{M}^{\expval{1}}_{\Delta\nu}}}}
\newcommand{\boldmtwo}{{\boldsymbol{\mathcal{M}^{\expval{2}}_{\Delta\nu}}}}

\newcommand{\boldmtworho}{{\boldsymbol{\mathcal{M}^{\expval{2}}_{\rhobar}}}}
\newcommand{\bmzero}{{\bar{\mathcal{M}}^{\expval{0}}_{\Delta\nu}}}
\newcommand{\bmone}{{\bar{\mathcal{M}}^{\expval{1}}_{\Delta\nu}}}
\newcommand{\bmtwo}{{\bar{\mathcal{M}}^{\expval{2}}_{\Delta\nu}}}
\newcommand{\bmany}{{\bar{\mathcal{M}}^{\expval{n}}_{\Delta\nu}}}

\newcommand{\bmonerho}{{\bar{\mathcal{M}}^{\expval{1}}_{\rho}}}
\newcommand{\bmonerhoq}{{\bar{\mathcal{M}}^{\expval{1}}_{\rho^q}}}
\newcommand{\bmtworho}{{\bar{\mathcal{M}}^{\expval{2}}_{\rho}}}
\newcommand{\bmanyrho}{{\bar{\mathcal{M}}^{\expval{n}}_{\rho}}}

\newcommand{\momone}[1]{\bar{\mathcal{M}}_{\bar{\rho}}^{\expval{1}}({#1})}
\newcommand{\momtwo}[1]{\bar{\mathcal{M}}_{\bar{\rho}}^{\expval{2}}({#1})}
\newcommand{\nmomone}[1]{\bar{\mathcal{M}}_{\Delta\nu}^{\expval{1}}({#1})}
\newcommand{\nmomtwo}[1]{\bar{\mathcal{M}}_{\Delta\nu}^{\expval{2}}({#1})}

\newcommand{\Deltas}{\Delta^s}
\newcommand{\sppr}[1][m]{s_{#1}^{+}}
\newcommand{\smpr}[1][m]{s_{#1}^{-}}

\newcommand{\fouriernoarg}{\mathfrak{F}}

\newcommand{\fourierspace}[1]{\mathfrak{F}_s\left[#1\right]}
\newcommand{\fourierspacetime}[1]{\mathfrak{F}_{st}\left[#1\right]}
\newcommand{\fouriertime}[1]{\mathfrak{F}_t\left[#1\right]}

\newcommand{\bs}{\vec{s}}
\newcommand{\rvec}{\vv{r}}
\newcommand{\kvec}{\vv{\kappa}}

\newcommand{\rhat}{\hat{r}}

\newcommand{\rvecsub}[1]{\vv{r}_{\!#1}}
\newcommand{\rsub}[1]{r_{\!#1}}
\newcommand{\Rvecsub}[1]{\vv{R}_{\!#1}}

\newcommand{\dvecsub}[1]{\vv{d}_{\!#1}}

\newcommand{\Rvec}{\vv{R}}
\newcommand{\Avec}{\vv{A}}

\newcommand{\wx}{\mathcal{X}}
\newcommand{\wy}{\mathcal{Y}}
\newcommand{\integer}{\mathbb{Z}}
\newcommand{\realone}{\mathbb{R}}
\newcommand{\complex}{\mathbb{C}}
\newcommand{\integerpos}{\mathbb{Z}^+}
\newcommand{\integerneg}{\mathbb{Z}^-}
\newcommand{\integernonneg}{\mathbb{Z}^+_0}
\newcommand{\integernonpos}{\mathbb{Z}^-_0}
\newcommand{\realpos}{\mathbb{R}^{+}}
\newcommand{\realnonneg}{\mathbb{R}^{+}_0}
\newcommand{\realneg}{\mathbb{R}^{-}}
\newcommand{\realnonpos}{\mathbb{R}^{-}_0}

\newcommand{\hrealone}{\fts{\realone}}
\newcommand{\domain}{\mathbb{X}}
\newcommand{\domaindual}{\domain^\ast}
\newcommand{\ydomain}{\mathbb{Y}}
\newcommand{\deltaO}{\delta_{\scriptscriptstyle \obs}}

\newcommand{\ddomain}{\mathbb{D}}

\newcommand{\odomain}{\mathbb{O}}

\newcommand{\image}{\mathbb{I}}

\newcommand{\realtwo}{\mathbb{R}^2}
\newcommand{\realthree}{\mathbb{R}^3}

\newcommand{\bulk}{{\mathfrak{B}}}
\newcommand{\surface}{{\mathfrak{S}}}
\newcommand{\surfaceL}{{\mathfrak{S}_L}}
\newcommand{\surfaceR}{{\mathfrak{S}_R}}
\newcommand{\material}{{\mathcal{M}}}
\newcommand{\plane}{{\mathfrak{P}}}

\newcommand{\fts}[1]{\grave{#1}}

\newcommand{\weight}{w}
\newcommand{\ftsdomain}{\fts{\domain}}

\newcommand{\ftt}[1]{\acute{#1}}
\newcommand{\ftst}[1]{\check{#1}}
\newcommand{\ftsvarphi}{\fts{\varphi}}
\newcommand{\ftspsi}{\fts{\psi}}

\newcommand{\lebesgue}{\mathcal{L}^2}
\newcommand{\lebesgone}{\mathcal{L}^1}

\newcommand{\lebesguep}{\mathcal{L}^p}
\newcommand{\seqlebesgue}{\ell^2}

\newcommand{\onetorus}{\mathbb{T}}

\newcommand{\bulksize}{S_\bulk}

\newcommand{\intdomain}{\int_{\domain}}
\newcommand{\intftsdomain}{\int_{\ftsdomain}}

\newcommand{\Epsilon}{\mathcal{E}}

\newcommand{\svec}{\vv{s}}
\newcommand{\uvec}{{\vv{u}}}

\newcommand{\br}{\vec{r}}

\newcommand{\bk}{\vv{k}}
\newcommand{\bolds}{\vec{s}}
\newcommand{\size}{S}
\newcommand{\hbulksize}{h_{g}}
\newcommand{\hreciplatt}{h_{G}}

\newcommand{\by}{\vec{y}}
\newcommand{\bz}{\vec{z}}
\newcommand{\bx}{\vec{x}}
\newcommand{\bxo}{\vec{x}_1}
\newcommand{\bxt}{\vec{x}_2}
\newcommand{\dbx}{{\boldsymbol{\mathrm{d}}\vec{x}}}

\newcommand{\absdbx}{\abs{\dbx}}

\newcommand{\mx}{\vec{x}}
\newcommand{\dmx}{\boldsymbol{\mathrm{d}}\vec{x}}

\newcommand{\Obs}{\mathrm{O}}
\newcommand{\tObs}{\tilde{\mathrm{O}}}

\newcommand{\obs}{\mathcal{O}}

\newcommand{\hObs}{\hat{\mathrm{O}}}
\newcommand{\expOp}{\overline{\Obs}}

\newcommand{\mxb}{\vec{x}_{b}}
\newcommand{\mxl}{\vec{x}_{L}}
\newcommand{\mxr}{\vec{x}_{R}}
\newcommand{\mxbl}{\vec{x}_{bL}}
\newcommand{\mxbr}{\vec{x}_{bR}}

\newcommand{\tdelta}{\tilde{\delta}}

\newcommand{\dnu}[1]{\nu^{(#1)}}

\renewcommand{\implies}{\Rightarrow}
\newcommand{\interval}{\mathfrak{I}}

\newcommand{\normal}{{\hat{\mathrm{n}}}}

\newcommand{\intmax}{{\ell}}
\newcommand{\amax}{{\mathfrak{a}}}

\newcommand{\Lequiv}{\overset{\tiny L}{\sim}}

\newcommand{\precNu}{{\varepsilon_{\mathcal{V}}}}
\newcommand{\stdNu}{{\sigma_{\mathcal{V}}}}

\newcommand{\precphi}{{\varepsilon_{\Phi}}}

\newcommand{\precRho}{{\varepsilon_{\varrho}}}

\newcommand{\prectheo}{{\varepsilon_x}}
\newcommand{\precmom}{{\varepsilon_p}}

\newcommand{\fca}{f^*_\alpha}
\newcommand{\bca}{b^*_\alpha}

\newcommand{\gca}{g^*_\alpha}
\newcommand\roth{\mathrel{\raisebox{-0.1em}{\rotatebox[origin=c]{180}{$h$}}}}
\newcommand\adbmal{\mathrel{\raisebox{-0.1em}{\rotatebox[origin=c]{180}{$\lambda$}}}}

\newcommand\recham{\hat{\roth}}

\newcommand{\N}{\mathscr{N}}

\newcommand{\uhat}{\hat{u}}

\newcommand{\tx}{\tilde{x}}

\theoremstyle{plain}
\makeatletter
\newcommand{\leqnomode}{\tagsleft@true}
\newcommand{\reqnomode}{\tagsleft@false}
\makeatother

\newcommand{\interior}[1]{%
  {\kern0pt#1}^{\mathrm{o}}%
}
\DeclareMathOperator{\Interior}{int}

\newcommand{\mtop}{resta-1993, %
kingsmith-vanderbilt-prb-1993-1, %
kingsmith-vanderbilt-prb-1993-2, %
resta-rmp-1994, %
resta-vanderbilt-2007%
}
\newcommand{\snl}[1][]{%
\ifthenelse{\equal{#1}{}}{\tau l}{\tau_{#1} l}%
}
\newcommand{\sn}[1][]{%
\ifthenelse{\equal{#1}{}}{\tau}{\tau_{#1}}%
}
\newcommand*\overbar[1]{%
   \hbox{%
     \vbox{%
       \hrule height 0.2pt 
       \kern0.5ex
       \hbox{%
         \kern-0.1em
         \ensuremath{#1}%
         \kern 0.1em
       }%
     }%
   }%
} 

\makeatletter
\let\save@mathaccent\mathaccent
\newcommand*\if@single[3]{%
  \setbox0\hbox{${\mathaccent"0362{#1}}^H$}%
  \setbox2\hbox{${\mathaccent"0362{\kern0pt#1}}^H$}%
  \ifdim\ht0=\ht2 #3\else #2\fi
  }
\newcommand*\rel@kern[1]{\kern#1\dimexpr\macc@kerna}
\newcommand*\widebar[1]{\@ifnextchar^{{\wide@bar{#1}{0}}}{\wide@bar{#1}{1}}}
\newcommand*\wide@bar[2]{\if@single{#1}{\wide@bar@{#1}{#2}{1}}{\wide@bar@{#1}{#2}{2}}}
\newcommand*\wide@bar@[3]{%
  \begingroup
  \def\mathaccent##1##2{%
    \let\mathaccent\save@mathaccent
    \if#32 \let\macc@nucleus\first@char \fi
    \setbox\z@\hbox{$\macc@style{\macc@nucleus}_{}$}%
    \setbox\tw@\hbox{$\macc@style{\macc@nucleus}{}_{}$}%
    \dimen@\wd\tw@
    \advance\dimen@-\wd\z@
    \divide\dimen@ 3
    \@tempdima\wd\tw@
    \advance\@tempdima-\scriptspace
    \divide\@tempdima 10
    \advance\dimen@-\@tempdima
    \ifdim\dimen@>\z@ \dimen@0pt\fi
    \rel@kern{0.6}\kern-\dimen@
    \if#31
      \overline{\rel@kern{-0.6}\kern\dimen@\macc@nucleus\rel@kern{0.4}\kern\dimen@}%
      \advance\dimen@0.4\dimexpr\macc@kerna
      \let\final@kern#2%
      \ifdim\dimen@<\z@ \let\final@kern1\fi
      \if\final@kern1 \kern-\dimen@\fi
    \else
      \overline{\rel@kern{-0.6}\kern\dimen@#1}%
    \fi
  }%
  \macc@depth\@ne
  \let\math@bgroup\@empty \let\math@egroup\macc@set@skewchar
  \mathsurround\z@ \frozen@everymath{\mathgroup\macc@group\relax}%
  \macc@set@skewchar\relax
  \let\mathaccentV\macc@nested@a
  \if#31
    \macc@nested@a\relax111{#1}%
  \else
    \def\gobble@till@marker##1\endmarker{}%
    \futurelet\first@char\gobble@till@marker#1\endmarker
    \ifcat\noexpand\first@char A\else
      \def\first@char{}%
    \fi
    \macc@nested@a\relax111{\first@char}%
  \fi
  \endgroup
}
\makeatother

\theoremstyle{remark}

\theoremstyle{plain}

\theoremstyle{definition}

\newtheorem*{definition*}{Definition}
\newtheorem*{notation*}{Notation}
\newtheorem*{nomenclature*}{Nomenclature}

\theoremstyle{remark}
\newtheoremstyle{assumption}
  {0.25cm}
  {0.25cm}
  {}
  {}
  {\itshape\bfseries\small}
  {\sf{:}\;\;}
  { }
  {}%
\theoremstyle{assumption}
\newtheorem*{assumption*}{Physical assumption}
\newtheorem{assumption}{Physical assumption}
\newtheorem*{conjecture*}{Conjecture/specification}

\newtheorem*{iassumption*}{Inconsequential assumption}

\newtheorem*{lemma*}{Lemma}
\theoremstyle{remark}

\newtheorem*{remark*}{Remark}
\newcommand{\closure}{{\mathfrak{cl}\,}}

\DeclareMathOperator{\dom}{dom}

\newcommand{\Lmin}{{\mathfrak{L}}}

\DeclareFloatingEnvironment[name=Homogenization Concept]{hconcept}
\mdfdefinestyle{mdfaside}{leftmargin=-0.2cm,rightmargin=-0.2cm,%
innerleftmargin=0.5cm,innerrightmargin=0.5cm, innertopmargin=0.3cm, 
innerbottommargin=0.3cm, roundcorner=1cm, backgroundcolor=black!5}
\DeclareMathAlphabet\mathbfcal{OMS}{cmsy}{b}{n}

\makeatletter
\def\l@subsection#1#2{}
\def\l@subsubsection#1#2{}
\makeatother
\makeatletter
\pretocmd{\section}{\addtocontents{toc}{\protect\addvspace{-50\p@}}}{}{}
\makeatother

\begin{document}
\author{Paul Tangney}
\date{\today}
\affiliation{Department of Physics and Department of Materials, Imperial College London}
\title{Electricity at the macroscale and its microscopic origins}
\begin{abstract}
\hspace{0.5cm} 
This work examines electrical structures, and the relationships between electrical structures
at the microscale and electrical structures at the macroscale.
By \emph{structures} I mean both
physical structures, such as spatial distributions of charge and potential, and the mathematical structures used
to specify physical structures and to relate them to one another.
I do not discuss magnetism and what little I say about energetics is incidental.

\hspace{0.5cm} 
I define the fields that specify electrical macrostructure, 
and their rates of change,
in terms of the microscopic charge density $\rho$, electric field $\me$, 
electric potential $\phi$, and their rates of change.
To deduce these definitions, I begin laying new foundations of a general theory of \emph{structure homogenization}, 
meaning a theory of how any observable macroscopic field $\Nu$ is related to
spatial averages of its microscopic counterpart $\nu$.
An integral part of structure homogenization theory is the definition of macroscopic
\emph{excess fields} in terms of microscopic fields. 
The excess field of ${\Nu:\realone^n\to\realone}$ on the boundary ${\partial\manifold}$ 
of a finite-measure subset $\manifold$ of $\realone^n$ is the field ${\bsigmaNu:\partial\manifold\to\realone}$
to which it is related by the generalized Stokes theorem,
${\int_\manifold \Nu\dd{\volumeform}=\int_{\partial\manifold}\bsigmaNu\volumeform}$;
where ${\volumeform}$ and ${\dd{\volumeform}}$ are volume forms on ${\partial\manifold}$
and ${\manifold}$, respectively, and 
${\Nu\dd{\volumeform}\equiv\dd{\left(\bsigmaNu\volumeform\right)}}$.
For example, the macroscopic volumetric charge density $\Rho$ in a material $\manifold$
is related to the areal charge density $\bsigma$ on its surface
by ${\int_{\manifold} \Rho\dd[3]{\br} = \int_{\partial\manifold}\bsigma\dd[2]{\bs}}$
and by
${\Rho\dd[3]{\br}\equiv\dd{\left(\bsigma\dd[2]{\bs}\right)}}$.
I derive an expression for ${\bsigmaNu[\nu]}$, which 
generalizes Finnis's expression~\citep{finnis} 
for excess fields at the surfaces of crystals (e.g., surface charge density ${\bsigma[\rho]}$)
to disordered microstructures.

\hspace{0.5cm} 
I use homogenization theory to define the macroscopic potential ${\bphi\equiv\bphi[\phi]}$, 
electric field ${\E\equiv\E[\me]}$, and charge density ${\Rho\equiv\Rho[\rho]}$, and I define
the macroscopic current density as ${\J\equiv \dbsigma[\drho]}$.
Using the microscopic theory, or \emph{vacuum theory}, of electromagnetism as my starting point, I deduce that the relationships
between these macroscopic fields are identical in form to the relationships between their microscopic counterparts.
Without invoking quantum mechanics, I use the definitions ${\J\equiv\dbsigma}$ and ${\bsigma\equiv\bsigma[\rho]}$ to 
derive the expressions for so-called 
\emph{polarization current} established by the \emph{Modern Theory of Polarization} (MTOP).
I prove that the bulk-average electric potential, or \emph{mean inner potential}, $\bphi$, vanishes in a macroscopically-uniform 
charge-neutral material, and I show that when a crystal lattice lacks inversion symmetry, it does not imply the existence 
of macroscopic $\E$ or $\pp$ fields in the crystal's bulk.

\hspace{0.5cm}
I point out that symmetry is scale-dependent. Therefore, if
anisotropy of the microstructure does not manifest as anisotropy of the macrostructure,
it cannot be the origin of a macroscopic vector field.
Only anisotropy of the \emph{macrostructure} can bestow directionality at the macroscale. 
The macroscopic charge density ${\Rho}$ is isotropic in the bulks of most materials, because it vanishes at every point.
This implies that, regardless of the microstructure $\rho$, a macroscopic electric field 
cannot emanate from the bulk.
I find that all relationships between \emph{observable} macroscopic fields can be expressed
mathematically without introducing the polarization ($\pp$) and electric displacement ($\D$) fields, neither
of which is observable.
Arguments for the existence of $\pp$ and $\D$, and interpretations of them, have varied since they were
introduced in the 19th century.
I argue that 
none of these arguments and interpretations are valid, and that macroscale isotropy prohibits the existence of $\pp$ and $\D$ fields.

\hspace{0.5cm}
Single-particle statistical states play prominent roles in the MTOP and in
most textbook descriptions of electrical microstructures. Therefore I discuss
several kinds of $1$-particle states in a many-particle system.
I derive exact expressions for the energy of a set of interacting indistinguishable particles in a
pure state in terms of the state's \emph{natural orbitals}, which are the 
eigenfunctions of its $1$-particle density matrix.
These expressions strengthen an already-strong
case against the traditional idea that a material's electron number density
has a substructure of \emph{localized} orbitals with almost integral 
occupancies.

\hspace{0.5cm}
I do not invoke quantum mechanics axiomatically in this work. Instead I
point out that statistical states of classical dynamical systems 
can be specified by wavefunctions and density matrices that
have the same basic properties as their quantum mechanical counterparts; 
and I include or append classical derivations of all
aspects of quantum mechanics that are relevant to this work.

\end{abstract}

\maketitle
\onecolumngrid
\clearpage
\twocolumngrid
\tableofcontents
\section{Introduction}
\label{section:introduction}
Most of the classical electromagnetic theory that is commonly described in textbooks
was established in the 19th century before electrons had been discovered 
or the existence of atoms had been confirmed~\citep{maxwell-1865,maxwell-book1,maxwell-book2,heaviside-book, lorentz, history_of_physics}.
The constitutive relations, ${\D = \varepsilon_0 \E + \pp}$ and ${\hh= \mu_0^{-1}\B - \M}$,
between the macroscopic electric and magnetic fields, ${\E}$ and ${\B}$, the
induced fields ${\pp}$ and ${\M}$, and the auxiliary fields ${\D}$ and ${\hh}$, are an important part of this theory.
To deduce them, materials were approximated as continua at the macroscale; and
the polarization ($\pp$) and magnetization ($\M$) densities were 
introduced to characterize how the state of the aether
was altered by their presence. 
When the concept of an aether was abandoned, $\pp$ and $\M$ were reinterpreted as
linear electromagnetic responses of materials. However this appears to have been
done \emph{ad hoc} and without due concern for consistency with 
the nascent theory of material microstructure.

The {\em microscopic theory} or {\em vacuum theory} of electromagnetism rightly underpins 
our microscopic theory of material structure, composition, and energetics. 
The purpose of {\em macroscopic} electromagnetism, which reduces to microscopic
electromagnetism when ${\pp}$ and ${\M}$ vanish, is to provide a unified 
description of materials and electromagnetic fields at the macroscale.
Therefore it should be underpinned by
our mutually-consistent 
microscopic theories of materials physics and vacuum electromagnetism;
and we should understand the microscopic origins of ${\pp}$ and ${\M}$ clearly.
However development of the macroscopic theory was completed 
before many of the discoveries on which we base our microscopic understanding
of materials were made, and it quickly became an established part of
physics doctrine.
Therefore it was not built on firm microscopic foundations and, unfortunately,
it has never been reconciled fully and satisfactorily with microscopic physics.

Inconsistencies between the microscopic and macroscopic theories were not apparent to most scientists until after 
crystallography had come of age and it had become possible to compute materials' 
{\em microstructures}, by which I mean the statistical 
distributions, on the nanoscale, of their constituent charges and magnetic moments.
It became obvious that we lacked precise and viable definitions of ${\pp}$ and ${\M}$
when attempts were made to define them in terms of
microstructures~\citep{martin-prb-1974,littlewood_1979,littlewood-1980,vogl_1978,tagantsev_1991,resta-1992,aizu_1962,landauer_1981,landauer_1960,larmor_1921,woo-prb-1971}.
Neither $\pp$ nor $\M$ is directly measureable, but definitions of observables 
attributed by classical electromagnetic theory to changes in their values
also proved elusive.
For example, it was not until the 1990s that researchers discovered 
how to calculate the so-called {\em polarization
current}, $\Jconv$, that flows through an insulating inversion-asymmetric crystal when it is uniformly
perturbed by a stimulus, such as a strain or a change in temperature.
It is $\Jconv$, and not ${\pp}$ or ${\D}$, that is measured directly 
in experiments that produce ${\pp-\E}$ or ${\D-\E}$ hysteresis loops.

An exact calculable expression for $\Jconv$ was eventually 
provided by a theory that became known as 
the \emph{Modern Theory of Polarization} (MTOP)~\citep{resta-1993,kingsmith-vanderbilt-prb-1993-1,kingsmith-vanderbilt-prb-1993-2,resta-rmp-1994,resta-vanderbilt-2007,Resta_2010}. 
According to the MTOP, a crystal's bulk polarization $\pp$ is a multivalued quantity that cannot be calculated
from the crystal's microscopic charge density $\rho$, because it is a property of the \emph{phase} of the
crystal's wavefunction, which has `nothing to do with' the charge density~\citep{Resta2018,Resta_2010,resta-1993}.
Therefore the MTOP deviates substantially from both the revised (20th century) definition of $\pp$ as a dipole
moment density, and the physical reasoning with which the use of $\pp$ as a measure
of dielectric response to a uniform field was justified.

Another elusive and hotly-debated definition was that of
the areal density of charge at a surface or interface, ${\bsigma[\rho]}$.
This problem was solved for crystals by Finnis in 1998~\citep{finnis}, and 
there appears to be agreement now that his definition is correct
~\citep{resta-vanderbilt-2007,goniakowski-rpp-2008,stengel-vanderbilt-prb-2009,stengel-prb-2011,
bristowe-JPCM-2011,noguera-chemrev-2013,goniakowski-2014,bristowe-JPCM-2014,vanderbilt_2018}. 
Unfortunately, the literature is far from clear on this point because multiple equivalent definitions
of ${\bsigma[\rho]}$ have been proposed, and some works continue to make the unnecessary distinction
between \emph{free charge} and \emph{bound charge}, and to express the latter's
contribution to $\bsigma$ as ${\bsigmabound=\pp\cdot\normal}$, where ${\normal}$ is the unit surface normal 
and $\pp$ is the polarization in the crystal's bulk.
The definition is complicated further by the multivaluedness of the MTOP definition of ${\pp}$.

A third illustration of the tension that exists
between the 20th century theory of material structure and 19th century electromagnetism, 
is the question of how to define and calculate the bulk macroscopic electric potential, $\bphi$, from 
the microscopic charge density, $\rho$.
This quantity, which is often called the {\em mean inner potential} (MIP), 
plays an important role in several areas of research, including theoretical electrochemistry 
and electron microscopy~\citep{miyake-1940, mip_sanchez_1985,mip_pratt_1987, mip_pratt_1988, 
mip_pratt_1989, mip_pratt_1992, gajdardziska-1993, spence-1993, mip_sokhan_1997, spence-1999, mip_leung_2010, 
mip_mundy_2011, mip_cendagorta_2015, mip_lars, mip_marzari, mip_water_2020, mip_madsen_2021,
mip_kathmann_2021,peng_1999,spence_1994,Gajdardziska-Josifovska1999,Ibers_1958,rez_1994,tildesley_1997,stillinger_1967}. 
An exact general definition of it has not previously been found, but 
several approximations to it have been proposed and are in use~\citep{mip_mundy_2011,tildesley_1997,saunders_1992,mip_pratt_1992}.
Bethe derived one such expression by approximating the microstructure as a
superposition of spherically-symmetric atomic charge densities~\citep{bethe-1928}.

\subsection{Motivations and objectives}
\label{section:motivation}
Considered individually, the three examples cited above suggest, at the very least, that
the connection between microscopic electromagnetism and macroscopic electromagnetism is subtle. 
However the situation appears more serious when they are considered collectively,  
because the fields ${\Jconv\equiv\dbsigma}$, ${\bsigma}$, and ${\bphi}$,
are all measureable elements of electricity at the macroscale. 
Therefore these are all examples of attempts to bridge, 
or to partially fill, the same hole in existing physical theory, namely:
We do not understand the relationship
between electricity at the macroscale and electricity at the microscale well
enough to express the fields that specify a material's {\em electrical macrostate}
in terms of the fields that specify its {\em electrical microstate}.

An {\em electrical microstate}, ${(\micro,\dmicro)}$, is an 
{\em electrical microstructure} $\micro$ and its time derivative, ${\dmicro\equiv\pdv*{\micro}{t}}$, at the same instant. 
An {\em electrical microstructure} is the most complete and detailed information pertaining to the instantaneous spatial distribution of
charge and electric potential that could, in principle, exist. It could be specified by the wavefunction,
density matrix, or position probability density function (pdf) of the set of all particles, 
but for many purposes the information required is contained in 
the microscopic electric potential $\phi$ and the microscopic charge 
density ${\rho\equiv-\laplacian\phi}$, in which case we say that the
electrical microstate is ${(\phi,\dphi)}$ or ${(\rho,\drho)}$.
An {\em electrical macrostate}, ${(\macro,\dmacro)}$,
is a specification of the spatial distributions of charge and electric
potential at the macroscale, $\macro$, and their time derivatives, $\dmacro$. 

In each of the three examples discussed above,
a different line of reasoning was 
followed to derive an expression for ${\Jconv=\Jconv[\dmicro]}$, ${\bsigma=\bsigma[\micro]}$, or 
${\bphi=\bphi[\micro]}$. However, none of these lines of reasoning were pursued
far enough to elucidate the relationship between macrostructure and microstructure fully, 
and with enough generality that what was learned could be applied, not only throughout electromagnetic
theory, but far beyond it: in elasticity theory, meteorology, astrophysics, and countless other
areas of research.
For example, the MTOP did not provide an expression
for ${\bphi[\micro]}$, and Finnis did not derive an expression for
${\Jconv[\dmicro]}$ from his expression for ${\bsigma[\micro]}$.

\subsubsection*{Objective 1}
My first objective is to reconcile the fundamental elements of our macroscale theory of 
electricity in materials with our mutually-compatible theories of electromagnetism
and material structure at the microscale. 
My focus is on the relationships between macroscopic fields and on how macroscopic fields 
can be defined in terms of microscopic fields.

I use the term macroscopic field to mean
\emph{uniform} field. A uniform field can be regarded as a field whose wavelength is orders of magnitude
larger than the material to which it is applied or from which it emanates. 
I say little about fields whose wavelengths are shorter than, or comparable to, 
a material's linear dimensions.  The physics of such fields is 
qualitatively different, in some respects, to the physics of macroscopic fields. 

\subsubsection*{Objective 2}
My second objective is to lay some groundwork for 
a comprehensive and rigorous theory of the relationship between physics at the 
microscale and physics at the macroscale. 

If you look around you, you will see surfaces, edges, and corners everywhere.
Everything you see is a feature of the macrostructure, meaning
that it is a blurred image of the microstructure at
a surface, edge, or corner. You do not see the microstructure in its full
horrendous complexity, and you do not notice that it changes from
one femtosecond to the next.
You see a relatively simple and relatively stable 
homogenized version of the microstructure. 

There are many sources of imprecision, such as the diffraction limit, 
and it would be impossible for you to be aware of the full microstructure because, for example,
a cubic molar sample of an element has ${\sim 10^{16}}$ atoms at each of its six faces, but the human
brain only has ${\sim 10^{11}}$ neurons. 
Therefore homogenization of microstructure to form macrostructure is
intrinsic to the act of observation. 

Nevertheless, given a microstructure $\nu$ and access to an
arbitrarily-powerful computer, it is not known how to calculate the macrostructure,
or even what mathematical form it would take.
The inconsistencies between Maxwell's macroscopic and microscopic
theories of electromagnetism are only one of many important 
consequences of this gap in our understanding.

Therefore I address the following question, which is of general
importance to mathematical physics: \\
\\
{\em How can a macrostructure
be expressed mathematically in terms of the microstructure underlying it?}

This question leads quickly to a more fundamental question: \\
\\
{\em If the microstructure
is a scalar field ${\nu:\realone^3\to \realone}$, what is the mathematical
form of the macrostructure?}

It turns out (see Sec.~\ref{section:homogenization}) 
that qualitative
differences exist between a macrostructure and a {\em base microstructure}, 
where I use the term {\em base microstructure} to mean a microstructure that is not itself 
the macrostructure arising from a structure on an even smaller length scale.

\subsubsection*{Objective 3}
An ancillary purpose of this work is to emphasize how little
of the physics of electricity in materials requires physical
assumptions that are incompatible with classical physics.

Most textbooks on solid state physics or electronic
structure theory do not clearly demarcate the features of 
mathematical representations of statistical microstates
that are peculiar to quantum mechanics
for fundamental physical reasons, 
from features that are consistent with classical statistical 
microstates.
For example, when we see a statistical state expressed as
${\Psi\equiv\sqrt{\pdf}e^{i\theta}}$, where ${\pdf=\pdf(\rvec_1,\rvec_2,\cdots)}$ is a position
pdf,
we often assume
that quantum mechanics is being `used'.
However, it is perfectly valid to express the
statistical state of a system of classical particles in this form, 
and it can be useful to do so.
Having done so, the classical many-particle state $\Psi$
can be expanded in a basis of single particle states, just as
in quantum mechanics.

We largely base our physical intuitions on what we observe
at the human scale. If the blurred lines between classical and 
quantum physics were made more clear, we would have a better
understanding of when we could apply our classical intuitions
to systems of quantum mechanical particles, and when our intuitions
were likely to fail us.

I begin to address this issue in the present work for two
reasons. The first is that fulfilling my first objective, 
and relating my findings to the MTOP, 
requires me to survey many parts of electronic
structure theory and solid state physics. Therefore I have the 
opportunity to point out that much of the mathematical infrastructure
that we usually associate with quantum mechanics is perfectly consistent with classical physics.

The second reason is that there are claims in the literature
on the MTOP that some of the observable quantities that I discuss
in this work have quantum mechanical origins and do
not have analogues within classical physics~\citep{resta-rmp-1994,resta-1993}.  
It is important to examine these claims carefully.

Single particle states play a prominent role in the MTOP.
Therefore I emphasize that there is nothing specific to quantum mechanics
about Bloch functions~\citep{bloch-1929} and Wannier functions~\citep{wannier}. If the bulk
of a crystal is represented in a torus, which is equivalent to 
using  \emph{Born-von K\'arm\'an boundary conditions}~\citep{born_von-karman}, and if
$\Psi$ is a stationary statistical state resulting from a classical
process that preserves the crystal's periodicity, it 
can be expanded in a basis of Bloch functions.
Each set of Bloch functions can be transformed into an infinite
number of sets of Wannier functions, which must
include a maximally localized set~\citep{ferreira_parada,marzari_mlwf}.

The MTOP approach to calculating $\Jconv$ gives exactly the
right result when the charge density can be expressed in the
form 
\begin{align*}
\rho(\rvec;\zeta)=\sum_i q_i \abs{\phi_i(\rvec;\zeta)}^2, 
\end{align*}
where  $\zeta$ is the stimulus whose rate of change ${\dot{\zeta}\equiv\dv*{\zeta}{t}}$ gives
rise to $\Jconv$; 
each ${\phi_i}$ is a function that varies smoothly with $\zeta$
while preserving its normalization; and each ${q_i}$ is independent of $\zeta$.

The MTOP is routinely applied to electrons in insulators by representing their
electron densities as sets of smoothly-evolving
Bloch or Wannier states of fixed occupancies; and it is trivial to apply it to 
classical nuclei or ions if they can be treated as point charges whose positions
as functions of $\zeta$ are known.
However it is probably not known how to adapt and apply the MTOP to the ${\zeta}$-dependent
statistical state of an arbitrary classical or quantum mechanical physical system.
I do not shed light on the answer to this representability problem, 
but I attempt to clarify the question and to emphasize its importance.

\subsection{Theoretical approach and outline of this work}
\label{section:intro_approach}
Section~\ref{section:notation} explains some of the notational 
conventions used in this work, and in Sec.~\ref{section:physical_assumptions} I explain
some of the physical assumptions about materials' microstructures 
that underpin many aspects of this work. The main body of this work begins in Sec.~\ref{section:aether}.

I define the {\em homogenization transformation} that turns an electrical microstructure
into an electrical macrostructure as a spatial averaging operation on a mesoscopic domain.
This obvious approach, which appears physically reasonable, 
has been attempted many times before by many authors~\citep{kaufman_1961,rosenfeld_1965,degroot_1965, schram_1960, russakoff-ajp-1970, robinson_1971, degroot_1964,mazur_1953,kirkwood_1936,mazur_1957,frias_2012,raab_2012,raab_2005,raab_2006,roche_2000}; and some of these attempts are presented in well known
textbooks~\citep{jackson-book, ashcroft_mermin_book}.
However none of these approaches have been adopted widely by
the research community as foundations for the development of rigorous theory, 
because they do not lead to Maxwell's macroscopic theory of electricity.

In Sec.~\ref{section:aether} I explain why we should not
be deterred by this: I outline the reasoning that led Maxwell 
to his macroscopic theory in order to demonstrate that
this reasoning has been invalidated by what has
since been learned about spacetime and the microstructures of materials.
Therefore we should not require unobservable elements of Maxwell's 
macroscopic theory, such as $\pp$, to be elements of a macroscopic
theory that is derived from, and consistent with, his vacuum theory of electromagnetism

In Sec.~\ref{section:definingP}, using the macroscopic polarization $\pp$ as an example, 
I briefly explain some of the ways in which definitions of macroscopic fields 
have failed in the past.

In Sec.~\ref{section:symmetry} I argue that many of my conclusions, 
and many elementary aspects of electricity at the macroscale, are demands of symmetry or asymmetry. 

For example, any stimulus changes the microscopic charge density in the bulk of a crystal,
to some degree. While $\rho$ is changing, microscopic polarization current ($\jconv$) flows, because ${\pdv*{\rho}{t}=-\div\jconv}$.
Whether or not a net, or \emph{macroscopic}, polarization current ($\Jconv$) flows
depends on the symmetry of the composite crystal+stimulus system:
The component of $\Jconv$ in direction $\uhat$
vanishes if there is a glide plane normal to $\uhat$, because then the
sum, 
\begin{align*}
\jconv(\rvec)\cdot\uhat+\jconv(\glide\rvec)\cdot\uhat, 
\end{align*}
of the contributions to $\Jconv\cdot\uhat$ from an arbitrary point $\rvec$ and
its image under the glide symmetry ${\glide\rvec}$ vanishes. 
On the other hand,  if symmetry does not demand that ${\Jconv\cdot\uhat}$ vanishes, it must be finite: 
There is a vanishing probability that, by chance, 
the positive contributions to ${\Jconv\cdot \uhat}$ 
cancel the negative contributions \emph{exactly} (i.e., to infinite precision).
Therefore either anisotropy demands that $\Jconv$ is finite or isotropy demands that it vanishes.

As another example, suppose that 
the average over the Cartesian ${y-z}$ plane of a 
microscopic charge density, ${\rho_k(x,y,z)}$, is 
\begin{align*}
\brho_k(x)=A_k\sin(kx+\theta_k), 
\end{align*}
for some real constants $A_k$, $\theta_k$, 
and ${k\gg \wavekmin\equiv 2\pi/\prectheo}$, where $\prectheo$ denotes the smallest macroscopic distance.
Then, since ${\brho_k(x)}$ is inversion symmetric, $\rho_k$
does not give rise to a macroscopic $\E$ field whose $x$-component, ${\E_x}$, is finite.
It follows by linearity that $\E_x$ vanishes for
any microscopic charge density whose average over the $y-z$ plane is of the form
\begin{align*}
\brho(x)= \sum_{k\gg\wavekmin} A_k\sin\left(kx+\theta_k\right).
\end{align*}
Since the planar average of the microscopic charge density of every perfect crystal 
can be expressed in this form, a macroscopic $\E$ field does not exist in the 
bulk of any isolated crystal whose surfaces are charge neutral - \emph{even} if 
its microstructure lacks inversion symmetry.

In Sec.~\ref{section:mtop} I discuss the {\em Modern Theory of Polarization}, and 
I derive the MTOP expression for ${\Jconv}$ without invoking quantum mechanics.

In Sec.~\ref{section:homogenization} I explain in more detail what I mean by
the prefixes {\em micro-} and {\em macro-}.
I outline some of the qualitative differences between a macrostructure
and a base microstructure and I explain the mathematical and physical origins of those differences.

In Sec.~\ref{section:excess_fields} I discuss the macroscopic \emph{excess
fields} that exist at surfaces,  interfaces, edges, and line and point defects.
Excess fields are the manifestations at the macroscale of \emph{abrupt} changes of the microstructure, meaning
changes that occur across microscopic distances.
For example, the difference in microstructure
between a material and vacuum manifests as an areal charge density $\bsigma$ on the material's surface.
I derive expressions for macroscopic excess fields in terms of microscopic volumetric fields, 
which generalize Finnis's expression for surface excesses to non-periodic microstructures.

In Sec.~\ref{section:average_charge} 
I use spatial averaging of the microscopic charge density $\rho$ to
calculate its  macroscopic counterpart $\Rho$, and in
Sec.~\ref{section:surface_charge} I define the surface charge
density $\bsigma$ as the integral of $\Rho$ along a path
that crosses the surface. This leads to Finnis's expression
for the surface charge of a crystal, ${\bsigma[\rho]}$, and to my generalization
of this expression to noncrystalline materials.

In Sec.~\ref{section:current} I derive the MTOP expression for $\Jconv$ again, but
this time I derive it by defining it as ${\Jconv\equiv\dbsigma=\dv*{\bsigma[\rho]}{t}}$, 
where ${\bsigma[\rho]}$ is Finnis's formula.

In Sec.~\ref{section:single_particle_states} I point out that 
there does not exist a theoretical justification for
interpreting the sets of single electron states that appear
in the MTOP definitions of polarization current as chemically meaningful
substructures of the electron density. 

In Sec.~\ref{section:average_potential} I prove that
the macroscopic potential, or mean inner potential, $\bphi$ vanishes in the bulk of
any isolated material whose surface is locally charge neutral.
Then I point out a flaw in the reasoning used by H. A. Lorentz to
deduce that a macroscopic electric field exists in 
the bulk of a crystal whose microstructure is anisotropic.
In Sec.~\ref{section:paradox} I discuss flaws in Bethe's derivation
of an approximate expression for $\bphi$, and I show that $\bphi$ vanishes
when these flaws are avoided.

This work comprises three interwoven strands, whose individual objectives
are the three objectives outlined above.
It concludes in Sec.~\ref{section:conclusions} with a summary of each strand.

\subsection*{Selectively reading this work}
I have tried to make this work as modular as possible, while preserving
the logic of the narrative as a whole and minimizing repetition.
My hope is that many of the sections, subsections, and appendices
are reasonably self-contained. 

Those interested only in the homogenization transformation that turns 
microstructure into macrostructure, should read Secs.~\ref{section:homogenization} and~\ref{section:excess_fields}, 
and Appendix~\ref{section:invariance_proofs}.
Those interested in everything except the homogenization transformation, and who are willing to 
trust the formulae derived in Sec.~\ref{section:excess_fields} and 
presented in Appendix~\ref{section:excess_formulae}, 
can safely skip those parts.

Those interested only in the \emph{mean inner potential}, $\bphi$, should read Sec.~\ref{section:symmetry} (particularly Sec.~\ref{section:potential_symmetry}),
Sec.~\ref{section:average_potential} and Sec.~\ref{section:paradox}.

Those interested only in polarization current, $\Jconv$, or the \emph{Modern Theory of Polarization} should read Secs.~\ref{section:symmetry},~\ref{section:mtop},
~\ref{section:surface_charge}, and
~\ref{section:current}.

Those interested only in the macroscopic electric field, $\E$, should read Sec.~\ref{section:symmetry} and Sec.~\ref{section:average_potential}.

Those interested only in surface charge, $\bsigma$, should read Secs.~\ref{section:what_is_macrostructure},~\ref{section:excess_fields} and~\ref{section:surface_charge}.

Those interested only in single particle states should 
read Secs.~\ref{section:bulk_subsystem},~\ref{section:representability}, 
and ~\ref{section:single_particle_states}, and Appendices~\ref{section:appendix_torus},~\ref{section:appendix_wannier}, and~\ref{section:appendix_natural}.

Those interested only in the relationship between quantum mechanics and classical statistical mechanics should
read Secs.~\ref{section:microstructures_of_materials} and ~\ref{section:fermi_dirac_derivation}, and Appendices
~\ref{section:appendix_states_as_vectors}, ~\ref{section:expectation_values}, ~\ref{section:appendix_measurement} and~\ref{section:smooth_evolution}.
 
I would be grateful for critical feedback on any part or aspect of this work (\texttt{p.tangney@imperial.ac.uk}).

\section{Notation}
\label{section:notation}
This section outlines some of the standard and non-standard notational conventions that are used
in this work. More notation is introduced in later sections, as and when it is needed.
Some of the appendices use notation that does not conform to the conventions introduced
in this section.

\subsection{Microscopic quantities vs macroscopic quantities}
Boldface type distinguishes macroscopic quantities from
microscopic quantities throughout this work. For example, the
macroscopic analogues of the
microscopic charge density $\rho$, the microscopic electric potential $\phi$, and the microscopic electric field $\me$, 
are denoted by ${\Rho}$, $\bphi$, and $\E$, respectively;
and $\br$ and ${\rvec}$ denote points, positions, or displacements at the macroscale
and the microscale, respectively.

In discussions of general features of the relationship between microscopic
and macroscopic fields, $\nu$ will denote an arbitrary microscopic field
and $\Nu$ will denote its macroscopic counterpart.

\subsection{Mathematical sets and spaces}
\subsubsection{Sets of numbers}
The symbols $\integer$, $\realone$, and $\complex$ denote the
sets of integers, real numbers, and complex numbers, respectively.

The sets of positive, negative, non-positive, and non-negative integers
or real numbers are defined and denoted as follows:
\begin{align*}
\integerpos
&\equiv\{x\in\integer:x>0\},
&
\integernonneg
&\equiv\{x\in\integer:x\geq0\},
\\
\integerneg &\equiv\{x\in\integer:x<0\},
&
\integernonpos&\equiv\{x\in\integer:x\leq0\},
\\
\realpos
&\equiv\{x\in\realone:x>0\},
&
\realnonneg
&\equiv\{x\in\realone:x\geq0\},
\\
\realneg &\equiv\{x\in\realone:x<0\},
&
\realnonpos&\equiv\{x\in\realone:x\leq0\}.
\end{align*}

\subsubsection{Lattices}
A one dimensional lattice with lattice spacing $a$ will be denoted by ${a\integer}$ 
if it includes the origin. If the origin is not one of the lattice points, it
will be denoted by ${x_0+a\integer}$, for some ${x_0\in\realone}$.
That is, 
\begin{align*}
x_0+a\integer\equiv \left\{x_0+ma: m\in\integer\right\}.
\end{align*}
This notation generalizes to lattices in higher dimensions. For example, 
a two dimensional lattice with lattice spacings ${a_1}$ and ${a_2}$, and with
one of the lattice points at a displacement ${\svec}$ from the origin, can 
be denoted by
\begin{align*}
\svec+\left(a_1\integer\right)\times\left(a_2\integer\right);
\end{align*}
and an $M$-dimensional lattice with lattice spacing $a$ in all directions
can be denoted by ${\left(a\integer\right)^M}$.

\subsubsection{Indexed sets}
${\{A_\alpha\}}$ denotes the set containing ${A_\alpha}$ for
every possible index $\alpha$; ${\{B_{\alpha\beta}\}}$ denotes the set containing ${B_{\alpha\beta}}$ for every possible
pair of indices, ${\alpha\beta}$; and ${\{B_{\alpha\beta}\}_\beta}$ denotes the set containing ${B_{\alpha\beta}}$
for every possible index $\beta$ and a fixed value of index $\alpha$.

\subsubsection{Intervals}
If $u$ is any real-valued quantity with a continuous range of possible values (e.g., an $x$ coordinate or an average of field $\nu$),
I use ${\interval(u,\Delta)}$, ${\interval[u,\Delta)}$ and ${\interval(u,\Delta]}$, and ${\interval[u,\Delta]}$ to denote interval
subsets of this range,
of width $\Delta$ and centered at $u$, which are open,
half-open, and closed, respectively.
I use the more conventional
notation ${(u_1,u_2)}$, ${[u_1,u_2)}$ and ${(u_1,u_2]}$, and ${[u_1,u_2]}$,
to specify intervals by their end points. For example,
${\interval(u,\Delta u]=(u-\Delta u/2,u+\Delta u/2]}$ is an interval
that is open at its lower boundary and closed at its upper boundary.

\subsubsection{Regions in the interiors of materials}
A simply-connected region in the interior of a material will
be denoted by ${V}$ and its measure will be denoted by ${\abs{V}}$.
In a one dimensional material ${\abs{V}}$ is a length; and in a three dimensional
material it is a volume.

A unit cell of a crystal will be denoted by $\unitcell$, and its length (1-d)
or volume (3-d) will be denoted by $\volume$.

\subsubsection{Tori}
I use the term \emph{torus} to mean a topological space.
A \emph{1-torus} of `circumference' ${C\in\realpos}$
is denoted by ${\onetorus(C)}$ or ${\onetorus}$ and defined as
\begin{align*}
\onetorus=\onetorus(C)\equiv \realone/(C\integer).
\end{align*}
Therefore if an arbitrary point in ${\onetorus(C)}$ is at position
${x\in\realone}$, it is also at 
all of the positions in the infinite set,
${\{x+qC: q\in\integer\}}$. In other words, each point in 
${\onetorus(C)}$ is its own image under a displacement by
an integer multiple of $C$.
${\onetorus(C)}$ is denoted by $\onetorus$ 
when the value of $C$ is clear from the context.

For any ${m\in\integerpos}$ and any set 
of $m$ circumferences, ${\{C_i\in\realpos:1\leq i \leq m\}}$, 
\begin{align*}
\onetorus^m(C_1\cdots C_m)\equiv \onetorus(C_1)\times\onetorus(C_2)\times\cdots\times\onetorus(C_m). 
\end{align*}
is an \emph{m-torus}.
Therefore, for any ${i\in\{1\cdots m\}}$ and
any ${q\in\integer}$, 
the coordinates
${(x_1\cdots,x_i+qC_i,\cdots x_m)}$ and
${(x_1\cdots x_m)}$ identify the same point in ${\onetorus^m(C_1\cdots C_m)}$.

If $f$ is a function whose domain is an interval ${[a,a+C)\subset\realone}$, such as if it is the restriction
to ${[a,a+C)}$ of a function whose domain is $\realone$,  it can be `rolled up'
into a function ${g}$ with domain ${\onetorus}$ by identifying a point ${\tilde{a}\in\onetorus}$ as
the counterpart in ${\onetorus}$ of ${a\in[a,a+C)}$; and defining 
\begin{align*}
g(\tilde{a}+(x-a))\equiv f(x),\forall x\in\closedopen{a,a+C}. 
\end{align*}
Conversely, any function ${f}$ whose domain is $\onetorus$  can be `unrolled' into $\realone$
by defining a $C$-periodic function ${g}$, with domain $\realone$,
as: ${g(x)\equiv f(x), \;\forall x\in\realone}$.

\subsection{Function spaces}
\label{section:function_spaces}
\subsubsection{$p-$norms}
If ${1\leq p \leq\infty}$, the \emph{p-norm} of a function,
\begin{align*}
f:\domain\to\image\in\{\realone,\complex\},
\end{align*}
from any \emph{measure space} $\domain$ to either ${\realone}$ or ${\complex}$, is
\begin{align*}
\norm{f}_p\equiv \left(\int_\domain \abs{f}^p \dd{\mu}\right)^{\frac{1}{p}},
\end{align*}
where $\mu$ is the \emph{measure} of $\domain$. The integral ${\int_\domain\cdots\dd{\mu}}$
is a \emph{Lebesgue integral}, but the difference between a Riemannn integral 
and a Lebesgue integral will rarely be relevant and will not be discussed.

\subsubsection{Lebesgue spaces}
A \emph{Lebesgue space} or \emph{$\lebesguep$-space} is a space of functions whose \emph{p-norm} is finite.
Therefore, for ${\image\in\{\realone,\complex\}}$, the set
\begin{align*}
\lebesguep(\domain,\image)\equiv\left\{f:\domain\to\image \;s.t.\; \norm{f}_p<\infty\right\}.
\end{align*}
is a ${\lebesguep}$-space.

\subsubsection{Hilbert-Lebesgue spaces}
A Hilbert-Lebesgue space is a $\lebesgue$-space. 
For example ${\lebesgue(\realone^m)}$ is the set of all functions ${f:\realone^m\to\complex}$ such that
\begin{align*}
\norm{f}_2\equiv \left(\int_{\realone^m}\abs{f(\rvec)}^2 \dd[m]{r}\right)^{\frac{1}{2}}<\infty.
\end{align*}
The norms of all of the Lebesgue spaces in this work are $2$-norms, ${\norm{\,\cdot\,}_2}$.
They will be denoted as ${\lebesgue(\domain)}$ or ${\lebesgue(\domain,\realone)}$, where $\domain$ is
usually ${\realone^m}$, ${\complex^m}$, or ${\onetorus^m}$ for some value
of ${m\in\integerpos}$. 
Omission of the second argument of ${\lebesgue(\domain,\image)}$
implies that ${\image=\complex}$.

It follows from ${\realone\subset\complex}$ that ${\lebesgue(\domain,\realone)\subset\lebesgue(\domain)}$.
Therefore a space may be denoted as ${\lebesgue(\domain)}$ despite the context suggesting that
its elements are real-valued or can be chosen to be real-valued. 
A Hilbert-Lebesgue space will only be denoted as ${\lebesgue(\domain,\realone)}$
when it is necessary to exclude complex valued-elements (e.g., because they
are unphysical).

\subsubsection{Hilbert-Lebesgue spaces of functions with countable domains}
\label{section:sequence_spaces}
The analogue of a $\lebesgue$-space for a set of functions whose domain
${\indexset}$ is a countable set, such as a lattice, is denoted by
${\seqlebesgue(\indexset,\image)}$ if ${\image\in\{\realone,\complex\}}$, or by
${\seqlebesgue(\indexset)}$ if ${\image=\complex}$.
It is defined as
\begin{align*}
\ell^2(\indexset,\image)\equiv\left\{ f:\indexset\to\image\;s.t.\; \norm{f}_2^2\equiv\sum_{i\in\indexset} \abs{f(i)}^2 < \infty\right\}.
\end{align*}

Strictly speaking, all of the $\lebesgue$-spaces
encountered in this work are really ${\seqlebesgue}$-spaces, because they are constructed as described in Appendix~\ref{section:states_as_vectors}.
Therefore the only difference between the notation
${\lebesgue(\domain)}$ and the notation ${\seqlebesgue(\indexset)}$ is that ${f\in\lebesgue(\domain)}$ emphasizes
that the domain of $f$ is continuous \emph{for all practical purposes}, whereas ${f\in\seqlebesgue(\indexset)}$ is used to 
emphasize that its domain is countable.

\subsubsection{Inner products}
\label{section:inner_product}
The inner product of space ${\lebesgue(\domain)}$ is
\begin{align*}
\braketD{f}{g}\equiv \intdomain f^* g \dd{\mu}.
\end{align*}
For example, the inner products of ${\lebesgue(\realone)}$, 
${\lebesgue(\onetorus)}$, ${\lebesgue(\unitcell)}$ are, respectively, 
\begin{align*}
\braketR{f}{g}&\equiv\int_\realone f^*(x)g(x)\dd{x},
\\
\braketT{f}{g}&\equiv\int_\onetorus f^*(x)g(x)\dd{x},
\\
\braketcell{f}{g}&\equiv\int_\unitcell f^*(x)g(x)\dd{x}.
\end{align*}
The inner products of all Hilbert-Lebesgue spaces may also be denoted
with the ambiguous notation ${\braket{f}{g}}$.

\subsubsection{Aside: The imaginary part of an `inner product'}
\label{section:imaginary_inner_product}
If the reason to use ${\lebesgue(\domain,\complex)}$ rather
than ${\lebesgue(\domain,\realone)}$ is to use
$\complex$ as a proxy for ${\realone\times\realone}$ (e.g., with
the real and imaginary parts of numbers representing Cartesian 
$x$- and $y$-components), then ${\braket{f}{g}}$ is not an inner 
product~\citep{gallier_2011,heil_2018,garling_2011,scharlau_2012,renteln_2013}.
Its real part, ${\Re\{\braket{f}{g}\}}$, does meet the definition of an inner product, but
its imaginary part does not have an obvious interpretation that is valid in all contexts.

If ${a,b\in\complex}$ represent vectors in ${\realone\times\realone}$, then
${\Re\{a^*b\}}$ is the vectors' inner product and, if $i$ is regarded
as a proxy for the unit pseudoscalar of the geometric algebra of 
${\realone\times\realone}$,  then
${i\Im\{a^*b\}}$ can be interpreted as their exterior product, which is a bivector
\citep{lounesto_2001,vaz_darocha_2016,doran_2003,garling_2011}.

\section{Assumptions about materials' structures}
\label{section:physical_assumptions}
This section outlines some of the physical assumptions about materials' structures that are made
in this work. More physical assumptions are introduced in later sections, as and when they are
needed. Most of the physical assumptions that are introduced in later sections are neither
strong nor significant.   
However 
Sec.~\ref{section:homogenization} introduces
physical assumptions that significantly narrow the range of materials to which parts of this work
apply. For example, they are not all valid  in materials whose textures are non-uniform 
on all length scales, such as wood.

\subsection{Length scales}
\label{section:length_scales}
The {\em microscale}, $a$, is the smallest length
scale of relevance to the physics of materials, and
the {\em macroscale}, $L$, is a much larger length scale, on
which materials appear continuous rather than particulate.

Whenever I use the prefixes {\em micro} and {\em macro} it is implicit that
there also exists a {\em mesoscale} $l$, where ${a\ll l \ll L}$.
The mesoscale is an intermediate length scale,
which is orders of magnitude larger than a bond length,
but small enough that all nonlinear contributions to the spatial variations of all macroscopic fields are negligible
on scale $l$.
The assumption that there exists a mesoscale is useful, and possibly necessary,
for understanding the relationship between microstructures and macrostructures.

The statements ${\Delta_a\sim a}$, ${\Delta_l\sim l}$, and ${\bDelta_L\sim L}$,
 mean that ${\Delta_a}$, ${\Delta_l}$, and ${\bDelta_L}$ are
distances or displacements which are microscopic, mesoscopic,
and macroscopic, respectively. I will explain precisely what I mean
by the terms {\em microscopic}, {\em mesoscopic}, and {\em macroscopic} 
in Sec.~\ref{section:homogenization}.

In Sec.~\ref{section:homogenization} I will define 
the {\em macroscale infinitesimal} ${\abs{\dbx}}$, which is a lower bound
on lengths and distances that are measurable at the macroscale.
At the microscale I denote ${\abs{\dbx}}$ by ${\prectheo}$

I denote the smallest measurable distance 
at the microscale by ${\abs{\dd{x}}\equiv\precmicro}$. Its value 
can be arbitrarily small, as long as it is finite and 
an inviolable lower bound.

\subsection{A {\em quasi-}one dimensional material}
\begin{figure}[h]
\includegraphics[width=8.5cm]{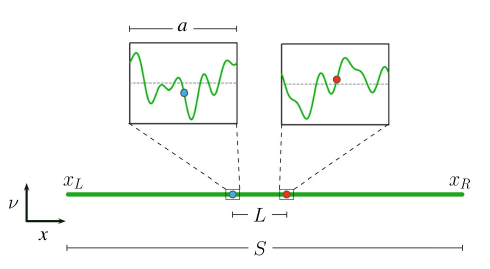}
\caption{Schematic of a {\em quasi} one dimensional material which
is macroscopically uniform but microscopically non-uniform. See Fig.~\ref{fig:crystal_dipole}
for a schematic of a three dimensional polarized charge-neutral material.
}
\label{fig:material}
\end{figure}
It may be useful to consider the macroscopically-uniform materials
depicted schematically in Fig.~\ref{fig:material} and Fig.~\ref{fig:crystal_dipole}. 
The material in Fig.~\ref{fig:material}
can be viewed as a three dimensional material with a large aspect
ratio and a microstructure, ${\nu:\realone\to\realone}$, which is
only one dimensional because its value ${\nu(x)}$ at $x$ is
really an average of its three dimensional microstructure in the plane
perpendicular to the page.

The planes perpendicular to the page at 
${x=\xl}$ and ${x=\xr=\xr+S}$ bound the material in the ${-\htx}$ and ${\htx}$ directions, respectively, 
where $S$ is the length of the material.
Therefore ${\nu(x)}$ is negligible when ${x\notin(\xl,\xr)}$.

The bulk macrostructure is depicted as uniform in Fig.~\ref{fig:material}, which
is not the general case. 
I do not make any assumptions about the macroscopic charge density $\Rho$, except
that it is differentiable {\em almost everywhere}. By this I mean that it
is differentiable everywhere except where it changes nonlinearly on an interval
of width ${\prectheo=\abs{\dbx}}$. 

For example, as I explain in Sec.~\ref{section:homogenization}, a surface is assumed to have a
width of less than ${\prectheo}$ at the microscale.  This is not a physical assumption about the surface, 
but a consequence of the definition of ${\prectheo}$.
It means that the microscopic
charge density $\rho$ changes from being characteristic of the material's bulk to
being characteristic of the vacuum above the surface (i.e., ${\rho=0}$) on 
an interval that is smaller than the macroscale infinitesimal.
Therefore, $\Rho$ cannot be assumed to be differentiable or continuous at a surface.

\subsubsection{Microstructure dimensionality}
Whenever the argument of a microscopic field $\nu$ has an arrow over it (e.g., ${\nu(\rvec)}$),
its domain is implicitly three dimensional. When it does not (e.g., ${\nu(x)}$) its domain is
implicitly one dimensional.  When it has one argument with an arrow and one without an arrow
(e.g., ${\nu(u,\svec)}$), its domain is three dimensional and the argument with the arrow ($\svec$)
denotes a vector in the plane perpendicular to the $x$ axis at the $x$ coordinate
specified by the argument without an arrow ($u$).

\subsubsection{The surface and bulk subsystems}
\label{section:boundaries}
At the microscale ${\plane(x)}$ denotes the set of all values of ${\svec}$ for which 
point ${\rvec=(x,\svec)}$ is within the material;
and ${\abs{\plane(x)}}$ denotes the area of the material's cross section at $x$.

As will be discussed in Sec.~\ref{section:homogenization},
homogenization of microstructure by spatial averaging is tantamount to spatial compression.
In particular, all points at the microscale that are within a distance
${\prectheo/2}$ of a material's boundary are mapped to the same locally-planar surface at the macroscale.
The sets of $x$-coordinates of points at the microscale that are 
mapped by homogenization to the material's left and right surfaces 
are ${\mxl\equiv\interval(\xl,\prectheo)}$
and ${\mxr\equiv\interval(\xr,\prectheo)}$, respectively. 

At the microscale
$\mxl$ and $\mxr$ are {\em coincidence sets} (see Sec.~\ref{section:homogenization}):
They are the sets of all $x$-coordinates that
are indistinguishable from ${\xl}$ and ${\xr}$, respectively, at the macroscale.
At the macroscale $\mxl$ and $\mxr$ can be regarded and treated as coordinates
of points because, if ${\eta\leq\prectheo}$, any interval ${\interval(x,\eta)}$
only contains points that indistinguishable from $x$ at the macroscale.

As shown in Fig.~\ref{fig:crystal_dipole}, the
material's bulk is
\begin{align*}
\bulk\equiv \left[\xl+\prectheo/2,\xr-\prectheo/2\right]
\end{align*}
at the microscale, and
the width of the bulk is 
\begin{align*}
\bulksize\equiv \abs{\bulk}=S-\prectheo\approx S.
\end{align*}
The left and right surfaces are the regions of finite widths, 
\begin{align*}
\surfaceL\equiv [\xl,\xl+\prectheo/2] 
\end{align*}
and
\begin{align*}
\surfaceR\equiv [\xr-\prectheo/2,\xr], 
\end{align*}
respectively. Therefore the entire material
is ${\bulk\cup\surface}$, where ${\surface\equiv\surfaceL\cup\surfaceR}$.

At the macroscale the material's bulk is the open interval ${(\mxl,\mxr)}$;
its left and right surfaces are ${\{\mxl\}}$ and ${\{\mxr\}}$, respectively;
and the entiire material is
${
\closure(\mxl,\mxr)=[\mxl,\mxr]}$, 
where the \emph{closure operator} $\closure$ closes a set by adding any
missing boundary points to it.

Note that the set of all points at the microscale that correspond to elements
of ${[\mxl,\mxr]}$ at the macroscale is not ${\bulk\cup\surface=[\xl,\xr]}$.
It is the larger set,
\begin{align*}
[\xl-\prectheo/2,&\, \xr+\prectheo/2]\\
&=[\xl,\xr]\cup
[\xl-\prectheo/2,\xl]\cup[\xr,\xr+\prectheo/2],
\end{align*}
It includes all points in vacuum that are within a distance ${\prectheo/2}$ of
either $\xl$ or $\xr$.

\subsection{Externally applied fields}
\label{section:external_fields}
I  will primarily be concerned with materials 
that are either isolated or under the influence of constant
or slowly-varying (${f\lesssim\;\text{GHz}\iff\lambda\gtrsim 1\,}$m) externally-applied 
electromagnetic waves.
I assume that the size ($S$) of the material object 
is much less than the wavelength ($\lambda$). Therefore, within the material, 
any external fields are effectively spatially-uniform. 
I also assume that the period of oscillation of the field,  ${1/f\gtrsim 1\;\text{ns}}$, 
is much longer than the charge density's relaxation time.

\subsection{Microstructures of materials} 
\label{section:microstructures_of_materials}
To facilitate discussing materials in general terms I use a
fairly general mathematical representation of a material. However I will
assume that the net charge of the material is zero and,
because I am not concerned with magnetism, I will
assume that all particles have zero spin.
For the purposes of this work, the only relevant characteristics of each particle are its charge and its mass.
The only relevance of particles' masses is that they determine how delocalized each particle is
on the time scales of interest, and that nuclei move much more slowly than electrons.

In quantum mechanics the state of an {\em isolated} thermally-disordered object at time $t$
can be specified completely by the position pdf
of its $\Nparticle$ constituent particles at time $t$. The set of all particles'
positions will often be referred to as their \emph{configuration} and their position
pdf will sometimes be referred to as the configuration pdf.
The time-dependent configuration pdf of the particles is the function
\begin{align}
\begin{split}
&\pdf:\realone^{3\Nparticle}\times\realone\to \realpos\,; \\ 
&(\rvecsub{1},\cdots,\rvecsub{\Nparticle},t)\mapsto \pdf(\rvecsub{1},\cdots,\rvecsub{\Nparticle},t).
\end{split}
\label{eqn:pdf}
\end{align}
The finite spatial and temporal precisions of all measurements, and the
fundamentally-perturbative nature of the act of observation,
mean that, even within classical physics,
an observer's knowledge of the state of any physical system is also a pdf, rather
than a set of precise values of positions and momenta.

When the only properties of interest to the observer are statistical properties
of functions of particles' \emph{positions}, such as the expectation value and variance of the electric
potential $\phi$ at a point, momenta can be integrated out of the pdf.
Momenta can also be integrated out
if it is the \emph{rates of change} of 
statistical properties of functions of positions that are of interest, 
as long as the partial time derivative of the particles' position pdf is known or calculable.
Therefore, both classically and quantum-mechanically the material's microstructure
can be specified by a time-dependent pdf of
the form specified by Eq.~\ref{eqn:pdf}.

Any discontinuous pdf can be approximated arbitrarily closely by a continuous pdf.
Therefore, 
since the precisions of all measurements are finite, $\pdf$ can be assumed to be continuous.
For example,
every {\em delta distribution}, by which I mean a weighted sum of Dirac delta functions,
is the ${\upsigma\to 0}$ limit of a weighted sum of Gaussians of variances $\upsigma^2$.
This means that there exists a smooth density arbitrarily close to any given delta distribution, from which
properties of the delta distribution can be calculated to arbitrary precision if smoothness
is required for the calculation.

\subsubsection{Symmetry of $\pdf$ under exchange of identical particles}
\label{section:pdf_symmetry}
Within both quantum mechanics and classical statistical mechanics 
the position pdf of a set of fast-moving identical 
particles must be invariant under exchange of the positions of any two particles. This \emph{exchange symmetry} 
reflects the fact that the particles' trajectories cannot be observed. 

For example, suppose that $\Nident$ identical particles were far enough apart
at time $t$ to allow a different label from the set ${\{1,2,\cdots,\Nident\}}$ to be assigned to each one, and to
express their pdf as the product, 
\begin{align*}
\pdf(t)=\pdf(\rvecsub{1},\cdots,\rvecsub{\Nident},t)=\prod_{i=1}^{\Nident} \pdf_i(\rvecsub{i},t),
\end{align*}
of $\Nident$ different single-position pdfs, ${\pdf_i(t)}$.
Now suppose that at time ${t+\tau}$ they are no longer widely separated and their direct
or indirect interactions
with one another mean that
${\pdf(t+\tau)}$ cannot be approximated as a product of single-position pdfs. 

Then, if $\tau$ is not so large that the particles have thermalized, the dependences of $\pdf$ on 
the particles' positions may not all be equivalent: For example, $\pdf$ may depend on the set 
${\{\pdf_i(\rvecsub{i},t)\}_{i=1}^{\Nident}}$ of statistical states that quantified the observer's knowledge
of the particles' positions at time $t$; and therefore the result of integrating out all but one of $\pdf$'s
position arguments may depend on which position was not integrated out.
However, if the particles continue to interact with one another while moving in the
same region of space, ${\pdf(t+\tau)}$ must become independent of 
set ${\{\pdf_i(\rvecsub{i},t)\}_{i=1}^{\Nident}}$ in the limit ${\tau\to\infty}$. 

In the limit ${\tau\to\infty}$ the identity of a particle observed at position
${\rvec}$ at time ${t+\tau}$ cannot be known. In other words, it cannot be known
which label the observed particle was assigned at time $t$; and
it is equally likely to have been assigned any label in the set ${\{1,2,\cdots,\Nident\}}$.
The function ${\pdf(t+\tau)}$ must reflect this loss of discernible identity, so 
it must assign the same probability density to each of the ${\Nident!}$ different arrangements of its 
$\Nident$ position arguments. 
This property of ${\pdf}$ is known as \emph{exchange symmetry}. 

Probability density function $\pdf$ is exchange symmetric if and only if it is invariant under exchange of any two of its position arguments.
For example, 
\begin{align*}
\pdf(\rvecsub{1},\rvecsub{2}\cdots\rvecsub{\Nident})
=\pdf(\rvecsub{2},\rvecsub{1}\cdots\rvecsub{\Nident})
=\pdf(\rvecsub{\Nident},\rvecsub{1}\cdots\rvecsub{2}).
\end{align*}

\subsubsection{Antisymmetry of $\Psi$ under exchange of identical particles}
\label{section:antisymmetry}
As discussed in Appendix~\ref{section:appendix_states_as_vectors}, the information possessed by $\pdf$ is also possessed
by any real- or complex-valued function of the form ${\Psi=\sqrt{\pdf}e^{i\theta}}$.
Since 
\begin{align*}
\pdf=\Psi^*\Psi = (-\Psi)^*(-\Psi), 
\end{align*}
the exchange symmetry of $\pdf$ does not require 
${\Psi}$ to be exchange symmetric: It could
also be exchange \emph{antisymmetric}.
Exchange antisymmetry would mean that $\Psi$ changes sign under exchange of any two of its position 
arguments. For example, 
\begin{align*}
\Psi(\rvecsub{1},\rvecsub{2},\cdots,\rvecsub{\Nident})=-\Psi(\rvecsub{2},\rvecsub{1},\cdots,\rvecsub{\Nident}).
\end{align*}

Now let us add the assumption that two or more particles cannot be at the same position
at the same time. Then $\pdf$ and $\Psi$ vanish at 
all \emph{coincidence points}, which are points
${(\rvecsub{1},\cdots,\rvecsub{\Nident})}$ in the particles' 
configuration space at which 
two or more of the particles' positions ${\rvecsub{i}}$ are the same.
It follows that if ${\precmicro\in\realpos}$ is a lower bound on the precisions with which 
lengths or distances can be measured, a necessary condition
for ${\Psi}$ to be 
(finite-difference) differentiable at coincidence point 
${(\rvec,\rvec,\rvecsub{3}\cdots\rvecsub{\Nident})}$ is
that forward differences equal backward differences at linear order in $\prectheo$
at this coincidence point. 
That is, for any unit vector ${\rhat\in\realone^3}$,
\begin{align}
\Psi(\rvec+\precmicro\rhat,\rvec\cdots\rvecsub{\Nident})
&-
\Psi(\rvec,\rvec\cdots\rvecsub{\Nident})
\nonumber
\\
=
\Psi(\rvec,\rvec\cdots\rvecsub{\Nident})
&-
\Psi(\rvec-\precmicro\rhat,\rvec\cdots\rvecsub{\Nident}) +\order{\precmicro^2}.
\nonumber
\end{align}
Since ${\Psi(\rvec,\rvec,\cdots\rvecsub{\Nident})=0}$, 
this can be expressed as
\begin{align}
\Psi(\rvec+\precmicro\rhat,\rvec\cdots\rvecsub{\Nident})
&=
-
\Psi(\rvec-\precmicro\rhat,\rvec\cdots\rvecsub{\Nident}) +\order{\precmicro^2}, 
\nonumber
\end{align}
which is satisfied if 
${\Psi}$ changes sign at ${(\rvec,\rvec\cdots\rvecsub{\Nident})}$.
It is also satisfied if
\begin{align*}
&\Psi(\rvec\pm\precmicro\rhat,\rvec\cdots,\rvecsub{\Nident})=\order{\precmicro^2}
\\
\implies
&\Psi(\rvec,\rvec\pm\precmicro\rhat\cdots,\rvecsub{\Nident})=\order{\precmicro^2}, 
\end{align*}
which is to say that it is satisfied if, at coincident points, the first derivatives
of $\Psi$ with respect to the coordinates of the coincident particles
vanish.

In summary, if the derivatives
of $\Psi$ with respect to the positions of coincident particles do
not vanish at coincident points, $\Psi$ must be
exchange antisymmetric if it is differentiable.
If its derivatives with respect to the positions
of the coincident particles vanish, it can be differentiable if it is either exchange
symmetric or exchange antisymmetric.
 
\subsubsection{Number density}
The exchange symmetry of the position pdf $\pdf$ of a set 
${\Nelec}$ electrons means that the electrons'
\emph{number density} at position $\rvec$ can be expressed as
\begin{align*}
n(\rvec)
&\equiv \Nelec\int\cdots\int
\pdf(\rvec,\rvecsub{2}\cdots\rvecsub{\Nelec})
\dd[3]{r_2}\cdots\dd[3]{r_{\Nelec}}
\\
&=    \Nelec\int\cdots\int
\abs{\Psi(\rvec,\rvecsub{2}\cdots\rvecsub{\Nelec})}^2
\dd[3]{r_2}\cdots\dd[3]{r_{\Nelec}}.
\end{align*}
In the limit ${\hilbv\to 0}$, 
${n(\rvec)\hilbv}$ is
the probability of there being an electron in a sphere of volume $\hilbv$ centered at ${\rvec}$.

\subsubsection{Adiabatic approximation}
Until I discuss currents in Sec.~\ref{section:current} I will
assume that, because nuclei move slowly, the electrons can respond adiabatically
to their motion. Therefore if, at a particular point in time, the subsystem of electrons
is close to either a stationary state, such as its ground state, or a metastable state,
it will remain close to this state as the state changes in response
to the slowly-evolving confining potential from the nuclei.
This means that, to a very good approximation,
the time dependence of the number density of a set of $\Nelec$ electrons, 
can be replaced by a parametric dependence on nuclear positions.
I will not usually make this parametric dependence explicit, but I will
omit $t$ as an argument to ${n(\rvec)}$ and to the microscopic charge
density $\rho$ whenever I am making this {\em adiabatic approximation}.

\subsubsection{Charge density}
The quantity of primary interest in electrostatics at the microscale is a
material's charge density function,
\begin{align}
\rho(\rvec) = \sum_{i=1}^{\Nparticle} q_i
\intthree
\cdots\intthree
&\delta(\rvec-\rvecsub{i}) \nonumber \\
\times \pdf(\rvecsub{1},&\cdots,\rvecsub{\Nparticle})
\dd[3]{r_1}\cdots\dd[3]{r_{\Nparticle}} \label{eqn:rho}
\end{align}
where $q_i$ is the charge of particle $i$, $\rvecsub{i}$ is its position, and $\delta$ is 
the Dirac delta distribution. 
From now on I will denote the position of the {\em nucleus} with index $i$ by $\Rvecsub{i}$, to distinguish
it from the positions of electrons; and
I will assume that, to a good approximation,
$\rho$ can be expressed in the form
\begin{align}
\rho(\rvec) =
\overbrace{\vphantom{\sum_{i\in\text{nuclei}}} -e\;n(\rvec)}^{\displaystyle \rhom(\rvec)}
+\overbrace{Ze\sum_{i\in\text{nuclei}}\tdelta(\rvec-\Rvecsub{i})  }^{\displaystyle \rhop(\rvec)}
\label{eqn:charge_density}
\end{align}
where $-e$ is the charge of an electron,
and $Z$ is the atomic number of the nuclei. For simplicity I will often assume that the material contains
only one species of nucleus.

The function $\tdelta(\rvec-\Rvecsub{i})$ is not quite the Dirac
delta distribution, but a highly localized smooth probability density function for
the position of nucleus $i$. In many situations, but not all, we can treat it
mathematically as we would treat the Dirac delta distribution.

The energy of attraction between the nuclei and the electrons can be expressed as
\begin{align*}
(n,\vext)\equiv \intthree n(\rvec)\vext(\rvec)\dd[3]{r},
\end{align*}
where ${\vext}$ is equal to ${-e}$ times the positive electric potential from the nuclei.
In studies of the electronic subsystem at fixed nuclear positions, it is
common to refer to $\vext$ as
as the {\em external potential}.

For a one dimensional material aligned with the $x-$axis, such as the one depicted in Fig.~\ref{fig:material}, 
the analogue of Eq.~\ref{eqn:charge_density}
is
\begin{align}
\rho(x)
& =\rhom(x)+\rhop(x) \nonumber \\
& =-e\,n(x)+Ze\sum_{i\in\text{nuclei}}\tdelta(x-X_i)
\label{eqn:charge_density1d}
\end{align}

For most purposes, I will specfiy the (electrical) microstructure of the material as $\rho$ or as ${(\rhop,\rhom)}$

\section{Maxwell's theory of the aether}
\label{section:aether}
I now summarize some of the reasoning that led
Maxwell to his macroscopic theory of electromagnetism.
My purpose is to show that almost all of his reasoning is inconsistent 
with what has since been learned about materials and spacetime and 
that, in hindsight, the existence of $\pp$ and $\D$ fields
appears not to have a sound logical basis.  
The sources I have relied on most heavily are
\REFS~\linecite{maxwell-1865},
\linecite{maxwell-book1},
\linecite{maxwell-book2},
\linecite{heaviside-book},
\linecite{lorentz},
and~\linecite{history_of_physics}.

\subsection{Maxwell's reasoning}
Maxwell believed that a {\em luminiferous aether} pervaded all matter
and was the domain in which all electromagnetic processes occurred. He
rejected the idea of electromagnetic action at a distance, believing instead
that the aether was the medium by which, and through which, all forces
between electrified bodies were exerted. 

In vacuum he regarded the aether as isotropic, homogeneous, and with 
properties characterised by only two scalar constants, 
such as its permittivity ${\epsilon_0}$ and its permeability ${\mu_0}$, or either
one and the speed of light $c$.
He believed that the aether's properties were altered in the presence of matter, 
but that the effects of matter on electromagnetic phenomena were indirect and
could, to a first approximation, be described by the 
changes 
${\epsilon_0\mapsto\tilde{\epsilon}(\br)}$ 
and 
${\mu_0\mapsto\tilde{\mu}(\br)}$
of the aether's characteristic constants
from uniform scalars to tensor fields. 

Maxwell used the term {\em electricity} in an abstract or vague sense and
he likened electricity in the aether to elasticity in a solid. He
regarded this analogy as so compelling that, on the basis of it, 
he was willing to impute to the aether the minimal set of physical properties 
necessary to make his theory internally consistent.
He reasoned that, just as a slack rope or an unstrained rod cannot transmit forces between its two ends,  
the aether must be in a state of mechanical stress when electric forces are transmitted through it. 
Therefore, just as an elastically-deformed solid has a density of stored energy at each point, which 
is released when the deforming force is removed and the body resumes its original shape, he speculated
that an electric force field was always accompanied by a displacement field {\em in the aether}, which 
stored potential energy.

As an alternative to specifying a deformed state of an elastic solid with a vector
field whose value at each point is the point's displacement from equilibrium, 
one can describe it by a flux density vector field that is parallel to the direction of material          
flow at each point and has a magnitude equal to the total quantity of material, per unit area, that
flowed through a small imaginary surface at the point during the deformation.
Maxwell chose this latter approach to describe the state of the aether and the motion of electricity 
within it. One of his reasons was that certain fluxes, namely electric currents, were measureable, 
and measured fluxes were spatial averages, which could not 
be calculated using the former approach unless much more detailed
information about the aether was available. For example, a rate of fluid flow cannot 
be calculated if one only knows the velocity of the fluid at each point;
knowledge of its density is also required.

Maxwell defined the confusingly-named 
\emph{electric displacement field} $\D$ as a flux density. It specified how much
electricity passed through each point, and in which direction, as 
an applied field that caused and maintained this displacement was `switched on'.
He regarded $\D$ as a specification of the electrically-deformed state of the aether, albeit
one that was a spatial average of a more detailed microscopic flux density. 
He regarded the current density $\J$ as a rate of motion or a velocity of the aether, that was 
driven by the electric force $\E$, and which changed the aether's displacement $\D$.

For reasons that remained mysterious to Maxwell, conductors lacked the restoring force that
returned the $\D$ field in an electrically-deformed dielectric to its original state 
when the electric field supporting it was switched off.
Therefore, instead of simply displacing, electricity flowed freely as a current.  
As it flowed, it dissipated some of the aether's energy into heat within the material.
Similarly, when a dielectric was placed 
in an electric field, a transient current $\bdot{\D}$ flowed and dissipated energy until the
equilibrium displacement was reached. 

For energy to be conserved the energy stored
in the aether by the displacement field had to be lower in the presence of a dielectric 
than it was in free space.
It followed that, for the same electric force $\E$, 
$\D$ was different within a dielectric to its value of ${\epsilon_0\E}$ in vacuum.
Its value in a dielectric was ${\D=\epsilon_0\E+\pp}$, where  $\pp$ was known
as the {\em electric polarization} of the dielectric.

Since $\D$ was different in dielectrics, it must change abruptly at a dielectric's boundaries.
Maxwell viewed charge, not as a substance that can accumulate, but as a spatial discontinuity of $\D$.
He did not understand a current to be a flow of charge but as
a state of motion of the aether, which changed the $\D$ field, creating
those discontinuities. So, although current did not transport charge, it 
caused it to exist.

\subsection{Heaviside's concern}
Although there are some similarities between Maxwell's conception 
of electric polarization and more modern viewpoints, overall
the physical picture described above bears little resemblance to 
modern conceptions of electromagnetism, spacetime, or the structures and compositions of
materials.
Maxwell's reasoning has become as obsolete as his conception of an aether is, and
both he and his contemporaries were alert to this eventuality.
They regarded the properties he imputed to the aether as conjectures which would, when more was
learned, either be confirmed and developed further, amended, or abandoned. 

Heaviside made his concern about the challenges the theory faced 
clear in 1893, more than a decade after Maxwell's death, when he wrote:
``{\em 
Whether Maxwell's theory will last, as a sufficient and satisfactory primary theory upon
which the numerous secondary developments required may be grafted, is a matter for the future
to determine.
Let it not be forgotten that Maxwell's theory is only the first step 
towards a full theory of the aether ; and, moreover, that no theory of the 
aether can be complete that does not fully account for the omnipresent force of gravitation}''
~\citep{heaviside-book}.

Maxwell's theory should not be expected to make sense \emph{conceptually} as a 
theory of material response, because 
he developed it as a theory of the aether. 
However, because he ensured that it reproduced all of the 
empirically-known mathematical relationships between
macroscale observables, its {\em accuracy} as a macroscale tool is undiminished by 
the historical peculiarities of its mathematical form.  

It became confusing conceptually when it was stripped of its logical foundations
by the concept of an aether becoming obsolete; and it fails as a conceptual scaffold
when it is used beyond the macroscale domain for which it was constructed:
At the microscale it conflicts with 20th century theories
of material structure and composition. This is one of the reasons why there has
been so much confusion and debate about how to calculate
macroscopic fields from microscopic ones.

\section{How is $\pp$ defined?}
\label{section:definingP}
Many attempts have been made to reconcile Maxwell's auxiliary electric fields, $\pp$ and $\D$, 
with modern conceptions of material structure and composition.
Most have proposed definitions of $\pp$ in terms of the microscopic charge density $\rho$.
Once definitions of $\E$ and $\pp$ are in hand, the definition of $\D$  follows from
the constitutive relation ${\D=\varepsilon_0\E+\pp}$.

None of the proposed definitions of $\pp$ are viable, to my knowledge, and
I briefly explain their shortcomings in this section.
I do not attempt to refute every paper directly, but I 
outline a few of the most common definitions of $\pp$ and the reasons why
they are unsatisfactory. 

The literature on the {\em Modern Theory of Polarization} (MTOP) (e.g., \REF~\linecite{resta-vanderbilt-2007}), 
which is discussed in Sec.~\ref{section:mtop}, can be consulted for more
discussion about shortcomings of pre-MTOP definitions of $\pp$. I outline
reasons to look beyond the MTOP definition of $\pp$ in Sec.~\ref{section:mtop}.

\subsubsection{Attempt 1}
\label{section:definingP_1}
$\pp$ has the dimensions of a dipole moment per unit volume and so it is natural to try to
define it as such. Let us consider a material that occupies and fills a space 
${\Omega\subset\realthree}$, whose volume is ${\abs{\Omega}}$.
Many authors have assumed, often tacitly, that $\Omega$ 
can be divided into microscopic partitions in some natural or `right' way. For example, 
in a molecular material there might be a separate partition for each molecule.
$\pp$ is then defined as the macroscopic spatial average of the partitions' dipole moments
divided by their volumes. This definition fails because, 
as Fig.~\ref{fig:surfcharge} and Fig.~\ref{fig:polar_surface} illustrate, 
there are an infinite number of ways to partition 
any material, which are equally justifiable theoretically, and 
each different set of partitions leads to a different magnitude 
and direction of $\pp$, in general.

\subsubsection{Attempt 2}
\label{section:definingP_2}
One could also define $\pp$ as the dipole moment of the entire material divided by its volume, i.e., 
\begin{align}
\pp\equiv \frac{1}{\abs{\Omega}}\int_\Omega \rho(\rvec)\,\rvec\dd[3]{r}.
\label{eqn:pdef1}
\end{align}
This definition is not satisfactory because it implies that $\pp$ 
is not a property of the bulk, in general. 
To understand why, consider the dipole moment of a charge-neutral crystalline rod 
of length $L$ and area of cross-section $A$, whose surfaces perpendicular to its long axis and
carry charges of $+q$ and $-q$. 
For simplicity let us suppose that the rod is carved from a perfect ionic crystal, 
and then isolated immediately so that its composition cannot change. 

If $L$ was large compared to the crystal's lattice constant, the rod's dipole moment would
be approximately equal to ${qL\,\normal}$, where $\normal$ is an outward unit normal to the surface
that carries
charge $q$. However, both the magnitude and the sign of $q$ are
determined by where along the rod's axis the bulk crystal was cleaved to form its surfaces.
As illustrated in Fig.~\ref{fig:polar_surface}, two surfaces formed by cleaving a crystal along 
relatively-shifted parallel planes have different charges, in general. Therefore, by this definition,
the value of 
\begin{align*}
\pp\approx \frac{qL}{\volume}\,\normal= \frac{q}{A}\,\normal
\end{align*}
is mostly determined by the areal density of charge
on the rod's surfaces,
${\bsigma\equiv q/A}$,
and it would vanish if the surfaces were neutralized.
Therefore, Eq.~\ref{eqn:pdef1} does not define a bulk property. 

\subsubsection{Attempt 3}
\label{section:definingP_3}
One could consider basing a definition on either 
${-\div\pp=\Rho}$ or ${-\div\mpp=\rho}$, where $\mpp$ is a microscale analogue of $\pp$.
This approach fails because, without boundary conditions, these equations only define $\pp$ and $\mpp$ up to 
arbitrary constants. With boundary conditions, their values are determined by the
charge at the material's surfaces, so they are not properties of the bulk.

\subsubsection{Attempt 4}
\label{section:definingP_4}
Finally, several well-known textbooks, including those by Jackson~\citep{jackson-book} and Ashcroft and Mermin~\citep{ashcroft_mermin_book}, 
use variants of a method refined by Russakoff~\citep{russakoff-ajp-1970} to define $\pp$. 
They assume that the microscopic charge density can be expressed in the 
form 
\begin{align*}
\rho(\rvec) = \sum_i\rho_i (\rvec-\rvecsub{i}), 
\end{align*}
where
each $\rho_i$ is a charge density that is localized around the origin and microscopic in extent, 
such that ${\rho_i(\rvec)\approx 0}$ when ${\abs{\rvec}\gg a}$.
For example, each $\rho_i$ might be the distribution of a different molecule's charge.

The next step is to find the spatial average of $\rho$ by  convolving it with a smooth 
spherically-symmetric {\em averaging kernel}, ${\mu(\epsilon):\realthree\to\realnonneg}$, whose
width is proportional to ${\epsilon}$ and which has an integral of one, as follows:
\begin{align*}
\expval{\rho;\mu}_\epsilon(\rvec) &\equiv
  \intthree \mu(\uvec;\epsilon) \, \rho(\rvec+\uvec)\dd[3]{u}  \\
& =   \sum_i \intthree \mu(\rvec-\rvecsub{i}-\uvec;\epsilon) \, \rho_i(\uvec)\dd[3]{u},
\end{align*}
where I have changed the variable of integration and made use of $\mu$'s spherical symmetry.
Taylor expanding ${\mu(\epsilon)}$ in each integrand to first order in $\uvec$ gives
\begin{align*}
\expval{\,\rho\,}(\rvec) 
& \approx   \sum_i \intthree \left[\mu(\rvec-\rvecsub{i};\epsilon) - \uvec\cdot\grad \mu(\rvec-\rvecsub{i};\epsilon)\right] \\
&\hspace{2cm} \times\rho_i(\uvec)\dd[3]{u} \\
& =   \sum_i \mu(\rvec-\rvecsub{i};\epsilon)\,q_i - \sum_i\grad \mu(\rvec-\rvecsub{i};\epsilon)\cdot\dvecsub{i}
\end{align*}
where 
${q_i\equiv\intthree \rho_i(\rvec)\dd[3]{r}}$ and 
${\dvecsub{i}\equiv\intthree \rho_i(\rvec+\uvec)\,\uvec\,\dd[3]{u}}$ are the net charge 
and the dipole moment of distribution $\rho_i$, respectively. 

If we identify ${\expval{\rho;\mu}_\epsilon}$ as the macroscopic
charge density $\Rho$ and express the second sum on the right
hand side as ${\div\left(\sum_i \mu(\rvec-\rvecsub{i};\epsilon)\,\dvecsub{i}\right)}$, we find that
\begin{align*}
\Rho & = \Rhofree - \div{\pp} = \Rhofree + \Rhobound
\end{align*}
where ${\Rhobound\equiv-\div\pp}$, and $\Rhofree$ and $\pp$
are volumetric densities of the molecules' net charges and dipole moments, respectively. 

There are multiple fatal flaws in these definitions of $\Rhobound$, $\Rhofree$, and $\pp$:
The value of each quantity depends sensitively on the value of $\epsilon$, which is arbitrary;
and both $\Rhobound$ and $\pp$ vanish in the limit ${\epsilon\to\infty}$. 
Each field also depends on how $\rho$ is partitioned
into localized distributions ${\rho_i}$. However, even if the set ${\{\rho_i\}}$ was given, 
if ${\Rhofree\neq 0}$ then the values of $\Rhobound$, $\Rhofree$, and $\pp$ would 
depend on the choice of origin.
This is because the dipole moment of any charge distribution is origin dependent unless its net charge is zero.

\section{Demands of symmetry and asymmetry}
\label{section:symmetry}
In this section I discuss some properties that we should expect
a macrostructure to have, on symmetry grounds,
if it is a spatial average of the microstructure.

I begin, in Sec.~\ref{section:averaging_commutes},  by
discussing consequences of the linearity of the spatial averaging operation 
${\nu\mapsto\expval{\nu;\mu}_\epsilon}$.
I discuss general macroscopic vector fields in Sec.~\ref{section:bulkfields}, and 
I discuss the macroscopic potential $\bphi$, the macroscopic polarization  $\pp$, 
and the polarization current ${\Jconv}$ in Secs.~\ref{section:potential_symmetry}, \ref{section:polarization_symmetry}, 
and~\ref{section:anisotropy}, respectively.

\subsection{Linearity and the superposition principle}
\label{section:averaging_commutes}
It is well known that,
given two microscopic charge densities, $\rho_1$ and $\rho_2$, from  
which the microscopic electric fields ${\me_1\equiv\hat{\mathcal{E}}_\rho[\rho_1]}$ and
${\me_2\equiv\hat{\mathcal{E}}_\rho[\rho_2]}$ emanate, where ${\hat{\mathcal{E}}_\rho}$ is a functional;
and given any two scalar constants ${\omega_1,\omega_2\in\realone}$; the following relation holds:
\begin{align}
\hat{\mathcal{E}}_\rho[\omega_1\rho_1+\omega_2\rho_2]=\omega_1\hat{\mathcal{E}}_\rho[\rho_1]+\omega_2\hat{\mathcal{E}}_\rho[\rho_2].
\label{eqn:superposition}
\end{align}
Analogous relations hold for other functionals, such as ${\hat{\phi}_\rho[\rho]}$, ${\hat{\mathcal{E}}_\phi[\phi]}$, and 
${\hat{\rho}_\phi[\phi]}$, which
relate ${\phi}$ to ${\rho}$, ${\me}$ to ${\phi}$ and ${\rho}$ to ${\phi}$, respectively.

The property of ${\hat{\mathcal{E}}_\rho}$ expressed by Eq.~\ref{eqn:superposition} is known as {\em linearity}, 
but in the context of electricity it is better known as 
the {\em principle of linear superposition} or simply the {\em superposition principle}.
The superposition principle follows from the fact that derivatives and integrals are linear operations and 
the fact that ${\me=-\grad\phi}$ and ${\rho/\varepsilon_0\equiv\div\me=-\laplacian\phi}$ are both
negative derivatives of ${\phi}$. 

A one dimensional spatial average has the general form
\begin{align}
\expval{\nu; \mu}_\epsilon(x) &\equiv \int_\realone \nu(x')\mu(x'-x;\epsilon)\dd{x'},
\label{eqn:average}
\end{align}
where $\epsilon$ is a parameter that is proportional to the 
width of the averaging kernel, ${\mu(\epsilon)}$.
It is straightforward to use Eq.~\ref{eqn:average} to show that this is 
also a linear operation, i.e., 
\begin{align*}
\expval{\omega_1\nu_1+\omega_2\nu_2; \mu}_\epsilon(x) = \omega_1\expval{\nu_1; \mu}_\epsilon(x)+\omega_2\expval{\nu_2; \mu}_\epsilon(x), 
\end{align*}
for any numbers ${\omega_1,\omega_2\in\realone}$ and any functions ${\nu_1=\nu_1(x)}$ and ${\nu_2=\nu_2(x)}$.
The spatial averages in two and three dimensions are also linear operations.

\subsection{Spatial averaging commutes with derivatives}
It can be shown from Eq.~\ref{eqn:average} that spatial averages and derivatives commute.
For example, 
\begin{align*}
\partial_x^n\expval{\nu;\mu}_\epsilon \equiv
\overbrace{\partial_x\partial_x\cdots\partial_x}^{n\;\text{times}}\expval{\nu;\mu}_\epsilon = \expval{\partial_x^n\nu;\mu}_\epsilon,
\end{align*}
where ${\partial_x}$ is the partial derivative with respect
to $x$. The analogous results for the gradient and laplacian in three dimensions, when ${\nu=\nu(\rvec)}$, 
are ${\grad\expval{\nu;\mu}_\epsilon=\expval{\grad\nu;\mu}_\epsilon}$
and ${\laplacian\expval{\nu;\mu}_\epsilon=\expval{\laplacian\nu;\mu}_\epsilon}$.

\subsubsection{Relationships between ${\bphi}$, ${\E}$, and ${\Rho}$}
Because spatial averaging commutes with derivatives, 
it would follow from defining macroscopic
fields as spatial averages of their microscopic counterparts that
the relationships between $\bphi$, $\E$, and $\Rho$ are the same as those
between $\phi$, $\me$, and $\rho$, i.e.,  ${\E=-\grad\bphi}$ and ${\Rho = -\varepsilon_0\laplacian\bphi=\varepsilon_0\div\E}$.

In Sec.~\ref{section:homogenization} we will identify ${\bphi}$, $\E$, and $\Rho$ with spatial averages
of ${\phi}$, $\me$, and $\rho$, respectively, but we will see that their
definitions are a bit more complicated than, for example, ${\bphi\equiv\expval{\phi;\mu}_\epsilon}$
for some averaging kernel $\mu$ and some width parameter $\epsilon$.
Nevertheless, they are spatial averages and the homogenization transformation
is linear, which means that the macroscopic counterpart of the derivative ${\dnu{1}}$
of $\nu$ is the derivative ${\dNu{1}}$ of the macroscopic
counterpart of $\nu$.

Therefore homogenization does not create new fields, $\pp$ and $\D$, as it does
in Maxwell's theory, and the linearity of the homogenization transformation
implies that
calculating $\bphi$ and $\E$ from $\Rho$ must be equivalent to
first calculating $\phi$ and $\me$ from $\rho$ and then spatially averaging them 
to find $\bphi$ and $\E$, respectively. 

\subsection{Symmetry is scale-dependent} 
\label{section:scale_symmetry}
It appears to follow 
from the fact that derivatives and spatial averaging commute
that symmetry is scale-dependent.
For example, a crystal with microstructure  ${\rho_\text{crystal}}$
and a glass with microstructure ${\rho_\text{glass}}$
can have exactly the same bulk macrostructure ${\Rho}$; and they
usually do because ${\Rho=0}$ in the bulk of any stable 
electromagnetically-isolated material whose surfaces are not charged.

The superposition principle implies that the macroscopic field 
emanating from the bulk of the crystal can be expressed as
\begin{align*}
\E^\text{bulk}_{\text{crystal}} 
& = \hat{E}_\rho[\rho_\text{crystal}]
= \expval{\hat{\me}_\rho[\rho_\text{crystal}]}
= \hat{\me}_\rho[\expval{\rho_\text{crystal}}]\\
& = \hat{\me}_\rho[\Rho] 
= \hat{\mathcal{E}}_\rho[\expval{\rho_\text{glass}}]
 = \hat{E}_\rho[\rho_\text{glass}]
= \E^\text{bulk}_{\text{glass}} 
\end{align*}
where ${\hat{E}_\rho}$
is a linear functional of $\rho$, which satisfies
${\hat{E}_\rho[\rho]=\hat{\me}_\rho[\Rho]}$,
and for simplicity I am denoting the spatial
average of each field ${\nu}$ simply as ${\expval{\nu}}$.

It follows that neither a crystal's symmetry, nor any other 
characteristic of its microstructure that differs from the glass, 
alters the macroscopic electric field or the macroscopic electric potential emanating from its bulk.
The only symmetries that manifest at the macroscale are
symmetries of the macrostructure.

\subsection{Nonlinear relationships and response functions}
It is important to note that the superposition principle applies only to {\em linear} physical
systems. A more accurate way to state this is that a linear physical system is defined to be
a system for which the superposition principle applies. 

The superposition principle could not apply to all of the quantities  
${\alpha(x)}$, ${\beta(x)}$ and ${\gamma(x)}$ if they were related by
${\alpha(x) = \beta(x)\gamma(x)}$ because $\beta$ and $\gamma$ are not linearly related.
For example, if ${\alpha=\alpha_1+\alpha_2}$ where
${\alpha_1=\beta_1\gamma_1}$ and 
${\alpha_2=\beta_2\gamma_2}$, then 
\begin{align*}
\alpha=
\alpha_1+\alpha_2 & = 
\beta_1\gamma_1 + \beta_2\gamma_2 \neq
\left( \beta_1+\beta_2\right) \left(\gamma_1+\gamma_2\right).
\end{align*}
This has important implications for material-specific response parameters, such 
as conductivities, which are not simply spatial averages of their
microscopic counterparts.

For example, if a material's macroscopic response to an arbitrarily-weak applied field $\Eext$ is ${\DRho}$, 
and if the change $\Dbsigma$ in its surface charge is the integral 
of ${\DRho}$ across the surface, then ${\Dbsigma\propto \Eext}$. However, the constant of 
proportionality does not have an analogue at the microscale to which it can be
related by spatial averaging. 

In fact, because the material is macroscopically uniform
at equilibrium before and after ${\Eext}$ is switched on, 
${\Rho}$ vanishes in the bulk both with and without ${\Eext}$.
Therefore the macroscopic response to $\Eext$ is not a response of the bulk, but
a change of the excesses of charge at surfaces and interfaces.
At the microscale, by contrast, the response of the material's equilibrium microstructure to $\Eext$ is a change
of the charge density at surfaces, interfaces, and at all points in the bulk.

The macrostructure of the bulk is indistinguishable from vacuum and
the macroscale response to a uniform field manifests only at surfaces and other 
macroscopic heterogeneities. 
Therefore it may be better to view a macroscopic response function, such as a conductivity or permittivity, 
as a property of a pair of surfaces, rather than as a
property of the material occupying the space between them.

For example, a macroscopic response function
${\bchi_{12}(\bsigma_1,\bsigma_2,\bsigmadot_1,\bsigmadot_2,\cdots)}$ for a pair of surfaces,
whose charge excesses are ${\bsigma_1}$ and ${\bsigma_2}$, 
might describe the (change in the) rate of change ${\bsigmadot_1 = -\bsigmadot_2}$ when a field is applied,
the rate of energy dissipation during this change, etc..
The bulk composition and microstructure would help to 
determine $\bchi_{12}$, but so would the microstructures and compositions
of both surfaces.

\subsection{Macroscopic vector fields}
\label{section:bulkfields}
As discussed in  Sec.~\ref{section:scale_symmetry}, and as 
Fig.~\ref{fig:material} illustrates,  symmetry is scale-dependent:
A material whose microstructure $\rho$ is highly inhomogeneous
and anisotropic can have a bulk macrostructure with local continuous translational symmetry, 
\begin{align}
\Rho(\bm{x+u})&=\Rho(\mx),\;\;\forall\, \mx\in\bulk\;\;\text{and}\;\;\forall\, \bm{u}:\bm{x+u}\in\bulk,
\label{eqn:homogeneity} \\
\intertext{which implies local isotropy,}
\Rho(\bm{x+u})&=\Rho(\bm{x-u}), \;\;\forall\, \mx,\bm{u}:\interval(\mx,2\abs{\bm{u}})\subset\bulk. 
\label{eqn:isotropy}
\end{align}
By {\em local} I mean that for any $\mx\in\bulk$ there are limits to the magnitudes
of $\bm{u}$ for which 
Eqs.~\ref{eqn:homogeneity} and~\ref{eqn:isotropy} hold. 
Eq.~\ref{eqn:homogeneity} holds for ${\bm{u}\in\left(-\abs{\mx-\mxl},\abs{\mx-\mxr}\right)}$
and Eq.~\ref{eqn:isotropy} holds for ${\abs{\bm{u}}<\bm{u_\text{max}}(\mx)\equiv\min\left\{\abs{\bx-\mxl},\abs{\bx-\mxr}\right\}}$.

As noted in Sec.~\ref{section:scale_symmetry}, the material cannot be stable
unless its bulk is charge-neutral on average, i.e., ${\Rho=0}$.
Therefore it is macroscopically uniform and locally isotropic at each point $\mx$. 
It follows that any observable directionality at $\mx$ must be a consequence
of the inequivalence of $\Rho$ at distances larger than ${\bm{u_\text{max}}(\mx)}$ in 
the two directions.
Therefore, a macroscopic vector field whose value at each point is a 
linear functional of $\Rho$ (or $\rho$), cannot emanate from the bulk of a 
macroscopically uniform material because the bulk macrostructure cannot bestow directionality. 
This implies that ${\E}$ vanishes and ${\bphi}$ is constant in the bulk.

Similarly, the existence of $\pp$ is attributed to the bulk microstructure
lacking inversion symmetry. However, regardless of the microstructure,
the bulk macrostructure has inversion symmetry and a macroscopic $\pp$ field 
appears incompatible with that.

\subsubsection{Current density}
There is one macroscopic vector field in a uniform material's bulk that is, in a 
certain sense at least, retained by the homogenized theory. 
This is the current density $\J$, which is not observable
at the macroscale {\em in the uniform bulk}, but whose consequences, namely, the
rates of change of accumulations of charge at macroscopic heterogeneities 
(surfaces, defects, etc.), are both observable and bestow directionality to it.

The rate of change of an accumulation of charge at a macroscopic defect or interface 
can be calculated from ${-\div\J}$ or from the difference $\Delta\J$ in the values of $\J$ on either side of an interface.
Therefore, instead of defining $\J$ at each point in the bulk, at the macroscale we need only concern ourselves with
$\Delta\J=\bsigmadot$ or ${-\div\J\abs{\dbx}=\bqdot}$, which are the rates
of change of the excesses of charge at an interface and at a point, respectively.

It is not possible, in general, to deduce the direction of $\J$ from the rate
of change of areal charge density $\bsigma$ at an interface.
However if, in a closed system, one knows the rates of change of accumulations of 
charge on {\em every} source of macroscopic heterogeneity, 
one could calculate the magnitude and direction of the current density everywhere
in the uniform regions surrounding and separating these heterogeneities.
This is the essence of electrical circuit theory.

My argument in this section is that symmetry is scale dependent and that, regardless of a material's microstructure, no
directionality should be observable at the macroscale if the macrostructure is isotropic.
$\J$ should not be regarded as either invalidating this argument or as an exception 
to this principle because $\J$ is not observable at the macroscale. Only 
its consequences, such as $\bsigmadot$ and the rate of change
of temperature ${\mathbf{\dot{T}}}$ are.

\subsection{Mean inner potential, $\bphi$}
\label{section:potential_symmetry}
The arguments of the previous section, and the linear relationship between 
the mean inner potential $\bphi$ 
and $\rho$ and $\Rho$, imply that $\bphi$  vanishes if a material's
surfaces are not charged.
This contradicts a great deal of existing literature, including 
textbooks and many recent research articles~\citep{bethe-1928, miyake-1940, mip_sanchez_1985,mip_pratt_1987, mip_pratt_1988, mip_pratt_1989, mip_pratt_1992, gajdardziska-1993, spence-1993, mip_sokhan_1997, spence-1999, mip_leung_2010, mip_mundy_2011, mip_cendagorta_2015, mip_lars, mip_marzari, mip_water_2020, mip_madsen_2021, mip_kathmann_2021}. 
However, those works often assume that ${\bphi}$ and the 
average potential {\em experienced by a particle}
moving through the bulk are the same quantity, or they approximate the latter as the former. 

Even if a particle spends a very long time in a material, it does not sample space uniformly. It samples
regions of positive electric potential, where the electron density is high, more than regions of negative potential.
Furthermore, an electron is a {\em perturbing} probe of electric potential. It does not
sample a material's equilibrium microscopic potential ${\phi(\rvec)}$ because its
presence at position ${\rvec}$ reduces the probability density of other electrons
being in a neighbourhood of ${\rvec}$, which 
reduces the negative potential it experiences from other electrons.

Therefore if the spatial average of ${\phi}$ is zero, meaning that a non-perturbing
probe would measure an average microscopic potential of zero if it sampled the entire space within a material
uniformly,
the average potential experienced by an electron would be positive. For the same reason,
the average potential experienced by a diffusing cation is negative because
it attracts electrons to it and repels nuclei and other cations.

The superposition principle helps to understand why ${\bphi}$ vanishes in a material's bulk  (${\implies-\grad\bphi=0}$).
It means that the potential emanating from any nucleus is the sum of the potentials emanating
from its constituent protons. Therefore the spatial average of the potential inside a charge-neutral material can be expressed as the sum of
the spatial averages of the potentials emanating from point particles of charge ${+e}$ (protons) and
point particles of charge ${-e}$ (electrons). The spatial average of the potential from an electron
is the negative of the spatial average of the potential from a proton.
Therefore, since there are equal numbers of protons and electrons in a charge-neutral material, $\bphi$ can be expressed as a
sum of vanishing contributions from proton-electron pairs.

\subsection{Bulk polarization, $\pp$}
\label{section:polarization_symmetry}
In this section I examine some consequences of assuming that $\pp$ 
can be expressed as some functional $\hat{P}_\rho$ of $\rho$. 
I do so under the assumption that ${\E\equiv\hat{E}_\rho[\rho]}$ and 
${\Rho\equiv\hat{\varrho}_\rho[\rho]}$ are both {\em linear} functionals of $\rho$.
If they were not linear, 
either the superposition principle would not apply at the macroscale, or the ${\rho\mapsto\Rho}$ 
homogenization transformation would not conserve net charge.
They are linear if $\E$ and $\Rho$ are spatial averages of $\me$ and $\rho$, respectively.

A polarization field is believed to exist in any crystal whose microstructure 
lacks inversion symmetry. Therefore ${\hat{P}_\rho[\rho]}$ must be a nonlinear functional 
because the superposition ${\rho = \rho_1+\rho_2}$ of two inversion-symmetric crystal structures 
does not have inversion symmetry, in general.
If ${\hat{P}_\rho[\rho_1]= 0}$, ${\hat{P}_\rho[\rho_2]=0}$, and ${\hat{P}_\rho[\rho]\neq 0}$, 
then 
\begin{align*}
\hat{P}_\rho[\rho_1+\rho_2]\neq \hat{P}_\rho[\rho_1]+\hat{P}_\rho[\rho_2]. 
\end{align*}

Let us assume that 
\begin{align*}
\D\equiv\hat{D}_\rho[\rho]=\hat{D}^{l}_\rho[\rho]+\hat{D}^{nl}_\rho[\rho] 
\end{align*}
and 
\begin{align*}
\pp\equiv\hat{P}_\rho[\rho]=\hat{P}^{l}_\rho[\rho]+\hat{P}^{nl}_\rho[\rho], 
\end{align*}
where the superscripts `$l$' and `$nl$' identify linear and nonlinear parts of the functionals, 
respectively.
It follows from the linearity of ${\hat{E}_\rho}$ 
and the relation ${\D = \varepsilon_0\E+\pp \implies \hat{D}_\rho = \varepsilon_0\hat{E}_\rho + \hat{P}_\rho}$
that ${\hat{D}_\rho^{nl}=\hat{P}_\rho^{nl}}$ and ${\varepsilon_0\hat{E}_\rho=\hat{D}^l_\rho-\hat{P}^l_\rho}$.

Any microscopic charge density $\rho$ can be written as a (possibly infinite) sum
of inversion-symmetric charge densities; the Fourier series of a periodic $\rho$ being one example.
It follows that ${\pp^l\equiv\hat{P}^l_\rho[\rho]=0}$ for any $\rho$,
that the linear part of $\D$ is
${\D^l\equiv\hat{D}_\rho^l[\rho]=\varepsilon_0\E}$, and that
its nonlinear part $\D^{nl}$ is simply ${\pp}$. 

With these constraints the set of relationships 
that constitute the macroscale theory of electricity can be expressed as
\begin{align*}
\Rho  &= \varepsilon_0\div\E,\;\; &\;\;
\div\J+\bdot{\Rho}  & = 0 \\
\curl\H  
&=\J+\varepsilon_0\bdot{\E}, \quad & \quad
\curl\E  & = -\bdot{\B}  
\end{align*}
where ${\J\equiv\Jcond +\bdot{\pp}}$, ${\Jcond}$ is the {\em conduction current}, and 
a dot denotes a partial time derivative at fixed position.
The linear and nonlinear parts of
$\bdot{\D}$ appear separately in these equations 
as ${\varepsilon_0\bdot{\E}}$ and $\bdot{\pp}$, respectively, 
and the only purpose served by $\pp$ is to define its time
derivative, ${\Jconv=\bdot{\pp}}$.

Let us consider two ways to proceed from here. The first is to follow convention by
finding a way to define $\pp$. This would lead to definitions of 
${\D}$, ${\Rhobound}$, ${\Rhofree}$, ${\Jbound}$, 
and ${\Jfree}$, where ${\Jbound\equiv\Jconv}$ and ${\Jfree\equiv\Jcond}$ are currents of 
bound charge density (${\Rhobound}$) and free charge density (${\Rhofree}$), respectively.
I denote them by ${\Jconv}$ and ${\Jcond}$ to avoid distinguishing between free and bound charges.
Introducing these six fields 
to the theory does not make it any more predictive or useful. Furthermore, $\pp$ and $\D$
are not observable; ${\Jfree}$ is only observable when ${\Jbound=0}$ and vice versa;
and free charge is only observable where the net bound charge vanishes and vice versa.

A much simpler and less conventional way to proceed is to {\em not} 
introduce any unobservable quantities into the theory, but 
to find a way to calculate ${\Jconv}$. 
We do not need to distinguish between different contributions to $\J$ in 
either Maxwell's equations or the contintuity equation if we 
are not distinguishing between $\Rhofree$ and $\Rhobound$.
Therefore ${\Jcond+\bdot{\pp}}$ can be replaced by $\J$
with the understanding that ${\J}$ 
is the net flow of charge from {\em all} mechanisms.

We are left with only the four equations above, which are identical in form 
to their counterparts at the microscale, and in which all three electrical
quantities that appear in them 
($\Rho$, $\J$, $\E$) are observables with clear and intuitive meanings.
There is no downside to scrapping $\pp$ and $\D$ and it
circumvents many problems, such as the fact that we do not have
a definition of $\pp$, and the fact that,  
because $\pp$ is nonlinear and $\E$ is 
linear, ${\pp\neq\varepsilon_0\boldsymbol{\chi}\E}$.

It is important to note that a key premise or conclusion of the {\em Modern Theory of Polarization} 
is that my central premise in this section, namely that $\pp$ can be calculated from $\rho$,  is false~\citep{resta-vanderbilt-2007}.
It is claimed, instead, that $\pp$ is a property of the phase $\theta$ of the material's 
wavefunction ${\Psi=\sqrt{\pdf}\,e^{i\theta}}$. 
I will discuss this claim in Sec.~\ref{section:mtop} 
and Sec.~\ref{section:current}.

\subsection{Polarization current as a demand of anisotropy}
\label{section:anisotropy}
\begin{figure}[!]
\includegraphics[width=0.48\textwidth]{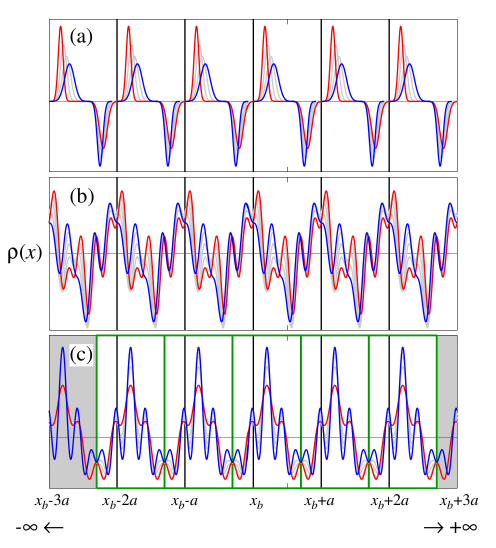}
\caption{Charge density $\rho$ as function of position $x$ in three one dimensional crystals. 
The crystals in (a) and (b) lack inversion symmetry, but the crystal in (c) has inversion symmetry, with
two inversion centers per primitive cell $\unitcell$. 
In (c) a choice of primitive unit cell whose dipole moment is zero is outlined in green.
If $\rho(x)$ changes continuously and uniformly between the densities
plotted in red and blue, a macroscopic current flows in crystals (a) and (b) 
because the probability that the net movement of charge in direction $\hat{x}$
equals the net movement of charge in the inequivalent direction ${-\hat{x}}$ is zero.
In (c) the symmetry of the crystal forbids a macroscopic flow of charge because the 
net movement of charge relative to an inversion
center cannot differ between the two equivalent directions $\hat{x}$ 
and $-\hat{x}$. One way to see this is to note that in (c) most of space can be tiled with 
unit cells $\unitcell$ whose dipole moment remains zero {\em throughout} the changing of the
density. There remains only the two shaded regions 
of combined width ${a=\volume}$ at the left and right boundaries of the chunk of 
bulk crystal comprised of ${\Nunitcell=6}$ primitive cells.
In the limit of large ${\Nunitcell}$
the change in the distance between $x_b$ and the center of charge
of the $\Nunitcell$ cells, divided by their combined width ${\Nunitcell a}$, vanishes.
In (a) and (b) the current cannot vanish because the ${\hat{x}}$
and ${-\hat{x}}$ directions are inequivalent. An important question, 
which the MTOP solved for quantum systems, is how the current
can be calculated from an evolving bulk microstructure, i.e., 
without knowing or calculating how much charge accumulates
at surfaces. If the integrals $q_1$ and $q_2$ of the two peaks per unit cell
in (a) remain constant, the current per unit length is simply
${\left(q_1\dot{x}_1+q_2\dot{x}_2\right)/a}$, where ${\dot{x}_1}$
and ${\dot{x}_2}$ are the velocities with which the peaks move.
However if the charge density is not organized
into packets of fixed charge, as in (b), the definition of current
is much less obvious.
}
\label{fig:current-symmetry}
\end{figure}

Figure~\ref{fig:current-symmetry} shows several stages in the
evolution of the equilibrium charge density $\rho$ in the bulks of three crystals as some stimulus $\zeta$
is applied uniformly to them. The stimulus might be a change in temperature, 
a strain, a displacement of one of the crystal's sublattices relative to 
the others, or anything else that changes a crystal's charge density 
uniformly throughout the bulk.

The charge density in Fig.~\ref{fig:current-symmetry}~(c) has inversion symmetry, with
two centers of inversion in each primitive unit cell; and it maintains this inversion symmetry as $\zeta$
changes and ${\rho(x;\zeta)}$ evolves. Clearly, the motion of charge relative to 
one of its centers of inversion must be the same in the ${-\hat{x}}$ and ${+\hat{x}}$ directions.
Therefore the existence of a net current is prohibited by symmetry.

On the other hand, the only symmetries possessed by the charge densities 
in  Figs.~\ref{fig:current-symmetry}(a) and~\ref{fig:current-symmetry}(b)
are their periodicities. Therefore it is impossible
for the motion of charge in the ${-\hat{x}}$ and ${+\hat{x}}$ directions to be equitable, because
those directions are inequivalent and two numbers within the same continuous range cannot be
\emph{exactly} equal, meaning equal to infinite precision, by chance.
If they are equal, it is by reason of symmetry; and if symmetry does not demand that
they are equal, the probability of
them turning out to be equal to $M$ significant figures vanishes in the ${M\to\infty}$ limit.

To tighten this argument let us assume that each of the three charge densities in Fig.~\ref{fig:current-symmetry}
can be expressed as a sum ${\rho(\zeta)=\rhop(\zeta)+\rhom(\zeta)}$ of the non-negative
charge density of the nuclei ($\rhop$) and the non-positive charge density
of the electrons ($\rhom$). Let $C$ denote
the center of electron charge of the six unit cell segment shown in Fig.~\ref{fig:current-symmetry}(b);
and let ${\dot{c}(x_b)\equiv\dv*{c(x_b)}{t}}$ 
be the rate of change of the center of charge of the $\Neleccell-$electron charge density in 
unit cell ${\unitcell\equiv(x_b,x_b+a)}$.
There is no symmetry reason to expect ${\dot{c}(x_b)}$ to vanish; therefore it does not.

The rate of change ${\dot{C}\equiv\dv*{C}{t}}$ of $C$ is the average of the rates of change of the centers of 
charge of the six unit cells. Since the cells are identical, ${\dot{C}}$ is also equal to ${\dot{c}(x_b)}$; 
and the current per unit length associated with this motion of electrons
is ${-\Neleccell\,e\,\dot{c}(x_b)/a}$. To find the total current density the contribution
from nuclei should be added to it.

Now let us turn to the material in Fig.~\ref{fig:current-symmetry}~(c).
However, instead of expressing ${\dot{C}}$ as the average value of ${\dot{c}}$ 
for the six complete cells delimited
by vertical black lines, let us express it as the average value of ${\dot{c}}$ for the five complete
green-bordered cells and the remaining cell, which is shaded in grey and divided into
two pieces on either side of the five.

The centers of charge of the green cells are time-invariant, by symmetry. 
Therefore ${\dot{C}=\dot{c}_g/6}$, 
where ${c_g}$ denotes the center of electron charge of the grey cell.
Now suppose that the number of primitive cells in the segment was not six, but of order ${l/a}$, such that the segment
was mesoscopic. Then the rate of change of its center of electron charge, and therefore the current, 
would be  ${\dot{C} \sim (a/l)\,\dot{c}_g}$, which is negligible.

\subsubsection{Anistropy introduced by the stimulus}
Just as a polarization current flows when a material without inversion symmetry is uniformly stimulated,
it also flows when the crystal has inversion symmetry, but the stimulus that changes $\rho$
breaks this symmetry. This is the case when, for example, the stimulus is an applied  electric field 
or a non-uniform strain (flexoelectricity).

Crystals can be categorized based on what sorts of stimuli are capable of
causing these transient polarization currents. An electric field induces
a polarization current in any crystal, whereas only crystals that lack inversion
symmetry tend to be pyroelectric, because temperature
does not reduce crystals' symmetries, in general, so they remain inversion
symmetric as they are heated.
A larger set of crystal symmetries are compatible with piezoelectricity 
because uniform uniaxial strain can break inversion symmetry.

\section{{\em Modern Theory of Polarization} }
\label{section:mtop}
\subsection{Introduction}
\label{section:mtop_intro}
The {\em Modern Theory of Polarization} (MTOP), which was developed in the 1990s by Resta and Vanderbilt and their
collaborators, solved the problem of how to define the polarization 
current ${\Jconv}$ of an insulating material in terms of the evolving microstructure {\em of its
bulk}, and provided a widely-applicable method of 
calculating it~\citep{resta-1993,kingsmith-vanderbilt-prb-1993-1,kingsmith-vanderbilt-prb-1993-2,resta-rmp-1994,resta-vanderbilt-2007}.

I emphasize that the MTOP defines $\Jconv$ in terms of the changing {\em bulk}
microstructure because the alternative, namely defining it as ${\Jconv\equiv\bsigmadot}$
where ${\bsigma}$ is the surface charge density,
would be of much less practical use: 
Calculations of $\Jconv$ would require knowledge of the time-dependent 
surface microstructure. Since the numbers of particles in surface regions tend to be very large, 
accurate calculations of $\Jconv$ would be intractable in many or most cases.
However, the MTOP can be used to calculate the electronic contribution
to ${\Jconv}$ from a simulation that only takes explicit account of the electrons in
a small region whose microstructure is representative of the entire bulk.
This makes it possible to calculate $\Jconv$ in a wide range of materials.

In subsections~\ref{section:illustrative_derivation}
and~\ref{section:bulk_subsystem} I present illustrative
derivations of the MTOP expressions for the electronic contribution
to $\Jconv$; and in Sec.~\ref{section:current} I present an 
explanation of the MTOP definition of $\Jconv$
that is based on the relation ${\J=\bsigmadot}$ 
and on Finnis's expression for the (macroscale) excess of charge on 
a surface, ${\bsigma[\rhop,\rhom]}$, in terms of the microscopic densities of
positive ($\rhop$) and negative ($\rhom$) charge. 
These derivations make it clear that the MTOP definition of the 
electronic polarization current density ${\Jconvm}$ in an insulator whose electrons are in their 
ground state is basically exact.

\subsection{How is the polarization current density $\Jconv$ defined?}
Suppose that the microstructure of a material is in a stationary or \emph{quasi}-stationary statistical
state and that $\zeta$ is one or more parameters on which the stationary state has a continuous 
and continuously-differentiable dependence. Then, 
as discussed in Sec.~\ref{section:anisotropy}, the polarization current density, 
\begin{align*}
\Jconv  = \Jconvp+\Jconvm,
\end{align*}
is the current that flows while the
stationary-state charge density, 
\begin{align*}
\rho(\zeta)=\rhop(\zeta)+\rhom(\zeta),
\end{align*}
is changing continuously in response to $\zeta$ changing quasistatically;
where \emph{quasistatically} means so slowly that deviations of the microstructure's
statistical state from its $\zeta$-dependent stationary state are negligible;
and ${\Jconvp}$ and ${\Jconvm}$ are the contributions to $\Jconv$ from the charge
densities of the nuclei (${\rhop}$) and electrons (${\rhom}$), respectively, changing.

A changing microscopic charge density implies
the existence of a microscopic current density, 
\begin{align*}
\jconv(\rvec,t)=\jconvp(\rvec,t)+\jconvm(\rvec,t), 
\end{align*}
which satisfies the charge conservation equation, 
\begin{align*}
\div\jconv &= -\pdv{\rho}{t}  = -\dot{\zeta}\pdv{\rho}{\zeta}
\\
&= 
-\dot{\zeta}
\pdv{\rhop}{\zeta}
-\dot{\zeta}
\pdv{\rhom}{\zeta}
= \div\jconvp+\div\jconvm.
\end{align*}
The macroscopic polarization current densities $\Jconv$, $\Jconvp$, and $\Jconvm$ are the mesoscale spatial averages 
of $\jconv$, $\jconvp$, and $\jconvm$, respectively, in the material's bulk. 

\subsubsection{Contribution of the nuclei to $\Jconv$}
Let us assume that ${\rho}$ has the form of Eq.~\ref{eqn:charge_density}.
Therefore the nuclei can be treated as point charges and, if $V$ is a
region of the bulk whose microstructure is representative of the bulk, 
their contribution to the polarization current density can be calculated
from the classical expression,
\begin{align*}
\Jconvp = 
\frac{e}{\abs{V}}\sum_{\Rvec_i\in V}Z_i\dv{\Rvec_i}{t}
=
\frac{e \dot{\zeta}}{\abs{V}}\sum_{\Rvec_i\in V}Z_i\dv{\Rvec_i}{\zeta},
\end{align*}
where ${Z_i}$ and ${\Rvec_i}$ are the ${i^\text{th}}$ nucleus's 
atomic number and position, respectively, and ${\abs{V}}$ is the volume of $V$.

\subsubsection{Contribution of the electrons to ${\Jconv}$}
This subsection states one form of the MTOP expression for the electronic polarization
current density ($\Jconvm$) that flows while $\zeta$ is changing. 

The electrons' charge density at position $\rvec$ is 
\begin{align*}
\rhom(\rvec;\zeta)=-e n(\rvec;\zeta), 
\end{align*}
where ${n(\zeta)}$ is the number density of electrons, or simply the `electron density'.
Let us assume that the electron density has a ${\zeta}$-dependent decomposition, 
\begin{align*}
n(\rvec;\zeta)=  -\rhom(\rvec;\zeta)/e= \sum_i n_i(\rvec;\zeta),
\end{align*}
where each density `packet' ${n_i(\zeta)}$ 
evolves smoothly with $\zeta$, without its integral changing
and therefore without the charge,
\begin{align*}
q_i =-e\int n_i(\rvec;\zeta)\dd[3]{r}, 
\end{align*} 
that it carries changing.
In subsection~\ref{section:decomposing_the_density} I will outline how
such a decomposition can be found in practice.

Assuming that it can be found, the electronic contribution to $\Jconv$ is 
\begin{align}
\Jconvm = 
\frac{1}{\abs{V}}\sum_{\rvecsub{i}\in V} q_i \dv{\rvecsub{i}}{t}= \frac{\dot{\zeta}}{\abs{V}} \dv{\zeta} \left(\sum_{\rvecsub{i}\in V} q_i \rvecsub{i}\right),
\label{eqn:mtopJ}
\end{align}
where the sum is over all
${\rvecsub{i}}$ within a region $V$ of the bulk that is 
representative of the entire bulk, and 
\begin{align*}
\rvecsub{i}(\zeta) \equiv \frac{\intthree \rvec \,n_i(\rvec;\zeta)\dd[3]{r}}{\intthree n_i(\rvec';\zeta)\dd[3]{r'}}, 
\end{align*}
is the \emph{center} of packet ${n_i}$.
Equation~\ref{eqn:mtopJ} is the MTOP definition of $\Jconvm$.

In practice, $\Jconvm$ is almost always calculated after the ground state electron density ${n(\zeta)}$ 
has been calculated with methods based on the density 
functional theory (DFT) of Hohenberg, Kohn, and Sham~\citep{hohenberg_kohn,vanleeuwen_2003,kohn_sham,dreizler_gross_1990}.
At the end of subsection~\ref{section:mtop_intro} I described the
MTOP definition of $\Jconvm$ as `exact' for an insulator in its ground state. 
I meant two things by this. The first is that
if $V$ is \emph{exactly} representative of the bulk,
Eq.~\ref{eqn:mtopJ} is an \emph{exact} expression for ${\Jconv}$. 
The second is that a natural byproduct of the most widely used DFT-based methods
of calculating the ground state density of
an insulator is a decomposition of ${n(\zeta)}$ into packets whose
centers are in $V$ and whose integrals are independent of $\zeta$.
Equation~\ref{eqn:mtopJ} cannot be used without such a decomposition.

If the material is a crystal, $V$ 
can be a single primitive unit cell, $\Omega$. In an amorphous
material $V$ must be large enough to sample all relevant features of the microstructure 
and, by expanding it to improve the sampling of the bulk, 
${\Jconv}$ can be calculated to any desired precision.

\begin{figure}[!]
\includegraphics[width=0.48\textwidth]{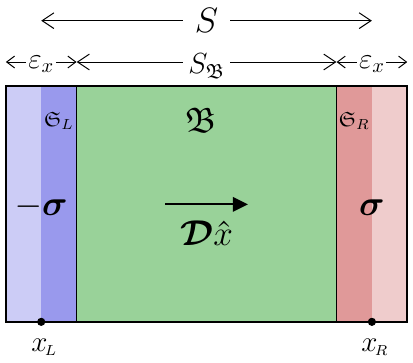} 
\caption{Schematic of a polarized three dimensional material, which complements the
schematic of the quasi-one dimensional material in Fig.~\ref{fig:material}.
The material is charge neutral, and its bulk (shaded green) is macroscopically 
charge neutral. 
The material has a finite dipole moment, ${\mbd\hat{x}}$, and
on the macroscale the surfaces with outward normals ${\hat{x}}$
and ${-\hat{x}}$ carry equal and opposite areal charge densities, $\bsigma$ and $-\bsigma$, respectively.
See Sec.~\ref{section:illustrative_derivation} for more details.}
\label{fig:crystal_dipole}
\end{figure}

\subsection{Illustrative derivation of Eq.~\ref{eqn:mtopJ}}
\label{section:illustrative_derivation}
The physical idea underpinning Eq.~\ref{eqn:mtopJ} is simple, and can be illustrated
using the isolated nonconducting material depicted in Fig.~\ref{fig:crystal_dipole}, 
which we will assume to be a crystal for the purposes of this section.

The crystal's dipole moment is $\mbd\hat{x}$ and the surfaces
perpendicular to $\hat{x}$ carry equal and opposite excesses of charge, in the form 
of areal charge densities of ${\pm\bsigma}$.
Between ${\xl}$ and ${\xr}$ 
the crystal's shape and cross-sectional area ${\abs{\plane}}$ are uniform:
The region ${\plane(x)}$ of each plane perpendicular to $\hat{x}$ that the crystal
occupies is the same for all ${x\in[\xl,\xr]}$. 

Although the width of each surface is zero at the macroscale, each macroscopically-planar surface
corresponds to a region of finite thickness $\prectheo$ at the microscale, half
of which is in vacuum and half of which is within the material. 
In Fig.~\ref{fig:crystal_dipole} 
the two surface regions
are coloured blue and red/pink, with lighter shades indicating the
parts of them that are in vacuum.
The parts that are within the material are indicated by both
darker shades and the labels ${\surfaceL}$ and ${\surfaceR}$.
The widths of the surface regions are exaggerated.
In reality $\prectheo$ would be many orders of magnitude smaller than $\size$; therefore,
\begin{align*}
\bulksize\equiv\abs{\bulk}=\size-\prectheo\approx\size.
\end{align*}

To understand Eq.~\ref{eqn:mtopJ}, let us choose the representative region of the bulk, $V$, 
to be an arbitrary unit cell in the crystal's bulk, $\unitcell$, and let
us derive the equation,
\begin{align}
\Jconvm\cdot\hat{x}
=
\left(\frac{1}{\volume}\sum_{\rvecsub{i}\in \unitcell} q_i \dv{\rvecsub{i}}{t}\right)\cdot\hat{x},
\label{eqn:mtopJ2}
\end{align}
which is the part of Eq.~\ref{eqn:mtopJ} that pertains to the $x$-component of $\Jconvm$.
Note that, in this subsection, the unit cell ${\unitcell}$ is three dimensional and ${\volume}$ is its \emph{volume}, 
but in subsection~\ref{section:bulk_subsystem}, 
in later chapters, and in Appendices~\ref{section:appendix_fourier},~\ref{section:appendix_torus}, and ~\ref{section:appendix_wannier}, 
each primitive unit cell $\unitcell$ is (quasi-) one dimensional and $\volume$ denotes its width.

Let us begin by assuming that the electron number density in the bulk can be expressed as
a superposition of ${\NUnitcell}$
identical electron densities, whose integrals are independent
of $\zeta$, such that each bulk primitive unit cell contains the center
of exactly one of them.  
Let ${\nunitcell(\zeta)}$ denote the density packet whose
center is in $\unitcell$, and which carries a total charge,
\begin{align*}
q^-= -e\intthree \nunitcell(\rvec;\zeta)\dd[3]{r}.
\end{align*}
Now let us choose an arbitrary point in unit cell $\unitcell$ to be the origin, 
which means that
the density in the bulk is
\begin{align*}
n(\rvec;\zeta) = \sum_{\Avec\in\bulk} \nunitcell(\rvec-\Avec;\zeta),\;\forall \rvec\in\bulk,
\end{align*}
where the sum is over all distinct lattice vectors ${\Avec}$ such that
the point displaced from the origin by $\Avec$ is in ${\bulk}$.

The charge neutrality of the crystal's bulk implies that 
${\rhocell(\zeta)}$, defined as the sum of 
${\rhomcell(\zeta)=-e\,\nunitcell(\zeta)}$
and the delta charge distribution of the nuclei in 
the cell $\unitcell$, 
is a charge neutral distribution
with a well defined (i.e., origin independent) dipole moment. 
The $x$-component of this dipole moment will be denoted by $\dipcell$. 

When the sum 
of all $\Nunitcell$ bulk unit cells' charge densities is
subtracted from the charge density of the entire crystal, $\rho$, the result is
\begin{align*}
\Drho(\rvec;\zeta)=\rho(\rvec;\zeta)-
\sum_{\Avec\in\bulk} \rhocell(\rvec-\Avec;\zeta).
\end{align*}
This vanishes everywhere in the bulk, but not everywhere in the 
surface regions, ${\surfaceL}$ and ${\surfaceR}$. 
Therefore if $\Drhobar(x;\zeta)$ denotes the average of ${\Drho(\rvec;\zeta)=\Drho(x,\svec;\zeta)}$ on the plane
perpendicular to $\hat{x}$ at $x$, 
the areal charge density,
\begin{align*}
\sigma_s(\zeta)
\equiv
-\int_{-\eta}^{\eta}\Drhobar(\xl+u;\zeta)\dd{u}=
\int_{-\eta}^{\eta}\Drhobar(\xr+u;\zeta)\dd{u},
\end{align*}
is finite and independent of $\eta$ as long as ${\prectheo/2<\eta\ll\bulksize}$.
The integrals of ${\Drhobar}$ across the left and right surfaces are equal in magnitude
and opposite in sign because we have assumed that the material is charge-neutral overall.
Note, however, that ${\bsigma}$ does not equal ${\sigma_s}$ unless ${\dipcell=0}$. This will be 
explained and discussed in Sec.\ref{section:excess_fields} and Sec.~\ref{section:surface_charge}.

The volume of the crystal can be expressed as
\begin{align}
\size\abs{\plane}=\left(\bulksize+\prectheo\right)\abs{\plane}
=\NUnitcell\volume+\prectheo\abs{\plane}, 
\label{eqn:crystal_volume}
\end{align}
where
the second term on the right hand side 
is much smaller than the first, because ${\NUnitcell\volume=\abs{\plane}\bulksize}$;
which also means that 
\begin{align*}
\prectheo\abs{\plane}/(\NUnitcell\volume)=\prectheo/\bulksize \sim a/S < a/L \ll a/l \ll 1. 
\end{align*}

The $x$-component of the crystal's dipole moment can be expressed as
\begin{align*}
\mbd=\NUnitcell \dipcell + \sigma_s \abs{\plane}(\NUnitcell\volume/\abs{\plane}+\prectheo),
\end{align*}
where the first term on the right hand side is the sum of the $x$-components of the 
dipole moments of the bulk unit cells; and the second term is the product of the net 
charge ${\sigma_s\abs{\plane}}$ in ${\surfaceR}$ and the distance between the surfaces.
It follows that the average, over the entire crystal, of the $x$-component
of the current density flowing within it is
\begin{align}
\Jconv\cdot\hat{x}& =\bsigmadot\cdot\hat{x}
=\mbpdot\cdot\hat{x}
=\frac{\mbddot}{\NUnitcell\volume+\prectheo\abs{\plane}}   
\nonumber
\\
& =\frac{\NUnitcell \dipcelldot + \dot{\sigma}_s (\NUnitcell\volume+\prectheo\abs{\plane})}{\NUnitcell\volume+\prectheo\abs{\plane}}   
\nonumber
\\
&= \frac{\dipcelldot}{\volume} + \dot{\sigma}_s + \order{a/S} 
 = \frac{\dipcelldot}{\volume} + \order{a/S},
\label{eqn:mtop01}
\end{align}
where ${\mbp\equiv \mbd/(\NUnitcell\volume+\prectheo\abs{\plane})}$ is the $x$-component
of the crystal's dipole
moment divided by the crystal's volume; and $\dot{\sigma}_s$ vanishes
because we have assumed that there is no conduction current and that the crystal is isolated.

The term of order ${a/S}$ on the right hand side of 
Eq.~\ref{eqn:mtop01} is negligible.
Therefore, 
\begin{align*}
\Jconv&\cdot\hat{x} 
= \frac{\dipcelldot}{\volume}
\\
&=\frac{1}{\volume}\dv{t}\left(-e \intthree \rvec \nunitcell(\rvec;\zeta)\dd[3]{r}  + e\sum_{\Rvecsub{i}\in\Omega} Z_i\Rvecsub{i}\right)\cdot\hat{x}
\\
&=\underbrace{\vphantom{\sum_{\Rvecsub{i}\in\Omega}}\frac{q^-}{\volume}  \dv{\rvecsub{\unitcell}}{t}\cdot\hat{x}}_{\displaystyle \Jconvm\cdot\hat{x}} 
+ \underbrace{\frac{e}{\volume} \sum_{\Rvecsub{i}\in\Omega} Z_i\dv{\Rvecsub{i}}{t}\cdot\hat{x}}_{\displaystyle \Jconvp\cdot\hat{x}}.
\end{align*}
The electronic contribution to the polarization current is
\begin{align}
\Jconvm\cdot\hat{x} &= \frac{1}{\volume}  q^- \dv{\rvecsub{\unitcell}}{t},
\label{eqn:mtop02}
\end{align}
which has the same form as Eq.~\ref{eqn:mtopJ2} when only one density packet has its center
in $\Omega$.

The derivation of Eq.~\ref{eqn:mtop02} neglected contributions to
${\mbddot}$ from redistributions of charge within the surface regions ${\surfaceL}$ and ${\surfaceR}$, 
because changes of \emph{surface} polarization should not be counted as contributing
to the \emph{bulk} polarization current, and because these surface contributions
vanish in the limit ${\prectheo/\bulksize\to 0}$.

\subsubsection{Merging and subdividing density packets}
Now let us suppose that density packet 
${\nunitcell(\zeta)}$ can be expressed as
\begin{align*}
\nunitcell(\rvec;\zeta) =  \sum_{i} n_i(\rvec;\zeta),
\end{align*}
where packet ${n_i(\zeta)}$ is centered at
${\rvecsub{i}(\zeta)}$ and carries a charge, 
\begin{align*}
q_i^- = -e\intthree n_i(\rvec;\zeta)\dd[3]{r},
\end{align*}
that is independent of $\zeta$.
Then,
\begin{align*}
q^- \rvecsub{0}(\zeta) 
&= -e\intthree \rvec\,\nunitcell(\rvec;\zeta)\dd[3]{r}
\\
&= -e\sum_i \intthree \rvec\,n_i(\rvec;\zeta)\dd[3]{r}
= \sum_i q^-_i \,\rvecsub{i}(\zeta).
\end{align*}
Therefore Eq.~\ref{eqn:mtop02} can be expressed as
\begin{align}
\Jconvm\cdot\hat{x} &= \left(\frac{1}{\volume}\sum_i q_i^-\dv{\rvecsub{i}}{t}\right)\cdot\hat{x}.
\label{eqn:mtop03}
\end{align}
This  demonstrates that decomposing density packets into smaller packets does not
change ${\Jconv}$ as long as the smaller packets'
integrals are independent of ${\zeta}$.
It can also be shown that ${\Jconv}$ is invariant
under mergers of multiple representative regions $V$ into
larger representative regions, and under mergers of multiple
$\zeta$-independent density packets into larger density packets.

Note that Eq.~\ref{eqn:mtop03} is identical to Eq.~\ref{eqn:mtopJ2} if the centers of the packets ${n_i(\zeta)}$ are
all in cell $\unitcell$. 
However if, for example, the center ${\rvecsub{j}}$ of ${n_j(\zeta)}$ is not in $\unitcell$, 
there exists an image ${\nunitcell(\rvec-\Avec;\zeta)}$ of ${\nunitcell(\rvec;\zeta)}$,
under a translation by a lattice vector $\Avec$, whose counterpart ${n_j(\rvec-\Avec;\zeta)}$ 
of ${n_j(\rvec;\zeta)}$ has its center in $\unitcell$. Therefore, Eq.~\ref{eqn:mtopJ2} could
be derived by repeating the derivation that led to Eq.~\ref{eqn:mtop02} with the definition
of ${\rhomcell(\zeta)}$ changed to
\begin{align*}
\rhomcell(\rvec;\zeta) \equiv -e\sum_{\Avec\in\bulk}\sum_{\Avec+\rvecsub{i}\in\unitcell} n_i(\rvec-\Avec;\zeta),
\end{align*}
i.e.,  by defining it as the sum of all of the smaller charge density packets 
whose centers are in $\unitcell$.

\subsection{Decomposing particle number densities into `packets'}
\label{section:decomposing_the_density}
It should be clear from the derivation above
that the partitioning of the electron density into packets (e.g., $\nunitcell$)
does not constitute an approximation; and that, if the density can be decomposed
into ${\zeta}$-dependent packets with $\zeta$-independent integrals,
and if region $V$ represents the bulk microstructure exactly,
Eq.~\ref{eqn:mtopJ} is an exact expression for $\Jconv$. 

\subsubsection{Applying the MTOP to electrons in their ground state}
\label{section:applying_the_mtop}
The derivation above does not assume that the material
is an insulator or that the electrons are in their ground state, 
but both of these assumptions are made when 
the MTOP is used to calculate $\Jconvm$.
It is also assumed that the electrons' \emph{external potential} ${\vext(\zeta)}$
is changing smoothly and quasistatically with $\zeta$. 
The external potential
${\vext(\rvec;\zeta)\in\realone}$ is the energy
of interaction of an electron at $\rvec$ with the quasistatic
nuclei and any externally-applied quasistatic fields~\citep{kaxiras_2019,vanleeuwen_2003}.

In most applications of the MTOP, the ground state density is calculated using the density functional 
theory (DFT) of Hohenberg, Kohn, and Sham~\citep{hohenberg_kohn,vanleeuwen_2003,kohn_sham,dreizler_gross_1990}.
In most DFT calculations the ground state density is calculated as
\begin{align}
n(\rvec;\zeta) = \sum_{i=1}^\infty f_i\abs{\psi_i(\rvec;\zeta)}^2,
\label{eqn:density_decomposition1}
\end{align}
where the infinite orthonormal set,
\begin{align*}
\densityset\big(\zeta;\hamfrakr\big)
\equiv
\bigg\{\psi_i(\zeta)\in&\lebesgue(\realone^3): \;1\leq i\leq \infty, 
\\
\hamfrakr\psi_i=\epsilon_i\psi_i,\;&\int_{\realone^3}\psi^*_i(\rvec;\zeta)\psi_j(\rvec;\zeta)\dd[3]{r}=\delta_{ij}\bigg\},
\end{align*}
is the set of eigenfunctions of the Hamiltonian,
\begin{align*}
\hamfrakr(\zeta):\lebesgue(\realone^3)\to\lebesgue(\realone^3),
\end{align*}
of a fictitious system of non-interacting electrons that has
the same ground state density ${n(\zeta)}$ as the real system of mutually-repulsive electrons;
and the eigenfunctions'
ground state `occupation numbers', $f_i$, satisfy,
\begin{align*}
\sum_{i=1}^\infty f_i = \Nelec, 
\end{align*}
where ${\Nelec}$ is the total number of electrons.

Let us assume that the eigenfunctions $\psi_i$ are indexed in order of increasing 
eigenvalue, $\epsilon_i$, i.e., such that,
\begin{align*}
i\leq j \iff \epsilon_i\leq\epsilon_j.
\end{align*}
Then, in an insulator, Eq.~\ref{eqn:density_decomposition1} simplifies to
\begin{align}
n(\rvec;\zeta) = 2\sum_{i=1}^{\Nelec/2} \abs{\psi_i(\rvec;\zeta)}^2 = \sum_{i=1}^{\Nelec/2} n_i(\rvec;\zeta), 
\label{eqn:density_decomposition2}
\end{align}
where it is assumed that each eigenstate is either empty (${f_i=0}$) or occupied by two electrons (${f_i=2}$)
with opposite spins, and that occupation numbers do not change as $\zeta$ changes.
Therefore the integral of each density `packet' $n_i(\zeta)$ is two.

The assumption that $\zeta$ and ${\vext(\zeta)}$ change \emph{quasistatically}
and the assumption that the material has a finite band gap, ${\bandgap>0}$, are necessary
for the assumption that electrons are in their ground state while the polarization 
current is flowing to be valid. 
The assumption that the material is an insulator,
rather than a semiconductor with a very small band gap, ${0<\bandgap\approx 0}$,  is necessary
for the assumption that electrons are in their ground
state to be approximately true at all temperatures that are 
low relative to ${\bandgap/k_B}$ (e.g., ${\SI{1}{\electronvolt}/k_B\approx \SI{11605}{\kelvin}}$);
and a large gap ensures that the charge of ${-2e}$ carried 
by each smoothly-evolving packet of density, ${\abs{\psi_i(\zeta)}^2}$, does not change as its center moves.

It is known that any insulating material's ground
state electron density ${n(\rvec;\zeta)}$ can be approximated
arbitrarily closely by Eq.~\ref{eqn:density_decomposition2}
for \emph{some} $1$-electron Hamiltonian of the form, 
\begin{align*}
\hamfrakr(\rvec;\zeta) = \hat{t} + \vext(\rvec;\zeta) 
+ \Delta v(\rvec;\zeta), 
\end{align*}
where ${\hat{t}}$ is the $1$-electron kinetic energy
operator; and ${\Delta v(\zeta)}$ adjusts 
${\vext(\zeta)}$ to account exactly, or arbitrarily closely to exactly, 
for treating the electrons as if they do not interact with one another~\citep{kohn_sham,vanleeuwen_2003,dreizler_gross_1990}.
A ground state electron density that is also the ground state 
density of a fictitious set of noninteracting electrons, which are confined by a potential
${v=\vext+\Delta v}$, it is said to be \emph{noninteracting $v$-representable}.

Not only is it known that \emph{in principle} every ground state electron density
is arbitrarily close to a noninteracting $v$-representable density, it is known
how to approximate ${\hamfrakr(\rvec;\zeta)}$ \emph{in practice}
~\citep{hohenberg_kohn,vanleeuwen_2003,kohn_sham,dreizler_gross_1990}.
Therefore a set ${\{n_i(\zeta)\}}$ of $\zeta$-dependent density packets with $\zeta$-independent integrals
can be calculated using one of many widely available and widely used software packages, 
such as Quantum ESPRESSO~\citep{espresso} or ABINIT~\citep{abinit}, both of which 
calculate ${\Jconvm}$ if fed the appropriate keywords and parameters.
These methods and codes approximate the ground state density, rather than calculating
it exactly, but the MTOP method of calculating $\Jconv$ from set ${\{n_i(\zeta)\}}$
does not, in principle, introduce further approximations: Equation~\ref{eqn:mtopJ} is an exact
expression if region $V$ represents the bulk microstructure exactly. 

Most calculations of set ${\{n_i(\zeta)\}}$ use {\em Born-von K\'arm\'an} boundary conditions, which will be discussed in subsection~\ref{section:bulk_subsystem},
and which does constitute an approximation, albeit one that seems likely to be accurate.

\subsubsection{Decomposing an arbitrary stationary state number density}
\label{section:decompose}
Although the integral of each density packet tends to be one or two 
in standard applications of the MTOP to electrons, it is clear from the derivation of Eq.~\ref{eqn:mtopJ2} above, and it will be clear from 
the derivation of Eq.~\ref{eqn:mtopJ} in Secs.~\ref{section:discrete} and~\ref{section:current}, 
that any method of partitioning a stationary state number density ${n(\zeta)}$ 
into $\zeta$-dependent packets {\em of fixed charge}
would give the same result, regardless of whether or not the
packets' charges are integer multiples of ${e}$. 

Furthermore, neither the derivation above nor
the derivation in Secs.~\ref{section:discrete} and~\ref{section:current},
which is based on the homogenization theory of
Secs.~\ref{section:homogenization} and~\ref{section:excess_fields}, 
require the particles whose number density is ${n(\zeta)}$ to be \emph{electrons}, 
or require ${n(\zeta)}$ to be a \emph{ground state} number density.
There is only an implicit assumption that ${n(\zeta)}$ is the number density
of a set of particles that are in a \emph{stationary} statistical state.

To my knowledge, it is not currently known how to decompose an arbitrary 
stationary-state number density into $\zeta$-dependent packets with $\zeta$-dependent integrals;
and it is not even known which densities admit such a decomposition \emph{in principle}.
However the literature on non-interacting $v$-representability
of electron number densities does clarify two important issues, albeit not necessarily explicitly
~\citep{vanleeuwen_2003,engel_and_dreizler,Corso_2025,Sutter_2024,v-rep-review_2023}:
First, there exists a broad range of particle number densities for which
such a decomposition is possible \emph{in principle}; and second, it is not necessary for
a number density to be a property of a \emph{quantum mechanical} stationary 
state for it to admit such a decomposition.

Therefore, if the number densities of other kinds
of charged particles in other kinds of physical systems can be decomposed
\emph{in practice},
Eq.~\ref{eqn:mtopJ} is more generally applicable than its developers 
claimed it to be in at least some of their presentations of it~\citep{\mtop}.

\subsubsection{Transforming between density decompositions}
Here, and in the remainder of this work, particles' spins, and the
spin-degeneracies of eigenstates, will be ignored. 
It will be assumed that an insulator
in its electronic ground state has ${\Nelec}$ singly-occupied eigenstates ${\ket{\psi_i(\zeta)}}$ of
the operator,
\begin{align*}
\hamfrak(\zeta)\equiv \intthree\dd[3]{r} \hamfrakr(\rvec;\zeta)\dyad{\rvec},
\end{align*}
which span a ${\Nelec-}$dimensional Hilbert space
$\hilbert_\zeta$ that changes smoothly with $\zeta$. 
This means that, as $\zeta$ changes, ${\hilbert_\zeta}$ rotates within the infinite-dimensional
Hilbert-Lebesgue space, ${\lebesgue(\realone^{\dimension \Nelec})}$, of which it is a subspace, 
where ${\dimension}$ is the dimensionality of the system. 
We will return to this concept in Sec.~\ref{section:H-representability}, and it is illustrated
schematically for ${\hilbert_\zeta\equiv\Span\left\{\ket{\varphi_1},\ket{\varphi_2},\ket{\varphi_3}\right\}}$
in Fig.~\ref{fig:rotating-vectors}.

The charge density packet ${-e\abs{\psi_i(\rvec;\zeta)}^2}$ of 
each occupied eigenfunction ${\psi_i(\rvec;\zeta)\equiv\braket{\rvec}{\psi_i(\zeta)}}$
of ${\hamfrakr(\rvec;\zeta)}$ carries a charge of ${-e}$
and contributes to the polarization current via
the motion of its center ${\rvecsub{i}(\zeta)}$ caused by ${\hilbert_\zeta}$'s rotation.

The remainder of Sec.~\ref{section:mtop}
will focus on the case of a (quasi-)one dimensional material.
Therefore ${\psi_i(\zeta)=\psi_i(x;\zeta)}$ is an eigenfunction of
\begin{align*}
\hamfrakx:\lebesgue(\realone)\to\lebesgue(\realone), 
\end{align*}
and  the center of $\psi_i$ will be denoted
by $x_i$ instead of $\rvecsub{i}$. 
In Sec.~\ref{section:bulk_subsystem}, 
it will be assumed that the bulk of the material is in a torus, and 
${\varphi_i(\zeta)}$
will denote the counterpart ${\psi_i(\zeta)}$ in that context.
In Sec.~\ref{section:single_particle_states}, Appendix~\ref{section:appendix_torus}, 
and Appendix~\ref{section:appendix_wannier} the counterpart
of ${\hamfrakx(\zeta)}$ in $\onetorus$ will be denoted by ${\hamsmallx(\zeta)}$.

As will be discussed in Sec.~\ref{section:current} and in the caption of Fig.~\ref{fig:rotating-vectors}, 
it is the space ${\hilbert_\zeta}$ that determines ${n(\zeta)}$, not its basis
of eigenstates. Therefore we are free to use a different basis to calculate
${\Jconvm}$; and it is common to transform the set ${\{\psi_i(\zeta)\}_{i=1}^{\Nelec}}$ to
one for which the density packets ${\{\abs{\psi_i(\zeta)}^2\}_{i=1}^{\Nelec}}$
are highly localized. However, this step is unnecessary, in principle:
the centers of the (delocalized) eigenstates can be used directly in the equation, 
\begin{align}
\Jconvm = -\frac{e}{\abs{V}}\sum_{x_i\in V} \dot{x}_i = -\frac{e\dot{\zeta}}{\abs{V}}\sum_{x_i\in V}\dv{x_i}{\zeta}
\label{eqn:mtopJx}
\end{align}
which is the one dimensional version of Eq.~\ref{eqn:mtopJ} with ${q_i=-e}$, 
and where $V$ represents a one dimensional region of the bulk of \emph{width} ${\abs{V}}$.

An alternative is to Fourier transform
the eigenfunctions with respect to their position arguments, 
and to compute each $x_i$ as the expectation value of
the Fourier transformed position operator ${\hat{x}^\ksuper\equiv i\!\pdv*{k}}$.
For example, the center of ${n_i(x)=\abs{\psi_i(x)}^2}$, when
${\braket{\psi_i}=1}$,  is
\begin{align*}
x_i &\equiv \int\dd{x} n_i(x) x  = \int \dd{x} \psi_i^*(x)\left(x \psi_i(x)\right) \\
& = \fourierconst^2\!\int \dd{x} \int \dd{k} 
e^{ikx}\fourierspace{x \psi_i}(k)
\int\dd{q} \fts{\psi}_i^*(k+q) e^{-i(k+q)x}\\
& = \fourierconst^2\!\int\dd{k} i\pdv{\fts{\psi}_i(k)}{k}\int\dd{q}\left(\int\dd{x} e^{-iqx}\right) \fts{\psi}_i^*(k+q) \\
& =  \int\dd{k} \fts{\psi}_i^*(k)\left(i\pdv{k}\right)\fts{\psi}_i(k)
 =  \int\dd{k} \fts{\psi}_i^*(k)\,\hat{x}^\ksuper\,\fts{\psi}_i(k), 
\end{align*}
where ${\fourierconst\equiv 1/\sqrt{2\pi}}$.
Despite my use of suggestive notation, there is no quantum mechanical content in these
lines of mathematics.
\emph{Any} classical probability density ${n_i(x)}$ can be expressed as
the square modulus of a Hilbert space vector, 
\begin{align*}
\psi_i(x)\equiv
\sqrt{n_i(x)}e^{i\theta(x)}\in\lebesgue(\realone).
\end{align*}
The phase ${\theta(x)}$ is often irrelevant
and arbitrary when studying the stationary states of an isolated system, 
but it becomes important when dividing a system into subsystems 
(e.g., nuclei and electrons or surface and bulk)
or when studying the time dependence of a nonequilibrium probability density.

\subsection{The bulk subsystem}
\label{section:bulk_subsystem}
In theoretical solid state physics it is common to study
a material's \emph{bulk subsystem}, meaning the material without its surfaces, 
by placing it in a torus. 
The bulk of a 1-d material would be represented in a 1-torus $\onetorus$, whose circumference
is the width, ${\bulksize}$, of the material's bulk. 
As discussed in Appendix~\ref{section:appendix_torus}, representing bulk wavefunctions in $\onetorus$ is equivalent to 
enforcing $\bulksize$-periodic {\em Born-von K\'arm\'an} boundary conditions on wavefunctions in ${\realone}$~\citep{born_von-karman}.

In textbooks on solid state physics, Born-von K\'arm\'an boundary conditions are usually
discussed in the context of crystalline materials~\citep{cohen_louie,ashcroft_mermin_book,ibach_and_luth,kittel}, but 
in computational research they
are also commonly used for amorphous materials~\citep{trave_2002,pasquarello_1992,tangney_silica_2002}.
The bulks of amorphous materials can be represented in $\onetorus$
by ensuring that $\bulksize$ is large enough to contain a representative sample of the material, where
`representative' means that the sample's microstructure shares, or approximates,
whatever statistical characteristics of the true material 
are relevant to the properties of it that are being studied.

The purpose of this subsection is to derive the MTOP expression for 
${\Jconvm}$ in an insulator in terms of a set ${\varphiset}$ of smoothly-evolving functions, which 
are $\bulksize$-periodic and therefore respect the Born-von K\'arm\'an boundary conditions.

\subsubsection{Calculating $\Jconvm$ in $\onetorus$}
\label{section:bvkany}
This subsection derives an expression for ${\Jconvm}$ in 
the limit ${\bulksize\to\infty}$ by assuming the existence of a set 
of functions,
\begin{align*}
\varphiset\equiv
\left\{\varphi_i(\zeta)\in\lebesgue(\onetorus): 1\leq i \leq \Nelecbulk
\right\},
\end{align*}
whose elements vary smoothly with $\zeta$ while 
reproducing the bulk electron density ${n(\zeta)}$ via 
\begin{align*}
n(x;\zeta)=\sum_{i=1}^{\Nelecbulk} \abs{\varphi_i(x;\zeta)}^2
,\;\;\forall x\in\onetorus;
\end{align*}
and while satisfying the orthonormality condition, 
\begin{align}
\braketT{\varphi_i(\zeta)}{\varphi_j(\zeta)}\equiv
\int_\onetorus\varphi_i^*(x;\zeta)\varphi_j(x;\zeta)\dd{x}=\delta_{ij},
\label{eqn:orthonormality}
\end{align}
for all ${i,j \in\{1\cdots\Nelecbulk\}}$, 
where ${\int_\onetorus \equiv \int_{x_0}^{x_0+\bulksize}}$ for any ${x_0\in\onetorus}$.

In a crystal the elements of ${\{\varphi_i(\zeta)\}}$ are usually
\emph{Bloch functions}, which are discussed in Appendix~\ref{section:appendix_torus}.
However, before we assume that ${n(\zeta)}$ is $\volume$-periodic, we will 
derive an expression for $\Jconvm$ in the more general case of a density
${n(\zeta)}$ that is ${\bulksize}$-periodic but not $\volume$-periodic.

The $\bulksize$-periodicity of $\onetorus$ allows each element of $\varphiset$ to be expressed
as a Fourier series,
\begin{align}
\varphi_i(x;\zeta)  \equiv \sum_{g\in \reciplattg} \ftsvarphi_i(g;\zeta) e^{igx}, 
\label{eqn:varphi1}
\end{align}
where $\reciplattg$ is the set of all wavevectors that are integer multiples
of ${\hbulksize\equiv 2\pi/\bulksize}$ (see Appendix~\ref{section:appendix_fourier})
and
\begin{align}
\ftsvarphi(g;\zeta)   \equiv \frac{1}{\bulksize}\int_\onetorus \varphi_i(x;\zeta) e^{-ig x}\dd{x}.
\nonumber
\end{align}

It follows from Eq.~\ref{eqn:orthonormality} that the normalization of ${\ftsvarphi_i}$ is 
\begin{align*}
\sum_g \abs{\ftsvarphi_i(g;\zeta)}^2=\frac{1}{\bulksize}, 
\end{align*}
where ${\sum_g}$ is an abbreviation of ${\sum_{g\in\reciplattg}}$.

Equation~\ref{eqn:varphi1} can be used to express 
the center of ${\varphi_i(\zeta)}$ as
\begin{align}
x_i(\zeta) &\equiv 
\int_\onetorus x \,\abs{\varphi_i(x;\zeta)}^2 \dd{x} 
\nonumber
\\
& =
\sum_{g}\ftsvarphi^*_i(g;\zeta)\sum_{g'}\ftsvarphi_i(g+g';\zeta)\int_\onetorus x e^{ig'x}\dd{x}.
\label{eqn:centerg}
\end{align}
Now let us assume that ${x_i(\zeta)}$ is being calculated
in the ${\bulksize\to \infty}$ limit, which is 
the limit in which $\hbulksize$, the smallest difference between
elements of $\reciplattg$, vanishes. 
In this limit, 
\begin{align}
\int_\onetorus x e^{i g' x}\dd{x}
&= -\frac{i}{\hbulksize}\int_\onetorus
\left(e^{i(g'+\hbulksize)x}-e^{ig'x}\right)\dd{x}.
\label{eqn:xeint}
\end{align}
The integrals ${\int_\onetorus e^{i(g'+\hbulksize)x}\dd{x}}$
and ${\int_\onetorus e^{ig'x}\dd{x}}$ 
vanish for all values of
${g'}$ in $\reciplattg$
except ${g'=-\hbulksize}$ and ${g'=0}$,
respectively, in which cases their values
are ${\int_\onetorus\dd{x}=\bulksize}$.
Therefore when Eq.~\ref{eqn:xeint} is substituted
into Eq.~\ref{eqn:centerg}, it simplifies to 
\begin{align*}
x_i(\zeta) 
& =
-i\sum_{g}\ftsvarphi^*_i(g;\zeta)\left(\frac{\ftsvarphi_i(g-\hbulksize;\zeta)-\ftsvarphi(g;\zeta)}{\hbulksize}\right)
\\
& = i\sum_g \ftsvarphi^*_i(g;\zeta)\partial_g\ftsvarphi_i(g;\zeta).
\end{align*}
In the limit ${\bulksize\to \infty}$ the sum over $g$ can be expressed as an integral 
over the set of all real wavenumbers, ${\hrealone}$; i.e., 
\begin{align*}
\sum_g=\frac{1}{\hbulksize}\sum_g \hbulksize \to \frac{\bulksize}{2\pi}\int_{\hrealone} \dd{g}.
\end{align*}
Therefore,
\begin{align}
x_i(\zeta) & = i\frac{\bulksize}{2\pi}\int_{\hrealone} \ftsvarphi^*_i(g;\zeta)\partial_g\ftsvarphi_i(g;\zeta)  \dd{g}
\nonumber
\\
\implies \dv{x_i}{\zeta} & =  i\frac{\bulksize}{2\pi}\int_{\hrealone}\bigg[\partial_\zeta\ftsvarphi^*_i(g;\zeta)\partial_g\ftsvarphi_i(g;\zeta)
\nonumber
 \\
&\qquad\qquad\qquad + \ftsvarphi^*_i(g;\zeta)\partial_\zeta\partial_g\ftsvarphi_i(g;\zeta)\bigg]\dd{g} 
\nonumber
\end{align}
If the second term in the integrand is integrated by parts, after
swapping the order of the derivatives with respect to $\zeta$ and $g$, 
the boundary term
vanishes because ${\lim_{g\to \infty} \varphi_i(\pm g;\zeta)=0}$.
Therefore we are left with
\begin{align}
\dv{x_i}{\zeta} 
& = -\frac{\bulksize}{2\pi} \times 2\im\left\{\int_{\hrealone}  \partial_\zeta\ftsvarphi^*_i(g;\zeta)\partial_g\ftsvarphi_i(g;\zeta)\dd{g}\right\}.
\label{eqn:dcenter}
\end{align}
By substituting this expression into Eq.~\ref{eqn:mtopJx} and choosing $V$ to be the entire
bulk (i.e., ${[x_0,x_0+\bulksize]}$, where ${x_0\in\onetorus}$ is arbitrary), we find that 
\begin{align}
\Jconvm
& =  \frac{e\dot{\zeta}}{\pi} \sum_i \im 
\left\{ \int_{\hrealone}  \partial_\zeta\ftsvarphi^*_i(g;\zeta)\partial_g\ftsvarphi_i(g;\zeta)\dd{g}\right\}.
\label{eqn:mtopJtwo}
\end{align}

\subsubsection{Calculating $\Jconv$ in $\onetorus$ for a crystal}
\label{section:mtopJ_crystal}
Now let us assume that the material whose bulk is represented in $\onetorus$ is a crystal
that comprises ${\NUnitcell}$ unit cells of widths $\volume$.
Then ${\bulksize=\NUnitcell\volume}$ and ${\Nelecbulk=\NUnitcell\Neleccell}$, 
where ${\Neleccell}$ is the number of electrons per unit cell. 

Since ${n(\zeta)}$ is ${\volume}$-periodic, 
$\varphiset$ can be chosen to be closed under translations by lattice vectors. In other words, 
if $\tvarphi(x;\zeta)$ is any element of $\varphiset$, and if $m$ is any integer, 
then the image ${\tvarphi(x-m\volume;\zeta)}$ of  ${\tvarphi(x;\zeta)}$ under
translation by the lattice vector ${m\volume}$ is also an element of ${\varphiset}$.
If ${m}$ is an integer multiple of $\Nunitcell$, translating ${\tvarphi(\zeta)}$ 
by ${m\volume}$ maps it onto itself; otherwise it maps it onto an identical 
function whose center is in a different unit cell.

It follows that $\varphiset$ is the union, 
\begin{align}
\varphiset = \bigcup_{m=0}^{\Nunitcell-1}
\left\{\varphi_{i}(x-m\volume;\zeta):1\leq i \leq\Neleccell \right\},
\label{eqn:varphi_subset}
\end{align}
of ${\Nunitcell}$ subsets, where the indices have been chosen such
that the center of every element of 
\begin{align*}
\varphisetcell\equiv
\left\{\varphi_i(\zeta):1\leq i \leq \Neleccell\right\}\subset\varphiset
\end{align*}
is in a particular unit cell, $\unitcell$.
Therefore the centers of all elements of each subset in the union 
on the right hand side of Eq.~\ref{eqn:varphi_subset} are in the 
same cell; and any one of the subsets can be transformed into any other one 
by translating all of its elements by the same lattice vector. 

Equation~\ref{eqn:varphi_subset} implies that subset ${\varphisetcell}$
can represent set $\varphiset$ exactly, and the electronic structure of the
bulk of the crystal exactly.
Therefore the sum in Eq.~\ref{eqn:mtopJtwo} can be expressed as ${\Nunitcell}$ times
the sum over the $\Neleccell$ elements of ${\varphisetcell}$. 

It is shown in Appendix~\ref{section:appendix_torus} that each element
of ${\varphisetcell}$ is a \emph{Bloch function} ${\bloch_{\alpha k}(\zeta)}$, which 
is associated with a particular element $k$ of
the subset ${\BZ}$ of ${\reciplattg}$ whose elements are in the first Brillouin zone
(see Appendix~\ref{section:appendix_fourier}). Therefore, let us replace
index ${i}$ with ${k\alpha}$, where $\alpha$ distinguishes between
different elements of ${\varphisetcell}$ that are associated with the same
element of $\BZ$. Let us also make all dependences on $\zeta$ implicit, instead
of denoting them implicitly.

Then Eq.~\ref{eqn:mtopJtwo} can be expressed as 
\begin{align}
\Jconvm
& =  \frac{e\dot{\zeta}}{\pi} \sum_{\alpha k} \im 
\left\{ \int_{\hrealone}  \partial_\zeta\ftsbloch^*_{\alpha k}(g)\partial_g\ftsbloch_{\alpha k}(g)\dd{g}\right\},
\label{eqn:mtopJthree}
\end{align}
where ${\ftsbloch_{\alpha k}}$ is the Fourier transform of Bloch function ${\bloch_{\alpha k}}$,
and where $\sum_i$  has been replaced by
${
\Nunitcell\sum_{\alpha=1}^{\Neleccell}\sum_{k\in\BZ}
}$. The factor ${\Nunitcell}$ disappears because $V$ is now
a single unit cell, which is ${1/\Nunitcell}$ times the
width of the entire bulk.

Bloch functions have the form 
\begin{align}
\bloch_{\alpha k}(x)=e^{ikx}\pbloch_{\alpha k}(x),
\label{eqn:bloch_local}
\end{align}
where ${\pbloch_{\alpha k}}$ is $\volume$-periodic. This implies that
\begin{align*}
\ftsvarphi_i(g)=\fts{\bloch}_{\alpha k}(g) = \fts{\pbloch}_{\alpha k}(g-k), 
\end{align*}
where ${\fts{\pbloch}_{\alpha k}}$ is the Fourier transform of ${\pbloch_{\alpha k}}$.
The $\volume$-periodicity of ${u_{\alpha k}}$ implies that
its Fourier transform ${\ftspbloch_{\alpha k}(g)}$ vanishes unless
${g}$ is a reciprocal lattice vector, which means that
 ${\fts{\bloch}_{\alpha k}(g)}$ 
vanishes unless ${g-k}$ is a reciprocal lattice vector.
Therefore if
${\int_{\hrealone}\dd{g}=\hbulksize\sum_g}$ 
is expressed as 
${\sum_{G}\int_{\BZ}\dd{k'}}$, where ${\sum_G=\sum_{G\in\reciplatt}}$
is a sum over all reciprocal lattice vectors 
and ${\int_{\BZ}\dd{k'}= \hbulksize\sum_{k'\in\BZ}}$, 
we find that the integral in Eq.~\ref{eqn:mtopJthree} can be
expressed as
\begin{align*}
\int_{\hrealone} & \partial_\zeta\ftsbloch^*_{\alpha k}(g)\partial_g\ftsbloch_{\alpha k}(g)\dd{g}
\\
&= \sum_G\int_{\BZ}\dd{k'} \partial_\zeta\ftsbloch^*_{\alpha k}(G+k')\partial_{k'}\ftsbloch_{\alpha k}(G+k)\dd{g}.
\end{align*}
This vanishes unless ${k=k'}$ because
${\ftsbloch_{\alpha k}(G+k')}$ vanishes unless ${G+k'-k}$ is a reciprocal lattice vector, 
and because the largest difference between two elements of $\BZ$ is less than the magnitude ${\hreciplatt}$
of the smallest finite reciprocal lattice vectors. 
Therefore Eq.~\ref{eqn:mtopJthree} can be expressed as
\begin{align*}
\Jconvm 
& =  \frac{e\dot{\zeta}}{\pi} 
\sum_{\alpha=1}^{\Neleccell}
 \im \left\{ \int_{\BZ} \dd{k}\sum_{G} \partial_\zeta\fts{u}^*_{\alpha k}(G)
\partial_k\fts{u}_{\alpha k}(G)\right\}.
\end{align*}
Now, since 
${\partial_\zeta u_{\alpha k}}$ and 
${\partial_k u_{\alpha k}}$ are ${\volume}$-periodic, 
we can express them as the Fourier series
\begin{align*}
\partial_\zeta u_{\alpha k}(x) 
& = \sum_G \partial_\zeta\fts{u}_{\alpha k}(G)e^{iGx}
\end{align*}
and
\begin{align*}
\partial_k u_{\alpha k}(x)
&= \sum_G \partial_k\fts{u}_{\alpha k}(G)e^{iGx},
\end{align*}
respectively.
By substituting these series, it is straightforward
to show that
\begin{align*}
\braketT{\partial_\zeta u_{\alpha k}}{\partial_k u_{\alpha k}}
& =\int_\onetorus \partial_\zeta u^*_{\alpha k}(x) \partial_k u_{\alpha k}(x)\dd{x}
\\
& =\sum_G \partial_\zeta\fts{u}^*_{\alpha k}(G)
\partial_k\fts{u}_{\alpha k}(G)
\end{align*}
Therefore the polarization current can be expressed as 
\begin{align}
\Jconvm 
& = \frac{e\dot{\zeta}}{\pi} \sum_{\alpha=1}^{\Neleccell} \im \bigg\{
\int_{\BZ} \dd{k}
\braketT{\partial_\zeta u_{\alpha k}}{\partial_k u_{\alpha k}}
\bigg\}.
\label{eqn:mtopJfour}
\end{align}
Finally, note that the normalization of each periodic Bloch function ${u_{\alpha k}}$
is the same as its non-periodic counterpart, ${b_{\alpha k}}$, i.e., 
\begin{align*}
\braketT{\pbloch_{\alpha k}}{\pbloch_{\alpha k}}=\braketT{\bloch_{\alpha k}}{\bloch_{\alpha k}} = 1.
\end{align*}
However it is more common to express $\Jconvm$ in terms of
periodic Bloch functions ${\tpbloch_{\alpha k}\equiv \sqrt{\Nunitcell}\pbloch_{\alpha k}}$ 
that are normalized on a single unit cell, i.e.,
\begin{align*}
\braketT{\tpbloch_{\alpha k }}{\tpbloch_{\alpha k}}&=\Nunitcell 
\\
\implies \braketcell{\tpbloch_{\alpha k}}{\tpbloch_{\alpha k}}&\equiv \int_\unitcell\dd{x}\abs{\tpbloch_{\alpha k}(x)}^2 = 1.
\end{align*}
Then,
\begin{align}
\Jconvm 
& = \frac{e\dot{\zeta}}{\pi} \sum_{\alpha=1}^{\Neleccell} \im \bigg\{
\int_{\BZ} \dd{k}
\braketcell{\partial_\zeta \tpbloch_{\alpha k}}{\partial_k \tpbloch_{\alpha k}}
\bigg\},
\label{eqn:mtopJfive}
\end{align}
which is one of the most commonly-quoted forms
of the MTOP definition of
$\Jconvm$.
%
%
%
%

I emphasize, again, that quantum mechanics has not been used in this derivation,
or anywhere in Sec.~\ref{section:mtop}.
It was assumed that the density could be expressed 
as ${n(x;\zeta)=\sum_{i=1}^{M}\abs{\varphi_i(x;\zeta)}^2}$, where the functions ${\varphi_i}$ are mutually
orthogonal and normalized to one, and ${M=\Nelecbulk}$. However they could also have 
different normalizations, and ${M}$ could be bigger or smaller than $\Nelecbulk$, 
as long as ${n(x;\zeta)}$ was normalized to $\Nelecbulk$
and the normalization of each $\varphi_i$ did not vary with $\zeta$.
Such a representation of ${n(\zeta)}$ is possible for at least some
classically-generated probability densities. 
For example, when the nuclei are
treated as classical particles, their delta distribution can be represented
in this way.

\subsection{Interpretation of the MTOP}
\label{section:interpretation}
In Sec.~\ref{section:illustrative_derivation} and Sec.~\ref{section:bulk_subsystem} 
I have derived the MTOP expression for $\Jconv$ via routes 
that are very different to the one originally used to derive it; 
and I have shown that it is more widely applicable than
previously believed.
Previous derivations have used quantum mechanical perturbation theory, 
which is nothing more than a Taylor expansion of a microstate about
a stationary state (i.e., an eigenstate) of a Hermitian operator. 
The applicability of such an expansion is not restricted to quantum mechanical
systems.

I have shown that the MTOP can be applied to a system of classical 
identical particles in an evolving
steady state $n(x;\zeta)$ that can be expressed as a sum of densities from
orthogonal functions $\varphi_i$ whose normalizations do not change as $\zeta$ changes.
Mathematically, this means that the MTOP is applicable whenever $n$ can be expressed as
a sum of contributions from eigenstates of a self-adjoint one-particle 
operator, such as the solutions of any Sturm-Liouville equation,  and when
the subset of eigenstates contributing to $n$ does not change suddenly
as $\zeta$ changes: Each contributing eigenstate must evolve
smoothly with $\zeta$, while preserving its normalization.

An alternative derivation of Eq.~\ref{eqn:mtopJ}, based on the homogenization
theory proposed in Secs.~\ref{section:homogenization} and~\ref{section:excess_fields},
is presented and discussed in Sec.~\ref{section:discrete}
and Sec.~\ref{section:current}.

Although I have concluded that the MTOP definition of polarization 
current is correct, and exact, I am not satisfied that the MTOP solves the 
problem discussed in Sec.~\ref{section:definingP} and I do
not agree with one interpretation of the MTOP; namely, that polarization is 
a fundamentally quantum mechanical phenomenon~\citep{resta-1993, resta-rmp-1994} and 
a property of the phase of a material's wavefunction.

The wavefunction of an isolated material at equilibrium does not have a
relevant phase. It is real, apart from an arbitrary and irrelevant
constant phase factor.
Therefore if $\pp$ were {\em exclusively} a property
of a wavefunction's phase, $\pp$ could only exist in an isolated material when it 
was being observed.
This particular strain of quantum weirdness may be palatable to some, but it should 
be noted that delving more deeply into it leads to difficult philosophical questions.

Note that, by applying the kinetic energy operator to a wavefunction ${\Psi = \sqrt{\pdf}e^{i\theta}}$
whose phase ${\theta(\{x_i\})}$ is nontrivial, 
the phase can be shown to be equivalent
to a potential ${\propto \sum_i\abs{\pdv*{\theta}{x_i}}^2}$. 
Although this potential is positive, potentials 
are only defined up to a constant applied to the entire physical system of interest.
This means that any positive or negative potential is either equivalent to, 
or arbitrarily close to, the combination of a phase and this irrelevant constant.
Therefore, when studying the bulk subsystem, the interaction between the bulk and surface subsystems
could be expressed as a phase. Then ${\Jconv=\bsigmadot}$ would
manifest as a time-dependence of this phase, which is equivalent to a changing
interaction with the surface. However this is not the rationale used to
relate $\Jconv$ to a phase in the MTOP. 

Furthermore, it has sometimes been implied or stated
that $\pp$ `is', rather than `can be', a property of the phase and that this means
that it does not have an analogue within classical physics. This interpretation 
of the mathematics cannot be correct because the only assumption that I made
in Sec.~\ref{section:bulk_subsystem} to derive Eq.~\ref{eqn:mtopJthree}, which is 
identical to the MTOP  expression for $\Jconv$, is that the electron 
density $n$ can be represented by a sum of contributions from
smoothly-evolving mutually-orthogonal functions in 
an ${\lebesgue}$ Hilbert-Lebesgue space.

The `quantum' MTOP definition of $\pp$ was developed 
to solve the problem discussed in Sec.~\ref{section:definingP} and the problem of 
how to calculate the currents demanded by asymmetry (Sec.~\ref{section:anisotropy}).
However, I did not invoke quantum mechanics to explain those problems because they 
are problems that exist within classical physics. 
This implies that, despite the development of the MTOP, there was more 
to understand because classical mechanics and classical statistical mechanics have 
never been derived from quantum mechanics. 
The {\em correspondence principle} is conjecture: Quantum mechanics is not known to be more 
general than classical statistical mechanics; it is assumed to be.

Classical physics is arguably no less internally-consistent than quantum physics. 
The problem with classical physics is not internal inconsistency but that, in its current state of 
development, it does not agree with experimental observations of very small or sensitive systems.
Therefore, with the exception of a disagreement with experiment, 
any physical problem that can be stated within the classical realm, and which 
does not expose an internal inconsistency in classical physics, 
must have a solution within the classical realm.

The well-defined problem of how to calculate the polarization current from the changing charge
densities in Fig.~\ref{fig:current-symmetry} is an excellent example. There exist classical
processes that result in the equilibrium distribution of charges  (e.g., ions in solution)
changing. In most cases the charges are either too numerous or too sensitive to the act of observation 
to allow the net flow of charge to be calculated within classical deterministic mechanics, and one must turn
to statistical mechanics. How can the current resulting from an adiabatic evolution of
a classical equilibrium statistical state be calculated?
The MTOP, as it was originally derived and interpreted,  does not answer this question directly.
But if a classically evolving 
charge density can be decomposed into packets whose integrals are constant, 
it answers it indirectly: $\Jconv$ can be calculated from Eq.~\ref{eqn:mtopJ}.

My approach to deriving Eq.~\ref{eqn:mtopJ} in Sections~\ref{section:surfstability} and~\ref{section:current} 
is conceptually very simple.
In Secs.~\ref{section:homogenization} and~\ref{section:excess_fields} I
use a systematic and unbiased approach to structure homogenization 
to derive expressions for interfacial excesses and changes 
in macroscopic fields across interfaces. From these I derive an expression
for $\bsigma$ in terms of the microstructure (a generalization of Finnis's expression), which leads to 
an expression for ${\J\equiv\bsigmadot}$ that is slightly more general than the MTOP expression, but
otherwise equivalent to it.

I do not find any reason to retain
the concept of a $\pp$ field, and it appears to violate macroscale symmetry.
However, as I will discuss in Sec.~\ref{section:quantized}, 
if one chooses to retain it, and if one also demands that ${\pp}$ determines the excess
charge on a pristine surface of a perfect crystal,  which is devoid of extrinsic charges,
then, even within classical physics, $\pp$ must be quantized in the same way as
the MTOP prescribes.

\section{Macrostructure as homogenized microstructure} 
\label{section:homogenization}
\subsection{Introduction}
In this section I present elements of a classical theory of 
\emph{structure homogenization}, meaning a theory of 
how microstructures determine macrostructures, and of
the resulting natures of macrostructures.
In later sections I will use this theory to elucidate the relationships between 
the microscopic fields $\phi$, ${\me}$, and ${\rho}$
and their macroscopic counterparts $\bphi$, $\E$, and $\Rho$.

The most obvious and intuitive approach to deriving 
electromagnetic theory at the macroscale from Maxwell's vacuum theory is
to define macroscopic fields as spatial averages of their microscopic counterparts.
However, despite this having been tried many times and in 
many ways~\citep{rosenfeld_1965,kaufman_1961,degroot_1965, schram_1960, degroot_1964,mazur_1953,mazur_1957, robinson_1971, russakoff-ajp-1970,jackson-book, ashcroft_mermin_book,kirkwood_1936,vanvleck-1937,kirkwood-1940,bethe-1928,miyake-1940,mip_pratt_1987,vinogradov-PRE-1999,kamenetskii-PRE-1998}, 
we lack a fundamental understanding of the relationship between microstructure and macrostructure.

There are two main reasons why previous attempts have not succeeded, or have not succeeded fully.
The first is that more fields appear in Maxwell's macroscopic theory than appear
in his vacuum theory. Therefore, for example, the fields $\pp$ and $\D$ cannot be defined as spatial
averages of their counterparts at the microscale because they do not have counterparts at the microscale.

The second reason is that, to define one field as the spatial
average of another, it is necessary to introduce
one or more parameters specifying the size and shape of the 
region of space that is averaged over, and the distribution of 
weights with which points in this region contribute to the average.
The dependences of macroscopic fields
on these parameters has been interpreted as a fatal flaw
in their definitions as spatial averages.

However, this non-uniqueness should not be interpreted as a fatal flaw, but as intrinsic
to the nature of macrostructure, which is not determined by microstructure alone.
It is determined, in part, by the relationship between the observer and the microstructure:
A macroscopic field is a microscopic field observed on a large length scale. What is observed depends on
how large that scale is. Therefore any definition of a macroscopic field in terms of its
microscopic counterpart must depend on a parameter that specifies it.

In addition to the scale on which the underlying microscopic field is observed, a
macroscopic field depends on the perspective from which the microstructure is observed, the apparatus with
which it is observed, and the fields that mediate the observation.
Therefore, when defining reproducibly-measureable macrostructure it 
is necessary to choose which of these influences are incorporated into the definition, and which of them 
are left to observers as apparatus-specific corrections.
For simplicity and generality, I will assume that the only parameter on which the definition of macrostructure
depends is the smallest distance, $\abs{\dbx}$ across which changes in macroscopic fields are observable at the macroscale.
All other observer-specific influences are left as corrections to be applied when a specific observation is
compared to the theory built on this one-parameter definition.

Perhaps unsurprisingly, there is a trade-off between the precisions to which
gradients of macroscopic fields are defined and the spatial precision at the macroscale, $\abs{\dbx}$.
When the uncertainty in position is small, the uncertainty in the gradient
of a macroscopic field is large, and vice-versa, with the product of these
uncertainties being proportional to the uncertainty in the value of the macroscopic field
itself. This uncertainty principle is derived in Sec.~\ref{section:uncertainty},
but I begin discussing its consequences for the nature of macrostructure in Sec.~\ref{section:what_is_macrostructure},
so that readers understand why defining a macroscopic field in terms of
its microscopic counterpart is only one of two primary objectives
of this structure homogenization theory.
An equally important objective, which is arguably more important from a practical perspective, 
is to define macroscopic \emph{excess fields},
such as surface charge densities, $\bsigma$.

I begin by outlining the simplifying assumptions that I make 
about the nature of the microstructure.

\subsection{Assumed properties of the microstructure, $\nu$}
\label{section:microstructure}
Here, and throughout this work, I will assume that the microstructure
only varies significantly on the microscale and on the macroscale, and
that the macroscale is many orders of magnitude larger than the microscale.
For example,  the microscale might be the nanometer scale, and the smallest distance across which 
variations of macroscopic fields are observed might 
be ${\abs{\dbx}\sim\SI{1}{\milli\meter}}$.

I will assume that the physical system of interest does not
contain any material with a microstructure that varies
significantly on intermediate length scales.
For example, the microstructure
of wood varies on every length scale between ${\sim\SI{1}{\nano\meter}}$
and the scale (${\sim\SI{10}{\meter}}$) of the tree from which 
it was harvested~\citep{wood}. 
When it is observed on any length scale in this 
ten order-of-magnitude range, its surface is observed to have texture.
Therefore the theory presented herein does not apply to wood. 
It applies when texture is only observable on the microscale and on the macroscale.
The texture observed on the macroscale is caused by
\emph{excess fields} at interfaces between regions with different microstructures, 
or by changes of the microstructure that occur gradually across macroscopic distances.

For simplicity, I will assume that the microstructure is specified
by a single scalar field, ${\nu:\realone^n\to\realone}$, and I will
denote its counterpart on the macroscale by $\Nu$.
Although the value of $n$ is three, 
the results derived for the ${n=1}$ case can be
applied in three dimensions by defining ${\nu(x)}$ to 
be the average of a function of ${(x,y,z)}$ on the ${y-z}$ plane at $x$.
Therefore, to make discussions of the main physical ideas simpler, 
in this section I will mainly focus on the one dimensional microstructure ${\nu:\realone\to\realone}$.

I make the following assumption about how observations and measurements
occur at the macroscale.
\begin{assumption}
At the macroscale, all positions, distances, and displacements are deduced
from measurements of $\Nu$. 
\end{assumption}
Generalizing the theory to the case of multiple pairs ${(\nu_i,\Nu_i)}$
of microscopic fields and their macroscopic counterparts is straightforward.
Therefore, in the context of electricity, this is the assumption that all measurements
and observations at the macroscale are measurements and observations of the electric
potential, $\bphi$, its derivatives ${\dbphi{1}=-\E}$ and ${\dbphi{2}=-\Rho}$, 
and/or their excesses (see Sec.~\ref{subsection:excess_fields} and Sec.~\ref{section:excess_fields}).

A microstructure $\nu$ may fluctuate, to some degree, on every length scale.
However, as discussed above, the theory proposed in this section applies under the 
assumption that $\nu$ may vary significantly on 
the macroscale $L$ and on the microscale $a$, which are widely separated, 
but its variations on any intermediate scale, or {\em mesoscale}, $l$ are negligible.
Generalizing the theory to microstructures that vary on three or more widely-separated length scales appears
straightforward, but generalizing it to microstructures that vary on all length
scales does not.

To define the microscale and the macroscale more precisely, 
I introduce the distances ${\amax}$ and ${\Lmin}$.
Roughly-speaking, $\Lmin$ is the largest distance such that nonlinearities in the variation of $\Nu$ across
distances less than $\Lmin$ are negligible; and $\amax$ is orders of magnitude
smaller than $\Lmin$, but large enough
that every interval in the set 
\begin{align*}
\left\{\interval(x,\amax):x\in\dom\nu\right\},
\end{align*}
whose center $x$ is not in vacuum, contains many local extrema of $\nu$.

These conditions do not define $\amax$ and $\Lmin$ uniquely, but it will not be necessary
to define them uniquely.
Their primary purpose will
be to define a mesoscopic distance ${\abs{\dbx}}$,
which is the macroscale infinitesimal, and
is much larger than any reasonable choice of $\amax$ and
smaller than any reasonable choice of $\Lmin$. 
At the microscale I will denote ${\abs{\dbx}}$ by $\prectheo$.

I define the microscale, the mesoscale, and the macroscale in terms of $\amax$ and $\Lmin$ as 
follows:
\begin{align*}
\eta_a\sim a &\iff \eta_a < \amax \\
\eta_l\sim l &\iff \amax < \eta_l < \Lmin \\
\eta_L\sim L &\iff \eta_L > \Lmin.
\end{align*}
My physical assumption that all fluctuations of the microstructure whose wavelengths are
between $\amax$ and $\Lmin$ have negligible amplitudes
can be expressed in terms of the Fourier transform ${\ftsnu(k)}$ of ${\nu(x)}$ as follows.
\begin{assumption}
\label{assumption:one}
\begin{align}
\int_{0}^{k_L}\abs{\ftsnu(k)}^2\dd{k}
\gg 
\int_{k_L}^{k_a}\abs{\ftsnu(k)}^2\dd{k} 
\ll 
\int_{k_a}^\infty \abs{\ftsnu(k)}^2\dd{k},
\nonumber
\end{align}
where 
${k_L\equiv 2\pi/\Lmin\lll k_a\equiv 2\pi/\amax}$.
\end{assumption}
Obviously, if this assumption holds true for particular values of $\amax$ and $\Lmin$, it
also holds true for larger values of $\amax$ and smaller values of $\Lmin$, as long as
${\amax<\Lmin}$.

\subsection{The nature of macrostructure}
\label{section:what_is_macrostructure}
I now outline some important features of macrostructure, which 
are consequences of finite spatial precision at the macroscale.

At the macroscale we observe homogenized microstructure. Under the two simplifying
assumptions presented in Sec.~\ref{section:microstructure}, the macrostructures
of the bulks of materials are uniform, except, possibly, for the existence of macroscopic
point, line, or locally-planar defects.
If there are no such defects, all materials appear uniform (textureless) at the macroscale
and we only perceive a difference between the bulks of two materials
on either side of an interface via observable properties of 
their \emph{microstructures}, such as their colours.

A material's colour is determined by the microscopic wavelengths of electromagnetic radiation
with which its microstructure interacts. Since the wavevectors of this radiation are normal
to the plane on which the spatial average that produces the observed macrostructure from the microstructure is performed,
we observe the radiation, rather than the averages of its electric and magnetic
fields along its axis of propagation, which would vanish.

We have assumed, via Physical Assumption 2, that differences
in texture are not observable at the macroscale.  
Physical Assumption 1 implies that differences in colour are irrelevant
to macroscale electricity.
For example, colour differences cannot be detected as changes in the
distributions of electric potential or charge at the macroscale.
Therefore the only observable structure at the macroscale is the network of interfaces
separating otherwise-indistinguishable regions of uniformity, and any macroscopic
defects within these otherwise-uniform regions. 
Interfaces and other macroscopic heterogeneities are observable because, and only if, they carry
observable excess fields.

\subsubsection{Excess fields}
\label{subsection:excess_fields}
At the macroscale, an excess field can be understood from Stokes' theorem. For example, 
the net charge within a material $\material$ is
\begin{align}
\int_\material \Rho(\br)\dd[3]{\br} = \int_{\partial\material}\bsigma(\bs)\dd[2]{\bs},
\label{eqn:stokes_charge}
\end{align}
where the integral on the right hand side is an integral of the areal charge density, 
$\bsigma$, on the material's surface, ${\partial\material}$.
In this case, $\bsigma$ is an excess field: it is the \emph{surface excess} of field ${\Rho}$.

Stokes' theorem also holds at the microscale if point charges are treated
as singularities and the sets of zero measure in which they reside are omitted
from the integrals. However, in that case both integrals in the microscale analogue of Eq.~\ref{eqn:stokes_charge}
vanish, and we must take a different route to understand surface excesses.

A further complication is that surfaces are ill-defined at the microscale: A surface's
microstructure differs, to some degrees, from the microstructures of both
the bulk and the atmosphere or vacuum above the surface. Its difference with respect
to the bulk microstructure lessens \emph{gradually}
with increasing depth below the surface, so there does not exist a depth below which
the material is bulk-like and above which it is not.
Therefore surfaces are ill-defined regions of indeterminate widths
at the microscale.

The surface ${\partial \material}$ of material $\material$ that appears in Eq.~\ref{eqn:stokes_charge}
is well defined because,
although the thickness of the surface region is indeterminate at the microscale, it
is less than ${\prectheo\equiv\abs{\dbx}}$. Therefore the depths of
any two points within it differ by less than ${\prectheo}$, 
which implies that, at the macroscale, the distance between the atmosphere
outside a material and the material's bulk is $\abs{\dbx}$. 
Surfaces are well-defined at the macroscale because 
every curve that crosses the surface exactly once contains
exactly one surface point. In other words, locally, surfaces are
\emph{literally} planar at the macroscale.

\begin{figure}[!]
\includegraphics[width=8cm]{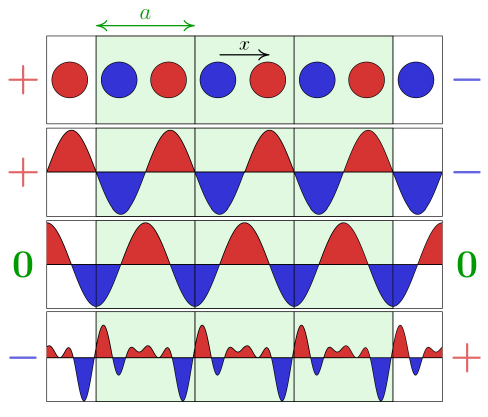} 
\caption{
{\em Excess fields}:
Each of the four vertically-stacked
panels is a schematic plot of the
microscopic charge density ${\rho(x)}$ of a different one dimensional material, 
with positive and negative charge(s) coloured red and blue, respectively. 
If, at each value of $x$, we calculate the average, ${\expval{\rho}_a(x)}$, of all charge 
within a distance ${a/2}$ of $x$, we find that it vanishes everywhere
in the green-shaded `bulk' of each material, but is finite
in the white surface regions. At each surface, the integral of ${\expval{\rho}_a(x)}$
over all points that are not in the bulk, but are within a distance $a$ of it,
is the surface's excess of charge, $\bsigma$.
The symbols $+$, $-$, and $0$ next to each surface indicate whether $\bsigma$
is positive, negative, or zero, respectively. 
The macroscopic analogue, $\Rho$, of $\rho$  is defined, to a finite 
precision ${\precRho}$, 
as its mesoscale average $\bar{\rho}$. 
${\Rho}$ vanishes everywhere in the bulk, but
not at interfaces, in general. Therefore, because
spatial averaging conserves charge, the excess charge at an interface, $\bsigma$, is simply
the integral of $\Rho$ across it.}
\label{fig:excess_fields}
\end{figure}

Figure~\ref{fig:excess_fields} illustrates, from a microscopic perspective, why surfaces
carry excess fields:
The average of a microscopic charge density, $\rho$, vanishes
in the green-shaded bulk of the one-dimensional materials depicted, but not at points near a surface, in general.
For example, consider a point very close to the left-hand edge of one of the unshaded surface regions
on the left-hand side. 
The macroscopic charge density, $\Rho$, vanishes in the green-shaded bulk because
the contribution to the spatial average at a point $x$ from all points ${\{x+u:0<u<a/2\}}$
to its right is cancelled exactly by the contribution from all points ${\{x-u:0<u<a/2\}}$
to its left. However, this cancellation cannot happen at a point near the left-most
edge of one of the surfaces on the left-hand side,
because most of the points to its left are in vacuum, where $\rho$ vanishes.

Therefore, $\Rho$ vanishes in the green-shaded bulk, but does not necessarily vanish
in the unshaded surface regions. Since the widths of these regions
are much less than ${\prectheo=\abs{\dbx}}$, each one corresponds
to a single point at the macroscale. 
This one-dimensional example illustrates that $\Rho$ vanishes everywhere in the bulk and it vanishes 
in the vacuum outside the material, but it is finite, in general, 
at a surface.

\begin{figure}[h]
\includegraphics[width=8cm]{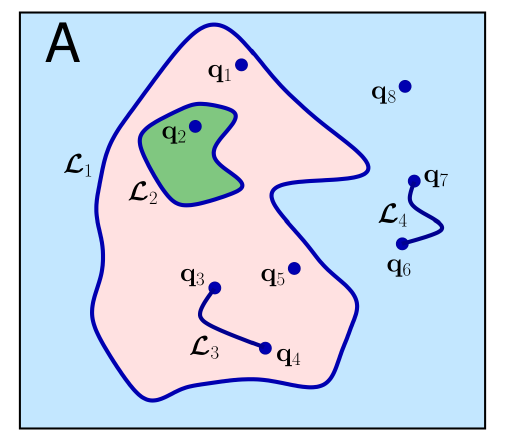}
\caption{
{\em A 2D macrostructure}:
An electrical microstructure in  ${A\subset\realtwo}$ is an areal 
charge density ${\sigma:A\to\realone; (y,z)\mapsto \sigma(y,z)}$, which
may be the excess of a volumetric charge density ${\rho(x,y,z)}$ on a surface or interface
normal to the ${x}$-axis. An excess field (see Fig~\ref{fig:excess_fields} and Sec.~\ref{section:excess_fields}) is created at an interface whenever 
a higher dimensional microstructure is homogenized along an axis normal to the interface.
The {\em macrostructure} of $A$ consists of continua that are punctuated with, and separated by, 
subspaces ${s_i}$ of dimensions zero or one, on which excess fields 
are defined.  
If $s_i$ has dimension one its excess field is a linear charge density ${\mathbfcal{L}_i}$;
and if it has dimension zero its excess field is a point charge ${\qfig{i}}$.
}
\label{fig:macrostructure}
\end{figure}
Therefore, in general, the value of a macroscopic field $\Nu$ on any curve that intersects a surface
or interface has a jump discontinuity or a removable discontinuity at the point of intersection, $\bx_s$.
The easiest way to deal with $\Nu$ being discontinuous at $\bx_s$ is by treating all surface
points separately: they can be omitted from the domain of $\Nu$ and an excess
field, ${\bsigmaNu(\bx_s)\equiv\Nu(\bx_s)\abs{\dbx}}$, can be defined on each surface, interface, 
or locally-planar defect.

More generally, excess fields describe macroscopically-observable accumulations whose manifestations
at the macroscale are best described by
distributions whose domains are manifolds of dimension 
${n<3}$. For example, the excess of charge on 
the $i^\text{th}$ surface or interface, which is a two dimensional manifold (or simply {\em 2-manifold}), $\manifoldn{2}_i$, 
is an areal density of charge, ${\bsigma_{i}:\manifoldn{2}_i\to\realone}$;
the $j^\text{th}$ linear charge density, ${\linecharge_{j}:\manifoldn{1}_j\to\realone}$,
is a charge excess defined on a curve (the 1-manifold, $\manifoldn{1}_j$); and the $k^\text{th}$ macroscopic point charge,
${\pointcharge_k:\manifoldn{0}_k\to\realone}$, is a charge excess defined at a point (the 0-manifold, $\manifoldn{0}_k$).
The macrostructure arising from a two-dimensional microscopic charge density
is illustrated in Fig.~\ref{fig:macrostructure}.

What this means is that, despite the microstructure being defined by a single field $\nu$,
there is more to the macrostructure than the single field ${\Nu}$. 
The macrostructure comprises $\Nu$, where 
\begin{align*}
\dom\Nu\equiv\realone^3\setminus \left(\{\manifoldn{2}_i\}\cup\{\manifoldn{1}_j\}\cup\{\manifoldn{0}_k\}\right),
\end{align*}
and the set 
\begin{align*}
\{\bsigmaNuind{i}\}\cup\{\linechargeNuind{j}\}\cup\{\pointchargeNuind{k}\} 
\end{align*}
of excess fields. 

We will see that $\Rho$ vanishes in the bulk of any stable material, and that $\bphi$ and $\E$ vanish
in a material's bulk if the material is isolated and its surfaces are charge-neutral. 
Therefore, 
defining a material's electrical macrostructure entails finding a way to define
excess fields in terms of the microstructure, $\nu$. Finnis solved this problem for periodic microstructures~\citep{finnis}, 
such as those plotted in Fig.~\ref{fig:excess_fields}, and I generalize his work to non-periodic
microstructures in Sec.~\ref{section:excess_fields}.

\subsection{Assumed properties of $\Nu$}
\label{section:gp}
Defining $\Nu$ in terms of $\nu$ is trickier than it first appears, so 
it is useful to list and discuss the properties that $\Nu$ is assumed
to have. There are three of them, which I list below and discuss 
in Secs.~\ref{section:gp1}, \ref{section:gp2}, and~\ref{section:gp3}.
\begin{assumption}
\label{assumption:two}
$\Nu$ is reproducibly measurable.
\end{assumption}
\begin{assumption}
\label{assumption:four}
$\nu$ fluctuates microscopically about ${\Nu(\bx)}$ at each ${\bx\in\dom\Nu}$.
\end{assumption}
\begin{assumption}
\label{assumption:three}
$\Nu$ is differentiable, except, possibly, on a set of zero measure.
\end{assumption}

\subsubsection{Reproducible measurability of $\Nu$}
\label{section:gp1}
$\Nu$ is measurable by a blunt probe or as an average of the values of $\nu$ measured 
by many sharp probes whose locations cannot be controlled or known to microscopic precisions.

For example, when you look at a surface, light enters your eye from 
many closely-spaced points of the surface's microstructure, but each ray enters at a different angle and with a 
different intensity and a different frequency, in general. The contributions from 
all of the rays merge to produce an image of homogenized microstructure in your mind. This is the macrostructure.
The merger that homogenizes the microstructure occurs in many stages,
involves many different mechanisms, and occurs at many different locations
along the path from the surface to your eye to your brain.

When I say that $\Nu$ is {\em reproducibly} measurable, I mean that when a particular
spatial resolution $\abs{\dbx}$ is chosen, and shared by all repeated measurements,
the value of $\Nu$ at each point can be defined independently of 
any measuring technique or apparatus. This means that, although a measured value of $\Nu$ at a point 
always contains artefacts
of the method used to measure it, if the magnitudes of these artefacts
could be made sufficiently small, or if corrections could be applied to remove them, any 
two measurements of $\Nu$ at the same point would both yield values that were \emph{both}
consistent with the microstructure $\nu$, \emph{and} with the definition of $\Nu$
in terms of $\nu$.

Despite the complexity of the processes that turn microstructure
into macrostructure, I base the homogenization theory presented
herein on the following assumption.
\begin{assumption}
\label{assumption:five}
Any accurate measurement of $\Nu$ at precisely the point $\bx$
is a measurement of a weighted spatial average of $\nu$ on a mesoscopic
domain centered at a point
that is macroscopically-indistinguishable from $\bx$, i.e., 
a point in the interval ${\interval(\bx,\abs{\dbx})}$.
\end{assumption}

\subsubsection{Definition of ``{\em fluctuates microscopically}''}
\label{section:gp3}
When I say that ${\nu}$ {\em fluctuates microscopically}, or that $\nu$ is a {\em microscopic quantity}
or a {\em microscopic function}, I mean that $\nu$ fluctuates on length scale $a$.
This means that every extremum of ${\nu}$ is within a distance ${\amax}$ of another extremum of ${\nu}$.
Defining the statements 
``{\em $\nu$  fluctuates microscopically about ${\valnu}$ at $\bx$}''
and
``{\em $\nu$  fluctuates microscopically about ${\Nu}$ at $\bx$}''
is more difficult, 
so I defer discussing them until Sec.~\ref{section:finding_precision}.

\subsubsection{Macroscale differentiability}
\label{section:gp2}
The purpose of a macroscopic field theory is to describe changes over macroscopic 
distances. Therefore if $\Nu$ fluctuated microscopically it would, effectively, be
nondifferentiable. For example, if the value of ${\DNu_1\equiv\Nu(\bx+\bh_l +\bh_a)- \Nu(\bx)}$ differed
significantly from the value of ${\DNu_2\equiv\Nu(\bx+\bh_l)- \Nu(\bx)}$, when ${\abs{\dbx}<\abs{\bh_l}<\Lmin}$
and ${\abs{\bh_a}<\amax}$, then it would not be possible to approximate
${\Nu(\bx+\bh_l)}$ with a truncated Taylor expansion of ${\Nu}$ about 
$\bx$ containing few terms.

$\Nu$ being differentiable at the macroscale means that,
given any point ${\bx\in\dom \Nu}$,
the values of ${\left(\Nu(\bx+\bh)-\Nu(\bx)\right)/\bh}$
and ${\left(\Nu(\bx)-\Nu(\bx-\bh)\right)/\bh}$
are equal in the limit ${\abs{\bh}\to\abs{\dbx}}$, \emph{to within the precisions to which they are defined}.

\subsection{Spatial averages}
I will define $\Nu$ \emph{in terms of} (not \emph{as}) spatial averages of $\nu$ of the form
\begin{align*}
\expval{\nu;\mu}_\epsilon(x)\equiv \left(\nu\ast\mu(\epsilon)\right)(x)\equiv\int_\realone\nu(x')\mu(x'-x;\epsilon)\dd{x'},
\end{align*}
where the parameter $\epsilon$ of the {\em averaging kernel}, ${\mu(\epsilon)}$, 
is twice its standard deviation, i.e., 
\begin{align*}
\int_\realone u^2\,\mu(u;\epsilon)\dd{u}&=
\left(\frac{\epsilon}{2}\right)^2.
\end{align*}
I will assume that $\mu(\epsilon)$ has three other properties for every value of $\epsilon$.
The first property is
\begin{align*}
\int_\realone \mu(u;\epsilon)\dd{u} &= 1,
\end{align*}
which implies that the homogenization
of $\nu$ is conservative.
The second property is
\begin{align*}
\int_\realone u\,\mu(u;\epsilon)\dd{u} &= 0,
\end{align*}
which implies that the value 
of ${\expval{\nu;\mu}_\epsilon}$ at $x$ is a weighted
average of $\nu$ from points whose weighted-average
position is $x$.
The third property is
\begin{align}
\mu(u;s\epsilon)&=s^{-1}\mu(u/s;\epsilon),
\label{eqn:scaling}
\end{align}
which simply means that the effect of changing $\epsilon$ is
to scale $\mu$ without changing its shape or its integral.

These properties do not place strong or unphysical constraints on the form of 
${\mu(\epsilon)}$, because any function with a well-defined mean and standard deviation 
can be normalized and translated 
to give a function whose integral is one and whose mean is zero.
That function can be identified as ${\mu(1)}$ and Eq.~\ref{eqn:scaling}
can be used to define the narrower or wider
function ${\mu(u;\epsilon)\equiv \mu(u/\epsilon;1)/\epsilon}$.

The reason for giving $\mu$ a parametric dependence on its width 
is that it makes it easier to discuss separately the effects on ${\expval{\nu;\mu}_\epsilon}$ of varying the width 
of ${\mu(\epsilon)}$
and of varying the shape of $\mu(\epsilon)$.
For example, it follows from Eq.~\ref{eqn:scaling} that the ${n^\text{th}}$ derivative
of $\mu$ satisfies
\begin{align}
\mu^{(n)}(su;s\epsilon)=\frac{\mu^{(n)}(u;\epsilon)}{s^{n+1}}.
\label{eqn:kernel2}
\end{align}
Therefore, whereas the average magnitude of ${\mu(\epsilon)}$ 
scales as ${1/\epsilon}$, the average magnitude of its first
derivative scales as ${1/\epsilon^2}$, and higher-order
derivatives decay even faster as $\epsilon$ increases.
An important implication of this is that the shape of $\mu$ has
less of an influence on the value of ${\expval{\nu;\mu}_\epsilon}$
as $\epsilon$ increases.

Note that if ${\mu(u;\epsilon)}$ has all of the properties discussed above, 
then so does the function ${\mu(-u;\epsilon)}$. Therefore
the general form of the spatial averages considered
in this work may also be expressed as
\begin{align}
\expval{\nu;\mu}_\epsilon(x)\equiv \int_\realone\nu(x+u)\mu(u;\epsilon)\dd{u}.
\label{eqn:average1}
\end{align}

\subsubsection{Schwartz and non-Schwartz averaging kernels}
For some averaging kernels, ${\mu(u;\epsilon)}$, there exist
values of $m$ such that their rates of decay in the limits
${u\to\pm\infty}$ are slower than ${1/\abs{u}^m}$.
Other kernels, such as Gaussians, decay faster than
any power law. 
A smooth function that decays faster than any power law is known as a {\em Schwartz function},
so I will refer to kernels of the first and second types as {\em non-Schwartz kernels}
and {\em Schwartz kernels}, respectively.

Non-Schwartz kernels tend to describe relatively-direct and 
physical weightings of microstructures, 
such as the shape of a blunt probe or
the decay of light intensity with distance.
Schwartz kernels tend to arise
from disorder and uncertainty: They describe limiting cases of homogenizing
physical processes, such as the ${N\to\infty}$ limit of
the combined effects of $N$ 
homogenizing influences, each of which can be described by non-Schwartz kernels.

For example, the {\em central limit theorem} 
demonstrates how a Gaussian
distribution arises from a very large number of contributions from
independent random variables whose distributions do not necessarily decay faster than 
a power law
(see, for example, ~\linecite{Riley_Hobson_Bence_2006}, Chapter~30).

Non-Schwartz kernels make it easier to illustrate the complications that arise from defining macroscopic
fields as spatial averages of their microscopic counterparts, and {\em top-hat kernels} are among the simplest
of non-Schwartz kernels. Therefore
I will introduce top-hat kernels 
in Sec.~\ref{section:tophat}
and use them to illustrate an important consequence of Eq.~\ref{eqn:kernel2}; namely, the fact that 
${\expval{\nu;\mu}_\epsilon}$ depends less and less on the shape of $\mu$ as $\epsilon$ increases.
Then, in Sec.~\ref{section:finite_precision}, I will use top-hat kernels to illustrate
why precision is finite at the macroscale.

\subsubsection{Top-hat kernels}
\label{section:tophat}
A very simple non-Schwartz kernel is the top-hat function,
\begin{align}
\mu\left(u;\epsilon/\sqrt{3}\right)=
T(u\,;\epsilon)\equiv
\begin{cases}
    0   & \text{if  $\abs{\,u\,} > \epsilon/2$} \\
   1/(2\epsilon) & \text{if $\abs{\,u\,} = \epsilon/2$}\\
   1/\epsilon & \text{if $\abs{\,u\,} < \epsilon/2$}.
\end{cases}
\end{align}
This function is discontinuous at ${u=\pm\epsilon/2}$, but I will sometimes use it when I
require $\mu$ to be differentiable. In those cases I use it with 
the understanding that I am using a differentiable function that approximates the top-hat function
arbitrarily closely.

I will refer to the average with a top-hat kernel as a {\em simple spatial average}, 
I will denote it by ${\expval{\nu}_\epsilon (x)}$, and although I will express it as
\begin{align}
\expval{\nu}_\epsilon (x) \equiv 
\frac{1}{\epsilon}\int_{-\epsilon/2}^{\epsilon/2}\nu(x+u)\dd{u},
\label{eqn:simple_average}
\end{align}
I do so with the understanding that
whenever ${\expval{\nu}_\epsilon(x)}$ is required to be a differentiable function of either $x$ or $\epsilon$, it is implied
that the spatial average is defined by Eq.~\ref{eqn:average1}, with a top-hat kernel ${\mu(\epsilon/\sqrt{3})}$ 
whose corners are arbitrarily sharp, but differentiable.

\begin{figure}[h]
\includegraphics[width=8.0cm]{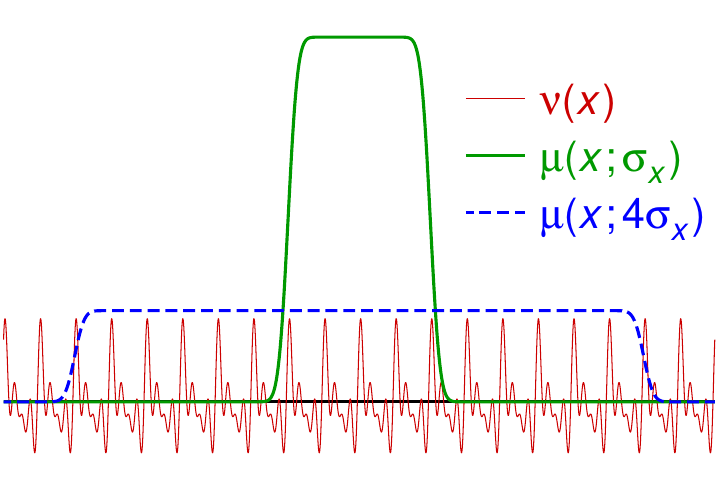}
\caption{A periodic function $\nu(x)$ and two smooth `top-hat' averaging kernels, ${\mu(\sigma_x)}$ and ${\mu(4\sigma_x)}$, 
which differ only by their widths, which are ${\epsilon=\sigma_x\sqrt{3}}$ and ${4\epsilon=4\sigma_x\sqrt{3}}$, respectively.
Increasing the kernel's width increases the number of periods of $\nu$ that contribute to
the average. The derivative of $\mu$ is only non-zero at the edges. Therefore, the rate at which it decays to zero 
becomes less and less significant to the average of $\nu$ as  $\mu$ is widened.
This illustrates a result that applies to a much wider 
class of kernels than top-hat kernels: the average is independent of the kernel's shape
in the ${\sigma_x\to \infty}$ limit.
}
\label{fig:kernel}
\end{figure}

Figure~\ref{fig:kernel} depicts a periodic microstructure and two differentiable top-hat
averaging kernels that might be used to find its spatial average.
One of the kernels is four times the width of the other, but both are almost constant
almost everywhere: their derivatives are only finite near where they decay
to zero. 

The wider kernel averages four times more periods
of the microstructure than the narrower kernel, but the weighting
it applies to each one is smaller by a factor of four.
Therefore, any difference between the averages calculated with the two kernels
is a result of them applying different non-uniform weights to points near where they decay to zero.

As ${\epsilon}$ increases, the contribution to the average of points where the derivative ${\mu^{(1)}(\epsilon)}$ is
non-negligible becomes smaller relative to the contribution from points where
${\mu^{(1)}(\epsilon)}$ almost vanishes. This illustrates the fact that the average magnitudes of derivatives of $\mu(\epsilon)$ 
decay faster as $\epsilon$ increases than the average of $\mu(\epsilon)$ does (Eq.~\ref{eqn:kernel2}).
Therefore it illustrates the fact that the shape of $\mu$ becomes increasingly irrelevant to the value of ${\expval{\nu;\mu}_\epsilon(x)}$ 
as ${\epsilon}$ increases.

In much of what follows I will assume that $\epsilon$ is large enough that
the shape of $\mu$ is irrelevant and, for simplicity, I will
only consider simple spatial averages.

\subsection{Why ${\Nu\equiv\expval{\nu}_\epsilon}$ fails as a definition}
\label{section:finite_precision}
\begin{figure}[h]
        \centering
\includegraphics[width=8.5cm]{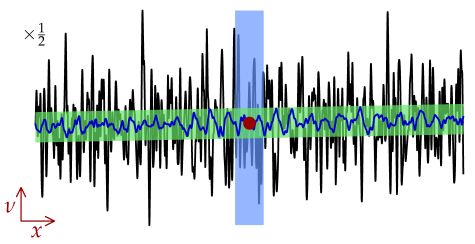} 
\caption{
The black curve is a microstructure, ${\nu(x)}$, after its fluctuations have been scaled 
by ${\frac{1}{2}}$ for visibility. The blue line within the green band is its simple spatial 
average ${\expval{\nu}_\epsilon}$ when ${\epsilon}$ is equal to the horizontal width of the vertical blue band.
The red dot is the average of ${\nu(x)}$ over all points $x$ within the blue band. 
The green band is the range of values that ${\expval{\nu}_\epsilon}$ takes on the segment of its domain shown.
Increasing the width, $\epsilon$, of the blue band reduces the width of the green band, but
its width only vanishes in the limit ${\epsilon\to\infty}$. 
}
\label{fig:homogenization1a}
\end{figure}
Figure~\ref{fig:homogenization1a} is a plot of a microstructure $\nu$ and its spatial
average, ${\expval{\nu}_\epsilon(x)}$. The range of values taken by ${\expval{\nu}_\epsilon}$
in a microscopic neighbourhood of the red dot is indicated by the almost-horizontal
green band. 
As $\epsilon$ increases, the microscopic fluctuations of ${\expval{\nu}_\epsilon(x)}$,
and therefore the width of the green band, reduce in magnitude as ${1/\epsilon}$ (Eq.~\ref{eqn:scaling}).
However they do not vanish.
They vanish only in the limit ${\epsilon\to \infty}$, which is 
the limit in which the average is performed over the entirety of $\nu$'s domain.
Therefore it is the limit in which ${\expval{\nu}_\epsilon}$
has the same value at every point, meaning that all structure has been lost.

Now let us assume that ${\Nu(x)\equiv\expval{\nu}_\epsilon(x)}$, for some finite value of $\epsilon$, 
so that the reasons why this definition fails become clear.

One reason why it fails is that two points $x_1$ and $x_2$, which are separated
by a microscopic distance ${\abs{x_1-x_2}<\amax}$, would be indistinguishable
at the macroscale. Measurements of ${\Nu(x_1)}$ and ${\Nu(x_2)}$ would
differ, despite appearing to have been performed at the same macroscale point.
Therefore $\Nu$ is not reproducibly-measureable \emph{at the macroscale}.

\begin{figure}[h]
\includegraphics[width=8.5cm]{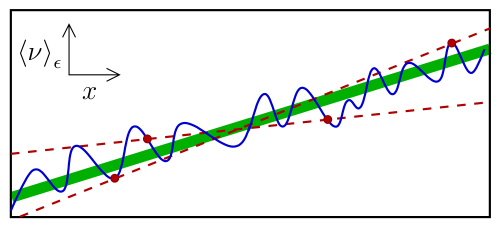} 
\caption{The blue curve, which is the simple average, ${\expval{\nu}_\epsilon}$, of $\nu$, 
fluctuates microscopically about the set of values plotted with a green band.  The magnitudes of these
fluctuations can be reduced by increasing $\epsilon$, 
but no matter how small
the fluctuations are, if they are finite, the finite-difference derivative of the blue curve differs from the 
slope of the green band, to some degree, for most choices of the two red points used to calculate it.}
\label{fig:differentiability}
\end{figure}

Another reason why it fails is that the finite difference derivative
${\left(\Nu(x+h)-\Nu(x)\right)/h}$ depends sensitively
on $x$ and $h$ and fluctuates microscopically as a function of each one, as illustrated
in Fig.~\ref{fig:differentiability}.
Therefore $\Nu$ is not differentiable \emph{at the macroscale}, because
its derivative does not converge with respect to $h$ 
while $h$ is still macroscopic or mesoscopic. It does not converge until $h$
is much smaller than the microscopic distances between successive
extrema of ${\Nu}$.

Both of these problems can be resolved by defining ${\Nu(\bx)}$ to 
be the set of \emph{all} values that would be measured
at the same macroscale point, $\bx$. 
This is a set of all spatial averages of ${\nu}$ centered at points in
an interval whose width is the lower bound, ${\prectheo=\abs{\dbx}}$ on distances
that are observable at the macroscale. 

Since $\abs{\dbx}$ is the limit of spatial precision
at the macroscale, the most precise measurements of $\Nu$ are either performed with
microscopically-blunt probes (radii ${\gtrsim \prectheo}$),
or with sharper probes whose positions can only be controlled or known to
within an interval of width ${\prectheo}$.

Therefore if $\prectheo$ is large enough that ${\expval{\nu;\mu}_\prectheo}$ is independent
of the shape of $\mu$,  the set of all measured values of ${\Nu(\bx)}$ is
the set of values of ${\expval{\nu}_\prectheo(x)}$ at microscale
points $x$ that are within an interval of width $\abs{\dbx}$ centered at $\bx$.
This set is an interval, ${\interval(\bbNu(\bx),\precNu(\bx))}$. Therefore 
${\Nu(\bx)}$ is only defined to a precision, $\precNu(\bx)$, that is finite.

\subsection{The macroscale infinitesimal, $\abs{\dbx}$}
\label{section:infinitesimal}
\begin{figure}[h]
        \centering
\includegraphics[width=8.5cm]{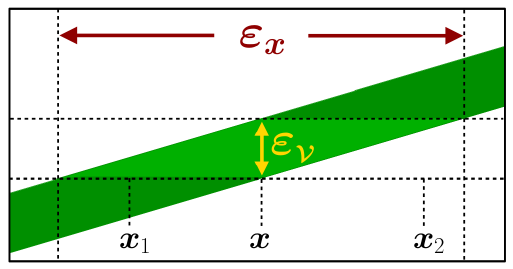}
\caption{
Plot of $\Nu$ versus $\bx$ when the precision $\precNu$ to which 
$\Nu$ is defined is finite. $\precNu$ is finite if 
the results of repeated accurate measurements of $\Nu$ at the same
macroscale point are not all equal, but only within $\precNu$ of one another.
It means that if the slope $\dNu{1}$ was
known, measurements of $\Nu$ could not be used to distinguish between two points, 
$\bx_1$ and $\bx_2$, reliably and conclusively, or to measure the distance between
them. Therefore, when measurements and observations at the macroscale are mediated
by macroscopic fields, 
there is an unavoidable imprecision ${\prectheo\equiv\abs{\dbx}}$
in positions, distances, and displacements.
}
\label{fig:homogenization1b}
\end{figure}
In this section I illustrate the fact that if $\Nu$ is only defined
to a finite precision, $\precNu$, the value of $\precNu$ imposes
a lower bound on the macroscale spatial precision, $\abs{\dbx}$.
I will then make the following assumption.
\begin{assumption}
The only limit on spatial precision at the macroscale, $\abs{\dbx}$, is 
the limit imposed by the finite precision, $\precNu$, to which $\Nu$ can be defined.
\end{assumption}
In other words, I will neglect all other sources of spatial imprecision in order to isolate and 
investigate imprecisions and uncertainties that are intrinsic to acts of observation 
in which the observer inhabits a length scale that is orders of magnitude larger than $\amax$.

Figure~\ref{fig:homogenization1b} illustrates why 
measurements of $\Nu$ cannot conclusively distinguish between 
$\bx_1$ and $\bx_2$ if ${\abs{\bx_1-\bx_2}< \precNu/\abs{\dNu{1}}}$.
In Sec.~\ref{section:uncertainty} I will present a discussion, for an arbitrary
choice of the averaging kernel $\mu$,  of the relationship
between $\prectheo$, $\precNu$, and the precision
$\precmom$ to which the derivative ${\dNu{1}}$ of $\Nu$ is
defined.

The macroscale infinitesimal ${\abs{\dbx}}$ 
is the smallest distance between empirically-distinguishable points, i.e.,
\begin{align*}
\abs{\dbx}\equiv\inf\bigg\{\abs{\bDx} :
\abs{\bx_1-\bx_2}>&\abs{\bDx}/2 \implies \bx_1\neq\bx_2,
\\
&\forall \bx_1,\bx_2\in\dom\Nu\bigg\}.
\end{align*}
This definition implies that distances smaller than 
${\abs{\dbx}}$ do not have meaning at the macroscale.
However, they do have meaning at the microscale, where 
${\abs{\dbx}}$ is denoted by ${\prectheo}$.
For simplicity, this definition also assumes that the value of ${\abs{\dbx}}$ 
is the same everywhere in ${\dom\Nu}$. 

At the microscale, ${\abs{x_1-x_2}<\prectheo/2=\abs{\dbx}/2}$ does not imply that 
${x_1=x_2}$. Therefore each point $\bx$ at the macroscale corresponds to an interval
of width $\prectheo$ at the microscale. 
I denote the midpoint of this interval by ${\barx(\bx)}$ and I refer
to the interval as the
{\em coincidence set} of ${\barx(\bx)}$. Mathematically, it is defined as
\begin{align*}
\left[\barx(\bx)\right]_{\Lequiv}\equiv
\left\{ x: x\Lequiv \barx(\bx)\right\}
=\interval(\barx(\bx),\prectheo),
\end{align*}
where
{\em macroscale coincidence}, $\Lequiv$, which is nontransitive and therefore
not an equivalence, is defined by
\begin{align*}
x_1\Lequiv x_2 \iff \abs{\,x_1-x_2\,} < \prectheo/2.
\end{align*}

The one-to-many relationship between points at the macroscale
and points at the microscale has important implications for the nature of macrostructure, 
which have already been discussed in Sec.~\ref{section:what_is_macrostructure}.
It implies that the transition from the microscale to the macroscale can be viewed
as a compression of space, which shrinks all microscopic distances to zero, 
resulting in surfaces and interfaces becoming literally planar, locally.

\subsection{Mutually-consistent values of $\precNu$ and $\prectheo$}
\label{section:finding_precision}
\begin{figure*}[!]
  \includegraphics[width=\textwidth]{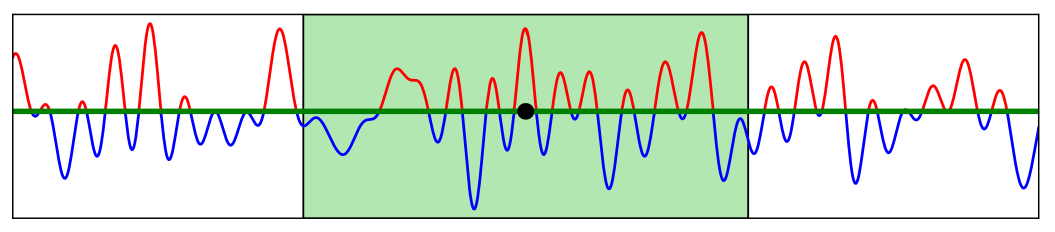}
\caption{Schematic. The red and blue curve represents a microstructure and the horizontal green
line is its average on the green-shaded interval whose center is marked by a black spot.
Although it seems natural to say that a microstructure {\em fluctuates microscopically} about its average,
its average is a microscopic function of both the width and the position of the averaging interval.
This means that the green line moves up and down
as either one of them changes, and that
the distances between successive extrema of these fluctuations of the average are microscopic.
There does not exist a unique 
value about which a microstructure fluctuates microscopically, in general.
because there is no reason to choose one interval width over another, or to choose to center
the interval at a particular point instead of one a microscopic distance away, 
}
\label{fig:fluctuates_microscopically}
\end{figure*}
It seems natural to say that $\nu$ fluctuates microscopically about its spatial average.
However the dependence of ${\expval{\nu}_\epsilon(x)}$ 
on $\epsilon$ means that the spatial average of $\nu$ at $x$ is not unique, and 
its dependence on $x$ means that the sets of all spatial averages on intervals of width less than $\epsilon$
centered at different macroscopically-coincident points are different. In other words, if ${\abs{x_1-x_2}>0}$, 
then, in general,
and notwithstanding the fact that ${x_1\Lequiv x_2}$, 
\begin{align*}
\left\{\expval{\nu}_\eta(x_1):0<\eta<\epsilon\right\}
\neq
\left\{\expval{\nu}_\eta(x_2):0<\eta<\epsilon\right\}.
\end{align*}

As discussed in Sec.~\ref{section:finite_precision}, 
the set of all possible accurately- and precisely-measured values of ${\Nu(\bx)}$ is
\begin{align}
\Nu(\bx)&\equiv 
\left\{\expval{\nu}_\prectheo(x): x\in\coincidence{\barx(\bx)}\right\}.
\label{eqn:Nuset}
\end{align}
However, on its own, this does not constitute a definition of ${\Nu(\bx)}$ because $\precNu$ determines $\prectheo$, 
so we cannot define $\precNu$ in terms of $\prectheo$.

Furthermore, we must take care to satisfy the requirement that $\nu$ fluctuates
microscopically about ${\Nu(\bx)}$ at $\bx$. If we choose an arbitrary mesoscopic
value of $\prectheo$, and then use Eq.~\ref{eqn:Nuset} as the definition
of the set of values about which ${\nu}$ fluctuates
microscopically at $\bx$, this requirement may not be satisfied.
For example, as Fig.~\ref{fig:homogenization1b} illustrates, if ${\abs{\dNu{1}}}$ is large enough, the sets 
\begin{align}
\left\{\expval{\nu}_\prectheo(x-\prectheo/2+u): 0<u<\amax\right\}
\end{align}
and
\begin{align}
\left\{\expval{\nu}_\prectheo(x+\prectheo/2-u): 0<u<\amax\right\}
\end{align}
do not intersect. Therefore, although ${x_1\in\coincidence{\barx(\bx)}}$ is required to imply that ${\expval{\nu}_\prectheo(x_1)}$ is
among the set of values of $\Nu$ that might be measured at $\bx$, if $\prectheo$ is not chosen carefully, 
and if the phrase
`\emph{$\nu$ fluctuates microscopically about $\upsilon$ at $x$}' is defined to mean
\begin{align*}
\upsilon\in\left\{\expval{\nu}_\prectheo(x+u): -\amax/2<u<\amax/2\right\},
\end{align*}
${\expval{\nu}_\prectheo(x_1)}$ may not be a value about which $\nu$ 
fluctuates microscopically at another point ${x_2\in\coincidence{\barx(\bx)}}$.

To remedy this problem we should define this phrase 
without referring to $\prectheo$, and then choose $\prectheo$ such that
$\Nu(\bx)$, as defined by Eq.~\ref{eqn:Nuset}, 
is a {\em subset} of the set of values about which $\nu$ fluctuates microscopically at 
\emph{every} ${x\in\coincidence{\barx(\bx)}}$.

I now propose possible definitions of the phrases 
`\emph{fluctuates microscopically about $\upsilon$ at $x$}'
and
`\emph{fluctuates microscopically about $\upsilon$ at $\bx$,}'
which I have not 
justified rigorously.
I present them to illustrate the difficulties with circular definitions, 
and because they might
be useful as a starting point for the development of rigorously-justified 
definitions that seamlessly link microstructure to what is measured
and observed at the macroscale.

\begin{definition*}
\emph{$\nu$ fluctuates microscopically
about $\upsilon$ at $x$} if and only if there exists a microscopic interval centered at $x$
on which the average of $\nu$ is $\upsilon$, i.e., if and only if 
\begin{align}
\upsilon\in A_\amax [\nu](x) & \equiv \Interior{\left\{ \expval{\nu}_{\eta}(x): \eta < \amax \right\}},
\label{eqn:spread}
\end{align}
\end{definition*}
In this expression `$\Interior$' denotes the interior 
of the set, meaning the set without its boundary points. 
Although 
it is not necessary to define
${A_\amax}$ to be an open set here, 
in more rigorous investigations of its properties
I have found it useful or necessary to define it as open.

\begin{definition*}
\emph{$\nu$ fluctuates microscopically about $\upsilon$ at $\bx$} if and only if
\begin{align}
\upsilon\in B_\amax^\prectheo[\nu](\barx(\bx))& \equiv
\bigcap_{x'\in\interval(\barx(\bx),\prectheo)} A_\amax[\nu](x').
\nonumber
\end{align}
\end{definition*}
\begin{definition*}
\emph{$\nu$ fluctuates microscopically about $\Nu$} if and only if
\begin{align*}
\Nu(\bx)\subseteq B^\prectheo_\amax[\nu](\barx(\bx)),\;\;\forall\bx\in\dom\Nu.
\end{align*}
\end{definition*}

Note that Eq.~\ref{eqn:Nuset} implies that
\begin{align*}
\Nu(\bx)\subseteq \bigcup_{x'\in\interval(\barx(\bx),\prectheo)} \closure A_\prectheo[\nu](x'),
\end{align*}
where ${A_\prectheo[\nu](x')\equiv\Interior\{\expval{\nu}_\eta(x'):\eta<\prectheo\}}$, and
the \emph{closure operator}, $\closure$, closes a set by adding its boundary points to it.

The following assumption specifies the domain
of validity of the three definitions proposed above.
\begin{assumption}
\label{assumption:eight}
There exist values
of $\amax$ and ${\prectheo}$ such that ${\amax\ll\prectheo<\Lmin}$,
and such that 
\begin{align*}
\Nu(\bx)
&\equiv
\left\{\expval{\nu}_\prectheo(x): x\in\coincidence{\barx(\bx)}\right\}
\subseteq B^\prectheo_\amax[\nu](\barx(\bx)),
\end{align*}
at every point ${\bx\in\dom\Nu}$.
\end{assumption}
Note that microstructures with perfect periodicities are pathological in various ways, but they are also 
unphysical because there always exists some degree of disorder. Even the periodicity of the
time average of a crystal's microstructure on an interval of length $\tau$
only has perfect periodicity in the limit ${\tau\to\infty}$.
Therefore I propose, as a conjecture, that Physical Assumption~8 
holds true for a useful subset of physical (disordered) microstructures, which
satisfy the first seven physical assumptions stated earlier in Sec.~\ref{section:homogenization}.

\subsection{Uncertainty principle}
\label{section:uncertainty}
I have discussed the case of a simple average, with 
a top-hat kernel, in some detail. The purpose
of this section is to discuss, in more general terms, how
the shape and width of the averaging kernel determine
unavoidable uncertainties in measured values of ${\Nu}$ and its
derivative, ${\dNu{1}}$. By analogy with Eq.~\ref{eqn:Nuset}, 
the set of values of ${\dNu{1}}$ that could be measured
at a point $\bx$ is
\begin{align}
\dNu{1}(\bx)&\equiv 
\left\{\expval{\partial_x\nu}_\prectheo(x): x\in\coincidence{\barx(\bx)}\right\},
\end{align}
and I denote the width of this interval by $\precmom$, where the subscript `$p$'
is intended to be reminiscent of momentum in quantum mechanics.

I will now derive a relationship between ${\precNu}$, $\precmom$, and $\prectheo$. 
For simplicity I will denote the spatial
average, ${\expval{\nu;\mu}_\prectheo}$, simply as ${\expval{\nu}}$,
I will denote the average of the average 
as ${\expval{\expval{\nu}}\equiv\expval{\expval{\nu;\mu}_\prectheo;\mu}_\prectheo}$, 
etc..

To a first approximation, the uncertainty in the value of  ${\dNu{1}}$ can be quantified
by the variance of ${\expval{\partial_x\nu}}$.
Let us use the fact that spatial derivatives commute with
spatial averaging to write
\begin{align*}
\expval{\partial_x\nu}&-\expval{\expval{\partial_x\nu}}
= \partial_x\left[\expval{\nu}-\expval{\expval{\nu}}\right]
= \partial_x\expval{\nu-\expval{\nu}}
\\
&=\partial_x\expval{\Dnu}
=
\partial_x\left(\mu(\prectheo)\ast\Dnu\right) = \mu(\prectheo)\ast\left(\partial_x\Dnu\right),
\end{align*}
where ${\Dnu(x)\equiv\nu(x)-\expval{\nu}(x)}$. 
In more explicit notation, this can be expressed as
\begin{align*}
\expval{\partial_x\nu}(x)&-\expval{\expval{\partial_x\nu}}(x) 
= \int_\realone \mu(u;\prectheo)\partial_x\Dnu(x+u)\dd{u}.
\end{align*}
Let us replace ${\int_\realone}$ with ${\int^\intmax_{-\intmax}}$, where the value of
${\intmax=\intmax(\prectheo)}$ has been chosen such that
\begin{align*}
\Bigg|\int^{\intmax}_{-\intmax} \mu(u;\prectheo) &\partial_x\Dnu(x+u)\dd{u}
\\
&-
\lim_{\intmax\to\infty} \int^{\intmax}_{-\intmax} \mu(u;\prectheo) \partial_x\Dnu(x+u)\dd{u}\Bigg|
\end{align*}
is negligible. 
If we also replace ${\mu(u;\prectheo)}$ with its Taylor expansion 
about ${u=0}$, we find     
\begin{align*}
\expval{\partial_x\nu}-\expval{\expval{\partial_x\nu}}
&=
\mu(0;\prectheo)\int^{\intmax}_{-\intmax} \partial_x\Dnu(x+u)\dd{u}
\\
&+
\sum_{m=1}^\infty \frac{\mu^{(m)}(0;\prectheo)}{m!}\int^{\intmax}_{-\intmax} u^m \partial_x\Dnu(x+u)\dd{u}.
\end{align*}
The integrals appearing in the sum on the right hand side 
can be expressed as
\begin{align*}
\int_{-\intmax}^\intmax u^m &\partial_x\Dnu(x+u)\dd{u}
\\
&=\int_0^\intmax u^m \left[\partial_x\Dnu(x+u)\pm\partial_x\Dnu(x-u)\right]\dd{u},
\end{align*}
where $\pm$ is $+$ when $m$ is even and $-$ when $m$ is odd.
In both cases there is partial cancellation, which 
reduces  the magnitudes of the integrals by a factor of about ${1/\sqrt{2}}$.
We know from Eq.~\ref{eqn:kernel2} that when $\prectheo$
is large the $m^\text{th}$ derivative of the kernel, ${\mu^{(m)}(0;\prectheo)}$, 
scales as ${1/\prectheo^m}$.  Furthermore, if $\mu$ is symmetric, then ${\mu^{(1)}(0;\prectheo)=0}$
and the ${m=1}$ term vanishes.

Therefore, to a first approximation, or in the limit of large $\prectheo$, 
the variance of the slope of $\Nu$ is
\begin{align*}
\left(\frac{\precmom}{2}\right)^2&\equiv\expval{\left(\dDNu{1}\right)^2} 
 =
\expval{\left(\expval{\partial_x\nu}-\expval{\expval{\partial_x\nu}}\right)^2}
\\
&\approx \mu(0;\prectheo)^2\expval{\left(\Dnu(x+\intmax)-\Dnu(x-\intmax)\right)^2}
\\
&=\mu(0;\prectheo)^2
\bigg[
\expval{\Dnu(x+\intmax)^2}
+
\expval{\Dnu(x-\intmax)^2}
\\
&\qquad\qquad-2\expval{\Dnu(x+\intmax)\Dnu(x-\intmax)}
\bigg]
\end{align*}
If we now assume that, for the purpose of
calculating ${\left(\precmom/2\right)^2}$,
the values of
${\Delta\nu(x+\intmax)}$ and
${\Delta\nu(x-\intmax)}$ can be treated as independent random variables with means
of zero and variances of ${\left(\precNu/2\right)^2}$, then
${\expval{\Dnu(x+\intmax)\Dnu(x-\intmax)}}$ vanishes and
we get
\begin{align}
\left(\frac{\precmom}{2}\right)^2
\approx 
2\mu(0;\prectheo)^2\left(\frac{\precNu}{2}\right)^2
\implies
\prectheo\precmom &\approx r_\mu\precNu,
\label{eqn:uncertainty1}
\end{align}
where 
${r_\mu\equiv\sqrt{2} \mu(0;\prectheo)\prectheo\sim 1}$, 
is dimensionless and with a value that depends
on the shape of $\mu$. If $\mu$ is Gaussian, then
${\mu(0;\prectheo)= (1/\prectheo)\sqrt{2/\pi}}$ and
${r_\mu= 2/\sqrt{\pi}}$.
If $\mu$ is a top-hat, then ${\mu(0;\prectheo)= 1/\left(\prectheo\sqrt{3}\right)}$
and ${r_\mu=\sqrt{2/3}}$.
If ${\sigma_x\equiv\prectheo/2}$, ${\sigma_{\mathcal{V}}\equiv\precNu/2}$,
and ${\sigma_p\equiv\precmom/2}$, Eq.~\ref{eqn:uncertainty1} can
be expressed as
\begin{subequations}
\begin{align}
\sigma_x\sigma_p & = r_\mu\stdNu/2\qquad\text{(General kernel)}
\\
\sigma_x\sigma_p & = \stdNu/\sqrt{\pi}\qquad\text{(Gaussian kernel)}
\\
\sigma_x\sigma_p & = \stdNu/\sqrt{6}\qquad\text{(Top-hat kernel)}
\end{align}
\label{eqn:uncertainty_principles}
\end{subequations}
These relations imply that 
there is a trade-off between macroscale spatial precision and
the uncertainty in ${\dNu{1}}$. When microscopic
fluctuations of $\nu$ are large, ${\precNu=2\stdNu}$ is large, and
${\prectheo\precmom=4\sigma_x\sigma_p}$ is large.

\color{black}
\subsection{Summary of the fundaments of homogenization theory} 
In this section I have discussed some of the fundamental
features of the homogenization transformation that turns
microstructure into macrostructure.
I have pointed out that macrostructure cannot be defined
uniquely, because it depends on 
the scale, ${\prectheo=\abs{\dbx}}$, at which the microstructure is observed, 
and which defines the smallest distance, ${\abs{\dbx}/2}$, between mutually-distinguishable
points at the macroscale.

However the value of $\abs{\dbx}$ cannot be chosen to be 
arbitrarily small if distances and displacements are measured with 
macroscopic fields.
This is because $\prectheo$ both determines, and is bounded from below by, 
the finite precisions, $\precNu$ and $\precmom$, to which macroscopic fields
and their derivatives, respectively, are defined. 
Therefore $\prectheo$, $\precNu$, and $\precmom$ are all interrelated, and can be interpreted either
as unavoidably-finite precisions or as measures of unavoidable uncertainty.
In Sec.~\ref{section:uncertainty} I derived uncertainty relations which imply that 
reducing $\prectheo$ increases the uncertainty $\precmom$ in derivatives of the macroscopic
field used to measure distances and displacements.

I have not presented a rigorously-justified relationship between
$\prectheo$ and $\precNu$, for an arbitrary microstructure, $\nu$, 
which satisfies my physical assumptions.
In part, this is because any such definition would have to be accompanied
by further physical assumptions, which specified more precisely the set of microstructures to which it would
apply. However, I have highlighted some of the difficulties that must
be overcome to devise rigorously-justified definitions, and I have
proposed a definition that I have found 
to be viable for a useful subset of microstructures that satisfy
my physical assumptions.
I will present these numerical and theoretical findings elsewhere.

The domain of validity of my proposed definitions is not
relevant to the two most fundamental conclusions of Sec.~\ref{section:homogenization}.
The first of these is that homogenization introduces unavoidable uncertainty 
at the macroscale, making spatial precision, 
and the precisions to which macroscopic fields are defined, finite.
The second is that finite spatial precision has important implications
for the nature of macrostructure, which
I discussed in Sec.~\ref{section:what_is_macrostructure}, 
and will discuss further in Sec.~\ref{section:excess_fields}.

Briefly, it means that surfaces and interfaces, which
do not exist in a well-defined sense at the microscale,  are created by the homogenization
transformation. When they are created they carry
excess fields, in general, and these fields are an integral component of macrostructure.
In fact, because the macroscopic charge density $\Rho$ vanishes in the bulks
of stable materials, in the context of electricity it can be the case that the excess fields, and the 2-, 1-, and 0-manifolds
they inhabit, \emph{are} the macrostructure.

Therefore the task of laying foundations for a homogenization theory that defines macrostructure in terms of 
microstructure, $\nu$, is far from complete. Completing the foundations entails defining excess fields in terms of $\nu$.
This is the subject of Sec.~\ref{section:excess_fields}.

It is straightforward to generalize the theory presented
in this section to systems in which the microstructure varies significantly
on three or more widely-separated length scales. In that case homogenization
proceeds in stages from the \emph{base microstructure}, on the smallest length scale, 
to the {\em apex macrostructure}, on the largest length scale.
The base microstructure is the only microstructure that is not
also a macrostructure determined by a microstructure on a smaller length scale, and
the apex macrostructure is the only macrostructure
that is not also a microstructure which determines a macrostructure on a larger
length scale.

It may not be straightforward to adapt the theory presented in 
this section to materials whose structures
vary significantly on every length scale, such as wood~\citep{wood}. 
However, such an adaptation may not be useful because, 
unlike most artificial materials,
wood does not appear to be locally homogeneous when observed with
either the naked eye or a microscope at any level of magnification.

\section{Excess fields}
\label{section:excess_fields}
\subsection{Introduction}
As discussed in Sec.~\ref{section:homogenization}, 
macrostructure essentially consists of 
smoothly-varying continua interspersed with
heterogeneitites such as defects, interfaces, and inclusions.

An inclusion is an embedded region whose microstructure differs 
from that of its host.
If all of its dimensions are smaller than ${\prectheo=\dbx}$, an inclusion 
is a singular point in the volumetric macroscopic field $\Nu$ and therefore
a macroscale point defect. 
Similarly, a macroscale
line [planar] defect is an inclusion that is larger than ${\prectheo}$ along only one [two] of 
its dimensions. Macroscale defects may require special treatments when applying
the macroscale theory, but their macrostructures can be calculated from the microstructures
by reasonably-straightforward application of the three-dimensional mesoscale averaging
operation.

Macroscale defects should not be confused with their microscale counterparts.
For example, consider vacancies and impurities in crystals, which are microscale point defects.
Although they may be charged, and therefore may contribute directly to the microscopic charge density $\rho$,
and they may perturb the arrangement of atoms, thereby indirectly changing $\rho$, their concentrations
are usually high enough and/or their effects on averages of $\rho$ small enough, that they can be regarded
as just another feature of the microstructure. Their presence in a crystal does not
alter the relationship between $\Rho$ and $\rho$ in most cases.

Exceptionally, microscopic defects might increase the upper bound, $\amax$, on distances regarded
as microscopic so much that it becomes comparable to $\Lmin$, thereby rendering Physical Assumpion~\ref{assumption:eight},
and much of the theory presented in Sec.~\ref{section:homogenization} invalid. However, I restrict attention
to systems in which ${\amax\ll\Lmin}$. 

If all of an inclusion's dimensions are much larger than ${\prectheo}$, the curvature
of its macroscale boundary with the host material will be negligible on the mesoscale.
Therefore, the inclusion is simply another macroscale material whose boundary with its host 
is locally flat. 
On either side of that boundary $\Nu$ is differentiable
and the boundary itself can be treated like any other mesoscopically-planar interface.
As discussed in Sec.~\ref{section:homogenization}, not only does $\Nu$ tend to be discontinuous at 
interfaces, but interfaces carry excesses ${\bsigmaNu}$ of $\Nu$, in general, 
which play important roles in physics at the macroscale.

I am trying to emphasise that, for a large class of systems, and from a practical perspective, the only ingredient 
of a mutually-consistent description of macrostructure and microstructure that we lack is the relationship between microstructure $\nu$ and 
the macroscale interfacial excess $\bsigmaNu$.
The purpose of this section is to derive this relationship. 
Specifically, I derive expressions for excesses at surfaces (interfaces with a vacuum), which are
trivial to generalize to interfaces by treating them as adjoined surfaces.

\subsubsection{Notation}
In this section, and henceforth, I will assume that $\Nu$ is
a single-valued field at the macroscale, which is
a mesoscale spatial average of $\nu$ that I will
often denote by ${\bnu}$. At the microscale
I will denote
the midpoint of interval ${\Nu(\bx)\equiv\interval(\bbNu(\bx),\precNu)}$, 
by ${\bNu(\barx(\bx))}$, and I will assume that ${\bbNu(\bx)}$ is the
average of ${\bNu(x)}$ over all ${x\in\coincidence{\barx(\bx)}}$.

As in Sec.~\ref{section:uncertainty}, ${\intmax=\intmax(\prectheo)>\prectheo}$ will denote the mesoscopic width
of the domain of a mesoscale spatial average. Increasing its value makes
the approximation
\begin{align*}
\bnu(x) \equiv\expval{\nu;\mu}_{\prectheo}(x)&\equiv\int_\realone \nu(x+u)\mu(u;\prectheo)\dd{u}
\\
&\approx \int_{-\intmax/2}^{\intmax/2} \nu(x+u)\mu(u;\prectheo)\dd{u}
\end{align*}
an arbitrarily close one. I introduce this finite width 
to help with derivations and I do not attach physical meaning to it.

In Fig.~\ref{fig:homogenization1b}, while considering the example
of a top-hat kernel, I defined ${\prectheo=\abs{\dbx}}$ such 
that if the distance between two macroscale points was greater than 
${\abs{\dbx}/2}$, they could be distinguished from one another, 
\emph{with certainty}, by macroscale measurements.  However, real measurements
do not provide
certainty  - only probabilities and degrees of certainty.

In this section I will consider spatial averages
with an arbitrary kernel, but to avoid cluttering and complicating
the theory and discussion, I will not discuss probabilities.
I will continue to refer to 
$\prectheo$ as the macroscopic spatial precision, and as
the width of an interface at the microscale, but with the understanding
that ${\prectheo/2}$ is now the standard deviation of the
position probability density function, ${\mu(\prectheo)}$.
In other words, I will continue
to use precise non-probabilistic language and mathematics, while
cognisant of the fact that this preciseness is unjustified.
For example, I will continue to regard the coincidence set ${\coincidence{\barx(\bx)}}$
of $\bx$ as a well-defined set of points, and I will continue to discuss an interface as having 
the precisely-defined width, $\prectheo$.

If this sloppiness introduces 
doubt about the validities of the derivations that follow, this doubt can 
be removed by strengthening our
physical assumptions about the nature of microstructure: We can assume
that ${\amax/\prectheo}$ and ${\prectheo/\Lmin}$ are both so small that the shape of $\mu$ has
a negligible influence on the macrostructure (see Eq.~\ref{eqn:kernel2}).
Then $\mu$ is effectively a top-hat kernel. To achieve further comfort, by reverting
to perfect consistency with our discussion of top-hat kernels in Sec.~\ref{section:homogenization}, 
we could mentally replace every instance of ${\mu(\prectheo)}$
in what follows with ${\mu(\prectheo/\sqrt{3})}$.

\subsection{Surface excesses in three dimensions}
\label{section:micromacrointerface}
Far from an interface, the relationship between the macrostructure
and the microstructure is as described, but not rigorously and precisely defined, in 
Sec.~\ref{section:homogenization}.
Homogenizing the interface region presents new problems as a consequence
of the fundamental difference between interfaces at the macroscale
and interfaces at the microscale, which I briefly discussed in Sec.~\ref{subsection:excess_fields}.
These are that interfaces are ill-defined at the microscale, because their widths are indeterminate, 
whereas at the macroscale they are well-defined two dimensional manifolds which carry excess fields.

In Sec.~\ref{subsection:excess_fields} we considered excesses of
one dimensional microstructures. Let us begin our discussion of 
excesses of
three dimensional microstructures by considering 
an excess field on a surface that is perpendicular to the $\bx-$axis and whose
macroscale $\bx-$coordinate is $\mxl$ (see Fig.~\ref{fig:surfcharge}). Let ${x_L\equiv\barx(\mxl)}$.
Above the surface, by which I mean  ${\bx<\mxl}$ at the macroscale and ${x<x_L-\prectheo/2}$ at the microscale,
there is vacuum, meaning that both ${\nu}$ and ${\Nu}$ are zero. 
Let us assume that $\Nu$ is also zero far below the surface, 
but that the average $\yznu(x)$ of $\nu$ on the plane parallel to the surface at $x$ does not vanish for every $x$.
This means that the mesoscale average of $\nu$ only vanishes if 
\begin{align*}
\int_{0}^{\intmax/2}\yznu(x+u)\mu(u;&\prectheo)\dd{u} 
\\
&= -\int_{-\intmax/2}^0\yznu(x+u)\mu(u;\prectheo)\dd{u},
\end{align*}
and, in general, neither of these integrals is zero.
Therefore, unless the average of $\nu$ vanishes on all 
planes parallel to the surface, $\Nu$ can only vanish if the
contributions to it from different depths below the surface cancel one another.

It follows that if we create a surface perpendicular to the $\bx-$axis 
by removing all material from one side of an 
imaginary plane passing through the bulk, 
the removal of this material disrupts the cancellation that
causes $\Nu$ to vanish. Therefore, within ${\pm\prectheo/2}$ of the surface, ${\Nu}$
does not vanish.
The integral of $\Nu(\bx,\by,\bz)$ between 
any point above the surface,
${\bx_1<\mxl-\absdbx/2}$, and any point below it, 
${\bx_2>\mxl+\absdbx/2}$, is
\begin{align}
\int_{\bxo}^{\mxl-\absdbx/2}&\Nu(\bx,\by,\bz)\dbx  
+\int^{\mxl+\absdbx/2}_{\mxl-\absdbx/2}\Nu(\bx,\by,\bz)\dbx \nonumber \\
 +\int^{\bxt}_{\mxl+\absdbx/2}&\Nu(\bx,\by,\bz)\dbx  
 =  \Nu(\mxl,\by,\bz)\abs{\dbx} \neq 0 \nonumber
\end{align}
The first and the third integrals are zero because ${\Nu(\bx,\by,\bz)=0}$ if ${\abs{\bx-\mxl}>\abs{\dbx}/2}$.
Therefore, at position ${(\by,\bz)}$ on the surface plane, the excess of $\Nu$ is 
${\bsigmaNu(\by,\bz)\equiv \Nu(\mxl,\by,\bz)\abs{\dbx}}$. The surface average, $\barbsigmaNu$, of 
$\bsigmaNu$ is ${\yzNu(\mxl)\abs{\dbx}}$, where ${\yzNu}$ is the macroscale counterpart of 
${\yznu}$, meaning its average on a mesoscopic two dimensional domain.

\subsection{Calculating interfacial excesses from the microstructure}
\label{section:calculating_excess}
To address the question of how excess fields can be calculated from the microstructure, $\nu$.
let us continue to assume that the $x$-axis is normal to the surface and, to
simplify the notation by keeping the problem one-dimensional, let us assume 
that $\nu(x)$ 
is the average of some other microscopic quantity on the plane parallel to the surface at $x$.
As before, let us assume that ${\Nu=0}$ in the bulk.
An obvious starting point is to define the microscale surface excess, $\sigma_\nu$, as 
\begin{align}
\sigma_\nu(x_b) = \int_{x_L-\prectheo/2}^{x_b}\nu(x)\,\dd{x}
\label{eqn:surfxs1}
\end{align}
where ${\nu(x)=0}$ if ${x<x_L-\prectheo/2}$, and $x_b$ is a point 
deep below the surface (`b'=`bulk').

\begin{center}
\begin{figure*}
\includegraphics[width=16cm]{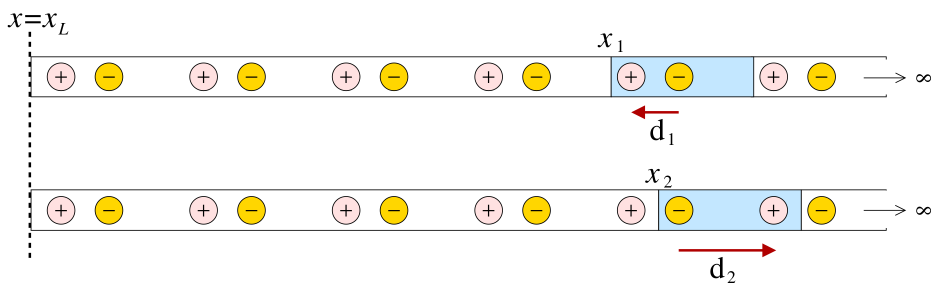}
\caption{
Cartoons depicting the surface (at $x=x_L$) of a one dimensional material.
The sums $\sigma_s(x_1)$ and $\sigma_s(x_2)$
of the charges between the surface at $x=x_L$ and $x_1$ and $x_2$, respectively, are very different.
Therefore neither can be identified as the surface charge.
The blue shaded regions are two different equally-valid choices of the unit cell of
the bulk crystal, which have very different
dipole moments $d(x_1+\frac{a}{2})$ and $d(x_2+\frac{a}{2})$, where
$d(x)$ is the dipole moment of the unit cell centered at $x$.
It was shown by Finnis~\citep{finnis} that the excess surface charge is
\begin{align*}
\sigma=\sigma_s(x_1) + \frac{d_1}{a}=\sigma_s(x_2) + \frac{d_2}{a}.
\end{align*}
Although the choice of unit
cell to describe the periodicity of the crystal changes
the ``dipole moment density'' $\mpp(x)=d(x)/a$ at every point,
the excess surface charge is independent of this choice and is well defined.}
\label{fig:surfcharge}
\end{figure*}
\end{center}

To see that ${\bsigmaNu\equiv\sigma_\nu(x_b)}$
is not a good definition of the macroscale surface excess, 
consider the example depicted in Fig.~\ref{fig:surfcharge}. In this example, the material 
could be a three dimensional crystal and $\nu$ 
the average of the charge density over planes parallel to the surface. The value of $\nu$ is zero everywhere
except at a discrete set of $x-$values, corresponding to lattice planes, on which it is either $+1$ or $-1$. 
Therefore, calculating $\sigma_\nu(x_b)$ is as simple as counting these charges from $x<x_L$ to 
$x=x_b$. By inspection, we find that ${\sigma_\nu(x_1)=0}$ and ${\sigma_\nu(x_2)=+1}$, where $x_1$ and 
$x_2$ are the positions indicated in Fig.~\ref{fig:surfcharge}.
If one continues counting beyond ${x=x_2}$, the value of $\sigma_\nu$ continues 
to jump between $0$ and $+1$ and it never converges.

The problem with defining ${\bsigmaNu\equiv\sigma_\nu(x_b)}$ is twofold. First, identifying ${\sigma_\nu(x_b)}$ as the surface
excess appears to imply that ${x<x_b}$ is the surface region and ${x>x_b}$ is the bulk.
However, as discussed in Sec.~\ref{subsection:excess_fields}, there is no clear boundary between surface and bulk
at the microscale and so the ``surface region'' is ill-defined.
Second, although  $\Nu$ vanishes in the bulk, the same
is not true of $\nu$, and any integral of a
 microscopic quantity is a microscopic function of its upper and lower
bounds of integration. Therefore, $\sigma_\nu(x_b)$ 
fluctuates microscopically as $x_b$ is varied.

This simple example, which is typical rather than pathological, illustrates how interfaces being ill-defined at the microscale
can be troublesome when one attempts to calculate macroscale properties
of interfaces from the microstructure. 
It also underscores the importance of a careful understanding of the relationship 
between microscale physics and macroscale physics.

To deduce the relationship between $\bsigmaNu$ and $\nu$,
consider the following two slightly-different
lines of reasoning.
The first is to define $\bsigmaNu$ as the mesoscale average
of $\sigma_\nu(x_b)$. This means that, instead of terminating
the integral at a 
single plane (at $x_b$) we take an average over an ensemble
of terminating planes. This was the approach taken by Finnis~\citep{finnis}, who
appears to have been the first to solve the problem of calculating 
what he called {\em thermodynamic excesses}
of charge and other quantities at interfaces. In \REF~\linecite{finnis} he reasoned that,
by averaging over terminating planes, ``{\em we can reconcile the atomistic picture, 
in which excesses appear to oscillate on the atomic length scale as a function of the region
size, with the thermodynamic picture}''. He used this approach
to derive an expression for the surface charge in crystals. One 
purpose of Sec.~\ref{section:excess_derivations} is to generalize
his result to noncrystalline materials.

The second line of reasoning,  which may seem more natural in the present context,
begins with the fact that surfaces, and therefore surface excesses, are only well-defined
at the macroscale.
Therefore, $\bsigmaNu$ must be the integral of $\Nu$ across 
the surface, i.e.,  
\begin{align*}
\bsigmaNu \equiv \int_{\bx_1}^{\bx_2}\Nu(\bx)\,\dbx,
\end{align*}
where ${\bx_1<\mxl}$ and ${\bx_2>\mxl}$.
This integral must converge to the surface excess 
because $\Nu=0$ in the bulk and above the surface, and because the spatial averaging operation 
is conservative, by virtue of $\mu$ being normalized to one.
It is straightforward to show that this viewpoint and Finnis's thermodynamic viewpoint
are equivalent, because the spatial average of an integral of $\nu$ is equal to
the integral of the spatial average of $\nu$.

\subsection{Changes of macroscale quantities across interfaces}
To calculate the change in $\Nu$ between a point ${\bx_2}$ on one
side of an interface and a point ${\bx_1}$ on the other, one 
could simply calculate ${\Nu(\bx_1)}$ and ${\Nu(\bx_2)}$ from
the microstructure, $\nu$. However, this is not always the easiest
approach. For example, if ${\dnu{1}}$ is known, 
but $\nu$ is not, it might be easier to recognise that the change of $\Nu$ across the interface is
the interfacial excess of $\dNu{1}$. Therefore, 
\begin{align}
\Nu(\bx_2)-\Nu(\bx_1)& = \bsigmadNu\equiv \int_{\bx_1}^{\bx_2}\dNu{1}(\bx)\,\dbx.
\label{eqn:surfxs2}
\end{align}
There are many important physical systems in which $\nu$ is related to a source function, $\psi$, 
by the Poisson equation, ${\dnu{2}=\psi}$. Since differentiation and spatial averaging
commute, their macroscale counterparts have the same relationship, ${\dNu{2}=\bpsi}$.
When $\Nu$ has different values
on either side of the interface, but is constant on both sides, 
the step change in its value across the interface can be calculated
from $\psi$ by integrating Eq.~\ref{eqn:surfxs2} by parts and substituting the Poisson equation to give
\begin{align}
\Nu(\bx_2)-\Nu(\bx_1)& = -\int_{\bx_1}^{\bx_2}\bx\,\bpsi(\bx)\,\dbx 
\label{eqn:surfxs3}
\end{align}
In Sec.~\ref{section:average_potential} we will find that Eq.~\ref{eqn:surfxs3} 
is useful way for calculating the change in the macroscopic potential $\bphi$  across
an interface, and therefore for calculating the 
\emph{mean inner potential}~\citep{bethe-1928,miyake-1940,mip_sanchez_1985,mip_pratt_1987, mip_pratt_1988, mip_pratt_1989, mip_pratt_1992, gajdardziska-1993, spence-1993, mip_sokhan_1997, mip_leung_2010, mip_mundy_2011,mip_cendagorta_2015,mip_water_2020,mip_kathmann_2021}.

For the purposes of calculating the interfacial excesses, and step-changes in macroscopic quantities across interfaces, 
that are required in this work about electricity, we only need to
deduce relationships between the right-hand-sides of Eqs.~\ref{eqn:surfxs2}
and~\ref{eqn:surfxs3} and the microstructures $\nu$ and $\psi$, respectively.
Mathematically, the problem at hand is to find simple and general expressions for
\begin{align}
\expval{\int_{x_1}^{x_2}\nu(x)\dd{x};\mu}_{\prectheo}
\;\;\text{and}\;\;
\expval{\int_{x_1}^{x_2}x\,\nu(x)\dd{x};\mu}_{\prectheo}
\label{eqn:integralave}
\end{align}
in terms of $\nu$, where the spatial average is performed over the upper bound, $x_2$, of the integrals at a
fixed value of $x_1$. Once these expressions are in hand, it will be straightforward to find expressions for the averages
of these integrals over $x_1$ or over both $x_1$ and $x_2$.

\subsection{Mesoscale averages of integrals}
\label{section:integral_averages}
The goal of this section is to deduce general relationships between the microstructure and
the integral averages in Expression~\ref{eqn:integralave}, which are
equivalent to the right-hand-sides of Eqs.~\ref{eqn:surfxs2} and~\ref{eqn:surfxs3}, apart from 
the appearance of $\bpsi$ instead of $\Nu$ in the latter.

Calculating the integral of ${\bNu}$, which is constant or linear on both sides
of an interface, is straightforward. Therefore, let us define ${\Dnu(x)\equiv \nu(x)-\bNu(x)}$, 
and instead calculate the mesoscale average over $x_b$ of
\begin{align}
\mss_r(x_b) & \equiv \int_{x_L}^{x_b} x^r\,\Dnu(x) \dd{x}
\label{eqn:mss}
\end{align}
for $r=0$ and $1$. I will denote these averages by ${\mbs_0}$ and ${\mbs_1}$.
The reasons for replacing $\nu$ by $\Delta\nu$ are that for ${r=0}$ the derivation
is made easier by the fact that $\Delta\nu$ fluctuates
microscopically about zero, and that the average of ${\mssx{\nu}_1(x_b)}$ does not converge with 
respect to $x_b$ unless $\nu$ fluctuates about zero.

Apart from those stated and discussed in Sec.~\ref{section:homogenization}, we will not 
make any assumptions about the microstructure. Therefore, our goal of 
deriving {\em generally-applicable} expressions for $\mbs_r$ in terms of $\nu$ is 
only possible if $\mbs_r$ can be related to some calculable characteristics
of the microstructure. Guided by Finnis's expression for the surface excess~\citep{finnis}, 
we will characterize the microstructure using moments and moment densities. 
This is explained in Sec.~\ref{section:partition} and Sec.~\ref{section:partitioning}.

\subsection{Partitioning space into microscopic intervals}
\label{section:partition}
To characterize the microstructure in the mesoscopic neighbourhood of $x_b$, let us assume
that the macrostructure is either uniform or linearly-varying in this neighbourhood.
Let us partition an interval of width $\intmax$ centered near $x_b$
into a set of $2 M$ contiguous microscopic subintervals, or {\em microintervals}, demarcated by
the set of points 
\begin{align*}
\Pi&(x_b,\intmax) \equiv \{x_m: m \in \mathbb{Z},\, \abs{m}\leq M,\,
x_0\equiv x_b,\, x_{m+1}>x_m,\nonumber \\
& x_{m+1}-x_m <\alpha,\, x_{-M}=x_b-\intmax/2,\, 
\abs{x_M-x_b-\intmax/2}<\amax\}
\end{align*}
Notice that, although ${\abs{x_{M}-x_{-M}}=\intmax(\prectheo)}$, the midpoint of ${(x_{-M},x_M)}$
is displaced from $x_b$ by a microscopic distance.  The reason for this will soon become clear.
The microinterval designated {\em `interval $m$'} and denoted by ${\interval_m\equiv\interval(\bar{x}_m,\Delta_m)}$ has midpoint 
\begin{align}
\bar{x}_m & \equiv \begin{cases} 
\frac{1}{2}\left(x_{m+1}+x_{m}\right) & \mbox{if } m<0 \\
\frac{1}{2}\left(x_{m-1}+x_{m}\right) & \mbox{if } m>0  ,
\end{cases} 
\end{align}
and width
\begin{align}
\Delta_m & \equiv \begin{cases} 
x_{m+1}-x_{m} & \mbox{if } m<0 \\
x_{m}-x_{m-1} & \mbox{if } m>0  .
\end{cases}
\end{align}
Now we can write the mesoscale average of $\nu$ as the following
 sum of integrals over microintervals:
\begin{align}
\bnu(x_b) 
& =\int_{-\intmax/2}^{\intmax/2} \mu(x_b-x;\prectheo)\,\nu(x)\dd{x} \nonumber \\
& = 
\sum_m
\int_{-\Delta_m/2}^{\Delta_m/2}\mu(x_b-\bar{x}_m-u;\prectheo)\,\nu(\bar{x}_m+u)\dd{u}  \nonumber \\
 & = 
\sum_m
\Delta_m \mu(x_b-\bar{x}_m;\prectheo)  \expval{\nu}_{\Delta_m}(\bar{x}_m),  \label{eqn:macroave3}
\end{align}
where $\sum_m$ denotes summation over all $2M$ microintervals
and we have used the fact that the change of $\mu$ across each 
microinterval is negligible when ${\amax/\prectheo}$ is sufficiently small.

Now let us place one further constraint on $\Pi(x_b,\intmax)$, which explains why $x_b$ is not the midpoint of
$(x_\mm,x_\m)$: The microinterval boundary points are chosen
such that ${\expval{\Dnu}_{\Delta_m}(\bar{x}_m)=0}$ for all $m$, which implies that
${\bnu(x)=\bNu(x)}$.
This choice is possible because $\Dnu$ fluctuates microscopically 
about zero everywhere in ${\interval(x_b,\intmax)}$. Therefore, starting from
$x_b$, the nearest point
$x_1>x_b$ such that the average of $\Dnu$ on ${[x_b,x_1]}$ is zero
must be a microscopic distance $\Delta_1$ away. The nearest point $x_2>x_1$ such that ${\expval{\Dnu}_{\Delta_2}(\bar{x}_2) = 0}$
is a microscopic distance $\Delta_2$ away, and so on.
Having chosen a set $\Pi(x_b,\intmax)$ for which the average of $\Dnu$ on each microinterval vanishes, 
Equation~\ref{eqn:macroave3} becomes
\begin{align}
\bar{\nu}(x_b) & =   
\sum_{m} \Delta_m \mu(x_b-\bar{x}_m;\prectheo) \bNu(\bar{x}_m)
\nonumber \\
& =  \,\intfull \mu(x_b-x;\prectheo)\bNu(x)\dd{x} = \bNu(x_b),
\nonumber
\end{align}
which is independent of $\mu$, as expected from Eq.~\ref{eqn:kernel2} in the
limit that ${\amax/\prectheo}$ vanishes.

\subsection{Characterising microstructure with moment distributions of microscopic intervals}
\label{section:partitioning}
Let us characterise the microstructure in interval $m$ by the set of moment densities 
\begin{align}
\many(\bar{x}_m,\Delta_m) 
\equiv 
\frac{1}{\Delta_m}
\int_{-\Delta_m/2}^{\Delta_m/2} \Dnu(\bar{x}_m+u) \, u^n \, \dd{u},
\nonumber
\end{align}
where $n=0,1,2,$ etc..
The zeroth moment density is simply the average of $\Dnu$ on
interval $[x_m,x_{m+1}]$, i.e., 
\begin{align}
\mzero(\bar{x}_m,\Delta_m)=\expval{\Dnu}_{\Delta_m}(\bar{x}_m).
\nonumber
\end{align}
Each moment density ${\many(\bar{x}_m,\Delta_m)}$ can be considered a microscopic quantity 
because its value fluctuates microscopically as a function of $\bar{x}_m$, at fixed $\Delta_m$, and
as a function of $\Delta_m$, at fixed $\bar{x}_m$.
Therefore, the set of all moment densities depends strongly on the choice of set $\Pi(x_b,\intmax)$, which 
is, to a large extent, arbitrary.

Let us define the mesoscale average ${\bmany(x_b)}$ of ${\many(\bar{x}_m,\Delta_m)}$ as
follows.
\begin{align}
\bmany(x_b) 
& \equiv   \expval{\mathcal{M}_\nu^{(1)};\mu}^*_\prectheo(x_b)   
 \equiv   \frac{1}{\intmax} \sum_m \Delta_m \mathcal{M}^{(n)}_\nu(\bar{x}_m,\Delta_m).
\nonumber
\end{align}
I have introduced the notation $\expval{\;.\;}^*$ to denote a particular kind of spatial
average - one which cannot be calculated by a continuous integral. 
It is a weighted average, over a discrete and finite set of values, each of which is calculated on 
a different microinterval from the set ${\left\{\interval_m\equiv\interval(\bar{x}_m,\Delta_m)\right\}}$ 
that partitions $\interval\left(x_{-M}+x_M)/2,\intmax\right)$.

In general, the average moment densities $\bar{\mathcal{M}}_\nu^{(n)}$  can
depend strongly on the choice of $\Pi(x_b,\intmax)$ and so 
they are not physically very meaningful. 
Nevertheless,  we will see that it is possible to derive useful expressions that relate them to
macroscopic observables, and which are valid for any choice 
of $\Pi(x_b,\intmax)$ that satisfies the conditions specified above.

In a crystal whose periodicity 
along the $x-$axis is $a$ (i.e., ${\nu(x+a)=\nu(x), \;\;\forall x}$), the $\expval{\;.\;}^*$ average
is unnecessary because, by choosing ${\Delta_m=a,\;\;\forall  m}$, all microintervals are identical
and so
\begin{align*}
\bmany(x_b) & = \many(x_b+a/2,a)
\\
&=\int_{-a/2}^{a/2}\Dnu(x_b+a/2)\,u^n\,\dd{u}.
\end{align*}

\subsection{Surface excess}
\label{section:excess_derivations}
Equation~\ref{eqn:mss} can be written as
\begin{align*}
\mss_r(x_b) & 
= \int_{x_L}^{\infty}x^r\,\Delta\nu(x)F(x-x_b)\dd{x}
\end{align*}
where ${F(x) = 1-H(x) = H(-x)}$ is one for ${x<0}$ and zero for ${x>0}$, and ${H(x)\equiv \dv{x}\max\{x,0\}}$ is the 
Heaviside step function. 
The mesoscale
average of $\mss_r(x_b)$ is
\begin{align}
\bsmsx{\Dnu}_r & \equiv 
\int_{-\infty}^{\infty}\mu(x'-x_b;\prectheo)\left(\int_{x_L}^{x'} x^r\,\Dnu(x)\dd{x}\right)\dd{x'}  \nonumber \\
& = \int_{-\intmax/2}^{\intmax/2}\mu(u;\prectheo)
\left(
\int_{x_L}^{x_b-\intmax/2} x^r\, \Dnu(x)\dd{x}\right)\dd{u} \nonumber\\
&+ \int_{-\intmax/2}^{\intmax/2}\mu(u;\prectheo)\left(\int_{x_b-\intmax/2}^{x_b+u} x^r\, \Dnu(x)\dd{x}\right)
\dd{u}  \nonumber \\
& = \bsmsx{\Dnu}_{r,s}(x_b) + \bsmsx{\Dnu}_{r,b}(x_b)
\nonumber
\end{align}
where we have assumed that ${x_b>x_L+\intmax/2}$ and 
we have split the mesoscale average, ${\bsmsx{\Dnu}_r}$, into the sum of a `surface' term, 
${\bsmsx{\Dnu}_{r,s}(x_b)}$, and a `bulk' term, ${\bsmsx{\Dnu}_{r,b}(x_b)}$, which can 
also be expressed as
\begin{align}
\bsmsx{\Dnu}_{r,s}(x_b) & \equiv
\int_{x_L}^{x_b-\intmax/2} x^r\, \Dnu(x)\dd{x}
\end{align}
and
\begin{align}
\bsms_{r,b}(x_b)
 & = \int_{-\intmax/2}^{\intmax/2}  \left(x_b+u\right)^r\Dnu(x_b+u) \mathcal{F}_\mu(u;\prectheo) \dd{u},
\label{eqn:a4}
\end{align}
where
${\displaystyle
{\mathcal{F}_\mu(u;\prectheo) \equiv 
\int_{-\infty}^{\infty} F(u-u')\mu(u';\prectheo)\dd{u'} }
}$
decays smoothly from a value of almost one at ${u=-\intmax/2}$
to almost zero at ${u=\intmax/2}$. Both its average value and its value at ${u=0}$
are one half and its derivative is ${\mathcal{F}^{(1)}_\mu(u;\prectheo)= -\mu(u;\prectheo)}$. 
The split of $\bsmsx{\Dnu}_r$ into bulk and surface terms is not unique:
both terms are microscopic functions of $x_b$, which is an arbitrarily-chosen 
point in the bulk. However, we will find that their sum is independent of $x_b$.

Now let us split the integral in Eq.~\ref{eqn:a4}
into a sum of integrals over the microintervals, ${\interval_m\equiv\interval(\bar{x}_m,\Delta_m)}$.
We can again exploit the slowness of the variation of $\mu$ and 
$\mathcal{F}_\mu$ on the microscale, when ${\amax/\prectheo}$ is very small, to 
replace $\mathcal{F}_\mu(x-x_b;\prectheo)$ in
each microinterval by its Taylor expansion about the 
microinterval midpoint. 
If ${\amax/\intmax}$ is sufficiently small, we can
discard the second and higher-order terms, which involve
first- and higher-order derivatives of $\mu$. 
Therefore, we get
\begin{widetext}
\begin{align}
\bsms_{r,b}(x_b) 
 = 
\sum_m
\bigg[
\mathcal{F}_\mu(\bar{x}_m-x_b;\prectheo) 
\int_{-\Delta_m/2}^{\Delta_m/2} 
(\bar{x}_m+u)^r
&
\Dnu(\bar{x}_m+u)\dd{u} 
\nonumber\\
&- \mu(\bar{x}_m-x_b;\prectheo) 
\int_{-\Delta_m/2}^{\Delta_m/2} 
u (\bar{x}_m+u)^r
\Dnu(\bar{x}_m+u)
\dd{u}
\bigg]
\label{eqn:sb3} 
\end{align}
\end{widetext}

\subsubsection{Case I: $\mbs_0$}
\label{section:s0ave}
Setting $r=0$ in Eq.~\ref{eqn:sb3} gives
\begin{align}
\bsms_{0,b}(x_b;\prectheo) 
& =\sum_m
\Delta_m \,\mathcal{F}_\mu(\bar{x}_m-x_b;\prectheo)\,\mzero(\bar{x}_m,\Delta_m) \nonumber \\
& - 
\sum_m
\Delta_m \,\mu(\bar{x}_m-x_b;\prectheo)\,\mone(\bar{x}_m,\Delta_m) \nonumber
\end{align}
Assuming that the microstructure is the same everywhere in 
a mesoscopic neighbourhood of $x_b$, the average of the 
microintervals'
$n^\text{th}$
moment density on every sufficiently-wide 
subinterval of ${[x_b-\intmax/2,x_b+\intmax/2]}$ should
be the same and equal to $\bmany(x_b)$
in the limit ${\amax/\prectheo\to 0}$.
Therefore the first term on the right hand side is simply
equal to ${(\intmax/2)\bmzero=(\intmax/2)\bDnu(x_b)}$, 
and
\begin{align}
\bsms_{0,b}(x_b;l)
& =
\frac{\intmax}{2}\,
\bDnu(x_b) - \expval{\mone;\mu}_\prectheo^*(x_b)   \nonumber \\
&= \int_{x_b-\intmax/2}^{x_b}\nu(x)\dd{x} - \bmone(x_b).
\label{eqn:bulkterm}
\end{align}
Adding $\bsms_{0,s}(x_b)$ and identifying the macroscopic quantity 
\begin{align}
\mbs_0(\mxb)  
& =  \int_{\mxl}^{\mxb}\,\DNu(\mx)\,\dmx \nonumber 
\intertext{as ${\bsms_0=\bsms_{0,s}(x_b)+\bsms_{0,b}(x_b)}$, we find that}
\mbs_0(\mxb)  
& = \int_{x_L}^{x_b}\Dnu(x)\dd{x}- \bmone(x_b)
\label{eqn:surfaverage0}
\end{align}
Note that ${\bsmsx{\Dnu}_{0,s}(x_b)=\mssx{\Dnu}_0(x_b)}$, which suggests that
${\bsmsx{\Dnu}_{0,b}(x_b) = - \bmone(x_b)}$ can be viewed as a correction 
to ${\mssx{\Dnu}_{0}(x_b)}$ that removes its sensitivity to $x_b$.

\subsubsection{Case II: $\mbs_1$ when $\Nu(\bm{x_b})=\bm{0}$}
\label{section:s1ave}
Returning to Eq.~\ref{eqn:sb3}, setting $r=1$, and using
the fact that, for all $m$,  
\begin{align}
\mzero(\bar{x}_m,\Delta_m)=
\frac{1}{\Delta_m}
\int_{x_m^-}^{x_m^+}\Dnu(x)\dd{x}=
\bDnu(x_b)=0,\nonumber
\end{align}
we find that
\begin{align}
&\bsms_{1,b}(x_b;\prectheo)   
= 
\sum_m
\Delta_m
\bigg\{
\mone(\bar{x}_m,\Delta_m)
\bigg[\mathcal{F}_\mu(\bar{x}_m-x_b;\prectheo) \nonumber \\
&- \bar{x}_m\,\mu(\bar{x}_m-x_b;\prectheo)\,\bigg]
 - 
\mtwo(\bar{x}_m,\Delta_m)\,
\mu(\bar{x}_m-x_b;\prectheo)
\bigg\} \nonumber 
\label{eqn:sb6} 
\end{align}
As in Sec.~\ref{section:s0ave}, 
the ${\many\,\mathcal{F}_{\mu}}$ term on the right hand side 
is equal to ${\left(\intmax/2\right)\bar{\mathcal{M}}^{(n)}_\nu(x_b)}$,
with ${n=1}$ in this case.
\begin{align}
\bsms_{1,b}&(x_b;\prectheo)  = \frac{\intmax}{2}\bmone(x_b) - \bmtwo(x_b) \nonumber \\
 -&\sum_m\,\Delta_m\,\mu(\bar{x}_m-x_b;\prectheo)\,\bar{x}_m\,\mone (\bar{x}_m,\Delta_m)
\end{align}
The third term on the right hand side is ${-\expval{x\,\mone\,;\mu}_{\prectheo}^*(x_b)}$. 
Subtracting ${x_b\expval{\mone;\mu}_{\prectheo}^*(x_b)}$ from the first term
and adding it to the third term gives
\begin{align}
\bsms_{1,b}&(x_b;\prectheo)  =
-\left(x_b-\frac{\intmax}{2}\right)\expval{\mone;\mu}_{\prectheo}^*(x_b) \nonumber \\
& -\expval{\mtwo\,;\mu}_{\prectheo}^*(x_b) 
-\expval{(x-x_b)\mone\,;\mu}_{\prectheo}^*(x_b) \nonumber
\nonumber 
\end{align}
It can be shown that the uniformity of the microstructure on the mesoscale
implies that the third term scales like $\amax/\prectheo$ when ${\prectheo\gg\amax}$.
This is because the distribution of microinterval moment densities 
is the same on either side of $x_b$, but the sign of $(x-x_b)$ is different.
Therefore, the contributions 
to this term from 
${(x_b-\intmax/2,x_b)}$ and ${(x_b,x_b+\intmax/2)}$ cancel one another.
Assuming that ${\prectheo\gg\amax}$, we get
\begin{align}
\bsms_{1,b}(x_b;\prectheo) & = -\left(x_b-\frac{\intmax}{2}\right)\bmone(x_b) - \bmtwo(x_b) 
\label{eqn:lastbutone}
\end{align}
Now, because ${\mzero(\bar{x}_m,\Delta_m)=0}$, we can write
\begin{align}
\int_{x_b-\intmax/2}^{x_b} \;x\;\Dnu(x)&\dd{x} 
 = \sum_{\bar{x}_m<x_b} \int_{-\Delta_m/2}^{\Delta_m/2} \, u \,\Dnu(\bar{x}_m+u) \dd{u} \nonumber \\
 = &\sum_{\bar{x}_m<x_b}\Delta_m \mone(\bar{x}_m,\Delta_m)
 =\frac{\intmax}{2} \bmone(x_b) \nonumber 
\end{align}
Therefore, adding ${\bsms_{1,s}(x_b;\prectheo)}$ to Eq.~\ref{eqn:lastbutone} gives 
\begin{align}
&\mbs_1(\mxb)  
 =  \int_{\mxl}^{\mxb}\mx\,\DNu(\mx)\,\dmx 
\nonumber \\
 &= \int_{x_L}^{x_b}\,x\,\Dnu(x)\dd{x} -\, x_b\, \bmone(x_b) - \bmtwo(x_b) 
\label{eqn:surfaverage1}
\end{align}
As with $\mbs_0$, we can write $\mbs_1$ as an $x_b$-independent sum of an $x_b$-dependent surface term, 
${\bsmsx{\Dnu}_{1,s}(x_b)}$, which is simply the original microscopically-varying integral ${\mssx{\Dnu}_1(x_b)}$, 
and an $x_b-$dependent bulk term, ${\bsmsx{\Dnu}_{1,s}(x_b)}$, which can be viewed as a correction that
removes the strong dependence on the arbitrarily-chosen position $x_b$.

\subsubsection{Idempotency of the mesoscale average}
\label{section:idempotency}
In Sec.~\ref{section:uncertainty} we assumed, implicitly, that 
the mesoscale averaging operation is not idempotent. This allowed us
to deduce that there is a trade-off between 
spatial precision/uncertainty and the precision/uncertainty
of macroscopic fields and their derivatives. However, the
uncertainty relations were derived under a `first approximation', and are
far from exact.
Throughout Sec.~\ref{section:excess_fields} we have assumed that we
are much closer to the limit ${\amax/\prectheo\to 0}$, and therefore closer
to the limit in which the averaging operation is idempotent.
Bearing this in mind, let us consider one important consequence of idempotency.

Idempotency of the averaging operation would allow the
following deduction to be made about
the mesoscale averages, $\boldmone$ and $\boldmtwo$, of
${\bmone}$ and ${\bmtwo}$, respectively. 
\begin{align}
\expval{\bsms_0}_{\prectheo} & = \bsms_0 \implies \boldmone(\mxb)\equiv \expval{\bmone}_{\prectheo}(x_b) = 0 \nonumber \\
\expval{\bsms_1}_{\prectheo} & = \bsms_1 \implies \boldmtwo(\mxb)\equiv \expval{\bmtwo}_{\prectheo}(x_b) = 0 \nonumber
\end{align}
The finding that both $\boldmone$ and $\boldmtwo$ are zero would have some very important
consequences. Therefore, guided by the knowledge that they 
vanish in the idempotent limit (${\amax/\prectheo\to 0,\; \prectheo/\Lmin\to 0}$), 
I show that they vanish without assuming
idempotency in Appendix~\ref{section:invariance_proofs}.
The importance of them vanishing will become clear in
Sec.~\ref{section:average_potential}.
The idempotency limit is the limit ${\amax/\prectheo\to 0,\; \prectheo/\Lmin\to 0}$, whereas
the limit in which they vanish is ${\amax/\prectheo\to 0}$.\\

\subsubsection{Mesoscale average over the lower limit of an integral}
Either by following similar procedures to those that led to Eq.~\ref{eqn:surfaverage0} and Eq.~\ref{eqn:surfaverage1}, 
or by invoking symmetry, one can find the following expressions for mesoscale averages of integrals
in which the average is performed over the lower bound, $x_b$, of the integrals from $x_b$ to $x_r$, where $x_r>x_b$.
\begin{widetext}
\begin{align}
\int_{\mxb}^{\mxr}\DNu(\mx)\,\dmx & =
\expval{\int_{x_b}^{x_r} \Dnu(x) \dd{x}; \mu}_{\prectheo}(x_b)  
 = \int_{x_b}^{x_r} \Dnu(x) \dd{x} +  \bmone(x_b)  \label{eqn:backsurfave0} \\
\int_{\mxb}^{\mxr}\mx\,\DNu(\mx)\,\dmx &=
\expval{\int_{x_b}^{x_r} \, x \,\Dnu(x) \dd{x}; \mu}_{\prectheo} (x_b)  
 =\int_{x_b}^{x_r} \, x \,\Dnu(x) \dd{x} +  x_b \bmone(x_b) + \bmtwo(x_b) \label{eqn:backsurfave1}
\end{align}
\end{widetext}

\section{Charge density ($\Rho$) and dipole moment density ($\mbp$)}
\label{section:average_charge}
In this section I consider the mesoscale averages of charge and dipole moment densities.
For simplicity I define the mesoscale average as the simple average introduced
and discussed in detail in Sec.~\ref{section:homogenization}.

\subsection{Charge density}
If we ignore the microscale variability of the mesoscale average, $\bar{\rho}$, of $\rho$, 
and the consequent uncertainty, $\precRho$, in the value of the macroscopic
charge density, its definition is simply
\begin{align}
\Rho(\bx) \equiv \bar{\rho}(x) = \frac{1}{\intmax}\int_{-\intmax/2}^{\intmax/2}\rho(x+u)\,\dd{u}
\label{eqn:Rhodef}
\end{align}
where ${\intmax \gg \amax}$. 

In the bulk of a crystal, 
$\intmax$ can be chosen to be an integer multiple of the periodicity, $a$, where
${\rho(x+a)=\rho(x),\;\;\forall\,x\in\bulk}$.
It is then easy to show that 
\begin{align}
\Rho(\mxb)= \bar{\rho}(x_b) = \frac{1}{a} \int_{-a/2}^{a/2}\rho(x_b + u)\,\dd{u}.
\nonumber
\end{align}
This vanishes if the crystal is charge-neutral, as expected of $\Rho$.
In amorphous materials, if $\rho$ fluctuates microscopically about zero, it is always possible to find 
microscopic displacements, ${\eta_1\sim a}$ and ${\eta_2\sim a}$,
such that 
\begin{align*}
\int_{-\intmax/2+\eta_1}^{\intmax/2+\eta_2}\rho(x_b+u)\,\dd{u}=0. 
\end{align*}
By expressing the integral in Eq.~\ref{eqn:Rhodef} as 
\begin{align*}
\int_{-\intmax/2}^{\intmax/2}=\int_{-\intmax/2}^{-\intmax/2+\eta_1}+ \int_{-\intmax/2+\eta_1}^{\intmax/2+\eta_2} -\int_{\intmax/2}^{\intmax/2-\eta_2}
\end{align*}
it is straightforward to show that 
\begin{align*}
\bar{\rho}(x_b)=0 + \order{\amax/\prectheo}\approx 0.
\end{align*}
Therefore, to within the finite precision $\precRho$ with which $\Rho$ can be defined, $\Rho$ vanishes
in the bulk of any material that is stable when it is electromagnetically isolated, and whose surfaces
are locally charge neutral.

\subsection{Dipole moment density}
$\pp$ has the dimensions of a dipole moment per unit volume, area, and length in 
three, two, and one dimensions, respectively.
Therefore, to define $\pp$ within the bulk of a one dimensional material, 
it seems natural to start from the quantity
\begin{align}
\mpp(x,\epsilon) \equiv \frac{1}{\epsilon} 
\int_{-\epsilon/2}^{\epsilon/2} \rho(x+u)\, u \, \dd{u}, \label{eqn:polden}
\end{align}
which is the 
dipole moment per unit length of ${\interval(x,\epsilon)\subset\bulk}$ 
with respect to an origin at $x$. 
$\mpp(x,\epsilon)$
is strongly dependent on both $x$ and $\epsilon$ and so it is 
difficult to attach physical meaning to it. However, it is clearly a microscopically-varying
quantity and its mesoscale average is
\begin{align}
\bar{\mpp}(x_b) & = 
\frac{1}{\intmax}\int_{x_b-\intmax/2}^{x_b+\intmax/2} 
\left(\frac{1}{\epsilon}\int_{-\epsilon/2}^{\epsilon/2} 
\rho(x+u)\,u \,\dd{u}\right) \;\dd{x}  \nonumber \\
& = 
\frac{1}{\epsilon}\int_{-\epsilon/2}^{\epsilon/2} 
u \left(\frac{1}{\intmax} \int_{x_b-\intmax/2}^{x_b+\intmax/2}  \rho(x+u)\, \dd{x}\right)\, \dd{u}  \nonumber \\
& = 
\frac{1}{\epsilon}\int_{-\epsilon/2}^{\epsilon/2} u 
\,\bar{\rho}(x_b+u) \dd{u} \nonumber \\
& = 
\frac{\bar{\rho}(x_b)}{\epsilon} 
\int_{-\epsilon/2}^{\epsilon/2} u \;\dd{u} 
+\order{\amax/\prectheo}
\approx 0, 
\label{eqn:zerodipole}
\end{align}
where, by
using  
${\bar{\rho}(x_b+u)=\bar{\rho}(x_b) + \order{\amax/\prectheo}}$, I am assuming that ${\bar{\rho}}$ fluctuates
microscopically but does not change systematically on length scale $\epsilon$.
Therefore, the mesoscale average $\bar{\mpp}$ of $\mpp$ is negligible when ${\amax/\prectheo}$ is sufficiently small, 
regardless of the value of ${\bar{\rho}}$. 

This result generalises to three dimensions, where it can be shown that each Cartesian component of
the mesoscale average of the dipole moment per unit volume of a region of arbitrary shape  scales
like ${\amax/\prectheo}$.
This is a generalisation to non-crystalline materials
of the well known result that, in a crystal, the average over all choices of unit cell of 
the dipole moment per unit cell is zero~\citep{resta-vanderbilt-2007}. 

These results suggest that ${\mbp\equiv\bar{\mpp}}$ is not a useful macroscopic quantity with which 
to characterise the bulk of a material because it 
does not distinguish between different mesoscopically-uniform materials, or even between a material and empty space.
We can only identify the macroscopic polarization $\pp$ as $\mbp$ if we are willing to accept that ${\pp=0}$
in {\em every} mesoscopically-uniform material, regardless of its microstructure.

\section{Interlude}
In the sections that follow I discuss several quantities 
that are commonly regarded as manifestations, or consequences, of either
the $\pp$ field itself or of its value changing. They include
surface charge $\bsigma$ and bound charge $\Rhobound$ 
(Sec.~\ref{section:surface_charge}), polarization current $\Jconv$(Sec.~\ref{section:current})
and the macroscopic (${\bk=0}$) electric field $\E$ (Sec.~\ref{section:macroscopicE}). 

Finnis's work~\citep{finnis} and Sec.~\ref{section:excess_fields} make it easy to write down an
expression for ${\bsigma=\bsigma[\rho]}$, which is a linear functional of ${\rho}$.
Its linearity means that, 
if $\rho$ can be decomposed as ${\rho=\sum_i\rho_i}$, where each $\rho_i$ is
either nonnegative or nonpositive, this becomes ${\bsigma[\rho] = \sum_i\bsigma[\rho_i]}$.
It follows immediately that,
when $\rho$ changes continuously in response to
a slowly varying stimulus, and if the set ${\{\rho_i\}}$ of charge packets
is chosen such that each one changes continuously as $\rho$ changes, but
its integral remains constant, then 
the polarization current can be expressed as the sum, 
${\dbsigma=\Jconv[\drho]=\sum_i\Jconv[\drho_i]}$. 

If the widths of the charge packets are microscopic, their shapes
are irrelevant to macroscale observables because homogenization transforms each packet into
a point charge. Therefore the contribution of each packet $\rho_i$
to $\Jconv$ can be calculated from the time derivatives of its integral,
${q_i=\int\rho_i\dd{x}}$, and its center, ${x_i = q_i^{-1}\int x \,\rho_i \dd{x}}$. 
If the packets can be chosen such that the integral of each one 
is time-invariant (${\dot{q}_i=0}$), we can use the MTOP to calculate $\Jconv$ 
from the evolving {\em bulk} microstructure.

It follows immediately from the results stated in 
Sec.~\ref{section:idempotency}, and proved in Appendix~\ref{section:invariance_proofs}, 
that the macroscopic potential $\bphi$ is zero in
an isolated macroscopically-uniform material whose surfaces are not charged.
It follows from this that a macroscopic $\E$ field cannot exist in such a material.
Nevertheless, in Sec.~\ref{section:macroscopicE} I prove this
by expressing ${\bphi}$ in terms of the microstructure $\rho$ using 
the results of Sec.~\ref{section:excess_fields}. In Sec.~\ref{section:macroscopicE}
I point out a fatal flaw in the cavity construction introduced by Lorentz
to relate the macroscopic $\E$ field to $\pp$, and in Sec.~\ref{section:paradox}
I refute Bethe's derivation of his approximate expression for the mean inner potential.

I conclude that neither $\pp$ nor the negative of its spatial derivative $\Rhobound$
are required elements of electromagnetic theory. I show that the quantization 
and multivaluedness of $\pp$ found within the MTOP are consequences of requiring that $\pp$ be a property
of the bulk and of defining the excess charge at a surface as ${\bsigmabound=\pp\cdot\normal}$.
As Fig.~\ref{fig:surfcharge} illustrates, the value of $\bsigmabound$ depends on how the surface is 
terminated (e.g., on a plane of net positive charge or on a plane of net negative charge). 
It follows that both ${\bsigmabound}$ and $\pp$ must be multivalued unless 
the excess surface charge is defined as ${\bsigma=\bsigmabound+\bsigmafree}$, where
$\bsigmafree$ takes full account of the dependence of $\bsigma$ on surface termination.

If it is accepted that $\pp$ is an unnecessary element of the theory, 
the importance of scrapping it should be obvious from its history: It has been 
interpreted in at least three different ways: as a property of the ether, as 
a dipole moment density, and as a property of the phase of a material's wavefunction. 
It can be misleading with regard to physical mechanisms; for example, expressing
the potential energy per unit volume as ${U=-\pp\cdot\E}$ suggests that $\E$ couples to the 
bulk, whereas expressing it as ${U=-\bsigma\E}$ makes clear that it only couples to 
charges at the surface, initially, and couples to the bulk indirectly by driving
charge through it. It can also lead to false conclusions, such as that
lack of inversion symmetry 
implies the existence of a uniform (${\bk=0 \notiff \bk\to 0}$) macroscopic $\E$ field 
in the bulk of a crystal.

\section{Surface charge ($\bsigma$)}
\label{section:surface_charge}
A charged surface or interface is not stable unless the electric potential from
it is compensated by, for example, an oppositely charged surface or interface.
The instability of isolated charged surfaces is due to the divergence of the electric
potential (see Sec.~\ref{section:average_potential}).
If a pristine isolated crystal surface is charged,
and therefore unstable unless neutralized by a change in
its composition with respect to the bulk, it is classified as {\em polar}.
An important question, about which a great deal has been written
\citep{tasker-1979, finnis, noguera-2000, goniakowski-rpp-2008, stengel-vanderbilt-prb-2009, bristowe-JPCM-2011, goniakowski-prb-2011, stengel-prb-2011,
noguera-chemrev-2013, bristowe-JPCM-2014,goniakowski-2014,goniakowski-2016}
is how to determine whether a particular surface is polar or non-polar and to
quantify its instability by calculating its surface charge.

\subsection{Calculating surface charge from $\rho$}
It does not seem difficult to intuit the meaning of the surface areal charge density $\bsigma$ when one first
encounters the concept. However, as soon as one tries to define it, 
in order to calculate it, difficulties become apparent.

The obvious way to calculate $\bsigma$ is simply to integrate the
volumetric charge density ${\rho(\rvec)}$ from a point above the surface to a point far beneath it.
Assuming that the surface is perpendicular to the $x-$axis, and that $\varrho(x)$ is the average of ${\rho(\rvec)}$ 
over the $yz-$plane at $x$, 
the obvious definition of the $yz-$averaged areal density of excess surface charge is
\begin{align}
\sigma_s(x_b) \equiv \int_{x_L}^{x_b}\varrho(x)\,\dd{x}
\label{eqn:surfchgeq1}
\end{align}
Fig.~\ref{fig:surfcharge} illustrates why this definition fails. It depicts
the `surface' of a one-dimensional crystal whose microscale charge distribution 
is a semi-infinite periodic array of alternating positive and negative 
point charges of magnitude one. For this simple case the integral in Eq.~\ref{eqn:surfchgeq1}
becomes the sum ${\sigma_s(x_b)=1-1+1-1+\cdots}$. Its value is either zero or
one, depending on the choice of $x_b$, and it continues to vary between these
values {\em ad infinitum} as $x_b$ increases.  
Therefore ${\sigma_s(x_b)}$ is a microscopic function of ${x_b}$ and, as a result, 
Eq.~\ref{eqn:surfchgeq1} fails as a definition of $\bsigma$.

Finnis presented an elegant solution to this problem 
in \REF~\linecite{finnis}, and a generalization of his result to amorphous materials 
is derived by a different route in Sec.~\ref{section:excess_fields} and Appendix~\ref{section:invariance_proofs}.
I quote and explain the more general result below. I then quote Finnis's result for crystals, which is 
simpler and easier to relate to the example depicted in Fig.~\ref{fig:surfcharge}.

\subsection{Macroscale surface charge}
\label{section:macrosurfcharge}
The problem of how to express $\bsigma$ in terms of $\rho$ is easy to resolve
once it is realised that $\bsigma$ only has meaning at the macroscale.
Microscopically, surfaces and interfaces are ill-defined entities because their widths
are indeterminate: in the vicinity of a surface, both structure and composition differ from the bulk, in general, but
they gradually become more bulk-like with depth. This gradual relaxation means that there is no clear boundary 
separating surface-like material from bulk-like material.

However, as Fig.~\ref{fig:surfcharge} illustrates, even if a material could be terminated abruptly at a 
plane and prevented from changing its local structure (bond lengths and angles) or composition, such 
that surface structure and composition were identical to the bulk, 
the concept of a surface charge density simply does not apply at the microscale: The microstructure is defined on a simply
connected subset of $\realthree$. One can define an areal charge density on any plane (e.g., ${\sigma(y,z;x)\equiv \rho(x,y,z)\dd{x}}$), but
no special surface plane exists.

However, as explained in Sec.~\ref{section:homogenization}, 
the spatial resolution $\prectheo$ at the macroscale is unavoidably finite and all 
points separated by microscopic distances coincide at the macroscale. As a result, the surface region
of indeterminate width is contracted to zero width by the homogenization transformation.
It becomes a two dimensional manifold.

\begin{figure*}[!]
\includegraphics[width=0.8\textwidth]{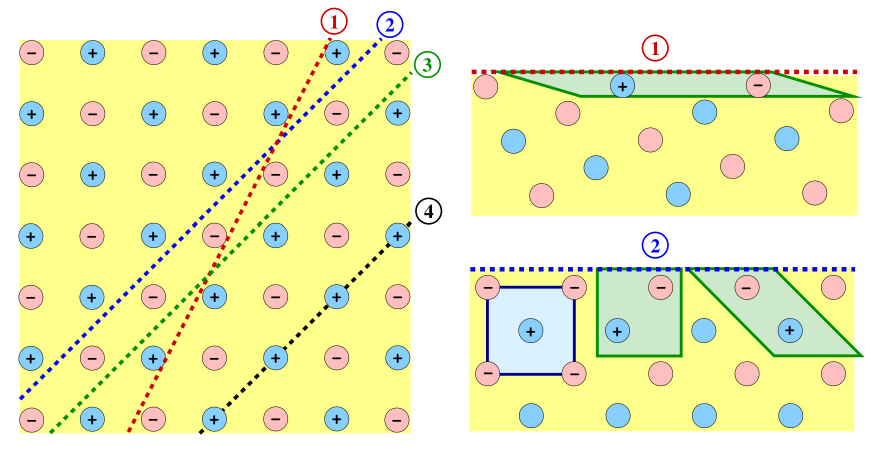}
\caption{A lattice of point charges can notionally be cleaved along an infinite number of lattice planes.
Here, four lattice planes are shown in cross section as dashed lines. Plane 4 
intersects a lattice of positive point charges and is included to illustrate a point made
in Sec.~\ref{section:macrosurfcharge}.
Cleaving along Planes
1, 2, and 3,  produce Surfaces 1, 2, and 3, respectively; Surfaces 1 and 2 are shown on the right. 
Deducing whether a surface is polar is as simple as
constructing a unit cell of the crystal (shown in green), which has two lattice vectors 
that are parallel to the surface plane and one face at the surface, and calculating the dipole moment of this cell.
The surface is polar if and only if the projection of the dipole moment onto the surface normal is non-zero. 
Surface 1 is non-polar but Surface 2 is polar. Surface 3 would be positively charged, despite Planes 2 and 3 being
parallel and Surface 2 being negatively charged.
Therefore, the orientation of the plane determines whether or not it is polar, but not the value of the surface charge density.
One could also construct the dipole-free unit cell (navy boundary) and, if this cannot be constructed with one of 
its faces at the surface, the integral of the charge density from
its uppermost face to the surface is the surface charge density, $\bsigma$. 
}
\label{fig:polar_surface}
\end{figure*}

The mesoscale average $\brho$ of $\rho$ at any microscale point 
whose macroscale image is on this manifold differs significantly (by more than ${\precRho/2}$), 
in general, from its value elsewhere. To understand why, consider the 
material depicted in Fig.~\ref{fig:polar_surface}. There exist planes (e.g., Plane 4)
on which the planar average of $\rho$ does not vanish. 
It follows that the three dimensional 
mesoscale average $\brho$ at any point (not just points on the charged planes) only vanishes
as a result of cancellation of positive and negative contributions whose displacements
from the point have components normal to those planes.
If all material from one side of such a plane is removed to create a surface, this
balance is disrupted and $\brho$ becomes finite, in general, at any point
within a distance ${\prectheo/2}$ of the plane.
The areal charge density $\bsigma$ at a point on the surface manifold 
is simply the integral of ${\brho}$ over the point's preimage
under the homogenization transformation. Therefore it is the integral 
of $\brho$ along on an interval of width ${\prectheo}$ on an axis normal
to the surface. The macroscopic volumetric charge density $\Rho$
is simply the average of $\brho$ on this interval.

It is logical, then, to define the surface charge as
\begin{align}
\bsigma = \int_{\mxl}^{\mxb}\Rho(\bx)\,\dbx .
\label{eqn:surfchgeq3}
\end{align}
This integral converges with respect to both of its limits because
${\Rho=0}$ in the bulk and in the vacuum above the surface.
By substituting the definition of $\Rho$ as the mesoscale average $\brho$ of $\rho$ (Eq.~\ref{eqn:Rhodef}), it 
is shown in Sec.~\ref{section:excess_fields} that
\begin{align}
\bsigma \equiv
\int_{x_L}^{x_b}\rho(x)\,\dd{x}-\bmonerho(x_b)\equiv\sigma_s(x_b) + \sigma_b(x_b),
\label{eqn:surfchgeq2}
\end{align}
where $x_b$ is {\em any} point deep below the surface and ${\bmonerho(x_b)\equiv -\sigma_b(x_b)}$ is defined
as follows: A mesoscopic neighbourhood of $x_b$ is partitioned
into a set of contiguous microscopic intervals ${\interval(\bar{x}_m,\Delta_m)}$ such that $x_b$ is at
a boundary between two of these {\em microintervals}, and such that the integral of $\rho$
on each microinterval is zero. The second condition is always possible because $\rho$ fluctuates microscopically about zero.
The dipole moment density of the $m^\text{th}$ interval is 
\begin{align*}
\monerho(\bar{x}_m,\Delta_m)
\equiv \frac{1}{\Delta_m}\int_{-\Delta_m/2}^{\Delta_m/2}\rho(\bar{x}_m+u)\,u\,\dd{u},
\end{align*}
and $\bmonerho(x_b)$ is defined as the average of this quantity over all microintervals in the discrete and finite
set that partitions the mesoscale neighbourhood of $x_b$.   It is shown in Appendix~\ref{section:invariance_proofs}
that the value of ${\bmonerho(x_b)}$ is the same for all sets of microintervals that satisfy the conditions
stated above. 

Although both ${\sigma_s(x_b)}$ and ${\sigma_b(x_b)}$ depend sensitively on $x_b$, their sum is independent of it.
This is easy to see in the special case of a periodic bulk charge density, ${\rho(x+a)=\rho(x)}$. 
The points $x_b+ma$, where $m\in\mathbb{Z}$, can be chosen as the microinterval boundary points. 
All microintervals are identical, in this case, and Eq.~\ref{eqn:surfchgeq2} simplifies to Finnis's result:
\begin{align}
\bsigma & = \int_{x_L}^{x_b}\rho(x)\,\dd{x}- \frac{1}{a} \int_0^a\rho(x_b+u)\,u\,\dd{u} \nonumber \\
& = \sigma_s(x_b) - \mpp(x_b+a/2;a).
\label{eqn:finnis}
\end{align}
Note that the definitions of $\monerho$ and $\mpp$ are identical. I use $\mpp$ when it is useful to make
clear that it is a dipole moment density. I use $\monerho$ when I favour
consistency with Sec.~\ref{section:excess_fields}
and with related quantities that will be introduced in Sec.~\ref{section:average_potential}.

Referring again to Fig.~\ref{fig:surfcharge}, and comparing the choices ${x_b=x_1}$ and ${x_b=x_2}$, we find 
that ${\sigma_s(x_1)=0}$ and ${\sigma_s(x_2)=1}$. If the minimum distance between positive and negative charges
is denoted by $b$, then ${\sigma_b(x_1)= b/a}$ and ${\sigma_b(x_2)=-(a-b)/a}$.
Therefore, 
\begin{align*}
\sigma_s(x_1)+\sigma_b(x_1)=\sigma_s(x_2)+\sigma_b(x_2)=\frac{b}{a}.
\end{align*}

It may be illuminating to note that applying the homogenization transformation 
is a lot like taking a thermodynamic limit; and we can think of macroscopic quantities as thermodynamic quantities which, 
in a non-ergodic system, can only be defined on macroscopic length scales.
Indeed, Finnis's reasoning when deriving Eq.~\ref{eqn:finnis} differed slightly from the reasoning outlined above.
He reasoned that one should average over an ensemble of terminating planes ($x_b$), in order to ``{\em reconcile the atomistic 
picture, in which} [surface] {\em excesses
appear to oscillate on the atomic length scale as  a function of the} [surface] {\em region size, with the thermodynamic picture.}''
In the language that I have chosen to use here and in Sec.~\ref{section:homogenization}, he found the mesoscale
average of the microscopic function $\sigma_s(x_b)$. The same result is found by substituting ${\Rho=\bar{\rho}}$
into Eq.~\ref{eqn:surfchgeq3} because, by changing the order of integration, the integral of a mesoscale average 
becomes the mesoscale average of an integral.

The subscripts of $\sigma_s$ and $\sigma_b$ 
are abbreviations of `surface' and `bulk', respectively.
$\sigma_s$ includes all contributions from
compositional differences between the surface and the bulk, including 
charged adsorbates, surpluses or deficits of electrons, charged impurities,
and non-stoichiometry associated with reconstructions or coordination defects.
On the other hand, $\sigma_b$, depends only on the charge density in the bulk
and is independent of the surface composition.
However, it is simplistic and wrong to view $\sigma_s$ as the contribution from
extrinsic surface charges and $\sigma_b$ as the contribution from the bulk 
charge distribution.  For example, it is always possible to choose $x_b$ such
that ${\sigma_b(x_b)=0}$ and ${\bsigma=\sigma_s(x_b)}$. 
Therefore, as well as containing all extrinsic contributions
to $\bsigma$, $\sigma_s$ can contain some, or all, of the contribution from the
bulk. Choosing ${\sigma_b(x_b)=0}$ is
equivalent to the ``dipole-free unit cell'' strategy used by Goniakowski {\em et al.} to deduce
surface charge and stability~\citep{goniakowski-rpp-2008}.

\subsection{Surface Stability}
\label{section:surfstability}
A pristine crystal surface is specified by the structure of the bulk crystal,
the surface-plane orientation, and the surface termination.
Consider the crystal depicted in Fig.~\ref{fig:polar_surface} and
the two examples given of surfaces of that crystal, which I'll refer to as Surface 1 and Surface 2.
These surfaces are defined by the planes  (Plane 1 and Plane 2, respectively)
at which one could imagine cleaving the perfect crystal. Each plane is defined by an orientation, which can be specified by the outward
surface normal  $\normal$, and by a position along an axis parallel to $\normal$.
The importance of the relative displacements of parallel surface planes is illustrated
by the fact that Surface 2 is negatively charged, whereas the surface created
by cleaving at Plane 3  would be positively charged.
In Fig.~\ref{fig:polar_surface} I am considering an unphysical frozen surface
where I have not allowed any relaxation from the bulk charge density
Nevertheless, this unphysical surface suffices to allow the surface polarity to be quantified
using Eq.~\ref{eqn:finnis}.

If we could prepare the frozen surfaces depicted in Fig.~\ref{fig:polar_surface}, when
we allowed them to relax in a vacuum they would relax and/or reconstruct.
The surface might even melt.
However, as long as the bulk crystal did not melt, or the surface didn't banish ions or electrons from it
(in the case of Surface 2 we would probably need to apply an electric field to prevent this), none
of the structural change near the surface would have any impact on the surface charge, $\bsigma$,
because the integral of the charge density between the surface and the crystalline bulk,
which is the first term in Eq.~\ref{eqn:finnis}, would be unchanged.

\subsection{Interface charge}
By treating an interface between two materials as a pair of adjoined surfaces
it is straightforward to show that the interface charge is
\begin{align}
\bsigma = \int_{x_1}^{x_2} \rho(x)\,\dd{x} + \momone{x_1}-\momone{x_2}
\end{align}
where ${x_1}$ is an arbitrary point in the bulk of the material on one side of the interface
and ${x_2>x_1}$ is an arbitrary point in the bulk of the material on the other side.

As a sanity check, let us imagine a plane perpendicular to ${\hat{x}}$ at position $x_b$ in the bulk
of a mesocopically-uniform material. This plane can be viewed as an interface between two
perfectly-aligned identical materials. The charge density at this imaginary interface is
\begin{align}
\bsigma = \int_{x_b-u}^{x_b+u}\rho(x)\,\dd{x} + \momone{x_b-u}-\momone{x_b+u}
\nonumber
\end{align}
where $u$ can have any value since all points on either side of $x_b$ are
in the bulk. For the sake of brevity, I choose the limit ${u\to 0^+}$, which leads immediately
to ${\bsigma=0}$, as should be the case. It is not too difficult to prove that
$\bsigma=0$ for an arbitrary value of $u$ in a non-periodic system.

\subsection{Consistency with the standard model of macroscale electrostatics}
\label{section:consistency}
Standard treatments of macroscale electrostatics tend to distinguish between a {\em free} charge density
${\Rhofree=\div{\D}}$ and a {\em bound} charge density ${\Rhobound=-\div{\pp}}$, where 
\begin{align*}
\Rho = \epsilon_0\div{\E} = \div{\D} -\div{\pp} = \Rhobound+\Rhofree.
\end{align*}
Substituting into Eq.~\ref{eqn:surfchgeq3} gives
\begin{align}
\bsigma & = 
\overbrace{\int_{\mx_L}^{\mxb} \Rhofree(\mx)\, \dbx}^{\displaystyle \bsigmafree} 
\overbrace{-\int_{\mx_L}^{\mxb} \div\pp \, \dbx}^{\displaystyle  \bsigmabound}  \nonumber \\
& = \int_{\mx_L}^{\mxb} \Rhofree(\mx)\, \dbx + \pp\cdot\vec{\hat{n}}
\end{align}
and consistency with Eq.~\ref{eqn:finnis} requires that
\begin{align}
\int_{\mx_L}^{\mxb} \Rhofree(\mx) \dbx & + \pp\cdot\vec{\hat{n}}  \nonumber \\
 = & \int_{x_L}^{x_b} \rho(x) \dd{x} + \mbp\left(x_b+a/2,a\right)\cdot\vec{\hat{n}}.
\label{eqn:macro-micro}
\end{align}
Both terms on the right hand side depend sensitively on $x_b$. However, because $x_b$ is arbitrary, 
both terms on the left hand side must be independent of it 
if physical meaning can be attributed to them independently of one other.
One way to resolve this is by defining $\Rhofree$ and $\pp$ to be
the mesoscale averages of $\rho$ and $\mbp$, respectively.
That is, ${\Rhofree \equiv \bar{\rho} = \Rho}$ and ${\pp \equiv \bar{\mbp} = 0}$ (see Eq.~\ref{eqn:zerodipole}), 
which implies that ${\Rhobound = 0}$ and ${\bsigmabound \equiv \pp\cdot\vec{\hat{n}} =0}$. 
These definitions preserve consistency between the standard model of electrostatics in dielectrics and
the apparently-reasonable definitions of $\Rho$ and $\bsigma$ presented herein; namely, $\Rho$ is 
the mesoscale average of $\rho$ and $\bsigma$ is its integral across a surface.
However, achieving consistency in this way entails discarding several quantities
from the standard model of electrostatics: $\D$, $\pp$, and $\Rhobound$ all vanish and ${\Rhofree}$
is simply ${\Rho}$.
The macroscopic Maxwell equations are now identical in form to their microscopic counterparts
because averaging commutes with differentiation; for example, 
\begin{align}
\epsilon_0 \div\me(x) = \rho(x) \Rightarrow \epsilon_0 \div\E(\mx) = \Rho(\mx).
\end{align}

\subsubsection{Quantization of $\pp$}
\label{section:quantized}
It is important to consider carefully whether or not the less drastic option of 
keeping quantities $\pp$, $\D$, $\Rhobound$, and $\Rhofree$ within the macroscale theory is logical or viable.
In this section I assume that $\pp$ remains an element of the theory and I 
show that consistency with the definitions of
$\Rho$ (Eq.~\ref{eqn:Rhodef}) and $\bsigma$ (Eq.~\ref{eqn:surfchgeq3}) 
requires it to be quantized. 

The fact that $\pp$ would be quantized for a classical crystal in the same way as it is quantized in the MTOP 
appears to have been appreciated from the beginning~\citep{kingsmith-vanderbilt-prb-1993-2,vanderbilt_2018}.
I explain it here for completeness, and also to emphasize it,  because it is easy to 
misinterpret the term \emph{quantum of polarization} as referring to something quantum mechanical.

As before I consider the surface of a pristine perfect crystal
whose outward unit normal is $\normal$ and whose structure and composition
have not been allowed to change after all the material on one side
of the surface plane was removed.
I choose the crystal's primitive
lattice vectors ${(\avec_1, \avec_2, \avec_3)}$ such that
${\avec_1\cdot\normal>0}$, 
${\avec_2\cdot\normal=\avec_3\cdot\normal=0}$, and ${\normal\cdot\left(\avec_2\times\avec_3\right)=A_\Omega>0}$. 
The volume of the bulk crystal's unit cell is ${\Omega\equiv\avec_1\cdot\left(\avec_2\times\avec_3\right)=\abs{\avec_1} A_\Omega}$. 
The surface, which is a two dimensional lattice, has primitive lattice vectors ${(\avec_2,\avec_3)}$
and the area of its primitive unit cell is ${A_\Omega}$.
Given the surface normal $\normal$, this choice of primitive unit cell of the bulk crystal allows all surface
terminations to be identified by a single parameter $\alpha$, which is the 
position along $\normal$ at which the uppermost primitive cell is sliced to form the surface.
For example, surfaces formed by cleaving at Planes 2 and 3 of Fig.~\ref{fig:polar_surface} differ only by
their values of $\alpha$. In this simple case the value of $\alpha$ determines only whether
the uppermost plane is a plane of cations or a plane of anions. In more general cases
the electron density would also be divided; however, it would be unphysical to remove
fractions of electrons by removing all density above the termination plane, 
so I assume that the integral of the 
density that remains in the uppermost cell is rounded up to a whole number. How this
density is distributed has no bearing on the arguments to follow.

The excess bound charge at the surface of the crystal is ${\bsigmabound=\pp_\perp\equiv\pp\cdot\normal}$.
Within the standard model of electrostatics $\pp$ is a bulk quantity and so it must be 
independent of surface termination $\alpha$. Therefore $\bsigmabound$ must be the same for 
all surfaces whose outward normal is $\normal$. 
However, as discussed in Sec.~\ref{section:surfstability}, and as Fig.~\ref{fig:polar_surface} illustrates, 
${\bsigma=\bsigmabound+\bsigmafree}$ is not the same for all values of $\alpha$.  
One could choose to include all of the $\alpha$-dependence of $\bsigma$  in $\bsigmafree$, which would
leave $\bsigmabound$ independent of surface termination. However, this is not the approach taken
within the MTOP~\citep{kingsmith-vanderbilt-prb-1993-2, stengel-prb-2011}. The MTOP assumes the standard
convention that $\Rhofree$ and $\bsigmafree$ only contain contributions from charges that are 
not intrinsic to the material~\citep{ashcroft_mermin_book, jackson-book}. As a consequence of preserving this old
convention, $\pp$ must be quantized~\citep{kingsmith-vanderbilt-prb-1993-2,vanderbilt_2018}. I now prove this.

There are no extrinsic charges in the idealized surfaces constructed; therefore 
${\bsigmafree=0}$ and ${\bsigma=\bsigmabound=\pp_\perp}$.
Now, because $\bsigma$ is known and single-valued, and because it can be changed to 
the value it would have for any other value of $\alpha$ by adding/removing
the same numbers and types of particles (nuclei and electrons) to/from
each unit cell of the surface lattice, $\bsigmabound$ must be multivalued.
Its set of values must be the set of values of $\bsigma$ for every possible 
choice of surface termination, $\alpha$. These values differ 
by integer multiples of ${e/A_\Omega}$. Therefore, $\pp_\perp$ is quantized such that
if ${\Delta\pp}$ is the difference between two values of
$\pp$ that are consistent with Eqs.~\ref{eqn:Rhodef} and~\ref{eqn:surfchgeq3}, then
\begin{align}
\Delta\pp_\perp \equiv \Delta\pp\cdot\normal 
&= \frac{m_1\, e}{\abs{\avec_2\times\avec_3}}
= \frac{m_1\, e}{\avec_1\cdot\left(\avec_2\times\avec_3\right)}\, \avec_1\cdot\normal \nonumber \\
\implies \Delta\pp & = \frac{m_1\,e}{\Omega}\,\avec_1 + \Delta\pp_{\parallel, 2}\,\avec_2 + \Delta\pp_{\parallel, 3}\,\avec_3
\nonumber
\end{align}
where ${\Delta\pp_\parallel \equiv \Delta\pp_{\parallel, 2}\,\avec_2 + \Delta\pp_{\parallel, 3}\,\avec_3 =\Delta\pp - \Delta\pp_\perp\,\normal}$, 
and $m_1$ is an integer.
By considering surfaces perpendicular to ${\avec_3\times\avec_1}$
and ${\avec_1\times\avec_{2}}$, the same logic would lead me to the following  general expression for the difference between any two
values of polarization that are consistent with Eqs.~\ref{eqn:Rhodef} and~\ref{eqn:surfchgeq3}.
\begin{align}
\Delta\pp = \frac{e\A}{\Omega}
\label{eqn:quantization}
\end{align}
where ${\A = m_1\,\avec_1+m_2\,\avec_2+m_3\,\avec_3}$ is an arbitrary lattice vector and ${m_1, m_2, m_3 \in\mathbb{Z}}$.
This is identical to the quantization of $\pp$ deduced within the MTOP, but this derivation makes clear
that the quantization of $\pp$ is not a quantum mechanical quantization because I have not invoked
quantum mechanics to deduce it. It is a consequence of adopting the apparently arbitrary and unnecessary convention that, in the absence
of extrinsic charges, $\bsigmabound$ is the total surface charge density, and of
preserving the macroscale theory's internal consistency while shoehorning $\pp$ into it.

\subsection{Mapping to a set of localized charge packets}
\label{section:discrete}
In this section I present a result that is pivotal for understanding 
the relationship between this work, which is founded on a
systematic approach to structure homogenization, 
and the MTOP's definition of polarization current, 
which is founded on quantum mechanical perturbation theory.

Let us express the charge density 
as the sum, ${\rho(x)=\sum_i\rho_i(x)}$, of a set of localized charge packets, ${\{\rho_i\}}$,
where each $\rho_i$ is either nonpositive or nonnegative.
The total charge in 
the $i^\text{th}$ packet is $q_i$ and its center of charge is $x_i$.
That is, 
\begin{align}
q_i  \equiv \int_{-\infty}^\infty \rho_i(x) \dd{x}, \quad 
x_i  \equiv \frac{1}{q_i}\int_{-\infty}^\infty \rho_i(x)\, x \,\dd{x}.
\nonumber
\end{align}
I assume that each ${\rho_i}$ can be chosen such that it is negligible outside 
interval ${\interval(x_i,\prectheo)}$.
For a system of nuclei and electrons whose
charge density is given by Eq.~\ref{eqn:charge_density1d}, the nonnegative
packets are ${\rho_i(x)\equiv Z_i\,e\,\delta(x-X_i)}$
and the nonpositive packets are ${\rho_i(x)\equiv-e\,n_i(x)}$, where ${n(x)\equiv\sum_{i}n_i(x)}$
is the electron number
density partitioned into a set of packets, ${\{n_i\}}$.

The localization transformation ${\rho_i(x)\to q_i\,\delta(x-x_i)}$
conserves charge and preserves $\rho_i$'s center of charge.
Therefore, the transformation of ${\rho}$ into the discrete
distribution of point charges ${\rho^q(x)\equiv\sum_i q_i\, \delta(x-x_i)}$ is
an isotropic spatial redistribution of charge. By `isotropic' I mean that 
it does not change the center of charge of either $\rho$ or 
${\trho\equiv\sum_{i\in\iset}\rho_i}$, 
where $\iset$ is {\em any} subset of the set of packet indices.
The equitable movement of charge in both directions
cannot change the macrostructure if charge is only moved 
across distances smaller than ${\prectheo=\dbx}$. 
Therefore, the macroscale counterpart ${\Rho^q}$ of ${\rho^q}$ cannot differ 
from $\Rho$. From this fact, and from Eq.~\ref{eqn:surfchgeq2}, it follows that 
\begin{align}
\bsigma & = \int_{\mxl}^{\mxb}\Rho^q(\bx)\,\dbx 
 = \sum_{i:x_i<x_b} q_i - \bmonerhoq(x_b)
\label{eqn:pointsigma_general}
\end{align}
Note that the derivation of Eq.~\ref{eqn:surfaverage0}  in Sec.~\ref{section:excess_fields}
assumed that space could be partitioned into microintervals whose net charges were equal.
This is always possible for a continuous charge density, but it is not possible for the $\rho^q$'s 
resulting from every possible partitioning of $\rho$ into packets because
the integral of $\rho^q$ on each microinterval is a sum of point charges.
If the magnitudes of these charges are irregular it is not possible, in general, to
partition space such that each interval's average charge density is precisely the same.
It seems likely that a more general derivation, which does not require each microinterval to have the same
average charge density, is discoverable.

In the bulk of a crystal with periodicity $a$, the charge packets can be chosen such that, 
for any packet $\rho_i$, whose center is $x_i$, there are identical packets with
centers at ${x_i+m a}$ for all ${m\in\mathbb{Z}}$. In this case, the  surface
charge is 
\begin{align}
\bsigma = \sum_{i:x_i<x_b} q_i - \frac{1}{a}\sum_{i:x_i\in(x_b,x_b+a)} q_i\,x_i,
\label{eqn:finnis2}
\end{align}
where I am assuming that ${\sum_{i:x_i\in(x_b,x_b+a)} q_i=0}$.
The time derivative of Eq.~\ref{eqn:finnis2} is the current $\J$. If the conduction current
vanishes, it is the polarization current $\Jconv$.

\section{Current ($\J$)}
\label{section:current}
As discussed in Sec.~\ref{section:anisotropy}, unless it is prohibited by symmetry, a 
polarization current $\Jconv$ flows through the bulk of
a material in response to any stimulus ${\zeta\to\zeta+\Delta\zeta}$, 
where $\zeta$ might be temperature, an externally-applied electric, magnetic, or stress field, 
or anything else that changes the material's equilibrium or steady-state-nonequilibrium
microstructure. 

\subsection{Polarization current as a rate of change of surface charge}
When all charge that flows through the bulk, by any mechanism or in response to any stimulus,
accumulates at surfaces, the current density deduced from Eq.~\ref{eqn:surfchgeq2} is simply
\begin{align}
\J = \bdot{\bsigma} = \int_{x_L}^{x_b}\pdv{\rho}{t}\eval_{(x,t)}\,\dd{x}-\pdv{\bmonerho}{t}\eval_{(x_b,t)}
\end{align}
Therefore, the current can be calculated if the time-dependent charge density ${\rho(x,t)}$ (or $\Rho$) 
is known everywhere. However, the amount of charge that can flow in an isolated material is limited,
and surface microstructures tend to be more difficult to calculate than 
microstructures in the bulks of crystalline materials. Therefore, we would like to be
able to calculate ${\Jconv}$ from the evolving equilibrium bulk microstructure.

If we can partition the electron density of a crystal into packets,
we can use Eq.~\ref{eqn:finnis2} to express the current
density as
\begin{align}
\J = \bdot{\bsigma} = 
\sum_{i:x_i<x_b} \dot{q}_i 
-\frac{1}{a}\sum_{i:x_i\in(x_b,x_b+a)} \left(\dot{q}_ix_i + q_i\dot{x}_i\right),
\end{align}
where each pair $(x_i,q_i)$ is either the position and charge of a nucleus
or the center and ${-e}$ times the integral of a packet $n_i$ of electron
density; and where ${x_b}$ has been chosen to not coincide with any of the $x_i$'s.  

Let us denote ${(x_b,x_b+a)}$, which is a primitive unit cell of the crystal, 
by $\unitcell$;
let us denote the sum of all $n_i$ whose centers 
are in $\unitcell$
by 
\begin{align*}
n_\unitcellscr(x,t)=\sum_{i:x_i\in\Omega} n_i(x,t);
\end{align*}
and let us denote the charge distribution of the nuclei in $\unitcell$ by ${\rhop_\unitcellscr(x,t)}$.
Finally, let us assume that the packets $n_i$ have been chosen such that
the integral of 
\begin{align*}
\rho_\unitcellscr(x,t)=\rhom_\unitcellscr(x,t)+\rhop_\unitcellscr(x,t)  = 
-e n_\unitcellscr(x,t)+\rhop_\unitcellscr(x,t), 
\end{align*}
is zero. Then, since the charge density in each bulk primitive unit cell 
is identical, the electron density in the bulk must be 
\begin{align*}
n(x,t) = \sum_{m\in\integer}n_\unitcellscr(x+ma,t),
\end{align*}
where ${a=\volume}$ is the lattice constant and $ma$ is a lattice vector.

We have realised the situation described in the discussion 
of Fig.~\ref{fig:crystal_dipole} in Sec.~\ref{section:mtop}: We have partitioned
the charge density of the electrons in the crystal's bulk into a set of identical
non-positive charge densities that are displaced from one another by lattice vectors;
and because the bulk must remain charge neutral, the
integral $\Neleccell$ of $n_\unitcellscr$ remains constant.
Therefore the polarization current is given by
\begin{align}
\Jconv & = -\frac{1}{a} \sum_{i:x_i\in\Omega}\left(\dot{q}_ix_i+q_i\dot{x}_i\right)
\label{eqn:currentA} 
\\
       & = -\frac{e\Neleccell}{a}\left(\dot{X}_\unitcellscr^+ - \dot{X}_\unitcellscr^-\right)
       = -\frac{\dot{d}_\unitcellscr}{a}
=\bdot{\mbp} (x_b+a/2,a)\cdot\vec{\hat{n}} \nonumber
\end{align}
where ${X_\unitcellscr^+}$ and 
${X_\unitcellscr^-}$ are the centers of charge of
${\rhop_\unitcellscr}$ and ${\rhom_\unitcellscr}$, respectively;
and I have used the fact that the sum of all the positive
$q_i$'s in $\unitcell$ and the sum of all the negative
$q_i$'s in $\unitcell$ are both time independent and equal to
${e\Neleccell}$ and ${-e\Neleccell}$, respectively.

If each packet $n_i$ has an integral that remains constant, the
polarization current can also be expressed as
\begin{align}
\Jconv & = -\frac{1}{a} \sum_{i:x_i\in\Omega}q_i\dot{x}_i;
\label{eqn:Jsimple}
\end{align}
and when the integral of each $n_i$ is one ($\times$ spin degeneracy), 
this is equivalent to the MTOP definition of $\Jconv$.

The generalization of Eq.~\ref{eqn:currentA} to amorphous 
materials is
\begin{align}
\Jconv = -\frac{1}{\ell}\sum_{i:x_i\in\interval(x_b,\ell)} \left(\dot{q}_i x_i + q_i\dot{x}_i\right).
\end{align}
Although the result for crystalline systems appears to be exact and precise, 
there are variations (${\sim a/\ell}$) in 
the value for amorphous systems with the choices of $x_b$ and $\ell$.
The source of these variations is 
differences in the averages of ${\dot{q}_ix_i+q_i\dot{x}_i}$ on different intervals.
Furthermore, because the net charge ${Q=\sum_{i:x_i\in \interval(x_b,\ell)}q_i}$ of
interval ${\interval(x_b,\ell)}$ is not zero, in general, the value
of $\Jconv$ calculated from this expression has an origin dependence unless
$Q$ is constant. In practice, it may be easier to find a set of
packets $n_i$ whose integrals are constant (${\implies\dot{q}_i=0}$)
and to calculate 
\begin{align*}
\Jconv\equiv  -\frac{1}{\ell}\sum_{i:x_i\in\interval(x_b,\ell)}  q_i\dot{x}_i.
\end{align*}

If we define the conduction current as ${\Jcond\equiv\J-\Jconv}$, 
and if ${\normal}$ is a surface's unit-magnitude outward normal, 
then ${\Jcond\cdot\normal}$ is simply
equal to $-e$ times the sum of the rates of change of the integrals of the packets $n_i$ 
whose centers are in the surface region, i.e.,
\begin{align}
\Jcond\cdot\normal = \sum_{i:x_i<x_b}\dot{q}_i.
\end{align}
Regardless of the choice of $x_b$, this does not contain
any contribution from bulk-like primitive cells, because
${\sum_{i:x_i\in\unitcell}\dot{q}_i}$ vanishes for each
bulk-like unit cell $\unitcell$.

\subsection{$\hilbert$-representability and ${\hilbert(t)}$-representability of $n$}
\label{section:H-representability}
I have established that $\Jconv$ can be calculated from any evolving 
bulk microscopic charge density ${\rho(x,t)}$ that can  
can be expressed as a sum ${\sum_i \rho_i}$ of moving packets of fixed amounts charge, each 
of which is either non-positive or non-negative.
When such a representation exists, it is
not unique because, for example, one can always add to any given
representation a co-moving pair of packets of equal and opposite charge, or
combine multiple packets into a single packet.

Clearly the distribution of nuclear charge admits such a representation, so in 
this section I focus on electrons.

\begin{figure}[!]
\includegraphics[width=0.45\textwidth]{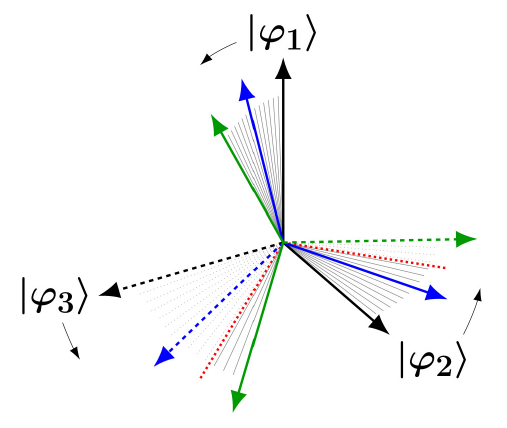}
\caption{
Consider the ground state number density ${n(x;\zeta)}$ 
of two spin-zero electrons in a static confining potential ${\vext(x;\zeta)=-e\varphi(x;\zeta)}$, where
${\zeta}$ is some physical parameter (e.g., $\Eext$). 
Since ${n(\zeta)}$ 
is arbitrarily close to a non-interacting $v$-representable density~\citep{vanleeuwen_2003},
it can be represented as a two dimensional Hilbert subspace 
${\hilbert_\zeta\equiv \SPAN\left\{\ket{\varphi_1(\zeta)},\ket{\varphi_2(\zeta)}\right\}}$, 
of ${\hilbert_\infty}$, which is
an abstract representation of
${\lebesgue(\realone)}$: If ${\P_\zeta}$ is a projector onto subspace 
$\hilbert_\zeta$,  the density is ${n(x;\zeta) =  \expval{\P_\zeta}{x}}$.
In the schematic above, ${\ket{\varphi_1(\zeta)}}$ and ${\ket{\varphi_2(\zeta)}}$
are depicted, for some value of $\zeta$, as solid black arrows; and ${\ket{\varphi_3(\zeta)}}$, 
which is in the orthogonal complement ${\hilbert_\zeta^\perp}$ of ${\hilbert_\zeta}$, is the 
black dashed arrow.
If ${\zeta}$ changes continuously, and while
${n(\zeta)}$'s response to this change is non-singular, ${\hilbert_\zeta}$
rotates continuously within $\hilbert_\infty$. Therefore
${\ket{\varphi_1(\zeta)}}$ and ${\ket{\varphi_2(\zeta)}}$ change continuously (from black to blue)
by mixing with vectors from ${\hilbert_\zeta^\perp}$. However, since the two states ${\ket{\varphi_i(\zeta)}}$
that contribute to the density are those with the lowest eigenvalues ${\epsilon_i(\zeta)}$ of a Hamiltonian
${\hamsmall(\zeta)}$, there may exist a critical
value ${\zeta_c}$ (dotted red lines) at which
${\epsilon_3}$ becomes lower than ${\epsilon_2}$. 
At ${\zeta=\zeta_c}$, 
${\hilbert_\zeta}$ changes \emph{abruptly} to ${\SPAN\left\{\ket{\varphi_1(\zeta)},\varphi_3(\zeta)\right\}}$, 
resulting in a discontinuous redistribution of electron density in $\realone$. 
Before the dotted red line is reached, the rotation of $\hilbert_\zeta$ manifests as a polarization current, 
\begin{align}
\Jconvm=-e\left(\dot{x}_1+\dot{x}_2\right), 
\label{eqn:currentfig}
\end{align}
where ${\dot{x}_i}$ denotes the time derivative of the center of ${\abs{\varphi_i(x;\zeta)}^2=\abs{\braket{x}{\varphi_i(\zeta)}}^2}$;
and $\Jconvm$ could be calculated from Eq.~\ref{eqn:currentfig} for \emph{any} 
basis ${\{\ket{\varphi_1(\zeta)},\ket{\varphi_2(\zeta)}\}}$ of ${\hilbert_\zeta}$.
However, the current that flows when ${\hilbert_\zeta}$
changes discontinuously at ${\zeta=\zeta_c}$ is not polarization current and cannot be calculated in this way.
When the density's response is singular, multiple basis
vectors can be exchanged between ${\hilbert_\zeta}$ and ${\hilbert_\zeta^\perp}$ in less
time than it takes for electrons to respond. In that case the MTOP
approach fails because, for example, if ${\ket{\varphi_i(\zeta)}}$ and ${\ket{\varphi_j(\zeta)}}$ are replaced in
${\hilbert_\zeta}$'s basis by
${\ket{\varphi_k(\zeta)}}$ and ${\ket{\varphi_l(\zeta)}}$, the value of $\Jconvm$ calculated
by assuming that the density at ${x_i}$ was displaced by ${x_k-x_i}$ to ${x_k}$ and
the density at ${x_j}$ was displaced by ${x_l-x_j}$ to ${x_l}$, would differ, in general, 
from the value calculated by assuming that
the densities at ${x_i}$ and ${x_j}$ were displaced to 
${x_l}$ and ${x_k}$, respectively. 
}
\label{fig:rotating-vectors}
\end{figure}

\subsubsection{Electrons}
\label{section:representability}
I say that a number density ${n(x)}$ is {\em $\hilbert$-representable} if there exists a projector $\P$ onto
a Hilbert space of dimension ${\Nelec=\int n}$ such that ${n(x)=\expval{\P}{x}}$. I say that a number density
${n(x,t)}$ is {\em ${\hilbert(t)}$-representable} (`{\em Ht representable}') if it is ${\hilbert}$-representable at all relevant 
times $t$ and if its time-dependent projector ${\P(t)}$ evolves smoothly with $t$. 

It is known that the ground state electron density of any material 
is either noninteracting $v$-representable or arbitrarily close to a 
noninteracting $v$-representable density~\citep{vanleeuwen_2003}.
This means that it can be represented as a set of packets ${n_i=\abs{\varphi_i}^2}$ 
of integral one (two for spin-degenerate packets), where the ${\varphi_i}$'s are the 
lowest-eigenvalue eigenstates of a single electron Hamiltonian, $\hamsmall$.

I make an adiabatic approximation by assuming that the ground state density's time dependence can be expressed as 
a parametric dependence on a slowly- and smoothly-varying stimulus  ${\zeta(t)}$. I express it as 
\begin{align}
n(x;\zeta) & =\sum_{i\leq \Nelec} \abs{\varphi_{i}(x; \zeta)}^2 
= \expval{\P_{\zeta}}{x}
\intertext{where ${\varphi_i(x;\zeta)\equiv \braket{x}{\varphi_i(\zeta)}}$ and}
\P_{\zeta}& \equiv\sum_{i\leq \Nelec} \dyad{\varphi_i(\zeta)} 
\end{align}
is a projector onto the Hilbert space ${\hilbert_\zeta}$ spanned by
the $\Nelec$ eigenvectors  ${\ket{\varphi_i(\zeta)}}$
of the single electron Hamiltonian ${\hamsmall(\zeta)}$ with the
lowest eigenvalues.
${\hilbert_\zeta}$ is an $\Nelec-$dimensional subspace of ${\lebesgue(\realone)}$, the infinite-dimensional
Lebesgue space of real- or complex-valued square integrable functions on ${\realone}$.
It changes as $\zeta$ changes and the eigenstates of ${\hamsmall(\zeta)}$ change.

To understand the representability problem, it may be useful to visualize it 
as it is depicted in Fig.~\ref{fig:rotating-vectors}.
In an insulator each vector in the basis ${\{\ket{\varphi_i(\zeta)}\}_{i=1}^{\Nelec}}$ of ${\hilbert_\zeta}$ changes
gradually with $\zeta$ as vectors from its orthogonal complement ${\hilbert_\zeta^\perp}$
are mixed into them. Therefore ${\hilbert_\zeta}$ rotates smoothly 
within ${\lebesgue(\realone)}$ as $\zeta$ changes. This is because there is a gap in the
eigenspectrum of ${\hamsmall(\zeta)}$ between the ${\Nelec^\text{th}}$ 
and the ${(\Nelec+1)^\text{th}}$ lowest eigenvalues, which never closes as $\zeta$ changes.
In a metal, by contrast, the $\Nelec^\text{th}$ eigenvalue is in a region 
of the spectrum where there is a quasicontinuum of eigenvalues. 
As $\zeta$ changes, the ordering of the eigenvalues is quasicontinuously 
changing, and each time the $\Nelec^\text{th}$ eigenvalue
and the ${(\Nelec+1)^\text{th}}$ eigenvalue cross, the ${\Nelec^\text{th}}$
basis vector ${\ket{\varphi_{\Nelec}(\zeta)}\in\hilbert_\zeta}$ is replaced with a vector
${\ket{\varphi_{\Nelec+1}(\zeta)}\in\hilbert_\zeta^\perp}$ to form ${\hilbert_{\zeta+\dd{\zeta}}}$.

The serene rotation of the basis vectors in a insulator is illustrated by the rotation of the
basis ${\{\ket{\varphi_1},\ket{\varphi_2}\}}$ in  Fig.~\ref{fig:rotating-vectors} {\em before}
${\ket{\varphi_2}}$ reaches the red dashed line, which indicates where
the second and third eigenvalues become equal in this two-electron example.
As soon as the variation of $\zeta$ rotates $\ket{\varphi_2}$ past the red line,
${\hilbert_\zeta}$ changes abruptly from ${\SPAN\{\ket{\varphi_1},\ket{\varphi_2}\}}$
to ${\SPAN\{\ket{\varphi_1},\ket{\varphi_3}\}}$.
The quasicontinuum of eigenvalues in a metal means that, instead of  ${\hilbert_\zeta}$
smoothly rotating, there is a rapid {\em click-clacking} of vectors in and out
of its basis. 

If $\zeta$ changes infinitely slowly, the electrons have time to reach, and settle at, each instantaneous $\hilbert$-representation
before it changes. The system can then be assumed to be close to equilibrium almost all of the time. However
immediately after each exchange of basis vectors between ${\hilbert_\zeta}$ and ${\hilbert_\zeta^\perp}$, 
it could be far from equilibrium.  This is likely to be the case if states are widely-separated spatially, 
such as when they are localized on opposite surfaces or on oppositely-charged electrodes
attached to different parts of the material.
At values of $\zeta$ at which the $\hilbert$-representation changes, the response
of ${n(x;\zeta)}$ to
changes of $\zeta$ is singular and occurs via a nonequilibrium dynamical process
involving many electrons, in general.

If the exchange of basis vectors between ${\hilbert_\zeta}$ and ${\hilbert_\zeta^\perp}$ occurs 
frequently, as is the case in a metal, the electron density does not have time to reach each new $\hilbert$-representation before it changes
again. Therefore the response of electrons in a metal to applied fields is singular and governed
by nonequilibrium dynamics.

In an insulator $\hilbert_\zeta$ rotates smoothly and its dimension is
${\gtrsim 10^{24}}$ for materials at the human scale. The set of vectors that span
it can always be transformed unitarily among themselves to localize
them or delocalize them.  These transformations do not change $\hilbert_\zeta$ or
the projector ${\P_\zeta}$, which means that
they do not change the density ${n(x;\zeta)=\expval{\P_\zeta}{x}}$ or the current
$\Jconv$. Therefore, although it is common to transform the 
eigenfunctions of ${\hamsmall(\zeta)}$ to a more localized set of basis functions
by linearly combining them, in principle this is not necessary.

\section{Single particle states}
\label{section:single_particle_states}
As mentioned above, one can change the minimal (${\Nelec}$-fold) basis of ${\hilbert_\zeta}$ by rotating
it. As shown in Sec.~\ref{section:mtop}, one can also change from a position ($x$) representation, or basis, of each
basis vector of ${\hilbert_\zeta}$ to a
wavevector ($k$) representation of each state, i.e.,  
\begin{align*}
\ket{\varphi_i(\zeta)} &= \intone \dd{x}\varphi_i(x;\zeta)\ket{x}
\end{align*}
where ${\varphi_i(x;\zeta)\equiv \braket{x}{\varphi_i(\zeta)}}$ and ${\braket{x}{x'}=\delta(x-x')}$, or
\begin{align*}
\ket{\varphi_i(\zeta)} &= \intone \dd{k}\ftsvarphi_i(k;\zeta)\ket{k}, 
\end{align*}
where ${\ftsvarphi_i(k;\zeta)\equiv \braket{k}{\varphi_i}}$,
${\braket{k}{k'}=\delta(k-k')}$, and ${\braket{x}{k}\equiv e^{ikx}/\sqrt{2\pi}}$.
One can also use
a mixed position/wavevector representation, as we have already seen
in Sec.~\ref{section:bvkany}, and as is shown in Appendix~\ref{section:appendix_torus}. This is
a common way to exploit the translational symmetry
of a crystal: Roughly speaking, real space is used to describe the electronic
structure in a single primitive unit cell $\Omega$ and reciprocal space is used
to describe variations of the structure between different primitive
cells. 

The polarization current can be calculated in 
any minimal basis ${\left\{\ket{\varphi_i(\zeta)}\right\}_{i=1}^{\Nelec}}$ of ${\hilbert_\zeta}$ and for any 
choice of the basis in which each vector ${\ket{\varphi_i(\zeta)}}$ is represented as a function.
All that is needed is the position operator for the chosen representation, 
which is simply $x$ when working with
${\varphi_i(x;\zeta)\equiv \braket{x}{\varphi_i(\zeta)}}$ 
and is ${i\pdv*{k}}$ when working with
${{\ftsvarphi}_i(k;\zeta)\equiv \braket{k}{\varphi_i(\zeta)}}$.

\subsection{Bloch and Wannier functions}
As discussed in Sec.~\ref{section:bvkany} and Appendix~\ref{section:appendix_torus}, when describing
the bulk of a material theoretically, or when simulating the bulk of a material,
it is common to use Born-von K\'arm\'an periodic boundary conditions~\citep{born_von-karman}.
This is equivalent to representing the material's bulk in 
a torus $\onetorus$ (or $\onetorus^m$ in $m$ dimensions), which 
obviates the need to deal with surfaces. 

If the material is a crystal, the absence of any surfaces means that 
the distributions of electrons and nuclei have the exact $\volume$-periodicity of the crystal
at mechanical equilibrium;
and at thermal equilibrium their time-averaged distributions are $\volume$-periodic.

In solid state physics it is common to use $\onetorus$ to
study the electronic subsystem in the presence of a $\volume$-periodic
distribution of static nuclei. In the limit of heavy nuclei, the
energy of interaction between electrons and nuclei is
\begin{align*}
(n,\vext)\equiv \int_{\onetorus}n(x)\vext(x)\dd{x},
\end{align*}
where
$\vext$ is the $\volume$-periodic {\em external potential}
from the positively charged nuclei.
It is also common to simplify the electronic structure
of the crystal by expressing the electron density $n(x)$ as a sum of contributions from the
eigenfunctions of an effective one-electron Hamiltonian $\hamsmallx$, which inherits
$\volume$-periodicity from ${\vext}$. These eigenfunctions are often interpreted physically
as real states that single electrons, or pairs of electrons with opposite spins, occupy.
However it is well known that this interpretation is not justified by rigorous theory.

As discussed in Appendix~\ref{section:appendix_torus},
the elementary eigenfunctions of a $\volume$-periodic operator,
${\hamsmallx:\lebesgue(\onetorus)\to\lebesgue(\onetorus)}$,
are known as {\em Bloch functions}.
They have the form
\begin{align*}
\bloch_{\alpha k}(x)\equiv\braket{x}{\bloch_{\alpha k}}=e^{ik x}u_{\alpha k}(x), 
\end{align*}
where the {\em Bloch state} ${\ket{\bloch_{\alpha k}}}$ is an eigenstate of
\begin{align*}
\hamsmall\equiv\int_\onetorus\dd{x}\hamsmallx\dyad{x}
\end{align*}
with eigenvalue ${\epsilon_{\alpha k}}$ and ${u_{\alpha k}(x)=\braket{x}{\pbloch_{\alpha k}}}$ has the crystal's $\volume$-periodicity.
The definition of $\hamsmall$ in terms of $\hamsmallx$ implies the inverse relation
${\hamsmallx\equiv\expval{\hamsmall}{x}}$.

Each Bloch function
${\bloch_{\alpha k}}$ can be chosen to be periodic in reciprocal space, and
such that the corresponding periodic
Bloch function ${u_{\alpha k}}$ is real-valued for all wavevectors
${k}$ in the first Brillouin zone.

Both ${\bloch_{\alpha k}(x)}$ and ${u_{\alpha k}(x)}$ are
delocalized over the entirety of $\onetorus$, 
and it follows from the eigenvalue equation, 
\begin{align*}
\hamsmall\ket{\bloch_{\alpha k}}=\epsilon_{\alpha k}\ket{\bloch_{\alpha k}}\iff \hamsmallx\bloch_{\alpha k}=\epsilon_{\alpha k}\bloch_{\alpha k},
\end{align*}
that ${u_{\alpha k}}$ is an eigenfunction of the $k$-dependent
Hamiltonian ${\hamsmallx_k\equiv e^{-ikx}\hamsmallx e^{ikx}}$.

In the large-torus limit, $\hamsmallx_k$ varies continuously with $k$. 
Therefore the eigenstates at different values of $k$ can be labelled such that ${u_{\alpha k}}$
and its eigenvalue ${\epsilon_{\alpha k}}$ vary continuously with $k$.
When the eigenvalues are plotted as functions of $k$, the set of points
${\{(k,\epsilon_{\alpha k})\}_k}$ forms a surface. Note that I use
a subscript $k$ on the parentheses to indicate that different elements of
the set correspond to different values of $k$; and the absence
of a subscript $\alpha$ means that $\alpha$ is the same for all
elements of the set. Therefore ${\{(k,\epsilon_{\alpha k})\}_k}$ 
and ${\{(k,\epsilon_{\beta k})\}_k}$ are different surfaces if ${\alpha\neq\beta}$.
Each surface, or set of intersecting surfaces, is usually referred to as a {\em band}
because its projection onto the energy (eigenvalue) axis is an interval, or `band', of energies.

Let us assume that the set of all Bloch functions, ${\{\bloch_{\alpha k}\}_{\alpha k}}$, has
been chosen to be orthonormal. Then, in an insulator, the electron density 
can be expressed as 
\begin{align*}
n(x) &=
\sum_{\alpha k}\abs{\bloch_{\alpha k}(x)}^2 
= \expval{\left(\sum_{\alpha k}\dyad{\bloch_{\alpha k}}\right)}{x}, 
\end{align*}
where the sum over ${\alpha k}$ is a sum over a finite number of `{occupied}'
states. 
For our purposes, the meaning of a state being occupied is simply that it contribute to $n$;
and will assume that each state is either vacant or occupied by one electron.

In Sec.~\ref{section:bvkany} it was shown how $\Jconv$ could be calculated
directly from the set of PBFs 
using Eq.~\ref{eqn:mtopJfour}.
Now let us try to calculate it using Eq.~\ref{eqn:Jsimple}; and let us also 
try to relate it more directly to the depiction, in Fig.~\ref{fig:rotating-vectors}, 
of the process by which a polarization current arises mathematically.

The Bloch states are an orthonormal basis for the $\hilbert$-representation
of $n$ and, by symmetry, there are the same number of Bloch state centers
in each primitive unit cell of the crystal.
Therefore we could use Eq.~\ref{eqn:Jsimple} to calculate $\Jconv$ from the 
velocities of their centers in one particular primitive cell.
Although this would yield the correct value of $\Jconv$, 
because the Bloch functions are delocalized, this mapping of 
the density onto a set of point charges is conceptually
inconsistent with the mapping envisaged in Sec.~\ref{section:discrete}.
There we assumed a mapping of the charge density, ${\rhom=-e n}$,
onto charge packets of microscopic widths, rather than delocalized
distributions.

We will now look for a microscopically-localized basis for the $\hilbert(t)$-representation
of ${n(x;\zeta(t))}$. Let us begin by expressing the density as
${n(x;\zeta) = \sum_\alpha n_{\alpha}(x;\zeta)}$, where
${n_\alpha(x;\zeta)\equiv \sum_k \abs{\bloch_{\alpha k}(x;\zeta)}^2}$
is the density from all Bloch states that contribute to band $\alpha$.
We know that each ${n_\alpha}$ is ${\hilbert(t)}$-representable because 
\begin{align*}
\P_\alpha(\zeta)\equiv\sum_k\dyad{\bloch_{\alpha k}(\zeta)} 
\end{align*}
is the projector onto its $\hilbert$-representation.
Therefore the polarization current can be 
calculated as the sum ${\Jconv\equiv \sum_\alpha \Jconv_\alpha}$, 
where ${\Jconv_\alpha}$ is the polarization current from the variation of ${n_{\alpha}(x;\zeta)}$
with $\zeta$. Matters become more complicated when bands cross one another~\citep{souza}, but I will
assume that, given any two bands, the band that is lower in energy at any given wavevector is also lower 
in energy at every other wavevector.

Let us focus on the contribution $\Jconv_\alpha$ of band $\alpha$
to $\Jconv$. We can transform the Bloch functions
to a more localized set with the generalized Fourier
transform,
\begin{align}
w_{\alpha X}(x;\zeta) \equiv 
\frac{1}{\sqrt{\Nunitcell}}\sum_{k} e^{-ikX}e^{i\vartheta_\alpha(k)}\bloch_{\alpha k}(x;\zeta),
\label{eqn:wannierdef}
\end{align}
where $\Nunitcell$ is the number of primitive unit cells in $\onetorus$; 
$X$ identifies a particular position ($x$) within the torus; and $\vartheta_\alpha(k)$ is any
$x$-independent constant or function of $k$.
The function ${w_{\alpha X}}$,  which is localized in real space, 
is known as a {\em Wannier function}~\citep{wannier, blount,ferreira_parada, kohn-prb-1973}.

Let ${\wannierxset(X_0)}$ denote the set 
of all images of a point ${X_0\in\onetorus}$ under translations by the crystal's 
lattice vectors, i.e.,
\begin{align*}
\wannierxset(X_0)\equiv\{X_0+m\volume: 0\leq m \leq \Nunitcell-1\}.
\end{align*}
Then it can be shown that, for any choice of ${X_0\in\onetorus}$, 
the set 
\begin{align*}
\wannierset(X_0)\equiv\{w_{\alpha X}: X\in\wannierxset(X_0)\},
\end{align*}
is both orthonormal and satisfies 
\begin{align*}
\sum_k \abs{\bloch_{\alpha k}(x;\zeta)}^2 
= \sum_X\abs{w_{\alpha X}(x;\zeta)}^2 = n_\alpha(x;\zeta).
\end{align*}
Therefore, the set 
${\{\ket{w_{\alpha X}(\zeta)}:X\in\wannierxset(X_0)\}}$ 
of $\Nunitcell$ {\em Wannier states}, 
\begin{align*}
\ket{w_{\alpha X}(\zeta)}\equiv\int_\onetorus \dd{x} w_{\alpha X}(x;\zeta) \ket{x},
\end{align*}
is a minimal orthonormal basis of the ${\hilbert}$-representation of $n_\alpha$, and
${\P_{\alpha}(\zeta)}$ can be expressed as
\begin{align*}
\P_{\alpha}(\zeta) = \sum_{X\in\wannierxset(X_0)} \dyad{w_{\alpha X}(\zeta)}.
\end{align*}

Because a finite integer multiple of $\volume$ separates any two of the points 
in ${\wannierxset(X_0)}$, 
each primitive unit cell contains exactly one of them. 
Furthermore, by substituting ${\bloch_{\alpha k}= e^{ikx} u_{\alpha k}}$ into Eq.~\ref{eqn:wannierdef}
and using the periodicity of ${u_{\alpha k}}$, it can be shown that 
any given Wannier function in set ${\wannierset(X_0)}$
transforms into any other under
a translation by an integer multiple of $\volume$.
Therefore each primitive cell contains the center of exactly one
of element of ${\wannierset(X_0)}$, where the center of ${w_{\alpha  X}}$ is
\begin{align*}
\bar{x}_{\alpha X}(\zeta)\equiv \int_\onetorus x \abs{w_{\alpha X}(x;\zeta)}^2\dd{x}.
\end{align*}
This means that we have decomposed ${n_\alpha}$ into a periodic array of identical localized packets of electron
density, ${n_{\alpha X}(x)\equiv\abs{w_{\alpha X}(x)}^2}$. It follows from Eq.~\ref{eqn:Jsimple} that 
\begin{align*}
\Jconv_\alpha = 
\frac{e}{\volume} \sum_{X}\dv{\bar{x}_{\alpha X}}{t}
=\frac{e\dot{\zeta}}{\volume} \sum_{X}\dv{\bar{x}_{\alpha X}}{\zeta}.
\end{align*}
The degree to which the Wannier functions are localized depends on the choice of $X_0$ and
on the choice 
of the function ${\vartheta_\alpha(k)}$ in Eq.~\ref{eqn:wannierdef}, but the most localized set, which is commonly known
as the set of {\em maximally-localized Wannier functions} (MLWF)~\citep{marzari_mlwf, mlwf_rmp}, is obtained
when ${\vartheta_\alpha}$ is a $k$-independent constant, in which case ${X_0}$ is one of the Wannier centers
and ${\wannierxset(X_0)}$ is the set of all of the Wannier centers
(see \REF~\linecite{ferreira_parada} and Appendix~\ref{section:appendix_wannier}).

\subsubsection{Interpretation of Wannier functions}
\label{section:wannier_interpretation}
Wannier functions, whether maximally localized or not, are not specific to
quantum mechanics and there is no obvious reason
to attach any particular physical meaning to them.

If a density is ${\hilbert}$-representable, its
${\hilbert}$-representation has an infinite number
of orthonormal minimal bases. Among those, there must
exist a maximally localized basis and either a maximally
delocalized basis. 
This is a mathematical observation which does not imply
that elements of these extreme sets have any further meanings or any physical meanings.
Therefore claims that MLWFs have greater physical significances
than elements of other minimal bases should be
substantiated and the precise physical meanings attributed
to them should be clarified. 

The Wannier states of band $\alpha$ are eigenstates of any operator of the form 
\begin{align*}
\Dop_1\equiv \sum_{X} d_{X} \dyad{w_{\alpha X}}, 
\end{align*}
which means that the 
Wannier functions are eigenfunctions of the generally-nonlocal integral operator whose
kernel is ${\Df_1(x';x)\equiv\mel{x'}{\Dop_1}{x}}$.
However, because electrons want to {\em delocalize} rather than localize,  Wannier functions are not, in general, 
either the eigenfunctions, or approximately equal to the eigenfunctions, of an operator that could reasonably be interpreted as the 
Hamiltonian of a real or idealized physical system. 
Therefore if, in a many-particle system, there existed single-particle states 
that could be regarded as `physical', in the sense that they resembled states
that individual particles would like to occupy in a system with simplified energetics 
(e.g., mean field interactions),
they would not be localized, in general, and certainly not by their mutual repulsion.
For example, changing a single-particle Hamiltonian, ${\hamsmall\equiv\hat{t}+\vextop}$, 
by adding a repulsive mean-field Coulomb potential from one or more
localized clouds of negative charge to ${\vextop<0}$, would not make its
eigenfunctions more localized, in general.

Furthermore, there is no `natural' or right way 
to partition the density into the same number of packets as there 
are electrons. Therefore $\vextop$, which together with $\Nelec$ determines the character
of chemical bonds, and which is usually the {\em only} localizing influence
on electrons, does not localize partitions of the density {\em individually}.
It localizes the density as a whole. There is nothing within rigorous physical or
chemical theory to suggest that it bestows a density with a substructure of localized partitions.

I emphasize this point because it has been claimed that Wannier functions, and MLWFs in particular, can
provide insight into chemical bonding by elucidating the substructure
of the electron density~\citep{marzari_mlwf,mlwf_rmp}. However, this
claim has not been justified theoretically, but by references to
the chemistry literature: It was claimed in \REFS~\linecite{marzari_mlwf} and~\linecite{mlwf_rmp} 
that chemists use {\em localized molecular orbitals}, which are the analogs of MLWFs 
for molecules, for this purpose.
However, the papers cited, namely \REFS~\linecite{boys},~\linecite{boys_fostera}, 
~\linecite{boys_fosterb}, and~\linecite{edmiston}, do not justify using localized
orbitals to analyse bonds, and they 
did not introduce them to represent the parts of the electron density that are 
most important to bonding. 
They introduced them to deal more efficiently with those parts of the electron density that are {\em least} important 
to bonding, or to changes in bonding. 

For example, when one is interested in a reaction that involves one reactive part of an otherwise-inert large molecule, 
it is unnecessary and computationally expensive to treat all parts of the molecule as reactive. One can freeze
the electronic structure of the inert part and calculate its effects on the
reactive part using methods that are much more computationally efficient
than treating the whole molecule as reactive would be. 

The same trick 
can be played when studying multiple large molecules, which are mostly the
same, but have different functional groups in one relatively-small region.
After calculating the electronic structure of one of the molecules, it should
not be necessary to recalculate it from scratch for another molecule: 
it is more efficient to reuse parts
of the density that are unaffected by the differences in functional groups.
Localized orbitals were introduced to facilitate this partitioning
of the electronic structure~\citep{boys, boys_fostera, boys_fosterb}.
For example, in calculations based on density functional theory, 
a boundary can be chosen between the reactive part of the molecule, $\mathcal{R}$, 
and the unreactive part, $\mathcal{U}$, and the density can be partitioned using
the centers $\bar{x}_\alpha$ of the localized functions ${w_{\alpha}}$ as
\begin{align*}
n(x) 
= \sum_{\bar{x}_\alpha \in \mathcal{R}} \abs{w_\alpha(x)}^2
+ \sum_{\bar{x}_\alpha \in \mathcal{U}} \abs{w_\alpha(x)}^2
\end{align*} 
The more localized the functions $w_\alpha$ are, the more well-defined the boundary is.
\vspace{0.5cm}

\subsection{The natural single particle substructure of the density}
\label{section:natural_substructure}
The electron density ${n(\rvec)}$ does possess a `natural'
substructure of single-particle states (${\varphi_\alpha}$)
and their `occupancies' ($\occ_\alpha$)~\citep{coleman_rmp,mcweeny_1960,lowdin_1955},
which satisfy
\begin{align*}
n(\rvec) = \sum_{\alpha}\occ_\alpha \abs{\varphi_\alpha(\rvec)}^2, 
&\quad 
\braket{\varphi_\alpha}{\varphi_\beta}=\delta_{\alpha\beta}, \\
\sum_\alpha\occ_\alpha = \Nelec, &\quad \occ_\alpha \leq 1,\; \forall \alpha, 
\end{align*}
and we will assume that they are indexed in order 
of decreasing occupation number, such that 
\begin{align*}
\alpha\leq \beta \iff\occ_\alpha\geq\occ_\beta.
\end{align*}
These {\em natural orbitals} are the normalized
eigenstates of the 1-particle reduced density matrix.
Their properties, some of which are discussed in Appendix~\ref{section:appendix_natural},
suggest that they are the only single-particle states 
that should be regarded as characteristics, or substructural components, 
of a many-particle state.

As an illustration of the physical meaning of natural orbitals, it
is proved in Appendix~\ref{section:appendix_natural} that
the energy of a normalized ${\Nelec}$-particle pure state, ${\Psi}$, 
can be expressed \emph{exactly} as
\begin{subequations}
\begin{align}
E&\equiv\expval{\hat{H}}{\Psi}
=\sum_\alpha
\occ_\alpha
\energy_{\alpha}
+
\sum_\alpha
\sum_{\beta\geq\alpha}\sqrt{\occ_\alpha\occ_\beta}
\,
w_{\alpha\beta}
\label{eqn:natural_energy01}
\\
&=\sum_\alpha\occ_\alpha
\left(
\energy_{\alpha}
+\vmf_\alpha\right)
+
\sum_\alpha
\sum_{\beta>\alpha}\sqrt{\occ_\alpha\occ_\beta}
\,
w_{\alpha\beta}
\label{eqn:natural_energy02}
\\
&=\sum_\alpha\occ_\alpha
\left(
\energy_{\alpha}
+
\frac{1}{2}\sum_{\beta}\sqrt{\frac{\occ_\beta}{\occ_\alpha}}w_{\alpha\beta}
\right),
\label{eqn:natural_energy03}
\end{align}
\label{eqn:natural_energy}
\end{subequations}
where the sums over $\alpha$ and $\beta$ are over the set of all natural orbitals, 
which is an infinite set.
Explanations of the symbols on the right hand side of Eq.~\ref{eqn:natural_energy02} will now be provided, 
and interpretations of them will be suggested.
It will be important to note that
the $\Nelec$-particle wavefunction ${{\Psi}}$ can be expressed
exactly as the infinite sum, 
\begin{align}
\Psi(\rvecsub{1}\cdots\rvecsub{\Nelec})
=\sum_\alpha c_\alpha\varphi_\alpha(\rvecsub{1})\Theta_\alpha(\rvecsub{2}\cdots\rvecsub{\Nelec}),
\label{eqn:natural_psi}
\end{align}
where 
${\displaystyle \abs{c_\alpha}^2=\lambda_\alpha=\frac{\occ_\alpha}{\Nelec}}$;
${\sum_\alpha\lambda_\alpha=1}$; and
${\Theta_\alpha}$ is an eigenfunction
of the ${(\Nelec-1)}$-particle reduced density matrix, 
with the same eigenvalue $\lambda_\alpha$ as ${\varphi_\alpha}$.
Functions
$\Theta_\alpha$ and ${\varphi_\alpha}$ are \emph{duals} of one another, in the
sense that the \emph{contraction} of ${\Psi}$ by
${\varphi_\alpha}$ is ${c_\alpha\Theta_\alpha}$ and the 
contraction of ${\Psi}$ by ${\Theta_\alpha}$ is ${c_\alpha\varphi_\alpha}$.
That is,
\begin{align*}
\varphi_\alpha\rfloor\Psi(\rvecsub{1}\cdots\rvecsub{N-1})
&\equiv \int\dd[3]{r}\varphi_\alpha^*(\rvec)\Psi(\rvec,\rvecsub{1}\cdots\rvecsub{N-1})
\\
&= c_\alpha\Theta_\alpha(\rvecsub{1}\cdots\rvecsub{N-1})
\end{align*}
and
\begin{align*}
\Theta_\alpha\rfloor\Psi(\rvec)
&\equiv \int\dd[3]{r_1}\cdots\int\dd[3]{r_{N-1}}\Theta_\alpha^*(\rvecsub{1}\cdots\rvecsub{N-1})
\\
&\times\Psi(\rvec,\rvecsub{1}\cdots\rvecsub{N-1})
= c_\alpha\varphi_\alpha(\rvec).
\end{align*}

Note that the derivation of Eqs~\ref{eqn:natural_energy} 
in Appendix~\ref{section:appendix_natural} applies to
any classical or quantum mechanical state ${\Psi}$ 
whose position probability density function is ${\Psi^*\Psi}$;
and the property ${\occ_\alpha\leq 1}$
arises from the normalization ${\braket{\Psi}=1}$: The fact that
${\{\varphi_\alpha\Theta_\alpha\}}$
is an orthonormal basis implies that 
\begin{align*}
\frac{\occ_\alpha}{N}=\abs{c_\alpha}^2=\abs{\braket{\varphi_\alpha\Theta_\alpha}{\Psi}}^2\leq 1.
\end{align*}

\subsubsection{Independent electron energy, $\varepsilon_\alpha$}
\label{section:independent_electron_energy}
The first term on the right hand side of 
Eq.~\ref{eqn:natural_energy01} is an occupation-weighted sum of 
\begin{align*}
\energy_\alpha\equiv\expval{\hamsmall}{\varphi_\alpha}=
\underbrace{\expval{\kinetic}{\varphi_\alpha}}_{\displaystyle \vphantom{\vext_\alpha}t_\alpha}+\underbrace{\expval{\vextop}{\varphi_\alpha}}_{\displaystyle \vext_\alpha},
\end{align*}
where 
${\hamsmall\equiv \kinetic + \vextop}$
would be the Hamiltonian of a single electron if the remaining ${\Nelec-1}$ electrons were not present:
$\kinetic$ is the single electron kinetic energy operator and
${\vextop}$ is the operator for the energy
of interaction between an electron and the nuclei. The nuclei are assumed to be static.

\paragraph*{Non-interacting electrons:} 
If all of the orbital coupling energies, $w_{\alpha\beta}$, vanished, the
occupation numbers at the energy minimum (i.e., the ground state) would be
\begin{align*}
\occ_\alpha
=
\begin{cases}
1,\;\text{if}\;\alpha\leq \Nelec,
\\
0,\;\text{if}\;\alpha> \Nelec,
\end{cases}
\end{align*}
and the occupied orbitals would be those with the lowest energies.
In words, the $\Nelec$ orbitals with the lowest energies would be occupied
and all other orbitals would be vacant.

The $\Nelec$ occupied natural orbitals of the ground state 
are those that minimize
\begin{align}
\sum_{\alpha=1}^{\Nelec} \energy_\alpha=
\sum_{\alpha=1}^{\Nelec} \expval{\hamsmall}{\varphi_\alpha},
\label{eqn:ground_state_energy}
\end{align}
while preserving
orthonormality. 

Orthonormality of the ground state natural orbitals implies that each term in Eq.~\ref{eqn:ground_state_energy}
is a stationary value of
\begin{align*}
\frac{\expval{\hamsmall}{\varphi}}{\braket{\varphi}}, 
\end{align*}
which means that each natural orbital ${\varphi_\alpha}$ is an eigenstate of
${\hamsmall}$.

\subsubsection{Interaction energy}
\label{section:natural_orbital_coupling}
The interaction energy, which is responsible for electrons moving between 
orbitals, is
\begin{align}
\sum_{\alpha, \beta\geq\alpha}\! \sqrt{\occ_\alpha\occ_\beta}w_{\alpha\beta}
=
\sum_\alpha \!\occ_\alpha\left(\vmf_\alpha + \sum_{\beta>\alpha}\sqrt{\frac{\occ_\beta}{\occ_\alpha}}w_{\alpha\beta}\right),
\label{eqn:interaction_energy}
\end{align}
where
${w_{\alpha\beta}\equiv 2\Re\left\{\mel{\varphi_\alpha}{\what_{\alpha\beta}}{\varphi_\beta}\right\}}$,
\begin{align*}
\what_{\alpha\beta}(\rvec)
\equiv \int\dd[3]{r_1}&\cdots\int\dd[3]{\rsub{\Nelec-1}}\bar{\Theta}_\alpha(\rvecsub{1}\cdots \rvecsub{\Nelec-1})
\\
\times
&\left(\sum_{j=1}^{\Nelec-1}\hamtwo(\rvec,\rvecsub{j})\right)
\Theta_\beta(\rvecsub{1}\cdots \rvecsub{\Nelec-1}),
\end{align*}
and ${\hamtwo(\rvec,\rvecsub{j})}$ 
denotes the energy of interaction between an electron at $\rvec$ and an electron at ${\rvecsub{j}}$.

\paragraph*{Mean-field interaction:}
The quantity ${\vmf_\alpha}$ on the right hand sides of Eq.~\ref{eqn:natural_energy02} and Eq.~\ref{eqn:interaction_energy} 
is 
\begin{align}
\vmf_\alpha = \frac{1}{2}\int\dd[3]{r}\abs{\varphi_\alpha(\rvec)}^2\what_{\alpha\alpha}(\rvec),
\label{eqn:vmf_explanation}
\end{align}
which is 
one half of the energy of interaction of an electron in orbital ${\varphi_\alpha}$ with the mean field
charge density of ${N-1}$ electrons occupying the dual state, ${\Theta_\alpha}$, of ${\varphi_\alpha}$.
That charge density is
\begin{align*}
\rho_\alpha^{\scriptscriptstyle (\Nelec-1)}(\rvec)\equiv -e\,\densityn_\alpha(\rvec),
\end{align*}
where ${\densityn_\alpha(\rvec)}$ is the number density of ${N-1}$ electrons
in state ${\Theta_\alpha}$.
Just as one electron's share of its energy of interaction with another electron is \emph{half}
of it, the share of the mean-field interaction energy ${2\vmf_\alpha}$ that belongs
to the electron in orbital ${\varphi_\alpha}$ is ${\vmf_\alpha}$. 

The expectation value of the interaction energy
of an electron in orbital $\varphi_\alpha$ with the ${N-1}$ electrons in a state ${\Theta_\alpha}$
would be ${2\vmf_\alpha}$
if the probability density of the single electron did not depend on the configuration of the ${N-1}$ electrons, 
and the probability density of the ${N-1}$ electrons did not depend on the position of the single electron.
In other words, ${2\vmf_\alpha}$ is an expectation value that is calculated without accounting for the correlation between the
motion of the electron in orbital $\varphi_\alpha$ and the motions of the electrons in state ${\Theta_\alpha}$.

\paragraph*{Correlation correction:}
One way to think of ${w_{\alpha\beta}}$ when ${\alpha\neq\beta}$ is as a correction to ${\vmf_\alpha}$ that accounts
for correlation: It accounts for the fact that ${\vmf_\alpha}$ has been calculated from a $N$-particle
pdf of the form 
\begin{align*}
\pdfarg{\varphi}_\alpha(\rvecsub{1})\pdfarg{\Theta}_\alpha(\rvecsub{2}\cdots\rvecsub{\Nelec})
\end{align*}
rather than one of the form
\begin{align*}
\pdfarg{\varphi|\Theta}_\alpha(\rvecsub{1}|\rvecsub{2}\cdots\rvecsub{\Nelec})&\pdfarg{\Theta}_\alpha(\rvecsub{2}\cdots\rvecsub{\Nelec})
\\
&=
\pdfarg{\varphi}_\alpha(\rvecsub{1})\pdfarg{\Theta|\varphi}_\alpha(\rvecsub{2}\cdots\rvecsub{\Nelec}|\rvecsub{1}),
\end{align*}
where ${\pdfarg{\varphi|\Theta}_\alpha}$ and ${\pdfarg{\Theta|\varphi}_\alpha}$ are \emph{conditional}
pdfs.
 
Another way to think of ${w_{\alpha\beta}}$ is as a coupling of orbital ${\varphi_\alpha}$ to orbital ${\varphi_\beta}$, which 
is \emph{mediated} by their dual states ${\Theta_\alpha}$ and ${\Theta_\beta}$, respectively. If it were a \emph{bare} (unmediated)
interaction, it would be ${\mel{\varphi_\alpha}{\what}{\varphi_\beta}}$ rather than ${\mel{\varphi_\alpha}{\what_{\alpha\beta}}{\varphi_\beta}}$.

\subsubsection{Hartree-Fock approximation}
In Appendix~\ref{section:appendix_natural} it is shown that, within the \emph{Hartree-Fock approximation}, the electron density is
\begin{align*}
n(\rvec)=\sum_{\alpha=1}^{\Nelec}\abs{\varphi_\alpha(\rvec)}^2, 
\end{align*}
and ${\what_{\alpha\beta}(\rvec)}$ 
simplifies to 
\begin{align*}
\what_{\alpha\beta}(\rvec)
= \int\what(\rvec,\rvec')\left[n(\rvec')-n_\alpha(\rvec')-n_\beta(\rvec')\right]\dd[3]{r'},
\end{align*}
where 
${n_\alpha(\rvec)=\abs{\varphi_\alpha(\rvec)}^2}$ and
${n_\beta(\rvec)=\abs{\varphi_\beta(\rvec)}^2}$. 
In other words, within the Hartree-Fock approximation, ${\what_{\alpha\beta}(\rvec)}$ is
the energy of interaction of an electron at ${\rvec}$ with the mean field charge
density from all of the electrons except those occupying orbitals ${\varphi_\alpha}$
and ${\varphi_\beta}$. The `share' of each of the electrons in orbitals
${\varphi_\alpha}$ and ${\varphi_\beta}$ of their coupling energy is
\begin{align*}
\frac{1}{2}w_{\alpha\beta}=\Re\left\{\mel{\varphi_\alpha}{\what_{\alpha\beta}}{\varphi_\beta}\right\}.
\end{align*}

\subsubsection{Interpreting natural orbitals and their energies}
Before discussing how natural orbitals and their occupation numbers 
should be interpreted, I wish to reiterate that
Eqs.~\ref{eqn:natural_energy}
are derived in Appendix~\ref{section:appendix_natural} without making
any non-classical assumptions. Therefore all of the mathematics in this section 
is compatible with the particles being identical and fast-moving (${\implies}$ indistinguishable)
billiard balls. 

I also wish to point out that 
it may not always be useful or appropriate to interpret the mathematics in this section
as meaning that the particles `occupy' orbitals. Sometimes it might be more appropriate
to regard the orbitals as nothing more than basis functions. 
Having said this, I will return to referring to the particles as electrons that occupy orbitals.

When electrons interact with one another, the natural orbitals are not eigenstates of the Hamiltonians
${\hamsmall_\alpha}$ or ${\hamsmall}$, or of any other Hamiltonian.
This is to be expected of `\emph{physical}' single-particle
states because Hamiltonian eigenstates are stationary states;
and the states that individual electrons occupy cannot be stationary 
if they interact with other electrons while occupying them.
Interactions perturb the electron occupying orbital ${\varphi_\alpha}$
and, sooner or later, displace it from that state to another one.

Each occupation number $\occ_\alpha$ can be
interpreted as either the fraction of time for which the
$\alpha^\text{th}$ natural orbital is occupied, or the probability 
that it is occupied at any given time. If it is possible
for an electron occupying natural orbital $\varphi_\alpha$ to be displaced from 
it, the fraction of time for which ${\varphi_\alpha}$ is occupied
must be less than one. 
This is one way to understand why 
the occupation numbers are less than one
for interacting particles.

The forms of Eqs.~\ref{eqn:natural_energy} suggest that
they can be interpreted within a quasi-independent-electron picture as follows:
When orbital ${\varphi_\alpha}$ is occupied, the energy of the electron occupying it is the sum of
the orbital energy, $\energy_\alpha$, and its share of its energies of interaction with 
electrons in other orbitals.
An equivalent statement is that its energy is the sum of $\energy_\alpha$ and its share of 
the energy of its interaction with the set of ${\Nelec-1}$ electrons occupying 
state ${\Theta_\alpha}$.  State ${\Theta_\alpha}$
is orthogonal to $\varphi_\alpha$, but is not orthogonal to all other natural orbitals.
Equation~\ref{eqn:natural_energy01} expresses the total energy as an occupation-weighted 
sum of orbital energies plus a sum over orbital pairs, ${\{\varphi_\alpha,\varphi_\beta\}}$, 
of the energy of mediated coupling between them,
weighted by the geometric mean of their occupation numbers, ${\sqrt{\occ_\alpha\occ_\beta}}$.

An important feature of Eq.~\ref{eqn:natural_energy01}, and the definition of $\energy_\alpha$, 
is that the total kinetic energy is exactly
\begin{align*}
\expval{\hat{T}}{\Psi}=\sum_\alpha\occ_\alpha\expval{\,\hat{t}\,}{\varphi_\alpha}, 
\end{align*}
which is simply the occupation-weighted sum of the natural orbitals' individual kinetic energies.
If ${\Psi}$ was expanded in terms of any other set of orbitals,  ${\{\psi_\alpha\}}$,
the kinetic energy would have the more complicated form,
\begin{align*}
\expval{\hat{T}}{\Psi}=
\sum_{\alpha}\sum_{\beta}C^*_{\alpha}C_\beta\mel{\psi_\alpha}{\,\hat{t}\,}{\psi_\beta}.
\end{align*}
The natural orbitals are the only orbitals for which the kinetic energy of the interacting many-electron system
can be expressed \emph{exactly} as a weighted sum of {\em single}-orbital contributions. 

Furthermore, 
if the magnitude of the electron-electron repulsion could be reduced gradually to zero, 
in the non-interacting limit
the energy would become
\begin{align*}
E=\sum_{\alpha=1}^\infty \occ_\alpha \expval{\hamsmall}{\varphi_\alpha}.
\end{align*}
As discussed in subsection~\ref{section:independent_electron_energy}, 
the occupied eigenstates of the noninteracting Hamiltonian, $\hamsmall$, 
are natural orbitals of the electronic ground state, $\Psigs$.

\subsubsection{How localized are natural orbitals?} 
As discussed in Secs.~\ref{section:H-representability} and~\ref{section:wannier_interpretation}, 
if an $\Nelec$-electron density ${n(\rvec)}$ is $\hilbert$-representable, it can be expressed as a sum of contributions
from $\Nelec$ maximally localized orbitals, or as a sum of contributions from $\Nelec$ maximally delocalized
orbitals.
When a crystal's bulk is represented in ${\onetorus^3}$, the density can be expressed as a sum of
contributions from ${\Nelec}$ Bloch functions,
which are \emph{infinitely} delocalized in the sense that each
one has an equal presence in every unit cell.

Therefore, unless the counterparts of Bloch functions in the bulks of real crystals (i.e., with surfaces, and not in a torus)
are qualitatively different, 
almost all elements $\psi_\alpha$, of the maximally delocalized basis of the $\hilbert$-representation of a crystal's electron density,
are delocalized throughout the entire crystal. There might be exceptions near surfaces or
other macroscopic heterogeneities, but, in a macroscopic crystal, the overwhelming majority of them 
would be spread throughout the entire crystal. 

Let us disregard the possible exceptions; and, on the basis of what is known theoretically about the bulks of crystals under Born-von K\`arman boundary conditions, 
let us assume that, in any macroscopic material, there exists a \emph{$\Nelec$-fold} set ${\{\psi_\alpha: 1\leq\alpha\leq\Nelec\}}$, 
almost all of whose elements are delocalized throughout the entire material, such that the material's electron density is
${n=\sum_{\alpha=1}^{\Nelec}\abs{\psi_\alpha}^2}$.

If that is the case, there must exist \emph{infinite} sets of equally delocalized or more delocalized 
orbitals ${\{\tpsi_\alpha\}}$, and their occupation numbers
\begin{align*}
\left\{\occ_\alpha\in[0,1]:\sum_\alpha\occ_\alpha=\Nelec\right\},
\end{align*}
such that 
\begin{align*}
n(\rvec)=\sum_{\alpha=1}^{\infty}\occ_\alpha\big|\tpsi_\alpha(\rvec)\big|^2=\sum_{\alpha=1}^{\Nelec}\abs{\psi(\rvec)}^2 
\end{align*}
and
\begin{align*}
\sum_{\alpha=1}^{\infty}\occ_\alpha\myexpval{\kinetic}{\tpsi_\alpha}\leq
\sum_{\alpha=1}^{\Nelec}\expval{\kinetic}{\psi_\alpha}.
\end{align*}

The reason to point this out is that, for noninteracting `electrons' (${\hamtwo=0\implies w_{\alpha\beta}=0, \;\forall \alpha,\beta}$), 
the energy is simply 
\begin{align*}
E=\sum_\alpha\occ_\alpha\energy_\alpha = \underbrace{(n,\vext)}_{\sum_\alpha\occ_\alpha\vext_\alpha}+\sum_\alpha\occ_\alpha t_\alpha.
\end{align*}
Since ${(n,\vext)}$ is independent of the density's substructure of orbitals and occupation numbers, 
if the ground state density $n_0$ were known, the natural orbitals and occupation numbers 
of the ground state would be those that minimized ${\sum_\alpha\occ_\alpha t_\alpha}$ under the 
constraint ${\sum_\alpha\occ_\alpha\abs{\varphi_\alpha}^2=n_0}$.

Delocalizing an orbital lowers its kinetic energy. Therefore, if a material (composed
of \emph{noninteracting} electrons) has a ground state 
density that can be expressed as a weighted sum of contributions from delocalized orbitals, 
its natural orbitals are no less delocalized than those orbitals.

Since the ground state natural orbitals of a material's noninteracting electrons would be delocalized
throughout the entirety of the material, a real material's ground state natural
orbitals would only be localized if their energy of mutual repulsion could be
lowered by them localizing. 
I have not been able to rule this possibility out, but it seems
unlikely to be the norm: Localization of orbitals would have to reduce the electrons' 
energy of mutual repulsion by more than it increased their kinetic energies. This seems
quite hard to achieve given the possibly-na\"ive expectation that, 
if all of the orbitals were spread across ${\Navogadro\approx 6.02\times 10^{23}}$
unit cells of a crystal, their kinetic energy would be lowered by a factor
of ${\sim 10^{8}\approx \Navogadro^{\frac{1}{3}}}$ by reducing their average
width to the length of a primitive lattice vector.

The literature on \emph{many body localization} and related
phenomena might
shed some light on this issue~\citep{mbloc_1,mbloc_2,mbloc_3,mbloc_4,mbloc_5,mbloc_6}. Conversely, 
Eq.~\ref{eqn:natural_energy01} or one of its variants might prove useful in studies of those phenomena.

\subsubsection{How large are the largest occupation numbers, and how rapidly do they decay?}
The questions of how close the largest occupation numbers ($\occ_1$, $\occ_2$, $\occ_3$, etc.) are to unity, 
and of how slowly ${\occ_\alpha}$ decays with increasing $\alpha$, 
appear to be of enormous fundamental importance for a variety of reasons~\citep{natural_occupations, cioslowski,cioslowski_2021,cioslowski_2019,tognetti_2016,helbig_2010,umrigar_1998}.
They might even be related to the question of which number densities are $\hilbert$-representable. 
However, although some calculations of natural orbitals and their occupation numbers have been 
performed for simple molecules and small contrived systems~\citep{tognetti_2016,natural_occupations,cioslowski,cioslowski_2021,cioslowski_2019}, 
little seems to be known about the occupation numbers in a macroscopic material.

One way to formulate the set of questions is to define the function,
\begin{align*}
F_\occ(s)\equiv
\frac{1}{\Nelec}\sum_{\alpha< s\Nelec}\occ_\alpha;
\end{align*}
and to ask what a plot of ${F_\occ}$ versus $s$ would look like; how it
would depend on the value of $\Nelec$; and how it would depend on other characteristics of the physical system.

It is known that \emph{ground state} occupation numbers become equal to one when interactions
between particles are turned off, or when they
are replaced with mean-field interactions, as in the drastic Hartree-Fock
approximation discussed in Appendix~\ref{section:hartree_fock} and at the end of
subsection~\ref{section:natural_orbital_coupling}. Therefore, in those cases, the
value of ${F_\occ}$ would be one for ${s\leq 1}$ and zero for ${s>1}$.

However, as discussed in subsection~\ref{section:natural_orbital_coupling}, it is \emph{interactions} that allow particles to pass between orbitals. 
Therefore, the fact that all occupation numbers are either $1$ or $0$ in a system without interactions should not be extrapolated
to the conclusion that they are close to either $1$ or $0$ in a system of interacting particles.
There does not appear to be any theoretical justification for such an extrapolation.
One reason for emphasizing this point is that I use the assumption that I object to, or that I am wary of, 
as the foundation on which the derivation in the next subsection is built.

\subsubsection{Classical derivation of the Fermi-Dirac distribution}
\label{section:fermi_dirac_derivation}
As mentioned above, the derivation of Eqs.~\ref{eqn:natural_energy} in 
Appendix~\ref{section:appendix_natural}
did not require either the $1$-particle Hamiltonian ${\hamsmallx=\expval{\hamsmall}{x}}$, 
or the interaction ${\what(x,x')=\mel{x}{\what}{x'}}$ 
to have any particular forms. They were only required to 
be functions of one and two positions, respectively;
or to be operators that act on one and two positions, respectively.
Therefore they are general expressions, which are perfectly consistent with classical physics.

In this subsection 
it will again be assumed that interactions are so weak that they are negligible. This 
assumption is not justified in the present context, but some of its consequences
can provide insight.
Under this assumption  the expectation value of the energy is simply
\begin{align*}
E&=\sum_\alpha\occ_\alpha\energy_\alpha.
\end{align*}
Then, since the energy is determined by the $1$-particle density matrix, 
${\Df_1(\rvec;\rvec')\equiv \mel{\rvec}{\Dop_1}{\rvec'}}$, 
which can be expressed in terms of its eigenfunctions and eigenvalues as
\begin{align*}
\Df_1(\rvec;\rvec')
&=\sum_\alpha\lambda_\alpha \varphi^*_\alpha(\rvec)\varphi_\alpha(\rvec') 
\\
&= \frac{1}{N}\sum_\alpha\occ_\alpha \varphi^*_\alpha(\rvec)\varphi_\alpha(\rvec'), 
\end{align*}
both $\Df_1$ and ${E}$ are determined by the sets ${\{\varphi_\alpha\}}$ 
and ${\{\occ_\alpha\}}$.

Now let us assume that the statistical state reflects a state
of knowledge/ignorance; and that the only thing known about the physical system
is that the value of $E$, the expectation value of its energy, is ${\tilde{E}}$. 
Then the natural orbitals and their occupation numbers are those
that maximize uncertainty subject to two constraints:  ${E=\tilde{E}}$ and 
the number of particles is $\Nelec$.
The logical basis for this approach, which is due to Jaynes and Shannon~\citep{jaynes1,jaynes2,shannon}, 
is discussed in some detail in~\linecite{tangney_bose_einstein}.

The uncertainty associated with the occupancy of orbital $\varphi_\alpha$ (i.e., the uncertainty about whether
it is occupied or empty)
can be quantified by the Shannon entropy, which
is the expectation value of the Shannon information~\citep{shannon,gu_murray_tangney}, i.e., 
\begin{align*}
\entropy_\alpha = \occ_\alpha(-\log\occ_\alpha) + (1-\occ_\alpha)(-\log(1-\occ_\alpha)).
\end{align*}
The expression on the right hand side of this equation 
is the probability that the state is occupied, 
multiplied by the quantity of information 
that would be revealed by the discovery that it is occupied
(${-\log\occ_\alpha}$), 
plus
the probability that it is unoccupied times the quantity  of information 
that would be revealed by the discovery that it is unoccupied.

The expectation value of the quantity of information that would be revealed by 
discovering which 
$\Nelec$ of the orbitals are occupied is
\begin{align}
\entropy\left[\{\occ_\alpha\}\right]=-\sum_\alpha\left[\occ_\alpha\log\occ_\alpha + (1-\occ_\alpha)\log(1-\occ_\alpha)\right].
\label{eqn:occupation_entropy}
\end{align}
The set ${\{\varphi_\alpha\}}$ is also unknown. Therefore, 
once the occupation numbers were known, a quantity of uncertainty that will be denoted
by ${\tentropy[\{\varphi_\alpha\}]}$ would remain. 
This quantity will not play much of a role in what follows.

To find ${\{\occ_\alpha\}}$ and ${\{\varphi_\alpha\}}$, 
we must maximise uncertainty subject to the constraints that what is known about
the state is true~\citep{tangney_bose_einstein,shannon,jaynes1,jaynes2}. 
The logic underpinning this approach to deriving empirically unfalsifiable
statistical models is discussed in detail in~\linecite{tangney_bose_einstein}.
In the present context the derivation entails finding a set ${\{\occ_\alpha\}}$ and
a set ${\{\varphi_\alpha\}}$ for which
\begin{align*}
\delta\bigg\{&\entropy\left[\{\occ_\alpha\}\right] + \tentropy\left[\{\varphi_\alpha\}\right]
-\beta\left(\sum_\alpha \occ_\alpha\energy_\alpha[\varphi_\alpha]-\tilde{E}\right)
\\
+&\beta\upmu\left(\sum_\alpha\occ_\alpha - N\right)  - \sum_{\alpha\beta}\upeta_{\alpha\beta}\left(\braket{\varphi_\alpha}{\varphi_\beta}-\delta_{\alpha\beta}\right)\bigg\}=0, 
\end{align*}
where ${\beta}$, ${\beta\upmu}$, and ${\{\upeta_{\alpha\beta}\}}$ are Lagrange multipliers to enforce, respectively, 
the information constraints ${E=\tilde{E}}$ and ${\sum_\alpha\occ_\alpha=N}$, and the set 
of orthonormality constraints, 
${\left\{\braket{\varphi_\alpha}{\varphi_\beta}=\delta_{\alpha\beta}, \forall\;\alpha,\beta\right\}}$.

The partial derivatives with respect to ${\occ_\alpha}$ 
of ${\energy_\alpha}$, ${\tentropy}$, and ${\braket{\varphi_\alpha}{\varphi_\beta}}$ all vanish. Therefore, at
stationarity, the occupation numbers satisfy
\begin{align*}
\pdv{\occ_\alpha}\left[\entropy\left[\{\occ_\alpha\}\right] - \beta \sum_\alpha \occ_\alpha\energy_\alpha+\beta\upmu\sum_\alpha\occ_\alpha\right]=0.
\end{align*}
It is straightforward to show from this expression and Eq.~\ref{eqn:occupation_entropy}
that the occupation numbers of this possibly-classical system are
\begin{align*}
\occ_\alpha = \frac{1}{1+e^{\beta(\energy_\alpha-\upmu)}}, 
\end{align*}
which means that the distribution of occupation numbers among the orbitals' energies is
a \emph{Fermi-Dirac} distribution.

This derivation is only valid in the non-interacting limit, which is the limit that 
must be assumed to derive the Fermi-Dirac distribution for quantum mechanical particles.
A classical derivation of the \emph{Bose-Einstein} distribution is presented in \linecite{tangney_bose_einstein}.

\subsection{Localized orbital based models of bonding}
Since the early days of quantum theory it has been common
to approximate the expectation value of the energy of a set of 
electrons as ${E\approx \Esingleapprox+\Emf+\Erestapprox}$, where 
${\Esingleapprox}$ approximates the  expectation value $\Esingle$ of the
sum of the electrons' $1$-particle energies;
$\Emf$ is a $1$-particle mean field approximation to the expectation value of
the sum of $2$-particle interaction energies;  and
\begin{align*}
\Erestapprox\approx E-\Esingleapprox-\Emf\approx \Erest\equiv E-\Esingle-\Emf.
\end{align*}
The true $1$-particle contribution, $\Esingle$, 
is the expectation value of the
sum of the  electrons' kinetic energies and their energies of interaction with any 
external fields, such as the electric field from nearby nuclei.

The main reason to express $E$ as a sum of $1$-electron contributions
plus a correction, and to make the correction as small as possible
by augmenting the $1$-electron contributions with a mean field 
contribution, $\Emf$,  is that $1$-electron energies
and orbitals are relatively easy to calculate, understand, and 
relate to one another. 

Their ease of use and relative conceptual simplicity
are partly responsible for orbitals 
playing a pervasive role in the teaching of chemistry, and in
the interpretation and communication of chemical research.
Another reason for their widespread use is that 
orbital-based reasoning can undoubtedly be predictive; and 
orbital-based calculations of bond lengths, excitation
energies, and other measurables, can be very accurate.

The prominence of orbitals within chemistry's lexicon might also be, 
to some small degree, an artefact of the history of quantum theory, including some 
very early and wildly inaccurate speculations about the natures of chemical bonds~\citep{lewis_1916,gillespie_robinson_2007}.
Notable contributors to the proposal and subsequent refinement, or abandonment, of early models of electronic
structure include Gilbert Lewis~\citep{lewis_1916}, Irving Langmuir~\citep{langmuir_covalency}, Hund~\citep{hund_1926}, Mulliken~\citep{mulliken_1928}, Lennard-Jones~\citep{lennard-jones_1929}, 
and Linus Pauling~\citep{pauling_1926,pauling_1928,pauling_1931_1, pauling_1931_2,pauling_1931_3,pauling_1932_1, pauling_1932_2,pauling_1960}. 
Dramatic progress in our understanding of chemical bonding was made in the ${\sim 15}$ years following the publication of \linecite{lewis_1916}, 
and many of the discoveries made remain useful and justified by theory. However, some
artefacts of early models of bonding, which have never been justified experimentally or by 
rigorous theory, are still being presented in introductory chemistry textbooks as established features of reality. 

An example is the idea that the instantaneous electronic structure
of an atom can approximately, but realistically, be specified as
a set of highly-localized nucleus-centered orbitals and their occupation states. In the simplest
versions, only four occupation states are possible:
If an orbital is not empty ($-$), it can be occupied by
one `spin up' electron ($\spinup$), one `spin down' electron ($\spindown$), or both (${\spinupdown}$).

The premise that an atom's electron cloud has this substructure of integer-occupied localized orbitals is 
then used as a basis for simplistic descriptions of chemical bonds. For example, 
an ionic bond is often explained as a Coulomb attraction between two ions, which are formed from
two atoms when  one or more of one atom's \emph{electrons} occupy one or more of another atom's orbitals.
However the charges of ions in a real ionic bond are not integer multiples of
an electron's charge, because the bonded ions are not formed
by atoms transferring electrons between them. They are formed by the atoms transferring
probability, or number density, between them,

Therefore, although the electron cloud of an isolated charge-neutral atom 
has a number density ${n}$ whose integral is an integer (i.e., the atom's atomic number), 
the integral of ${n}$ over the electron
cloud of a bonded ion is not an integer, either usually or in general, which implies
that the charges of bonded ions are not integer multiples of $e$. For example, even in
the canonically-ionic compound NaCl, the magnitudes of the ions' charges 
are ${\sim 0.8\, e}$~\citep{nacl_1,nacl_2,nacl_3,nacl_4}.

The traditional concept of a covalent bond is an 
example of a failing of the integer-occupied-orbital model of bonding
that cannot be fixed by allowing the orbitals to have fractional occupancies.
Covalency is often regarded as a qualitatively-distinct type of bond, or mechanism of bonding;
and  a covalent bond is often described as the attraction of a pair
of atoms to electrons that they share between them, and which 
occupy orbitals that are hybrids of orbitals from both atoms.
However, if a pair of atoms shared electrons \emph{between them}, the electron 
density would have a local maximum between them; but it never does: Maxima of the electron
density coincide with maxima of the electric potential~\citep{density_maxima}, which coincide with 
the positions of the nuclei.

Both of these misconceptions have been perpetuated by
the widespread use of approximations that are based on simplifying
the mathematical form of the many-electron wavefunction  
(${\Psi}$) to make calculations tractable.
For example, the \emph{Hartree-Fock approximation}, which
is probably the simplest useful mean-field approximation, is based on
restricting ${\Psi}$ to be a Slater determinant.
When $\Psi$ has this form, integer-occupied single particle states appear to
have clear physical meanings. When the single
particle states in the determinant are linear combinations of 
atom-centered basis functions, such as the orbitals in
the Hartree-Fock wavefunctions of isolated atoms,
each basis function  `belongs' to one of
the atoms. Therefore, there appears to be a clear and meaningful 
qualitative distinction between a covalent bond and an ionic bond.
However this is an artefact of $\Psi$'s simplified form.

The primary reason for approximating $\Psi$ as a
determinant, or as a sum of few determinants, 
is not to simplify bonding \emph{conceptually}, but
to make \emph{calculations} tractable.
Therefore it seems valid to question whether localized atomic or molecular
orbitals help to simplify bonding conceptually, 
or whether they complexify and obfuscate it.
The artificial qualitative distinction between covalent bonding and ionic bonding illustrates that,
at least when building the most basic understanding of bonding from the most
computationally tractable form of wavefunction, they can be misleading.

Although most research scientists understand this, and also understand that the terms
\emph{ionic}, \emph{covalent}, and \emph{metallic} refer to varying
degrees of localization of the electron density around nuclei,
students are still being taught more traditional and misleading ideas~\citep{bacskay_reimers_1997,zurcher_2018,grundmann_2016,chemistry1,sutton_2024,chemistry_for_IB}.
Therefore the purpose of this subsection is to emphasize that, in some ways and for some purposes,
the essence of bonding is simpler than it appears from descriptions of it in terms of localized orbitals.

Within rigorous theory, there appears to be only two `types' of bonding, 
which are really two opposing idealized limits: One is
the limit in which all of the
electron density is strongly localized around nuclei by the electrons' Coulomb attraction
to them, and the other is the limit in which some of the electron density is localized around nuclei, 
and the rest of it is uniformly distributed.
The former is the ionic limit, the latter is the metallic limit, and there is no
theoretical or empirical reason to believe that
covalent bonding is anything other than bonding that does
not conform closely to either limit.

\subsubsection{Summary of chemical bonding in terms of electron density}
The electron density ($n$) is high where the microscopic electric potential from the nuclei ($\phinuc$) 
is high, and it only has maxima at the positions of the nuclei. If two nuclei shared
electrons between them, there would be a local maximum of the density between them, but 
there never is.

The shape of $n$ is determined by the shape of $\phinuc$; and $\phinuc$ is determined by the charges
and positions of the nuclei.  Localizing density where $\phinuc$ is high makes the potential energy of attraction between
electrons and nuclei more negative, but delocalizing
density makes both the electrons' kinetic energy and their mutual repulsion less positive.
The ground state density is the lowest-energy compromise between these localizing and delocalizing
influences. 

Most of the density is localized around nuclei, 
and most of it is at points where its gradient ${\grad n}$ is directed towards the nearest nucleus. 

\paragraph*{Ionic bonding:}
The bonding for which a 
superposition of spherically symmetric electron densities 
most closely approximates the true electron density 
is referred to as \emph{ionic} bonding; and the \emph{ionic limit} of bonding
is the limit in which an approximation of this form becomes exact. 

If the time average of the net charge of each nucleus and its almost-spherical electron cloud was zero, 
the attraction between atoms would be very weak and arise from electron correlation, rather
than electrostatics (see Appendix~\ref{section:nonoverlapping}). 
Atoms bind chemically by becoming ions, thereby lowering the potential energy via their mutual attraction. 
However they do not become ions
by donating or accepting electrons, but by donating or accepting electron \emph{density}.
When atoms are close enough to one another to bond chemically, and on the shortest time scales 
relevant to atomic motion ($\sim 10^{-15}$ seconds),
there is no theoretical reason why ions' average charges are likely to be integers or close to integers.

\paragraph*{Covalent and metallic bonding:}
If more of the electron density is in regions where ${\grad n}$ is not 
directed towards the nearest nucleus,
we describe the bonding as either \emph{covalent} or \emph{metallic}.
Bonding is metallic if the density in these regions is so delocalized that 
a significant fraction of the electrons are mobile. Otherwise
we refer to it as covalent, for historical reasons.
There is not a clear boundary between ionic and covalent bonding. 

Large atomic numbers, delocalized electrons, and high coordination numbers go hand in hand because,
when an atom's radius is large, the interactions of its nucleus with electrons on the outskirts
of its electron cloud are weak, and comparable in magnitude to its interactions with
electrons on the outskirts of neighbouring atoms.
Therefore energy is lowered by atoms arranging such that each one has many neighbours, 
whose electrons its nucleus interacts with. 
The energy is lowered further by the electron density on the outskirts of atoms delocalizing, 
so that more electrons have interactions of comparable strengths with multiple nuclei.
This delocalization worsens the approximation of the density as a superposition of spherical densities.

The \emph{metallic limit} is the limit in which the electron density becomes
a superposition of spherically-symmetric densities \emph{and} a uniform density.


\section{Macroscopic potential ($\bphi$) and  field ($\E$)}
\label{section:average_potential}
It is well known that the electric potential is a relative quantity and that,
when its value at a point is quoted, this value is always
the difference between the potential at the point and a reference potential. 
In theoretical work the reference point is often taken to be a notional 
point in the vacuum at infinity, in experimental work it may be a particular
electrode, and in engineering it is common to reference potentials to the `ground' or `earth'.
The difference in meaning between the terms {\em  mean inner potential} and {\em macroscopic
potential} is of little relevance to this work. The MIP is the average of $\phi$ over all
points in a material; therefore it is a scalar constant. The macroscopic
potential is a scalar field, defined at all points in a material, but defined only
to a finite precision ${\precphi}$. To within this precision its value at each point in the bulk 
is equal to the MIP.  I mostly refer to $\bphi$ as the macroscopic potential in this section 
and as the MIP when discussing Bethe's approximate expression for it in
Sec.~\ref{section:paradox}.

It is very important to be able to calculate changes in macroscopic potential.
From a conceptual and theoretical viewpoint, our understanding of the relationship between
macroscale and microscale electrostatics cannot be considered complete if we do not know 
how, in principle, to calculate the potential at the macroscale from the charge density or potential
at the microscale.
From a practical perspective, the MIP, which is believed to be positive, is a key
quantity in several areas of experimental and computational science.
For example, computer simulations
can be used to calculate microscopic charge densities and, up to an unknown constant, microscopic potentials.
To improve the designs of devices, such as batteries, fuel cells, chemical sensors, and solar cells, 
computational scientists simulate their constituent materials independently and,
from those simulations, try to calculate the change in the average potential that an electron or
ion would experience if it moved from one device component to another~\citep{mip_lars,mip_marzari}. 
The MIP is also used in electron microscopy to analyse and interpret 
electron diffraction images~\citep{mip_water_2020}.
In both of these contexts, the distinction between the MIP and the {\em time average} of the 
potential {\em felt by} the charge carrier is rarely made, but it is the latter
that is of interest and the MIP is used as an approximation to it.

In this section I show how to calculate the change ${\Delta\bphi}$ in the macroscale 
electric potential $\bphi$ between the vacuum above a material's surface and its bulk
from the microscopic charge density $\rho$.
The results are easy to generalize to the change 
in potential across an interface between two materials by treating it as a pair of adjoined surfaces. 
They also allow the value of $\bphi$ in an isolated material to be calculated
relative to a distant point in the vacuum surrounding it.

Some of my results, such as the macroscale potential within a \emph{macroscopically}-thin film whose 
surfaces are equally- and oppositely-charged, are well known and serve as a 
sanity check on my theory and reasoning. However, the main result, which I justified
on symmetry grounds in Sec.~\ref{section:symmetry}, contradicts most of the
literature on this subject over the past century. This result is that ${\bphi=0}$
in the bulk $\bulk$ of an isolated material unless it has charged surfaces or unless
its bulk contains charged macroscopic heterogeneities.

It is important to note that ${\Delta\bphi}$ is the
difference in potential across a surface \emph{at the macroscale}.
Therefore the derivation that follows is valid in the limit
${a/l\to 0}$, which is the limit in which surfaces become truly planar.
Many similar derivations and calculations in existing literature do not
assume this limit~\citep{baldereschi_1988,colombo_1991,junquera_2007}, 
and this limit is not appropriate for many purposes. For example, it would not
be appropriate to assume this limit when calculating
variations in the average potential on the nanoscale~\citep{junquera_2003,bernardini_1998}, 
because it is the limit in which all such variations vanish.

\subsection{Change in potential across a surface (${\Delta\bphi}$)}
\label{section:surfacepot}
To calculate  ${\Delta\bphi}$, I will consider the microscopic
potential at an arbitrary position ${\rvec_b\in\bulk}$ deep
below the surface at ${x=x_L}$. I will calculate the potential
within a finite chunk of material and then take
the large size limit. It is important, when doing this,
to order the limits appropriately. To illustrate the possible
pitfalls, consider the well-known example of the macroscopic potential $\bphi$ 
on the plane ${\bx=\mxb\approx (\mxl+\mxr)/2}$ from 
equally- and oppositely-charged surfaces at $\mxl$ and $\mxr$.
If I calculate $\bphi$ for an isolated material that is finite in all directions
and take the limit ${\abs{\mxl-\mxr}\to\infty}$ before I take
the limit of large size in the lateral ($yz$) directions, I find that $\bphi$
vanishes. However, if I take the limit of large cross-sectional area
first, I find that $\bphi$ is linear in $\bx$ and that
$\E$ is constant.
The appropriate order to choose for the limits
depends on the aspect ratio of the material and on the position within 
the material at which the potential is being calculated.

I want to calculate the average potential in a bulk-like region
of the material that is much closer to one surface than any other.
The plane ${x=x_L}$ is parallel to this surface and in the vacuum just beyond it.
Because all other surfaces are further away, it is appropriate to assume that
$\abs{x-x_L}$ is much smaller than the material's lateral dimensions.
I will first calculate the microscopic
potential due to the charge within a cylindrical region of the material, of radius
$R$, whose axis is normal to the surface.
I will use the cylindrical
coordinates $\rvec=(x,s,\phi)$ or $\rvec=(x,\svec)$, where $\svec\in\realtwo$ is
a vector in the plane parallel to the surface, $s=\abs{\svec}$, and $\phi$ is the azimuth.

The potential at ${\rvec_b\equiv(x_b,s_b,\phi_b)=(x_b,\svec_b)}$ due to the charge density
within a cylinder bounded by the surfaces ${\abs{\svec-\svec_b}=R}$, ${x=x_L}$, and
${x=x_b}$ is
\begin{align*}
 \Phi_L&(x_b,\svec_b;R) \nonumber \\
  = &
\kappa\,
\int_{x_L}^{x_b}
\left(
\iint_{\abs{\svec-\svec_b}<R}
\frac{\rho(x,\svec)}{\sqrt{(x-x_b)^2+\abs{\svec-\svec_b}^2}}\dd[2]{s}\right)\dd{x} 
\end{align*}
where ${\kappa\equiv \left(4\pi\epsilon_0\right)^{-1}}$.
I will first cast this expression into a more convenient form. Then I
will calculate its mesoscale average over a range of positions $(x_b,\svec_b)$
in the limit of large $R$.
Finally, I will add the mesoscale average of the potential, $\Phi_r$, from charge 
deeper below the surface than $\rvec_b$. 
The right-hand boundary of the region whose charge density 
contributes to $\Phi_r$ is the plane ${x=x_r}$, where ${x_b<x_r<x_R}$;
I use a lower case subscript for $\Phi_r$ to 
distinguish the plane ${x=x_r}$ from the right hand surface plane ${x=x_R}$.
I will be considering the cases ${x_r=x_R}$ and 
${\abs{x_r-x_b}\ll\abs{x_R-x_r}}$ separately.

The average volumetric charge density on a disc of radius $R$, which is parallel to the surface
and centered at $(x,\svec)$, is
\begin{align*}
\bar{\rho}(x,\svec\,;R)  & \equiv \frac{1}{\pi R^2} 
\iint_{\abs{\uvec}<R}
\rho(x,\svec+\uvec) \dd[2]{u}
\end{align*}
Defining
$\displaystyle \Delta\rho(\uvec;x,\svec,R)  \equiv \rho(x,\svec+\uvec)-\bar{\rho}(x,\svec;R)$
allows $\Phi_L$ to be split into two terms:
\begin{widetext}
\begin{align}
\Phi_L(x_b,\svec_b;R)
 = 
\overbrace{\frac{1}{2\epsilon_0}
\int_{x_L}^{x_b}
\bar{\rho}(x,\svec_b;R)
\left(
\int_0^R
\frac{u}{\sqrt{(x-x_b)^2+u^2}}\dd{u}\right)\dd{x}}^{\displaystyle \Phi_L^{[\rhobar]}(x_b,\svec_b;R)}
+
\overbrace{
\vphantom{\left(\int_0^R\frac{u}{\sqrt{(x-x_b)^2+u^2}}\dd{u}\right)}
\kappa
\int_{x_L}^{x_b}
\iint_{\abs{\uvec}<R}
\frac{\Delta\rho(\uvec;x,\svec_b,R)}{\sqrt{(x-x_b)^2+\abs{\uvec}^2}}\dd[2]{u}\dd{x}}^{\displaystyle 
\Phi_L^{[\Delta\rho]}(x_b,\svec_b;R)}
\label{eqn:Phidrho}
\end{align}
\end{widetext}
and similarly for $\Phi_r$.
It is easy to see that the surface excess of ${\Delta \rho(\uvec;x,\svec_b,R)}$ vanishes when
averaged over the plane parallel to the surface. 
The planar averages of 
${\Phi^{[\Delta\rho]}_L}$ and ${\Phi^{[\Delta\rho]}_r}$
also vanish in the large $R$ limit, i.e., 
\begin{align}
\lim_{\depth/R\to 0} \left\{\frac{1}{\pi R^2}\iint_{\abs{\uvec}<R} \Phi_L^{[\Delta\rho]}(x_b,\svec_b+\uvec;R)\dd[2]{u}\right\}  & =0, \nonumber \\
\lim_{\depth/R\to 0} \left\{\frac{1}{\pi R^2}\iint_{\abs{\uvec}<R} \Phi_r^{[\Delta\rho]}(x_b,\svec_b+\uvec;R)\dd[2]{u}\right\} & = 0,\nonumber
\end{align}
where ${\depth\equiv\abs{x_L-x_b}}$ is the depth of ${\rvec_b}$ below the surface.
Therefore, the only contributors 
to the bulk average of the microscopic potential are $\Phi^{[\rhobar]}_L$ and $\Phi_r^{[\rhobar]}$.
\\

\subsection{Mesoscale average of ${\Phi_L^{[\rhobar]}}$}\label{section:phi1ave}
Integrating over $u$ in the expression for ${\Phi_L^{[\rhobar]}}$, choosing ${R>\depth}$,
and using a Taylor expansion 
gives
\begin{widetext}
\begin{align}
\Phi_L^{[\rhobar]}(x_b,\svec_b;R)
& =
\frac{1}{2\epsilon_0}
\int_{x_L}^{x_b}
\bar{\rho}(x,\svec_b;R)
\left(
\sqrt{(x-x_b)^2+R^2}+(x-x_b)\right)\dd{x} \nonumber \\
& =
\frac{1}{2\epsilon_0}
\int_{x_L}^{x_b}
\bar{\rho}(x,\svec_b;R)
(x-x_b)
\dd{x}
+
\frac{R}{2\epsilon_0}
\int_{x_L}^{x_b}
\bar{\rho}(x,\svec_b;R)
\left[ 1 
+ \frac{1}{2}\left(\frac{x-x_b}{R}\right)^2 
+\order{\left(\frac{x-x_b}{R}\right)^4}
\right]
\dd{x} 
\label{eqn:expansion1}
\end{align}
\end{widetext}
I assume that there exists a well-defined macroscopic 
average of the volumetric charge density on every plane parallel 
to the surface. 
By this I mean that, although $\bar{\rho}(x,\svec_b;R)$ may exhibit microscopic fluctuations
as $R$ increases,
it converges to a well-defined value rather than
systematically growing or shrinking. Furthermore, as $\svec_b$ is varied at fixed $R$, 
${\bar{\rho}(x,\svec_b;R)}$
fluctuates microscopically about the value to which it converges in the large $R$ limit.
If I also assume that the bulk of the material is charge-neutral, 
only the first term of the series expansion in Eq.~\ref{eqn:expansion1} 
can survive the large $R$ limit.
In anticipation of this limit, and with the understanding 
that `$=$' means `$\approx$' until the limit is taken, I write
\begin{align}
\Phi_L(x_b,\svec_b;R)
= 
\frac{1}{2\epsilon_0} \bigg[
\left(R-x_b\right)
&\int_{x_L}^{x_b} \bar{\rho}(x,\svec_b\,;R)\dd{x}\nonumber \\
+
\int_{x_L}^{x_b}
\, x \, &\bar{\rho}(x,\svec_b\,;R)  \dd{x} 
\bigg]
\label{eqn:phip1}
\end{align}
$\Phi_L^{[\rhobar]}(x_b,\svec_b;R)$ depends sensitively on $x_b$ and so do both of its constituent terms on 
the right hand side.  Therefore, I will 
average over $x_b$. Before doing so, let us consider the average over $\svec_b$, which we 
should perform to calculate the three dimensional macroscopic bulk average. 
Because we will be taking the limit of large $R$, we can assume that at every value of ${\svec_b}$
the planar averages of ${\bar{\rho}(x,\svec_b;R)}$ and ${\Phi_L(x_b,\svec_b;R)}$
are converged, to our desired precisions, 
on an area much smaller than ${\pi R^2}$. 
Because they are now insensitive to $\svec_b$, 
I will drop it from their list of arguments.
The mesoscale average over $x_b$ of 
$\int_{x_L}^{x_b}\bar{\rho}(x\,;R)\dd{x}$ 
is the average areal surface charge density, ${\bsigma_L(R)}$.

The mesoscale average over $x_b$ of 
$\int_{x_L}^{x_b}x\,\bar{\rho}(x\,;R)\dd{x}$ 
is $X_{\bsigma}^L \,\bsigma_L(R)$, where
\begin{align}
X_{\bsigma}^L  &\equiv 
\frac{\expval{\int_{x_L}^{x_b} \,x\, \bar{\rho}(x;R)\, \dd{x}}_\prectheo}{\expval{\int_{x_L}^{x_b} \bar{\rho}(x;R)\, \dd{x}}_\prectheo}, \nonumber
\end{align}
is the center of the microscale distribution of excess surface charge.
The microscale charge density is bulk-like at mesoscopic depths and so $X_\bsigma^L$ is within ${\prectheo/2}$
of $x_L$, which means that
${X_\bsigma^L\in\mxl\equiv\interval(x_L,\prectheo)}$.  When working
at the macroscale, $\mxl$ and $\mxb$ are treated as having precise values.
Therefore, taking the mesoscale average
of both sides of Eq.~\ref{eqn:phip1} gives
\begin{align}
\bphi_L(R) & = \frac{1}{2\epsilon_0}\bsigma_L(R) \left(R - \abs{\mxb-\mxl}\right) \label{eqn:potmacro}
\end{align}

\subsection{Mesoscale average of $\Phi_r$}
By the same approach that led to Eq.~\ref{eqn:phip1}, we can find the 
potential at ${(x_b,\bolds_b)}$ from charge within the cylinder
bounded by the surfaces ${\abs{\svec-\svec_b}=R}$, ${x=x_b}$, and ${x=x_{r}}$, where ${x_{r}>x_b}$ 
is any position
to the right of $x_b$ such that ${\abs{x_r-x_b}\ll R}$. This potential is 
\begin{align}
\Phi_r(x_b,x_r;R) 
  = 
\frac{1}{2\epsilon_0}\bigg[(R+x_b)
\int_{x_b}^{x_r}&\bar{\rho}(x;R)\dd{x} \nonumber \\
-
\int_{x_b}^{x_r}
\, x \, &\bar{\rho}(x;R)  \dd{x}
\bigg]  \label{eqn:phiright}
\end{align}
Except in the case of thin material films, if 
${\abs{x_L-x_b}\ll R}$ then ${\abs{x_R-x_b}\not\ll R}$, which invalidates the derivation 
of Eq.~\ref{eqn:phiright} for $x_r=x_R$.
Therefore, I will separately treat two cases. First, I will consider the case of a thin film, for which $\abs{x_L-x_R}\ll R$. I will set $x_r=x_R$, and
calculate the mesoscale average of the potential in the center of the film from the charge density of the entire film. 
In the second case, I will assume that the surface at ${x_R}$ 
is far away and that its macroscopic surface charge density $\bsigma_R$ is zero.
In this case I will add to $\bphi_L$ the contribution to the mesoscale average of the potential from charge density at
positions ${x>x_b}$ which are still within the bulk of the material.

\subsubsection{Case I: Thin film, ${\abs{x_L-x_R}\ll R}$}
Defining ${\bsigma_R(R)}$ as the average areal charge density
of the surface at $\mxr$, and ${X_\bsigma^R\in\mxr}$ as
the center of the microscale distribution of excess 
charge at the right-hand surface, allows me to express
the mesoscale average of Eq.~\ref{eqn:phiright}, when ${x_r=x_R}$, as
\begin{align}
\bphi_R(R)=\frac{1}{2\epsilon_0}\bsigma_R(R)\left(R-\abs{\mxr-\mxb}\right)
\end{align}
The total potential at $\mxb$ is the sum of $\bphi_L$ and $\bphi_R$ in the large $R$ limit. That is, 
\begin{align}
\bphi(x_b)& 
 = 
\lim_{R\to \infty} \bigg\{\frac{R}{\epsilon_0}\left(\frac{\bsigma_L(R)+\bsigma_R(R)}{2}\right) \nonumber \\
&- \frac{1}{2\epsilon_0}\left[\sigma_L(R)\,\abs{\mxb-\mxl}
+ \bsigma_R(R)\abs{\mxb-\mxr}\right]\bigg\}
\label{eqn:totpot0}
\end{align}
The term proportional to $R$ becomes infinite in this limit unless ${\bsigma_L+\bsigma_R=0}$. Therefore, 
I assume that the surfaces have equal and opposite average areal charge densities and I 
define ${\bsigma\equiv \bsigma_L=-\bsigma_R}$. Then Eq.~\ref{eqn:totpot0} becomes
\begin{align}
\bphi(\mxb)& = \frac{\bsigma}{\epsilon_0}\left[\frac{1}{2}\left(\mxl+\mxr\right)-\mxb\right]
\label{eqn:filmmacro}
\end{align}
From Eq.~\ref{eqn:filmmacro}, we can immediately calculate 
the change in the macroscopic potential between the charged surfaces at $\mxl$ and $\mxr$ as
\begin{align}
\Delta\bphi\equiv\bphi(\mxl)-\bphi(\mxr) = \frac{\bsigma}{\epsilon_0}\left(\mxr-\mxl\right)
\end{align}
For very large finite values of $R$ we can write this as
${\Delta\bphi = Q/(\epsilon_0 A)}$,
where ${Q\equiv A\,\bsigma\left(\mxr-\mxl\right)}$ and ${A\equiv \pi R^2}$.  
This is the familiar formula for the magnitude of the potential 
difference between parallel plates carrying equal and opposite charges.
Therefore an important sanity check on the theory has been passed.

\subsubsection{Case II: Macroscopic sample, $\abs{x_L-x_R}\not\ll R$}
When the surface at $x_R$ is not charged and is sufficiently far away that it does not
contribute to the potential at $x_b$, the total microscopic potential at $x_b$ 
from all charge at greater depths ($x>x_b$) can be assumed to emanate from bulk regions where the charge density
is macroscopically uniform and neutral.

If the bulk of the material is charge neutral, there is no contribution to the potential in the vicinity 
of $x_b$ from points $x_r>x_b$ sufficiently far from it. However, the potential at $x_b$ will depend on the precise
choice of $x_r$ because the integrated charge between $x_b$ and $x_r$ depends sensitively on its value.
Therefore, as before, I will take the mesoscale average over $x_r$ of the microscopic potential, $\Phi_r(x_b,x_r;R)$, at $x_b$.
Eqs.~\ref{eqn:excessone} and~\ref{eqn:excessthree}, which are derived in Appendix~\ref{section:excess_fields},
can be used in Eq.~\ref{eqn:phiright} to write the following expression for 
the mesoscale average of $\Phi_r(x_b,x_r;R)$ over $x_r$. 
\begin{align}
&\overline{\Phi_r(x_b;R)}  =
\frac{1}{2\epsilon_0}
\Bigg[
(R+x_b)\left(
\int_{x_b}^{x_r}\bar{\rho}(x;R)\dd{x}
- \momone{x_r}
\right)  \nonumber \\
& \qquad -
\left(
\int_{x_b}^{x_r}
 x \, \bar{\rho}(x;R)  \dd{x} 
- x_r\momone{x_r} - \momtwo{x_r}\right)
\Bigg] \nonumber \label{eqn:phiaver}
\end{align}
This is independent of the choice of $x_r$, and so $x_r$ may be chosen such that 
$\int_{x_b}^{x_r}\bar{\rho}(x;R)\dd{x} = 0$.
The reason for making this choice is that it implies the following relationships.
\begin{align}
\momone{x_r} & =\momone{x_b} \nonumber \\
\momtwo{x_r} & =\momtwo{x_b} \nonumber \\
\int_{x_b}^{x_r}\,x\,\bar{\rho}(x;R)\dd{x} & = (x_r-x_b) \momone{x_b} \nonumber
\end{align}
Using these formulae, Eq.~\ref{eqn:phiaver} simplifies to
\begin{align}
\overline{\Phi_r(x_b;R)} = \frac{1}{2\epsilon_0}\left(\momtwo{x_b} - R\momone{x_b}\right)
\end{align}
and so the total microscopic potential at $x_b$ is
\begin{align}
\Phi&(x_b;R)  = \Phi_L(x_b;R)+\overline{\Phi_r(x_b;R)} \nonumber  \\
& =
\frac{1}{2\epsilon_0}
\Bigg[
(R-x_b)\int_{x_L}^{x_b}\bar{\rho}(x_b;R)\dd{x} \nonumber \\
& + \int_{x_L}^{x_b}\,x\,\bar{\rho}(x_b;R)\dd{x} + \momtwo{x_b} - R\momone{x_b}
\Bigg] \label{eqn:totalpot}
\end{align}
It is straightforward to show that the average of $\momone{x_b}$ over $x_b$ is zero for any macroscopically-uniform charge density 
(see Eq.~\ref{eqn:zerodipole} of Sec.~\ref{section:average_charge}). 
Making use of Eq.~\ref{eqn:potmacro} we can write down the mesoscale average of $\Phi(x_b;R)$, which is
\begin{align}
\bphi(R) & = \frac{1}{2\epsilon_0} \left[\bsigma_L(R) \left(R - \abs{\mxb-\mxl}\right) + \boldmtworho(R)\right] \nonumber \\
 & = \frac{\bsigma_L(R)}{2\epsilon_0} \left(R - \abs{\mxb-\mxl}\right) 
\label{eqn:potmacro2} 
\end{align}
where ${\boldmtworho(R)}$ is the mesoscale average over $x_b$ of $\momtwo{x_b}$, which is
shown in Appendix~\ref{section:average_momdensity} to be zero.
The remaining term diverges in the limit ${R\to\infty}$, which demonstrates that charged
surfaces are not stable. 

Now let us assume that ${\bsigma_L=0}$. Eq.~\ref{eqn:potmacro2} becomes ${\bphi=0}$, which 
implies that the macroscopic potential, which is the mesoscale average of the microscopic
potential, is zero in the bulk of any material that is mesoscopically charge-neutral in the bulk
and which does not have charged surfaces or contain any charged macroscale heterogeneities.
There are two very important points to note about this result. 

The first is that it
contradicts the prevailing view that the MIP 
is finite and positive~\citep{bethe-1928,mip_water_2020}.
It also contradicts a view commonly expressed or implied in textbooks on electromagnetism~\cite{jackson-book}
and solid state physics~\citep{ashcroft_mermin_book, kittel}, namely, that it is possible
for the symmetry of a crystalline microstructure to endow a material with
a macroscale electric field.

The second important point is that, because $\Phi[\rho]$ is
a linear functional of $\rho$, this result is the {\em only} result that could emerge from 
an internally-consistent theory of structure homogenization.
Mathematically, the spatial average ${\Rho}$ of $\rho$ is simply the
weighted sum (integral) of the infinite set of charge densities
${\{\rho_u: \rho_u(x+u)\equiv \rho(x), \forall u\in\realone 
\;\;\text{and}\; \forall x \in \realone\}}$. It follows from linearity that
${\bphi=\expval{\Phi[\rho]}_\prectheo = \Phi[\expval{\rho}_\prectheo] = \Phi[\Rho]}$. 
Therefore, if ${\Rho=0}$ everywhere, as is the case for an isolated
uniform material whose surfaces are uncharged, then ${\bphi=0}$ everywhere.

\subsection{Lorentz's fallacy: the macroscopic local field}
\label{section:macroscopicE}
As discussed in Sec.~\ref{section:bulkfields}, unless there exist sources of macroscopic fields
that are external to the material's bulk (e.g., an applied field $\Eext$ or a net charge at one or more of 
its surfaces) the isotropy and homogeneity of its macrostructure $\Rho$, 
which vanishes everywhere in the bulk, preclude the existence of a non-vanishing $\E$ field.
Isotropy is incompatible with the existence of a vector field.
As discussed in Sec.~\ref{section:potential_symmetry} and Sec.~\ref{section:average_potential}, 
if the macroscopic charge density is zero in the bulk of an isolated material whose surfaces
are uncharged, there is no source of macroscopic potential $\bphi$. If $\bphi=0$ throughout
the bulk, $\E=0$ throughout the bulk.

Nevertheless, most textbooks posit the existence of a non-vanishing $\E$ emanating from the bulks
of crystals that lack inversion symmetry at the microscale. 
Furthermore, it is commonly believed that the net field
acting at each point $x$ in a material's bulk is ${\E + \pp/3\epsilon_0 + \me(x)}$, where
$\me$ is a microscopic field  and ${\pp/3\epsilon_0}$ is another macroscopic (${\bk=0}$) contribution.
The purpose of this section is to critically examine the reasoning used to infer the
existence of the contribution ${\pp/3\epsilon_0}$. Almost all derivations of this term are
based on a construction and line of reasoning first presented by Hendrik A. Lorentz
in a series of lectures given at Columbia University in 1906, which were subsequently
published in book form~\citep{lorentz} (p137). 
His construction, which is illustrated in Fig.~\ref{fig:macrofields}, 
is sometimes known as the {\em Lorentz cavity}.
This construction has been used in many textbooks~\citep{griffiths-book, jackson-book, born-huang-book, kirkwood_1936, vanvleck-1937, kirkwood-1940}, 
but as I will now explain, both Lorentz's original argument and all of its descendents that I am aware of are fatally flawed.


\begin{figure}[h]
\includegraphics[width=0.45\textwidth]{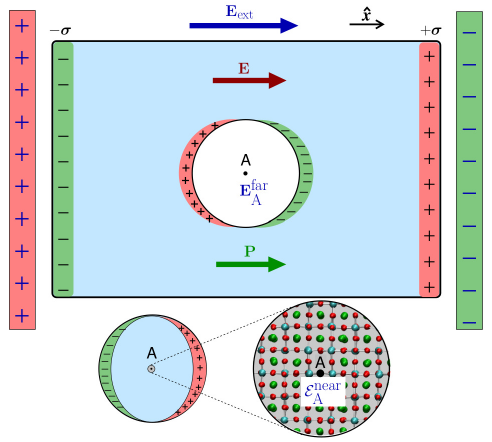}
\caption{The {\em Lorentz cavity}~\citep{lorentz}. See Sec.~\ref{section:macroscopicE}.} 
\label{fig:macrofields}
\end{figure}

Lorentz set out to calculate the average force or electric field acting on 
a microscopic particle $A$ in the bulk of a dielectric in which there exists
a macroscopic  electric field $\E$.
The particle could be a molecule, an atom, or an electron. 
He expressed the electric field at $A$ as  ${\me_A = \EAfar + \EAnear}$, where $\EAfar$ is the
field emanating 
from all material beyond a mesoscopic spherical region of 
radius $R$ centered at $A$
and $\EAnear$ is the field emanating from all charges within the region, except $A$ 
itself.
As the boldface notation suggests, $\EAfar$ is calculated 
by treating the material as a continuum; therefore it is a macroscopic quantity. This is
reasonable because the length scale on which fluctuations of the
microscopic charge density occur is much smaller than the distance ($>R \sim l$) to $A$.
On the other hand $\EAnear$ can be expressed as
${\EAnear = \EAnearmacro + \Delta\EAnearmicro}$, 
where ${\EAnearmacro}$ and ${\Delta\EAnearmicro}$
are macroscopic and microscopic contributions, respectively.
It is a microscopic quantity.

In some presentations of this approach, and as illustrated in Fig.~\ref{fig:macrofields}, 
these two separate contributions are imagined in different and separated material systems. 
To calculate $\EAfar$ a continuous material 
with a cavity in its bulk is imagined, with $A$ at the center of the cavity.
To calculate $\EAnear$ a microscopically-varying spherical 
charge distribution is imagined, with $A$ at its center. This is the charge 
that was evacuated to form the cavity and it is frozen in the arrangement 
it had prior to being evacuated.

From here, different authors have derived the
term ${\pp/3\epsilon_0}$ in different ways, but it tends to be thought of as arising either
from the  charge on the cavity's surfaces or from the dipole moment of the material
evacuated from it. It is interesting that the version of the argument that appears in
the first edition of Jackson's book~\citep{jackson_firsted} differs substantially
from the one appearing in its 1975 second edition~\citep{jackson-seconded} and that the latter is very similar
to the one in Ashcroft and Mermin's 1976 book~\linecite{ashcroft_mermin_book}.
However, it is not necessary to go in detail into 
these differences because we have already 
introduced the fatal flaw in Lorentz's reasoning and, to my knowledge, all 
variants of his derivation suffer from it.

Just as the charges at the surfaces of the materials
depicted in Figs.~\ref{fig:surfcharge} and~\ref{fig:polar_surface}
depend sensitively, in magnitude and sign, on how the surfaces are terminated, 
so too does the charge on the surface of the cavity and the dipole moment of the 
evacuated material. They both depend
sensitively and microscopically on the cavity radius $R$ and vanish
when averaged over a continuous mesoscopic range of radii.
Therefore the true value of the macroscopic field at $A$ is the sum
of only two contributions: the applied field and the field
from charge at the material's surfaces.

\subsection{LO-TO splitting}
I have argued that inversion asymmetry of a crystal's microstructure
does not endow it with a macroscopic $\E$ field.
This implies that a macroscopic field is not created
when the sublattices of an inversion symmetric crystal are relatively displaced.
An oscillating rigid relative displacement of a crystal's sublattices 
can be regarded as a ${\bk=0}$ phonon, so another way of saying that
${\E}$ vanishes is to say that a ${\bk=0}$ phonon does not have an intrinsic electric field.

However, it is well known that the frequency of a ${\bk\to 0}$ 
longitudinal optical (LO) phonon
is increased by the electric field that is intrinsic to it, 
and which opposes its motion~\citep{stengel_2021,pick_1970,cochran_cowley_1962,born-huang-book,ashcroft_mermin_book, Jones_and_March1, coiana_mgo,lyddane_sachs_teller}.
Were it not for this field, the frequencies of some crystals'
LO and TO phonons would be equal, by symmetry, 
in the long wavelength limit (${\bk\to 0}$).
The breaking of this degeneracy by the LO phonon's intrinsic field 
is commonly referred 
to as {\em LO-TO splitting}~\citep{born-huang-book,ashcroft_mermin_book,Jones_and_March1}.

Therefore I am claiming that, in the ${\bk\to 0}$ limit,
an LO phonon of wavevector $\bk$ creates an electric field
of wavevector $\bk$, but that a ${\bk=0}$ phonon does not
create an electric field in a crystal whose surfaces are earthed.

A ${\bk=0}$ phonon does create a uniform (i.e., ${\bk=0}$) electric field if the surfaces are not earthed, because
the polarization current that flows during the rigid relative motion of sublattices
changes the areal charge densities on parallel opposing surfaces at equal 
and opposite rates. If this charge accumulates, a field emanates from it.

\subsubsection{Why finite-wavevector LO phonons create electric fields}
\label{section:finite_k_field}
\begin{figure}[h]
\includegraphics[width=0.49\textwidth]{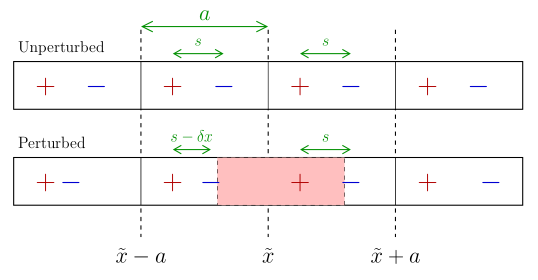}
\caption{See text of Sec.~\ref{section:finite_k_field}. The net charge within
the pink-shaded interval of width $a$ is ${+q}$. In the unperturbed crystal, the net charge
in every interval of width $a$ is zero.}
\label{fig:macrofields}
\end{figure}

Suppose that the crystal's microstructure ($\rho$) is modulated
along $\hat{x}$ by an LO phonon, of {\em finite} wavelength $\lambda$,
propagating along $\hat{x}$.
At each instant, this modulation creates regions of excess positive
charge and regions of excess negative charge, which alternate
along $\hat{x}$ with a wavelength of $\lambda$. 
This excess charge density wave, in turn, creates an electric field of 
the same wavelength, $\lambda$, which opposes the LO mode's motion. 

To help understand {\em why} excesses of charge are created, it 
is instructive to consider the perfect crystal, without any perturbation, 
and to calculate the excess charge, ${\bsigma(\tx)}$,
on plane ${\plane(\tx)}$, which is perpendicular to ${\hat{x}}$ at ${\tx\in\bulk}$.
To do so, we can treat the plane  as a pair of
adjoined surfaces, use Eq.~\ref{eqn:finnis} to 
calculate the excess charge on each one, and add them to give ${\bsigma(\tx)}$.
Because ${\tx}$ is in the bulk, and the bulk of an unperturbed crystal is uniform, 
we can choose ${x_L=x_b=\tilde{x}}$. We find that the 
excess charge at $\tx$ is
\begin{align}
\bsigma(\tx)=\bsigma_+(\tx) +\bsigma_-(\tx) & = 
\frac{1}{a}\int_{-a}^{0}\rho(\tx+u)\,u\,\dd{u}  
\nonumber \\
& -
\frac{1}{a}\int_{0}^{a}\rho(\tx+u)\,u\,\dd{u} = 0,
\end{align}
where ${\bsigma_+}$ and ${\bsigma_-}$ are the areal charge densities on the 
`surfaces' at $\tx$ whose outward normals
are ${\hat{x}}$ and ${-\hat{x}}$, respectively, and their cancellation follows
from the periodicity of the crystal. 

Now let us consider what happens when an LO phonon breaks periodicity by modulating 
the structure along the $x$ axis. 
When this happens, ${\bsigma_+(\tx)}$ and ${\bsigma_-(\tx)}$ are no longer
exactly equal in magnitude, in general, which means that
${\bsigma(\tx)}$ does not vanish. Calculating its value is more
complicated than in the periodic case because the crystal is no longer
uniform. Therefore it is no longer valid to regard
the point $\tx$ as both defining the position of our imaginary surfaces
and as points in the `bulk' beneath them. However, as an illustration, let us 
calculate the net charge in the interval
${\interval(\tx+u,a)}$ averaged over all ${u}$ between ${-a/2}$ and ${a/2}$.

For the purpose of this illustration, let us suppose that
each unit cell contains a single anion-cation pair and
that the distance between the pair in interval ${[\tx-a,\tx]}$ is
smaller by ${\delta x}$ than the distance between the pair in interval
${[\tx,\tx+a]}$, such that the difference between the 
dipole moments of these unit cells is ${\Delta d = q\delta x}$, where ${q}$
is the cation's charge. Then, the average over ${u\in(-a/2,a/2)}$ 
of the net charge in interval ${\interval(\tx+u,a)}$ is 
${q\delta x/ a =\Delta d/a}$. 

Now suppose that we have calculated the same quantity for
every pair of adjacent unit cells in an interval ${\interval(\tx,\ell)}$,
where ${a\ll \ell\ll\lambda}$, and then repeated this calculation
for  a continuous range of values of $\tx$.
Let us denote ${1/\ell}$ times the sum of all net
charges in interval ${\interval(x,\ell)}$ by
${\expval{\rho}^*(x)}$ and ${1/\ell}$ times the sum of
the cells' dipole moments by ${\expval{\mpp}^*(x)}$.
Then, it can be shown 
that ${\expval{\rho}^*(x)=-\div \expval{\mpp}^*(x)}$.
The similarity of this expression to
the relation ${\rho=-\div\pp}$ is not coincidental:
Maxwell used a similar line of reasoning to deduce it, 
albeit with displacements of charges replaced by 
displacements of the ether.

This example illustrates that ${\bsigma(\tx)}$ does not vanish in the presence
of an LO perturbation because the symmetry reason for it vanishing no longer exists. 
Furthermore,  because a more realistic charge density $\rho$ would be 
a smooth function of position, ${\bsigma(\tx)}$
would be a smooth function of $\tx$, with the same
periodicity $\lambda$ as the LO perturbation that created it.
Therefore, there would be an excess charge density wave of periodicity 
$\lambda$, from which would emanate an electric field of periodicity $\lambda$.

\subsubsection{Zero-wavevector LO phonons} 
If ${\melo(\bk,u)}$
denotes the electric field created by displacing a crystal by $u$ along the eigenvector
of an LO phonon of wavevector $\bk$, my claim about the difference between
the point ${\bk=0}$ in reciprocal space and
the ${\bk\to 0}$ limit can be stated as follows:
\begin{align*}
0\neq \lim_{\bk\to 0} \melo(\bk,u)\neq \melo(0,u)=0.
\end{align*}
In other words, the limit ${\bk\to 0}$ is a \emph{singular limit}
of ${\melo(\bk,u)}$.

The fact that ${\melo(\bk,u)}$ is discontinuous at ${\bk=0}$ is
well known when expressed in a different way:
Squared phonon frequencies are eigenvalues of a crystal's
dynamical matrix. Therefore, if ${\melo(\bk,u)}$ vanishes
suddenly at ${\bk=0}$, causing LO phonon frequencies
at ${\bk=0}$ to be smaller than their values in the
${\bk\to 0}$ limit, the dynamical 
matrix must be discontinuous at ${\bk=0}$.
It is very well known that it is discontinuous, and 
\emph{non-analytic
corrections} are commonly applied to the ${\bk=0}$ dynamical
matrix to calculate the ${\bk\to 0}$ dynamical 
matrix~\citep{pick_1970,cochran_cowley_1962,cochran_1960,giannozzi_1991,gonze_1997,baroni_rmp,born-huang-book,Jones_and_March1}.
The ${\bk\to 0}$ dynamical matrix is not one matrix, in general, because both it, and the LO phonon frequency, depend
on the direction in reciprocal space from which the point ${\bk=0}$ is approached.

To understand why the ${\bk\to 0}$ limit is singular, it is easier to
think about the LO phonon's wavelength in the ${\lambda\to\infty}$ limit
than its wavevector in the ${\bk\to 0}$ limit:
Imagine a microscopic or mesoscopic neighbourhood of a point in
the bulk of an arbitrarily large perfect crystal, and then 
imagine that the crystal is perturbed by displacing it from 
equilibrium along the eigenvector of an LO phonon of wavelength $\lambda$.
Now imagine increasing $\lambda$.

As ${\lambda}$ becomes much larger than the size of the neighbourhood, 
and continues to increase, 
the microstructure within the neighbourhood looks more and more like
it would look if
the crystal's sublattices had been displaced rigidly relative to one another.
Therefore it looks more and more like a crystal that 
has been perturbed by displacing it along the eigenvector of a ${\bk=0}$
phonon.
Nevertheless, no matter how large $\lambda$ becomes, if one moves
a distance ${\lambda/2}$ in the direction of ${\bk}$, the relative
displacements of the atoms in the direction of $\bk$ are reversed.

In other words, \emph{microscopically}, increasing $\lambda$ brings the structure closer to
the ${\bk=0}$ structure, but \emph{macroscopically} it does not; and it is the macroscopic
structure that determines whether or not there is a macroscopic $\E$ field.

\section{A potential paradox}
\label{section:paradox}
In this section I highlight some subtleties in the meaning and definition of the microscopic potential $\phi$
and its relationships with its macroscopic counterpart ${\bphi}$ and
the microscopic charge density $\rho$.
As mentioned at the beginning of Sec.~\ref{section:average_potential}, the value of the MIP 
is believed to be positive~\citep{mip_sanchez_1985,mip_pratt_1987, mip_pratt_1988, mip_pratt_1989, mip_pratt_1992, mip_sokhan_1997, mip_leung_2010, mip_mundy_2011, mip_cendagorta_2015, mip_lars, mip_marzari, mip_water_2020, mip_madsen_2021, mip_kathmann_2021}.
This contradicts my finding that it is zero.
Therefore, to illustrate the subtleties, I use the example of Hans Bethe's 1928 derivation of an approximate expression 
for the MIP, which is sometimes known as the {\em Bethe potential} (${\bphibethe}$), from several different perspectives.
I begin, in Sec.~\ref{section:bethederivation}, by outlining his derivation and line of reasoning.

My focus is on the `paradox' referred to in the section title and 
I do not address the question(s) of most relevance to those using 
the Bethe potential, or one of its descendants, as a 
parameter in the analysis and/or interpretation of their calculations (e.g., theoretical electrochemistry) 
or experiments (e.g., electron holography).
In most of these applications $\bphibethe$ is used as an approximation
to the average potential {\em experienced by an electron} as it passes through 
the material. This quantity is likely to depend heavily on the electron's
energy as it enters the material and the time that it spends inside the material.
Furthermore, one should not calculate it from the probability density
${n(\rvec)}$ of an electron being at ${\rvec}$, but on the
conditional probability density ${n_c(\rvec_1|\rvec_2)}$ of there being
an electron at ${\rvec_1}$ {\em given} that the probe electron is at ${\rvec_2}$.

As an illustration of the importance of basing calculations on ${n_c}$
rather than $n$, consider the example of a neutral atom meeting a stray electron in a vacuum. One might
deduce from the atom's electron density ${n(\rvec)}$ that they would not be attracted
to one another; but by considering how the distribution of the atom's electrons
is changed by their interaction with the stray electron, one can quickly deduce
that they do attract one another and that {\em all} singly-charged anions are stable in vacuum
when the electrons' temperature is sufficiently low.

\subsection{Bethe's fallacy: the mean inner potential}
\label{section:bethederivation}
Bethe assumed that the charge densities of materials are not too dissimilar from
a superposition of atomic charge densities.
For a crystal with one spherically-symmetric atom in its
primitive unit cell $\unitcell$, the expression he derived is
\begin{align}
\bphibethe = \frac{2\pi e}{3 \epsilon_0\volume}\int_0^\infty n(r) r^4 \dd{r} > 0, 
\label{eqn:bethe0}
\end{align}
where ${\volume}$ is the volume of $\unitcell$; and ${n(r)}$ is the number of
electrons per unit volume in each atom's electron cloud
at a distance $r$ from its nucleus.
Bethe deduced from Eq.~\ref{eqn:bethe0} that $\bphi$ is positive, which contradicts my finding 
that it is zero.
I will now rederive Eq.~\ref{eqn:bethe0} via a more explicitly-careful 
mathematical route than Bethe chose to present, but using
his starting point and physical reasoning.
His starting point was the expression 
\begin{align}
\phir(r) \equiv
 \frac{1}{\epsilon_0 r}\int_0^r \rho(u) u^2\dd{u} 
+ 
\frac{1}{\epsilon_0}\int_r^\infty \rho(u) u\dd{u}, 
\label{eqn:sphericalpot}
\end{align}
for the electric potential ${\phir}$
at a distance $r$ from the center of an isolated spherically-symmetric charge distribution ${\rho(r)}$.
Eq.~\ref{eqn:sphericalpot} can be derived from Gauss's law by assuming that the electric
field inherits spherical symmetry from $\rho$ and by expressing the potential at a distance $r$ from the 
nucleus, $\phi_r(r)$, as the integral of the spherically-symmetric field 
from an infinitely-distant point to one whose distance from the nucleus is $r$,
along an axis passing through the nucleus. 

Bethe reasoned that the average potential in the crystal 
is the potential emanating from one atom, integrated over all
points in space, and divided by the volume per atom, i.e., 
the ${R\to \infty}$ limit of
\begin{align}
\phibar(R)
& = 
\frac{4\pi}{\volume}\int_0^R \phir(r) r^2 \dd{r}.
\label{eqn:bethe1}
\end{align}
His reason for integrating $\phir$ {\em over all space}, but dividing
by the volume of {\em only one unit cell}, was that the potential in each cell has contributions 
from atoms in all other cells and, either by symmetry, or when averaged over all other cells,
the sum of the contributions of the atom in a given cell to the averages of the electrostatic potential in
all other cells must equal the sum of the contributions of atoms in all other cells to the average
electrostatic potential in the given cell. 

Bethe expressed each atom's charge density (charge per unit volume at a distance $r$ from its center) 
as ${\rho(r)=\rhop(r)+\rhom(r)}$, where ${\rhop}$ is the density of nuclear charge and
${\rhom(r) = -e \, n(r)}$ is the density of electron charge. He expressed
${\rhop}$ as a delta distribution, i.e., 
\begin{align}
\rho(r) = Z\, e\, \delta(r) - e\,n(r), 
\label{eqn:betherho}
\end{align}
but I will keep it more general for now.  Substituting Eq.~\ref{eqn:sphericalpot} into 
Eq.~\ref{eqn:bethe1}  and 
simplifying leads to
\begin{align}
\phibar(R) & = -\frac{2\pi}{3\epsilon_0\volume}\int_0^R \rho(r) r^4 \dd{r}  \nonumber \\
&+ \frac{\frac{4}{3}\pi R^3}{\volume}\left[\frac{3}{2}\frac{\kappa Q(R)}{R} 
 + \frac{1}{\epsilon_0}\int_R^\infty \rho(r)  r\dd{r}\right]
\label{eqn:bethe2}
\end{align}
where 
\begin{align*}
Q(R)\equiv 4\pi\int_0^R \rho(r) r^2 \dd{r} 
\end{align*}
is the net
charge in a sphere of radius $R$.
When $R$ is chosen large enough that the total charge outside
this sphere is negligible, only the first term on the right
hand side remains, i.e., 
\begin{align*}
\phibar(R) & = -\frac{2\pi}{3\epsilon_0\volume}\int_0^R \rho(r) r^4 \dd{r}  
\end{align*}
Since ${\rho(r)<0}$ when ${r}$ exceeds the spatial extent of ${\rho_+}$, 
${\phibar(R)}$ is positive and we recover Eq.~\ref{eqn:bethe0}
in the limit of large $R$ if ${\rhop}$ is localized at a point.

If, instead, we assume that ${\rhop}$ has a finite width and denote the total nuclear charge by ${Q_+}$, we
can use the atom's overall charge neutrality, i.e., 
\begin{align*}
Q_+\equiv 4\pi\int_0^\infty u^2\rhop(u)\dd{u} = -4\pi\int_0^\infty u^2\rhom(u)\dd{u}, 
\end{align*}
to express $\bphibethe$ as
\begin{align}
\bphibethe& = \frac{Q_+}{6\epsilon_0\volume}\left(s_-^2-s_+^2\right) 
\label{eqn:bethe3}
\end{align}
where $s_+^2$ and $s_-^2$ are the mean squared distances of positive and 
negative charges, respectively, from
the atom's center, i.e.,
\begin{align*}
s_\pm^2 & \equiv \pm \frac{4\pi}{Q_+}\int_0^\infty \rhopm(u) u^4 \dd{u}
\end{align*}
From Eq.~\ref{eqn:bethe3} it seems clear that the MIP being
positive is a consequence of the electrons being more 
delocalized than the nuclei (${s_-^2>s_+^2}$).
For example, if we assume that ${\rhop}$ and ${\rhom}$ are constant
within concentric spheres of radii ${r_+}$ and ${r_-}$, respectively, 
and zero outside them, then Bethe's derivation would lead to
\begin{align*}
\bphibethe = \frac{Q_+}{10\epsilon_0\volume}\left(r_-^2 - r_+^2\right) > 0.
\end{align*}
It seems reasonable to interpret this expression as follows: Gauss' Law implies
that ${\phir(r)}$ 
vanishes if ${r>r_-}$ because
the net charge within a sphere of radius $r_-$ centered at the atom's center 
is zero. The potential is positive at the atom's center and it decreases monotonically
to its value of zero at ${r=r_-}$. Therefore the potential is positive 
in a sphere of radius ${r_-}$ and zero everywhere else.
It seems obvious, then, that ${\bphi}$ must be positive.

I will now show, by illustration, why this obviously-right result must be wrong. 
Then I will explain the flaw in Bethe's reasoning and  show that
a more careful treatment of the problem leads to the conclusion that ${\bphi}$ is zero.
I illustrate the flaw from several different perspectives to highlight some of
the many pitfalls that exist when working with the electric potential.
Readers who have already spotted the flaw, or who don't like whodunnits, 
might want to skip the illustrations and proceed directly to Sec.~\ref{section:solution}.

\begin{figure}[!]
\includegraphics[width=0.5\textwidth]{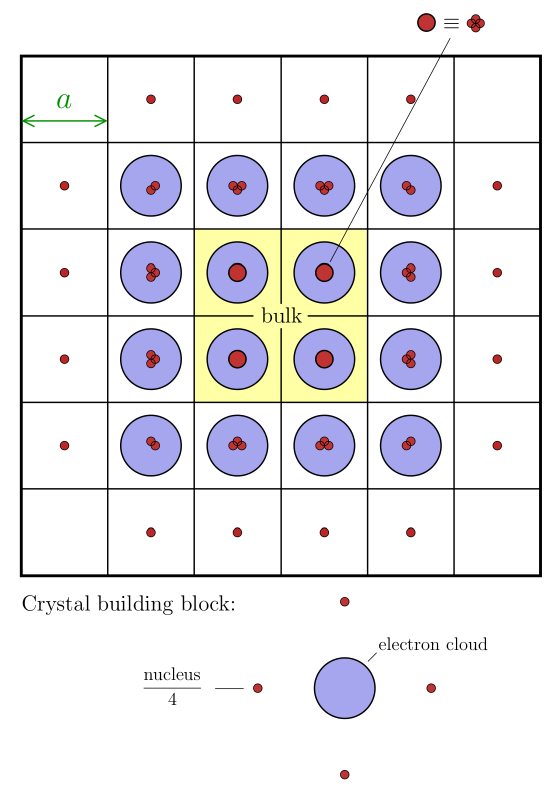} 
\caption{
If a crystal is built from a superposition of atoms (Crystal 1),
Bethe's derivation leads to the result ${\bphibethe_1>0}$. Crystal 2, shown schematically above,
is built from a charge-neutral building block comprised of an electron cloud
surrounded by fractions of nuclei (bottom). 
Bethe's derivation leads to the result ${\bphibethe_2<0}$ for Crystal 2.
Both crystals are identical in the bulk (yellow region), but differ
at their surfaces; however the surfaces
of both crystals are charge-neutral (${\bsigma=0}$).
Crystal 2 can be transformed into Crystal 1 
by cancelling each of the outermost ${+Ze/4}$ charges, where $Z$ is the atomic number,
with a charge of ${-Ze/4}$ at the same position and
adding a charge of ${+Ze/4}$ to the centers of the 
outermost electron clouds.
If we assume this to be equivalent to adding a dipole moment 
density (dipole moment per unit length, in this 2-d example)
of ${-\left(Ze/4\right)\,\hat{n}}$ to each surface, where ${\hat{n}}$ is
the surface's outward unit normal vector, it is easy to calculate
the effect of such a layer on the MIP for the 3-d analogue
of the crystal shown.
It would shift the average potential below the surface
relative to its value in the vacuum above the surface  by
exactly ${\bphibethe_1-\bphibethe_2}$, thereby making
the values of the MIP calculated for Crystal 1
by two different routes equal.
However, at the macroscale, adding this dipole moment density
should have no effect on the value of ${\bphi}$ because
a microscopic distance is indistinguishable from
a distance of zero at the macroscale (see Sec.~\ref{section:homogenization}).
Therefore an isolated dipole moment shrinks to a 
point of no net charge under the homogenization transformation
and the addition of a plane of such dipoles
to a surface does not change either the net charge ${\bsigma}$
at the surface or the macroscopic charge density ${\Rho}$ at
any point inside or outside the material. 
}
\label{fig:building_blocks}
\end{figure}

\subsubsection{Existence of a flaw - Illustration 1}
\label{section:illustration1}
Bethe chose to build his material from a
spherically-symmetric charge density, with a localized
distribution of 
positive charge at its center and
relatively delocalized distribution of negative charge surrounding it.
However, just as there is no `right' way to partition the charge density 
of a material for the purpose of defining its average dipole
moment density (see Sec.~\ref{section:definingP_1}), there is no right way to partition
it for the purpose of calculating its average potential.
It is no less justified to build a crystal from
a superposition of charge densities of the form
\begin{align}
\rho(\rvec) = -e\,n(r) + 
\frac{Q_+}{N_A}\sum_{i: |\Rvec_i|=A} \delta(\rvec-\Rvec_i)
\label{eqn:star}
\end{align}
where, as before, the origin is chosen to coincide with a nucleus;
${r=\abs{\rvec}}$; each ${\Rvec_i}$ is a lattice vector and
therefore a relative displacement of two nuclei (for simplicity I assume that
${T\to 0}$); and the sum is over all ${\Rvec_i}$'s of a given finite magnitude ${A}$, of 
which there are a total of ${N_A=\sum_{i:|\Rvec_i|=A}  1}$.

An example of a building block of this form is shown schematically for a 2-d crystal in
Fig.~\ref{fig:building_blocks}. It has negative charge at its center, positive 
charge further away, and it is charge neutral overall, meaning that 
the net flux of the electric field through any surface that encloses it is zero, 
as it is for an atom. The flux
from an atom is zero at all points on a surface enclosing it, whereas
the flux from the charge density of Eq.~\ref{eqn:star} is finite almost
everywhere on a surface enclosing it, but with regions of the surface where it is positive
and regions where it is negative. Nevertheless, Gauss's law implies
that the net potential outside the surface 
from charge within it is zero, as it is for an atom. 

Using the same physical reasoning with which Bethe deduced that ${\bphibethe>0}$ for
a material built from atoms, it can be shown that ${\bphibethe<0}$ in
a material built from this charge distribution, because
${s_+=A > s_-}$.
Furthermore, the magnitude of ${\bphibethe}$ 
depends on the value of $A$. For example, consider a simple cubic crystal with 
lattice spacing ${a}$ and let us build it from Eq.~\ref{eqn:star} with the
choice ${A=a}$ (${\implies N_A=6}$). I will refer to the crystal built in this way as
{\em Crystal 2}, I will refer to the crystal built by Bethe from atoms 
as {\em Crystal 1}, and I will denote their MIPs, as derived
using Bethe's approach, by ${\bphibethe_2}$
and ${\bphibethe_1}$,  respectively.
Then if, following Bethe, we assume the nuclei to be localized
at a point, we find that 
\begin{align*}
\bphibethe_2=\bphibethe_1-Q_+ \frac{a^2}{6\epsilon_0\volume}.
\end{align*}

Crystal 1 and Crystal 2 
are identical in the bulk; they differ only near surfaces. 
However, because
all surfaces of each crystal are charge-neutral (${\bsigma=0}$), 
and because ${\Rho=0}$ in the bulk in each case, 
the macroscale theory would not be internally consistent if ${\bphi>0}$ in one
case and ${\bphi<0}$ in the other.
If the value of ${\bphi}$ is defined it must be the same in each
case because the macrostructures of the two crystals
are identical: electrostatically, each one is indistinguishable from empty space.

Crystal 1 and Crystal 2 could be made identical by
adding a pair of charges of opposite signs, of magnitudes ${Q_+/6}$,
and separated by a distance $a$, to each surface unit cell of one of
the crystals. For example, to make Crystal 2 into Crystal 1
we would have to add a charge of ${Q_+/6}$ to the center of each of the
electron clouds closest to its surface and a charge of ${-Q_+/6}$
at a displacement ${a\,\hat{n}}$ from the first charge, 
where ${\hat{n}}$ is the surface's outward unit normal.
Let us temporarily assume that,
from the perspective of a point whose depth below the surface
is much greater than $a$, this is equivalent to adding to the surface
an approximately-uniform areal density of dipole moments, 
\begin{align*}
\sigma_\mpp=-\left(Q_+ a/6 a^2\right) \hat{n} = -\left(Q_+ a^2/6\volume\right)\hat{n}.
\end{align*}
Then the result would be an upward shift of the potential
in the crystal relative to the vacuum near the surface of
\begin{align*}
\Delta\phi = \sigma_\mpp/\epsilon_0=Q_+ a^2/6\epsilon_0\volume.
\end{align*}
This cancels the difference between ${\bphibethe_2}$ and ${\bphibethe_1}$.
Therefore, although the values of ${\bphibethe}$ derived by Bethe's method
differ, it appears possible to correct the difference between them by changing the
surface structure of either crystal to make the two crystals identical.

Unfortunately, although we have corrected the difference between
the two values of ${\bphibethe}$, this does not solve our problem. 
We still do not have any reason to prefer
one building block over another; therefore we do not
have any reason to prefer one of the
resulting crystal surface structures over the other.
We have two derivations,  which appear equally valid, and from which 
we deduce two different values of the MIP. 
This appears to imply that there is a flaw in the construction that
Bethe used for his derivation.

Bethe did not involve surfaces in his derivation because, when ${\bsigma=0}$, he regarded
the MIP as a property of the bulk. However, the fact that
${\bphibethe_1}$ and ${\bphibethe_2}$ differ suggests that 
the MIP is a surface property, which can be changed by adding
or removing equal amounts of positive and negative charge
at each surface. 
This is problematic if we wish to identify ${\bphibethe}$ as the
macroscopic potential ${\bphi}$
because,
as mentioned above, the addition of a layer of  {\em microscopic} 
dipoles to a surface
should not change ${\bphi}$ in an internally-consistent linear macroscale theory.
This is because, at the macroscale, a microscopic distance is equivalent to ($\Lequiv$)
a distance of zero (see Sec.~\ref{section:homogenization} and Sec.~\ref{section:excess_fields}), 
and because a layer of microscopic dipoles, ${qa\hat{n}}$, is equivalent
to two layers with equal and opposite charges per unit area that are separated
by a distance $a$.
Since ${a\Lequiv 0}$ these two layers are equivalent, at the macroscale, to a single charge-neutral layer,
which would not change ${\bphi}$ inside the crystal.
Therefore if Bethe's derivation is right, and if ${\Rho}$ is a linear
spatial average of $\rho$, the MIP cannot be identified as
the macroscopic potential because that would be tantamount to saying
that the same macroscale distribution of charge can give
rise to different values of ${\bphi}$.
This would imply that, even when $\Rho$ and the macroscale boundary conditions
are known, the value of $\bphi$ 
cannot be calculated; its value depends, 
in some way, on certain microscopic details of $\rho$ that are lost by
the ${\rho\mapsto\Rho}$ homogenization transformation.

If we can assume that ${\phi}$ is a linear functional ${\phi[\rho]}$ of $\rho$,
 and that $\Rho$ is a linear spatial average $\expval{\rho}$ of $\rho$, then the linearity of 
both operations implies  that 
${\phi[\Rho]= \phi[\expval{\rho}]=\expval{\phi[\rho]}}$. 
Therefore, if there exist two microscopic charge
densities, $\rho$ and  ${\rho+\Delta\rho}$ , with the same macroscopic
charge density ${\Rho}$ (${\implies\expval{\Delta\rho}=0}$) and different macroscopic
potentials, $\bphi$ and ${\bphi+\Dbphi}$, then ${\bphi}$ does not equal ${\phi[\Rho]}$
and is a nonlinear functional ${\bphi[\rho]}$ of $\rho$. Linearity would imply that
\begin{align*}
\Dbphi\equiv\bphi[\rho+\Delta\rho]-\bphi[\rho]=\bphi[\Delta\rho]\neq 0; 
\end{align*}
and it would also imply that the mesoscale spatial average,
\begin{align*}
\expval{\Dbphi}=\expval{\bphi[\Delta\rho]}=\bphi[\expval{\Delta\rho}],
\end{align*}
is zero because ${\expval{\Delta\rho}=0}$.
Therefore ${\Delta\bphi}$ would be a harmonic function (${\mathbf{\laplacian}\,\Dbphi=0}$)
that fluctuates microscopically about zero. It would follow that ${\Delta\bphi}$ and $\bphi$ are
microscopic quantities, not a macroscopic ones.
On the other hand, nonlinearity of ${\bphi[\rho]}$ implies that
a material's macroscopic potential can depend on its history;
for example, if the material's microstructure ${\rho=\rho_1+\Delta\rho_1=\rho_2+\Delta\rho_2}$ is built
by superimposing the charge densities ${\rho_1}$ and ${\Delta\rho_1}$, its
macroscopic potential would differ, in general, from its value if it was
built by superimposing the two different charge densities, ${\rho_2}$ and ${\Delta\rho_2}$.
There are many problems that arise if we are tempted to assume that
${\bphi}$ depends on microscopic details of $\rho$ that are washed
away by the homogenization transformation; I have only mentioned a few of them.

Returning to the example of Crystal~1 and Crystal~2:
if we rigidly shift the MIP of Crystal~2 by ${\Delta \phi>0}$ by coating
its surfaces with a layer of dipoles, 
the same layer would shift the average potential in the vacuum just outside the crystal 
by ${-\Delta \phi}$.
Outside Crystal~1, $\phi$ appears to be zero 
because the field from each atom is zero.
Outside Crystal~2, the {\em average} of $\phi$ appears to be zero because
the average electric field emanating from each building block is zero. 
Adding a dipole layer to Crystal~2 to turn it into Crystal~1 appears to
shift the {\em mean vacuum potential} (MVP) in a layer surrounding the crystal up, while
shifting $\bphibethe$ down by the same amount. In the limit of large distance from the 
crystal the potential vanishes, because the crystal is charge neutral overall, but it does
not begin to decrease in magnitude significantly until the distance to the closest surface
is comparable to one or more of the surface's linear dimensions; therefore the MVP is
shifted by ${-\Delta \phi}$ in a {\em macroscopic} layer of vacuum surrounding the crystal.
So if Bethe's derivation was correct, and if a macroscopic layer of microscopic dipoles could shift
the average potential in a macroscopic region, the MVP would be zero in a macroscopic layer
of vacuum surrounding the crystal both before and after it had 
been shifted by a finite amount ${-\Delta \phi}$!
Clearly, this is absurd.

\begin{figure}[!]
\includegraphics[width=0.5\textwidth]{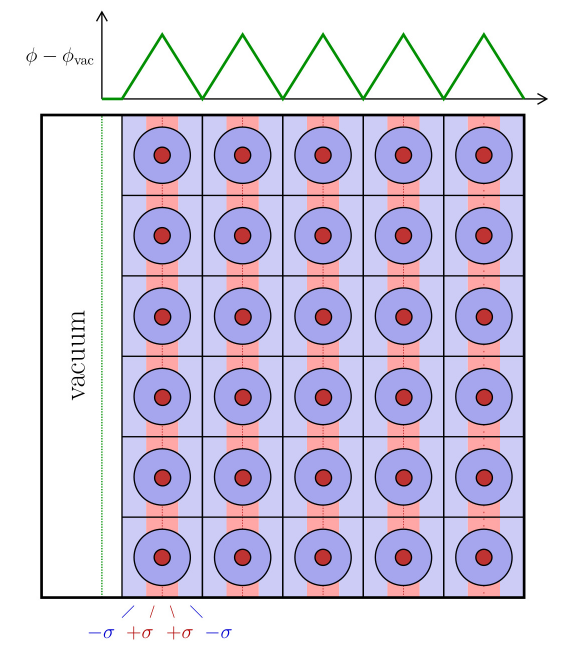} 
\caption{
One can think of each of the coloured layers parallel to
the surface as planes of charge, with the layers coloured
pink carrying charge densities of ${\sigma>0}$ and those
coloured blue carrying charge densities of ${-\sigma}$.
Alternatively, one can think of each blue layer and 
the red layer next to it as a plane of dipoles. 
The plane of dipoles closest to the surface is pointing
into the surface, causing an upward step in $\phi$, which in this example denotes the planar average
of the microscopic potential, relative
to its value of ${\phi_{\text{vac}}}$ in the vacuum 
close to the surface (dotted green line).
The next
layer causes a downward step of $\phi$, and so on.
The average potential is clearly positive if ${\phi_{\text{vac}}}$
is regarded as the zero of potential. It 
would be negative if the layer closest to the vacuum was
positive instead of negative, as in Fig.~\ref{fig:building_blocks}, 
and if the value of ${\phi_{\text{vac}}}$ was again
set equal to zero, despite its value differing
from its value above the surface of the
crystal pictured above.
}
\label{fig:surface_dipoles}
\end{figure}

Now consider Fig.~\ref{fig:surface_dipoles}, which uses 
the concept of a dipole layer to illustrate one argument for why the MIP
is positive. The crystal in question is identical to Crystal~1, 
so this construction appears to validate Bethe's result. However, 
there is no justification for dividing the material's microstructure
into the blue and pink layers.
If, for example, we combined each adjacent pair of pink and blue layers into a single
charge-neutral layer, we would find that the MIP vanishes.
Therefore, as with the construction Bethe used in his derivation, two equally-justified
ways to partition and spatially-average the microstructure leads to 
two different values of the MIP.

The superposition principle, on which Bethe's derivation and much
of electromagnetic theory are based, allows us to do the following: Let
us partition the space $\Vregion$ occupied by an electron cloud into
$M$ partitions of volume ${\Volume/M}$ and let us divide
the nucleus's charge into $M$ `pieces', such that for each 
partition there is a piece of nucleus with the same magnitude of charge.
Now, after taking the large-$M$ limit, let us displace the pieces to the
partitions so that each one becomes charge neutral.
The atom's spherical symmetry initially ensures that, after displacing
all of the pieces, it is again spherically symmetric.
After this redistribution of charge, the MIP must be zero
because the nucleus and every partition have become charge neutral.
If we view the displacement of nuclear charge as the
superposition of the negative of an atom's charge density
on each atom, this makes sense. We have simply superimposed 
a crystal's charge density and its negative, so of course
the MIP of the superposition vanishes.
However, we could also view the displacement of each
piece of nucleus as the placement of a negative charge
at the nucleus and a positive charge in the partition. 
Placing a dipole at a point in space changes the potential
everywhere but, by symmetry, it cannot change the spatial average
of the potential. Therefore placing all of these dipoles inside
the crystal should not change the MIP.
It appears that the superposition principle does not apply.

\subsubsection{Existence of a flaw - Illustration 2}
\label{section:illustration2}
Another way to see that there must be problem with Bethe's result is to treat electrons as point particles 
instead of expressing ${\rhom}$ as a smooth and delocalized density.
Using a line of reasoning similar to Bethe's
we could say that the total potential in each unit
cell from all electrons and nuclei outside the cell is
approximately equal to the sum of the total
potentials emanating from the particles inside the cell.
Then we could calculate the total potential
from each point particle at all points within a distance
$R$ of it, add together the total potentials from all particles 
within each unit cell of the crystal, and take the limit
of large $R$ to get the total potential
emanating from each unit cell. This total would vanish because
the potential emanating from a point charge ${Ze}$ is the negative of ${Z}$ times
the total potential emanating from a point charge ${-e}$.
Cancellation is obvious when ${Z=1}$ (hydrogen), but Bethe's construction
does not exclude this case, either explicitly or implicitly. Therefore his
derivation leads to the conclusion that the magnitude of the
spatial average of the potential from a proton is greater than the magnitude
of the spatial average of the potential from an electron.

This suggests that the problem in Bethe's derivation might be related to
his use of a continuous charge density for electrons and a (discrete)
delta distribution of charge for nuclei.
Usually this form of $\rho$ is regarded 
as a time average of the true charge distribution: 
electrons are whizzing around the more massive nuclei
so fast that their charge, when observed on a timescale
of about ${10^{-16}-10^{-15}}$ seconds, appears to be
smeared into a continuous charge density.
This timescale is too short for nuclei to move significantly, but 
long enough for each electron to trace out a very long trajectory.
Nevertheless, because the integral of 
\begin{align*}
\phi\left(\uvec \right) \equiv 
\overbrace{\sum_{i\in\text{nuclei}}\frac{\kappa Z e}{\abs{\uvec-\Rvec_i}}}^{\displaystyle \phi_n(\uvec)}
\overbrace{- \sum_{j\in\text{electrons}}\frac{\vphantom{Z}\kappa e}{\vphantom{\Rvec}\abs{\uvec-\rvec_j}}}^{\displaystyle \phi_e(\uvec)}
\end{align*}
over all space vanishes, which implies that its average over all space, ${\expval{\phi(\uvec)}_{\uvec}}$,
vanishes, one might expect the spatial average ${\expval{\expval{\phi(\uvec)}}_{\uvec}}$
of its expectation value 
with respect to the material's wavefunction $\Psi$,
\begin{align*}
\expval{\phi(\uvec)}\equiv\frac{\expval{\phi(\uvec)}{\Psi}}{\braket{\Psi}},
\end{align*}
to also vanish.
However, it is easy to reduce this expectation value and its average over all $\uvec$ to the forms
\begin{align}
\expval{\phi(\uvec)} = &\frac{\expval{\phi(\uvec)}{\Psi}}{\braket{\Psi}} =\phi_n(\uvec)  -\kappa\,e\, \int \frac{n(\rvec)}{\abs{\uvec-\rvec}}\dd[3]{r}, \nonumber \\
\expval{\expval{\phi(\uvec)}}_{\uvec} & = \expval{\phi_n}_{\uvec} + \expval{-\kappa\,e\, \int \frac{n(\rvec)}{\abs{\uvec-\rvec}}\dd[3]{r}}_{\uvec},
\label{eqn:likebethe}
\end{align}
where ${n(\rvec)}$ is the probability density that there is an electron at ${\rvec}$.
At first glance, Eq. ~\ref{eqn:likebethe} might appear to validate Bethe's approach, 
because it appears to be the spatial average of the potential from 
a charge distribution whose form becomes equivalent to the one he used (Eq.~\ref{eqn:betherho})
when there is spherical symmetry.
It would be very strange if it were equivalent:  it would mean
that ${\expval{\expval{\phi(\uvec)}}_{\uvec}}$ is finite but 
that the expectation value ${\expval{\expval{\phi(\uvec)}_{\uvec}}}$
vanishes. Therefore, it would mean that its value is changed simply by changing the order of 
integration such that the integral with respect to ${\uvec}$ is performed first.
We need to understand this better - both physically and mathematically.

\subsection{The flaw}
\label{section:solution}
The flaw in Bethe's derivation is that he calculated the
electrons' contribution to the potential from
a volumetric density of negative charge, ${\rhom(\rvec) = -e\,n(\rvec)}$, which is 
defined at all points in space. Then, because the electron density
is spread over a greater volume than the nuclear density on femtosecond
time scales, the spatial average of the potential does not vanish.

To understand why it does not vanish, let us again 
consider the potential ${\phi_r(r;\eta)}$
at a distance $r$ from the center of a
spherically-symmetric nonpositive or nonnegative charge density ${\rho(u;\eta)}$, where
the width of ${\rho}$ is proportional to the value of $\eta$, which is a parameter.
Let us denote the integral of $\rho$ within a sphere of radius $r$ by ${Q(r;\eta)}$ and
its integral over all space by 
\begin{align*}
Q_\infty\equiv\lim_{r\to \infty}Q(r;\eta)= \lim_{\eta\to 0}Q(r;\eta).
\end{align*}
Using Eq.~\ref{eqn:sphericalpot} we can express 
the magnitude of the potential at a distance ${r}$ from the center of $\rho$ as 
\begin{align*}
\abs{\phi_r(r;\eta)} & = \abs{\frac{\kappa Q(r;\eta)}{r} + \frac{1}{\epsilon_0}\int_r^\infty \rho(u;\eta) u\dd{u}} \\
& < \abs{\frac{\kappa Q(r;\eta)}{r} + \frac{1}{\epsilon_0}\int_r^\infty \rho(u;\eta) u\left(\frac{u}{r}\right)\dd{u} } \\
\therefore \abs{\phi_r(r;\eta)} & < \frac{\kappa\abs{Q_\infty}}{r}.
\end{align*}
At very large distances (${r\gg \eta}$) the potential is approximately equal to
${\kappa Q_\infty/r}$, and it gets closer to this value as $r$ increases.
At short distances (${r \sim \eta}$) the magnitude of the potential
is significantly smaller than ${\kappa \abs{Q_\infty}/r}$ and the 
ratio ${\phi_r(r;\eta)/\left(\kappa Q_\infty/r\right)}$ gets smaller
as $r$ decreases.

Now let us consider the average of $\phi_r$ over all
points within a fixed distance $R$ of the center of $\rho$
as the value of $\eta$ changes.
When ${\eta/R}$ is very small, the magnitude of $\phi_r$ at
almost all points is approximately ${\kappa \abs{Q_\infty}/r}$
and it is only significantly smaller than that value
in a volume fraction
${\sim \left(\eta/R\right)^3}$ of the sphere.
In the limit ${\eta/R\to 0}$, the average potential
is the same as it would be if $\rho$ was the delta distribution 
of a point charge.
However, as $\eta$ increases, the magnitude of the average
potential in the sphere of radius $R$ decreases because
the fraction of the volume occupied by points at which $\abs{\phi_r}$ is
significantly less than  ${\kappa \abs{Q_\infty}/r}$ increases.
Therefore, the average potential in the sphere reduces in 
magnitude as $\eta$ increases.

This is why the potential from the electrons does not cancel the
potential from the nuclei in Bethe's derivation: the value
of ${\eta}$ is finite for the electrons, but vanishingly small for nuclei, 
which makes the magnitude of the average potential from the nuclei
greater than that from the electron cloud.
 
I will now explain, from three different perspectives, what is wrong
with Bethe's derivation and with his use of the charge density in 
Eq.~\ref{eqn:betherho}.

\subsubsection{Perspective 1}
Bethe's use of a continuous electron charge density ${\rhom(\rvec)=-e\,n(\rvec)}$ 
suggests that he interpreted it as the time average of
the electrons' instantaneous delta distribution of charge.
However ${n(\rvec)}$ is not a time average of the electrons' 
positions, it is a probability density that an electron (any one of them)
is at position $\rvec$ at any precisely-specified time.
The average, over a time interval ${\interval(t,\Delta t)}$, of the delta distribution 
of a set of moving particles is not a volumetric charge density, 
but a set of linear charge densities defined only
along the segments of the trajectories followed during ${\interval(t,\Delta t)}$.
Therefore the time average of the electrons' delta distribution 
is a set of linear charge densities defined on a set of curves and
there is no charge at points that do not lie along these curves.

This means that the set of points at which the time-average of the
electrons' charge distribution is nonzero is a set whose
measure in $\realthree$ is zero, regardless of the magnitude of ${\Delta t}$; 
therefore electrons are not more delocalized than nuclei because both occupy zero volume.
Using the fact that ${Ze/r}$ is cancelled by ${Z\times -e/r}$, it
is easy to show that the potential from the true time
average of the electrons' charge distribution exactly cancels
the potential from the nuclei.

\subsubsection{Perspective 2}
\label{section:perspective2}
Although Eq.~\ref{eqn:likebethe} appears to be the spatial
average of the potential from a set of point nuclei
and a continuous density of negative charge ${\rho_-(\rvec)=-e\,n(\rvec)}$, 
it is not.
The quantity
\begin{align}
\expval{\phi_e(\uvec)} =
-\kappa e \int \frac{n(\rvec)}{\abs{\uvec-\rvec}}\dd[3]{r}
\label{eqn:phie}
\end{align}
is not the potential at $\uvec$ from ${\rhom(\rvec)=-e\,n(\rvec)}$ because
Coulomb's law does not hold for the charge from a continuous charge density
within a region of infinitesimal volume.
To see that it does not hold, consider a spherical Gaussian surface of radius $\Delta r$
centered at the point $\rvec$. The charge within the sphere is ${\rho(\rvec)\times \frac{4}{3}\pi(\Delta r)^3}$, 
but the electric field, $\me$, on its surface is not directed radially outward from $\rvec$ because it
has a contribution from the nucleus.
This would not present a problem for point charges because one could choose ${\Delta r}$
to be arbitrarily small without changing the amount of charge enclosed by it; and,
as ${\Delta r}$ got smaller, the direction of the field passing through the surface would become arbitrarily close 
to radially outward.

However this reasoning does not apply to a continuous charge density because
the magnitude of the charge enclosed by the surface scales like ${\left(\Delta r\right)^3}$
in the small ${\Delta r}$ limit.
A correct application of Gauss's law for a spherically-symmetric charge density
leads to Eq.~\ref{eqn:sphericalpot}, which does not give the same result
as Eq.~\ref{eqn:phie}. Equation~\ref{eqn:phie} is the correct expression
for the expectation value of the potential at ${\uvec}$ from the electrons, 
but it is not the correct expression for the potential from charge
density ${\rhom}$; when there is spherical symmetry, the latter is Eq.~\ref{eqn:sphericalpot}.

Equation~\ref{eqn:likebethe} can be expressed in the slightly more general form
\begin{align*}
\expval{\expval{\phi(\uvec)}}_{\uvec} & = \kappa
\expval{\int \frac{\rho(\rvec)}{\abs{\uvec-\rvec}}\dd[3]{r}}_{\uvec} 
\end{align*}
where ${\rho=\rhop+\rhom}$ and $\rhop$ is the distribution
of the nuclei.
Assuming that $\rho$ has spherical symmetry, and choosing its center as the origin, 
the integral of ${\expval{\phi(\uvec)}}$ over all points within a distance $R$ of its center is
\begin{align*}
\int_{\abs{\uvec}<R} \expval{\phi(\uvec)} \dd[3]{\uvec}
& =
\kappa
\int_{\abs{\uvec}<R} \dd[3]{u}\int_{\realthree} \dd[3]{r}\left(\frac{\rho(\rvec)}{\abs{\uvec-\rvec}}\right)  \\
& =
\kappa
\int_{\realthree} \dd[3]{r} \rho(\rvec)\int_{\abs{\uvec}<R} \frac{1}{\abs{\uvec-\rvec}}\dd[3]{u} .
\end{align*}
In the limit of large $R$ the right hand side becomes arbitrarily close to
\begin{align*}
\kappa \left(\int_{\realthree} \rho(\rvec) \dd[3]{r} \right) \left(\int_0^R 4 \pi u^2 \times \frac{1}{u} \dd{u}\right),
\end{align*}
which vanishes because ${\int_{\realthree} \rho(\rvec)\dd[3]{r}}$ vanishes.
Therefore the total potential from the atom, divided by the volume of a primitive
unit cell, which is the quantity calculated by Bethe, is zero.
Furthermore, to deduce that it is zero I have not assumed anything about the degrees
to which the nuclear density and the electron density ${n(\rvec)}$ are localized.

\subsubsection{Perspective 3}
The relations ${\me=-\grad \phi}$ and ${\rho/\epsilon_0 = -\laplacian\phi}$
are preserved by the homogenization transformation because $\grad$, $\laplacian$, 
and the spatial average are all linear operations, which   commute
when they are applied
in a mutually-consistent manner.
For example, if $\expval{\;}_x$ and $\expval{\;}_y$
denote averages along the $x$-axis and $y$-axis, respectively, 
then 
\begin{align*}
\expval{\rho}_x/\epsilon_0 = -\expval{\laplacian\phi}_x=-\laplacian\expval{\phi}_x.
\end{align*}
However, it cannot generally be true that ${\expval{\rho}_x/\epsilon_0 = -\laplacian\expval{\phi}_y}$.
Calculating the average potential, along an axis normal to a surface,
from the average of the charge density on planes parallel to the surface does not, 
in general, lead to a meaningful result.

This is why it is not physically reasonable to calculate the MIP from 
the average of the green curve in Fig.~\ref{fig:surface_dipoles} and
it is why the value of the MIP deduced by averaging the charge distribution
in layers parallel to the surface depends on the choice of the layers' positions and thicknesses.
For example, if, instead of the division into the pink (P) and blue (B) layers depicted in Fig.~\ref{fig:surface_dipoles}, 
each layer was chosen to be a layer of atoms, the green curve would be flat because
each layer would be charge neutral. 
A layer of atoms comprises two pink layers and
two blue layers in the order BPPB, so I will call it
a BPPB-layer. One could choose the first layer
at the surface to be a negatively charged
B-layer and all others to be charge-neutral PPBB-layers.
In that case the green curve would be a straight line with a negative slope;
therefore, not only would the MIP appear to be negative, there would be a 
macroscopic electric field in the material emanating
from the plane of negative charge.
If the first layer is a BPP-layer and subsequent layers
alternate between B- and BPP-layers, the MIP appears
to be negative, with the potential as a function of depth
resembling a skewed version of 
the negative of the green curve in Fig.~\ref{fig:surface_dipoles}.

The electron charge density used by Bethe in his derivation
can be regarded as the result of performing the following sequence
of temporal and spatial averages:
first, the time average of the electron charge distribution  is calculated
to give a set of curves carrying linear charge densities; next, this set of
linear charge densities are turned into a volumetric charge density
by averaging in the radial direction over a small width ${\dd{r}}$;
finally, the resulting distribution is given spherical symmetry
by setting the charge density at distance $r$ equal to its spatial
average on the spherical surface of radius $r$.
The resulting charge density
is then used to calculate the potential as a function 
of position along the radial direction, i.e., along 
an axis that, at its point of intersection with the surface
on which the final spatial average is taken, is perpendicular
to this surface.
There is no reason to expect this procedure to produce
results that are any more meaningful than those derived
by partitioning the
surface in Fig.~\ref{fig:surface_dipoles} into artificial layers
using an arbitrary, unjustified, and mutually-inconsistent
sequence of partial spatiotemporal averages.

I avoided the problems with Bethe's derivation in 
Sec.~\ref{section:average_potential} by not making
any assumptions about the microscopic charge
density, except that it is mathematically smooth;
and
by only averaging along the $x$-axis; and by
using the same mesoscopic interval 
width for all spatial averages.
Because I used a general form of $\rho$, 
the derivations of Sec.~\ref{section:average_potential}
apply to charge distributions that 
are arbitrarily close to delta distributions;
and Coulomb's law can be applied to
a smooth charge distribution in this limit.

\section{Summary}
\label{section:conclusions}
This work concludes now with three summaries, followed by a brief discussion of
context and outlook. 
Each of the three summaries addresses
a different one of the three objectives stated in Sec.~\ref{section:introduction}.

\subsection*{Structure homogenization}
This work lays some foundations of a 
theory of the relationship between a \emph{microstructure} and
its \emph{macrostructure}. 
The microstructure is assumed to consist of one or more 
differentiable fields, ${\nu_i:\realone^n\to\realone}$, which fluctuate
on the \emph{microscale} $a$. The microstructure's macrostructure 
is the set of all observable manifestations 
on the \emph{macroscale}, ${L\ggg a}$, of the microstructure,  ${\{\nu_i\}}$.

For the purposes of this summary, it will be assumed that
the microstructure comprises a single scalar field, ${\nu:\realone^3\to\realone}$.

\subsubsection{Premises of structure homogenization theory}
Structure homogenization theory, in its most basic form, is
founded on two premises.

The first premise is that the microstructure fluctuates on the  microscale $a$
and the macroscale $L$, but there exists an intermediate \emph{mesoscale} $l$,
where ${a\ll l \ll L}$,  on which its fluctuations are negligible. 
Therefore, on any mesoscopic domain, the average of the microscopic fluctuations of $\nu$ 
almost vanish, and non-linear contributions to its macroscopic variations are negligible.

A more mathematical statement of this premise is that the 
Fourier transform ${\ftsnu\equiv\fourierspace{\nu}}$ of 
${\nu}$ satisfies
\begin{align}
\int_{\abs{\kvec}<k_L}  \abs{\ftsnu(\kvec)}^2\dd[3]{\kappa}
\gg &\int_{\abs{\kvec}\in(k_L,k_a)} \abs{\ftsnu(\kvec)}^2\dd[3]{\kappa}
\nonumber
\\
&\ll
\int_{\abs{\kvec}>k_a} \abs{\ftsnu(\kvec)}^2\dd[3]{\kappa},
\label{eqn:scale_separation}
\end{align}
where ${k_a\equiv 2\pi/\amax}$ and ${k_L\equiv 2\pi/\Lmin}$;
where the microscale $a$  and the macroscale $L$ are defined by
${\epsilon\sim a\iff \epsilon<\amax}$ and 
${\epsilon\sim L\iff \epsilon>\Lmin}$;
and where ${\amax}$ is a property of the microstructure ($\nu$)
and ${\Lmin}$ is determined by both the microstructure and 
the scale on which the microstructure is observed. 

The second premise of structure homogenization theory is that when $\nu$ is observed or measured with 
a probe of macroscopic dimensions, such as the pupil of an eye,
what is observed is a weighted spatial average of $\nu$ on a mesoscopic domain.

\subsubsection{Observable artefacts of structure homogenization}
It turns out that the homogenization transformation introduces 
qualitative differences between a simple microstructure $\nu$ and any 
macrostructure defined by it.
These mathematical peculiarities and observable artefacts of structure homogenization 
have some very important physical consequences.

They exist because perfect homogenization, meaning a total elimination of
$\nu$'s microscopic fluctuations, is only possible in the limit in which 
the macrostructure is the average of $\nu$ over \emph{all} points in its domain.
This is the limit in which the counterpart of $\nu$ at the macroscale, ${\Nu:\realone^3\to\realone}$, 
is flat and featureless. For example, the Earth's surface macrostructure is close
to this limit in Voyager~1's famous `\emph{pale blue dot}' photograph~\citep{sagan_1994}.

Away from this limit,  homogenization is imperfect, and spatial
averages of $\nu$ are not uniquely defined: The average of $\nu$
on a mesoscopic domain is only defined to a precision that is finite.
Therefore $\Nu$ can only be defined to a finite precision, $\precNu$. 

The fact that $\Nu$ is only defined to a finite precision means that if 
the only way to distinguish between two points ${\bx_1, \bx_2 \in\realone^3}$
is to observe the difference in its value at those points, ${\Nu(\bx_1)-\Nu(\bx_2)}$,
there is a limit, $\prectheo$, to the precisions with which
positions and displacements can be measured or observed.
An approximate relationship between $\precNu$, $\prectheo$,
and the uncertainty $\precmom$ in the gradient of $\Nu$ 
is ${\prectheo\precmom\propto\precNu}$. 

The uncertainty in positions and
displacements implies a one to many relationship between
points $\bx$ at the macroscale and points ${x}$ 
at the microscale. Effectively, microstructure homogenization
is a compression of space which causes all microscopic distances to vanish. 
This spatial compression causes surfaces and interfaces, which 
are ill-defined at the microscale because their widths are indeterminate,
to become well-defined and locally planar (zero width) at the macroscale:
Since the domain of $\nu$ is $\realone^3$, 
they become two dimensional manifolds which carry excess fields, in general.
For example, homogenizing a material's microscopic volumetric charge density $\rho$ not only
defines a macroscopic analogue $\Rho$ of $\rho$ within the material, 
it also turns the material's boundary into a two dimensional manifold (the surface),
which carries an areal charge density, $\bsigma$.

I have derived expressions that relate boundary excess fields  
to the microscopic fields whose homogenization created them. For example, I have derived
an expression ${\bsigma[\rho]}$ relating the areal charge density at a surface to the microscopic
volumetric charge density, $\rho$. This expression generalizes Finnis's expression~\citep{finnis} for
the surface charge density of a crystal to amorphous microstructures.

\subsection*{Electrical macrostructure}
I used the basic elements of the theory of structure homogenization 
to deduce how the microscopic fields $\rho$, $\me$, and $\phi$, that  
appear in Maxwell's vacuum theory of electricity
manifest as macroscopic fields, and to deduce the relationships
between those macroscopic fields.

The set ${\{\rho,\me,\phi\}}$ does not only define
a set ${\{\Rho,\E, \bphi\}}$ of macroscopic-counterpart fields. 
It also defines macroscopic excess fields on lower-dimensional 
manifolds, such as surfaces, interfaces, edges, line defects, and point defects.
These manifolds and fields are created by the spatial compression that
is intrinsic to structure homogenization.

The linearity of the spatial averaging operation that turns
microstructure into macrostructure
means that the relationships between 
$\rho$, $\phi$, and $\me$ are preserved by
the homogenization transformation. Therefore 
${\Rho = -\laplacian\bphi}$ and ${\E=-\grad\bphi}$. 

It is a well-known and obvious stability requirement that ${\Rho=0}$ in the bulk of every material. 
It follows that, in the bulk of a macroscopically-uniform material whose surfaces
are charge-neutral, either $\bphi$ is constant and ${\E=0}$ or $\bphi$ is a linear 
function of position and $\E$ is constant.

Both the $\pp$ and $\D$ fields that appear in macroscopic electromagnetic theory
have been interpreted, and their existences justified, in multiple mutually-inconsistent
ways since Maxwell introduced them in the 19th century. 
I have pointed out that none of these interpretations or justifications are valid
and that $\pp$ and $\D$ appear within physical theory for historical reasons only:
$\pp$ is not observable, is not a necessary element of electromagnetic theory, 
cannot be defined uniquely, and its existence is prohibited by macroscale symmetry.
Furthermore, it continues to cause a great deal of confusion, 
without adding to the utility of electromagnetic theory.
Scrapping it removes the distinction between $\E$ and the electric 
displacement $\D$, so $\D$ should also be scrapped. 

The only volumetric fields that are required at the macroscale 
are $\bphi$ and its derivatives $\E$ and $\Rho$; but the linearity
of their interrelationships facilitates the decomposition of each one into components
with distinct origins and effects.
For example, when studying dielectric response it might be useful to write
${\E=\Eext + \DE}$ and ${\Rho=\Rho_0 + \DRho}$, where $\Eext$ is
an externally-applied electric field, ${\DRho}$ is the change that it induces
in the charge density, and $\DE$ is the field emanating from ${\DRho}$.

When studying the long wavelength electric fields that emanate from modulations of
the structure by optical lattice vibrations, it makes more sense to express these
modulations directly as changes in charge density (${\Delta \rho}$ and/or ${\Delta\Rho}$)
than as a diverging polarization field. The electric field can be calculated directly
from the former, whereas the latter must be translated into a charge density to deduce
its field. Furthermore, expressing these modulations as a charge density makes the qualitative
difference between the long wavelength limit (${\bk\to 0}$) and a rigid relative
displacement of sublattices (${\bk=0}$) clearer: a longitudinal 
optical phonon of wavelength ${\lambda_L\sim L}$ creates an electric field 
of wavelength ${\lambda_L}$; but if the material's surfaces are earthed,
a rigid relative displacement of sublattices does not create any macroscopic field. 
Therefore if a material is at equilibrium, ${\Rho=0}$ implies that there cannot be 
any macroscopic electric field emanating from its bulk.
The electric potential also vanishes unless it has a source. Therefore, 
if all surfaces of an electromagnetically-isolated (${\E^\mathrm{ext}=0}$) material are neutral, 
${\E}$  and ${\bphi}$ both vanish in its bulk. This has important implications for materials physics.

The absence of a macroscopic field can also be understood as a demand of symmetry: symmetry is scale-dependent and the bulks of all compositionally- 
and structurally- uniform materials are isotropic at the macroscale, regardless of their microstructures. 
A vector field that has a linear relationship with ${\Rho}$
cannot exist if ${\Rho}$ is uniform because all directions are equivalent. 
Therefore if ${\E}$ does not vanish in the bulk of a homogeneous material, it is either externally applied 
or it emanates from an accumulation of charge at surfaces, interfaces, or 
other {\em macroscopic} heterogeneities.

On the macroscale, a material's response to an external field ${\Eext}$ 
is the changing of the areal densities of charge at all points on surfaces and interfaces
whose tangent planes are not parallel to ${\Eext}$. When ${\Eext}$ is perpendicular
to two opposing surfaces, and parallel to all others, 
the net change in
the macroscopic field in the material at equilibrium is ${\Eext-\Delta\bsigma/\epsilon_0}$
where ${\Delta\bsigma>0}$ is the magnitude of the changes in the surface charges 
induced by ${\Eext}$.

When a crystal possesses a spontaneous polarization field, by which I mean {\em only} that its
microstructure lacks inversion symmetry, any surface perpendicular to an axis of anisotropy
would carry an areal surface charge density $\bsigma$ unless neutralized by extrinsic charges.
A charged surface is unstable unless stabilized by another source of potential, such as an oppositely charged surface.

The definition of surface charge density $\bsigma$ as the integral 
of $\Rho$ across the surface is equivalent to Finnis's definition, which
I have generalized to non-crystalline microstructures.
By relating currents to changes of surface charge, Finnis's result
can be used to calculate the normal component
of the current density $\J$ at any interface if the time dependence
of $\rho$ at the interface is known.

The current density in an insulator can also be calculated using 
the main practical result of the {\em Modern Theory of Polarization} (MTOP), 
which is a definition of the polarization current density $\Jconv$ in terms of the time-dependent
microstructure of a material's bulk. 
I have shown that this result follows from Finnis's result and that quantum mechanics
is not required to derive it. My derivations make clear that the original
MTOP definition of polarization current~\citep{resta-1993,kingsmith-vanderbilt-prb-1993-1,kingsmith-vanderbilt-prb-1993-2} is exact: 
The fact that $\Jconv$ is expressed in terms of single particle states does not constitute an approximation.

\subsection*{Mathematical representations of classical microstructures}
I have shown that all features of the mathematical structure
of the quantum mechanical theory of electricity in materials that are relevant
to this work are compatible with, or required features of,
an internally-consistent statistical theory of a 
\emph{deterministic} dynamical system of charged particles.

If a classical material comprised a large number of particles whose charges and masses were comparable in magnitude to those of
electrons and nuclei, those particles would move so rapidly, and respond so sensitively to the act of observing them, 
that it would be impossible to observe their instantaneous positions or to follow their individual trajectories.

Therefore, as in quantum mechanics,
the \emph{observable} microstructure would not be the particles' instantaneous configuration (set of positions), 
but their joint position probability distribution, ${\pdf=\pdf(\rvecsub{1}\cdots\rvecsub{\Nelec})}$. 
If a subset of the fast-moving classical particles were
identical, the impossibility of following their trajectories would make them indistinguishable.
Therefore any empirically-unfalsifiable pdf $\pdf$ would be symmetric with respect to the 
exchange of any two of the particles' positions.

A \emph{coincidence point} in the particles' configuration space $\configspace$
is a configuration in which two or more particles occupy the same point in space.
Since two massive particles cannot occupy precisely the same point in space, 
$\pdf$ must vanish at all coincidence points. This requirement 
implies that $\pdf$ has non-differentiable cusps at coincidence points unless 
its derivatives with respect to the coincident particles' positions vanish
sufficiently rapidly as coincidence points are approached.
If $\pdf$ is non-differentiable at coincidence points, the information it possesses can be specified
by an exchange-antisymmetric function ${\Psi=\sqrt{\pdf}e^{i\theta}\in\lebesgue(\realone^{3\Nelec})}$, 
which is differentiable for all practical purposes because it changes sign at coincidence points.

Any set of single-particle functions ${\{\varphi_i\}}$, 
which is a complete orthonormal basis of ${\lebesgue(\realone^3)}$,
defines an infinite set of ${\Nelec}$-particle Slater determinants. The set of Slater determinants is a 
complete orthonormal basis of the subspace of ${\lebesgue(\realone^{3\Nelec})}$ that comprisess its anti-symmetric elements.
Therefore the statistical state $\Psi$ of the microstructure of a set of mutually-repulsive and indistinguishable classical particles
can be expressed exactly as a weighted sum of an infinite number of Slater determinants, or approximated by
a weighted sum of a finite number of Slater determinants.

I have discussed various kinds of single-particle statistical states, or \emph{orbitals}, in a many-particle system. 
Orbitals play important roles in both the MTOP and in commonly-used and commonly-taught 
models of chemical bonding.

If a material's statistical microstructure changes as some stimulus $\zeta$ varies,
it may be possible to express the number density of each of its constituent sets of 
identical particles exactly as ${n(\rvec;\zeta)=\sum_{i=1}^{\Nelec} n_i(\rvec;\zeta)}$,
where ${n_i(\rvec;\zeta)\equiv\abs{\varphi_i(\rvec;\zeta)}^2}$, and
${\{\varphi_i(\zeta)\in\lebesgue(\realone^3)\}_{i=1}^{\Nelec}}$
is an orthonormal set whose elements vary continuously as $\zeta$ changes.
Without invoking quantum mechanics I have shown that
when ${n(\zeta)}$ admits such a representation,
the polarization current that flows as $\zeta$ changes
is given exactly by the MTOP expression~\citep{resta-1993,kingsmith-vanderbilt-prb-1993-1,kingsmith-vanderbilt-prb-1993-2},
\begin{align*}
\Jconv=q\dot{\zeta}\sum_{i=1}^{\Nelec}\dv{\zeta}\left(\int_{\realone^3}\dd[3]{r}\rvec n_i(\rvec;\zeta)\right), 
\end{align*}
where $q$ is the charge of each particle in the identical set.
If the number density and its continuous dependence on $\zeta$ 
can be represented by one $\Nelec$-fold set of single-particle states, it can be represented by
an infinite number of such sets, which are related to one another by rotations within the $N$-dimensional
subspace of ${\lebesgue(\realone^3)}$ that they span.

When the bulk of a crystal is represented in a torus, delocalized sets whose elements have the 
crystal's periodicity, such as the eigenfunctions of single-particle effective Hamiltonians,
are known as \emph{Bloch functions}. 
Each set of Bloch functions can be transformed into any number of sets of
localized \emph{Wannier functions}. The most localized set is known as the set of
\emph{maximally localized Wannier functions} (MLWFs)~\citep{ferreira_parada,marzari_mlwf,souza,mlwf_rmp}.
I point out that MLWFs do not have an obvious physical interpretation, and that using them to analyse the
natures of chemical bonds is not justified theoretically.

However I point out that good reasons exist to attach physical meaning to so-called \emph{natural states} and \emph{natural orbitals}.
Many of these reasons were summarized by Coleman~\citep{coleman_rmp} and others~\citep{lowdin_1955,davidson_1972,mcweeny_1960,ando_1963,helbig_2010,umrigar_1998,natural_occupations}, 
while others are illustrated by results derived in Appendix~\ref{section:appendix_natural} and discussed in Sec.~\ref{section:single_particle_states}.

For example, I show that the expected energy, $E$, of a classical or quantum mechanical 
system of $\Nelec$ identical particles in a pure state can be expressed exactly as
\begin{align}
E&=\sum_\alpha
\occ_\alpha
\energy_{\alpha}
+
\sum_\alpha
\sum_{\beta\geq\alpha}\sqrt{\occ_\alpha\occ_\beta}
\,
w_{\alpha\beta},
\label{eqn:energy_conclusions}
\end{align}
where the sums are over the set ${\{\varphi_\alpha\}}$ of all natural orbitals; the orbital
`occupations' satisfy ${\occ_\alpha\in[0,1]}$ and ${\sum_\alpha\occ_\alpha=\Nelec}$; ${\energy_\alpha\equiv\expval{\hamsmall}{\varphi_\alpha}}$
is the expectation value of a single-particle Hamiltonian $\hamsmall$ for a particle in natural orbital ${\varphi_\alpha}$;
and ${w_{\alpha\beta}}$ is the energy of a \emph{mediated} coupling between particles in natural orbitals
${\varphi_\alpha}$ and ${\varphi_\beta}$.

If $E$ is expressed in terms of an \emph{arbitary} orthonormal set of orbitals, the expression is
much more complicated than Eq.~\ref{eqn:energy_conclusions} and does not give a clear meaning to the concept 
of orbitals being \emph{occupied}.
Equation~\ref{eqn:energy_conclusions} shows that, when interactions between particles are sufficiently weak, 
$E$ is approximately an occupation-weighted sum of the expected energies of particles that
are statistically almost independent of one another. 

The approximation ${E\approx\sum_\alpha\occ_\alpha\energy_\alpha}$ becomes an equality
in the limit of infinitesimally weak interactions; and it is shown in Sec.~\ref{section:natural_substructure} that, in that limit, 
the relationship ${\occ_\alpha\equiv\occ(\energy_\alpha)}$ 
between a natural orbital's occupation number at thermal equilibrium and its energy ${\energy_\alpha}$ is
a Fermi-Dirac distribution. Quantum mechanics is not invoked to draw this conclusion, so it applies
equally to classical and quantum mechanical systems of indistinguishable particles.

\subsection{Outlook}
The main body of this work began in Sec.~\ref{section:aether}
with a brief description of the gross features of Maxwell's theory of a 
luminiferous aether that occupies space and pervades all matter.
The purpose of Sec.~\ref{section:aether}
was to contrast Maxwell's understanding of matter and materials with 
the current state of physical theory; and, on the basis of that contrast,  
to encourage the abandonment of unobservable and redundant features of his theory; namely,
$\pp$ and $\M$.

Although Sec.~\ref{section:aether} may have appeared critical of 
Maxwell's work, the sequel to this work will argue that he was on 
the right track. It will argue that, in some important and fundamental ways, 
the physical picture he espoused is a better likeness of reality 
than the one painted by modern textbooks. Its major flaw is that it is a theory
of fields whose common domain is three dimensional \emph{space}.
Therefore science and technology might be in more 
advanced states today if its development had continued after 
Poincar\'e, Minkowski, Einstein and others had elevated time's status from a parameter to one of 
the dimensions of spacetime~\citep{walter_2014,miller_1981,darrigo_2006}.

Throughout this work, and in many of its appendices, it is shown
that most or all of the `quantum mechanical' aspects of our theory of electricity in materials 
are consistent with electrons and nuclei being as classical as billiard balls.
Their `quantum mechanical' natures are artefacts of
them occupying time and length scales that are many orders of magnitude
shorter than human time and length scales.
In light of this, it is interesting to wonder how science might have progressed 
had physics not been diverted quite so radically by the experimental discovery of 
quantum mechanics early in the 20th century.

For generations, physicists and chemists have sought to understand the quantum world 
\emph{from within} quantum mechanics. Relatively little effort has
been put into the development of classical statistical mechanics; and even less
effort has been put into trying to understand quantum mechanics from within 
the classical realm. It seems valid to question the wisdom of this imbalance of effort, 
because the common belief that quantum mechanics is somehow more general than classical 
statistical mechanics lacks a rigorous theoretical justification.

When Bohr coined the term \emph{correspondence principle}~\citep{bohr_1920}, 
he is unlikely to have anticipated it becoming mistaken for a reference to 
a rigorously-derived result or set of results~\citep{falkenburg_2009,messiah_1961}.
Nevertheless, it is often used as an assertion that purports to summarize physical
theory near the classical/quantum boundary - an underdeveloped region of physical theory that is 
not ripe for such brief and breezy summarization.

The sequel to this work will further examine the observable 
artefacts of large differences in scale between an observer and their subject.
It will suggest that the observed structure of the 
microscopic world may not be what a microscopic observer would observe.
It will examine the possibility that particles with charge and spin are not 
the components of a \emph{base microstructure}.
They might be components of a macrostructure whose underlying microstructure occupies even smaller scales. 
Some of the qualitative differences between a macrostructure and a base microstructure
are discussed in Sec.~\ref{section:homogenization} and illustrated
by Fig.~\ref{fig:macrostructure}. 

If it turns out that differences in scale explain some or all of the differences
between what is observed of the microscopic world and what we see around us 
on human scales, it may be worth investigating whether there exist observable
artefacts of an observer occupying time and length scales that are orders of magnitude \emph{smaller}
than their subject. In other words, do cosmologists draw their conclusions from
observations of the apex macrostructure, and do they account for all artefacts
of the observables occupying such large time and length scales?

\onecolumngrid
\newpage
\appendix
\addtocontents{toc}{\protect\setcounter{tocdepth}{0}}
\appendixpage
\twocolumngrid
\section{Areal charge densities and other integrals of macroscopic fields across interfaces}
\label{section:excess_formulae}
Here I quote expressions 
for the excesses of microscopic scalar fields at macroscale surfaces
and interfaces.
Derivations of these expressions can be found in Sec.~\ref{section:excess_fields} and
a derivation of a special case of one of them can be found in \REF~\linecite{finnis}.

I use the microscopic charge density ${\rho}$ 
as my example, but the expressions are applicable to any scalar field $\nu$, including
scalar components of vector or tensor fields. They are valid when
there exists a mesoscale $l$ on which the statistical characteristics
of the microscopic ($\sim a$) fluctuations in $\nu$ do not vary appreciably
and the 
macroscale counterpart ${\Nu}$ of ${\nu}$ varies at most linearly.
When that is the case it is possible to calculate a mesoscale average ${\bnu}$
of ${\nu}$ at each point, such that ${\Delta\nu\equiv \nu-\bnu}$ fluctuates
microscopically about zero, and such that ${\bnu}$ equals ${\Nu}$
to within the finite precision ${\precNu}$ to which ${\Nu}$ is defined.

Since ${\rho}$ fluctuates microscopically about zero 
in a material's bulk, ${\brho=0}$ is among the values of its
mesoscale average at each point. Therefore, I define ${\Delta\rho\equiv\rho}$
and I will express the quoted formulae in terms of ${\rho}$, rather
than ${\Delta \rho}$. Before quoting them, it is necessary to 
explain the construction used to define the quantities appearing
within them.

Consider a material with two surfaces which, on the mesoscale, are locally planar
where they intersect the Cartesian $x$ axis. Let ${x_L}$ and ${x_R}$ be points on the $x$ axis
in the vacuum immediately outside the material at the surfaces with outward unit
normals ${-\hat{x}}$ and ${+\hat{x}}$, respectively.
Let ${x_b\in (x_L,x_R)}$ be a point on the $x$ axis that is arbitrary apart
from the requirement that it is far enough away from both surfaces that it
can be regarded as being in the bulk of the material.
In each of the expressions quoted below, ${x_b}$ should be regarded as a point
in the bulk below whichever surface (at $x_L$ or $x_R$) the expression pertains to.

Now consider a mesoscopic neighbourhood ${\interval(x_b+u,\ell)}$ of $x_b$, where
${u\sim a}$ and ${\ell\sim l}$.
Let us assume that it is partitioned into contiguous microscopic (${\sim a}$) intervals 
${\interval(\bar{x}_m,\Delta_m)}$, such that ${x_b}$ is at the boundary
point shared by two of them, and such that the average of ${\rho}$ on each interval
vanishes.
Let ${\manyrho(\bar{x}_m,\Delta_m)\equiv \Delta_m^{-1}\int_{-\Delta_m/2}^{\Delta_m/2}\rho(\bar{x}_m+u)\,u^n\,\dd{u}}$
be the $n^\text{th}$ moment of the ${m^\text{th}}$ interval divided by its width ${\Delta_m\sim a}$, 
and let $\bmanyrho(x_b)$ denote the average of ${\manyrho(\bar{x}_m,\Delta_m)}$ over
all intervals in the set that partitions ${\interval(x_b+u,\ell)}$.

Then, if ${\Rho}$ denotes the macroscopic counterpart of ${\rho}$,  and ${\mxl}$ and ${\mxr}$
denote macroscale points in the vacuum beyond the surfaces normal to ${-\hat{x}}$
and ${\hat{x}}$, respectively, 
the areal densities of charge at these surfaces can be expressed as 
\begin{align}
\bsigma_L  & = \int_{\mxl}^{\mxb}\Rho(\bx)\dbx 
\nonumber
\\
 &= \int_{x_L}^{x_b}\rho(x) \dd{x} - \bmonerho(x_b),
\label{eqn:excessone}
\\
\bsigma_R  &= \int_{\mxb}^{\mxr}\Rho(\bx)\dbx
\nonumber 
\\
&=
-\int_{x_b}^{x_R}\rho(x) \dd{x} + \bmonerho(x_b),
\label{eqn:excesstwo}
\end{align}
and the integrals of ${\bx\Rho(\bx)}$ across the surfaces 
can be expressed as
\begin{align}
\int_{\mxl}^{\mxb} \bx \Rho(\bx)\dbx = 
\int_{x_L}^{x_b} x \rho(x)\dd{x} 
&- x_b\bmonerho(x_b)
\nonumber
\\
&-\bmtworho(x_b)
\label{eqn:excessthree}
\\
\int_{\mxb}^{\mxr} \bx \Rho(\bx)\dbx = 
\int_{x_b}^{x_r} x \rho(x)\dd{x} &+ x_b\bmonerho(x_b)
\nonumber
\\
&+\bmtworho(x_b)
\label{eqn:excessfour}
\end{align}
The expression for the charge $\bsigma$ at an interface between two materials can be
deduced from these expressions by assuming that ${\mxbl}$ and ${x_{bL}}$ are macroscale
and microscale points, respectively, in the bulk of one of the materials, and that
${\mxbr>\mxbl}$ and ${x_{bR}>x_{bL}}$ are points in the bulk of the other.
\begin{align}
\bsigma &=  \int_{\mxbl}^{\mxbr} \Rho(\bx)\dbx 
\nonumber
\\
&= 
\int_{x_{bL}}^{x_{bR}}\rho(x)\dd{x} + 
\bmonerho(x_{bL})
-\bmonerho(x_{bR})
\label{eqn:excessfive}
\end{align}
It follows from this expression that the excess of charge at any plane in the bulk
of a macroscopically-uniform material vanishes. The plane can be treated
as an interface and, since the plane itself is in the bulk, the values 
of ${x_{bL}}$ and ${x_{bR}}$ can both be chosen to be the point at which the $x$ axis
intersects the plane. Then the right hand side of Eq.~\ref{eqn:excessfive} vanishes.

\onecolumngrid
\newpage
\twocolumngrid
\section{Fourier transforms, Fourier series, and sets of wavevectors} 
\label{section:appendix_fourier}
This appendix outlines some notation and normalization 
conventions that are used for Fourier transforms and Fourier series.

One simple convention that is used throughout this work
is that if $f$ is a function of space (i.e., one or more positions)
and time, then 
its Fourier transforms in space, time, and both space and time
are denoted by
${\ftsf}$, ${\ftt{f}}$, and ${\ftst{f}}$, respectively.

If $\domain$ is the spatial domain of each component of the position
argument of $f$, the domain of each 
wavevector component of the argument of 
${\ftsf}$ or ${\ftst{f}}$ will be denoted by ${\ftsdomain}$.

It will be assumed that the time domain of $f$ is ${\realone}$;
and the frequency domain of ${\ftt{f}}$ or ${\ftst{f}}$
is also ${\realone}$ (or ${\realnonneg}$ if frequencies
are required to be positive) but
will be denoted by ${\ftt{\realone}}$.

To reduce clutter in expressions for unitary Fourier transforms, 
let ${\displaystyle \fourierconst\equiv 1/\sqrt{2\pi}}$.

\subsection{Fourier transforms with respect to position, time, and both position and time}
Let ${\domain\in\{\realone,\onetorus\}}$; let
${\image\in\{\realone,\complex\}}$; 
and let
\begin{align*}
f:\domain^m\times \realone\to \image; (u,t)\mapsto f(u,t),
\end{align*}
where each value of ${u\in\domain^m}$ specifies one or more positions
and ${t\in\realone}$ is a time variable.

\subsubsection{Fourier transform with respect to position}
The spatial Fourier transform exists  if the function
${f(t):\domain^m\to\image; u\mapsto f(u,t)}$ is an element of ${\lebesgone(\domain^m,\image)}$
at all values of ${t\in\realone}$, i.e., if
\begin{align*}
\int_{\domain^m}\dd{u} \abs{f(u,t)} < \infty,\;\forall t\in\realone.
\end{align*}

The unitary Fourier transform with respect to space is effected by
the operator, 
\begin{align*}
&\fouriernoarg_s:\lebesgone(\domain^m\times\realone)\to\lebesgone(\ftsdomain^m\times\realone); f\mapsto \ftsf\equiv \fourierspace{f},
\end{align*}
where ${\ftsdomain=\realone}$ if ${\domain=\realone}$, and ${\ftsdomain= (2\pi/C)\integer}$ if ${\domain=\onetorus(C)}$.

The unitary Fourier transform of $f$ with respect to position is
the function
\begin{align}
&\fourierspace{f}\equiv\ftsf: \ftsdomain^m\times\realone\to\image,
\nonumber
\\
&(q,t)\mapsto \ftsf(q,t)\equiv \fourierconst^m\int_{\domain^m} \dd{u} f(u,t)e^{-i q\cdot u},
\label{eqn:fts}
\end{align}
where ${q\in\ftsdomain^m\cong \domain^m}$; and
the inverse of Eq.~\ref{eqn:fts} is
\begin{align*}
f(u,t) \equiv \fourierconst^m\int_{\domain^m} \dd{q} \fts{f}(q,t)e^{i q\cdot u}.
\end{align*}

\subsubsection{Fourier transform with respect to time}
The Fourier transform of $f$ with respect to time exists 
if the function
${f(u):\realone\to\image; t\mapsto f(u,t)}$ is an element of ${\lebesgone(\realone,\image)}$
at all values of ${u\in\domain^m}$, i.e., if
\begin{align*}
\int_{\realone}\dd{t} \abs{f(u,t)} < \infty,\;\forall u\in\domain^m.
\end{align*}

The unitary Fourier transform with respect to time is effected 
by the operator
\begin{align*}
&\fouriernoarg_t:\lebesgone(\domain^m\times\realone)\to\lebesgone(\domain^m\times\ftt{\realone}); \; f\mapsto \ftt{f}\equiv \fouriertime{f},
\end{align*}
where ${\ftt{\realone}\cong\realone}$.

The unitary Fourier transform of $f$ with respect to time is the function
\begin{align}
&\fouriertime{f}\equiv\ftt{f}: \domain^m\times\ftt{\realone}\to\image,
\nonumber
\\
&(u,\omega)\mapsto \ftt{f}(u,\omega)\equiv \fourierconst\int_{\realone} \dd{t} f(u,t)e^{i \omega t},
\label{eqn:ftt}
\end{align}
where ${\omega\in\ftt{\realone}\cong \realone}$.
The inverse of Eq.~\ref{eqn:ftt} is
\begin{align*}
f(u,t) \equiv \fourierconst\int_{\ftt{\realone}} \dd{\omega} \ftt{f}(u,\omega)e^{-i \omega t}.
\end{align*}

\subsubsection{Fourier transform with respect to position and time}
The Fourier transform of $f$ with respect to position and time exists 
if ${f\in\lebesgone(\domain^m\times\realone,\image)}$, i.e., if
\begin{align*}
\int_{\domain^m}\dd{u} \int_{\realone}\dd{t} \abs{f(u,t)} < \infty.
\end{align*}

The unitary Fourier transform with respect to position and time is effected 
by the operator
\begin{align*}
&\fouriernoarg_{st}:\lebesgone(\domain^m\times\realone)\to\lebesgone(\ftsdomain^m\times\ftt{\realone}); \; f\mapsto \ftst{f}\equiv \fourierspacetime{f},
\end{align*}
where ${\ftt{\realone}\cong\realone}$;
and ${\ftsdomain=\realone}$ if ${\domain=\realone}$, and ${\ftsdomain= (2\pi/C)\integer}$ if ${\domain=\onetorus(C)}$.

The unitary Fourier transform of $f$ with respect to position and time is the function
\begin{align*}
&\fourierspacetime{f}\equiv\ftst{f}: \ftsdomain^m\times\ftt{\realone}\to\image,
\\
&(q,\omega)\mapsto \ftst{f}(q,\omega)\equiv \fourierconst^{m+1}\int_{\domain^m}\dd{u}\int_{\realone} \dd{t} f(u,t) e^{-i (q\cdot u-\omega t)},
\end{align*}
where ${\omega\in\ftt{\realone}}$; and the 
expression for the inverse transform is
\begin{align*}
f(u,t) \equiv \fourierconst^{m+1}\int_{\ftsdomain^m}\dd{q}\int_{\ftt{\realone}} \dd{\omega} \ftst{f}(q,\omega)e^{i (q\cdot u-\omega t)}.
\end{align*}

\subsection{Wavevectors compatible with $\volume$ and $\bulksize$ periodicities}
Definitions of the wavevector differentials, ${\hbulksize}$ and ${\hreciplatt}$, and the sets of wavevectors, ${\reciplattg}$, $\BZ$, and ${\reciplatt}$, follow.
Although all five are determined by the values of one or both of ${\bulksize}$ and $\volume$,
these dependences will not usually be made explicit.
For example, to make mathematical expressions less cluttered, ${\hbulksize(\bulksize)}$ will usually be denoted as ${\hbulksize}$.

It will sometimes be important to be conscious of the implicit dependences of ${\hbulksize}$, ${\hreciplatt}$, 
${\reciplattg}$, $\BZ$, and ${\reciplatt}$ on ${\bulksize}$ and/or $\volume$.

The smallest wavevectors that are compatible with $\bulksize$-periodicity have magnitude
\begin{align*}
\hbulksize=\hbulksize(\bulksize)\equiv 2\pi/\bulksize,
\end{align*}
and the set of all wavevectors that are compatible with
${\bulksize}$-periodicity is
\begin{align*}
\reciplattg=\reciplattg(\bulksize)\equiv \hbulksize\integer\equiv\left\{g=m_g\hbulksize:m_g\in\integer\right\}.
\end{align*}

The subset of ${\reciplattg}$ that comprises all 
wavevectors in the first \emph{Brillouin zone} that are
compatible with $\bulksize$-periodicity is
\begin{align*}
\BZ=\BZ(\bulksize,\volume)\equiv \left\{ k \in \reciplattg : -\frac{\pi}{\volume} < k \leq \frac{\pi}{\volume}\right\}.
\end{align*}

The magnitude of the smallest wavevectors that are compatible with
$\volume$-periodicity is 
\begin{align*}
\hreciplatt=\hreciplatt(\volume)\equiv 2\pi/\volume;
\end{align*}
and the \emph{reciprocal lattice},
\begin{align*}
\reciplatt=\reciplatt(\volume)\equiv \hreciplatt\integer,
\end{align*}
is the set of all wavevectors that are compatible with
${\volume}$-periodicity.

Every element of ${\reciplattg}$ can be expressed as the sum of
a reciprocal lattice vector and an element of $\BZ$, i.e., 
\begin{align*}
\reciplattg = \left\{g=G+k : G\in\reciplatt,\; k \in\BZ\right\}.
\end{align*}

\onecolumngrid
\vspace{0.8cm}
\PRLsep
\vspace{1cm}
\twocolumngrid
\section{Statistical states as vectors}
\label{section:appendix_states_as_vectors}
This appendix shows how the position probability density 
function ${\pdf}$ 
of a set of particles can be represented as an element
of a Hilbert-Lebesgue space. It is straightforward to adapt it to
the probability density function $\pdfO$ of an arbitrary observable, ${\Obs}$.

This appendix deviates from the convention 
that ${x}$ represents a single coordinate of a single particle's position vector
(see Sec.~\ref{section:notation}).
In this appendix ${x\in\domain}$ is a set of ${\Ndof}$ of coordinates. Therefore it is a specification of
the microstructure of a system with $\Ndof$ degrees of freedom.

For simplicity, it will be assumed that
${\domain\subset\realone^{\Ndof}}$, where ${0<\abs{\domain}<\infty}$.
It will also be assumed that the precisions of all measurements are finite.

\subsection{Statistical states as rays}
\label{section:states_as_rays}
The position pdf of an arbitrary system of ${\Nparticle=\Ndof/\dimension}$ classical
particles confined to ${\domain}$ is 
\begin{align*}
\pdf:\domain\to\realpos; x\mapsto \pdf(x), 
\end{align*}
where ${x\in\domain}$ and the integral of ${\pdf(x)}$ over the set $\domain$ of all microstructures is unity. 

The information possessed by $\pdf$
can be specified by a square-integrable function,
\begin{align*}
\Psi(x) = r e^{i\theta(x)}\sqrt{\pdf(x)}\in \lebesgue(\domain),
\end{align*}
where ${r\in\realpos}$ is arbitrary and finite, and ${\theta:\domain\to\realone}$ is arbitrary 
and may be constant.
The position pdf can be retrieved from $\Psi$ as follows:
\begin{align}
\pdf(x) = \frac{\Psi^*(x)\Psi(x)}{\int_\domain\Psi^*(x')\Psi(x')\dd{x'}} = r^{-2}\abs{\Psi(x)}^2.
\label{eqn:wfn_to_pdf}
\end{align}
Since $r$ is an arbitrary positive constant and ${\theta}$ is an arbitrary
function, there is a one-to-one map  between $\Nparticle$-particle position pdfs $\pdf$ and
equivalence classes of elements $\Psi$ of ${\lebesgue(\domain)}$.

\subsubsection{Exchange symmetry or antisymmetry of $\Psi$}
\label{section:appendix_symmetry}
As discussed in Sec.~\ref{section:antisymmetry}, if ${x}$ specifies 
the configuration (positions) 
of a set of identical particles, and if the particles
move so fast within the same region of space that it is impossible to 
follow their individual trajectories, then their identicality
makes them indistinguishable. To reflect this indistinguishability, 
$\pdf$ must be invariant under exchange of any two of the particles' positions.
When that is the case, $\Psi$ must 
be either exchange symmetric or exchange antisymmetric
to preserve the exchange symmetry of $\pdf$.

Under the additional assumption that 
two or more particles cannot occupy the same point in space, $\Psi$
must vanish at \emph{coincidence points}, which are points in $\domain$ that
represent configurations in which two or more particles coincide. 

If $\Psi$ is exchange symmetric, it is not differentiable at a coincidence point
unless its derivatives with respect to the positions of the coincident particles
vanish. However it can be finite-difference differentiable, while preserving 
the exchange symmetry of $\pdf$, if it is exchange-antisymmetric
and changes sign at coincidence points.

\subsection{Statistical states as vectors}
\label{section:states_as_vectors}
A one-dimensional lattice, with spacing ${a\in\realpos}$, is denoted by ${\alattx\integer}$. 
Therefore ${\discreal\equiv\left(\alattx\integer\right)^{\Ndof}}$ is a ${\Ndof}$-dimensional hypercubic 
lattice whose `volume' per lattice
site will be denoted by ${\hilbv\equiv \alattx^{\Ndof}}$. That is, $\hilbv$ is the measure in ${\realone^{\Ndof}}$
of the set ${\N_x\subset\realone^{\Ndof}}$ of points that are closer in ${\realone^{\Ndof}}$
to the element ${x}$ of ${\discreal}$ than to any other element of $\discreal$.

Now let ${\discreal\equiv (\alattx\integer)^{\Ndof}\cap\domain}$, and let us define a function
${\sqhilbv\tPsiv:\discreal\to\complex}$ 
whose square modulus divided by $\hilbv$ is
\begin{align}
\abs{\tPsiv(x)}^2 = \frac{1}{\hilbv}\int_{\N_x} \abs{\Psi(u)}^2 \dd{u} 
= \frac{1}{\hilbv}\int_{\N_x} \pdf(u) \dd{u},
\end{align}
where $\Psi$ is defined as in subsection~\ref{section:states_as_rays}. 
This definition of ${\tPsiv}$ implies that ${\pdf(x)=\lim_{\hilbv\to 0}\abs{\tPsiv(x)}^2}$ and 
that 
${\abs{\sqhilbv\tPsiv(x)}^2}$ is the probability that the true configuration of the system, $x_t$,
is closer to $x$ than to any other element of ${\discreal}$.
Therefore, ${\sqhilbv\tPsiv}$ is a square root of a \emph{probability mass function}, and ${\abs{\sqhilbv\tPsi(x)}^2}$
becomes ${\hilbv \pdf(x)}$ in the limit ${\hilbv\to 0}$.

Since $\discreal$ is a finite lattice, ${\sqhilbv\tPsiv}$ can 
be specified by a finite set of complex numbers, ${\left\{\sqhilbv\tPsiv(u)\right\}_{u\in\discreal}}$. 
This set is square-summable, because ${\hilbv\abs{\tPsiv(x)}^2}$ is a probability mass function, i.e., 
\begin{align*}
\sum_{u\in\discreal} \abs{\tPsiv(u)}^2\hilbv =\sum_{u\in\discreal}\tPsiv^*(u)\tPsiv(u)\hilbv =1< \infty,
\end{align*}
which implies that ${\sqhilbv\tPsiv\in\seqlebesgue(\discreal)}$,
where ${\seqlebesgue(\discreal)=\seqlebesgue(\discreal,\complex)}$ is the Hilbert-Lebesgue space of
square-\emph{summable} functions ${\discreal\to\complex}$.

Any probability mass function, ${\abs{\sqhilbv f}^2}$, for the identity
${x\in\discreal}$ of the neighbourhood ${\N_x}$ containing ${x_t}$ 
can be represented by one of its square roots, 
${\sqhilbv f\in\seqlebesgue(\discreal,\complex)}$. This square
root can be represented by
a vector ${\ket{\sqhilbv f}\equiv \sqhilbv\ket{f}}$ in a vector space ${\hilbertv}$ whose
`inner product' (see Sec.~\ref{section:inner_product} to interpret the quotes) is defined by
\begin{align*}
\braket{\sqhilbv f}{\sqhilbv g} \equiv \sum_{u\in\discreal}f^*(u)g(u)\epsilon.
\end{align*}
One reason for choosing this form for the inner product is that in
the dense-sampling limit ${\alattx\to 0 \implies \hilbv\to 0}$, it becomes
the Riemann integral,
\begin{align*}
\lim_{\hilbv\to 0}\braket{\sqhilbv f}{\sqhilbv g}=\int_{\domain} f^*(u)g(u)\dd[\Ndof]{u}.
\end{align*}
To achieve this form for the inner product, let us introduce the 
set 
\begin{align*}
\left\{\phix(\hilbv)\in\seqlebesgue(\discreal): x\in\discreal\right\}, 
\end{align*}
where 
\begin{align*}
\sqhilbv\phix(\hilbv):\discreal\to\complex; u\mapsto \sqhilbv\phix(u;\hilbv), 
\end{align*}
and ${\abs{\sqhilbv\phix(\hilbv)}^2}$ is a probability mass function 
which vanishes everywhere except at ${x\in\discreal}$, i.e., 
${\abs{\sqhilbv\phix(u;\hilbv)}^2=\Pr(x_t\in\N_u)}$ vanishes
if ${u\neq x}$  and is unity if ${u=x}$.
This implies that ${\phix(u;\hilbv)}$ vanishes if ${u\neq x}$ 
and that ${\phix(x;\hilbv)=e^{i\theta_x}}$, for some ${\theta_x\in\realone}$.
For simplicity we choose ${e^{i\theta_x}}$ to be unity (${\implies\theta_x=0}$) for every ${x\in\discreal}$.
To simplify notation, I will not make the dependence of $\phix$ on parameter $\hilbv$ explicit in much of what follows, and
I will use ${\ket{\sqhilbv x}}$ rather than ${\ket{\sqhilbv \phix(\hilbv)}}$ to denote the element of $\hilbertv$ that
represents ${\sqhilbv \phix(\hilbv)}$.
I will refer to 
${\{\sqhilbv\ket{u}\}_{u\in\discreal}}$ as the
\emph{microstructure basis set}. 

The element ${\ket{\sqhilbv f}}$ of ${\hilbertv}$ representing a square-summable function ${\sqhilbv f}$ can be 
expressed in terms of this basis set as follows,
\begin{align}
\ket{\sqhilbv f} = \sum_{x\in\discreal}\sqhilbv f(x)\left(\sqhilbv\ket{x}\right) = \sum_{x\in\discreal} f(x)\hilbv\ket{x},
\label{eqn:hilbertvec}
\end{align}
where ${\ket{\sqhilbv x}}$ is the element of ${\hilbertv}$ representing ${\sqhilbv\phix}$, i.e., 
\begin{align*}
\ket{\sqhilbv x} = \sum_{u\in\discreal} \phix(u)\sqhilbv\left(\sqhilbv\ket{u}\right) = \sum_{u\in\discreal}\phix(u)\hilbv\ket{u}.
\end{align*}
This implies that the inner product of ${\ket{\sqhilbv x}}$ and ${\ket{\sqhilbv x'}}$ can be expressed as
\begin{align*}
\braket{\sqhilbv x}{\sqhilbv x'}=\hilbv\braket{x}{x'}=\sum_{u,v\in\discreal}\phi_x^*(u)\phi_{x'}(v)\hilbv^2\braket{u}{v},
\end{align*}
where ${\phi_x^*(u)\phi_{x'}(v)}$ vanishes unless ${u=v=x=x'}$, in which
case its value is ${1/\hilbv}$. Therefore ${\braket{x}{x'}=0}$ if
${x\neq x'}$, and 
\begin{align*}
\braket{x}{x}=
\sum_{u\in\discreal}\abs{\phix(u)}^2\hilbv\braket{u} = \abs{\phix(x)}^2\hilbv\braket{x} =\braket{x}.
\end{align*}
These choices do not determine the norm of ${\ket{x}}$, so let us choose it to be
\begin{align*}
\norm{\ket{x}}_2\equiv\sqrt{\braket{x}}=\frac{1}{\sqrt{\hilbv}} \implies \braket{x}=\frac{1}{\hilbv},
\end{align*}
which means that the norm of ${\ket{\sqhilbv x}\equiv\sqhilbv\ket{x}}$ is one and that the microstructure
basis, ${\{\sqhilbv\ket{u}\}_{u\in\discreal}}$, is orthonormal.
In other words, ${\ket{x}}$ is not a unit vector, but ${\sqrt{\hilbv}\ket{x}}$ is a unit vector.

We can express ${\ket{f}}$ in terms of the microstructure basis as follows:
\begin{align*}
\ket{f} = \frac{1}{\sqhilbv}\ket{\sqhilbv f} &= 
\frac{1}{\sqhilbv}\sum_{x\in\discreal}\braket{\sqhilbv x}{\sqhilbv f}\sqhilbv\ket{x}
\\
&=\sum_{x\in\discreal}\braket{x}{f}\hilbv\ket{x}.
\end{align*}
By comparison with Eq.~\ref{eqn:hilbertvec}, this implies that 
${\braket{x}{f} = f(x)}$.
Therefore, the element of $\hilbertv$ that represents 
${\ket{\sqhilbv\tPsiv}\equiv\sqhilbv\ket{\tPsiv}}$ is
\begin{align}
\ket{\sqhilbv \tPsiv}\equiv \sqhilbv \sum_{x\in\discreal}\tPsiv(x)\hilbv\ket{x};
\nonumber
\label{eqn:hilbertpsi}
\end{align}
and the element that represents ${\ket{\tPsiv}}$ is
\begin{align}
\ket{\tPsiv}=  \sum_{x\in\discreal}\tPsiv(x)\hilbv\ket{x}.
\end{align}

From this point forward let us assume that ${\hilbv}$ is small enough that the variations of
${\Psi}$ and $\pdf$ between any two neighbouring points of $\discreal$ are negligible; and 
the effects on all theoretical calculations of making it even smaller are negligible.
When it is this small, 
let us denote ${\ket{\tPsiv}}$, ${\tPsiv}$, and ${\hilbertv}$ by ${\ket{\psi}}$, ${\psi}$, and ${\hilbert}$, respectively;
and let us express ${\ket{\tPsiv}\equiv \sum_{x\in\discreal}\tPsiv(x)\hilbv\ket{x}}$ as
\begin{align}
\ket{\psi} \equiv \int_\domain \psi(x)\ket{x}\dd{x},
\label{eqn:wfn_to_vector}
\end{align}
where ${\dd{x}\equiv\hilbv}$ and ${\psi(x)\equiv \braket{x}{\psi}}$. In other  words, from now on it is implicit that ${\int_\domain \ldots\dd{x}}$ really
means ${\sum_{x\in\discreal}\ldots\hilbv}$.
Under this convention, we can express the identity in $\hilbert$ as 
\begin{align*}
\identity = \sum_{x\in\discreal} \dyad{\sqhilbv x}=\int_\domain \dyad{x}\dd{x}.
\end{align*}
That is,
\begin{align*}
\identity\ket{\psi} = \left(\int_\domain\dyad{x}\dd{x}\right)\ket{\psi} = \int_\domain \braket{x}{\psi}\ket{x} \dd{x} = \ket{\psi}
\end{align*}
and
\begin{align*}
\identity\ket{x} = \left(\int_\domain\dyad{x'}\dd{x'}\right)\ket{x} = \int_\domain \braket{x'}{x}\ket{x'} \dd{x'} = \ket{x}.
\end{align*}
All of the Hilbert-Lebesgue spaces in this work that are denoted by $\lebesgue$ rather than $\seqlebesgue$, 
should be regarded as having been constructed in the same way as $\hilbert$. 
The notation $\lebesgue$ is used despite the fact that the construction of ${\hilbert}$ above
implies that ${\hilbv}$ is finite, albeit arbitrarily small; and despite the fact that
${\intdomain\cdots\dd{x}}$ is defined as a Riemann integral. 
These aspects of the definition of $\hilbert$ imply that, 
strictly speaking, it is a space of functions whose domain is countable.
Therefore it is (an abstract representation of) a ${\seqlebesgue}$-space.
Nevertheless, the notation $\lebesgue$ is used throughout this work for spaces defined in this way. A function space with a $2$-norm 
is only denoted by ${\seqlebesgue}$ when it is important to emphasize that its elements have a countable domain, as was the
case for ${\seqlebesgue(\discreal)}$ above.

It may be troubling to some that subtleties 
appear to have been swept under a rug by using
the finite precisions of all measurements
to justify quantizing the domain of $\pdf$.
A more careful justification of this step is presented 
in Section~3 of \linecite{tangney_bose_einstein}.

\subsection{The space of Fourier transforms of elements of ${\lebesgue(\domain)}$}
Any function ${\psi\in\lebesgue(\domain)}$ can be Fourier transformed with respect to each of the $\Ndof$ coordinates
specified by points ${x\in\domain}$, as follows:
\begin{align*}
\ftspsi(k)&\equiv \fourierconst^N \intftsdomain\dd{x} \psi(x) e^{-ikx}
\implies \psi(x)= \fourierconst^N \intdomain\dd{k} \ftspsi(k) e^{ikx},
\end{align*}
where ${\fourierconst\equiv 1/\sqrt{2\pi}}$, and  ${\ftsdomain\equiv \domaindual/\fourierconst^2}$, where
\begin{align*}
\domain^\ast\equiv \left\{ \adbmal\equiv k/(2\pi):k\in\ftsdomain\right\}
\end{align*}
is the \emph{dual space} of ${\domain}$. In other words, ${\domain}$ is a vector
space over field ${\realone}$, and ${\domain^\ast}$ is
the set of all linear maps, 
\begin{align*}
(\adbmal,\cdot):\domain\to\realone;\; x\mapsto (\adbmal,x)\equiv \;\adbmal x.
\end{align*}

There is nothing within the construction of ${\lebesgue(\domain)}$ and ${\hilbert}$ 
in subsection~\ref{section:states_as_vectors} that would make it inapplicable
to space $\ftsdomain$ or space ${\domaindual}$. Therefore let us use an analogous 
construction to define
the Hilbert-Lebesgue space ${\lebesgue(\ftsdomain)}$.
Then, since the Fourier transform ${\ftsf}$ of any  ${f\in\lebesgue(\domain)}$ 
exists, and is uniquely defined by $f$; and since \emph{Plancherel's theorem}~\citep{decay_of_fourier,strichartz_distribution_theory} states that
\begin{align*}
\intftsdomain \ftsf^*(k)\ftsf(k)\dd{k}=
\intdomain f^*(x) f(x)\dd{x} = 
\norm{f}_2^2<\infty,
\end{align*}
${\ftsf}$ is an element of ${\lebesgue(\ftsdomain)}$ and 
there is a one to one correspondence between elements of ${\lebesgue(\domain)}$ and elements of 
${\lebesgue(\ftsdomain)}$, which implies
that there is a one to one correspondence between
elements of ${\lebesgue(\ftsdomain)}$ and elements of ${\hilbert}$.

Furthermore, \emph{Parseval's theorem}~\citep{decay_of_fourier,strichartz_distribution_theory} states that
the Fourier transforms ${\ftsf,\ftsg\in\lebesgue(\ftsdomain)}$ of
any ${f,g\in\lebesgue(\domain)}$ satisfy
\begin{align*}
\braket{f}{g}\equiv\intdomain f^*(x)g(x) \dd{x} = \intftsdomain\ftsf^*(k)\ftsg(k)\dd{k}.
\end{align*}
Therefore, instead of introducing a second abstract Hilbert space, ${\hilbert}$
can play the same role for ${\lebesgue(\ftsdomain)}$ that it plays for ${\lebesgue(\domain)}$:

Recall that each of the elements
elements ${\ket{x}}$ of ${\hilbert}$ represents a function ${\phix\in\lebesgue(\domain)}$
which is localized in a neighbourhood of the point ${x\in\domain}$, whose measure is ${\hilbv}$.
By following the same procedure in spaces $\ftsdomain$ and ${\lebesgue(\ftsdomain)}$, 
let us define a set of functions ${\phik\in\lebesgue(\ftsdomain)}$, 
each of which is localized within a different element of a partition of ${\ftsdomain}$. Let us choose the partition 
such that the measure
of each of its elements is ${\ftshilbv\equiv 1/(\fourierconst^{2\Ndof}\hilbv)}$; and let us denote the element of $\hilbert$ that
represents the function ${\phik}$ that is localized
within the partition centered at ${k\in\ftsdomain}$ by ${\ket{k}}$, 
where 
\begin{align*}
\norm{\ket{k}}_{\hilbert}=\norm{\phik}_2=\frac{1}{\sqrt{\ftshilbv}} = \fourierconst^\Ndof\sqrt{\hilbv}.
\end{align*}
Since set ${\{\ket{k}\}}$ is a complete orthonormal basis of ${\ftsdomain}$, 
any ${\ftsf\in\lebesgue(\ftsdomain)}$ is represented  in $\hilbert$ 
by a vector that can be expressed as
\begin{align}
\myket{\ftsf}
&\equiv \intftsdomain \dd{k} \ftsf(k)\ket{k} 
= \fourierconst^\Ndof \intftsdomain\dd{k}\intdomain\dd{x} f(x) e^{-ikx} \ket{k}
\nonumber
\\
& = \intdomain\dd{x} f(x) \left(\fourierconst^\Ndof\intftsdomain\dd{k} e^{-ikx}\ket{k}\right).
\label{eqn:ftsf1}
\end{align}
Now, if ${\ftsf}$ and ${f}$ are to be represented by the same element of $\hilbert$,
this must be equal to
${\ket{f}  = \intdomain\dd{x} f(x) \ket{x}}$.
Therefore, 
\begin{align*}
\ket{x} 
&=\fourierconst^\Ndof\intftsdomain\dd{k}e^{-ikx}\ket{k},
\end{align*}
which implies that
\begin{align*}
\braket{k}{x} = \fourierconst^\Ndof e^{-ikx} = \frac{e^{-ikx}}{\sqrt{(2\pi)^\Ndof}} = \braket{x}{k}^*.
\end{align*}

\onecolumngrid
\vspace{0.8cm}
\PRLsep
\vspace{1cm}
\twocolumngrid
\section{Expectation values of observables}
\label{section:expectation_values}
This appendix deviates from the convention introduced in Sec.~\ref{section:notation}
by using $x$ to denote a set of $\Ndof$ coordinates.

\subsection{Preliminaries}
Let $\configspace$ denote the \emph{configuration space}
of a physical system with $\Ndof$ degrees of freedom.
In other words, ${\configspace}$ is the set of all possible
microstructures of the system, $x$.

Let ${\pdf:\configspace\to\realpos}$, ${x\mapsto\pdf(x)}$ be a probability
density function. Then, 
if ${\varphi\in\lebesgue(\configspace)}$ is any function
\begin{align*}
\varphi:\configspace\to\complex, x\mapsto \varphi(x)\equiv \sqrt{\pdf(x)}e^{i\theta(x)},
\end{align*} 
where ${\theta:\configspace\to\realone}$ is arbitrary for now, 
it specifies ${\pdf=\varphi^*\varphi}$. Therefore $\varphi$ specifies
the statistical state specified by $\pdf$. 

The state can also be represented by the element, 
\begin{align*}
\ket{\varphi}\equiv\intconfig \dd{x}\varphi(x)\ket{x}\in\hilbert, 
\end{align*}
of an abstract Hilbert space ${\hilbert}$, whose elements are in
one-to-one correspondence with elements of ${\lebesgue(\configspace)}$.

Vector space ${\hilbert}$ and vector 
${\ket{\varphi}}$ are defined by a construction analogous to the one
outlined in Appendix~\ref{section:appendix_states_as_vectors}:
Briefly, the integral ${\intconfig\cdots\dd{x}}$ is really 
the discrete Riemann sum ${\hilbv\sum_{x\in\discreal}}$ over a partition 
of $\configspace$. The elements of the partition are all simply-connected subsets of $\configspace$,
with the same arbitrarily-small \emph{finite} measure $\hilbv$, 
centered at points on a lattice ${\discreal=\discreal(\hilbv)}$.
Therefore ${\hilbert}$ is a finite dimensional vector space, but its dimension is
arbitrarily large: It increases monotonically as ${\hilbv\in\realpos}$
decreases, and it diverges in the limit ${\hilbv\to 0^+}$.

It will be assumed in this appendix that all functions and vectors that specify statistical
states are normalized to one, i.e., 
\begin{align*}
\norm{\varphi}_2^2\equiv\braket{\varphi}=1.
\end{align*}
The vector ${\ket{x}}$ does not represent statistical state. It represents
a smooth function that (almost) vanishes everywhere
except in the element of the partition of $\configspace$ that is centered at ${x\in\discreal}$, 
where its (almost) constant value is ${1/\sqrt{\hilbv}}$.
Therefore, for any function ${f:\configspace\to\complex}$, 
\begin{align*}
\intconfig\dd{x} f(x)\braket{x}{x'} = \hilbv\sum_{x\in\discreal} f(x)\braket{x}{x'} = f(x').
\end{align*}
For all practical purposes, ${\braket{x}{x'}}$ is the \emph{Dirac delta function}, ${\delta(x-x')}$.

\subsection{An observable and its expectation value}
Let ${\Obs:\configspace\to\realone,\;x\mapsto\Obs(x)}$
be an observable whose value is determined by the microstructure.
Its expectation value in the statistical state specified
by $\pdf$ or $\varphi$ is
\begin{align*}
\expval{\Obs}=\expOp[\pdf]
&\equiv \intconfig \pdf(x)\Obs(x)\dd{x}
\\
& =
\intconfig \varphi^*(x) \Obs(x)\varphi(x)\dd{x}.
\end{align*}
Since ${O(x)\in\realone, \;\forall x\in\configspace}$, the operator 
${\hO:\hilbert\to\hilbert}$  defined by 
\begin{align*}
\hO \equiv \intconfig\dd{x} \Obs(x)\dyad{x}
\end{align*}
is a \emph{Hermitian} or \emph{self-adjoint} operator,
i.e., 
\begin{align*}
\hO^\dagger 
= \intconfig\dd{x} \Obs^*(x)\left(\dyad{x}\right)^\dagger
= \intconfig\dd{x} \Obs(x)\dyad{x} = \hO.
\end{align*}
The result of its action on ${\ket{\varphi}}$ is
\begin{align*}
\hO\ket{\varphi} 
&= \left(\intconfig\dd{x}\Obs(x)\dyad{x}\right)\left(\intconfig\dd{x'}\varphi(x')\ket{x'}\right)
\\
& = \intconfig\dd{x}\Obs(x)\ket{x}\left(\intconfig\dd{x'}\varphi(x')\braket{x}{x'}\right) 
\\
&=\ket{\Obs\varphi} \equiv \intconfig\dd{x} \Obs(x)\varphi(x)\ket{x} = \bra{\Obs\varphi}^\dagger.
\end{align*}
Therefore ${\expval{\Obs}}$ can be expressed as
\begin{align*}
\expval{\Obs}
&=\expval{\hO}{\varphi}=\braket{\varphi}{\Obs\varphi}=\braket{\Obs\varphi}{\varphi}
\\
&= \left(\intconfig\dd{x}\varphi^*(x)\bra{x}\right)\left(\intconfig\dd{x'}\Obs(x')\varphi(x')\ket{x'}\right)
\\
& = \intconfig\dd{x}\varphi^*(x)\Obs(x)\varphi(x).
\end{align*}
In other words, ${\expval{\Obs}}$ is the real-valued
inner product ${\braket{\varphi}{\Obs\varphi}=\braket{\Obs\varphi}{\varphi}}$ of ${\varphi}$ and
the function
\begin{align*}
\Obs\varphi:\configspace\to\complex; \;x\mapsto \Obs(x)\varphi(x).
\end{align*}

\subsection{Stationary states}
In this subsection, with generalizations that will be introduced in later
subsections in mind, I will sometimes denote ${\Obs(x)}$ by ${\hOx}$;
and the expectation
value of ${\Obs}$ in a unit-normalized state ${\ket{\varphi}}$ will 
sometimes be denoted by
${\obs[\varphi]}$. i.e., 
\begin{align*}
\obs[\varphi]\equiv\expval{\hO}{\varphi} = \mybraket{\varphi}{\hOx\varphi}.
\end{align*}
A necessary condition for this expectation value to be stationary with respect to
variations of $\varphi$ that conserve total probability
is ${\delta\functionalarg{\Obs}[\varphi;\eta]=0, \forall\eta\in\lebesgue(\configspace)}$, where
\begin{align*}
\functionalarg{\Obs}[\varphi]\equiv \obs[\varphi]+\lagrangearg{\Obs}\braket{\varphi};
\end{align*}
${\lagrangearg{\Obs}}$ is a Lagrange multiplier; and 
${\delta\functionalarg{\Obs}[\varphi;\eta]}$ is the
Gateaux derivative of ${\functionalarg{\Obs}}$ in direction $\eta$ at
${\varphi}$. That is, 
\begin{align*}
\delta\functionalarg{\Obs}[\varphi;\eta]
&= \dv{\zeta}\functionalarg{\Obs}[\varphi+\zeta\eta]\eval_{\zeta=0}
\\
&=\lim_{\zeta\to 0}\frac{\functionalarg{\Obs}[\varphi+\zeta\eta]-\functionalarg{\Obs}[\varphi]}{\zeta}=0.
\end{align*}

\subsubsection{Stationary states are eigenstates}
Let ${\varphi_r\equiv\Re\{\varphi\}}$ and ${\varphi_i\equiv\Im\{\varphi\}}$;
and let use choose ${\eta}$ to be an element of a real-valued complete
basis of ${\lebesgue(\configspace)=\lebesgue(\configspace;\complex)}$.
Then,
\begin{align}
2\Re\left\{\delta\functionalarg{\Obs}[\varphi;\eta]\right\}
&=
\mel{\varphi}{\hO}{\eta}
+
\mel{\eta}{\hO}{\varphi}
-\lagrangearg{\Obs}\left[\braket{\varphi}{\eta}+\braket{\eta}{\varphi}\right] 
\nonumber
\\
& = 2\mel{\eta}{\hO}{\varphi_r} - 2\lagrangearg{\Obs}\braket{\eta}{\varphi_r}=0
\nonumber
\\
\implies &\mel{\eta}{\hO}{\varphi_r} =\lagrangearg{\Obs}\braket{\eta}{\varphi_r};
\label{eqn:real_stationary}
\end{align}
and
\begin{align}
2i\Im\left\{\delta\functionalarg{\Obs}[\varphi;\eta]\right\}
&= 
\mel{\varphi}{\hO}{\eta}
-
\mel{\eta}{\hO}{\varphi}
-\lagrangearg{\Obs}\left[\braket{\varphi}{\eta}-\braket{\eta}{\varphi}\right] 
\nonumber
\\
& = 2i\mel{\eta}{\hO}{\varphi_i} - 2i\lagrangearg{\Obs}\braket{\eta}{\varphi_i}=0
\nonumber
\\
\implies
& \mel{\eta}{\hO}{\varphi_i} = \lagrangearg{\Obs}\braket{\eta}{\varphi_i}=0.
\label{eqn:imag_stationary}
\end{align}
Since ${\eta\in\lebesgue(\configspace;\realone)\subset\lebesgue(\configspace)}$ 
is an arbitrary element of a complete basis,
Eqs.~\ref{eqn:real_stationary} and~\ref{eqn:imag_stationary} must 
hold for every element of the basis. Therefore
the functions ${\hOx\!\varphi_r,\hOx\!\varphi_i\in\lebesgue(\configspace;\realone)}$ 
can be expressed as
\begin{align*}
\hOx\!\varphi_r & = \sum_\eta \mel{\eta}{\hO}{\varphi_r}\eta
 = \sum_\eta \braket{\eta}{\varphi_r}\eta 
= \lagrangearg{\Obs}\varphi_r,
\\
\hOx\!\varphi_i 
& = \sum_\eta \mel{\eta}{\hO}{\varphi_i}\eta
 = \sum_\eta \braket{\eta}{\varphi_i}\eta 
= \lagrangearg{\Obs}\varphi_i,
\end{align*}
where the sums are over all elements of the basis.
It follows that
\begin{align*}
\hOx\!\varphi=
\hOx(\varphi_r+i\varphi_i)=\lagrangearg{\Obs}\varphi
\implies\hO\ket{\varphi}=\lagrangearg{\Obs}\ket{\varphi}.
\end{align*}
Therefore $\varphi$ and ${\ket{\varphi}}$ are normalized eigenstates
of ${\hOx}$ and ${\hO}$, respectively, if and only if they
are states at which the expectation value ${\obs[\varphi]=\expval{\hO}{\varphi}}$
is stationary with respect to variations that preserve their normalizations.


\subsection{Symmetry}

\subsubsection{Degeneracies imply symmetries}
\label{section:degeneracies_imply_symmetries}
Let us assume that the set 
\begin{align*}
\odomain\equiv\left\{\expval{\hO}{\varphi}:\ket{\varphi}\in\hilbert,\;\braket{\varphi}=1\right\}
\end{align*}
of all possible values of ${\expval{\Obs}}$
is $\realone$ or a connected subset of $\realone$, such as an interval
or ${\realnonneg}$.

Let $\varphi_i$ and ${\varphi_j\neq\varphi_i}$, be two different stationary states
of $\Obs$, whose stationary values are 
${\obs_i \equiv \obs[\varphi_i]}$
and
${\obs_j \equiv \obs[\varphi_j]}$, respectively. 
Let us also assume that ${\obs_i}$ and ${\obs_j}$ are not \emph{both }
boundary points of ${\odomain}$.
Then, if ${\deltaO>0}$ is sufficiently small, the probability
\begin{align*}
\Pr\left(\abs{\obs_i-\obs_j}< \deltaO\right), 
\end{align*}
decreases as ${\deltaO}$ decreases, and it vanishes in the limit ${\deltaO\to 0^+}$. 
In other words, it is improbable that, by chance, two stationary values turn out to be very close
in value; and it is impossible that they turn out to be exactly equal (i.e., to infinite precision)
by chance.

Since ${\obs_i}$ and ${\obs_j}$ cannot be equal by chance, 
${\obs_i=\obs_j}$ implies that the stationary states specified by $\varphi_i$ and $\varphi_j$ 
are equivalent to one another by a symmetry of the system.

\color{red}
\vspace{2cm}
\begin{center}
SEVERAL SUBSECTIONS ARE MISSING \\ (mid rewrite)
\end{center}
\vspace{2cm}
\color{black}

\color{black}

\onecolumngrid
\vspace{0.8cm}
\PRLsep
\vspace{1cm}
\twocolumngrid
\section{Slow measurements of small or sensitive systems}
\label{section:appendix_measurement}
The thesis on which most of this work is built is that what is observed at the macroscale is 
homogenized \emph{microstructure}. The microstructure, in its most detailed and specific form, 
can be expressed as a pdf, 
\begin{align*}
\cfgdist:\configspace\to\realnonneg; \; x\mapsto \cfgdist(x), 
\end{align*}
where $x$ specifies the \emph{configuration} of 
an unimaginably-large number of degrees of freedom, $\Ndof$. For almost
all purposes, the degrees of freedom can be regarded as 
the ${\dimension=3}$ coordinates of ${\Nparticle=\Ndof/\dimension}$ particles.
Therefore each configuration $x$ is a precise specification of all
of the particles' positions, and ${\configspace\subseteq\realone^{\Ndof}}$ is the set of all accessible
configurations.

Some might argue that a classical \emph{microstate} has 
the mathematical form of a point ${\Gammat\equiv (\pi_t, x_t)}$ 
in the particles' phase space, ${\phasespace\cong\momspace\times\configspace}$,
where ${\pi_t\in\momspace\subseteq\realone^{\Ndof}}$ is a specification of the $\Ndof$ momenta 
conjugate to the coordinates specified by $x_t$, and $\momspace$ is the set of accessible momenta. 
This is true, strictly speaking, but if classical particles had
charges, masses, and separation distances comparable to those of electrons 
and nuclei, the average magnitudes of their accelerations would be enormous. 
Therefore they would move so fast that it would be impossible to observe the particles
in a particular microstate;  or even to observe, or measure properties of, 
short segments of their trajectory. 

For example, in diamond, the average number density of electrons 
is approximately ${\SI{1}{\per\angstrom\cubed}=\SI{e30}{\per\meter\cubed}}$, and 
the shortest phonon period is approximately ${\SI{25}{\femto\second}=\SI{2.5e-14}{\second}}$~\citep{peckham_1967}.
If electrons were classical particles, the Coulomb repulsion between 
two electrons separated by a distance of ${\SI{1}{\angstrom}\equiv 10^{-10}\;\unit{\meter}}$ would 
give them accelerations of ${\sim \SI{250}{\angstrom\per\femto\second\squared}}$. Moreover, 
if $\Nelec$ particles are \emph{identical}, exchange symmetry reduces the set of symmetry-inequivalent 
points in their phase space by a factor of ${\Nelec!}$. For example, if they were confined
to a region ${\Vregion\subset\realone^3}$, the time that they would take to sample their
configuration space would be less, by a factor of ${\Nelec!}$, than the time that each individual
particle would take to explore ${\Vregion}$.

These considerations justify the assumption that, during each `moment of 
consciousness'~\citep{consciousness1,consciousness3}, and during most indirect observations (i.e., those not
limited by our brains and senses because they are performed by artificial devices), the particles
sample $\phasespace$ comprehensively.
Consequently, observations of the set of particles, and measurements of their properties, are not observations and measurements
of the particles in a microstate or in a small region of their phase space. They are observations and measurements
of the particles' \emph{statistical} state.
\color{red}
\vspace{2cm}
\begin{center}
SUBSECTION MISSING (mid rewrite)
\end{center}
\vspace{2cm}

\color{black}

\subsection{Slow perturbative measurements of fast-moving classical systems}
\label{section:slow_measurements}
The outcome of a measurement of an observable $\Obs$ is determined
by an interaction between a \emph{probe} $\probe$ and the measurement \emph{subject}, $\subject$.
This section assumes that the duration $\tau$ of this
interaction is much longer than the time scale on which 
the true instantaneous configuration $x_t$ of $\subject$ changes.
It is so long, which is to say that $x_t$ changes so fast, that it is impossible
to measure properties of short segments of the trajectory, ${x_t(t)}$;
and the fraction of $\tau$ for which ${x_t}$ is in any substantial
subset of the configuration space $\configS$ of $\subject$ is approximately
equal to the fraction of time it would spend there in
an infinite number of identical repetitions of the measurement.

In this context, `identical' means that the information
available to the observer
(prior information, their readings, etc.)
is the same in each repetition. For simplicity, let us assume that 
only one pdf is consistent with this information. In other words, 
no symmetries exist that make multiple symmetry-equivalent pdfs consistent with it.

Let us denote the configuration space of $\probe$ by $\configP$, and
let us denote its true instantaneous configuration by $y_t$. 
Let us denote the composite system  whose instantaneous microstate is ${(x_t,y_t)}$
by ${\combined\equiv\subject\cup\probe}$.
As discussed above, for simplicity, let us assume that what is being measured does not depend on 
the momenta conjugate to $x_t$ and $y_t$. Therefore
the measured value is the expecation value of some function,
\begin{align*}
\Obs:\configS\times\configP\to\realone;\;(x,y)\mapsto \Obs(x,y).
\end{align*}

Let us assume that the observer either has no information about $x_t$
prior to the measurement, or that they only possess a statistical description of $x_t$.
Therefore, they do not know $x_t$ precisely immediately before $\probe$ and $\subject$ begin to interact, which
implies that they cannot know ${(x_t,y_t)}$ precisely during the interaction. 
For the purposes of this section, it 
is unnecessary to assume that $y_t$ is not known precisely before or after the interaction, 
but of course nothing can be known to infinite precision. 

These assumptions mean that everything that can be known about
$\subject$, $\probe$, and $\combined$ before, during, and after the measurement, can be expressed
as the time-dependent probability density functions, ${\prs(x;t)}$, $\prp(y;t)$, and ${\prcm(x,y;t)}$
for ${x_t}$, ${y_t}$, and ${(x_t,y_t)}$, respectively.
For simplicity, time ($t$) will sometimes be omitted as a parameter of these pdfs.

The probe is \emph{perturbative} because it interacts with $\subject$. If it did not interact with $\subject$, 
nothing could be deduced about $\subject$ from ${\prp(y;t_0)}$ and ${\prp(y;t_0+\tau)}$, where
${t_0}$ is the earliest time at which the interaction between $\probe$ and $\subject$ becomes non-negligible, 
and ${t_0+\tau}$ is the earliest later time at which it becomes negligible again.
The interaction implies that 
\begin{align*}
\prcm(x,y) &=\prsp(x|y)\prp(y)
\\
&=\prps(y|x)\prs(x)\neq \prs(x)\prp(y), 
\end{align*}
where
$\prsp$ is the conditional pdf for $x_t$ given $y_t$; and $\prps$ is the conditional pdf for $y_t$
given $x_t$. 

The measurement is \emph{slow} in the sense that the fraction of interval ${\intervalmeas\equiv(t_0,t_0+\tau)}$ 
for which $\prcm$ depends on the configuration ${(x_t(t_0),y_t(t_0))}$ of $\combined$
when ${\subject}$ and ${\probe}$ begin to interact
is negligible. In other words, the \emph{transient} is short-lived and has a negligible influence
on the measurement outcome. For almost the entire duration of the measurement, 
$\combined$ is either in a steady state (the time dependence of $\prcm$ is periodic and its period is much less than $\tau$), 
a stationary state (the time dependence of $\prcm$ is negligible), or
$\subject$ is responding adiabatically to $\probe$. 

Let us treat cases in which the measurement is performed while $\combined$ is in a steady, but non-stationary, state
by redefining the time-periodic pdf, $\prcm$, as the average of itself over one of its periods.
For present purposes, this removes the distinction between steady states and stationary states, so steady states will not 
be discussed further.

Adiabatic response of $\subject$ to $\probe$
means that the influence of $\probe$ on $\prs$ changes so slowly that, on the timescale 
on which $x_t$ changes, it appears not to change at all. This
does not imply that $y_t$ is not changing rapidly, but that, during
the measurement interval ${\intervalmeas}$, the joint pdf ${\prcm(t)}$
changes so slowly that, instantaneously, it is almost the same as it would
be if the probe's influence was not changing.

For example, if $\probe$ was the moving tip of a scanning tunnelling
microscope (STM), its constituent atoms and electrons would move rapidly.
However we are assuming that the tip, as a whole, moves so slowly that, at each of its positions,
$\prcm$ is the same as it would be if the tip was static and had been at that position for a long time.
This assumption of \emph{adiabatic decoupling}~\citep{ott_1979,spohn_2001,Tangney_JCP_2006} is like the \emph{Born-Oppenheimer approximation}~\citep{born_oppenheimer}
used to deduce the instantaneous statistical state
of electrons from the instantaneous positions of the much slower-moving nuclei.
However, for an STM tip it is likely to be a much better approximation, because the timescale of seconds on which 
an STM tip moves is about fifteen orders of magnitude larger than the 
timescale on which nuclei move.  

On the other hand, if $\probe$ is an emitted wave pulse, its frequency would have
to exceed the highest frequencies of visible electromagnetic waves for its adiabatic
decoupling from classical electron-like particles to be less effective
than the decoupling between nuclei and electrons that the Born-Oppenheimer approximation exploits.

It follows from all of these physical assumptions and considerations that
the measurement is not a measurement of $\subject$ in
a particular configuration, $x_t$. It is not even a measurement or observation
of $\subject$ in one of the stationary statistical states it can be in when it is isolated.
It is \emph{either} a measurement or observation pertaining to a stationary state of $\combined$, 
\emph{or} it is a property of the \emph{stationary state trajectory}, ${\{\prcm(t):t\in\interval_\text{meas}\}}$,
where each pdf ${\prcm(t)=\prcm(x,y;t)}$ along the stationary state trajectory is almost exactly the
stationary state that $\combined$ would be in if the influence of $\probe$ on $\prcm$ was not changing.

\subsection{Why measured values are eigenvalues}
Let us consider the case in which what is measured is
not determined by a stationary state \emph{trajectory} of $\combined$, but by
a single stationary state of $\combined$. Let
${\{\prcm_\gamma\}}$
be the set of all stationary states of $\combined$ 
that might be measured, where 
index $\gamma$ distinguishes between different stationary states.
Then the $\gamma^\text{th}$ possible measured value of observable $\Obs$ is
\begin{align}
\obs_\gamma
&\equiv \int_{\configS}\dd{x}\int_{\configP}\dd{y} \prcm_\gamma(x,y) \Obs(x,y)
\nonumber\\
& = \int_{\configS}\dd{x}\prs_\gamma(x)\overbrace{\left(\int_{\configP}\dd{y} \prps_\gamma(y|x) \Obs(x,y)\right)}^{\displaystyle \tObs_\gamma(x)} 
\nonumber \\
& =  \int_{\configS}\dd{x} \prs_\gamma(x) \tObs_\gamma(x).
\label{eqn:observable1}
\end{align}
Equation~\ref{eqn:observable1} implies that, under the physical assumptions stated above, 
${\obs_\gamma}$ is not a property of $\subject$, but a property of 
the $\gamma^\text{th}$ stationary state \emph{of the act of measurement of $\Obs$}.

However, the purpose of most measurements is to reveal information about $\subject$, not $\combined$. Therefore most 
observables of interest are functions 
\begin{align*}
\Obs:\configS\to\realone;\; x\mapsto \Obs(x), 
\end{align*}
rather than functions
\begin{align*}
\Obs:\configS\times\configP\to\realone;\; (x,y)\mapsto \Obs(x,y). 
\end{align*}
The instantaneous value of such an observable 
is ${\Obs(x_t)}$, rather than ${\Obs(x_t,y_t)}$, and what is measured is its average, $\obs_\gamma$, in one of the stationary
states of ${\combined}$. 

When ${\Obs=\Obs(x)}$, Eq.~\ref{eqn:observable1} simplifies, because 
${\int_{\configP}\dd{y}\prps_\gamma(y|x)=1}$ implies that
${\tObs_\gamma(x)=\Obs(x), \;\forall \gamma}$. Therefore,
\begin{align}
\obs_\gamma = \int_{\configS}\dd{x} \prs_\gamma(x)\Obs(x).
\label{eqn:observable2}
\end{align}
Equation~\ref{eqn:observable2} means that the result of the slow perturbative measurement of property $\Obs$ of the fast-moving 
system $\subject$ is an average, over a very large number of configurations of $\subject$, while $\combined$ is in one of the 
stationary states \emph{of the act of measurement of $\Obs$}.

Note that, by definition of a stationary state, $\prcm_\gamma$ is independent of time. Therefore, ${\prs_\gamma(x)\equiv \int_{\configP} \prcm_\gamma(x,y)\dd{y}}$
is also independent of time, and is a stationary state of subsystem
$\subject$ of $\combined$ during the act of measurement of $\Obs$.
As discussed in Appendix~\ref{section:appendix_states_as_vectors}, we can specify this stationary state pdf
with a function ${\psicm_\gamma\in\lebesgue(\configspace\times\ydomain)}$ such that ${\big|\psicm_\gamma\big|^2=\prcm_\gamma}$.

Let us specify $\prs$, which is an arbitrary pdf for $x_t$, with the function ${\psis\in\lebesgue(\configspace)}$, 
where ${\big|\psis\big|^2=\prs}$. Let us represent ${\psis}$
as an element ${\ket{\psis}}$ of an abstract Hilbert space ${\hilbert}$, and let ${\hO:\hilbert\to\hilbert}$
denote the Hermitian operator for which
\begin{align}
\expval{\Obs}\equiv \int_{\configS}\dd{x}\prs(x)\Obs(x) = \frac{\expval{\hO}{\psis}}{\braket{\psis}}.
\end{align}
The set of eigenstates of a Hermitian operator is a complete basis of the space on
which it acts and it is always possible to choose this set such that they are 
mutually orthogonal. Therefore we can express ${\ket{\psis}}$ as
\begin{align*}
\ket{\psis}=\sum_{\alpha} a_\alpha \ket{\uppsis_\alpha}, 
\end{align*}
where each ${\ket{\uppsis_\alpha}}$ is one of the mutually-orthogonal
eigenstates of ${\hO}$; and ${a_\alpha\in\complex}$.
Let us denote the eigenvalue of $\hO$ corresponding to eigenstate ${\ket{\uppsis_\alpha}}$ 
by $\obs_\alpha$. 
That is,
\begin{align*}
\hO\ket{\uppsis_\alpha}&=\obs_\alpha\ket{\uppsis_\alpha}
\Rightarrow \obs_\alpha = \frac{\expval{\hO}{\uppsis_\alpha}}{\braket{\uppsis_\alpha}}
\end{align*}
Therefore, when $\subject$ is in state ${\ket{\psis}}$, the expectation value of $\Obs$ 
can be expressed as 
\begin{align*}
\expval{\Obs}&=
\frac{\expval{\hO}{\psis}}{\braket{\psis}}=\frac{\sum_\alpha \abs{a_\alpha}^2 \obs_\alpha}{\sum_\beta \abs{a_\beta}^2}.
\end{align*}
Now let us consider an arbitrary change of state, 
\begin{align*}
\ket{\psis}\mapsto \ket{\psis+\zeta\delta\psis} \equiv \ket{\psis}+\zeta\ket{\delta\psis},
\end{align*}
where ${\zeta\in\realone}$ and
\begin{align*}
\ket{\delta\psis}=\sum_\alpha \delta a_\alpha\ket{\uppsis_\alpha}\in\hilbert.
\end{align*}
The corresponding change in ${\expval{\Obs}}$ as a function of $\zeta$ is
\begin{align*}
\delta\expval{\Obs}(\zeta) &= 
\frac{\expval{\hO}{\psis+\zeta\delta\psis}}{\braket{\psis+\zeta\delta\psis}}
-
\frac{\expval{\hO}{\psis}}{\braket{\psis}}
\\
&=
\frac{\sum_\alpha \abs{a_\alpha+\zeta\delta a_\alpha}^2 \obs_\alpha}{\sum_\beta \abs{a_\beta+\zeta\delta a_\beta}^2}
-
\frac{\sum_\alpha \abs{a_\alpha}^2 \obs_\alpha}{\sum_\beta \abs{a_\beta}^2}.
\end{align*}
After a little algebra, we find 
that the lowest-order contribution to ${\delta\expval{\Obs}(\zeta)}$ is linear in
$\zeta$ and proportional to 
\begin{align}
\sum_{\alpha\beta}
\obs_\alpha
&\bigg[
\abs{a_\beta}^2\left(a_\alpha\delta a_\alpha^*+a_\alpha^*\delta a_\alpha\right)
\nonumber
-\abs{a_\alpha}^2\left(a_\beta\delta a_\beta^*+a_\beta^*\delta a_\beta\right)
\bigg]
\nonumber
\\
&=
2\sum_{\alpha\beta}
\Re\{a_\alpha^*\delta a_\alpha\}
\abs{a_\beta}^2
\left(\obs_\alpha-\obs_\beta\right)
\nonumber
\\
&=
2\sum_{\alpha}
\Re\{a_\alpha^*\delta a_\alpha\}
\left(\obs_\alpha\sum_\beta\abs{a_\beta}^2-\sum_\beta 
\abs{a_\beta}^2
\obs_\beta\right)
\nonumber
\\
&=
2\left(\sum_\beta\abs{a_\beta}^2\right)\sum_{\alpha}
\Re\{a_\alpha^*\delta a_\alpha\}
\left(\obs_\alpha-\expval{\Obs}\right)
\label{eqn:expval_derivative}
\end{align}
The set of values of ${\expval{\Obs}}$ in states
of $\subject$ that are stationary \emph{during the act of measurement of $\Obs$}
is the set of possible measured values of $\Obs$. 
These values depend only on the time-independent function ${\Obs(x)}$
and on whichever stationary state $\subject$ is in during the measurement. 
The $\gamma^\text{th}$ such state can be represented by $\prs_\gamma$ or by an element
${\ket{\uppsis_\gamma}}$ of $\hilbert$.

If $\zeta$ is sufficiently small, 
the condition that our arbitrary state ${\ket{\psis}}$ must satisfy
to be one of these stationary states is that
${\zeta\ket{\delta \psis}}$ does not change
${\expval{\Obs}}$ to linear order in $\zeta$.
Therefore ${\dv*{\expval{\Obs}}{\zeta}=0}$, which implies that
the set of possible measured values of ${\expval{\Obs}}$ is
the set of its values for which
\begin{align*}
\sum_{\alpha}
\Re\{a_\alpha^*\delta a_\alpha\}
\left(\obs_\alpha-\expval{\Obs}\right)
=0
\end{align*}
for \emph{any} choice of ${\ket{\delta\psis}\in\hilbert}$, and therefore for
any set ${\{\delta a_\alpha\}}$. 
This implies that, for each $\alpha$, \emph{either} ${\expval{\Obs}=\obs_\alpha}$ 
or ${a_\alpha=0}$.

This concludes the demonstration that the set of possible measured values of any observable $\Obs$ is the set
of eigenvalues of a Hermitian operator ${\hO}$.

\color{black}

\onecolumngrid
\vspace{0.8cm}
\PRLsep
\vspace{1cm}
\twocolumngrid
\section{Time dependences of statistical states}
\label{section:smooth_evolution}
A probability distribution function (pdf) for the microstructure or microstate of
a classical Hamiltonian system is a state of knowledge. If the deterrministic
system it describes is left undisturbed, the pdf evolves smoothly until
new information about the physical system is revealed, such as the result of an earlier measurement.
New information would change the pdf abruptly in the same way that, when the first lottery number is drawn, 
the probability that your ticket is a winning ticket abruptly changes.

In classical statistical mechanics, the smooth evolution of the pdf for a system's microstate
is described by the \emph{Liouville equation}, which is the \emph{continuity equation} for the
flow of probability in phase space expressed in terms of the system's Hamiltonian~\citep{evans_morriss_2008}. 
If the Hamiltonian does not depend explicitly on time, 
the flow of probability in phase space is smooth and 
conserves probability \emph{locally} in phase space. This means that the probability of
the system's microstate being in any region of phase space can only change
via a flow of probability density through the region's boundaries.

\subsection{Evolution of a statistical microstate}
Let ${\phasespace\equiv\momspace\times\configspace}$ denote the phase
space of a classical system, where $\momspace$ and $\configspace$ are
the spaces of all possible momenta and configurations (microstructures), respectively.
Let ${\Gammat\equiv(\pi_t,x_t)\in\phasespace}$ denote the system's true microstate, 
and its Hamiltonian is 
\begin{align*}
\smallham:\phasespace\equiv\momspace\times\configspace\to\realone;\; (\pi,x)\mapsto\smallham(\pi,x), 
\end{align*}
where 
${\pi=(\pi_1,\pi_2,\cdots)}$,
${x=(x_1,x_2,\cdots)}$, and $x_\alpha$ and $\pi_\alpha$ are the coordinate
and momentum, respectively, of the ${\alpha^\text{th}}$ degree of freedom.

The partial derivatives with respect to $x_\alpha$ and ${\pi_\alpha}$ 
are denoted by ${\gradx^\alpha}$ and ${\gradp^\alpha}$, respectively, and we will
assume that all second partial derivatives of ${\smallham}$ 
are continuous. This implies that~\citep{hormander}
\begin{align*}
\gradx^\alpha\gradp^\alpha\smallham=\gradp^\alpha\gradx^\alpha\smallham
\implies\gradx\cdot\gradp\smallham=\gradp\cdot\gradx\smallham.
\end{align*}

To define a probability distribution for the location of $\Gammat$ in $\phasespace$, 
let us use a construction similar to that used in Appendix~\ref{section:states_as_vectors}:
Let ${\G\subset\phasespace}$ denote a regular lattice, and let 
${\{\N_\Gamma:\Gamma\in\G\}}$ be a partition of $\phasespace$ such 
that all points in $\phasespace$ that are closer to ${\Gamma\in\G}$
than to any other element of $\G$ are elements of ${\N_\Gamma}$.
Since $\G$ is a regular lattice, the measure ${\abs{\N_\Gamma}}$ of ${\N_\Gamma}$ in $\phasespace$
is the same for all ${\Gamma\in\G}$, and is denoted by ${\hilbv>0}$.

Our state of knowledge of $\Gammat$ at time $t$ can be expressed by
the probability \emph{mass} function, 
\begin{align*}
\psdist(t)\hilbv:\G\to[0,1];\;\Gamma\mapsto \psdist(\Gamma;t)\hilbv\equiv\Pr(\Gammat\in\N_\Gamma), 
\end{align*}
where ${\hilbv}$ is small enough that variations of the probability
\emph{density} function, 
\begin{align*}
\psdist(t):\phasespace\to\realpos; \;\Gamma\mapsto \psdist(\Gamma;t)=\psdist(\pi,x;t),
\end{align*}
across every subset ${\N_\Gamma}$ in the partition of $\phasespace$ are negligible.

Let us assume that no new information about the physical system is revealed during the
interval ${(t,t+\tau)}$, so that 
${\psdist(t+\tau)}$ is determined by $\psdist$ evolving smoothly
between times $t$ and ${t+\tau}$ as described by
the Liouville equation~\citep{evans_morriss_2008},
\begin{align}
\gradt\psdist  & + \gradG\cdot\left(\psdist\dot{\Gamma}\right)  = 0
\label{eqn:liouville_general}
\\
\implies
\gradt\psdist  & + \dot{\Gamma}\cdot\gradG\psdist = -\psdist\gradG\cdot\dot{\Gamma} = -\compress\psdist,
\label{eqn:liouville0}
\end{align}
In Eq.~\ref{eqn:liouville0}, 
${\compress=\compress(\Gamma)\equiv \gradG\cdot\dot{\Gamma}(\Gamma)}$
is known as the 
\emph{phase space compression factor}~\citep{evans_morriss_2008,fluctuation_theorem}.

If the physical system is isolated, its Hamiltonian $\smallham$ is time-independent.
Therefore using Hamilton's equations, 
${\dot{x}_\alpha=\gradp^\alpha\smallham}$
and
${\dot{\pi}_\alpha=-\gradx^\alpha\smallham}$, we find that
\begin{align*}
\compress&=\gradG\cdot\dot{\Gamma}
=\sum_\alpha\left(\gradx^\alpha\dot{x}_\alpha+\gradp^\alpha\dot{\pi}_\alpha\right)
\\
&=\sum_\alpha\left(\gradx^\alpha\gradp^\alpha\smallham-\gradp^\alpha\gradx^\alpha\smallham\right)
=\gradx\cdot\gradp\smallham-\gradp\cdot\gradx\smallham
=0.
\end{align*}
Since the phase space compression factor vanishes, we are left with
the simplified Liouville equation,
\begin{align}
\gradt\psdist  & + \dot{\Gamma}\cdot\gradG\psdist = 0,
\label{eqn:liouville}
\end{align}
which can also be expressed as 
\begin{align}
\gradt\psdist &+ \gradx \psdist\cdot\gradp \smallham -\gradp \psdist\cdot\gradx \smallham= 0.
\label{eqn:liouville1}
\end{align}

Probability must always be conserved, but Eqs.~\ref{eqn:liouville} and~\ref{eqn:liouville1}
state that there is \emph{local} conservation of probability. In other words, the rate of change of the 
probability density at a point in $\phasespace$ equals the rate
at which probability density is flowing into
the point from neighbouring points.

If the phase space compression factor were finite, which would be the
case if ${\partialt\smallham}$ was finite, 
an increasing probability of ${\Gammat}$ being 
in one region of $\phasespace$ could be counterbalanced by a decreasing probability
of it being in another region of ${\phasespace}$, with which the first region 
had no overlap or common boundary points.

There is local conservation of probability \emph{if and only if} the phase space compression
factor vanishes.

\subsection{\emph{Reduced} Liouville equations}
Let us express ${\psdist(t)}$ as 
\begin{align*}
\psdist(\pi,x;t)=\cfgdist(x;t)\momcfgdist(\pi|x;t), 
\end{align*}
where
${\momcfgdist(\pi|x;t)}$ is the conditional probability density function for ${\pi_t(t)}$ when
it is known that ${x_t(t)=x}$.
Then Eq.~\ref{eqn:liouville1} can be expressed as
\begin{align*}
\momcfgdist\gradt \cfgdist &+ \cfgdist\gradt \momcfgdist + \momcfgdist\gradx \cfgdist\cdot\gradp \smallham 
\\
&+ \cfgdist\gradx \momcfgdist\cdot\gradp \smallham 
-\cfgdist\gradp \momcfgdist\cdot\gradx \smallham = 0.
\end{align*}
If we integrate over the set ${\momspace}$ of accessible
momenta, and use the fact that
${\intmom \momcfgdist\dd{\pi}=1}$,
which implies that
\begin{align*}
\gradt\intmom \momcfgdist\dd{\pi}=\intmom\gradt \momcfgdist \dd{\pi}=0,
\end{align*}
we get
\begin{align}
\gradt \cfgdist + \gradx \cfgdist\cdot&\left(\intmom \dd{\pi} \momcfgdist\gradp \smallham\right)
+ \cfgdist\left(\intmom\dd{\pi}\gradx \momcfgdist\cdot\gradp \smallham\right) 
\nonumber
\\
&- \cfgdist\left(\intmom\dd{\pi}\gradp \momcfgdist\cdot \gradx \smallham\right) = 0.
\label{eqn:liouville2}
\end{align}
The identities
\begin{align*}
\intmom\dd{\pi} \gradp\cdot\left(\momcfgdist\partial_x \smallham\right)
&=
\intmom\dd{\pi} \gradp \momcfgdist\cdot \gradx \smallham
\\
&+
\intmom\dd{\pi}  \momcfgdist \gradp\cdot\gradx \smallham,
\\
\intmom\dd{\pi}\gradx\cdot\left(\momcfgdist\gradp\smallham\right)
&=
\intmom\dd{\pi}\gradx\momcfgdist\cdot\gradp\smallham
\\
&+\intmom\dd{\pi}\momcfgdist\gradx\cdot\gradp\smallham, 
\end{align*}
allow Eq.~\ref{eqn:liouville2} to be expressed in the form
\begin{align}
\gradt\cfgdist+\gradx\cdot\bigg(\cfgdist\intmom&\dd{\pi}\momcfgdist\gradp\smallham\bigg) 
\nonumber
\\
&=\cfgdist\intmom\dd{\pi}\gradp\cdot\left(\momcfgdist\gradx\smallham\right).
\label{eqn:liouville2}
\end{align}
The generalized Stokes theorem~\citep{baez_1994,doran} implies that
the integral on the right hand side can be expressed as 
\begin{align*}
\intmom\dd{\pi}\gradp\cdot\left(\momcfgdist\gradx\smallham\right)  
= \int_{\partial\momspace}\momcfgdist\dd{s_\pi}\left(\hat{s}_\pi\cdot\gradx\smallham\right),
\end{align*}
where ${\partial\momspace}$ is the boundary of
set ${\momspace}$; ${\dd{s_\pi}}$ is a volume form
of ${\partial\momspace}$; and ${\hat{s}_\pi}$ is a unit vector that is normal
to ${\partial\momspace}$ and directed away from set $\momspace$.

By definition of the set ${\momspace}$ of \emph{accessible} (${\implies}$ \emph{possible}) 
momenta, ${\momcfgdist}$ vanishes on ${\partial\momspace}$. Therefore
the boundary integral vanishes and Eq.~\ref{eqn:liouville2} becomes
\begin{align}
\gradt \cfgdist + \gradx\cdot \left( \cfgdist\intmom\dd{\pi}\momcfgdist\gradp \smallham \right) &=  0,
\label{eqn:liouville3}
\end{align}
which can also be expressed as
\begin{align*}
\gradt \cfgdist +  \left(\intmom\dd{\pi}\momcfgdist\gradp \smallham \right)\cdot\gradx\cfgdist &=  -\left(\gradx\cdot\intmom\dd{\pi}\momcfgdist\gradp \smallham\right)\cfgdist.
\end{align*}
The quantity in parentheses on the left hand side is
the expectation value of ${\dot{x}}$
\emph{given that} ${x_t=x}$. It will be denoted as
\begin{align}
\barv(x;t)
\equiv
\intmom\dd{\pi}\momcfgdist\gradp \smallham  = \intmom\dd{\pi}\momcfgdist(\pi|x;t)\dot{x}(\pi,x).
\label{eqn:velocity}
\end{align}
Therefore we can express Eq.~\ref{eqn:liouville3} in the forms
\begin{subequations}
\begin{align}
\gradt \cfgdist +\gradx\cdot\left( \cfgdist \barv\right) &= 0,
\label{eqn:liouville4a}
\\
\gradt\cfgdist + \barv\cdot\gradx\cfgdist &= -\left(\gradx\cdot\barv\right)\cfgdist.
\label{eqn:liouville4b}
\end{align}
\label{eqn:liouville4}
\end{subequations}
Equations~\ref{eqn:liouville4} are forms of 
the \emph{reduced Liouville equation} in configuration space, $\configspace$.

\subsubsection{Reduced Liouville equation in momentum space}
If we express $\psdist$ as 
\begin{align*}
\psdist(\pi,x;t)=\momdist(\pi;t)\cfgmomdist(x|\pi;t), 
\end{align*}
a derivation analogous to the derivation of Eq.~\ref{eqn:liouville4} from
Eq.~\ref{eqn:liouville} would lead us to
\begin{subequations}
\begin{align}
\gradt \momdist + \gradp\cdot\left(\momdist\barF\right) & =0,
\label{eqn:liouvillepa}
\\
\gradt\momdist + \barF\cdot\gradp\momdist & =-\left(\gradp\cdot\barF\right)\momdist,
\label{eqn:liouvillepb}
\end{align}
\label{eqn:liouvillep}
\end{subequations}
where,
\begin{align*}
\barF=\barF(\pi;t)\equiv 
\int_\configspace\dd{x} \momdist(x|\pi;t)\dot{\pi}(\pi,x).
\end{align*}
Equations~\ref{eqn:liouvillep} are forms of the reduced Liouville 
equation in momentum space, $\momspace$.

\subsubsection{Configuration space and momentum space compression factors}
Let us define the
\emph{configuration space compression factor},
\begin{align*}
\compressx=\compressx(x;t)\equiv \gradx\cdot\barv(x;t),
\end{align*}
and the
\emph{momentum space compression factor},
\begin{align*}
\compressp=\compressp(\pi;t)\equiv \gradp\cdot\barF(\pi;t).
\end{align*}
Then we can express the reduced Liouville 
equations (Eqs.~\ref{eqn:liouville4b} and~\ref{eqn:liouvillepb}) as
\begin{align}
\gradt \cfgdist +\barv\cdot\gradx\cfgdist &= -\compressx\cfgdist,
\label{eqn:liouville_x}
\\
\gradt \momdist +\barF\cdot\gradp\momdist &= -\compressp\momdist.
\label{eqn:liouville_p}
\end{align}
Recall that the flow of probability described by Eq.~\ref{eqn:liouville0} 
conserves probability locally if and only if the phase space compression factor ${\compress}$ 
vanishes everywhere in $\phasespace$.
Similarly, the flows of probability described by 
Eqs.~\ref{eqn:liouville_x} and~\ref{eqn:liouville_p} conserve
probability locally if and only if ${\compressx}$ and ${\compressp}$, respectively, 
vanish everywhere in ${\configspace}$ and ${\momspace}$, respectively.

However $\compressx$ and $\compressp$ are finite, in general, so
Eqs.~\ref{eqn:liouville_x} and~\ref{eqn:liouville_p} do
not conserve probability locally in $\configspace$ and $\momspace$, respectively.
This makes them unsuitable for calculating
the \emph{simultaneous} evolutions of ${\cfgdist}$ and ${\momdist}$,
because ${\bar{v}}$ and ${\bar{F}}$ vary in time if
$\momcfgdist$ and ${\cfgmomdist}$, respectively, vary in time.

For example, 
Eq.~\ref{eqn:liouville_x} can only be used to describe
the evolution of $\cfgdist$ in a given time interval 
if ${\barv}$ is known at all times in that interval.
However $\momcfgdist$ must be known to calculate $\barv$;
and ${\momcfgdist=\psdist/\cfgdist}$ cannot be calculated without knowing
${\psdist}$ and ${\cfgdist}$.
Similarly, $\barF$ cannot be known unless $\psdist$ and ${\momdist}$
are known.
Therefore Eq.~\ref{eqn:liouville_x} and Eq.~\ref{eqn:liouville_p}
could only be used to calculate the simultaneous evolutions of ${\cfgdist}$ and ${\momdist}$
if those evolutions had already been calculated (e.g., from Eq.~\ref{eqn:liouville}).

Note that Eqns~\ref{eqn:liouville_x} and~\ref{eqn:liouville_p} must conserve total probability. Therefore,
\begin{align*}
\intconfig\gradt\cfgdist(x;t)\dd{x} &= \dv{t}\left(\intconfig\cfgdist(x;t)\dd{x}\right) = 0
\\
\implies \intconfig\barv(x)\cdot\gradx\cfgdist(x)\dd{x}&= -\intconfig\compressx(x)\cfgdist(x)\dd{x},
\end{align*}
and
\begin{align*}
\intmom\gradt\momdist(\pi)\dd{\pi} &= \dv{t}\left(\intmom\momdist(\pi)\dd{\pi}\right) = 0,
\\
\implies \intmom\barF(\pi)\cdot\gradx\momdist(\pi)\dd{\pi}&= -\intmom\compressp(\pi)\momdist(\pi)\dd{\pi}.
\end{align*}

\color{red}
\vspace{2cm}
\begin{center}
SUBSECTION MISSING (mid rewrite)
\end{center}
\vspace{2cm}
\color{black}

\subsection{Evolutions of microstructures in Hilbert spaces}
\label{section:evolution}
Let ${\hilbert_x}$ denote the Hilbert space whose rays correspond to
microstructure pdfs, such as ${\cfgdist(t)}$.
Let ${\ket{\psi(t)}}$ and ${\ket{\psi(t+\tau)}}$ be elements
of ${\hilbert_x}$ which possess the same information as ${\cfgdist(t)}$ 
and ${\cfgdist(t+\tau)}$, respectively. 
That information could be retrieved via Eqs.~\ref{eqn:wfn_to_pdf} and~\ref{eqn:wfn_to_vector}, 
as follows:
\begin{align*}
\cfgdist(x;t) & = \frac{\braket{\psi(t)}{x}\braket{x}{\psi(t)}}{\braket{\psi(t)}} 
 = \frac{\psi^*(x;t)\psi(x;t)}{\int_\configspace \psi^*(x';t)\psi(x';t)\dd{x'}}.
\end{align*}
Let us choose the set 
\begin{align*}
\left\{\ket{\psi(t')}:t\leq t'\leq t+\tau\right\}
\end{align*}
of state vectors at times between $t$ and ${t+\tau}$
such that they all have
unit norms (${\braket{\psi}=1}$) and such that
the evolution of
${\ket{\psi(t)}}$ into ${\ket{\psi(t+\tau)}}$
is differentiable with respect to time.

Let us also choose each state ${\ket{\psi(t')}}$ along this \emph{state trajectory}
such that the \emph{wavefunction}, 
\begin{align*}
\psi(x;t')\equiv\braket{x}{\psi(t')}=\abs{\braket{x}{\psi(t')}}e^{2\pi i\theta(x;t')},
\end{align*} 
is a differentiable function of $x$.
As discussed in Sec.~\ref{section:antisymmetry} and Appendix~\ref{section:appendix_states_as_vectors}, 
this is even possible at a point ${x_0\in\configspace}$ where ${\cfgdist}$ vanishes, but its
gradient ${\gradx \cfgdist(x)}$ remains finite as ${x\to x_0}$.
In other words, even if $\cfgdist$ has a non-differentiable cusp at $x_0$, 
${\psi}$ can be differentiable at $x_0$
as long as the phase factor  ${e^{2\pi i\theta(x;t')}}$ changes sign at $x_0$.

Because the state trajectory is smooth and norm-preserving, 
${\ket{\psi(t)}}$ and ${\ket{\psi(t+\tau)}}$ 
are related by a rotation.
In other words, for every ${\tau\in\realpos}$,
there exists a unitary operator 
${
\hU(\tau):\hilbert_x\to\hilbert_x}$,
which varies
smoothly with $\tau$, such that
\begin{align}
\ket{\psi(t+\tau)} = \hU(\tau)\ket{\psi(t)}.
\label{eqn:unitary_evolution}
\end{align}
${\hU(\tau)}$ is independent of $t$ because 
$\cfgdist$ evolves according to Eq.~\ref{eqn:liouville4} and
its evolution from time $t$ depends only on ${\cfgdist(t)}$ and the system's Hamiltonian, ${\smallham}$, which 
is time-independent. 
Therefore, given any time ${\tilde{t}\neq t}$, 
if ${\cfgdist(\tilde{t})=\cfgdist(t)}$,
then ${\cfgdist(\tilde{t}+\tau)=\cfgdist(t+\tau)}$. 
This implies that the information shared by ${\ket{\psi(t)}}$, ${p(t)}$, and ${p(\tilde{t})}$
could also be expressed as a unit vector ${\ket{\phi(\tilde{t})}\in \hilbert_x}$, and 
${\hU(\tau)\ket{\phi(\tilde{t})}}$ would possess the same physical information
as ${\cfgdist(\tilde{t}+\tau)}$.

Another way to say this is that time is homogeneous 
for the physical system, and for the evolution of what is known about it, 
because the system is isolated. Its isolation implies that $\smallham$ does
not depend explicitly on time, and that the evolution of $\cfgdist$
between time $t$ and time ${t+\tau}$ depends only on ${\cfgdist(t)}$ and on $\smallham$.

Every unitary operator can be expressed as the exponential of an anti-Hermitian operator. Therefore, 
let 
\begin{align*}
\hU(\tau)\equiv e^{-2\pi i\hB(\tau)/h}, 
\end{align*}
where ${\hB(\tau)}$ is Hermitian, which means that ${-i\hB(\tau)}$ is anti-Hermitian, 
and $h$ is the constant with dimensions ${\text{energy}\times\text{time}}$ 
that was introduced in Section~\ref{section:CMQM_boundary}.
Smooth evolution of ${\ket{\psi}}$ implies
that ${\hB(\tau)/h}$ must vanish in the limit ${\tau\to 0}$. 
Therefore, in this limit we can express Eq.~\ref{eqn:unitary_evolution}
as
\begin{align}
\ket{\psi(t+\tau)} &=  e^{-2\pi i\hB(\tau)/h}\ket{\psi(t)}
\nonumber \\
& =\left(1-2\pi i\hB(\tau)/h\right)\ket{\psi(t)}+\order{\hB^2/h^2}
\nonumber
\\
\implies \gradt\ket{\psi(t)}&\equiv \lim_{\tau\to 0}\frac{\ket{\psi(t+\tau)}-\ket{\psi(t)}}{\tau} 
\nonumber \\
& = -\frac{i}{\hbar}\left(\lim_{\tau\to 0}\frac{\hB(\tau)}{\tau}\right)\ket{\psi(t)},
\label{eqn:pre_tdse}
\end{align}
where ${\hbar\equiv h/(2\pi)}$.
We have not made any assumptions about the state ${\ket{\psi(t)}}$ to derive this equation, which means that it is generally valid
and that ${\hU(\tau)}$, and hence ${\hB(\tau)}$, are $t$-independent properties of the physical system.

If we define the system-dependent operator, 
\begin{align*}
\hH_x\equiv  \lim_{\tau\to 0} \left(\frac{\hB(\tau)}{\tau}\right),
\end{align*}
then
Eq.~\ref{eqn:pre_tdse} becomes 
\begin{align}
\hH_x\ket{\psi}=i\hbar\gradt\ket{\psi}.
\label{eqn:tdse}
\end{align}
What this shows is that, when a classical statistical microstate, ${\cfgdist(x)}$, is expressed
as an element ${\ket{\psi}}$ of a Hilbert space, the equation governing 
its time-evolution has the same mathematical form as the \emph{time-dependent Schr\"odinger equation}.
The analogous equation, 
\begin{align}
\hH_p\ket{\psi_p}=i \hbar\gradt\ket{\psi_p},
\label{eqn:tdse_p}
\end{align}
could be derived from Eq.~\ref{eqn:liouvillep} to describe the evolution
of an element ${\ket{\psi_p}}$ of a Hilbert space $\hilbert_p$, which 
contains the same physical information as the statistical momentum state, ${\tilde{\momcfgdist}}$.

Note that I have used suggestive notation to make the similarity 
between Eqs.~\ref{eqn:tdse} and~\ref{eqn:tdse_p} and the time-dependent
Schr\"odinger more obvious. 

\subsection{Schr\"odinger equation (generic form, position representation)}
Now let us try to find a more explicit form for ${\hH_x}$, when it
is represented in the basis of microstructures, $\ket{x}$,
by attempting to recover Eq.~\ref{eqn:liouville4} from Eq.~\ref{eqn:tdse}.
After projecting both sides of Eq.~\ref{eqn:tdse} onto ${\ket{x}}$, and
inserting the identity, we get
\begin{align*}
\mel{x}{\hH_x\hone}{\psi_x}
&= \bra{x}\hH_x\left(\int_\configspace \dyad{x'}\dd{x'}\right)\ket{\psi_x}
= i\hbar \gradt\braket{x}{\psi_x}
\\
\implies &
\int_\configspace H_x(x,x')\psi_x(x')\dd{x'} = i \hbar \gradt\psi_x(x)
\end{align*}
where ${\psi_x(x)\equiv\braket{x}{\psi_x}}$
and ${H_x(x,x')\equiv\mel{x}{H_x}{x'}}$.
Let us use ${\smallhamop}$ to denote the integral operator that acts on $\psi_x$ on the left
hand side of this equation, so that we can write it 
as 
\begin{align*}
\smallhamop\psi_x = i\hbar \gradt \psi_x.
\end{align*}

\onecolumngrid
\vspace{0.8cm}
\PRLsep
\vspace{1cm}
\twocolumngrid
\section{The bulk of a crystal in a torus}
\label{section:appendix_torus}
In solid-state physics it is common to 
use \emph{Born-von K\'arm\'an (BvK) periodic boundary conditions}
when studying the bulk of a crystal theoretically~\citep{born_von-karman}.
This dispenses with surfaces and is equivalent to placing the bulk of the crystal in
a torus. This appendix assembles some of the theoretical infrastructure
that is commonly used to simplify theory under BvK boundary conditions.
Its main focus is on the mathematical origins and mathematical properties of Bloch functions, and the 
assumptions under which they enter physical theory.

Bloch functions are single-particle statistical states that are usually introduced 
in the context of \emph{quantum mechanical} descriptions of crystals.
For example, at the time of writing the 
Wikipedia page on `\emph{Bloch's Theorem}' introduces them as `\emph{solutions
to the Schr\"odinger equation in a periodic
potential}' and uses the example of electrons in a crystal as the context for 
most of its discussion of them~\citep{bloch_wiki}. However it does note their
greater generality, and it refers readers to the mathematical field of
spectral geometry.

One purpose of this appendix is to emphasize that Bloch functions are
not inherently quantum mechanical entities, and do not originate from
any particular kind of energetics. Their generality is made clear 
by introducing them, and deducing 
their properties, under very loose physical assumptions.

\subsection{Physical assumptions}
For simplicity I consider a one dimensional crystal whose
bulk is represented in a $1$-torus, $\onetorus$, whose
circumference is the thickness $\bulksize$ of bulk
crystal that occupies it. 
If $\volume$ is the width of a primitive unit cell of the crystal,
${\bulksize=\Nunitcell\volume}$ for some large positive integer $\Nunitcell$.

The crystal whose bulk is represented in $\onetorus$ consists of a set of particles
whose statistical state is ${\volume}$-periodic, in the sense that the particles'
position pdf is invariant under any translation of all of the particles' positions
by the same lattice vector, ${m\volume}$, where $m$ is any integer.
The set of particles comprises comprises multiple subsets of identical
particles. 

We will focus on only one of the crystal's sets of identical particles. 
The $\volume$-periodic number density of those particles is
\begin{align*}
n:\onetorus\to\realnonneg;\;x\mapsto n(x),
\end{align*}
where ${n(x+m\volume)=n(x),\;\forall m\in\integer}$, and
\begin{align*}
\int_\onetorus n(x)\dd{x}=\Nident\in\integerpos.
\end{align*}
Let us assume that the $\Nident$ particles are not only identical; they are also indistinguishable.
They could be electrons, but none of the assumptions made are inconsistent with
them being identical billiard balls that are rendered indistinguishable by the very
high speeds at which they move: They move too fast for their individual trajectories
to be observed. Therefore, because it is impossible to keep track of which particle is which, their
position pdf ${\pdf}$ must be \emph{exchange symmetric}. If $\pdf$ is specified
by ${\Psi\in\lebesgue(\onetorus^\Nident)}$, where ${\Psi^*\Psi=\pdf}$, then
${\Psi}$ must be either exchange antisymmetric or exchange
symmetric (see Sec.~\ref{section:pdf_symmetry} and Appendix~\ref{section:appendix_states_as_vectors}).

Let us assume that ${n}$ is continuous and has continuous derivatives. 
The number of derivatives that are required to exist and be continuous 
is context dependent, but is usually one or two.  
This is not a stringent requirement given that \emph{physical} number densities, such as the
steady state number densities in a reaction-diffusion system, tend to be
smooth. 

\subsection{A family of `Hamiltonians', $\hamsmallx$}
\label{section:hamiltonian}
This subsection demonstrates that, for any given $\volume$-periodic number density $n$, 
any number of operators can be constructed that have all of the properties of the $1$-particle
Hamiltonian operator ${\hamsmallx}$ that will be assumed in later subsections of this appendix. 
The operators introduced in this section are contrived. They are not Hamiltonians, in general.
They are introduced simply because their existence demonstrates that, 
when ${\hamsmallx}$ is used later in this appendix, it does not
constitute a non-classical assumption.

\subsubsection{A function that specifies the density per particle}
As our starting point, consider the function
\begin{align*}
\densityfunction(x)\equiv \sqrt{\frac{n(x)}{\Nident}}e^{i\theta(x)}\in\lebesgue(\onetorus), 
\end{align*}
where ${\theta:\onetorus\to\realone}$ is a smooth function whose purpose is to ensure that
${\densityfunction}$ is differentiable. It may be necessary because ${\densityfunction}$ may
not be differentiable, or may not be
\emph{practically} differentiable, at points where ${n}$ vanishes and 
its gradient ${\partial_x n}$ does not vanish.

Although \emph{physical} number densities, such as those at the steady states of
reaction-diffusion systems, do not vanish \emph{identically}, they may become so small that
the magnitude of the derivative,
\begin{align*}
\partial_x\densityfunction = \left(\frac{1}{2}\partial_x\log n + i\partial_x\theta\right)
\densityfunction,
\end{align*}
becomes too large for all practical purposes. Therefore $\densityfunction$
may become practically non-differentiable wherever ${n}$ becomes very small,
in the following sense:

Let ${\precmicro\in\realpos}$ denote the smallest measurable distance at the microscale. In other words,
it is the counterpart at the microscale of ${\prectheo=\abs{\dbx}}$ at the macroscale (see Sec.~\ref{section:homogenization}).
Then an unequality,
\begin{align*}
\frac{1}{\precmicro}\big[\densityfunction(x+\precmicro)&-\densityfunction(x)\big]
\\
&\neq
\frac{1}{\precmicro}\big[\densityfunction(x)-\densityfunction(x-\precmicro)\big] + \order{\precmicro^2}, 
\end{align*}
of the left and right finite-difference derivatives would
occur wherever ${n}$ becomes very small \emph{and} its derivative
does not vanish more rapidly than ${\sqrt{n}}$.

Wherever this problem arises, $\theta$ is necessary to ensure
that, for practical purposes, all required derivatives of ${\densityfunction}$ exist.
It can ensure this if it is such that ${\exp(i\theta)}$ changes sign wherever $n$ either
vanishes or becomes so small that it would otherwise be \emph{practically} non-differentiable.

Since $n$ determines where ${\exp(i\theta)}$ changes sign, and since $n$ is $\volume$-periodic, 
let us assume that ${\exp(i\theta)}$ is also ${\volume}$-periodic.
Therefore ${\densityfunction}$ is ${\volume}$-periodic and differentiable.

\subsubsection{A family of $1$-particle decompositions of the density}
Now let us assume that there exists a complete set 
of differentiable phase factors which are orthogonal under the
density-weighted inner product~\citep{tang_2006,decay_of_fourier},
\begin{align*}
\braketT{f}{ng}\equiv\int_\onetorus f^*(x)n(x)g(x)\dd{x}= \braketT{nf}{g}.
\end{align*}
That is, there exists a set
\begin{align*}
\phaseset\equiv
\bigg\{
\vartheta_\alpha:\onetorus\to\realone\;\bigg|  
&\braketT{e^{i\vartheta_\alpha}}{n e^{i\vartheta_\beta}}
=\Nident\delta_{\alpha\beta}, \;\alpha,\beta\in\integernonneg
\bigg\},
\end{align*}
which can be used to define the set,
\begin{align*}
\densityset\equiv\bigg\{ \psi_\alpha\equiv 
\densityfunction e^{i\vartheta_\alpha}:\theta_\alpha\in\phaseset
,\;\alpha\in\integernonneg
\bigg\}.
\end{align*}
of $\phaseset$, ${\exp(i\vartheta_\alpha)}$ is $\volume$-periodic.
The (unweighted) inner product of ${\psi_\alpha,\psi_\beta\in\densityset}$ is
\begin{align*}
\braketT{\psi_\alpha}{\psi_\beta} 
&= \int_\onetorus \psi_\alpha^*(x)\psi_\beta(x)\dd{x}
\\
&= \frac{1}{N}\int_\onetorus e^{-i\vartheta_\alpha(x)}
n(x)e^{i\vartheta_\beta(x))}\dd{x}
\\
&=\braketT{e^{i\vartheta_\alpha}}{ne^{i\vartheta_\beta}} = \delta_{\alpha\beta}.
\end{align*}
Therefore, for any choice of function,
\begin{align*}
\occ:\integernonneg\to [0,1];\; \alpha\mapsto\occ_\alpha,
\end{align*}
such that ${\sum_\alpha \occ_\alpha =\Nident}$, 
\begin{align*}
\sum_\alpha \occ_\alpha\abs{\psi_\alpha(x)}^2 
= \sum_\alpha\occ_\alpha\abs{\densityfunction(x)}^2 = \frac{n(x)}{N}\sum_\alpha\occ_\alpha=n(x).
\end{align*}

For an arbitrary function $\occ$, let us define a function
\begin{align*}
&\epsilon:\integernonneg\to\realone;\; \alpha\mapsto \epsilon_\alpha, 
\end{align*}
and let us assume that ${\alpha \geq\beta\iff \epsilon_\alpha\leq\epsilon_\beta}$.
Then we can define a `Hamiltonian' operator, 
\begin{align*}
\hamsmall\equiv \sum_{\alpha}\epsilon_\alpha\dyad{\psi_\alpha},
\end{align*}
and a number density operator,
\begin{align*}
\denop\equiv \sum_{\alpha} \occ_\alpha \dyad{\psi_\alpha}\implies n(x)=\expval{\denop}{x}.
\end{align*}

The ${\lebesgue(\onetorus)\to\lebesgue(\onetorus)}$
counterpart of ${\hamsmall:\hilbert\to\hilbert}$ is the integral
operator $\hamsmallx$ with kernel ${\hamsmallx(x,x')\equiv \mel{x}{\hamsmall}{x'}}$. Its action on
an arbitrary function ${f\in\lebesgue(\onetorus)}$ is
\begin{align*}
\left(\hamsmallx f\right)(x)=
\int_\onetorus \hamsmallx(x,x')f(x')\dd{x'};
\end{align*}
and its action on any element ${\psi_\alpha}$ of ${\densityset}$ is 
\begin{align*}
\int_\onetorus \hamsmallx&(x,x')\psi_\alpha(x')\dd{x'}
\\
&= \sum_{\beta}\epsilon_{\beta} \psi_\beta(x)
\int_\onetorus\psi_\beta^*(x')\psi_\alpha(x')\dd{x'}=\epsilon_\alpha\psi_\alpha(x), 
\end{align*}
where use has been made of the orthonormality of set ${\densityset}$.
Therefore each element ${\psi_\alpha}$ of ${\densityset}$ is
an eigenfunction of ${\hamsmallx}$ with eigenvalue ${\epsilon_\alpha}$.
It is straightforward to show that ${\hamsmallx}$ is Hermitian.

\subsubsection{Energy distribution of occupation numbers}
For any choices of the functions ${\occ}$ and ${\epsilon}$, any number of `energy distributions', 
\begin{align*}
\occ_\epsilon:\realone\to[0,1];\; e\mapsto \occ_\epsilon(e), 
\end{align*}
can be defined, such that ${\occ_\epsilon(\epsilon_\alpha)=\occ_\alpha,\;\forall \alpha\in\integernonneg}$.

Therefore, given any number density $n$, an infinite number of Hamiltonian operators ${\hamsmallx}$ 
and energy distributions $\occ$ can be defined such that
\begin{align*}
n(x) &= \sum_\alpha \occ_\epsilon(\epsilon_\alpha) \abs{\psi_\alpha(x)}^2, 
\;&\;
\sum_\alpha\occ_\epsilon(\epsilon_\alpha)&=\sum_\alpha\occ_\alpha=\Nident.
\end{align*}

\subsubsection{A family of $\hilbert$-representations of ${n(x)}$}
Let ${\indexsetN}$ denote \emph{any} subset of ${\integernonneg}$ with
exactly ${\Nident}$ elements.
Then it is straightforward to show that the ${\Nident}$-dimensional Hilbert space, 
\begin{align*}
\Span\bigg\{\ket{\psi_\alpha}\in\hilbert:\alpha\in\indexsetN, \; \psi_\alpha\in\densityset\bigg\},
\end{align*}
is a ${\hilbert}$-representation of the density, $n$;
where ${\ket{\psi_\alpha}}$ is the counterpart in the abstract
Hilbert space ${\hilbert}$ of ${\psi_\alpha\in\lebesgue(\onetorus)}$.

\subsection{Fourier series expansion of an arbitrary eigenfunction, $\varphi$}
\label{section:fourier_series_expansion}
Let us forget the simplifications and contrivances of subsection~\ref{section:hamiltonian};
but let us assume that the $\volume$-periodic number density ${n}$ can be expressed as
\begin{align*}
n(x)=\sum_i \occ_i\abs{\varphi_i(x)},
\end{align*}
where ${\{\varphi_i\}}$ is a complete orthonormal set of eigenfunctions of some $\volume$-periodic
operator $\hamsmallx$; and ${\occ_i\in[0,1],\;\forall i}$.

Let ${\varphi}$ be an arbitrary eigenfunction of $\hamsmallx$, with eigenvalue
${\epsilon}$. That is, 
\begin{align}
\hamsmallx\varphi= \epsilon\varphi.
\label{eqn:eigenvalue_equation}
\end{align}
Since $\hamsmallx$ is both ${\volume}$-periodic and 
$\bulksize$-periodic, it will be useful to express the Fourier series
expansion of $\varphi$ as the nested sum, 
\begin{align}
\varphi(x) 
= \!\!\sum_{g\in\reciplattg} \!\ftsvarphi(g)e^{igx}
= \!\sum_{k\in\BZ}\sum_{G\in\reciplatt\vphantom{\BZ}} \!\ftsvarphi(G+k)e^{i(G+k)x}.
\label{eqn:eigenfunction1}
\end{align}
Because it is an infinite lattice, $\reciplatt$ is closed under addition and its
elements have additive inverses. Therefore,
\begin{align*}
\left\{G+\tG: \tG\in\reciplatt\right\}=\reciplatt,
\end{align*}
for any ${G\in\reciplatt}$. In words, $\reciplatt$ is 
not changed by adding the same reciprocal lattice vector $G$ 
to each of its elements.
It follows that, for any choice of ${G\in\reciplatt}$, 
Eq.~\ref{eqn:eigenfunction1} can be expressed as
\begin{align}
\varphi(x)   &= 
\sum_{k} e^{i(G+k) x}\left(\sum_{\tG} \ftsvarphi(\tG+G+k) e^{i\tG x}\right),
\label{eqn:eigen0}
\end{align}
where $\sum_k$ denotes ${\sum_{k\in\BZ}}$ and 
$\sum_{\tG}$ denotes ${\sum_{\tG\in \reciplatt}}$.

By its definition as a Fourier series, the
function 
\begin{align}
&u:\reciplattg\times\onetorus\to\complex, \; (g,x)\mapsto u(g,x)\equiv  \sum_{\tG} 
\ftsvarphi(\tG+g) e^{i\tG x}
\nonumber
\end{align}
is ${\volume}$-periodic in its second argument.
Let us also define the function,
\begin{align*}
&b:\reciplattg\times\onetorus\to \complex;\;
(g,x)\mapsto b(g,x)\equiv e^{igx}u(g,x),
\end{align*}
so that, for any ${G\in\reciplatt}$, Eq.~\ref{eqn:eigen0} can be expressed as
\begin{align}
\varphi(x)
&= \sum_k e^{i(G+k)x}u(G+k,x)  = \sum_k b(G+k,x),
\label{eqn:eigenfunction2-0}
\end{align}
where,
\begin{align}
&\bloch(G+k,x)\equiv e^{i(G+k)x}\pbloch(G+k,x).
\label{eqn:bloch_functionG}
\end{align}
The functions
\begin{align*}
b_g\equiv b(g):\onetorus\to\complex; \; x\mapsto b_g(x)\equiv b(g,x) = e^{igx}u(g,x)
\end{align*}
are known as \emph{Bloch functions}.
They are discussed in most textbooks on condensed matter physics~\citep{cohen_louie,ashcroft_mermin_book,ibach_and_luth,kittel}.
The $\volume$-periodic functions, 
\begin{align*}
u_g\equiv u(g):\onetorus\to\complex; \; x\mapsto u_g(x)\equiv u(g,x), 
\end{align*}
are often referred to as \emph{periodic Bloch functions}
or as \emph{cell-periodic Bloch functions}~\citep{vanderbilt_2018}.
The latter term avoids confusion between 
$\bulksize$-periodicity, which the functions ${b(g)}$ and ${u(g)}$ both possess, 
and ${\volume}$-periodicity, which ${u(g)}$ possesses but which ${b(g)}$ does not
possess unless ${g\in\reciplatt}$.

Bloch functions and their periodic counterparts will sometimes be 
denoted as ${b_g(x)}$ and ${u_g(x)}$ 
and sometimes as ${b(g,x)}$ and ${u(g,x)}$. 
These notations can be used interchangably, but 
when the fact that they are \emph{functions} of wavevector $g$ is not relevant, $g$ will
usually be a subscript.

\subsection{Properties of Bloch functions}
\label{section:bloch_functions}
Since $G$ is an arbitrary element of $\reciplatt$,
$\varphi$ can be expressed in the form
of Eq.~\ref{eqn:eigenfunction2-0} for any
choice of ${G\in\reciplatt}$, including ${G=0}$.
Therefore, 
\begin{align}
\varphi(x) 
&= \sum_k b(G+k,x) = \sum_k b(k,x),
\label{eqn:sumk0}
\end{align}
where ${G}$ is an arbitrary finite element of $\reciplatt$
in the first expression, and the zero element in the second expression.
It follows that
\begin{align}
\sum_k &\big[b(G+k,x)-b(k,x)\big]
\nonumber
\\
&= \sum_k e^{ikx}\left[e^{iGx}u(G+k,x)-u(k,x)\right] =0.
\label{eqn:sumk}
\end{align}
Since this equation is valid at any point ${x\in\onetorus}$, let us replace $x$
by ${x-m\volume}$ and use 
the ${\volume}$-periodicities of ${u(G+k)}$, ${u(k)}$ and ${e^{iGx}}$
to express it as
\begin{align*}
\sum_k e^{ikx}e^{imk\volume}&\left[e^{iGx}u(G+k,x)-u(k,x)\right] 
\\
&=
\sum_k e^{imk\volume}\left[b(G+k,x)-b(k,x)\right]  = 0,
\end{align*}
where $m$ could be any integer.
If this equation is multiplied by
${e^{-imk'\volume}}$,
for any ${k'\in\BZ}$, 
and summed over all values of $m$ between $0$ and ${\Nunitcell-1}$,
the result is
\begin{align*}
\sum_k\left(\sum_{m=0}^{\Nunitcell-1}e^{im(k-k')\volume}\right)\left[b(G+k,x)-b(k,x)\right]=0.
\end{align*}
The sum in parentheses vanishes unless ${k=k'}$, in which case its
value is ${\Nunitcell}$. Therefore, since $G$ and $k'$ are
arbitrary elements of ${\reciplatt}$ and $\BZ$, respectively, 
\begin{align}
b(G+k,x)=b(k,x),\;\;\forall G\in\reciplatt,\;\forall k\in\BZ.
\label{eqn:bloch_periodicity1}
\end{align}
Since $x$ is an arbitrary point in $\onetorus$, 
it follows that ${b}$ has the
periodicity of the reciprocal lattice, i.e., 
\begin{align}
b(g,x)=b(g+G,x),
\;\forall g\in\reciplattg,\;\forall G\in\reciplatt,\;\forall x\in\onetorus.
\label{eqn:ham_bloch_periodic}
\end{align}
It follows from this that
\begin{align*}
b(g+G,x) &= e^{iGx}e^{igx}u(g+G,x)  = e^{igx}u(g,x),
\end{align*}
which means that,
\begin{align*}
u(g+G,x) & = e^{-iGx}u(g,x), 
\end{align*}
for all ${g\in\reciplattg}$, for all ${G\in\reciplatt}$, 
and for all ${x\in\onetorus}$.

\subsubsection{Eigenvalue equation}
By replacing $x$ with ${x-m\volume}$ in the eigenvalue equation, 
\begin{align*}
\left(\hamsmallx\varphi\right)(x)&=\epsilon\varphi(x),
\end{align*}
and
expressing $\varphi$ as in Eq.~\ref{eqn:sumk0}, 
the 
$\volume$-periodicities of
${e^{iGx}}$ and ${u(k)}$ can be used to
show that
\begin{align*}
\sum_k \big[\hamsmallx b(k,x-m\volume) - \epsilon b(k,&x-m\volume)\big] 
\\
= \sum_k e^{-ikm\volume}\big[\hamsmallx b(k,x) &- \epsilon b(k,x)\big] = 0.
\end{align*}
If we multiply this by ${e^{ik'm\volume}}$ and sum over $m$ we get
\begin{align*}
\sum_k \left(\sum_{m=0}^{\Nunitcell-1}e^{i(k'-k)m\volume}\right)&\big[\hamsmallx b(k,x) - \epsilon b(k,x)\big] = 0;
\end{align*}
and, once again, the sum over $m$ vanishes unless ${k'=k}$.
Therefore, since ${k'\in\BZ}$ is arbitrary, we have
found that
\begin{align*}
\hamsmallx b(k,x) & = \epsilon b(k,x).
\end{align*}
It follows from Eq.~\ref{eqn:ham_bloch_periodic} and the $\hreciplatt$-periodicity
of ${b(k,x)}$ that
\begin{align}
\hamsmallx b(k+G,x) & = \epsilon b(k+G,x), \; \forall G\in\reciplatt.
\label{eqn:bloch_eigenfunctionG}
\end{align}

\subsubsection{Summary of subsections~\ref{section:fourier_series_expansion} and~\ref{section:bloch_functions}}
Subsection~\ref{section:fourier_series_expansion} showed that
an \emph{arbitrary} eigenfunction $\varphi$ of $\hamsmallx$ could be
expressed as the sum of Bloch functions, 
\begin{align*}
\varphi=\sum_k b(k+G),
\end{align*}
where ${G}$ is an arbitrary reciprocal lattice vector and 
each Bloch function ${b(k+G)}$ in the sum is associated with a different element $k$ of ${\BZ}$.

Subsection~\ref{section:bloch_functions} has shown that Bloch functions are ${\hreciplatt}$-periodic (${b(G+k)=b(k), \forall G\in\reciplatt,\;\forall k\in\BZ}$)
and that each term in the sum satisfies Eq.~\ref{eqn:bloch_eigenfunctionG}. Therefore, each
Bloch function
in the sum either vanishes or is itself an eigenfunction of ${\hamsmallx}$ with
the same eigenvalue $\epsilon$ as $\varphi$.

\subsection{Elements of ${\{b(G+k):k\in\BZ\}}$ are not related by lattice translations}
\label{section:lattice_translations}
As discussed in Appendix~\ref{section:degeneracies_imply_symmetries}, two
eigenfunctions of ${\hamsmallx}$ have exactly the same eigenvalue if and only if
symmetry demands it. In other words, if two Bloch eigenfunctions share an eigenvalue, 
they are equivalent to one another by symmetry.

It will now be shown that translating a Bloch eigenfunction by
a lattice vector does not change the element of ${\BZ}$ with which
it is associated. Therefore the invariance of $\hamsmallx$ under translations by
elements of the set ${\volume\integer}$ of lattice vectors would not explain
multiple elements of ${\{b(G+k):k\in\BZ\}}$ having the same finite eigenvalue.

The translation operator ${\trans_m}$, where ${m\in\integer}$, is defined by its action on an arbitrary
function ${f\in\lebesgue(\onetorus)}$, as follows:
\begin{align*}
\trans_m f(x) = f(x-m\volume).
\end{align*}
To see that Bloch eigenfunctions, ${b(k)}$ and ${b(k')}$, at different wavevectors ${k,k'\in\BZ}$ 
are not images of one another under lattice translations, let us translate 
${b(k)}$ by an arbitrary lattice vector ${m\volume\in\volume\integer}$.
\begin{align*}
\trans_m b(k,x)
&\equiv b(k,x-m\volume) 
= e^{ik(x-m\volume)}u(k,x-m\volume).
\\
&= e^{-ikm\volume}\left(e^{ikx}u(k,x)\right)
=e^{-ikm\volume}b(G+k,x).
\end{align*}
Therefore translating an arbitrary Bloch function ${b(G+k)}$ is
equivalent to multiplying it by the factor ${e^{-ikm\volume}}$;
and the only element of $\BZ$ on which ${e^{-ikm\volume}}$ depends
is $k$. 

In summary, 
for any ${m\in\integer}$, 
the Bloch function ${b(k)}$ (${=b(G+k),\;\forall G\in\reciplatt}$)
and its image
under translation by lattice vector ${m\volume}$
are both associated with the same element ${k}$ of $\BZ$.
They differ only by the $x$-independent phase factor, ${\exp(-ikm\volume)}$.

\subsection{Parity}
\label{section:parity}
It has been shown above that if there exist multiple
Bloch eigenfunctions with the same eigenvalue, and if they are associated with different elements of $\BZ$,
the symmetries responsible for them having the same eigenvalue are not lattice translations.
Since $\volume$-periodicity is the only symmetry that we explicitly 
assumed the crystal to have, either our construction of the crystal in $\onetorus$ implicitly
assumes the existence of one or more other symmetries, or
the sum over ${k}$ in Eq.~\ref{eqn:eigenfunction2-0}  only has
one non-vanishing term.

Assuming that Bloch functions are complex-valued, in general, 
and have $x$-dependent phases,
we can dismiss the second possibility. This is because, if we take the complex conjugate of 
both sides of Eq.~\ref{eqn:bloch_eigenfunctionG}, we get
\begin{align}
\hamsmallx b^*(k,x)=\epsilon b^*(k,x).
\label{eqn:bloch_conjugate_eigen}
\end{align}
Therefore, if ${b(k)}$ is an eigenfunction of
$\hamsmallx$ with a finite eigenvalue, then
\begin{align*}
b^*(k) &= e^{-ikx}u^*(k,x) 
= e^{-ikx}u^*(k,x)
\end{align*}
is an eigenfunction of ${\hamsmallx}$ with the same eigenvalue.

Note that, since ${u(k)}$ is ${\volume}$-periodic, its
real and imaginary parts are ${\volume}$-periodic. Therefore
${u^*(k)}$ is $\volume$-periodic, which means that
${b^*(k)}$ has exactly the form expected of the Bloch function, 
\begin{align*}
b(-k,x) = e^{-ikx}u(-k,x), 
\end{align*}
which is the Bloch function associated with element $-k$ of $\BZ$.

Also note that
\begin{align*}
k\in\BZ\cup\left\{-\pi/\volume\right\}\iff -k\in\BZ\cup\left\{-\pi/\volume\right\}
\end{align*}
and
\begin{align*}
k\in\BZ\setminus\left\{\pi/\volume\right\}\iff -k\in\BZ\setminus\left\{\pi/\volume\right\}.
\end{align*}
In words: with the exception of the wavevector ${\pi/\volume}$ 
on the boundary of the first Brillouin zone (for which ${b(\pi/\volume)=b(-\pi/\volume)}$ due to 
$\hreciplatt$-periodicity), the negative
of every element of $\BZ$ is also an element of $\BZ$.

Therefore, assuming that ${b(k)}$ does not vanish, and that it differs from ${b^*(k)}$ by more than
a constant, which implies that the real and imaginary 
parts of ${b(k)}$ differ by more than a constant, then ${b^*(k)=b(-k)}$;
and ${b(k)}$ and ${b(-k)}$ are eigenfunctions of $\hamsmallx$ that have the
same eigenvalue. 

Since ${b(k)}$ and ${b(-k)}$ have the same eigenvalue, and since we have already
ruled out the possibility that the crystal's $\volume$-periodicity could be
the symmetry responsible for this, there must be another symmetry of the system that explains it.

\subsubsection{Complex Bloch functions imply higher dimensions}
If ${b(k)}$ is real-valued, then ${b^*(k)=b(k)}$. In that case, 
Eq.~\ref{eqn:bloch_conjugate_eigen} does not imply the existence
of degenerate eigenvalues, and does not imply that 
an additional symmetry of the system has implicitly been assumed.

\begin{figure}[!]
\centering
\includegraphics[width=0.50\textwidth]{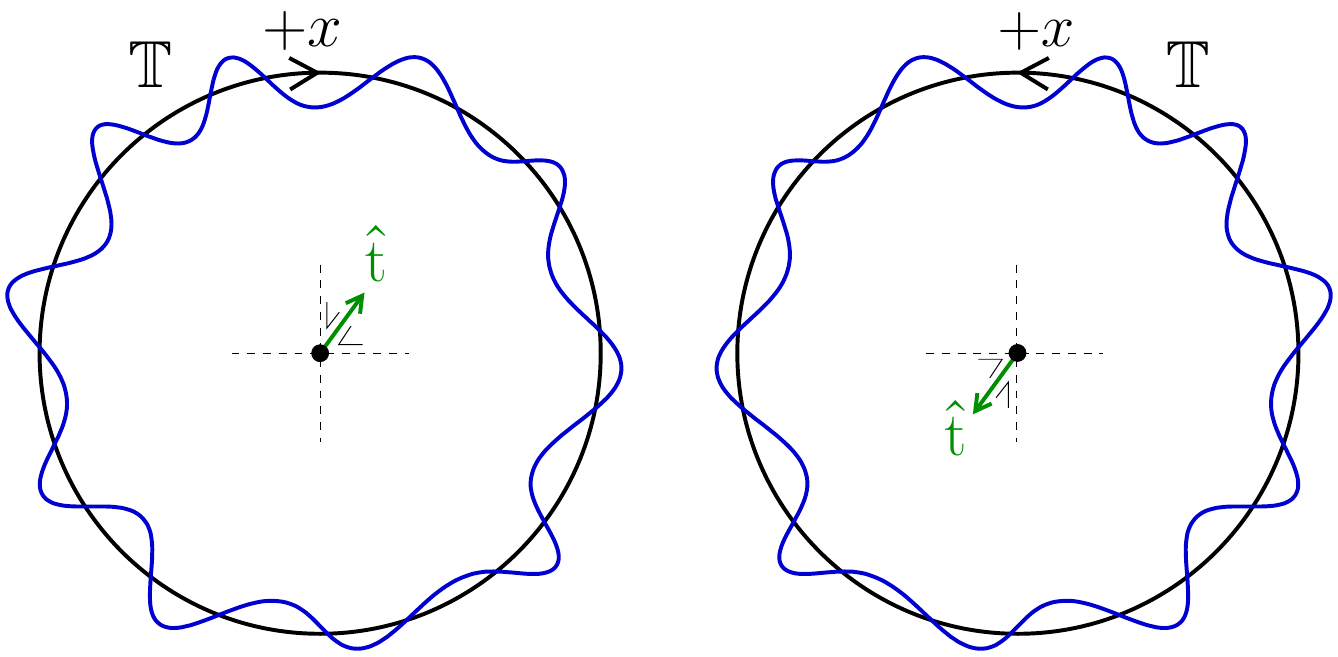}
\caption{The black circles on the left and right are the
$1$-torus $\onetorus$, which inhabits the
Cartesian ${y-z}$ plane.  The radial displacement of
the blue curve from the black circle is the value ${b_k(x)}$ of a $\bulksize$-periodic
function ${b_k:\onetorus\to\realone}$, ${x\mapsto b_k(x)}$.
The only difference between the two figures is that, on the left
the torus is viewed along the positive $t$ direction, whereas
on the right it is viewed along the negative $t$ direction.
}
\label{fig:parity}
\end{figure}

Let us focus on the case in which ${b(k)}$ has both real and imaginary parts, 
which means that there exists an unexplained degeneracy.
It will be useful to consider Fig.~\ref{fig:parity}: In the schematic on the left
hand side, each point on the black circle represents a different point ${x\in\onetorus}$;
and the radial displacement of the blue curve from the black circle
represents the real part of ${b_k=b(k)\in\lebesgue(\onetorus)}$ at $x$.

Recall that 
\begin{align*}
b_k(x)=\braket{x}{b(k)}=\Re\left\{\braket{x}{b(k)}\right\}+i\Im\left\{\braket{x}{b(k)}\right\}, 
\end{align*}
where ${\ket{x}}$ denotes the counterpart in $\hilbert$ of an element of ${\lebesgue(\onetorus)}$
that is localized around a point ${x\in\onetorus}$ (see Appendix~\ref{section:appendix_states_as_vectors}).
Also recall from Sec.~\ref{section:imaginary_inner_product} that the inner product of 
${\ket{x}}$ and ${\ket{b(k)}}$ is not ${\braket{x}{b(k)}}$, but 
${\Re\left\{\braket{x}{b(k)}\right\}}$; and that ${\Im\left\{\braket{x}{b(k)}\right\}}$
does not have any meaning unless the space ${\Span\left\{\ket{x},\ket{b(k)}\right\}}$ is embedded in a higher
dimensional space.

In Fig.~\ref{fig:parity}, 
${\ket{b(k)}}$ represents the blue curve, 
${\ket{x}}$ represents a function localized at a point $x$ on the black
circle, 
and ${\Re\{\braket{x}{b(k)}\}}$ is the 
radial displacement of the blue curve from the black circle at that point.
Therefore, if ${b_k(x)}$ has an imaginary part, 
the plane in which the black circle resides (the plane of the page) must be embedded
in a higher dimensional space. This is because, as noted in Sec.~\ref{section:imaginary_inner_product},  
${i\Im\{b_k(x)\}}$
represents the outer product of ${\ket{x}}$ and ${\ket{b(k)}}$, which is a bivector.
The vector ${\hat{\mathrm{t}}}$ resides in that higher dimensional space, and is perpendicular
to the plane of the page.

The axis parallel to $\hat{\mathrm{t}}$ might be the time axis, or it might be a spatial axis, or
it might have temporal and spatial components, or it might not be an axis in spacetime. 
If it is an axis in spacetime, and 
since real crystals are three dimensional, it is usually regarded as the time axis.
Regardless of what axis it is, the schematic on the right hand side of Fig.~\ref{fig:parity} shows the plane of the circle
from the opposite side. 

Now, since the blue curve in Fig.~\ref{fig:parity} is an eigenfunction ${b_k(x)}$ of ${\hamsmallx}$;
and since ${\hamsmallx}$ does not depend on the `side' of the plane that the torus
is on (i.e., on whether 
${\hat{\mathrm{t}}}$ points into or out of the plane), both sides of the plane must be equivalent.
In other words, if all of the eigenfunctions of ${\hamsmallx}$ with a given eigenvalue were plotted, the
set of curves would appear the same from both sides of the plane. This can only be the
case if there exists an eigenfunction that looks just like ${b_k(x)}$ when it is viewed
from the opposite side of the plane of $\onetorus$.

\begin{figure}[!]
\centering
\includegraphics[width=0.30\textwidth]{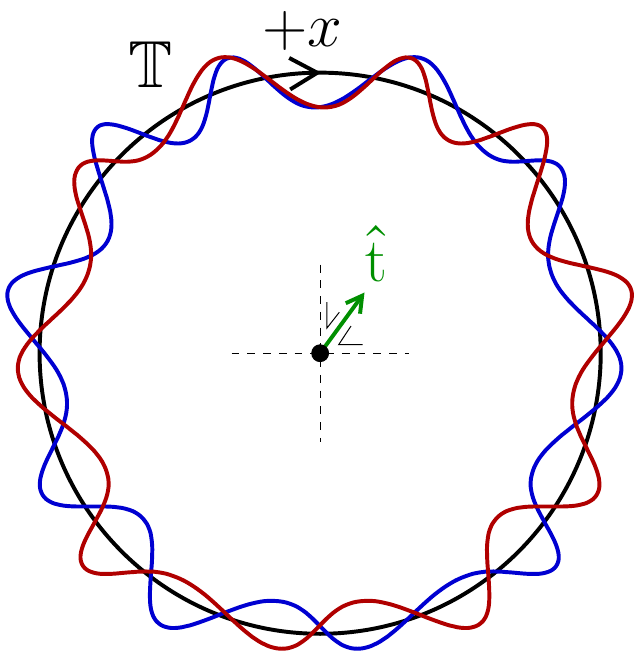}
\caption{The same as the left panel of Fig.~\ref{fig:parity}, but with 
the addition of the red curve. The red curve 
is identical to the blue curve in the right panel of Fig.~\ref{fig:parity}.
Since the blue curve is a eigenfunction 
of a Hamiltonian that is independent of whether
${\hat{\mathrm{t}}}$ is directed into the page or out of the page, 
there must exist another eigenfunction, with the same eigenvalue, 
whose shape is the same as blue curve when it is viewed
from the opposite side of the plane. That solution is the red curve.
}
\label{fig:parity2}
\end{figure}

That eigenfunction is the red curve in Fig.~\ref{fig:parity2}, which is
identical to the blue curve on the right of Fig.~\ref{fig:parity}.
The symmetry that relates the red curve and blue curve is usually referred
to as \emph{time inversion symmetry}. However, it is clear from Fig.~\ref{fig:parity}, 
that inverting ${\hat{\mathrm{t}}}$ would be equivalent to changing the positive $x$ direction
from clockwise to anticlockwise, i.e., 
\begin{align*}
x&\mapsto -x 
\\
\implies b(k,x)&\mapsto b(k,-x) 
\\
\implies 
e^{ikx}u(k,x)&\mapsto e^{-ikx}u(k,-x)  
\end{align*}
Therefore, 
\begin{align*}
u^*(k,x)=u(-k,x)=u(k,-x),
\end{align*}
and
\begin{align*}
b^*(k,x)=b(-k,x)=b(k,-x).
\end{align*}
This demonstrates that any Bloch eigenfunction, which is associated
with any wavevector ${k\in\BZ}$, is doubly degenerate as a consequence
of $\hamsmallx$ being independent of the \emph{parity} of 
${\hat{\mathrm{t}}}$ and the positive ${x}$ direction. 
In simple terms, if you close your fists
and extend your thumbs, and imagine that your thumbs 
are parallel to $\hat{\mathrm{t}}$, 
the fingers of only one of your hands go around its thumb in the positive $x$ direction.
The Hamiltonian $\hamsmallx$ does not depend on which hand this is. This independence
is the symmetry
responsible for ${b(k,x)}$ and ${b(-k,x)}$ having the same eigenvalue.
We may call it \emph{parity inversion symmetry}, but it is more
commonly referred to as \emph{time inversion symmetry}.


\subsection{Notation, normalization, and orthogonality}
In general, $\hamsmallx$ has an infinite number of
different eigenvalues. Since no assumptions
were made about $\varphi$, everything in this 
appendix from subsection~\ref{section:fourier_series_expansion}
onwards applies to each eigenfunction independently.
Therefore, for every different eigenvalue $\epsilon$ of $\hamsmallx$ 
there are two Bloch eigenfunctions that are equivalent to one
another under parity inversion symmetry, each of which is associated with
a different element of $\BZ$. If one of them is associated with ${k\in\BZ}$, 
the other is associated with ${-k\in\BZ}$.

Since ${\hamsmallx}$ has an infinite number of eigenfunctions, whereas
${\BZ}$ is a finite set, there are an infinite number of eigenfunctions
associated with each element $k$ of $\BZ$.
Therefore, each eigenfunction is usually identified
by a double index ${\alpha k}$, where index $\alpha$ distinguishes
between different eigenfunctions at the same wavevector $k$.
In other words, the elementary eigenfunctions of ${\hamsmallx}$, 
can be expressed as
\begin{align}
b_{\alpha k}(x) 
\equiv b_\alpha(k,x)
= e^{ikx} u_{\alpha k}(x)\equiv e^{ikx}u_\alpha(k,x),
\label{eqn:blochfunction}
\end{align}
where ${u_{\alpha k}(x)\equiv u_\alpha(k,x)}$ is ${\volume}$-periodic.

Here and in Sec.~\ref{section:single_particle_states} and Appendix~\ref{section:appendix_wannier} it will
sometimes be convenient to denote the Bloch functions as ${b_\alpha(k,x)}$ and the periodic Bloch functions as ${u_\alpha(k,x)}$.
At other times it will be convenient to denote them as ${b_{\alpha k}(x)}$ and ${u_{\alpha k}(x)}$, respectively.
Therefore these notations will be used interchangeably.
Similar interchangeable notations will be used for eigenstates and eigenvalues, namely,
${\epsilon_{\alpha k}\equiv \epsilon_\alpha(k)}$
and
\begin{align*}
\ket{b_{\alpha k}}\equiv \int_\onetorus \dd{x} b_{\alpha k}(x)\dyad{x}\equiv \ket{b_\alpha(k)}.
\end{align*}

Since ${\braketT{b_{\alpha k}}{b_{\alpha k}}=\braketT{u_{\alpha k}}{u_{\alpha k}}}$, each Bloch function ${b_{\alpha k}}$ bestows its unit 
normalization on its periodic counterpart, ${u_{\alpha k}}$. However, the set of periodic Bloch functions does not
inherit orthogonality from the set of Bloch functions, in general, i.e., 
\begin{align*}
\braketT{b_{\alpha k}}{b_{\beta k'}}
&=\int_{\onetorus}b_{\alpha k}^*(x) b_{\beta k'}(x)\dd{x}=\delta_{\alpha\beta}\delta_{kk'}
\\
&=\int_{\onetorus}u_{\alpha k}^*(x)e^{i(k'-k)x} u_{\beta k'}(x)\dd{x}\neq\braketT{u_{\alpha k}}{u_{\beta k'}}.
\end{align*}
However, when ${k=k'}$, the exponential in the integrand is unity, and this equation becomes
\begin{align*}
\braketT{b_{\alpha k}}{b_{\beta k}}=
\braketT{u_{\alpha k}}{u_{\beta k}}=\delta_{\alpha\beta}.
\end{align*}

\onecolumngrid
\vspace{0.8cm}
\PRLsep
\vspace{1cm}
\twocolumngrid
\section{Wannier functions of minimal width}
\label{section:appendix_wannier}
\subsection{Introduction}
{\em Wannier function of minimal width} is simply another term for {\em maximally localized Wannier function}~\citep{marzari_mlwf}.
The title of this section acknowledges the work of 
Ferreira and Parada, on which it is based~\citep{ferreira_parada}. 
My presentation differs from theirs in several ways, the most deliberate of which 
is my avoidance of quantum mechanical perturbation theory.
This is to demonstrate that there is nothing specific to quantum mechanics
in the theory of Wannier functions and their relationships with Bloch functions.

All of this section would apply to the eigenfunctions,
\begin{align*}
b_{\alpha k}(x)\equiv \braket{x}{b_{\alpha k}} \equiv b_{\alpha}(k,x),
\end{align*}
of any bounded and self-adjoint $\volume$-periodic operator, 
\begin{align*}
\hamsmallx:\lebesgue(\onetorus)&\to\lebesgue(\onetorus),
\\
\hamsmallx(x)\equiv \expval{\hamsmall}{x}&=\hamsmallx(x+m\volume), \;\forall m\in\integer, 
\end{align*}
where ${\ket{b_{\alpha k}}=\ket{b_\alpha(k)}}$ is an eigenstate of 
\begin{align*}
\hamsmall = \int_\onetorus \dd{x} \hamsmallx(x) \dyad{x},
\end{align*}
and the circumference of $\onetorus$ is ${\bulksize=\Nunitcell\volume}$, where
${\Nunitcell\in\integerpos}$.

For example, consider a classically-modelled process in a crystal whose bulk
is represented in $\onetorus$. If ${\pdf(x)}$ is a 
a one-particle position probability density function 
that has the crystal's $\volume$-periodicity, it could be specified 
by a smooth function ${\psi(x)=\sqrt{\pdf(x)}e^{i\theta(x)}}$;
and $\psi$ could be expanded in either a Bloch basis or a Wannier basis. Either basis 
could be used to build a basis of many-particle states to represent a function
$\Psi$ whose square modulus is a classical many-particle
position probability density function.

\subsection{Theoretical setup}
In this appendix it will be assumed that ${\{b_{\alpha k}\}}$
is an orthonormal set of single-particle Bloch functions, 
\begin{align*}
b_{\alpha k}(x)\equiv b_\alpha(k,x) \equiv e^{ikx}u_{\alpha k}(x)\equiv e^{ikx}u_\alpha(k,x),
\end{align*}
which is a complete basis of ${\lebesgue(\onetorus)}$. 
The subscripts $\alpha$ and $k$ of the set ${\{b_{\alpha k}\}}$ indicate that its elements include those
with all band indices $\alpha$, and all wavevectors $k$ that are elements of ${\BZ}$ (see Appendix~\ref{section:appendix_fourier}).
Orthonormality of the set of Bloch functions can be expressed as 
\begin{align*}
\braketT{b_{\alpha k}}{b_{\beta k'}}\equiv\int_\onetorus\dd{x} b_{\alpha k}^*(x) b_{\beta k'}\dd{x}=\delta_{\alpha\beta}\delta_{kk'}.
\end{align*}

As discussed in Appendix~\ref{section:appendix_torus},
at each point ${x\in\onetorus}$, the set ${\{b_{\alpha k}(x)\}_k}$ 
of all Bloch functions whose band indices are $\alpha$
can be regarded as a function ${b_\alpha(k,x)}$ of
$k$, which is ${\hreciplatt}$-periodic, meaning that
\begin{align*}
b_\alpha(g+G,x)=b_\alpha(g,x),
\end{align*}
for any ${g\in\reciplattg}$ and any
reciprocal lattice vector ${G\in\reciplatt}$; and it will be assumed that the periodic Bloch functions ${u_{\alpha k}(x)=u_\alpha(k,x)}$
are real valued for all ${k\in\BZ}$. 
These assumptions are discussed and justified
in Appendix~\ref{section:appendix_torus}, and the ${\hreciplatt}$-periodicity of ${b_\alpha(g)}$ is 
shown to imply that
\begin{align*}
u_\alpha(g+G,x)=e^{-iGx}u_{\alpha}(g,x), 
\end{align*}
for any ${g\in\reciplattg}$ and any reciprocal lattice vector ${G\in\reciplatt}$.

This appendix considers
a general Wannier transformation of the 
set ${\{b_{\alpha k}\}_k\subset\{b_{\alpha k}\}_{\alpha k}}$ 
to a Wannier function ${w_\alpha(x)}$.
The general Wannier transformation will be 
used to deduce the specific Wannier transformation that minimizes the spread of ${w_\alpha(x)}$.

We have assumed that the set of Bloch functions is orthonormal and complete. 
Therefore there exist an infinite
number of Hermitian operators ${\expval{\hamsmall}{x}}$, where
\begin{align*}
\hamsmall\equiv\sum_{\alpha k}\epsilon_{\alpha k}\dyad{b_{\alpha k}}, 
\end{align*}
of which they are eigenfunctions. These operators 
differ only  by their sets of eigenvalues, ${\{\epsilon_{\alpha k}\}_{\alpha k}}$. 

One of them will be denoted by $\hamsmallx$,
and it will be assumed that
${\hamsmallx}$ varies smoothly with $x$.
This facilitates the assumption that all required partial derivatives of each Bloch
function with respect to $k$ and/or $x$ exist, where, as in Sec.~\ref{section:bvkany}, 
\begin{align*}
\partial_k b_\alpha(k,x) \equiv \lim_{\bulksize\to\infty} \hbulksize^{-1} 
\left[b_\alpha\left(k+\hbulksize,x\right)-b_\alpha\left(k,x\right)\right].
\end{align*}

\subsection{The most localizing Wannier transformation}
Consider the general Wannier transformation,
\begin{align}
w_\alpha(x) = \intbz f_\alpha(k) b_\alpha(k,x)\dd{k},
\label{eqn:wannier}
\end{align}
where, as in Sec.~\ref{section:bvkany}, ${\intbz\dd{k}}$ denotes
the sum ${\hbulksize\sum_{k\in\BZ}}$;
and ${f_\alpha:\reciplattg\to\complex}$ is ${\hreciplatt}$-periodic
and normalized on $\BZ$. That is, 
for any ${g\in\reciplattg}$ and any ${G\in\reciplatt}$, 
${f_\alpha(g+G) = f_\alpha(g)}$ and
\begin{align*}
\intbz f^*_\alpha(k) f_\alpha(k)\dd{k} &= 
\intbz f^*_\alpha(g+k) f_\alpha(g+k)\dd{k} =  1.
\end{align*}
As we are interested in finding the most localized Wannier function, 
let us assume that $f_\alpha(g)$ is a smooth function of $g$, and
that ${b_\alpha(g,x)}$ is a smooth function of both $g$ and $x$.

We will be assuming that we are in the limit ${\bulksize\to\infty}$, 
which means that the set of points in $\BZ$ is quasicontinuous. It also
means that we can assume that the width of $w_\alpha$ is much smaller
than $\bulksize$. 
Therefore when we calculate the spread,
\begin{align*}
W_\alpha(X) = \int_\onetorus \abs{w_\alpha(x)}^2(x-X)^2\dd{x},
\end{align*}
of $w_\alpha$ about an arbitrary point $X\in\onetorus$, or when we calculate
any integral of a \emph{localized} function of position in $\onetorus$, 
$\int_\onetorus$ should always be taken to mean
${\int_{x_0}^{x_0+\bulksize}}$, where $x_0$ is chosen such
that the integrand is negligible at the point ${x_0=x_0+\bulksize\in\onetorus}$. 

Let us define a generating function ${G_\alpha(s,X)}$ from which $W_\alpha(X)$ 
can be calculated, as follows:
\begin{align}
 G_\alpha(s,X) &\equiv \int_\onetorus \abs{w_\alpha(x)}^2 e^{-is(x-X)} \dd{x}
\label{eqn:G}
\\
\implies 
W_\alpha(X)  
 &= \lim_{s\to 0}\int_\onetorus \abs{w_\alpha(x)}^2 \left(x-X\right)^2 e^{-is\left(x-X\right)}\dd{x}  
\nonumber
\\
&= -\lim_{s\to 0} \partial_s^2 G_\alpha(s,X) 
\label{eqn:W}
\end{align}
${G_\alpha(s,X)}$ is a Fourier transform of ${\abs{w_\alpha(x)}^2}$ after 
it has been displaced by ${-X}$. Therefore, if $X$ was the center of $w_\alpha$, 
${G_\alpha(s,X)}$ would be the Fourier transform of $w_\alpha$ after 
its center had been moved to the origin.

Inserting Eq.~\ref{eqn:wannier} into Eq.~\ref{eqn:G} gives 
\begin{align}
G_\alpha(s,X) 
 = 
&
\intbz\dd{k'} 
\intbz\dd{k} 
\fca(k')
f_\alpha(k+s)
e^{isX}
\nonumber \\
&\times
\int_\onetorus\dd{x}
\bca(k',x)
b_\alpha(k+s,x)
e^{-isx},
\label{eqn:G2}
\end{align}
where the ${\hreciplatt}$-periodicities of $b_\alpha$ and $f_\alpha$
have been used
to shift the domain of the integration over $k$ from $\BZ$ to $\BZ+s$.

The set 
${\left\{u_{\alpha k}\right\}_{\alpha}}$ of all 
periodic Bloch functions at $k$ is orthonormal
because we have chosen the set of all Bloch functions to be orthonormal, 
i.e., 
\begin{align*}
\braket{b_{\alpha k}}{b_{\beta k'}}
&=
\int_\onetorus\dd{x} b_{\alpha k}^*(x) b_{\beta k'}(x)=
\delta_{\alpha\beta}\delta_{kk'} 
\\
\implies 
\braket{b_{\alpha k}}{b_{\beta k}}&=
\braket{u_{\alpha k}}{u_{\beta k}}=\delta_{\alpha\beta}.
\end{align*}
Furthermore, the set ${\{u_{\alpha k}\}_\alpha}$ of periodic Bloch functions at $k$ is 
a complete basis of ${\lebesgue(\onetorus)}$ because it is
the set of eigenfunctions of an operator ${e^{-ikx}\hamsmallx e^{ikx}}$, which is
Hermitian because $\hamsmallx$ is Hermitian. 

The completeness of ${\{u_{\alpha k}\}_\alpha}$ allows us to express ${u_\alpha(k+s,x)}$ as
\begin{align}
u_\alpha(k+s,x) & = \sum_{\beta} C_{\alpha\beta}(k,s)u_\beta(k,x)
\label{eqn:uks}
\\
\implies
b_\alpha(k+s,x)&=e^{isx}\sum_\beta C_{\alpha\beta}(k,s)b_\beta(k,x),
\label{eqn:bks}
\end{align}
where
\begin{align}
C_{\alpha\beta}(k,s) &\equiv \braket{u_\beta(k)}{u_\alpha(k+s)} \nonumber \\
& \equiv \int_{\onetorus} u^*_\beta(k,x)u_\alpha(k+s,x)\dd{x}.
\label{eqn:C2}
\end{align}
Equation~\ref{eqn:C2} implies that the orthonormality of the set of periodic Bloch functions at wavevector $k$ can be expressed
as ${C_{\alpha\beta}(k,0)=\delta_{\alpha\beta}}$.

Inserting Eq.~\ref{eqn:bks} into Eq.~\ref{eqn:G2} gives
\begin{align}
G_\alpha(s,X) = 
e^{isX}
\sum_{\beta}
&
\intbz\dd{k'} 
\intbz\dd{k} 
\fca(k')
f_\alpha(k+s)\nonumber \\
&\times C_{\alpha\beta}(k,s) 
\braket{b_{\alpha k'}}{b_{\beta k}},
\label{eqn:G3}
\end{align}
and the orthogonality of the set of Bloch functions allows
this to be simplified to
\begin{align}
G_\alpha(s,X)& = 
e^{isX}
\intbz\dd{k} \fca(k) f_\alpha(k+s) C_{\alpha\alpha}(k,s).
\label{eqn:G4}
\end{align}
It follows that
\begin{widetext}
\begin{align}
\partial^2_s G_\alpha(s,X)
= 
e^{isX}
\intbz\dd{k} \fca(k) 
\bigg[
&C_{\alpha\alpha}(k,s) 
\partial^2_s f_\alpha(k+s)
+
2\partial_s f_\alpha(k+s) \partial_s C_{\alpha\alpha}(k,s) 
+
2iX f_\alpha(k+s) \partial_s C_{\alpha\alpha}(k,s)
\nonumber
\\
+
&2iX C_{\alpha\alpha}(k,s) \partial_s f_\alpha(k+s) 
-X^2 f_\alpha(k+s) C_{\alpha\alpha}(k,s)
+ f_\alpha(k+s) \partial_s^2 C_{\alpha\alpha}(k,s)
\bigg]
\label{eqn:d2G}
\end{align}
\end{widetext}
Before calculating ${W_\alpha(X)=-\lim_{s\to 0}\partial_s^2 G_\alpha(s,X)}$ some useful
results will be deduced and derived. 

The first thing to note is that, because
${u_{\alpha k}}$ inherits smoothness from ${b_{\alpha k}}$,
\begin{align*}
\lim_{s\to 0}\partial_s^n u_\alpha(k+s,x)= 
\partial_k^n u_\alpha(k,x), \;\forall n\in\integerpos.
\end{align*}
Therefore the Taylor expansion of ${u_\alpha(k+s,x)}$ about ${s=0}$
is
\begin{align}
u_\alpha(k+s,x) 
= 
u_\alpha(k&,x)+s\partial_k u_\alpha(k,x) 
\nonumber
\\
&+\frac{1}{2} s^2 \partial^2_k u_\alpha(k,x) + \order{s^3}.
\label{eqn:utaylor}
\end{align}
Inserting this into Eq.~\ref{eqn:C2} gives
\begin{align*}
C_{\alpha\beta}(k,s) = \delta_{\alpha\beta}&+s\braket{u_{\beta k}}{\partial_k u_{\alpha k}} 
\\
&+\frac{1}{2}s^2\braket{u_{\beta k}}{\partial_k^2 u_{\alpha k}}
+\order{s^3}
\\
\implies C_{\alpha\alpha}(k,s) & = 1 + 
\frac{1}{2}s^2\braket{u_{\alpha k}}{\partial_k^2 u_{\alpha k}}+\order{s^3},
\end{align*}
where the term ${s\braket{u_{\alpha k}}{\partial_k u_{\alpha k}}}$ vanishes
because the only way for the normalization ${\braket{u_{\alpha k}}=1}$ to be preserved as $k$ varies is if 
${\partial_k u_{\alpha k}}$ is orthogonal to ${u_{\alpha k}}$. 
It follows that the first derivative of ${C_{\alpha\alpha}(k,s)}$ with respect to $s$ is
\begin{align*}
\partial_s C_{\alpha\alpha}(k,s) 
& =  s\braket{u_{\alpha k}}{\partial_k^2 u_{\alpha k}} 
+\order{s^2}, 
\end{align*}
which vanishes in
limit ${s\to 0}$  because all terms on the right hand side 
are proportional to powers of $s$.
The second derivative of ${C_{\alpha\alpha}(k,s)}$ with respect to $s$ is
\begin{align*}
\partial^2_s C_{\alpha\alpha}(k,s) & =  \braket{u_{\alpha k}}{\partial_k^2 u_{\alpha k}} 
+\order{s},
\end{align*}
and all terms with powers of $s$ as prefactors vanish in the limit ${s\to 0}$.
By integrating by parts it is found that
\begin{align*}
\braket{u_{\alpha k}}{\partial_k^2 u_{\alpha k}}
&=-\braket{\partial_k u_{\alpha k}}{\partial_k u_{\alpha k}},
\end{align*}
where the boundary term vanishes because the limits of integration, $x_0$ 
and ${x_0+\bulksize}$, are the same point in $\onetorus$. 
It follows that
\begin{align*}
\lim_{s\to 0} \partial^2_s C_{\alpha\alpha}(k,s) & = -\braket{\partial_k u_{\alpha k}}{\partial_k u_{\alpha k}}.
\end{align*}
The following results will help us to take the ${s\to 0}$ limit of Eq.~\ref{eqn:d2G}:
\begin{align*}
&\lim_{s\to 0} C_{\alpha\alpha}(k,s)=1,
\\
&\lim_{s\to 0} \partial_s C_{\alpha\alpha}(k,s)=0,
\\
&\lim_{s\to 0}\partial_s^2 C_{\alpha\alpha}(k,s)
=-\braket{\partial_k u_{\alpha k}}{\partial_k u_{\alpha k}}, 
\\
&\lim_{s\to 0} \partial^n_s f_\alpha(k+s)=\partial^n_k f_\alpha(k).
\end{align*}
We find that
\begin{align*}
W_\alpha&(X) =-\lim_{s\to 0} G_\alpha(s,X)
\\
=&
\intbz\dd{k} \fca(k) 
\bigg[
-
\partial^2_k 
-
2iX \partial_k 
+X^2  
+ t_\alpha(k)
\bigg]
f_\alpha(k),
\end{align*}
where
${t_\alpha(k)\equiv
\braket{\partial_k u_{\alpha k}}{\partial_k u_{\alpha k}}}$.

By defining the function
${g_\alpha(k)\equiv e^{ikX} f_\alpha(k)}$, we can express ${W_{\alpha}(X)}$
as
\begin{align}
W_\alpha(X) = \intbz\gca(k)
\left[
-\partial_k^2
+
t_\alpha(k)
\right]g_\alpha(k)\dd{k}.
\label{eqn:Wfinal}
\end{align}
This is stationary with respect to norm-preserving variations of $g_\alpha$ when $g_\alpha$ is
any eigenfunction of 
\begin{align*}
\recham_\alpha(k)\equiv\expval{\recham_\alpha}{k}\equiv-\partial_k^2+t_\alpha(k), 
\end{align*}
and the stationary values of $W_\alpha(X)$ are the eigenvalues of $\recham_\alpha(k)$.
Note that the relationship in reciprocal space between ${u_\alpha(k,x)}$ and $t_\alpha(k)$ is the same, 
up to a multiplicative constant, as the
relationship in real space between the Bloch function ${b_{\alpha k}}$ and 
what would be referred to within quantum mechanics as its 
kinetic energy.   Also note that 
the operator $\recham_\alpha(k)$ is like a Hamiltonian in reciprocal space; and that
${t_\alpha(k)}$ plays the role of a positive potential in reciprocal space
for ${g_{\alpha}}$. 

If we now set ${g_\alpha(k)\equiv \hreciplatt^{-1} e^{i\theta_\alpha(k)}}$, 
${f_\alpha}$ has the required normalization, Eq.~\ref{eqn:wannier} becomes
\begin{align}
w_\alpha(x) 
&= \hreciplatt^{-\frac{1}{2}}\intbz e^{-ikX} e^{i\theta_\alpha(k)} b_\alpha(k,x)\dd{k}
\end{align}
and Eq.~\ref{eqn:Wfinal} becomes
\begin{align}
W_\alpha(X) = \frac{1}{\hreciplatt}\intbz
\left[
\abs{\partial_k\theta_\alpha}^2
+
t_\alpha(k)
\right]\dd{k}.
\label{eqn:Wfinal2}
\end{align}
This has its minimum value when $\theta_\alpha$ is a constant, and this minimum value is
\begin{align}
W_\alpha^{\textrm{min}}(X) = \frac{1}{\hreciplatt} \intbz \braket{\partial_ku_{\alpha k}}{\partial_ku_{\alpha k}} \dd{k}.
\end{align}
Finally, let us return to equations~\ref{eqn:G} and~\ref{eqn:G4} to find the center
${\bar{x}_{\alpha X}}$ of the Wannier function whose spread about point $X$ is minimal:
\begin{align}
\bar{x}_{\alpha X}&= X + i\lim_{s\to 0}\partial_s G_\alpha(k,s)
\nonumber \\
& = X + i\intbz f^*_\alpha(k)\left[\partial_k + i X\right] f_\alpha(k)\dd{k}
\nonumber \\
& =  \intbz f^*_\alpha(k)\left(i\partial_k f_\alpha(k)\right)\dd{k} = X
\end{align}
I emphasize that this section does not have any quantum mechanical content.
I have presented a derivation that is as applicable
within classical statistical mechanics as it is within quantum mechanics.

\onecolumngrid
\vspace{1cm}
\PRLsep
\vspace{1cm}
\twocolumngrid
\section{Energies of pure states in their natural bases}
\label{section:appendix_natural}
One purpose of this appendix is to derive some illuminating, and possibly useful, expressions 
for the expectation values of the energies of two kinds of physical systems. 
The first comprises multiple interacting indistinguishable particles, such as
electrons, confined by a potential, such as the electrostatic  potential 
from a set of quasi-static nuclei. The second physical system
is a non-overlapping interacting pair of charge-neutral composite particles, 
such as nanoparticles or noble gas atoms.

A second purpose of this appendix is to use these examples to illustrate
how much can be learned about the structure of 
a generic pure statistical state $\Psi$ of a physical system or subsystem $\subject$
by expressing the expectation values of $\subject$'s observables 
in basis sets that are intrinsic properties of state ${\Psi}$, 
rather than eigenstates of the observables' operators.

The eigenstates of the operator $\hObs$ of an observable $\Obs$ 
inherit their characteristics from three sources: Namely, 
the observable $\Obs$, the physical system $\subject$ of which they are statistical
states, and the apparatus or probe to which $\subject$ is coupled during
the measurement of $\Obs$. Therefore, they are not necessarily
representative of an arbitrary pure state of $\subject$. However, even if they were, 
there is 
nothing general to be learned about pure states from an expression 
for the expectation value ${\expval{\Obs}}$ of $\Obs$ that is derived by expanding
an arbitrary pure state $\Psi$ in a basis of eigenstates of ${\hObs}$:
For every observable, this procedure leads to the classical expression,
\begin{align*}
\expval{\Obs}=\sum_\alpha \Pr(\Obs=\obs_\alpha)\obs_\alpha,
\end{align*}
where each ${\obs_\alpha}$ is one of the possible results of measuring
${\Obs}$.

This work demonstrates by example that it is possible to 
gain insight into characteristics  
of an arbitrary pure state $\Psi$ 
by expressing expectation values of its observables in terms of its
\emph{natural basis sets}, which are basis sets whose elements are 
\emph{natural states}. 
A pure state's natural states are the eigenstates of its
reduced density matrices; and their eigenvalues are their occupation numbers~\citep{coleman_rmp,lowdin_1955,davidson_1972,ando_1963,mcweeny_1960}.

All of the theory that is presented or developed in this appendix is applicable to 
a pure state of a set of classical particles whose energy expectation 
value can be expressed as the expectation value of a sum of $1$-particle
energies and $2$-particle interaction energies. Although, I briefly express the $1$-particle
energy in the form that it would take for a set of quantum mechanical particles in
an external potential, my derivations do not require it to have this form.

\subsection{Pure states and mixed states}
\subsubsection{Pure states}
A pure state of a classical or quantum mechanical system $\subject$ that
comprises $N$ indistinguishable particles is a function 
${\Psi\in\lebesgue(\domain^N)}$
such that
\begin{align*}
\pdf(x_1\cdots x_N) =
\abs{\Psi(x_1\cdots x_N)}^2 
\end{align*}
is the probability distribution for the system's microstructure, ${(x_1\cdots x_N)}$, 
where ${x_i\in\domain}$ specifies the coordinates of the ${i^\text{th}}$ particle;
and ${\domain}$ is the set of all possible coordinates of a single particle. For example, if ${\domain=\realone^3}$, 
${\pdf(x_1\cdots x_N)}$ is a position probability density function and
each element of ${\domain}$ is a point in physical space and a 
possible location of a particle.

The pure state specified by $\Psi$ will sometimes be represented by
an element ${\ket{\Psi}}$ of an abstract Hilbert space ${\hilbert_N}$, 
whose elements are in one to one correspondence with the
elements of ${\lebesgue(\domain^N)}$.
It can also be specified by an idempotent density operator, 
\begin{align*}
\Dop\equiv\dyad{\Psi} \implies \Dop\Dop = 
\ket{\Psi}\braket{\Psi}\bra{\Psi}
=\dyad{\Psi}=\Dop.
\end{align*}
Idempotency of the density operator can be regarded as the defining property of a pure state.

\subsubsection{Mixed states}
A \emph{mixed state} is a more general kind of statistical state. A mixed state density
operator is not idempotent and has the mathematical form,
\begin{align*}
\Dop =\sum_\alpha \Pr(\Psi=\Psi_\alpha)\Dop_\alpha =\sum_\alpha\Pr(\Psi=\Psi_\alpha)\dyad{\Psi_\alpha},
\end{align*}
where ${\Pr(\Psi=\Psi_\alpha)}$ is the probability that the physical
system is in pure state ${\Psi_\alpha}$,  and
\begin{align*}
\sum_\alpha \Pr(\Psi=\Psi_\alpha)=1.
\end{align*}

\subsubsection{Pure states define the set of all states}
\begin{figure}
    \includegraphics[width=0.49\textwidth]{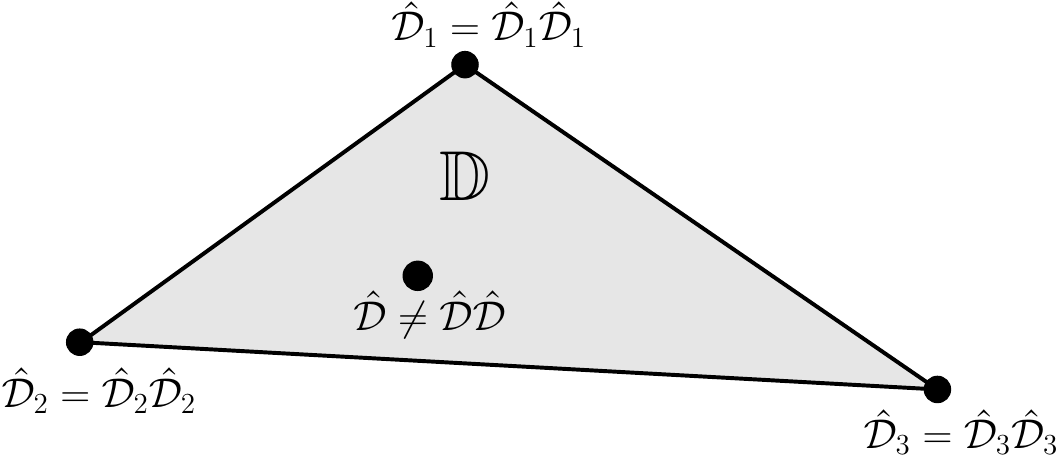}
\label{fig:density_operator}
\caption{Schematic depiction of the set $\ddomain$ of all density operators. Its 
vertices are idempotent, and therefore pure states; and every interior point ${\Dop}$
is a mixed state, which can be expressed 
as ${\Dop=\weight_1\Dop_1+\weight_2\Dop_2+\weight_3\Dop_3}$, where
${\weight_1,\weight_2,\weight_3\in(0,1)}$ and
${\weight_1+\weight_2+\weight_3=1}$. Set $\ddomain$ is defined by the pure states, just
as a triangle is defined by its vertices.
}
\end{figure}
The set of all density operators, $\ddomain$, is the set of all positive Hermitian operators
${\hilbert_N\to\hilbert_N}$ with unit trace. It is a \emph{convex} set~\citep{coleman_rmp,von_neumann_new,von_neumann_old},
which means that
\begin{align*}
\weight\Dop_1+(1-\weight)\Dop_2\in\ddomain, \;\forall \Dop_1,\Dop_2\in\ddomain, \;\forall \weight\in[0,1].
\end{align*}
Convexity of ${\ddomain}$ implies
that if ${\ddomain}$ is visualized as a simplex, as it is depicted in Fig.~\ref{fig:density_operator}, 
an element of ${\ddomain}$ is a pure state if and only if it is at a vertex of the simplex.
Therefore the set of pure state density operators is the set of \emph{extreme points} of $\ddomain$;
and convexity implies that the set of extreme points (vertices) defines the entire set.
Therefore, as pointed out by~\citet{coleman_rmp}, 
we can hope to understand a great deal about an arbitrary mixed state of $\subject$
by understanding an arbitrary pure state of ${\subject}$.

\subsection{Notation, definitions, and assumptions}
\label{section:natural_notation}
The notation in this appendix deviates from the convention,
introduced in Sec.~\ref{section:natural_notation} and used throughout the 
rest of this work, that ${x}$ or ${x_i}$ represents a single
coordinate of a particle's position vector.
In this appendix a single variable, such as 
$x$, $y$, $x_i$ or $y_j$ denotes all of a single particle's coordinates;
and in subsection~\ref{section:orbital_energy} I will often use $i$ to denote $x_i$.

For simplicity it is assumed in subsection~\ref{section:nonoverlapping} that all particles
are either spin-0 electrons or spin-0 nuclei, whose coordinates are simply their positions.
However, no assumptions are made that are inconsistent with classical physics.
Therefore the particles referred to as
`electrons' represent any set of indistinguishable charged light particles, 
and those referred to as `nuclei' represent any set of oppositely-charged indistinguishable heavy particles.

In subsection~\ref{section:natural_states} and subsection~\ref{section:orbital_energy}, 
if the particles have spins the $i^\text{th}$ particle's coordinates will be ${i\equiv x_i\equiv(\rvecsub{i},\spin_i)}$, where
$\rvecsub{i}$ is its position and $\spin_i$ is its spin;
and I will often use the abbreviation ${\dmeasure{1\cdots p}\equiv\dd{x_1}\cdots\dd{x_p}}$.
I do not discuss magnetization, particles' spins,
or magnetic interactions explicitly; and all of the theory presented is applicable to charged
spin-$0$ particles.  However, if the theory is applied to particles with spins, 
integrals over the coordinates of one or more particles 
denote a sum over all possible (sets of) spins 
of the integral over all possible (sets of) positions.
For example, 
\begin{align*}
\int f(i,j)\dmeasure{i,j}
&\equiv
\int f(x_i,x_j)\dd{x_i}\dd{x_j}
\\
&\equiv \sum_{\spin_i,\spin_j}\int f(\rvecsub{i},\spin_i,\rvecsub{j},\spin_j)\dd[3]{r_i}\dd[3]{r_j}.
\end{align*}

\subsubsection{Basis bras and kets}
The meaning of ${\ket{x}}$ was explained in Appendix~\ref{section:appendix_states_as_vectors}
the general case in which $x$ represents $\Ndof$ degrees of freedom, but no
account was taken of the
degrees of freedom of a single particle it can be thought of as the tensor product
of the state vectors of those degrees of freedom. 

For example, if ${\rvec\in\realone^3}$, and if $x$ momentarily represents the $x$-coordinate of $\rvec$, 
then ${\ket{\rvec}}$ should be thought of as
\begin{align*}
\ket{\rvec}=\ket{x}\otimes\ket{y}\otimes\ket{z}
\end{align*}
or as
\begin{align*}
\ket{\rvec}=\ket{z}\otimes\ket{x}\otimes\ket{y}
\end{align*}
or as a tensor product, in any other order, of the three states that represent the \emph{same} particle's
Cartesian position coordinates. The order does not matter as long as the same order is used consistently for
all particles.

Therefore, from now on, ${\ket{x}}$ is an element of ${\hilbert_1}$ that represents a possible configuration
of a single particle; and just as ${\hilbert_N}$ is an abstract representation of ${\lebesgue(\domain^N)}$, 
${\hilbert_1}$ is an abstract representation of ${\lebesgue(\domain)}$. Therefore if, for example, 
${\domain\equiv\onetorus}$, then ${\ket{x}}$ also represents a square integrable function that is 
localized around ${x\in\onetorus}$. If the width $\dwidth$ of that function vanished, it would be
the Dirac delta distribution. However its width is arbitrarily small, but \emph{finite}.

The vector ${\ket{x_1\cdots x_p}\in\hilbert_p}$, where ${\hilbert_p}$ is an abstract
representation of ${\lebesgue(\domain^p)}$, is defined as
\begin{align*}
\ket{x_1\cdots x_p}\equiv \antisymmetrizer\left\{\ket{x_1}\otimes\ket{x_2}\cdots\otimes\ket{x_p}\right\},
\end{align*}
where I will be using operator $\antisymmetrizer$ in multiple vector spaces to denote the norm-preserving antisymmetriser.

Note that a tensor product of multiple states is only antisymmetrized if each state refers to a different particle.
It was not needed for the definition ${\ket{\rvec}=\ket{x}\otimes\ket{y}\otimes\ket{z}}$.

Since the space $\domain$ has not been specified, it would be unnecessarily complicated to 
be specific and rigorous about the normalizations of states like ${\ket{x_1\cdots x_p}}$. It suffices
to consider the expression,
\begin{align*}
\braket{x_1\cdots x_p}{x'_1\cdots x'_p} = \delta(x_1-x'_1)\cdots \delta(x_p-x'_p),
\end{align*}
where ${\delta}$ plays the same role within integrals over ${\domain}$ that is played
by Dirac's delta function, when it is in the hands of physicists. This expression 
is inappropriate and problematic in several ways, but it is fine for present purposes.
It is implied by the equally-problematic expression ${\braket{x}{x'}=\delta(x-x')}$ 
and the fact that $\antisymmetrizer$ preserves normalizations.

\subsubsection{Contractions}
\label{section:contraction}
If ${\ket{f}}$ is a $p$-state 
and ${\ket{F}}$ is a $q$-state, then
the \emph{contraction}~\citep{lounesto_2001,chisolm_2012,dorst_contraction,vaz_darocha_2016,doran_2003} of ${\ket{F}}$ by ${\ket{f}}$ is
the ${(q-p)}$-state, 
\begin{align*}
\lcontract{f}{F}
\equiv 
\int \bar{f}(1\cdots p) F(1\cdots p\cdots q)\ket{p+1\cdots q} \dmeasure{1\cdots q},
\end{align*}
and the contraction of ${\bra{F}}$ by ${\bra{f}}$ is the dual of $\lcontract{f}{F}$,
\begin{align*}
\rcontract{F}{f}
\equiv 
\int 
\bar{F}(1\cdots p\cdots q)
f(1\cdots p) 
\bra{p+1\cdots q} 
\dmeasure{1\cdots q}.
\end{align*}
Note that ${\ket{f\rfloor F}}$ and ${\bra{f\rfloor F}}$
can also be denoted as ${\ket{F\lfloor f}}$
and ${\bra{F\lfloor f}}$, respectively.

It will be useful to denote the unit-normalized contractions  with a `$1$' subscript, 
i.e., 
\begin{align*}
\lcontractN{f}{F}\equiv \frac{\lcontract{f}{F}}{\sqrt{\braket{F\lfloor f}{f\rfloor F}}} = \rcontractN{F}{f}^\dagger,
\\
\rcontractN{F}{f}\equiv \frac{\rcontract{F}{f}}{\sqrt{\braket{f\rfloor F}{F\lfloor f}}} = \lcontractN{f}{F}^\dagger,
\end{align*}
where, for example, 
\begin{align*}
\norm{\lcontract{f}{F}}=\sqrt{\braket{F\lfloor f}{f\rfloor F}}.
\end{align*}

\subsection{Natural states}
\label{section:natural_states}
A {\em natural $p$-state} ${\mathcal{X}_\alpha(x_1,\cdots,x_p)}$ of an isolated system of ${N=p+q}$ identical particles
in a pure state, ${\Psi(x_1,\cdots,x_{N})\in\hilbert_N}$,
is an eigenstate of its
$p^\text{th}$-order reduced density matrix (or simply ${p}${\em-matrix}). 
That is, 
\begin{align*}
\int \dmatrixarg{\Psi}_p(x_1\cdots x_p; x_1'\cdots x_p') \mathcal{X}_\alpha&(x'_1,\cdots,x'_p)\dd{x'_1}\cdots\dd{x'_p}\\
&= \lambda_\alpha \mathcal{X}_\alpha(x_1,\cdots,x_p),
\end{align*}
where $\lambda_\alpha$ is a nonnegative real number and
\begin{align*}
\dmatrixarg{\Psi}_p(x_1\cdots x_p;& x_1'\cdots x_p') 
\equiv
\int \Psi(x_1\cdots x_p,x_{p+1}\cdots x_N)
\\
&\times\bar{\Psi}(x'_1\cdots x'_p, x_{p+1}\cdots x_N) \dd{x_{p+1}}\cdots \dd{x_N}.
\end{align*}
Natural states have many nice properties.  For example, if ${\{\tilde{\mathcal{X}}_\alpha\}}$
and ${\{\tilde{\mathcal{Y}}_\beta\}}$ are not sets of natural states, but are any other
complete orthonormal bases of the ${p}$-particle and ${q}$-particle Hilbert spaces, respectively, then
${\Psi}$ can be expressed exactly as the double infinite sum
\begin{align}
\Psi(x_1\cdots &x_{N}) \nonumber \\
&= \sum_{\alpha,\beta} \tilde{C}_{\alpha\beta} 
\tilde{\mathcal{X}}_\alpha(x_1\cdots x_p) \tilde{\mathcal{Y}}_\beta(x_{p+1}\cdots x_{N}),
\label{eqn:unnatural1}
\end{align}
for some set of constants ${\tilde{C}_{\alpha\beta}\in \mathbb{C}}$.
However if ${\{\mathcal{X}_\alpha\}}$ and ${\{\mathcal{Y}_\beta\}}$ are the
sets of natural $p$-states and $q$-states, this expression simplifies to the single infinite sum,
\begin{align}
\Psi(x_1\cdots &x_{N}) \nonumber \\
&= \sum_{\alpha} C_{\alpha} \mathcal{X}_\alpha(x_1\cdots x_p) \mathcal{Y}_\alpha(x_{p+1}\cdots x_{N}),
\label{eqn:natural1}
\end{align}
where ${C_\alpha\in\complex}$ and ${\mathcal{X}_\alpha}$  and ${\mathcal{Y}_\alpha}$ are eigenstates 
of the $p$-matrix and the $q$-matrix, respectively, with the same eigenvalue,
${\lambda_\alpha\equiv \abs{C_{\alpha}}^2}$. Furthermore, 
\begin{align}
C_\alpha &\mathcal{Y}_\alpha(x_{p+1}\cdots x_{N})   =  
\left(\bra{x_{p+1}\cdots x_N}\otimes\bra{\mathcal{X}_\alpha}\right)\ket{\Psi} \nonumber \\
& \equiv
\int \bar{\mathcal{X}}_\alpha(x_1\cdots x_p) \Psi(x_1\cdots x_{N}) \dd{x_1}\cdots\dd{x_p},
\label{eqn:contraction}
\end{align}
which means that ${C_\alpha\mathcal{Y}_\alpha}$ is the contraction of ${\bar{\mathcal{X}}_\alpha}$
onto ${\Psi}$. 

Therefore ${\mathcal{Y}_\alpha}$ resides in the Hilbert
subspace that is orthogonal to $\mathcal{X}_\alpha$.  Equation~\ref{eqn:contraction} also means that both 
${\mathcal{X}_\alpha}$ and ${\mathcal{Y}_\alpha}$ inherit from $\Psi$ its symmetry or antisymmetry with respect to exchange 
of positions. 

I refer the reader to~\linecite{coleman_rmp} for a clear explanation of many 
of the nice properties of natural states.
These properties suggest that natural $p$-states are the only ${p}$-particle states
to which physical meaning should be attached in a system comprised of more than $p$ particles.
I state only two of these properties here. 

\paragraph*{Property 1:} It can be shown (see Coleman's Theorem 3.1)
that if ${\Phi}$ is restricted to the mathematical form
\begin{align*}
\Phi(x_1\cdots &x_N)
=\sum_{\alpha\leq u,\, \beta\leq v} 
\tilde{C}_{\alpha\beta}\tilde{\mathcal{X}}_\alpha(x_1\cdots x_p)\tilde{\mathcal{Y}}_\beta(x_{p+1}\cdots x_N), 
\end{align*}
where ${u\leq v<\infty}$, and if ${\norm{\Psi-\Phi}^2}$ is minimized
with respect to the set of coefficients ${\{\tilde{C}_{\alpha\beta}\}}$ and the
sets of functions, ${\{\tilde{\mathcal{X}}_{\alpha}\}_{\alpha\leq u}}$ and ${\{\tilde{\mathcal{Y}}_{\beta}\}_{\beta\leq v}}$, 
the minimum is obtained by the following truncation of the sum in Eq.~\ref{eqn:natural1}:
\begin{align*}
\Phi(x_1\cdots &x_N) 
=\sum_{\alpha\leq u} 
C_{\alpha}\mathcal{X}_\alpha(x_1\cdots x_p)\mathcal{Y}_\alpha(x_{p+1}\cdots x_N),
\end{align*}
where the coefficients are indexed such that ${\alpha<\beta\implies \abs{C_{\alpha}}\geq \abs{C_\beta}}$.

\paragraph*{Property 2:} It can also be shown (see Coleman's Theorem 3.3)
that if ${p}$ is odd and ${2p<N}$, then
\begin{align}
\int \bar{\mathcal{X}}_\alpha(x_1\cdots x_p) \mathcal{Y}_\alpha(x_1\cdots x_{N}) \dd{x_1}\cdots\dd{x_p}= 0.
\label{eqn:orthogonality}
\end{align}

As mentioned in Sec.~\ref{section:natural_notation}, 
I will often use $j$ as shorthand for $x_j$. Therefore,
\begin{align*}
\ket{1\cdots N} &= \ket{1\cdots p}\otimes\ket{p+1\cdots N}\\
&\equiv 
\ket{x_1\cdots x_p}\otimes\ket{x_{p+1}\cdots x_{N}}=
\ket{x_1\cdots x_N}.
\end{align*}

\subsection{Natural orbitals}
\label{section:orbital_energy}
There is a long history of simplifying many-particle states and many-particle energetics
by treating the particles as quasi-independent; and approximations based on this simplification are widely used. 
The purpose of this subsection is to present a rigorous theoretical justification of the concept 
of a quasi-independent particle state in some many-particle systems; and to provide insight into the conditions
under which this concept ceases to be meaningful or justified.

It is hoped that this may lead to a better understanding of how Bloch functions, Wannier functions,
and other kinds of single particle states should be interpreted; and a better
understanding of the validity of the assumption that the electron densities of atoms 
and chemical bonds have substructures of atomic and molecular orbitals.

The justification that is presented consists of a derivation of
a few closely-related exact expressions for the energy 
\begin{align*}
E=\expval{\Ham}{\Psi}
\end{align*}
of a set of $N$ interacting indistinguishable particles with wavefunction $\Psi$.
One of those expressions is
\begin{align}
E& = \sum_\alpha \occ_\alpha \energy_\alpha
+ \sum_{\{\alpha,\beta\}} \sqrt{\occ_\alpha \occ_\beta}\,w_{\alpha\beta}, 
\label{eqn:two}
\end{align}
where the first sum is over the set ${\{\varphi_\alpha\}}$ of all of
$\Psi$'s \emph{natural orbitals} (natural $1$-states);
the second sum is over all distinct pairs of natural orbitals;
the set ${\{\occ_\alpha\}}$, which is uniquely determined by $\Psi$,
has the properties ${\occ_\alpha\in[0,1]}$ and
${\sum_\alpha \occ_\alpha =N}$ that would be required
of orbital occupation probabilities; ${\energy_\alpha}$ denotes ${\expval{\hamsmall}{\varphi_\alpha}}$,
where ${\hamsmall}$ is a $1$-particle Hamiltonian; and
\begin{align*}
w_{\alpha\beta}\equiv 2\Re\left\{\mel{\varphi_\alpha}{\what_{\alpha\beta}}{\varphi_\beta}\right\},
\end{align*}
where ${\what_{\alpha\beta}}$ can be viewed as
a coupling between natural orbitals ${\varphi_\alpha}$ and ${\varphi_\beta}$ that is mediated
by the ${(N-1)}$-particle states ${\ket{\Theta_\alpha}\equiv\lcontractN{\varphi_\alpha}{\Psi}}$ and 
${\ket{\Theta_\beta}\equiv\lcontractN{\varphi_\beta}{\Psi}}$.

Equation~\ref{eqn:two} is an exact expression for the energy 
of a set of $N$ interacting indistinguishable particles in a pure state
as a weighted sum of the energies ${\{\energy_\alpha\equiv\expval{\hamsmall}{\varphi_\alpha}\}}$ of
independent particles whose wavefunctions are the \emph{natural orbitals}
(natural $1$-states) ${\{\varphi_\alpha\}}$ of $\Psi$, plus an interaction term.

The interaction term is a weighted sum of the 
terms ${\left\{w_{\alpha\beta}\right\}}$, 
which have the appearance of pairwise couplings between orbitals.
However $\what_{\alpha\beta}$ 
depends on the natural ${(N-1)}$-states ${\Theta_\alpha}$ and ${\Theta_\beta}$
that are the dual states of ${\varphi_\alpha}$ and ${\varphi_\beta}$, respectively.
Therefore ${\mel{\varphi_\alpha}{\what_{\alpha\beta}}{\varphi_\beta}}$
is not a $2$-particle term, but an $N$-particle term, which it might be appropriate
to interpret as a \emph{mediated} coupling between orbitals $\varphi_\alpha$ and $\varphi_\beta$,
or as a mediated interaction between two particles occupying them.

If all of the interaction energies were sufficiently small, 
$E$ could be interpreted as approximately a weighted sum of the
energies of independent particles occupying different orbitals, where the weight given
to each energy is both the probability of the corresponding orbital being occupied 
at a particular instant and the fraction of time for which it is occupied.
Then ${\{w_{\alpha\beta}\}}$ could be interpreted as the set of
energies of the interactions responsible for moving particles
between orbitals.

Therefore, when interaction energies are small, Eq.~\ref{eqn:two} is consistent 
with the physical picture of each particle occupying an orbital for a long
period until, eventually, its weak or rare interactions with other particles 
move it to a different orbital.
On the other hand,  the fact
that ${w_{\alpha\beta}}$ is an $N$-particle energy means that 
when interaction energies are large, 
$E$ is not approximately a sum of single particle energies. In that case, 
Eq.~\ref{eqn:two} is consistent with the residence times of particles in orbitals 
being too short for the concept of orbital occupation to be valid.

The derivation of Eq.~\ref{eqn:two} makes use of properties
possessed only by the set natural orbitals. Therefore it strengthens the 
case for natural orbitals being the most `physical' $1$-particle
states in a many particle system, and suggests that 
comparisons with natural orbitals might shed light on how
other sets $1$-particle states should be interpreted.

\subsubsection{Theoretical setup}
Let us begin with the exact expression
\begin{align}
\Psi(1\cdots N) = \sum_\alpha c_\alpha \varphi_\alpha(1)\Theta_\alpha(2\cdots N),
\label{eqn:orbital_expansion}
\end{align}
where ${\{\varphi_\alpha\}}$ and ${\{\Theta_\alpha\}}$ are the sets
of natural orbitals and natural ${(N-1)}$-states, respectively, 
and
the functions in each set are mutually orthogonal and 
have been chosen to be
normalized to one, i.e., 
${\braket{\varphi_\alpha}{\varphi_\beta}=\delta_{\alpha\beta}}$
and
${\braket{\Theta_\alpha}{\Theta_\beta}=\delta_{\alpha\beta}}$. 
Let us also choose ${\Psi}$ to be normalized to one, 
which implies that ${\sum_\alpha\lambda_\alpha=1}$, 
where ${\lambda_\alpha\equiv\abs{c_\alpha}^2}$.

We will make use of the function 
\begin{align*}
\densityn_\alpha(1)\equiv (N-1)
\int\abs{\Theta_\alpha(1\cdots N-1)}^2 \dmeasure{2\cdots N-1},
\end{align*}
which would be the number density of ${N-1}$ particles
whose state was the
${\alpha^\text{th}}$ natural ${(N-1)}$-state, $\Theta_\alpha$.

The $N$-particle Hamiltonian is the following
sum of an independent-particle operator, $\Hamone$, 
and an interaction operator, ${\Hamtwo}$:
\begin{align}
\Ham \equiv \overbrace{\sum_{i=1}^N\hamone(i)}^{\displaystyle \Hamone} 
+ \overbrace{\sum_{i=1}^{N}\sum_{j=i+1}^N\hamtwo(i,j)}^{\displaystyle \Hamtwo}.
\label{eqn:hamiltonian}
\end{align}
Both $\Hamone$ and $\Hamtwo$ operate on $N$-particle states, 
but are sums of $1$-particle operators and $2$-particle operators, respectively.
The independent-particle Hamiltonian, $\Hamone$, is a sum over
$i$ of the single-particle Hamiltonian, ${\hamsmall(i)}$, 
which operates on the coordinates of the ${i^\text{th}}$ particle.

The interaction term has the form
${\Hamtwo=\sum_{i,j>i}\hamtwo(i,j)}$,
where ${\hamsmalltwo(i,j)}$ is the interaction between particles
with coordinates $i$ and $j$.
I will use ${\Hamtwo}$ more generally to denote the 
interaction operator of a system with $M$ particles, 
where $M$ is
the number of particles of the state on which ${\Hamtwo}$ acts.

Equation~\ref{eqn:two} and the other expressions 
for ${\expval{\Ham}{\Psi}}$ 
that will be derived in this section (Sec.~\ref{section:orbital_energy}), are quite general. 
They are valid for any set of indistinguishable 
classical or quantum mehcanical particles and any operator $\Ham$ that
can be expressed as a sum of $1$-particle and $2$-particle terms.
Nevertheless, I will sometimes refer to the particles
as electrons and I will assume that the $1$-particle operator 
has the form,
${\hamsmall(i)= \kinetic(i) + \vextop(i)}$, 
where $\kinetic$ is the $1$-particle kinetic energy operator; and $\vextop$
is the operator for the energy of a single particle in
an external potential.

\subsubsection{Single particle energy}
Equation~\ref{eqn:orbital_expansion} can be used to express
the expectation value of the $1$-particle energy as
\begin{align*}
E_1\equiv\expval{\Hamone}{\Psi}
& =
\sum_{\alpha,\beta} \bar{c}_\alpha c_\beta 
\int
\bar{\varphi}_\alpha(1)\bar{\Theta}_\alpha(2\cdots {N})
\nonumber \\
&\times
\left(\sum_i \hat{h}(i)\right)
\varphi_\beta(1)\Theta_\beta(2\cdots {N})
\dmeasure{1\cdots N}.
\end{align*}
Using
the orthonormality 
of the natural $(N-1)$-states
and the antisymmetry of $\Psi$, this can be simplified to
\begin{align}
E_1= \sum_\alpha \occ_\alpha\left(t_\alpha + \vext_\alpha\right) = \sum_\alpha\occ_\alpha\energy_\alpha,
\label{eqn:H0}
\end{align}
where 
${\occ_\alpha \equiv N\lambda_\alpha \equiv N\abs{c_\alpha}^2}$;
${\energy_\alpha\equiv t_\alpha+\vext_\alpha}$; and
I have introduced the $1$-particle energy expectation values, 
${t_\alpha \equiv \expval{\,\hat{t}\,}{\varphi_\alpha}}$
and 
\begin{align*}
\vext_\alpha 
&\equiv \expval{\vextop}{\varphi_\alpha}   
  =  \int \vext(x) n_\alpha(x)\dd{x},
\end{align*}
where ${n_\alpha(x)\equiv\abs{\varphi_\alpha(x)}^2}$;
${\vextop_\alpha\equiv \int \vext(1)\dyad{1}\dmeasure{1}}$;
and ${\vext(1)=\vext(x_1)}$ is the external potential
felt by a single particle whose coordinates are $x_1$.
\\

\subsubsection{Interaction energy - Expression 1}
\label{section:interaction_energy1}
The interaction energy is
\begin{align*}
\Energy
&\equiv\expval{\Hamtwo}{\Psi}
\\
&
= \int\Psi^*(1\cdots N)\left(\sum_{i,j>i}\hamtwo(i,j)\right)\Psi(1\cdots N)\dmeasure{1\cdots N};
\end{align*}
and if ${\Psi}$ is expressed as in Eq.~\ref{eqn:orbital_expansion}, it becomes
\begin{align}
\Energy
\equiv\sum_{\alpha\beta}&\bar{c}_\alpha c_\beta 
\int\bar{\varphi}_\alpha(1)\bar{\Theta}_\alpha(2\cdots N)
\nonumber
\\
\times&\left(\sum_{i,j>i}\hamtwo(i,j)\right)\varphi_\beta(1)\Theta_\beta(2\cdots N)\dmeasure{1\cdots N};
\label{eqn:Wexpr1}
\end{align}
The exchange symmetries of ${\Psi^*\Psi}$ and ${\bar{\Theta}_\alpha\Theta_\beta}$ allow the 
sum of interactions ${\hamtwo(i,j)}$ in parentheses to be 
replaced by any of the following three expressions:
\begin{align}
\sum_{i,j>i} \hamtwo(i,j) 
&= \sum_{j>1}\hamtwo(1,j) + \sum_{i>1,j>i}\hamtwo(i,j)
\nonumber
\\
&= (N-1)\hamtwo(1,2) + \sum_{i>1,j>i}\hamtwo(i,j)
\label{eqn:Wsimplification1}
\\
&= N\sum_{j>1}\hamtwo(1,j) = N(N-1)\hamtwo(1,2).
\label{eqn:Wsimplification2}
\end{align}
Equation~\ref{eqn:Wsimplification1} will be used in subsection~\ref{section:interaction_energy2}.
We will use Eq.~\ref{eqn:Wsimplification2} in this subsection.
Let us define the $1$-particle operator, 
\begin{align*}
\what_{\alpha\beta}(1)
&\equiv \int\bar{\Theta}_\alpha(2\cdots N)\left(\sum_{j>1}\hamtwo(1,j)\right)\Theta_\beta(2\cdots N)\dmeasure{2\cdots N}
\\
&= (N-1)\int\bar{\Theta}_\alpha(2\cdots N)\hamtwo(1,2)\Theta_\beta(2\cdots N)\dmeasure{2\cdots N}.
\end{align*}
This allows Eq.~\ref{eqn:Wexpr1} to be expressed as
\begin{align}
W &=N\sum_{\alpha,\beta}
\bar{c}_\alpha c_\beta\mel{\varphi_\alpha}{\what_{\alpha\beta}}{\varphi_\beta}.
\label{eqn:Wexpr2}
\end{align}

Now let us denote the argument of ${c_\alpha}$ by $\vartheta_\alpha$, i.e., ${c_\alpha=\abs{c_\alpha}e^{i\vartheta_\alpha}}$. 
Then we \emph{could} define
\begin{align}
w_{\alpha\beta}&\equiv 2\Re\left\{e^{i(\vartheta_\beta-\vartheta_\alpha)}\mel{\varphi_\alpha}{\what_{\alpha\beta}}{\varphi_\beta}\right\}, 
\label{eqn:firstwdef}
\end{align}
and this definition would lead to the expressions presented in Sec.~\ref{section:total_energy}.
However, because the magnitudes of the coupling energies ${\{w_{\alpha\beta}\}}$ determine
whether or not orbital occupation is a valid concept, let us choose a path by which it can be expressed
in a simpler mathematical form.
Without losing generality, let us choose the coefficients ${\{c_\alpha\}}$ 
to be real. This is possible because we can express ${c_\alpha\varphi_\alpha}$ as
\begin{align*}
c_\alpha\varphi_\alpha&=\abs{c_\alpha}e^{i\vartheta_\alpha}\varphi_\alpha = \abs{c_\alpha}\left(e^{i\vartheta_\alpha}\varphi_\alpha\right)
=c_\alpha \left(e^{i\vartheta_\alpha}\varphi_\alpha\right)
\intertext{or, if we want $c_\alpha$ to be negative, as}
c_\alpha\varphi_\alpha&=-\abs{c_\alpha}\left(e^{i(\vartheta_\alpha+\pi)}\varphi_\alpha\right) = c_\alpha\left(e^{i(\vartheta_\alpha+\pi)}\varphi_\alpha\right).
\end{align*}
Therefore the phase factors of the coefficients ${\{c_\alpha\}}$ can be merged into the natural orbitals.
Having done so, we can define the coupling energy between ${\varphi_\alpha}$ and ${\varphi_\beta}$ as
\begin{align*}
w_{\alpha\beta}&\equiv 2\Re\left\{\mel{\varphi_\alpha}{\what_{\alpha\beta}}{\varphi_\beta}\right\}.
\end{align*}

Now let us denote ${\frac{1}{2}w_{\alpha\alpha}=\frac{1}{2}\expval{\what_{\alpha\alpha}}{\varphi_\alpha}}$ as $\vmf_\alpha$
and express it as
\begin{align}
\vmf_\alpha
&= (N-1)\int \abs{\varphi_\alpha(1)}^2\hamtwo(1,2)\abs{\Theta_\alpha(2\cdots N)}^2\dmeasure{1\cdots N}
\nonumber
\\
&= \int\int n_\alpha(x)\hamtwo(x,x')\densityn_\alpha(x')\dd{x}\dd{x'}.
\label{eqn:vmf_definition}
\end{align}
This expression makes it clear that ${\vmf_\alpha}$ is the
mean field interaction between an electron in orbital ${\varphi_\alpha}$ 
and ${N-1}$ electrons in $\varphi_\alpha$'s dual ${(N-1)}$-state, $\Theta_\alpha$.

Now $W$ can be expressed as
\begin{align}
W&=\sum_\alpha\left(\occ_\alpha \vmf_{\alpha} + \frac{1}{2}\sum_{\beta\neq\alpha} \sqrt{\occ_\alpha\occ_\beta}\,w_{\alpha\beta}\right)
\nonumber
\\
&=\sum_\alpha\occ_\alpha\vmf_\alpha 
+ \sum_{\alpha,\beta>\alpha}\sqrt{\occ_\alpha\occ_\beta}\,w_{\alpha\beta}
\label{eqn:Eint}
\\
&=\sum_{\alpha}\sum_{\beta\geq\alpha}\sqrt{\occ_\alpha\occ_\beta} \,w_{\alpha\beta}
=N\sum_{\alpha}\sum_{\beta\geq\alpha}c_\alpha c_\beta \,w_{\alpha\beta}.
\nonumber
\end{align}
If the coefficients ${\{c_\alpha\}}$ are real, expressing ${N c_\alpha c_\beta}$ as ${\sqrt{\occ_\alpha\occ_\beta}}$ implies that they
are positive as well as real.
However, none of these expressions, except the final one in which ${\sqrt{\occ_\alpha\occ_\beta}}$ has been replaced by ${ N c_\alpha c_\beta}$
rather than by ${N \bar{c}_\alpha c_\beta}$, would be different if the coefficients ${\{c_\alpha\}}$ were complex and ${w_{\alpha\beta}}$
had been defined as in Eq.~\ref{eqn:firstwdef}.

Note that if the phase factors of the natural ${(N-1)}$-states were independent of particles'
positions and spins, they could be merged into the natural orbitals. Then
${\what_{\alpha\beta}}$ could be expressed as
\begin{widetext}
\begin{align}
\what_{\alpha\beta}(1)
&= (N-1)\int\sqrt{\pdfarg{\Theta}_\alpha(2\cdots N)\pdfarg{\Theta}_\beta(2\cdots N)} 
\hamtwo(1,2)
\dmeasure{2\cdots N}, 
\label{eqn:geometric_mean}
\end{align}
\end{widetext}
where 
\begin{align*}
\pdfarg{\Theta}_\alpha(x_2\cdots x_N)\equiv\abs{\Theta_\alpha(x_2\cdots x_N)}^2 
\end{align*}
is the probability density that a set of ${(N-1)}$ particles whose wavefunction is ${\Theta_\alpha}$
have configuration ${(x_2\cdots x_N)}$. The reason to express ${\what_{\alpha\beta}}$ in this form
is to show that the integrand on the right hand side of Eq.~\ref{eqn:geometric_mean} is the product of ${\hamtwo(x_1,x_2)}$ and
the geometric mean of ${\pdenarg{\Theta}_\alpha(x_2\cdots x_N)}$ and ${\pdenarg{\Theta}_\beta(x_2\cdots x_N)}$.

\subsubsection{Interaction energy - Expression 2}
\label{section:interaction_energy2}
Another expression for $W$ can be found by
inserting Eq.~\ref{eqn:Wsimplification1} into Eq.~\ref{eqn:Wexpr1}. 
Then the definition of ${\what_{\alpha\beta}}$ can be used to simplify
the first term, and ${\braket{\varphi_\alpha}{\varphi_\beta}=\delta_{\alpha\beta}}$ can be
used to simplify the second term, to give
\begin{align*}
\Energy  
&= \sum_{\alpha\beta}\bar{c}_\alpha c_\beta\mel{\varphi_\alpha}{\what_{\alpha\beta}}{\varphi_\beta}
\\
&+\sum_\alpha \lambda_\alpha \int\bar{\Theta}_\alpha(2\cdots N)\left(\sum_{i>1,j>i}\hamtwo(i,j)\right)\Theta_\alpha(2\cdots N)\dmeasure{2\cdots N}
\\
&= \sum_{\alpha\beta}\bar{c}_\alpha c_\beta \mel{\varphi_\alpha}{\what_{\alpha\beta}}{\varphi_\beta} + \sum_\alpha \lambda_\alpha \underbrace{\expval{\Hamtwo}{\Theta_\alpha}}_{\displaystyle \Walpha}
\end{align*}
Note that the first term on the right hand side only differs from right hand side of Eq.~\ref{eqn:Wexpr2} by a factor of $N$.
Therefore we can replace it with the right hand side of Eq.~\ref{eqn:Eint} divided by $N$, i.e., 
\begin{align}
W&= \sum_\alpha \lambda_\alpha\left(\vmf_\alpha + \Walpha\right)+ \frac{1}{2}\sum_{\alpha,\beta\neq\alpha}\sqrt{\lambda_\alpha\lambda_\beta} w_{\alpha\beta}
\label{eqn:Wexpr4}
\\
&= \sum_\alpha\left(\lambda_\alpha\Walpha + \sum_{\beta\geq\alpha}\sqrt{\lambda_\alpha\lambda_\beta}w_{\alpha\beta}\right).
\label{eqn:Wexpr5}
\end{align}

\subsubsection{Total energy}
\label{section:total_energy}
By combining Eqs.~\ref{eqn:H0},~\ref{eqn:vmf_definition} and \ref{eqn:Eint}, 
the total energy can be expressed as
\begin{align}
E
&=\sum_\alpha 
\occ_\alpha
\left(
\energy_{\alpha}
+\vmf_\alpha\right)
+
\sum_\alpha
\sum_{\beta>\alpha}\sqrt{\occ_\alpha\occ_\beta}
\,
w_{\alpha\beta}
\label{eqn:totalenergy1}
\\
&=\sum_\alpha
\occ_\alpha
\energy_{\alpha}
+
\sum_\alpha
\sum_{\beta\geq\alpha}\sqrt{\occ_\alpha\occ_\beta}
\,
w_{\alpha\beta}.
\label{eqn:totalenergy2}
\end{align}
Now let us define ${\hamsmall_\alpha\equiv \hamsmall + \vmfop}$, 
where
\begin{align*}
\vmfop(x)\equiv \int \what(x,x') \densityn_\alpha(x')\dd{x'}.
\end{align*}
This allows us to write
\begin{align*}
\expval{\hamsmall_\alpha}{\varphi_\alpha} = 
\expval{\hamsmall}{\varphi_\alpha}+\expval{\vmfop}{\varphi_\alpha}=
\energy_\alpha + \vmf_\alpha.
\end{align*}
Therefore, the following exact expressions are equivalent:
\begin{align*}
E 
&= \sum_\alpha\occ_\alpha\expval{\hamsmall_\alpha}{\varphi_\alpha}
+\sum_{\alpha}\sum_{\beta>\alpha}\sqrt{\occ_\alpha\occ_\beta}\mel{\varphi_\alpha}{\what_{\alpha\beta}}{\varphi_\beta}
\\
&= \sum_\alpha\occ_\alpha\expval{\hamsmall}{\varphi_\alpha}
+\sum_{\alpha}\sum_{\beta\geq\alpha}\sqrt{\occ_\alpha\occ_\beta}\mel{\varphi_\alpha}{\what_{\alpha\beta}}{\varphi_\beta}.
\end{align*}

If Eqs.~\ref{eqn:Wexpr4} and ~\ref{eqn:Wexpr5} are used instead of Eq.~\ref{eqn:Eint}, 
the total energy can be expressed as
\begin{align}
E 
 = \sum_\alpha \occ_\alpha\bigg[\energy_\alpha + \frac{1}{N}\big(\vmf_\alpha &+\Walpha \big)\bigg]
\nonumber
\\
&+ \sum_{\alpha,\beta>\alpha}\sqrt{\lambda_\alpha\lambda_\beta} w_{\alpha\beta},
\label{eqn:totalenergy3}
\end{align}
where ${\Walpha}$ is the total interaction energy of ${N-1}$ particles whose state is ${\Theta_\alpha}$.
The first part of this expression is an occupation-weighted sum of the
natural orbital energy $\energy_\alpha$ plus one electron's share of (i.e., ${\frac{1}{N}\times}$)
the mean field interaction between an electron in orbital $\varphi_\alpha$ and the ${N-1}$ 
remaining electrons in state ${\Theta_\alpha}$ plus one electron's share
of the energy ${\Walpha}$ of interaction between the electrons in state ${\Theta_\alpha}$.
The second part of the expression could also be expressed as
\begin{align*}
\sum_{\alpha,\beta>\alpha}\sqrt{\lambda_\alpha\lambda_\beta} w_{\alpha\beta}
=
\frac{1}{N}\sum_{\alpha,\beta>\alpha}\sqrt{\occ_\alpha\occ_\beta} w_{\alpha\beta}.
\end{align*}
It can be interpreted as a correlation term to correct the sum ${\sum_\alpha \lambda_\alpha \vmf_\alpha}$ of
mean field interactions between the electron in orbital $\varphi_\alpha$ and
the electrons in state ${\Theta_\alpha}$.

\subsubsection{Hartree-Fock approximation}
\label{section:hartree_fock}
It can be shown~\citep{coleman_rmp} that
\begin{align*}
\lambda_\alpha\leq \frac{1}{N}\implies \occ_\alpha \leq 1,
\end{align*}
with equality if and only if
$\Psi$ has the form,
\begin{align*}
\Psi(1\cdots {N}) &= \antisymmetrizer 
\left\{\varphi_\alpha(1) \Theta_\alpha(2\cdots {N})\right\}
\\
&= \frac{1}{\sqrt{N}}\bigg[\varphi_\alpha(1)\Theta_\alpha(2\cdots i-1, i, i+1 \cdots N)
\\
&\quad-\sum_{i=2}^N \varphi_\alpha(i)\Theta_\alpha(2\cdots i-1, 1, i+1 \cdots N)\bigg].
\end{align*}
The Hartree-Fock wavefunction is
the simplest wavefunction
with this form, as it can be expressed as 
\begin{align*}
\PsiHF(1\cdots N)=\antisymmetrizer\{\varphi_{\alpha_1}(1)\cdots\varphi_{\alpha_N}(N)\}.
\end{align*}
Therefore, in the Hartree-Fock approximation, ${\varphi_\alpha}$'s dual ${(N-1)}$-state is
\begin{align*}
\Theta_\alpha(1\cdots N-1)
&=
\antisymmetrizer\bigg\{\varphi_1(1)\cdots \breve{\varphi}_\alpha(\alpha)\cdots \varphi_{N}(N-1)\bigg\},
\end{align*}
where ${\breve{\varphi}_\alpha(\alpha)}$ denotes the absence of ${\varphi_\alpha(\alpha)}$
in the product, e.g., ${\varphi_1\breve{\varphi}_2\varphi_3=\varphi_1\varphi_3}$.

It follows that ${\what_{\alpha\beta}}$ is
\begin{align*}
\what_{\alpha\beta}(x) 
&= 
\sum_{\eta\notin\{\alpha,\beta\}}
\int \what(x,x')\abs{\varphi_\eta(x')}^2\dd{x'}
\\
&= \int\what(x,x')\left(\sum_{\eta\notin\{\alpha,\beta\}}n_\eta(x')\right)\dd{x'}
\\
&= \int\what(x,x')\left[n(x')-n_\alpha(x')-n_\beta(x')\right]\dd{x'},
\end{align*}
where 
\begin{align*}
n(x)\equiv\sum_{\alpha=1}^N\abs{\varphi_\alpha(x)}^2=\sum_{\alpha=1}^Nn_\alpha(x) 
\end{align*}
is the number density of $\Psi$. 

This demonstrates that, within the Hartree-Fock approximation,  ${\what_{\alpha\beta}(x)}$ is the
mean field potential at $x$ from the density ${n-n_\alpha-n_\beta}$, which is
the number density of $\Psi$ minus the contribution to it from 
orbitals $\varphi_\alpha$ and ${\varphi_\beta}$.
In other words, the coupling between state ${\ket{\varphi_\alpha}}$ 
and ${\ket{\varphi_\beta}}$ is mediated by
a mean field potential, which does not include a self interaction.

\subsubsection{Summary}
The theory presented in Sec.~\ref{section:orbital_energy} may be important
in many contexts, but it has been developed and presented with chemical bonding in mind, where
the term \emph{chemical} is intended to mean that the attraction between the bonded atoms
occurs due to a substantial redistribution of the atoms' electron densities.

The next section is concerned with attractions between
atoms, surfaces, nanoparticles, or other objects that can occur without substantial redistribution
of the objects' electron densities, because it is the attraction due to dynamical 
correlation of the objects' constituent particles.

\subsection{Non-overlapping bodies}
\label{section:nonoverlapping}
This section presents one way to understand the forces and torques exerted 
by two unmagnetized charge-neutral bodies on one another when they do not overlap spatially and do not
exchange particles.
The bodies could be  atoms,  molecules, nanoparticles, or any other objects composed
of more than one charged particle. Therefore they will be referred to as C-particles, where
`C' abbreviates {\em composite}, and they will be identified individually as
CP1 and CP2. 
The isolated system comprised {\em only} of CP1 and CP2 will be referred to as CP1+CP2.

For simplicity it will be assumed that each C-particle is composed of only two species
of more elementary particle; namely, nuclei of atomic number $Z$ and electrons.

The only approximation made in this section is the neglect of overlap of the CP's wavefunctions.

\begin{figure}
    \includegraphics[width=0.49\textwidth]{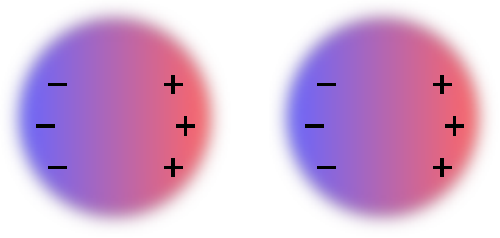}
\label{fig:CP1+CP2}
\caption{A pair of charge-neutral composite particles polarize to lower their potential energy.}
\end{figure}

\subsubsection{Macroscopic charge distributions and their ensembles}
Since CP1 and CP2 are not charged, the forces and torques they exert on one another
arise from non-uniformities of their charge distributions. 
For example, the C-particles depicted in Fig.~\ref{fig:CP1+CP2} attract one another 
because they have polarized such that each one has a dipole moment that is directed
from left to right. They would also attract one another if they both
had opposite (left to right) linear polarizations, or for any number of other
more complex charge distributions, ${\pden=\pdenup{1}+\pdenup{2}}$, of CP1+CP2.

In general, non-uniformities can be either static or dynamic.
Static non-uniformities are non-uniformities of their time-averaged charge
distributions, and dynamic non-uniformities are transient and arise
from interaction-biased quasi-random fluctuations of the CP's microstructures.
The focus of this section is on dynamic uniformities, and I will
refer to the energy and force of interaction between the CPs arising
from the dynamic uniformities as the \emph{correlation} energy and force, 
despite the fact that there also exist intra-CP correlations.

If the fluctuations of the microstructures
of CP1 and CP2 were independent of one another, the probabilities of
the net correlation force being repulsive and attractive at a given instant would be equal, 
and the time average of the net correlation force would vanish.

However, their charge distributions do not change independently of one another, 
and we will see that the energy of CP1+CP2 can be expressed as
\begin{align*}
E= E_1+E_2+\Eint,
\end{align*}
where ${\Eint}$ is the energy of interaction between them.

\subsubsection{Notation}
The variables $x_i$ and $y_j$ specify the cooordinates 
of the $i^\text{th}$ constituent particle of CP1 and the ${j^\text{th}}$ constituent particle of CP2, respectively,
and $\interact{x_i}{y_j}$ denotes the Coulomb repulsion between particles 
with coordinates $x_i$ and $y_j$ if their charges are both either $e$ or $-e$.

The vector ${X\equiv (x_1\cdots x_p)}$ 
specifies the coordinates of all $p$ constituent particles of CP1
and ${Y\equiv (y_1\cdots y_q)}$ specifies the coordinates of 
all $q$ constituent particles of CP2.

Integrals will continue to incorporate sums over spin configurations 
and the abbreviations 
${\dmeasureA{i_1\cdots i_m}\equiv\dd{x_{i_1}}\cdots\dd{x_{i_m}}}$
and
${\dmeasureB{i_1\cdots i_m}\equiv\dd{y_{j_1}}\cdots\dd{y_{j_n}}}$
will be used.
For example, 
\begin{align*}
\int f(x_1,x_2)\dmeasureA{1,2}  \equiv \int f(x_1,x_2)\dd{x_{1}}\dd{x_{2}}.
\end{align*}

\subsubsection{Wavefunction}
The wavefunction of CP1+CP2 is 
\begin{align*}
\Psi(s_1\cdots s_N)=\sum_\alpha C_\alpha\wx_\alpha(s_1\cdots s_p)\wy_\alpha(s_{p+1}\cdots s_N),
\end{align*}
where ${\{s_1\cdots s_N\}\equiv\{x_1,x_2\cdots x_p,y_1\cdots y_q\}}$;
${\{\wx_\alpha\}}$ and ${\{\wy_\alpha\}}$ are the sets of all natural
$p$-states and $q$-states, respectively; and the dual of $p$-state
$\wx_\alpha$ is the $q$-state $\wy_\alpha$.

Although $\Psi$ is antisymmetric, the C-particles do not overlap
significantly. Furthermore, 
as the distance between them increases, the rates
at which particles move between them decrease, while the characteristic
time and length scales of fluctuations of their charge
distributions that are capable of producing significant relative forces and torques increase.
Therefore let us make the physical assumption that the particles are at a separation $r$ for
which there exists a time scale $\tau$ such that
the average frequency with which particles travel between them  
is much smaller than ${1/\tau}$, and the time scale of the charge redistribution 
processes responsible for their relative forces and torques is
much smaller than $\tau$.

Under this assumption, and since the degree of overlap between CP1 and CP2 is negligible,  it does not change
the energy or the expectation value of any observable if $\Psi$ is chosen to not have the correct (anti-)symmetry with respect
to exchange of coordinates {\em between} CP1 and CP2.
Therefore $\Psi$ can be expressed as
(see subsection~\ref{section:natural_states} or~\linecite{coleman_rmp})
\begin{align}
\Psi(X,Y) & = \sum_{\alpha} C_\alpha \wx_\alpha(X)\wy_\alpha(Y), \label{eqn:state}
\end{align}
where each $\wx_\alpha$  is an eigenfunction of the integral operator with kernel
\begin{align*}
\dmatrixarg{\wx}_p(X;X')  &\equiv \int \Psi(X,Y) \Psi^*(X',Y) \dmeasureB{1\cdots q}
\end{align*}
and each $\wy_\alpha$ is an eigenfunction of the integral operator with kernel
\begin{align*}
\dmatrixarg{\wy}_q(Y;Y')  &\equiv \int \Psi(X,Y) \Psi^*(X,Y') \dmeasureA{1\cdots p}.
\end{align*}
That is, 
\begin{align*}
\int \dmatrixarg{\wx}_p(X;X')\wx_\alpha(X')\dmeasureA{1\cdots p}  &= \lambda_\alpha \wx_\alpha(X),
\\
\int \dmatrixarg{\wy}_q(Y;Y')\wy_\alpha(Y')\dmeasureB{1\cdots q}  &= \lambda_\alpha \wy_\alpha(Y), 
\end{align*}
where ${\lambda_\alpha =\abs{C_\alpha}^2}$.

The sets ${\{\wx_\alpha(X)\}}$ and
${\{\wy_\alpha(Y)\}}$ are orthonormal, meaning that
${\braket{\wx_\alpha}{\wx_\beta}=\delta_{\alpha\beta}}$ and
${\braket{\wy_\alpha}{\wy_\beta}=\delta_{\alpha\beta}}$, and 
their elements 
have the appropriate symmetry with respect to interchange of any two identical particles on the
same C-particle.
For example if $x_i$ and $x_j$ are the coordinates of electrons on CP1, then 
\begin{align*}
\wx_\alpha(x_1\cdots x_i \cdots x_j \cdots x_p)=
-\wx_\alpha( x_1\cdots  x_j \cdots  x_i \cdots x_p).
\end{align*}

\subsubsection{Energy}
The Hamiltonian of CP1+CP2 can be expressed as 
${\Ham = \Ham_1 + \Ham_2 + \Hamint}$, where
$\Ham_1$ and $\Ham_2$ are the Hamiltonians of
CP1 and CP2, respectively, and $\Hamint$ is the interaction between them. 
Using the notation
\begin{align*}
&\mel{\wx_\alpha\wy_\alpha}{\Ham}{\wx_\beta\wy_\beta}
\\
&\qquad\quad\equiv
\int\int 
\bar{\wx}_\alpha( X)\bar{\wy}_\alpha( Y)\,\Ham\,
\wx_\beta(X)\wy_\beta(Y) \dmeasureA{1\cdots p}\dmeasureB{1\cdots q},
\end{align*}
the energy  of CP1+CP2 can be expressed as
\begin{align*}
 E &= 
\expval{\Ham}{\Psi}
=
\sum_{\alpha\beta}
\bar{C}_\alpha C_\beta
\mel{\wx_\alpha\wy_\alpha}{\Ham}{\wx_\beta\wy_\beta}
\\
&= \sum_\alpha\lambda_\alpha\left(\CPenergy{1}_\alpha+\CPenergy{2}_\alpha\right) + \expval{\Hamint}{\Psi},
\end{align*}
where ${\Eint\equiv\expval{\Hamint}{\Psi}}$ is the energy 
of interaction between CP1 and CP2; and
 ${\CPenergy{1}_\alpha\equiv\expval{\Ham_1}{\wx_\alpha}}$ and
 ${\CPenergy{2}_\alpha\equiv\expval{\Ham_2}{\wy_\alpha}}$ are the expectation values
of $\Ham_1$ and ${\Ham_2}$, respectively, when the state $\ket{\wx_\alpha}$ of CP1 is the $\alpha^\text{th}$ natural
$p$-state of $\ket{\Psi}$ and the state ${\ket{\wy_\alpha}}$ of CP2 is the natural ${q}$-state
of ${\ket{\Psi}}$ that is ${\ket{\wx_\alpha}}$'s dual state, i.e., 
\begin{align*}
\lcontractN{\wx_\alpha}{\Psi}&=\ket{\wy_\alpha}, 
&
\lcontractN{\wy_\alpha}{\Psi}&=\ket{\wx_\alpha}.
\end{align*}

The remainder of Sec.~\ref{section:nonoverlapping} will focus on the interaction
energy, $\Eint$.

\subsubsection{Interaction energy}
In what follows, the interaction energy will be expressed as
\begin{align*}
\Eint\equiv \expval{\Hamint}{\Psi}\equiv \Eintcp{+-}+\Eintcp{-+}+\Eintcp{--}+\Eintcp{++}, 
\end{align*}
where ${\Eintcp{+-}}$ denotes the energy
of interaction between the nuclei of CP1 and the electrons of CP2; ${\Eintcp{-+}}$
denotes the energy of interaction between the electrons of CP1 and the nuclei of CP2;
and ${\Eintcp{--}}$ and ${\Eintcp{++}}$ denote the energies of interaction between
CP1 and CP2 that only involve electrons and nuclei, respectively.
An expression for ${\Eintcp{+-}}$ will now be derived, and expressions for ${\Eintcp{-+}}$, ${\Eintcp{--}}$, 
and ${\Eintcp{++}}$ that could be derived by a similar route will then be presented.

Let ${\sxn}$ and ${\sxe}$ denote the sets of all indices $i$ for which ${x_i}$ is
the coordinate of one of CP1's nuclei and electrons, respectively; and let ${\syn}$
and ${\sye}$ denote the sets of all indices $j$ for which $y_j$ 
is one of CP2's nuclei and electrons, respectively.
Then
the interaction between CP1's nuclei and CP2's electrons can be expressed as
\begin{widetext}
\begin{align}
\Eintcp{+-} 
&=-Z
\int\dd{x}\int\dd{y}
w(x,y)
\left(
\sum_{\substack{i\in \sxn \\ j\in\sye}}
\int \int \abs{\Psi(x_1\cdots y_q)}^2
\delta(x-x_i)\delta(y-y_j)
\dmeasureA{1\cdots p}\dmeasureB{1\cdots q}
\right)
\nonumber
\\
&=-Z
\int\dd{x}\int\dd{y}
w(x,y)
\left(
\numpx\nummy
\int \int \abs{\Psi(x,x_2\cdots x_p,y,y_2\cdots y_q)}^2
\dmeasureA{2\cdots p}\dmeasureB{2\cdots q}
\right),
\label{eqn:Eintfirst}
\end{align}
\end{widetext}
where ${\numpx}$ denotes the number of nuclei in CP1, 
${\nummy}$ denotes the number electrons in CP2;
and the symmetry of ${\abs{\Psi}^2}$ with respect to exchange
of identical particles belonging to the same CP has been used to reach the second expression from the first.

Equation~\ref{eqn:Eintfirst} will be simplified by 
expressing it in terms of position probability density functions (pdfs). First, the notation
used to identify pdfs, joint pdfs, and condition pdfs will be introduced.

\paragraph{Notation for probability density functions (pdfs):}
The joint probability density that one of CP1's nuclei is at $x$ and one of 
CP2's electrons is at $y$ is
\begin{align*}
&\denpm(x,y)\\
&\quad\equiv  \numpx\nummy\int\int 
\abs{\Psi(x,x_2\cdots y,y_2\cdots y_q)}^2\dmeasureA{2\cdots p}\dmeasureB{2\cdots q}.
\end{align*}
More generally, ${\den_{\scriptscriptstyle ss'}(x,y)}$, where ${s,s'\in\{+,-\}}$,
will denote the joint probability
density that one of CP1's particles of type ${s}$ is at $x$ and  one of CP2's
particles of type ${s'}$ is at $y$, where particles of type `$+$' are nuclei and
those of type
`$-$' are electrons.

Let ${\denargp{i}}$ and ${\denargm{i}}$ denote the
number densities of CPi's nuclei and electrons, respectively, where ${\text{CPi}\in\{\text{CP1},\text{CP2}\}}$.
For example, ${\denargm{2}(y)}$ is the probability density that one of CP2's electrons 
is at $y$, which implies that ${\int\dd{y}\denargm{2}(y)=\nummy}$.

Let ${\den^{\scriptscriptstyle (i)}_{\scriptscriptstyle s|s'}(u|v)}$, 
where ${s, s'\in \{-,+\}}$, denote the conditional
probability density that one of CPi's 
particles of type $s$ 
is at $u$ \emph{given} that one of the other CP's particles of type $s'$ is at $v$.
For example, ${\dencmp{2}(y|x)}$ is the conditional probability density
that one of CP2's electrons is at $y$, given that one of CP1's nuclei is at $x$;
and ${\dencmm{1}(x|y)}$ is the conditional probability density that one
of CP1's electrons is at $x$ given that one of CP2's electrons is at $y$.

These definitions imply the following relations:
\begin{align*}
\denpm(x,y) &= \dencpm{1}(x|y)\denargm{2}(y) = \denargp{1}(x)\dencmp{2}(y|x)
\\
\denmp(x,y) &= \dencmp{1}(x|y)\denargp{2}(y) = \denargm{1}(x)\dencpm{2}(y|x)
\\
\denpp(x,y) &= \dencpp{1}(x|y)\denargp{2}(y) = \denargp{1}(x)\dencpp{2}(y|x)
\\
\denmm(x,y) &= \dencmm{1}(x|y)\denargm{2}(y) = \denargm{1}(x)\dencmm{2}(y|x).
\end{align*}

\paragraph{Interaction energies in terms of pdfs}
We can now express Eq.~\ref{eqn:Eintfirst} as 
\begin{align*}
\Eintcp{+-}
&= -Z\int\dd{x}\int\dd{y} w(x,y)\denpm(x,y).
\\
&= -Z\int\dd{x}\int\dd{y} w(x,y)\dencpm{1}(x|y)\denargm{2}(y)
\\
&\equiv -Z\bbraket{\dencpm{1}}{\denargm{2}}
= -Z\bbraket{\denargp{1}}{\dencmp{2}},
\end{align*}
where I have introduced the shorthand
notation,
\begin{align*} 
\bbraket{f}{g}\equiv \int \dd{x}\int \dd{y} w(x,y) f(x) g(y), 
\end{align*}
which I will now begin to use extensively.
The function $f$ that occupies the first
slot of ${\bbraket{\,\cdot\,}{\,\cdot\,}}$ will always be 
a position pdf for CP1's nuclei or electrons, 
and the function $g$ in the second slot will always be a 
position pdf for CP2's nuclei or electrons. Either $f$ or $g$ may or may not 
be a conditional pdf, i.e., ${f=f(x|y)}$ or ${f=f(x)}$ and ${g=g(y|x)}$ or ${g=g(y)}$.

In this notation, the energy of interaction between CP1's electrons and CP2's nuclei
is 
\begin{align*}
\Eintcp{-+}\equiv 
-Z(\dencpm{1}|\denargm{2})=
-Z(\denargp{1}|\dencmp{2});
\end{align*}
the energy of interaction between CP1's electrons and CP2's electrons is
\begin{align*}
\Eintcp{--}\equiv (\dencmm{1}|\denargm{2})=(\denargm{1}|\dencmm{2});
\end{align*}
the energy of interaction between CP1's nuclei and CP2's nuclei is
\begin{align*}
\Eintcp{++}\equiv Z^2(\dencpp{1}|\denargp{2})=Z^2(\denargp{1}|\dencpp{2}); 
\end{align*}
and the total interaction energy is
\begin{align}
\Eint 
&= \Eintcp{++} + \Eintcp{--} +\Eintcp{+-}+\Eintcp{-+}
\nonumber
\\
&= 
Z^2(\denargp{1}|\dencpp{2})
+
(\denargm{1}|\dencmm{2})
\nonumber
\\
&\qquad\qquad\qquad-Z\left[(\denargp{1}|\dencmp{2})
+(\denargm{1}|\dencpm{2})\right]
\label{eqn:Eintshort1}
\\
&= 
Z^2(\dencpp{1}|\denargp{2})
+
(\dencmm{1}|\denargm{2})
\nonumber
\\
&\qquad\qquad\qquad-Z\left[(\dencpm{1}|\denargm{2})
+(\dencmp{1}|\denargp{2})\right]
\label{eqn:Eintshort2}
\end{align}
Now let us define a set of probability or number density
\emph{response}
functions, as follows:
\begin{align*}
&\ddencmm{i}\equiv \dencmm{i} - \denargm{i},&
&\ddencmp{i}\equiv \dencmp{i} - \denargm{i},&
\\
&\ddencpp{i}\equiv \dencpp{i} - \denargp{i},&
&\ddencpm{i}\equiv \dencpm{i} - \denargp{i}.&
\end{align*}
Then $\Eint$ can be expressed as
${\Eint=\Estatic + \Edyn}$,
where $\Estatic$ 
is the energy of the mean-field interaction between the
CPs' average charge densities, 
\begin{align*}
\rhoarg{1} &\equiv eZ\denargp{1}-e\denargm{1},
&
\rhoarg{2} &\equiv eZ\denargp{2}-e\denargm{2}; 
\end{align*}
and $\Edyn$ is the contribution to $\Eint$ from
dynamical correlation.

\subsubsection{Mean field interaction energy}
The mean-field interaction energy can be expressed as
\begin{align}
\Estatic&\equiv
\overbrace{
Z^2\bbraket{\denargp{1}}{\denargp{2}}
}^{\displaystyle \Estatarg{++}}
+
\overbrace{
\bbraket{\denargm{1}}{\denargm{2}}
}^{\displaystyle \Estatarg{--}}
\nonumber
\\
&\qquad\qquad\qquad
\underbrace{
-Z\bbraket{\denargp{1}}{\denargm{2}}
}_{\displaystyle \Estatarg{+-}}
\underbrace{-Z
\bbraket{\denargm{1}}{\denargp{2}}
}_{\displaystyle \Estatarg{-+}}
\\
& = 
Z\bbraket{\denargp{1}}{\pdenarg{2}}
-
\bbraket{\denargm{1}}{\pdenarg{2}}=\bbraket{\pdenarg{1}}{\pdenarg{2}}
\label{eqn:Emfshort}
\end{align}
where ${\pdenarg{1}}$, ${\pdenarg{2}}$ and ${\pden\equiv\pdenarg{1}+\pdenarg{2}}$ 
are the charge densities of CP1, CP2, and CP1+CP2, respectively.

If the separation $r$ between the CPs' centers is large relative to 
their linear dimensions (e.g., their diameters, if they are spherical) it may
be useful to express
${\pdenarg{1}}$ and ${\pdenarg{2}}$ as multipole expansions. Then $\Estatic$
can be expressed exactly as an infinite sum of multipole-multipole interactions,
or approximated by a truncation of the infinite sum.

By assumption, each CP is charge-neutral. Therefore the term
in the infinite sum that decays slowest as $r$
increases is the ${1/r^3}$ dipole-dipole term; the next slowest decaying terms are the 
${1/r^4}$ dipole-quadrupole and quadrupole-dipole terms; and, in general and in principle, there are terms that decay as
${1/r^m}$ for all integers ${m\geq 3}$.

Note that ${\denargp{1}}$, ${\denargm{1}}$, ${\denargp{2}}$, and ${\denargm{2}}$
are all non-negative. Therefore ${\Estatarg{++}}$ and ${\Estatarg{--}}$ are positive
contributions to $\Eint$, which always contribute repulsions to the inter-CP force;
and ${\Estatarg{+-}}$ and ${\Estatarg{-+}}$ are negative contributions to $\Eint$, which
always contribute attractions to the inter-CP force.

\subsubsection{Correlation interaction energy}
The correlation interaction energy can be expressed as
\begin{align}
\Edyn 
&\equiv
Z^2\bbraket{\denargp{1}}{\ddencpp{2}}
+
\bbraket{\denargm{1}}{\ddencmm{2}}
\nonumber
\\
&\quad\qquad\quad
-Z\left[
\bbraket{\denargp{1}}{\ddencmp{2}}
+
\bbraket{\denargm{1}}{\ddencpm{2}}
\right]
\label{eqn:Ecor1}
\\
&= 
Z^2\bbraket{\ddencpp{1}}{\denargp{2}}
+
\bbraket{\ddencmm{1}}{\denargm{2}}
\nonumber
\\
&\quad\qquad\quad-Z\left[\bbraket{\ddencpm{1}}{\denargm{2}}
+\bbraket{\ddencmp{1}}{\denargp{2}}\right]
\label{eqn:Ecor2}
\\
& = \Edynarg{++}+\Edynarg{--}+\Edynarg{+-}+\Edynarg{-+},
\nonumber
\end{align}
where
\begin{align*}
\Edynarg{++}\equiv Z^2\bbraket{\denargp{1}}{\ddencpp{2}}=Z^2\bbraket{\ddencpp{1}}{\denargp{2}}
\end{align*}
is the energy of correlation between CP1's nuclei and CP2's nuclei;
\begin{align*}
\Edynarg{--}\equiv \bbraket{\denargm{1}}{\ddencmm{2}}=\bbraket{\ddencmm{1}}{\denargm{2}}
\end{align*}
is the energy of correlation between CP1's electrons and CP2's electrons;
\begin{align*}
\Edynarg{+-}\equiv -Z\bbraket{\denargp{1}}{\ddencmp{2}}=-Z\bbraket{\ddencpm{1}}{\denargm{2}}
\end{align*}
is the energy of correlation between CP1's nuclei and CP2's electrons; and
\begin{align*}
\Edynarg{-+}\equiv -Z\bbraket{\denargm{1}}{\ddencpm{2}}=-Z\bbraket{\ddencmp{1}}{\denargp{2}}
\end{align*}
is the energy of correlation between CP1's electrons and CP2's nuclei.

\subsubsection{Density response functions}
The response functions ${\ddencpp{i}}$, ${\ddencmm{i}}$, ${\ddencpm{i}}$, and ${\ddencmp{i}}$ are
not non-negative everywhere because they are not pdfs. They are differences between pdfs.
For example,
\begin{align*}
\ddencpp{1}(x|y)\equiv \dencpp{1}(x|y)-\denargp{1}(x)
\end{align*}
is the difference between the probability density 
that one of CP1's nuclei is at $x$ when one of CP2's nuclei is at $y$ 
and the probability density that one of CP1's nuclei is at $x$ when 
nothing more specific than ${\denargp{2}}$ is known about the locations of CP2's nuclei.

Conservation of probability and conservation of the number of nuclei imply that 
\begin{align*}
\int\dd{x}\denargp{1}(x)=\int\dd{x}\dencpp{1}(x|y)\implies
\int\dd{x}\ddencpp{1}(x|y)=0. 
\end{align*}
Therefore they imply that if the discovery or revelation that
there is a nucleus is at $y$ causes
the number density of nuclei to decrease in one part of 
CP1, it must cause it to increase in another part of CP1.

\begin{figure}
    \includegraphics[width=0.49\textwidth]{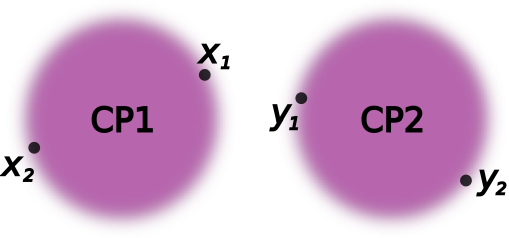}
\label{fig:CP1+CP2_4}
\caption{Schematic. See text.}
\end{figure}
Nuclei repel one another. Therefore, because the points $x_1$ and
$y_1$ shown in Fig.~\ref{fig:CP1+CP2_4} are relatively close to one another,
it seems reasonable to expect that the probability density of there being a nucleus at
$x_1$ would be reduced by Coulomb repulsion if there was a nucleus at $y_1$. 
If that  was the case,  ${\dencpp{1}(x_1|y_1)}$ would be less
than ${\denargp{1}(x_1)}$, so
${\ddencpp{1}(x_1|y_1)}$ be negative, and the contribution of 
points $x_1$ and $y_1$ to ${\Edynarg{++}}$ would be negative.

On the other hand, ${\ddencpp{1}(x_2|y_2)}$ might be positive, despite
the repulsion that would exist between nuclei at $x_2$ and $y_2$, because
probability density is conserved and because
the distance between $y_2$
and $x_2$ is larger than the distance between ${y_2}$ and 
most other points in CP1. 
Since
the electric potential 
from a nucleus at $y_2$ would be lower at $x_2$ than at most
other points in CP1, the presence of a nucleus at $y_2$ might increase
the nucleus number density at $x_2$ in order to reduce
it elsewhere.

Overall, however, if nuclei were sufficiently mobile,
and if other contributions to the correlation
energy 
(${\Edynarg{--}}$, 
${\Edynarg{+-}}$, and
${\Edynarg{-+}}$) had negligible effects on the dynamics of nuclei,
points in CP1 that are close to CP2 would be more likely
to contribute negatively to $\Edynarg{++}$, 
and points in CP1 that are further away from CP2 
would be more likely to contribute positively to ${\Edynarg{++}}$.

We can state this mathematically by defining
\begin{align*}
\edynarg{++}(x) &\equiv Z^2\int\dd{y}
w(x,y)
\ddencpp{1}(x|y)
\denargp{2}(y)
\\
\implies \Edynarg{++}&= \int\dd{x}\edynarg{++}(x).
\end{align*}
Then ${\edynarg{++}(x_1)}$ is likely
to be negative and 
${\edynarg{++}(x_2)}$ is likely
to be positive. 
Furthermore, since $x_1$ is
closer to CP2 than $x_2$ is, it
is likely that 
\begin{align*}
\abs{\edynarg{++}(x_1)}>\abs{\edynarg{++}(x_2)}.
\end{align*}
Therefore ${\Edynarg{++}}$ 
is more likely to be negative than positive.
Similar reasoning could be used to
argue that 
${\Edynarg{--}}$,
${\Edynarg{+-}}$, and
${\Edynarg{-+}}$ are all likely to be negative;
and it is known that the sum 
${\Edyn}$ of all contributions to the correlation energy is negative.

However, while arguing that $\Edynarg{++}$ is
likely to be negative, we have implicitly made some physical assumptions
that are not necessarily justified or valid, or even likely to be justified or valid.
For example, we have assumed that ${\Edynarg{++}}$ is not
changed significantly by correlations between electrons and nuclei.
However,  we will not discuss this possibility because we will
discuss another of our implicit assumptions, which 
but we will discuss another
of our implicit assumptions. Its dubious validity that we have made, and
which makes that assumption likely to be valid.

At first sight, and as it is presented above, the reasoning 
appears  to contain a strong and unjustified implicit
assumption. Namely, it appears that the fact that one of CP2's nuclei is at 
$y$ at time $\tau$ changes the probability density \emph{at time $\tau$}
that one of CP1's nuclei is at $x$.
If this assumption was being made, it would not be valid because the time taken
by nuclei in a neighbourhood of $x$ to respond to the arrival of
a nucleus at $y$ is finite. Their response is not instantaneous.

However, let us not forget that ${\Edynarg{++}}$
can be expressed as 
${Z^2\bbraket{\dencpp{1}}{\denargp{2}}}$ or as ${Z^2\bbraket{\denargp{1}}{\dencpp{2}}}$, 
and that the correlation is not the nuclei of CP1 responding to fluctuations
in the spatial distribution of CP2's nuclei, or vice-versa, but the nuclei
of each CP moving under the influence of the other CP's nuclei.
The motions of the nuclei and electrons of CP1 are correlated with one another and, 
to a lesser degree because they are further away, they are correlated
with the motions of CP2's nuclei and electrons.

Therefore it is not appropriate to interpret ${\ddencpp{1}(x|y)}$ as the
response of CP1's number density of nuclei at $x$, ${\denargp{1}(x)}$, to one of CP2's nuclei
suddenly appearing at $y$; and it is not even appropriate to interpret
it as a response to the entire history of the nucleus whose
position at time $t$ is $y$. It should be interpreted as an
average of the responses to the trajectories in the set
\begin{align*}
\big\{
\left\{(X(t),Y(t)): t\in (-\infty,\tau]\right\} :\exists j\in\syn \;s.t.\; y_j(\tau)=y\big\}
\end{align*}
of all possible trajectories ${(X(t),Y(t))}$ of CP1+CP2 that are consistent with 
one of CP2's nuclei being at $y$ at time $\tau$.

\subsubsection{Decoupled non-overlapping bodies}
It has been shown above that the energy of CP1+CP2
can be expressed as
\begin{align*}
E=\sum_\alpha\lambda_\alpha\left(\CPenergy{1}_\alpha + \CPenergy{2}_\alpha\right)
+ \Emf + \Edyn.
\end{align*}

\subsubsection{Separation of time scales}
We have found the following expression for the total energy of CP1+CP2, which is exact in the limit of zero overlap between CP1 and CP2
if CP1+CP2 is in a pure state $\Psi$.
\begin{align}
E[\Psi] & = \sum_\alpha\lambda_\alpha\left(\Epsilon_1^\alpha
+\Epsilon_2^\alpha\right) + \Eintmf[\pdensuper{1},\pdensuper{2}] \nonumber \\
 & +\frac{1}{2}\left[
\expval{\pdensuperm{1},\dbarvm{2}}
+
\expval{\pdensuperm{2},\dbarvm{1}}
\right]
\nonumber \\
&-\frac{Z}{2}\left[
\expval{\pdensuperp{1},\dbarvp{2}}
+
\expval{\pdensuperp{2},\dbarvp{1}}
\right]
\label{eqn:E12_2}
\end{align}
Let us assume that an isolated C-particle is approximately spherical but thermally disordered. 
When two C-particles approach one another the interaction between them can break their near-spherical symmetry.
If they are observed on a time scale that is short relative to the time scale on which they rotate about an axis
passing through their centers, and that is short relative to the time scale on which the internal structure of a C-particle can rearrange, 
it is reasonable to assume that they are observed in a pure state. This is because there are no relevant symmetries on such a time scale.

Let us now consider the different types of correlation described by the $\delta\bar{v}$ terms on the right hand side of Eq.~\ref{eqn:E12_2}.
The terms $-Z\expval{\pdensuper{1}_\p,\delta\bar{v}_\p^{(2)}}$ and $-Z\expval{\pdensuper{2},\delta\bar{v}_\p^{(1)}}$ account for the energy associated
with synchronicity between the motion of nuclei on one C-particle and the motion of nuclei and electrons on the other. If we assume that nuclei
move much more slowly that electrons and that electrons are free to move so that, on the time scale of nuclear motion, 
they perfectly screen any fields from nuclei on the other C-particle, then $\delta\bar{v}_\p{(1)}=\delta\bar{v}_\p{(2)}=0$ and only
the synchronous motion of electrons on different C-particles is relevant. Our assumption that electrons move freely also implies that 
$\Eintmf[\pdensuper{1},\pdensuper{2}] =0$, since both C-particles are globally charge-neutral and since on nuclear time scales electrons 
move rapidly to ensure local charge-neutrality. Therefore, it is expected that a very good approximation to the energy of CP1+CP2 is 
provided by 
\begin{align*}
E & \approx \sum_\alpha\lambda_\alpha\left(\Epsilon^\alpha_1+\Epsilon^\alpha_2\right) 
 +\frac{1}{2}
\expval{\pden_\subminus^{(1)},\delta\bar{v}_\subminus^{(2)}}
+\frac{1}{2}\expval{\pden_\subminus^{(2)},\delta\bar{v}_\subminus^{(1)}},
\end{align*}
or
\begin{align}
E[\{\lambda_\alpha,&\wx_\alpha,\wy_\alpha\} ] \approx \sum_\alpha\lambda_\alpha\left(E_1[\wx_\alpha]+E_2[\wy_\alpha]\right)  
\nonumber \\
& +\frac{1}{2}
\expval{\pden_\subminus^{(1)}[\{\lambda_\alpha,\wx_\alpha\}],\delta\bar{v}_\subminus^{(2)}[\{\lambda_\alpha,\wx_\alpha,\wy_\alpha\}]}\nonumber \\
&+\frac{1}{2}\expval{\pden_\subminus^{(2)}[\{\lambda_\alpha,\wy_\alpha\}],\delta\bar{v}_\subminus^{(1)}[\{\lambda_\alpha,\wx_\alpha,\wy_\alpha\}]}
\label{eqn:E12_3}
\end{align}
\vspace{1cm}

\subsubsection{Appendix to Appendix~\ref{section:appendix_natural}:Creation and annihilation operators}
Note that ${\Wn_\beta}$, where ${\beta\neq\alpha}$,
depends indirectly on ${\ket{\varphi_\alpha}}$, and this
dependence could be made explicit, but I will not do this.
However, I will draw out the dependence of
${\Delta\varepsilon_{\alpha\beta}}$ on orbitals other than 
${\ket{\varphi_\alpha}}$ and ${\ket{\varphi_\beta}}$.
These dependences enter ${\Delta\varepsilon_{\alpha\beta}}$ via
$\hat{\mathcal{V}}_{\alpha\beta}$, because
${\ket{\Theta_\alpha}}$ and ${\ket{\Theta_\beta}}$
both contain finite overlaps with {\em at least} ${N-1}$ natural $1$-states.

The overlap  of ${\ket{\Theta_\alpha}}$ with ${\ket{\varphi_\alpha}}$
vanishes by Eq.~\ref{eqn:orthogonality}; 
however ${\braket{\varphi_\alpha}{\Theta_\beta}}$ does
not vanish, in general, if ${\beta\neq\alpha}$. 
Therefore, 
let us express
${\ket{\Theta_\beta}}$ as the sum of a state with finite overlap
with ${\ket{\varphi_\alpha}}$ and a state $\ket{\Theta_{\beta\perp\alpha}}$ whose projection 
onto ${\ket{\varphi_\alpha}}$ vanishes.
To facilitate this decomposition, let us define the {\em annihilation operator }
${\hat{a}_\alpha}$ and the {\em creation operator} ${\hat{a}_\alpha^\dagger}$ 
by their actions on an $M$-particle state $\chi_M$ and an $(M-1)$-particle state $\chi_{M-1}$, respectively.
\begin{align*}
&\left(\hat{a}_\alpha \chi_M\right)(1\cdots M-1) 
 \equiv M^{\frac{1}{2}}\int \chi_M(1\cdots M)\bar{\varphi}_\alpha(M)\dd{x_M}
\\
&\left(\hat{a}^\dagger_\alpha \chi_{M-1}\right)(1\cdots M) 
 \equiv M^{-\frac{1}{2}}\hat{\mathcal{A}}\left\{\chi_{M-1}(1\cdots M-1)\varphi_\alpha(M)\right\}
\end{align*}
where ${\hat{\mathcal{A}}}$ is the antisymmetrization operator.
With a bit of algebra it can be shown that 
${\hat{a}_\alpha\hat{a}_\alpha^\dagger + \hat{a}_\alpha^\dagger\hat{a}_\alpha=\identity}$, 
where $\identity$ is the identity.
Note that this notation is a bit sloppy and, as a result,
this expression for the identity is misleading. We should really express it as
${\hat{a}_{M+1,\alpha}\hat{a}_{M+1,\alpha}^\dagger + \hat{a}_{M,\alpha}^\dagger\hat{a}_{M,\alpha}=\identity_M}$, 
where ${\hat{a}_{M,\alpha}}$ acts on $M$-particle states to produce ${(M-1)}$-particle
states, 
${\hat{a}_{M,\alpha}^\dagger}$ acts on $(M-1)$-particle states to produce ${M}$-particle
states, and ${\identity_M}$ is the identity in the $M$-particle Hilbert space.
With this in mind, let us proceed with the simpler sloppy notation.
We can write
\begin{align}
\Theta_\beta(2\cdots N) 
&= 
\hat{a}_\alpha^\dagger\hat{a}_\alpha\Theta_\beta(2\cdots N) +
\hat{a}_\alpha\hat{a}^\dagger_\alpha\Theta_\beta(2\cdots N)
\nonumber\\
&= \hat{a}_\alpha^\dagger 
\Theta_{\beta-\alpha}(2\cdots N-1) +\Theta_{\beta\perp\alpha}(2\cdots N)
\nonumber
\end{align}
where ${\braket{\varphi_\alpha}{\Theta_{\beta-\alpha}}}$
and
${\braket{\varphi_\alpha}{\Theta_{\beta\perp\alpha}}}$ both vanish.
Then, 
\begin{align}
N\left(\frac{c_\alpha}{c_\beta}\right)
&\Delta\varepsilon_{\alpha\beta}
\equiv 
\mel{\Theta_\alpha}{\hat{\mathcal{U}}_{\alpha\beta}}{\hat{a}_\alpha^\dagger\Theta_{\beta-\alpha}}
+ \mel{\Theta_\alpha}{\hat{\mathcal{U}}_{\alpha\beta}}{\Theta_{\beta\perp\alpha}}
\nonumber \\
& =
\int
\bar{\varphi}_\alpha(1)
\bar{\theta}_{\alpha\beta}(2)
\hat{w}(1,2)
\varphi_\alpha(2)
\varphi_\beta(1)
\dmeasure{1,2}
\nonumber \\
&+ 
\int
\bar{\varphi}_\alpha(1)
\bar{\Theta}_\alpha(2\cdots N)
\hat{w}(1,2)
\nonumber \\
&\qquad\times\Theta_{\beta\perp\alpha}(2\cdots N)
\varphi_\beta(1)
\dmeasure{1\cdots N}
\label{eqn:X}
\end{align}
where ${\ket{\theta_{\alpha\beta}}\equiv \braket{\Theta_{\beta-\alpha}}{\Theta_\alpha}}$
is a $1$-particle state that is orthogonal to $\ket{\varphi_\alpha}$
and, 
to reach the second equation from the first, I have
used the orthogonality of ${\Theta_\alpha}$ to ${\varphi_\alpha}$, as
follows: in the expression for ${\hat{a}_\alpha^\dagger\Theta_{\beta-\alpha}}$, 
I expanded the antisymmetrized product of ${\Theta_{\beta-\alpha}}$ and ${\varphi_\alpha}$
as a sum; then I used the fact that each integral for which 
the argument of ${\varphi_\alpha}$ is not $2$ vanishes.

\onecolumngrid
\vspace{1cm}
\PRLsep
\vspace{1cm}
\twocolumngrid
\section{Excess field invariance proofs}
\label{section:invariance_proofs}
Equations~\ref{eqn:surfaverage0},~\ref{eqn:surfaverage1}, 
~\ref{eqn:backsurfave0} and~\ref{eqn:backsurfave1}, 
which are expressions for
the macroscale interfacial excesses of $\Dnu(x)$ and ${x\,\Delta\nu(x)}$, 
are the most important results of Sec.~\ref{section:excess_fields}
and among the most important results of the homogenization theory
presented in this work.

On first examination these expressions appear to depend on $x_b$
and on how the mesoscale neighbourhood of $x_b$ 
is partitioned into microscopic intervals. Since any such 
dependence would make them ill-defined quantities, 
it is crucial to the importance and generality of these expressions
that all choices of $x_b$ and $\Pi(x_b,\ell)$, which 
satisfy the conditions stated in Sec.~\ref{section:partition}, 
give the same values for $\mbsx{\Dnu}_0$ and $\mbsx{\Delta\nu}_1$.
This section is devoted to proving that this is indeed the case.

\subsection{Derivatives of $\bmone$ and $\bmtwo$ with respect to $x_b$}
\label{section:xb_derivative}
The derivatives of ${\bmany(x_b)}$ with respect to $x_b$. 
will be used to demonstrate that the surface excesses 
calculated in Sec.~\ref{section:s0ave} and~\ref{section:s1ave} are independent
of $x_b$. 
This will demonstrate that $x_b$ is a parameter that determines the values of each term on
the right hand sides of Eqs.~\ref{eqn:surfaverage0} and~\ref{eqn:surfaverage1}, but not their sums - 
$\mbs_0(\mxb)$ and $\mbs_1(\mxb)$, respectively.
I will assume that ${\amax/\prectheo}$ is sufficiently small that
the kernel average ${\expval{\many;\mu}^*_\prectheo}$ can be replaced 
with a simple average, i.e., 
\begin{align}
\bmany(x_b)
&= \frac{1}{\ell}\;\sum_{m} \many(\bar{x}_m,\Delta_m) \Delta_m \label{eqn:moment2}
\end{align}
where
$
\many(\bar{x}_m,\Delta_m) \equiv 
{\Delta_m^{-1}} \int_{x_m^-}^{x_m^+} \left(x - \bar{x}_m\right)^n \Dnu(x) \dd{x} 
$.
For convenience I have denoted the left-hand and right-hand boundaries of $\interval_m$ by
${x_m^+\equiv\bar{x}_m+\frac{1}{2}\Delta_m\in\Pi(x_b,\ell)}$ and ${x_m^-\equiv \bar{x}_m-\frac{1}{2}\Delta_m\in\Pi(x_b,\ell)}$, 
respectively.

The simple average, ${\expval{\Dnu}_\intmax(x_b)}$, fluctuates microscopically
and continuously within the range ${\interval(0,\precNu)}$ as $x_b$ changes.
However, the definition of ${\expval{\;\cdot\;}^*_\ell}$ stipulates that ${x_b\in\Pi(x_b,\ell)}$ and
that the average of $\Delta\nu$ is the same on every microinterval, which means that, 
in general, ${x_b\pm\ell/2\neq x_{\pm M}}$ and that
${0\leq\abs{\expval{\Dnu}_\prectheo(x_b)-\bmzero}<\precNu}$.
I will preserve the constraint ${\bmzero=0}$ as $x_b$ changes, which means that
every element of ${\Pi(x_b,\ell)}$, including $x_M$ and $x_{-M}$, changes with $x_b$.
Therefore, ${\ell=x_M-x_{-M}}$ also changes.
I will assume that
${\nu(x_b)\neq\bnu(x_b)}$, leaving the special case ${\nu(x_b)=\bar{\nu}(x_b)}$ to the interested reader.
Using primes to denote total derivatives with respect to $x_b$, I can write
\begin{align}
\dv{x_b}\bmany(x_b) 
 & = 
 \frac{1}{\ell}\sum_{m} \dv{x_b}\left[\Delta_m\many(\bar{x}_m,\Delta_m)\right] \nonumber \\
& -\frac{\ell'}{\ell}\bmany(x_b)
\label{eqn:totalderiv0}  
\end{align}
where
\begin{align}
\dv{x_b}\bigg[&\Delta_m\many(\bar{x}_m,\Delta_m)\bigg] \nonumber \\
& = 
\left(\frac{\Delta_m}{2}\right)^n
\left[\dxppr\Dnu(\xppr)
+\left(-1\right)^{n+1}
\dxmpr\Dnu(\xmpr)\right] \nonumber \\
&\qquad\qquad -n\Delta_m\bar{x}'_m\mathcal{M}^{(n-1)}_\Dnu(\bar{x}_m,\Delta_m) \label{eqn:termderiv}
\end{align}
I will derive expressions for the derivatives of ${\bmone}$ and ${\bmtwo}$
below. Before doing so, I will simplify this task by deducing a relationship between $\dxmpr$ and $\dxppr$
from the constraints ${\expval{\Dnu}_{\Delta_m}(\bar{x}_m)=\bDnu(x_b)=0\implies \bDnu'(x_b)=0}$.
\begin{align}
 \bDnu\,'(x_b)  &=
\dv{x_b}
\left(\frac{1}{\Delta_m}\int_{\xmpr}^{\xppr}\Dnu(x)\dd{x}\right)=0   \nonumber \\
\implies
-\frac{\Delta'_m}{\Delta_m}&\bDnu(x_b) 
+ \frac{1}{\Delta_m}\left[\dxppr\Dnu(\xppr) - \dxmpr\Dnu(\xmpr) \right] =0
\nonumber 
\end{align}
${\bDnu=0}$ means that ${\dxppr\Delta\nu(\xppr) = \dxmpr \Delta\nu(\xmpr)}$. Recursively
applying this relationship and using the fact that ${x_0\equiv x_b\implies x'_0=1}$ we
find that
\begin{align}
x'_m \Delta\nu(x_m) = \Dnu(x_b), \;\;\forall\;x_m\in\Pi(x_b,\ell)
\label{eqn:constraint2}
\end{align}
Because a microscopic change of $x_b$ cannot change $\ell$ 
by more than $\amax$, the second term on the right
hand side of Eq.~\ref{eqn:totalderiv0} is negligible for our purposes. 
Therefore, by substituting Eq.~\ref{eqn:constraint2} into Eq.~\ref{eqn:totalderiv}
we find that
\begin{align}
\dv{x_b}\bmany(x_b) 
 & = 
 \frac{\Dnu(x_b)}{\ell}\left[1-\left(-1\right)^n\right]
\sum_{m} \left(\frac{\Delta_m}{2}\right)^n \nonumber \\
& - \frac{n}{\ell}\sum_m
\Delta_m\bar{x}'_m\mathcal{M}^{(n-1)}_\Dnu(\bar{x}_m,\Delta_m)
\label{eqn:totalderiv}  
\end{align}

\subsubsection{Case I: $\displaystyle \dv*{\bmone}{x_b}$}
For $n=1$, Eqs.~\ref{eqn:termderiv} and~\ref{eqn:constraint2} 
mean that
\begin{align}
\dv{x_b}&\left[\Delta_m\mone(\bar{x}_m,\Delta_m)\right]  \nonumber \\
& = \frac{\Delta_m}{2} \left[\dxppr\Delta\nu(\xppr) + \dxmpr\Delta\nu(\xmpr)\right]  
 = \Delta_m \Delta\nu(x_b) \nonumber 
\end{align}
Subsistuting this and ${\bmzero(x_b)=0}$ into Eq.~\ref{eqn:totalderiv} gives
\begin{align}
\dv{x_b}\bmone(x_b) & =  \Dnu(x_b)
\label{eqn:final0}
\end{align}
This result can be derived at greater length without
requiring that ${\bDnu\,'(x_b)=0}$ or ${\bDnu(x_b)=0}$. 

\subsubsection{Case II: $\displaystyle \dv*{\bmtwo}{x_b}$}
\label{section:derivatives_case2}
Inserting $n=2$ in Eq.~\ref{eqn:totalderiv} and using Eqs.~\ref{eqn:termderiv} 
and~\ref{eqn:constraint2} gives
\begin{align}
\dv{x_b}\bmtwo(x_b) 
  = 
-\frac{2}{\ell}\sum_m\bar{x}'_md_m
\label{eqn:dbmtwo}
\end{align}
where ${d_m \equiv \Delta_m \mone(\bar{x}_m,\Delta_m)}$ is
the first moment of $\Dnu$ in $\interval_m$.
I define $X_d$ to be the first-moment-weighted average of the interval midpoints, $\bar{x}_m$.
\begin{align}
X_d \equiv \frac{\sum_{m}\bar{x}_m d_m}{\sum_m d_m} 
\label{eqn:xd}
\end{align}
In the limit $a/l\to 0$ in an infinite macroscopically-uniform material, $X_d$ coincides both
with $x_b$ and with ${\frac{1}{2}\left(x_\mm+x_\m\right)}$.
Rearranging Eq.~\ref{eqn:xd} and taking the derivative with respect to $x_b$ gives
\begin{align}
\sum_m\bar{x}'_m d_m = X'_d\sum_{m}d_m + X_d\sum_m d'_m - \sum_m \bar{x}_m d'_m  \label{eqn:combo1}
\end{align}
Using Eq.~\ref{eqn:constraint2} and ${\int_{\xmpr}^{\xppr}\Dnu(x)\dd{x}=0}$, the 
derivative of $d_m$ can be expressed as 
\begin{align}
d\,'_m & = \left(\frac{\Delta_m}{2}\right)\left[\dxppr\Dnu(\xppr)+\dxmpr\Dnu(\xmpr)\right] \nonumber \\
 & = \Delta_m\Dnu(x_b) \implies 
  \sum_m d\,'_m  = \ell\, \Delta\nu(x_b) \label{eqn:combo2}
\end{align}
The last term on the right hand side of Eq.~\ref{eqn:combo1} is
\begin{align}
\sum_m \bar{x}_m d\,'_m & = \Delta \nu(x_b) \sum_m \Delta_m \bar{x}_m 
 \approx 
\Delta \nu(x_b) \int_{x_\mm}^{x_\m} \, x \, \dd{x} \nonumber \\ 
&= \Delta \nu(x_b) \, \frac{\ell}{2} \, \left(x_\mm+x_\m\right) \label{eqn:combo3}
\end{align}
Eqs.~\ref{eqn:dbmtwo}, ~\ref{eqn:xd}, ~\ref{eqn:combo1}, ~\ref{eqn:combo2}, 
and~\ref{eqn:combo3} can be combined with 
${\sum_m d_m  = \ell \bmone(x_b)}$
to show that
\begin{align}
\dv{x_b}\bmtwo(x_b) 
& = 
2\,\Dnu(x_b)\left[\left(\frac{x_\mm+x_\m}{2}\right)-X_d\right] \nonumber \\
& - 2\,X'_d\,\bmone(x_b) 
\end{align}
Now, because ${\frac{1}{2}(x_\mm+x_\m)}$ and 
${X_d}$ both get closer $x_b$ as $l$ increases, 
their difference vanishes and $X'_d$ becomes one in the
limit ${\amax/l\to 0}$. Therefore, in this limit,
\begin{align}
\dv{x_b}\bmtwo(x_b) = - 2\bmone(x_b) 
\label{eqn:final1}
\end{align}
To derive Eqs.~\ref{eqn:final0} and~\ref{eqn:final1} we assumed that $a/l$ could be brought to zero without 
straying into regions having different mesoscale averages $\bar{\nu}$. When this 
assumption is not valid, some of the terms that
were discarded should be considered more carefully. 

\subsubsection{${\dv*{\mbs_0}{\mx}}$ and 
${\dv*{\mbs_1}{\mx}}$}
It is straightforward to use Eqs.~\ref{eqn:surfaverage0}, ~\ref{eqn:surfaverage1}, 
~\ref{eqn:final0}, and~\ref{eqn:final1} to show that
\begin{subequations}
\label{eqn:derv12}
\begin{align}
\dv{\mbs_0}{x_b}  & = 0 \label{eqn:derv1}\\
\dv{\mbs_1}{x_b}  & = \bmone(x_b)\label{eqn:derv2}
\end{align}
\end{subequations}

\subsection{Macroscopic moment densities are independent of the choice of microscopic intervals}
\label{section:pi_dependence}
It is important to demonstrate that, in the limit $a/l\to 0$, our results
do not depend on the choice of the set of microintervals, $\Pi(x_b,\ell)$, that partition
the space around $x_b$. In this section it is demonstrated that all sets of 
points which satisfy the requirements explained in Sec.~\ref{section:partition} and 
and Sec.~\ref{section:partitioning} give the same
values of $\nmomone{x_b}$ and $\nmomtwo{x_b}$ and therefore the same values of $\mbs_0$ and $\mbs_1$. 

\subsubsection{$\nmomone{x_b}$}
\label{section:momone}
Let us assume that we are in the limit ${a/l\to 0}$ and that
for a particular choice, $\Pi_1(x_b,\ell)$, of the set of microinterval boundary points, we find
\begin{align}
\nmomone{x_b,\Pi_1} = \frac{1}{\ell}\sum_{m} \int_{\xmpr}^{\xppr}\,x\,\Dnu(x)\dd{x}
\end{align}
where ${\expval{\Dnu}_{\Delta_m}(\bar{x}_m)=0}$ has allowed each 
integrand ${(x-\bar{x}_m)\Dnu(x)}$ to be simplified to ${x\Dnu(x)}$.

Now suppose that a new set $\Pi_2(x_b,\ell_s)$ of boundary points $s_m\equiv x_m+\delta x_m$ is formed
by changing every point $x_m\in\Pi_1(x_b,\ell)$, {\em except} $x_0=x_b$, 
by an amount $\delta x_m$, such that the average of $\Dnu(x)$ on each of 
the new microscopic intervals remains equal to ${\bDnu(x_b)=0}$ and such that the ordering of the points 
does not change ($s_{m+1}>s_m,\;\;\forall\; m$). The new set of microintervals partitions the interval 
${[s_\mm,s_\m]}$, where ${s_\m-s_\mm = \ell_s}$, ${\abs{s_\m-x_b-\ell_s/2}<\amax}$, and 
${\amax/\ell_s\sim \amax/l \to 0}$.
I denote the midpoint, width, left-hand boundary, and right-hand boundary of the new $m^{th}$ interval by
${\bar{s}_m}$, ${\Deltas_m}$, ${\smpr}$, and ${\sppr}$, respectively. 
By construction, the average of $\Dnu(x)$ on each microinterval is zero. Therefore,
\begin{align}
\int_{\smpr}^{\sppr}\Dnu(s)\dd{s}&=
\int_{\xmpr}^{\xppr} \Dnu(x)\dd{x} =0 \nonumber \\
\implies  
\int_{\xppr}^{\sppr}\Dnu(x)\dd{x} &= \int_{\xmpr}^{\smpr}\Dnu(x)\dd{x}
\label{eqn:forty7}
\end{align}
The new average moment density is
\begin{align}
 \nmomone{x_b,\Pi_2}& 
   = \frac{1}{\ell_s}\sum_m \Bigg[\int_{\xmpr}^{\xppr}x\,\Dnu(x)\dd{x}   \nonumber \\
   &+  \int_{\xppr}^{\sppr} x \, \Dnu(x) \dd{x} 
 - \int_{\xmpr}^{\smpr} x \, \Dnu(x) \dd{x}  
\Bigg] 
\end{align}
After cancelling terms in the sum, this can be written as
\begin{align}
 \nmomone{x_b,\Pi_2}& 
   = \frac{1}{\ell_s}\sum_{m=-M}^{M-1} \int_{x_m}^{x_{m+1}}x\,\Dnu(x)\dd{x}   \nonumber \\
+\frac{1}{\ell_s}\int_{x_\m}^{s_\m} & x \; \Dnu(x) \dd{x}
+\frac{1}{\ell_s}\int_{s_\mm}^{x_\mm} x \; \Dnu(x) \dd{x}
\label{eqn:newmom}
\end{align}
Although the widths of all microintervals defined by ${\Pi_1(x_b,\ell)}$ and ${\Pi_2(x_b,\ell_s)}$ are less than $\amax$, 
I have not assumed that ${\abs{\bar{s}_m-\bar{x}_m}<\amax}$. 
If, for example, the width of each new interval was larger than each old interval, i.e.,
${0<\Delta_m<\Delta_m^s <\amax}$, this would imply that ${0<s_\m-x_\m\sim Ma \not\ll l}$.  
Therefore, we cannot immediately dismiss the second and third terms on the right hand side as negligible.
However, $\nmomone{x_b,\Pi_1}$ was assumed to be converged with respect to 
the magnitude of $\ell$. 
Therefore, without changing its value significantly, 
I can expand the set $\Pi_1$ to encompass the ranges $[x_\m,s_\m]$ and $[s_\mm,x_\mm]$ by dividing these
ranges into microintervals and adding their boundary points to $\Pi_1(x_b,\ell)$ to form 
a new set ${\Pi_1^{\text{new}}(x_b,\ell^\text{new})\supset\Pi_1(x_b,\ell)}$ containing
${M^\text{new}>M}$ microinterval boundary points on each side of $x_b$, and such that
${0<x_\mm-x_\mmnew<\amax}$, 
${0<x_\mnew-x_\m<\amax}$, and ${0<\ell^\text{new}-\ell_s\lesssim\amax}$.
Eq.~\ref{eqn:newmom} then becomes
\begin{align}
\nmomone{x_b,\Pi_2}& 
   = \frac{1}{\ell_s}\sum_{m=\mmnew}^{\mnew-1} \int_{x_m}^{x_{m+1}}x\,\Dnu(x)\dd{x}   \nonumber \\
-\frac{1}{\ell_s}&\int_{s_\m}^{x_\mnew} x \; \Dnu(x) \dd{x}
-\frac{1}{\ell_s}\int^{s_\mm}_{x_\mmnew} x \; \Dnu(x) \dd{x} \nonumber \\
& = 
\frac{\ell^\text{new}}{\ell_s}\,
\nmomone{x_b,\Pi_1^\text{new}} + \order{\frac{\amax}{l}} 
\end{align}
Therefore, if $\order{\amax/l}$ terms are neglected, 
\begin{align}
\nmomone{x_b,\Pi_2} 
&= \nmomone{x_b,\Pi_1} 
\label{eqn:random}
\end{align}
I have assumed that ${s_\m>x_\m}$ and ${s_\mm<x_\mm}$, but a similar procedure
can be followed to prove the same result for any other case, such as
${s_\m>x_\m}$ and ${s_\mm>x_\mm}$.

\subsubsection{$\nmomtwo{x_b}$}
I will now demonstrate that
$\nmomtwo{x_b}$ is independent of how the region around $x_b$ is partitioned. 
I will use the partitions $\Pi_1(x_b,\ell)$, $\Pi_2(x_b,\ell_s)$ and $\Pi_1^{\text{new}}(x_b,\ell_s)$, introduced in the previous section, where $\ell_s>\ell$, 
and I again assume the limit ${\amax/l\to 0}$, which implies that
$\nmomtwo{x_b,\Pi_1}=\nmomtwo{x_b,\Pi_1^{\text{new}}}$.
I will show that 
${\nmomtwo{x_b,\Pi_1}=\nmomtwo{x_b,\Pi_1^{\text{new}}}=\nmomtwo{x_b,\Pi_1}}$.
\begin{widetext}
\begin{align}
&\ell_s\nmomtwo{x_b,\Pi_2} 
 = \sum_{m} \int_{\smpr}^{\sppr}(x-\bar{s}_m)^2 \Dnu(x) \dd{x}   
  = \sum_{m} \left[\int_{\smpr}^{\sppr}(x-\bar{x}_m)^2 \Dnu(x) \dd{x}   
-2\,\left(\bar{s}_m-\bar{x}_m\right)\,\int_{\smpr}^{\sppr}x\,\Dnu(x)\dd{x}\right]
\nonumber \\
& = \sum_{m} \left[
\int_{\xmpr}^{\xppr}(x-\bar{x}_m)^2 \Dnu(x) \dd{x}  
+\int_{\xppr}^{\sppr}x^2 \Dnu(x) \dd{x}  
-\int_{\xmpr}^{\smpr}x^2 \Dnu(x) \dd{x}\right.  
-2\,\bar{s}_m\,\int_{\smpr}^{\sppr}x\,\Dnu(x)\dd{x}
\left.
+2\,\bar{x}_m\,\int_{\xmpr}^{\xppr}x\,\Dnu(x)\dd{x}
\right] \nonumber \\
& = \ell\nmomtwo{x_b,\Pi_1} 
+ \sum_{m=\m+1}^{\mnew}\int_{\xmpr}^{\xppr}(x-\bar{x}_m)^2 \Dnu(x) \dd{x} 
+ \sum_{m=\mmnew}^{\mm-1}\int_{\xmpr}^{\xppr}(x-\bar{x}_m)^2 \Dnu(x) \dd{x}  
+2\sum_{m=\m+1}^{\mnew}\bar{x}_m\int_{\xmpr}^{\xppr}x\,\Dnu(x)\dd{x}\nonumber \\
&+2\sum_{m=\mmnew}^{\mm-1}\bar{x}_m\int_{\xmpr}^{\xppr}x\,\Dnu(x)\dd{x}
-2\,\sum_{m=-M}^M\bar{s}_m\int_{\smpr}^{\sppr}x\,\Dnu(x)\dd{x}
+2\,\sum_{m=-M}^M\bar{x}_m\int_{\xmpr}^{\xppr}x\,\Dnu(x)\dd{x} \nonumber \\
&=\ell^\text{new}\nmomtwo{x_b,\Pi_1^\text{new}} +2 \sum_{m=\mmnew}^{\mnew}\bar{x}_m \int_{\xmpr}^{\xppr}x\,\Dnu(x)\dd{x}
-2\,\sum_{m=-M}^M\bar{s}_m\int_{\smpr}^{\sppr}x\,\Dnu(x)\dd{x}
\end{align}
\end{widetext}
Denoting the dipole weighted mean positions (Eq.~\ref{eqn:xd}) of sets $\Pi_1^{\text{new}}(x_b,\ell_s)$ and $\Pi_2(x_b,\ell_s)$
by $X_d(\Pi_1^{\text{new}})$ and $X_d(\Pi_2)$, respectively, using  Eq.~\ref{eqn:random}, and
neglecting $\order{\amax/l}$ terms, 
allows this to be written as
\begin{align}
& \nmomtwo{x_b,\Pi_2} 
  = \nmomtwo{x_b,\Pi_1^{\text{new}}}  \nonumber \\
&+ 2 \nmomone{x_b,\Pi_1^{\text{new}}} X_d(\Pi_1^{\text{new}})
- 2 \nmomone{x_b,\Pi_2} X_d(\Pi_2) \nonumber \\
& = \nmomtwo{x_b,\Pi_1} + 2 \nmomone{x_b,\Pi_1}\left[X_d(\Pi_1)-X_d(\Pi_2)\right] \nonumber
\end{align}
In the limit $a/l \to 0$, both $X_d(\Pi_1)$ and $X_d(\Pi_2)$ tend to $x_b$ and so 
\begin{align}
\nmomtwo{x_b,\Pi_2} = \nmomtwo{x_b,\Pi_1}
\end{align}

\subsection{Mesoscale averages of $\bmany(x_b)$}
\label{section:average_momdensity}
In general, both $\bmone(x_b)$ and $\bmtwo(x_b)$ vary microscopically
with $x_b$. It can be necessary to know their mesoscale
averages over $x_b$, which are denoted by ${\boldmone}$ and ${\boldmtwo}$, 
respectively.  In Sec.~\ref{section:idempotency} it was argued that
idempotency of the mesoscale-averaging operation applied to 
surface integrals requires both ${\boldmone}$ and ${\boldmtwo}$
to be zero in regions of mesoscale uniformity.
In this section I prove that this requirement is satisfied
in the ${a/l\to 0}$ limit.

I define ${\ell_1\sim l}$ and ${\ell_2\sim l}$ as the widths
of the intervals on which the mesoscale averages of 
$\many$ and $\bmany$, respectively, are calculated.
I consider the average, over all ${u\in\interval(-\ell_2/2,\ell_2/2)}$, of
$\bmone(x_b+u)$. As usual, when $\bmone$ is being evaluated
at ${x_b+u}$, I partition an interval of width $\ell_2$ centered at a microscopic distance
from  ${x_b+u}$ into a set of microintervals.
I denote the left-hand boundary, 
right-hand boundary, midpoint, and width of the $m^\text{th}$ microinterval, ${\interval_m(u)}$,
by ${\xmpr=\xmpr(u)}$, ${\xppr=\xppr(u)}$, 
${\bar{x}_m=\bar{x}_m(u)}$, and ${\Delta_m=\Delta_m(u)}$, respectively. 
For each value of $u$, the value of 
${\ell_1(u)=\ell_1(0)+\Delta\ell_1(u)}$ is chosen such that ${\bDnu(x_b+u)=0}$. 
It is always possible to choose it such that ${\Delta\ell_1(u)<\amax}$.
The values of ${\bar{x}_m}$
and ${\Delta_m}$ are chosen such that ${\expval{\Dnu}_{\Delta_m(u)}(\bar{x}_m(u))=\bDnu(x_b+u)=0}$.
To avoid clutter I will only make the dependences of $x_m$, $\xmpr$, $\xppr$, 
$\Delta_m$, $\bar{x}_m$, and $\ell_1$ on $u$ 
explicit in my notation when it is necessary for clarity.

The mesoscale average of  $\nmomone{x_b}$ is
\begin{align}
&\expval{\bar{\mathcal{M}}_\nu^{(1)}}_{\ell_2}(x_b) 
=\frac{1}{\ell_2}\int_{-\ell_2/2}^{\ell_2/2}\left(\frac{1}{\ell_1}\sum_{m}
\int_{\xmpr}^{\xppr} x \, \Dnu(x) \dd{x}\right) \dd{u} \nonumber \\
&=\frac{1}{\ell_2}\int_{-\ell_2/2}^{\ell_2/2}\left(\frac{1}{\ell_1}
\int_{-\ell_1/2}^{\ell_1/2} v \, \Dnu(x_b+u+v) \dd{v}\right) \dd{u} \nonumber 
\end{align}
Because terms of order ${\amax/\ell_1}$ are negligible, we can ignore the 
dependence of $\ell_1$ on $u$. This allows us to switch
the order of the integrations over $u$ and $v$.
\begin{align}
&\expval{\bar{\mathcal{M}}_\nu^{(1)}}_{\ell_2}(x_b) \nonumber \\
& = \frac{1}{\ell_2\,\ell_1}\int_{-\ell_1/2}^{\ell_1/2}u\;\left(\int_{\ell_2/2}^{-\ell_2/2}
\Delta\nu(x_b+u+v)\dd{v}\right)\dd{u} \nonumber
\end{align}
The inner integral vanishes in the limit ${\amax/\ell_2\to 0}$. Therefore, 
\begin{align}
\boldmone(\mxb) \equiv \expval{\bmone}_l(x_b) = 0
\end{align}

Because $\nu_B$ is constant on the microscale, it follows that
${\boldsymbol{\mathcal{M}^{\expval{1}}_{\nu}}}$ is also zero.
This is simply a consequence of mesoscale uniformity implying local mesoscale isotropy. By 
{\em local mesoscale isotropy} I mean that, for all $n$, 
\begin{align}
\int_{-\ell/2}^{\ell/2} u^n \,\nu(x+u)\dd{u} = 
\int_{-\ell/2}^{\ell/2} u^n \,\nu(x-u)\dd{u} + \order{a/l}\nonumber
\end{align}

I now want to prove that the mesoscale average $\boldmtwo$ of $\bmtwo$ is zero, where
\begin{align}
\nmomtwo{x_b+u}  
 &= \frac{1}{\ell_1}\sum_{m}
\int_{\xmpr}^{\xppr} \left(x-\bar{x}_m\right)^2 \Dnu(x)\dd{x} \nonumber \\
 &= \frac{1}{\ell_1}\sum_{m} \bigg[\int_{\xmpr}^{\xppr}\left(x-x_b-u\right)^2 \Dnu(x) \dd{x} \nonumber \\
&-2\left(\bar{x}_m-x_b-u\right)\int_{\xmpr}^{\xppr} x\;\Dnu(x)\dd{x}\bigg] \nonumber \\
& = \frac{1}{\ell_1}  \int_{x_\mm}^{x_\m} \left(x-x_b-u\right)^2 \Delta\nu(x)\dd{x} \nonumber \\
&-2\left[X_d-(x_b+u)\right]\bmone(x_b+u)
\end{align}
As defined in Sec.~\ref{section:derivatives_case2}, ${X_d=X_d(u)}$ is the first-moment-weighted average of the microinterval 
midpoints. Its value tends to ${x_b+u}$ in the ${a/\ell_1\to 0}$ limit; therefore, the second term on the right hand side 
vanishes in this limit.
Since $\order{a/l}$ terms can be disregarded, we can assume that ${x_{\pmm}=x_b+u\pm\ell_1/2}$
and express the average over $u$ as
\begin{align}
& \boldmtwo(\mx_b) \nonumber \\
&= 
\frac{1}{\ell_2}\int_{-\ell_2/2}^{\ell_2/2} 
\frac{1}{\ell_1}\left(\int_{-\ell_1/2}^{\ell_1/2} v^2 \,
\Delta\nu(x_b+u+v)\dd{v}\right)\dd{u}  \nonumber \\
&= 
\frac{1}{\ell_1}\int_{-\ell_1/2}^{\ell_1/2}
v^2 \expval{\Dnu}_{\ell_2}(x_b+v) \dd{v}
\end{align}
where, to reach the second line from the first, we 
reverse the order of integration, thereby
neglecting the $\order{\amax/\ell_1}$ error 
made by ignoring the $u-$dependence of $\ell_1$.
The value of ${\expval{\Dnu}_{\ell_2}(x_b+v)}$ fluctuates microscopically about zero as $v$ is varied, 
and its magnitude remains smaller than ${\delta_A^{[\bnu]}/2}$. Therefore, it is possible
to choose a value of ${\eta_1<\amax}$ for which ${\int_{-\ell_1/2}^{\ell_1/2+\eta_1}\expval{\Dnu}(x_b+v)\dd{v}=0}$.
Therefore, we can add $\eta_1$ to the upper limit of the integral  at the expense of a negligible $\order{a/l}$ error.
Integrating by parts gives
\begin{align}
 \boldmtwo&(\mx_b) \nonumber \\
& =
-\frac{2}{\ell_1}\int_{-\ell_1/2}^{\ell_1/2+\eta_1}
v \left(\int_{-\ell_1/2}^v \expval{\Dnu}_{\ell_2}(x_b+v') \dd{v'}\right) \dd{v}\nonumber
\end{align}
Let ${\eta_2>0}$ be the shortest distance for which ${\int_{-\ell_1/2}^{v-\eta_2}\expval{\Dnu}_{\ell_2}(x_b+v') \dd{v'} = 0}$.
Because ${\expval{\Dnu}_{\ell_2}(x_b+v')}$ fluctuates microscopically about zero, ${\eta_2<\amax}$ and
${\abs{\int_{v-\eta_2}^v\expval{\Dnu}_{\ell_2}(x_b+v') \dd{v'}} < \eta_2\,\delta_A^{[\bnu]}/2}$.
 Therefore, 
\begin{align}
 \abs{\boldmtwo(\mx_b)} 
 <
\frac{\amax}{\ell_1}\,\delta_A^{[\bnu]}\,\abs{\int_{-\ell_1/2}^{\ell_1/2+\eta_1} 
v  \dd{v}}
\sim 
\frac{1}{2}\frac{\amax^3}{\ell_2}\delta_B^{[\nu]} 
\end{align}
This vanishes in the ${\amax/\ell_2\to 0}$ limit and so
\begin{align}
\boldmtwo(\mx_b) = 0 + \order{a/l}
\end{align}\\

\onecolumngrid
\vspace{0.5cm}
\PRLsep
\vspace{0.5cm}

\twocolumngrid
\bibliography{\basefolder/references}

\begin{thebibliography}{219}%
\makeatletter
\providecommand \@ifxundefined [1]{%
 \@ifx{#1\undefined}
}%
\providecommand \@ifnum [1]{%
 \ifnum #1\expandafter \@firstoftwo
 \else \expandafter \@secondoftwo
 \fi
}%
\providecommand \@ifx [1]{%
 \ifx #1\expandafter \@firstoftwo
 \else \expandafter \@secondoftwo
 \fi
}%
\providecommand \natexlab [1]{#1}%
\providecommand \enquote  [1]{``#1''}%
\providecommand \bibnamefont  [1]{#1}%
\providecommand \bibfnamefont [1]{#1}%
\providecommand \citenamefont [1]{#1}%
\providecommand \href@noop [0]{\@secondoftwo}%
\providecommand \href [0]{\begingroup \@sanitize@url \@href}%
\providecommand \@href[1]{\@@startlink{#1}\@@href}%
\providecommand \@@href[1]{\endgroup#1\@@endlink}%
\providecommand \@sanitize@url [0]{\catcode `\\12\catcode `\$12\catcode
  `\&12\catcode `\#12\catcode `\^12\catcode `\_12\catcode `\%12\relax}%
\providecommand \@@startlink[1]{}%
\providecommand \@@endlink[0]{}%
\providecommand \url  [0]{\begingroup\@sanitize@url \@url }%
\providecommand \@url [1]{\endgroup\@href {#1}{\urlprefix }}%
\providecommand \urlprefix  [0]{URL }%
\providecommand \Eprint [0]{\href }%
\providecommand \doibase [0]{http://dx.doi.org/}%
\providecommand \selectlanguage [0]{\@gobble}%
\providecommand \bibinfo  [0]{\@secondoftwo}%
\providecommand \bibfield  [0]{\@secondoftwo}%
\providecommand \translation [1]{[#1]}%
\providecommand \BibitemOpen [0]{}%
\providecommand \bibitemStop [0]{}%
\providecommand \bibitemNoStop [0]{.\EOS\space}%
\providecommand \EOS [0]{\spacefactor3000\relax}%
\providecommand \BibitemShut  [1]{\csname bibitem#1\endcsname}%
\let\auto@bib@innerbib\@empty
\bibitem [{\citenamefont {Abanin}\ \emph {et~al.}(2019)\citenamefont {Abanin},
  \citenamefont {Altman}, \citenamefont {Bloch},\ and\ \citenamefont
  {Serbyn}}]{mbloc_1}%
  \BibitemOpen
  \bibfield  {author} {\bibinfo {author} {\bibnamefont {Abanin}, \bibfnamefont
  {D~A}}, \bibinfo {author} {\bibfnamefont {E.}~\bibnamefont {Altman}},
  \bibinfo {author} {\bibfnamefont {I.}~\bibnamefont {Bloch}}, \ and\ \bibinfo
  {author} {\bibfnamefont {M.}~\bibnamefont {Serbyn}}} (\bibinfo {year}
  {2019}),\ \bibfield  {title} {\enquote {\bibinfo {title} {Colloquium:
  Many-body localization, thermalization, and entanglement},}\ }\href {\doibase
  10.1103/RevModPhys.91.021001} {\bibfield  {journal} {\bibinfo  {journal}
  {Rev. Mod. Phys.}\ }\textbf {\bibinfo {volume} {91}},\ \bibinfo {pages}
  {021001}}\BibitemShut {NoStop}%
\bibitem [{\citenamefont {Aizu}(1962)}]{aizu_1962}%
  \BibitemOpen
  \bibfield  {author} {\bibinfo {author} {\bibnamefont {Aizu}, \bibfnamefont
  {K}}} (\bibinfo {year} {1962}),\ \bibfield  {title} {\enquote {\bibinfo
  {title} {Polarization, pyroelectricity, and ferroelectricity of ionic
  crystals},}\ }\href@noop {} {\bibfield  {journal} {\bibinfo  {journal} {Rev.
  Mod. Phys.}\ }\textbf {\bibinfo {volume} {34}},\ \bibinfo {pages}
  {550--576}}\BibitemShut {NoStop}%
\bibitem [{\citenamefont {Ando}(1963)}]{ando_1963}%
  \BibitemOpen
  \bibfield  {author} {\bibinfo {author} {\bibnamefont {Ando}, \bibfnamefont
  {T}}} (\bibinfo {year} {1963}),\ \bibfield  {title} {\enquote {\bibinfo
  {title} {Properties of fermion density matrices},}\ }\href {\doibase
  10.1103/RevModPhys.35.690} {\bibfield  {journal} {\bibinfo  {journal} {Rev.
  Mod. Phys.}\ }\textbf {\bibinfo {volume} {35}},\ \bibinfo {pages}
  {690--702}}\BibitemShut {NoStop}%
\bibitem [{\citenamefont {Ashcroft}\ and\ \citenamefont
  {Mermin}(1976)}]{ashcroft_mermin_book}%
  \BibitemOpen
  \bibfield  {author} {\bibinfo {author} {\bibnamefont {Ashcroft},
  \bibfnamefont {N~W}}, \ and\ \bibinfo {author} {\bibfnamefont {N.~D.}\
  \bibnamefont {Mermin}}} (\bibinfo {year} {1976}),\ \href@noop {} {\emph
  {\bibinfo {title} {Solid State Physics}}},\ HRW international editions\
  (\bibinfo  {publisher} {Holt, Rinehart and Winston})\BibitemShut {NoStop}%
\bibitem [{\citenamefont {Ayers}\ and\ \citenamefont
  {Parr}(2003)}]{density_maxima}%
  \BibitemOpen
  \bibfield  {author} {\bibinfo {author} {\bibnamefont {Ayers}, \bibfnamefont
  {P~W}}, \ and\ \bibinfo {author} {\bibfnamefont {R.~G.}\ \bibnamefont
  {Parr}}} (\bibinfo {year} {2003}),\ \bibfield  {title} {\enquote {\bibinfo
  {title} {Sufficient condition for monotonic electron density decay in
  many-electron systems},}\ }\href {\doibase https://doi.org/10.1002/qua.10622}
  {\bibfield  {journal} {\bibinfo  {journal} {International Journal of Quantum
  Chemistry}\ }\textbf {\bibinfo {volume} {95}}~(\bibinfo {number} {6}),\
  \bibinfo {pages} {877--881}}\BibitemShut {NoStop}%
\bibitem [{\citenamefont {Bacskay}\ \emph {et~al.}(1997)\citenamefont
  {Bacskay}, \citenamefont {Reimers},\ and\ \citenamefont
  {Nordholm}}]{bacskay_reimers_1997}%
  \BibitemOpen
  \bibfield  {author} {\bibinfo {author} {\bibnamefont {Bacskay}, \bibfnamefont
  {G~B}}, \bibinfo {author} {\bibfnamefont {J.~R.}\ \bibnamefont {Reimers}}, \
  and\ \bibinfo {author} {\bibfnamefont {S.}~\bibnamefont {Nordholm}}}
  (\bibinfo {year} {1997}),\ \bibfield  {title} {\enquote {\bibinfo {title}
  {The mechanism of covalent bonding},}\ }\href@noop {} {\bibfield  {journal}
  {\bibinfo  {journal} {J. Chem. Educ.}\ }\textbf {\bibinfo {volume}
  {74}}~(\bibinfo {number} {12}),\ \bibinfo {pages} {1494}}\BibitemShut
  {NoStop}%
\bibitem [{\citenamefont {Baez}\ and\ \citenamefont
  {Muniain}(1994)}]{baez_1994}%
  \BibitemOpen
  \bibfield  {author} {\bibinfo {author} {\bibnamefont {Baez}, \bibfnamefont
  {J~C}}, \ and\ \bibinfo {author} {\bibfnamefont {J.~P.}\ \bibnamefont
  {Muniain}}} (\bibinfo {year} {1994}),\ \href@noop {} {\emph {\bibinfo {title}
  {Gauge Fields, Knots And Gravity}}},\ Series On Knots And Everything\
  (\bibinfo  {publisher} {World Scientific Publishing Company})\BibitemShut
  {NoStop}%
\bibitem [{\citenamefont {Baldereschi}\ \emph {et~al.}(1988)\citenamefont
  {Baldereschi}, \citenamefont {Baroni},\ and\ \citenamefont
  {Resta}}]{baldereschi_1988}%
  \BibitemOpen
  \bibfield  {author} {\bibinfo {author} {\bibnamefont {Baldereschi},
  \bibfnamefont {A}}, \bibinfo {author} {\bibfnamefont {S.}~\bibnamefont
  {Baroni}}, \ and\ \bibinfo {author} {\bibfnamefont {R.}~\bibnamefont
  {Resta}}} (\bibinfo {year} {1988}),\ \bibfield  {title} {\enquote {\bibinfo
  {title} {Band offsets in lattice-matched heterojunctions: A model and
  first-principles calculations for {GaAs/AlAs}},}\ }\href {\doibase
  10.1103/PhysRevLett.61.734} {\bibfield  {journal} {\bibinfo  {journal} {Phys.
  Rev. Lett.}\ }\textbf {\bibinfo {volume} {61}},\ \bibinfo {pages}
  {734--737}}\BibitemShut {NoStop}%
\bibitem [{\citenamefont {Bao}\ \emph {et~al.}(2018)\citenamefont {Bao},
  \citenamefont {Verma},\ and\ \citenamefont {Truhlar}}]{nacl_2}%
  \BibitemOpen
  \bibfield  {author} {\bibinfo {author} {\bibnamefont {Bao}, \bibfnamefont
  {J~L}}, \bibinfo {author} {\bibfnamefont {P.}~\bibnamefont {Verma}}, \ and\
  \bibinfo {author} {\bibfnamefont {D.~G.}\ \bibnamefont {Truhlar}}} (\bibinfo
  {year} {2018}),\ \bibfield  {title} {\enquote {\bibinfo {title} {How well can
  density functional theory and pair-density functional theory predict the
  correct atomic charges for dissociation and accurate dissociation energetics
  of ionic bonds?}}\ }\href@noop {} {\bibfield  {journal} {\bibinfo  {journal}
  {Phys. Chem. Chem. Phys.}\ }\textbf {\bibinfo {volume} {20}},\ \bibinfo
  {pages} {23072--23078}}\BibitemShut {NoStop}%
\bibitem [{\citenamefont {Baroni}\ \emph {et~al.}(2001)\citenamefont {Baroni},
  \citenamefont {de~Gironcoli}, \citenamefont {Dal~Corso},\ and\ \citenamefont
  {Giannozzi}}]{baroni_rmp}%
  \BibitemOpen
  \bibfield  {author} {\bibinfo {author} {\bibnamefont {Baroni}, \bibfnamefont
  {S}}, \bibinfo {author} {\bibfnamefont {S.}~\bibnamefont {de~Gironcoli}},
  \bibinfo {author} {\bibfnamefont {A.}~\bibnamefont {Dal~Corso}}, \ and\
  \bibinfo {author} {\bibfnamefont {P.}~\bibnamefont {Giannozzi}}} (\bibinfo
  {year} {2001}),\ \bibfield  {title} {\enquote {\bibinfo {title} {Phonons and
  related crystal properties from density-functional perturbation theory},}\
  }\href@noop {} {\bibfield  {journal} {\bibinfo  {journal} {Rev. Mod. Phys.}\
  }\textbf {\bibinfo {volume} {73}},\ \bibinfo {pages} {515--562}}\BibitemShut
  {NoStop}%
\bibitem [{\citenamefont {Bernardini}\ and\ \citenamefont
  {Fiorentini}(1998)}]{bernardini_1998}%
  \BibitemOpen
  \bibfield  {author} {\bibinfo {author} {\bibnamefont {Bernardini},
  \bibfnamefont {F}}, \ and\ \bibinfo {author} {\bibfnamefont {V.}~\bibnamefont
  {Fiorentini}}} (\bibinfo {year} {1998}),\ \bibfield  {title} {\enquote
  {\bibinfo {title} {Macroscopic polarization and band offsets at nitride
  heterojunctions},}\ }\href {\doibase 10.1103/PhysRevB.57.R9427} {\bibfield
  {journal} {\bibinfo  {journal} {Phys. Rev. B}\ }\textbf {\bibinfo {volume}
  {57}},\ \bibinfo {pages} {R9427--R9430}}\BibitemShut {NoStop}%
\bibitem [{\citenamefont {Bethe}(1928)}]{bethe-1928}%
  \BibitemOpen
  \bibfield  {author} {\bibinfo {author} {\bibnamefont {Bethe}, \bibfnamefont
  {H}}} (\bibinfo {year} {1928}),\ \bibfield  {title} {\enquote {\bibinfo
  {title} {Theorie der beugung von elektronen an kristallen},}\ }\href@noop {}
  {\bibfield  {journal} {\bibinfo  {journal} {Ann. Phys. (Berl.)}\ }\textbf
  {\bibinfo {volume} {392}},\ \bibinfo {pages} {55--129}}\BibitemShut {NoStop}%
\bibitem [{\citenamefont {Bloch}(1929)}]{bloch-1929}%
  \BibitemOpen
  \bibfield  {author} {\bibinfo {author} {\bibnamefont {Bloch}, \bibfnamefont
  {F}}} (\bibinfo {year} {1929}),\ \bibfield  {title} {\enquote {\bibinfo
  {title} {{\"U}ber die quantenmechanik der elektronen in kristallgittern},}\
  }\href@noop {} {\bibfield  {journal} {\bibinfo  {journal} {Z. Phys.}\
  }\textbf {\bibinfo {volume} {52}},\ \bibinfo {pages} {555--600}}\BibitemShut
  {NoStop}%
\bibitem [{\citenamefont {Blount}(1962)}]{blount}%
  \BibitemOpen
  \bibfield  {author} {\bibinfo {author} {\bibnamefont {Blount}, \bibfnamefont
  {E~I}}} (\bibinfo {year} {1962}),\ \bibfield  {title} {\enquote {\bibinfo
  {title} {Formalisms of band theory},}\ }\href@noop {} {\bibfield  {journal}
  {\bibinfo  {journal} {Solid State Physics}\ }\textbf {\bibinfo {volume}
  {13}},\ \bibinfo {pages} {305--373}}\BibitemShut {NoStop}%
\bibitem [{\citenamefont {Blumenthal}\ \emph {et~al.}(2017)\citenamefont
  {Blumenthal}, \citenamefont {Kahk}, \citenamefont {Sundararaman},
  \citenamefont {Tangney},\ and\ \citenamefont {Lischner}}]{mip_lars}%
  \BibitemOpen
  \bibfield  {author} {\bibinfo {author} {\bibnamefont {Blumenthal},
  \bibfnamefont {L}}, \bibinfo {author} {\bibfnamefont {J.~M.}\ \bibnamefont
  {Kahk}}, \bibinfo {author} {\bibfnamefont {R.}~\bibnamefont {Sundararaman}},
  \bibinfo {author} {\bibfnamefont {P.}~\bibnamefont {Tangney}}, \ and\
  \bibinfo {author} {\bibfnamefont {J.}~\bibnamefont {Lischner}}} (\bibinfo
  {year} {2017}),\ \bibfield  {title} {\enquote {\bibinfo {title} {Energy level
  alignment at semiconductor–water interfaces from atomistic and continuum
  solvation models},}\ }\href@noop {} {\bibfield  {journal} {\bibinfo
  {journal} {RSC Adv.}\ }\textbf {\bibinfo {volume} {7}},\ \bibinfo {pages}
  {43660--43670}}\BibitemShut {NoStop}%
\bibitem [{\citenamefont {Bohr}(1920)}]{bohr_1920}%
  \BibitemOpen
  \bibfield  {author} {\bibinfo {author} {\bibnamefont {Bohr}, \bibfnamefont
  {N}}} (\bibinfo {year} {1920}),\ \bibfield  {title} {\enquote {\bibinfo
  {title} {\"uber die serienspektra der elemente},}\ }\href {\doibase
  10.1007/BF01329978} {\bibfield  {journal} {\bibinfo  {journal} {Z. Physik}\
  }\textbf {\bibinfo {volume} {2}},\ \bibinfo {pages} {423--469}}\BibitemShut
  {NoStop}%
\bibitem [{\citenamefont {Born}\ and\ \citenamefont
  {Huang}(1954)}]{born-huang-book}%
  \BibitemOpen
  \bibfield  {author} {\bibinfo {author} {\bibnamefont {Born}, \bibfnamefont
  {M}}, \ and\ \bibinfo {author} {\bibfnamefont {K.}~\bibnamefont {Huang}}}
  (\bibinfo {year} {1954}),\ \href@noop {} {\emph {\bibinfo {title} {Dynamical
  Theory of Crystal Lattices}}},\ International Series of Monographs on
  Physics\ (\bibinfo  {publisher} {Oxford University Press})\BibitemShut
  {NoStop}%
\bibitem [{\citenamefont {Born}\ and\ \citenamefont {von
  {K\'arm\'an}}(1912)}]{born_von-karman}%
  \BibitemOpen
  \bibfield  {author} {\bibinfo {author} {\bibnamefont {Born}, \bibfnamefont
  {M}}, \ and\ \bibinfo {author} {\bibfnamefont {T.}~\bibnamefont {von
  {K\'arm\'an}}}} (\bibinfo {year} {1912}),\ \bibfield  {title} {\enquote
  {\bibinfo {title} {{\"U}ber schwingungen im raumgittern},}\ }\href@noop {}
  {\bibfield  {journal} {\bibinfo  {journal} {Phys. Z.}\ }\textbf {\bibinfo
  {volume} {13}},\ \bibinfo {pages} {297--309}}\BibitemShut {NoStop}%
\bibitem [{\citenamefont {Born}\ and\ \citenamefont
  {Oppenheimer}(1927)}]{born_oppenheimer}%
  \BibitemOpen
  \bibfield  {author} {\bibinfo {author} {\bibnamefont {Born}, \bibfnamefont
  {M}}, \ and\ \bibinfo {author} {\bibfnamefont {R.}~\bibnamefont
  {Oppenheimer}}} (\bibinfo {year} {1927}),\ \bibfield  {title} {\enquote
  {\bibinfo {title} {Zur quantentheorie der molekeln},}\ }\href@noop {}
  {\bibfield  {journal} {\bibinfo  {journal} {Ann. Phys. (Berl.)}\ }\textbf
  {\bibinfo {volume} {389}}~(\bibinfo {number} {20}),\ \bibinfo {pages}
  {457--484}}\BibitemShut {NoStop}%
\bibitem [{\citenamefont {Boys}(1960)}]{boys}%
  \BibitemOpen
  \bibfield  {author} {\bibinfo {author} {\bibnamefont {Boys}, \bibfnamefont
  {S~F}}} (\bibinfo {year} {1960}),\ \bibfield  {title} {\enquote {\bibinfo
  {title} {Construction of some molecular orbitals to be approximately
  invariant for changes from one molecule to another},}\ }\href@noop {}
  {\bibfield  {journal} {\bibinfo  {journal} {Rev. Mod. Phys.}\ }\textbf
  {\bibinfo {volume} {32}},\ \bibinfo {pages} {296--299}}\BibitemShut {NoStop}%
\bibitem [{\citenamefont {Bristowe}\ \emph {et~al.}(2014)\citenamefont
  {Bristowe}, \citenamefont {Ghosez}, \citenamefont {Littlewood},\ and\
  \citenamefont {Artacho}}]{bristowe-JPCM-2014}%
  \BibitemOpen
  \bibfield  {author} {\bibinfo {author} {\bibnamefont {Bristowe},
  \bibfnamefont {N~C}}, \bibinfo {author} {\bibfnamefont {P.}~\bibnamefont
  {Ghosez}}, \bibinfo {author} {\bibfnamefont {P.~B.}\ \bibnamefont
  {Littlewood}}, \ and\ \bibinfo {author} {\bibfnamefont {E.}~\bibnamefont
  {Artacho}}} (\bibinfo {year} {2014}),\ \bibfield  {title} {\enquote {\bibinfo
  {title} {The origin of two-dimensional electron gases at oxide interfaces:
  Insights from theory},}\ }\href@noop {} {\bibfield  {journal} {\bibinfo
  {journal} {J. Phys.: Condens. Matter}\ }\textbf {\bibinfo {volume}
  {26}}~(\bibinfo {number} {14}),\ \bibinfo {pages} {143201}}\BibitemShut
  {NoStop}%
\bibitem [{\citenamefont {Bristowe}\ \emph {et~al.}(2011)\citenamefont
  {Bristowe}, \citenamefont {Littlewood},\ and\ \citenamefont
  {Artacho}}]{bristowe-JPCM-2011}%
  \BibitemOpen
  \bibfield  {author} {\bibinfo {author} {\bibnamefont {Bristowe},
  \bibfnamefont {N~C}}, \bibinfo {author} {\bibfnamefont {P.~B.}\ \bibnamefont
  {Littlewood}}, \ and\ \bibinfo {author} {\bibfnamefont {E.}~\bibnamefont
  {Artacho}}} (\bibinfo {year} {2011}),\ \bibfield  {title} {\enquote {\bibinfo
  {title} {The net charge at interfaces between insulators},}\ }\href@noop {}
  {\bibfield  {journal} {\bibinfo  {journal} {J. Phys.: Condens. Matter}\
  }\textbf {\bibinfo {volume} {23}}~(\bibinfo {number} {8}),\ \bibinfo {pages}
  {081001}}\BibitemShut {NoStop}%
\bibitem [{\citenamefont {Buchwald}\ and\ \citenamefont
  {Fox}(2013)}]{history_of_physics}%
  \BibitemOpen
  \bibinfo {editor} {\bibnamefont {Buchwald}, \bibfnamefont {J}}, \ and\
  \bibinfo {editor} {\bibfnamefont {R.}~\bibnamefont {Fox}},\ Eds. (\bibinfo
  {year} {2013}),\ \href@noop {} {\emph {\bibinfo {title} {The Oxford Handbook
  of The History of Physics}}}\ (\bibinfo  {publisher} {Oxford University
  Press},\ \bibinfo {address} {Oxford})\BibitemShut {NoStop}%
\bibitem [{\citenamefont {Cendagorta}\ and\ \citenamefont
  {Ichiye}(2015)}]{mip_cendagorta_2015}%
  \BibitemOpen
  \bibfield  {author} {\bibinfo {author} {\bibnamefont {Cendagorta},
  \bibfnamefont {J~R}}, \ and\ \bibinfo {author} {\bibfnamefont
  {T.}~\bibnamefont {Ichiye}}} (\bibinfo {year} {2015}),\ \bibfield  {title}
  {\enquote {\bibinfo {title} {The surface potential of the water–vapor
  interface from classical simulations},}\ }\href@noop {} {\bibfield  {journal}
  {\bibinfo  {journal} {J. Phys. Chem. B}\ }\textbf {\bibinfo {volume}
  {119}}~(\bibinfo {number} {29}),\ \bibinfo {pages} {9114--9122}}\BibitemShut
  {NoStop}%
\bibitem [{\citenamefont {Chisolm}(2012)}]{chisolm_2012}%
  \BibitemOpen
  \bibfield  {author} {\bibinfo {author} {\bibnamefont {Chisolm}, \bibfnamefont
  {Eric}}} (\bibinfo {year} {2012}),\ \href {https://arxiv.org/abs/1205.5935}
  {\enquote {\bibinfo {title} {Geometric algebra},}\ }\Eprint
  {http://arxiv.org/abs/1205.5935} {arXiv:1205.5935 [math-ph]} \BibitemShut
  {NoStop}%
\bibitem [{\citenamefont {Cioslowski}\ and\ \citenamefont
  {Pr\k{a}tnicki}(2019)}]{cioslowski_2019}%
  \BibitemOpen
  \bibfield  {author} {\bibinfo {author} {\bibnamefont {Cioslowski},
  \bibfnamefont {J}}, \ and\ \bibinfo {author} {\bibfnamefont {F.}~\bibnamefont
  {Pr\k{a}tnicki}}} (\bibinfo {year} {2019}),\ \bibfield  {title} {\enquote
  {\bibinfo {title} {Universalities among natural orbitals and occupation
  numbers pertaining to ground states of two electrons in central
  potentials},}\ }\href {\doibase 10.1063/1.5123669} {\bibfield  {journal}
  {\bibinfo  {journal} {J. Chem. Phys.}\ }\textbf {\bibinfo {volume}
  {151}}~(\bibinfo {number} {18}),\ \bibinfo {pages} {184107}}\BibitemShut
  {NoStop}%
\bibitem [{\citenamefont {Cioslowski}\ and\ \citenamefont
  {Strasburger}(2021{\natexlab{a}})}]{cioslowski}%
  \BibitemOpen
  \bibfield  {author} {\bibinfo {author} {\bibnamefont {Cioslowski},
  \bibfnamefont {J}}, \ and\ \bibinfo {author} {\bibfnamefont {K.}~\bibnamefont
  {Strasburger}}} (\bibinfo {year} {2021}{\natexlab{a}}),\ \bibfield  {title}
  {\enquote {\bibinfo {title} {From {Fredholm} to {Schr\"odinger} via
  {Eikonal}: A new formalism for revealing unknown properties of natural
  orbitals},}\ }\href@noop {} {\bibfield  {journal} {\bibinfo  {journal} {J.
  Chem. Theory Comput.}\ }\textbf {\bibinfo {volume} {17}}~(\bibinfo {number}
  {11}),\ \bibinfo {pages} {6918--6933}}\BibitemShut {NoStop}%
\bibitem [{\citenamefont {Cioslowski}\ and\ \citenamefont
  {Strasburger}(2021{\natexlab{b}})}]{cioslowski_2021}%
  \BibitemOpen
  \bibfield  {author} {\bibinfo {author} {\bibnamefont {Cioslowski},
  \bibfnamefont {J}}, \ and\ \bibinfo {author} {\bibfnamefont {K.}~\bibnamefont
  {Strasburger}}} (\bibinfo {year} {2021}{\natexlab{b}}),\ \bibfield  {title}
  {\enquote {\bibinfo {title} {From {Fredholm} to {Schr\"odinger} via
  {Eikonal}: {A} new formalism for revealing unknown properties of natural
  orbitals},}\ }\href {\doibase 10.1021/acs.jctc.1c00709} {\bibfield  {journal}
  {\bibinfo  {journal} {J. Chem. Theory Comput.}\ }\textbf {\bibinfo {volume}
  {17}}~(\bibinfo {number} {11}),\ \bibinfo {pages} {6918--6933}}\BibitemShut
  {NoStop}%
\bibitem [{\citenamefont {Cochran}(1960)}]{cochran_1960}%
  \BibitemOpen
  \bibfield  {author} {\bibinfo {author} {\bibnamefont {Cochran}, \bibfnamefont
  {W}}} (\bibinfo {year} {1960}),\ \bibfield  {title} {\enquote {\bibinfo
  {title} {Crystal stability and the theory of ferroelectricity},}\ }\href@noop
  {} {\bibfield  {journal} {\bibinfo  {journal} {Adv. Phys.}\ }\textbf
  {\bibinfo {volume} {9}}~(\bibinfo {number} {36}),\ \bibinfo {pages}
  {387--423}}\BibitemShut {NoStop}%
\bibitem [{\citenamefont {Cochran}\ and\ \citenamefont
  {Cowley}(1962)}]{cochran_cowley_1962}%
  \BibitemOpen
  \bibfield  {author} {\bibinfo {author} {\bibnamefont {Cochran}, \bibfnamefont
  {W}}, \ and\ \bibinfo {author} {\bibfnamefont {R.~A.}\ \bibnamefont
  {Cowley}}} (\bibinfo {year} {1962}),\ \bibfield  {title} {\enquote {\bibinfo
  {title} {Dielectric constants and lattice vibrations},}\ }\href@noop {}
  {\bibfield  {journal} {\bibinfo  {journal} {J. Phys. Chem. Solids}\ }\textbf
  {\bibinfo {volume} {23}}~(\bibinfo {number} {5}),\ \bibinfo {pages}
  {447--450}}\BibitemShut {NoStop}%
\bibitem [{\citenamefont {Cohen}\ and\ \citenamefont
  {Louie}(2016)}]{cohen_louie}%
  \BibitemOpen
  \bibfield  {author} {\bibinfo {author} {\bibnamefont {Cohen}, \bibfnamefont
  {M~L}}, \ and\ \bibinfo {author} {\bibfnamefont {S.~G.}\ \bibnamefont
  {Louie}}} (\bibinfo {year} {2016}),\ \href@noop {} {\emph {\bibinfo {title}
  {Fundamentals of Condensed Matter Physics}}}\ (\bibinfo  {publisher}
  {Cambridge University Press})\BibitemShut {NoStop}%
\bibitem [{\citenamefont {Coiana}\ \emph {et~al.}(2024)\citenamefont {Coiana},
  \citenamefont {Lischner},\ and\ \citenamefont {Tangney}}]{coiana_mgo}%
  \BibitemOpen
  \bibfield  {author} {\bibinfo {author} {\bibnamefont {Coiana}, \bibfnamefont
  {G}}, \bibinfo {author} {\bibfnamefont {J.}~\bibnamefont {Lischner}}, \ and\
  \bibinfo {author} {\bibfnamefont {P.}~\bibnamefont {Tangney}}} (\bibinfo
  {year} {2024}),\ \bibfield  {title} {\enquote {\bibinfo {title} {Breakdown of
  phonon band theory in {MgO}},}\ }\href {\doibase 10.1103/PhysRevB.109.014310}
  {\bibfield  {journal} {\bibinfo  {journal} {Phys. Rev. B}\ }\textbf {\bibinfo
  {volume} {109}},\ \bibinfo {pages} {014310}}\BibitemShut {NoStop}%
\bibitem [{\citenamefont {Coleman}(1963)}]{coleman_rmp}%
  \BibitemOpen
  \bibfield  {author} {\bibinfo {author} {\bibnamefont {Coleman}, \bibfnamefont
  {A~J}}} (\bibinfo {year} {1963}),\ \bibfield  {title} {\enquote {\bibinfo
  {title} {Structure of fermion density matrices},}\ }\href@noop {} {\bibfield
  {journal} {\bibinfo  {journal} {Rev. Mod. Phys.}\ }\textbf {\bibinfo {volume}
  {35}},\ \bibinfo {pages} {668--686}}\BibitemShut {NoStop}%
\bibitem [{\citenamefont {Colombo}\ \emph {et~al.}(1991)\citenamefont
  {Colombo}, \citenamefont {Resta},\ and\ \citenamefont
  {Baroni}}]{colombo_1991}%
  \BibitemOpen
  \bibfield  {author} {\bibinfo {author} {\bibnamefont {Colombo}, \bibfnamefont
  {L}}, \bibinfo {author} {\bibfnamefont {R.}~\bibnamefont {Resta}}, \ and\
  \bibinfo {author} {\bibfnamefont {S.}~\bibnamefont {Baroni}}} (\bibinfo
  {year} {1991}),\ \bibfield  {title} {\enquote {\bibinfo {title} {Valence-band
  offsets at strained si/ge interfaces},}\ }\href {\doibase
  10.1103/PhysRevB.44.5572} {\bibfield  {journal} {\bibinfo  {journal} {Phys.
  Rev. B}\ }\textbf {\bibinfo {volume} {44}},\ \bibinfo {pages}
  {5572--5579}}\BibitemShut {NoStop}%
\bibitem [{\citenamefont {Corso}(2025)}]{Corso_2025}%
  \BibitemOpen
  \bibfield  {author} {\bibinfo {author} {\bibnamefont {Corso}, \bibfnamefont
  {Thiago~Carvalho}}} (\bibinfo {year} {2025}),\ \bibfield  {title} {\enquote
  {\bibinfo {title} {v-representability and hohenberg-kohn theorem for
  non-interacting schrödinger operators with distributional potentials on the
  one-dimensional torus},}\ }\href {\doibase 10.1088/1751-8121/adc04c}
  {\bibfield  {journal} {\bibinfo  {journal} {J. Phys. A: Math. Theor.}\
  }\textbf {\bibinfo {volume} {58}}~(\bibinfo {number} {12}),\ \bibinfo {pages}
  {125203}}\BibitemShut {NoStop}%
\bibitem [{\citenamefont {D'Alessio}\ \emph {et~al.}(2016)\citenamefont
  {D'Alessio}, \citenamefont {Kafri}, \citenamefont {Polkovnikov},\ and\
  \citenamefont {Rigol}}]{mbloc_6}%
  \BibitemOpen
  \bibfield  {author} {\bibinfo {author} {\bibnamefont {D'Alessio},
  \bibfnamefont {L}}, \bibinfo {author} {\bibfnamefont {Y.}~\bibnamefont
  {Kafri}}, \bibinfo {author} {\bibfnamefont {A.}~\bibnamefont {Polkovnikov}},
  \ and\ \bibinfo {author} {\bibfnamefont {M.}~\bibnamefont {Rigol}}} (\bibinfo
  {year} {2016}),\ \bibfield  {title} {\enquote {\bibinfo {title} {From quantum
  chaos and eigenstate thermalization to statistical mechanics and
  thermodynamics},}\ }\href {\doibase 10.1080/00018732.2016.1198134} {\bibfield
   {journal} {\bibinfo  {journal} {Adv. Phys.}\ }\textbf {\bibinfo {volume}
  {65}}~(\bibinfo {number} {3}),\ \bibinfo {pages} {239--362}}\BibitemShut
  {NoStop}%
\bibitem [{\citenamefont {Darrigol}(2006)}]{darrigo_2006}%
  \BibitemOpen
  \bibfield  {author} {\bibinfo {author} {\bibnamefont {Darrigol},
  \bibfnamefont {O}}} (\bibinfo {year} {2006}),\ \enquote {\bibinfo {title}
  {The genesis of the theory of relativity},}\ in\ \href {\doibase
  10.1007/3-7643-7436-5_1} {\emph {\bibinfo {booktitle} {{Einstein, 1905--2005:
  Poincar{\'e} Seminar 2005}}}},\ \bibinfo {editor} {edited by\ \bibinfo
  {editor} {\bibfnamefont {T.}~\bibnamefont {Damour}}, \bibinfo {editor}
  {\bibfnamefont {O.}~\bibnamefont {Darrigol}}, \bibinfo {editor}
  {\bibfnamefont {B.}~\bibnamefont {Duplantier}}, \ and\ \bibinfo {editor}
  {\bibfnamefont {V.}~\bibnamefont {Rivasseau}}}\ (\bibinfo  {publisher}
  {Birkh{\"a}user {Basel}},\ \bibinfo {address} {Basel})\ pp.\ \bibinfo {pages}
  {1--31}\BibitemShut {NoStop}%
\bibitem [{\citenamefont {Davidson}(1972)}]{davidson_1972}%
  \BibitemOpen
  \bibfield  {author} {\bibinfo {author} {\bibnamefont {Davidson},
  \bibfnamefont {E~R}}} (\bibinfo {year} {1972}),\ \bibfield  {title} {\enquote
  {\bibinfo {title} {Properties and uses of natural orbitals},}\ }\href
  {\doibase 10.1103/RevModPhys.44.451} {\bibfield  {journal} {\bibinfo
  {journal} {Rev. Mod. Phys.}\ }\textbf {\bibinfo {volume} {44}},\ \bibinfo
  {pages} {451--464}}\BibitemShut {NoStop}%
\bibitem [{\citenamefont {{de Groot}}\ and\ \citenamefont
  {Vlieger}(1965)}]{degroot_1965}%
  \BibitemOpen
  \bibfield  {author} {\bibinfo {author} {\bibnamefont {{de Groot}},
  \bibfnamefont {S~R}}, \ and\ \bibinfo {author} {\bibfnamefont
  {J.}~\bibnamefont {Vlieger}}} (\bibinfo {year} {1965}),\ \bibfield  {title}
  {\enquote {\bibinfo {title} {Derivation of {Maxwell's} equations: The
  statistical theory of the macroscopic equations},}\ }\href@noop {} {\bibfield
   {journal} {\bibinfo  {journal} {Physica}\ }\textbf {\bibinfo {volume}
  {31}}~(\bibinfo {number} {3}),\ \bibinfo {pages} {254--268}}\BibitemShut
  {NoStop}%
\bibitem [{\citenamefont {Doran}\ and\ \citenamefont
  {Lasenby}(2003)}]{doran_2003}%
  \BibitemOpen
  \bibfield  {author} {\bibinfo {author} {\bibnamefont {Doran}, \bibfnamefont
  {C}}, \ and\ \bibinfo {author} {\bibfnamefont {A.}~\bibnamefont {Lasenby}}}
  (\bibinfo {year} {2003}),\ \href@noop {} {\emph {\bibinfo {title} {Geometric
  Algebra for Physicists}}}\ (\bibinfo  {publisher} {Cambridge University
  Press})\BibitemShut {NoStop}%
\bibitem [{\citenamefont {Doran}\ and\ \citenamefont {Lasenby}(2007)}]{doran}%
  \BibitemOpen
  \bibfield  {author} {\bibinfo {author} {\bibnamefont {Doran}, \bibfnamefont
  {C}}, \ and\ \bibinfo {author} {\bibfnamefont {A.}~\bibnamefont {Lasenby}}}
  (\bibinfo {year} {2007}),\ \href@noop {} {\emph {\bibinfo {title} {Geometric
  Algebra for Physicists}}}\ (\bibinfo  {publisher} {Cambridge University
  Press})\BibitemShut {NoStop}%
\bibitem [{\citenamefont {Dorst}(2002)}]{dorst_contraction}%
  \BibitemOpen
  \bibfield  {author} {\bibinfo {author} {\bibnamefont {Dorst}, \bibfnamefont
  {L}}} (\bibinfo {year} {2002}),\ \enquote {\bibinfo {title} {Applications of
  geometric algebra in computer science and engineering},}\ Chap.~\bibinfo
  {chapter} {2}\ (\bibinfo  {publisher} {Birkh\"auser},\ \bibinfo {address}
  {Boston, MA})\ pp.\ \bibinfo {pages} {35--46}\BibitemShut {NoStop}%
\bibitem [{\citenamefont {Dreizler}\ and\ \citenamefont
  {Gross}(1990)}]{dreizler_gross_1990}%
  \BibitemOpen
  \bibfield  {author} {\bibinfo {author} {\bibnamefont {Dreizler},
  \bibfnamefont {R~M}}, \ and\ \bibinfo {author} {\bibfnamefont {E.~K.~U.}\
  \bibnamefont {Gross}}} (\bibinfo {year} {1990}),\ \href {\doibase
  10.1007/978-3-642-86105-5} {\emph {\bibinfo {title} {Density Functional
  Theory: An Approach to the Quantum Many-Body Problem}}}\ (\bibinfo
  {publisher} {Springer Berlin Heidelberg},\ \bibinfo {address} {Berlin,
  Heidelberg})\BibitemShut {NoStop}%
\bibitem [{\citenamefont {Edmiston}\ and\ \citenamefont
  {Ruedenberg}(1963)}]{edmiston}%
  \BibitemOpen
  \bibfield  {author} {\bibinfo {author} {\bibnamefont {Edmiston},
  \bibfnamefont {C}}, \ and\ \bibinfo {author} {\bibfnamefont {K.}~\bibnamefont
  {Ruedenberg}}} (\bibinfo {year} {1963}),\ \bibfield  {title} {\enquote
  {\bibinfo {title} {Localized atomic and molecular orbitals},}\ }\href@noop {}
  {\bibfield  {journal} {\bibinfo  {journal} {Rev. Mod. Phys.}\ }\textbf
  {\bibinfo {volume} {35}},\ \bibinfo {pages} {457--464}}\BibitemShut {NoStop}%
\bibitem [{\citenamefont {Engel}\ and\ \citenamefont
  {Dreizler}(2011)}]{engel_and_dreizler}%
  \BibitemOpen
  \bibfield  {author} {\bibinfo {author} {\bibnamefont {Engel}, \bibfnamefont
  {E}}, \ and\ \bibinfo {author} {\bibfnamefont {R.~M.}\ \bibnamefont
  {Dreizler}}} (\bibinfo {year} {2011}),\ \href@noop {} {\emph {\bibinfo
  {title} {Density functional theory: An advanced course}}}\ (\bibinfo
  {publisher} {Springer Berlin, Heidelberg})\BibitemShut {NoStop}%
\bibitem [{\citenamefont {Evans}\ and\ \citenamefont
  {Morriss}(2008)}]{evans_morriss_2008}%
  \BibitemOpen
  \bibfield  {author} {\bibinfo {author} {\bibnamefont {Evans}, \bibfnamefont
  {D~J}}, \ and\ \bibinfo {author} {\bibfnamefont {G.}~\bibnamefont {Morriss}}}
  (\bibinfo {year} {2008}),\ \href {\doibase 10.1017/CBO9780511535307} {\emph
  {\bibinfo {title} {Statistical Mechanics of Nonequilibrium Liquids}}},\
  \bibinfo {edition} {2nd}\ ed.\ (\bibinfo  {publisher} {Cambridge University
  Press})\BibitemShut {NoStop}%
\bibitem [{\citenamefont {Evans}\ and\ \citenamefont
  {Searles}(2002)}]{fluctuation_theorem}%
  \BibitemOpen
  \bibfield  {author} {\bibinfo {author} {\bibnamefont {Evans}, \bibfnamefont
  {D~J}}, \ and\ \bibinfo {author} {\bibfnamefont {D.~J.}\ \bibnamefont
  {Searles}}} (\bibinfo {year} {2002}),\ \bibfield  {title} {\enquote {\bibinfo
  {title} {The fluctuation theorem},}\ }\href {\doibase
  10.1080/00018730210155133} {\bibfield  {journal} {\bibinfo  {journal} {Adv.
  Phys.}\ }\textbf {\bibinfo {volume} {51}}~(\bibinfo {number} {7}),\ \bibinfo
  {pages} {1529--1585}}\BibitemShut {NoStop}%
\bibitem [{\citenamefont {Falkenburg}(2009)}]{falkenburg_2009}%
  \BibitemOpen
  \bibfield  {author} {\bibinfo {author} {\bibnamefont {Falkenburg},
  \bibfnamefont {B}}} (\bibinfo {year} {2009}),\ \enquote {\bibinfo {title}
  {Correspondence principle},}\ in\ \href {\doibase
  10.1007/978-3-540-70626-7_39} {\emph {\bibinfo {booktitle} {Compendium of
  Quantum Physics}}},\ \bibinfo {editor} {edited by\ \bibinfo {editor}
  {\bibfnamefont {D.}~\bibnamefont {Greenberger}}, \bibinfo {editor}
  {\bibfnamefont {K.}~\bibnamefont {Hentschel}}, \ and\ \bibinfo {editor}
  {\bibfnamefont {F.}~\bibnamefont {Weinert}}}\ (\bibinfo  {publisher}
  {Springer Berlin Heidelberg},\ \bibinfo {address} {Berlin, Heidelberg})\ pp.\
  \bibinfo {pages} {125--131}\BibitemShut {NoStop}%
\bibitem [{\citenamefont {Ferreira}\ and\ \citenamefont
  {Parada}(1970)}]{ferreira_parada}%
  \BibitemOpen
  \bibfield  {author} {\bibinfo {author} {\bibnamefont {Ferreira},
  \bibfnamefont {L~G}}, \ and\ \bibinfo {author} {\bibfnamefont {N.~J.}\
  \bibnamefont {Parada}}} (\bibinfo {year} {1970}),\ \bibfield  {title}
  {\enquote {\bibinfo {title} {{Wannier} functions and the phases of the
  {Bloch} functions},}\ }\href@noop {} {\bibfield  {journal} {\bibinfo
  {journal} {Phys. Rev. B}\ }\textbf {\bibinfo {volume} {2}},\ \bibinfo {pages}
  {1614--1618}}\BibitemShut {NoStop}%
\bibitem [{\citenamefont {Finnis}(1998)}]{finnis}%
  \BibitemOpen
  \bibfield  {author} {\bibinfo {author} {\bibnamefont {Finnis}, \bibfnamefont
  {M~W}}} (\bibinfo {year} {1998}),\ \bibfield  {title} {\enquote {\bibinfo
  {title} {Accessing the excess: An atomistic approach to excesses at planar
  defects and dislocations in ordered compounds},}\ }\href@noop {} {\bibfield
  {journal} {\bibinfo  {journal} {Phys. Status Solidi A}\ }\textbf {\bibinfo
  {volume} {166}}~(\bibinfo {number} {1}),\ \bibinfo {pages}
  {397--416}}\BibitemShut {NoStop}%
\bibitem [{\citenamefont {Foster}\ and\ \citenamefont
  {Boys}(1960{\natexlab{a}})}]{boys_fostera}%
  \BibitemOpen
  \bibfield  {author} {\bibinfo {author} {\bibnamefont {Foster}, \bibfnamefont
  {J~M}}, \ and\ \bibinfo {author} {\bibfnamefont {S.~F.}\ \bibnamefont
  {Boys}}} (\bibinfo {year} {1960}{\natexlab{a}}),\ \bibfield  {title}
  {\enquote {\bibinfo {title} {Canonical configurational interaction
  procedure},}\ }\href@noop {} {\bibfield  {journal} {\bibinfo  {journal} {Rev.
  Mod. Phys.}\ }\textbf {\bibinfo {volume} {32}},\ \bibinfo {pages}
  {300--302}}\BibitemShut {NoStop}%
\bibitem [{\citenamefont {Foster}\ and\ \citenamefont
  {Boys}(1960{\natexlab{b}})}]{boys_fosterb}%
  \BibitemOpen
  \bibfield  {author} {\bibinfo {author} {\bibnamefont {Foster}, \bibfnamefont
  {J~M}}, \ and\ \bibinfo {author} {\bibfnamefont {S.~F.}\ \bibnamefont
  {Boys}}} (\bibinfo {year} {1960}{\natexlab{b}}),\ \bibfield  {title}
  {\enquote {\bibinfo {title} {A quantum variational calculation for {HCHO}},}\
  }\href@noop {} {\bibfield  {journal} {\bibinfo  {journal} {Rev. Mod. Phys.}\
  }\textbf {\bibinfo {volume} {32}},\ \bibinfo {pages} {303--304}}\BibitemShut
  {NoStop}%
\bibitem [{\citenamefont {Frias}\ and\ \citenamefont
  {Smolyakov}(2012)}]{frias_2012}%
  \BibitemOpen
  \bibfield  {author} {\bibinfo {author} {\bibnamefont {Frias}, \bibfnamefont
  {W}}, \ and\ \bibinfo {author} {\bibfnamefont {A.~I.}\ \bibnamefont
  {Smolyakov}}} (\bibinfo {year} {2012}),\ \bibfield  {title} {\enquote
  {\bibinfo {title} {Electromagnetic forces and internal stresses in dielectric
  media},}\ }\href@noop {} {\bibfield  {journal} {\bibinfo  {journal} {Phys.
  Rev. E}\ }\textbf {\bibinfo {volume} {85}},\ \bibinfo {pages}
  {046606}}\BibitemShut {NoStop}%
\bibitem [{\citenamefont {Gajdardziska-Josifovska}\ and\ \citenamefont
  {Carim}(1999)}]{Gajdardziska-Josifovska1999}%
  \BibitemOpen
  \bibfield  {author} {\bibinfo {author} {\bibnamefont
  {Gajdardziska-Josifovska}, \bibfnamefont {M}}, \ and\ \bibinfo {author}
  {\bibfnamefont {A.~H.}\ \bibnamefont {Carim}}} (\bibinfo {year} {1999}),\
  \enquote {\bibinfo {title} {Applications of electron holography},}\ in\
  \href@noop {} {\emph {\bibinfo {booktitle} {Introduction to Electron
  Holography}}}\ (\bibinfo  {publisher} {Springer US},\ \bibinfo {address}
  {Boston, MA})\ pp.\ \bibinfo {pages} {267--293}\BibitemShut {NoStop}%
\bibitem [{\citenamefont {Gajdardziska-Josifovska}\ \emph
  {et~al.}(1993)\citenamefont {Gajdardziska-Josifovska}, \citenamefont
  {McCartney}, \citenamefont {de~Ruijter}, \citenamefont {Smith}, \citenamefont
  {Weiss},\ and\ \citenamefont {Zuo}}]{gajdardziska-1993}%
  \BibitemOpen
  \bibfield  {author} {\bibinfo {author} {\bibnamefont
  {Gajdardziska-Josifovska}, \bibfnamefont {M}}, \bibinfo {author}
  {\bibfnamefont {M.~R.}\ \bibnamefont {McCartney}}, \bibinfo {author}
  {\bibfnamefont {W.~J.}\ \bibnamefont {de~Ruijter}}, \bibinfo {author}
  {\bibfnamefont {D.~J.}\ \bibnamefont {Smith}}, \bibinfo {author}
  {\bibfnamefont {J.~K.}\ \bibnamefont {Weiss}}, \ and\ \bibinfo {author}
  {\bibfnamefont {J.~M.}\ \bibnamefont {Zuo}}} (\bibinfo {year} {1993}),\
  \bibfield  {title} {\enquote {\bibinfo {title} {Accurate measurements of mean
  inner potential of crystal wedges using digital electron holograms},}\
  }\href@noop {} {\bibfield  {journal} {\bibinfo  {journal} {Ultramicroscopy}\
  }\textbf {\bibinfo {volume} {50}}~(\bibinfo {number} {3}),\ \bibinfo {pages}
  {285--299}}\BibitemShut {NoStop}%
\bibitem [{\citenamefont {Gallier}(2011)}]{gallier_2011}%
  \BibitemOpen
  \bibfield  {author} {\bibinfo {author} {\bibnamefont {Gallier}, \bibfnamefont
  {J}}} (\bibinfo {year} {2011}),\ \href@noop {} {\emph {\bibinfo {title}
  {Geometric Methods and Applications: For Computer Science and
  Engineering}}},\ Texts in Applied Mathematics\ (\bibinfo  {publisher}
  {Springer New York})\BibitemShut {NoStop}%
\bibitem [{\citenamefont {Garling}(2011)}]{garling_2011}%
  \BibitemOpen
  \bibfield  {author} {\bibinfo {author} {\bibnamefont {Garling}, \bibfnamefont
  {D~J~H}}} (\bibinfo {year} {2011}),\ \href@noop {} {\emph {\bibinfo {title}
  {Clifford Algebras: An Introduction}}},\ London Mathematical Society Student
  Texts\ (\bibinfo  {publisher} {Cambridge University Press})\BibitemShut
  {NoStop}%
\bibitem [{\citenamefont {Giannozzi}\ \emph {et~al.}(2017)\citenamefont
  {Giannozzi}, \citenamefont {Andreussi}, \citenamefont {Brumme}, \citenamefont
  {Bunau}, \citenamefont {Nardelli}, \citenamefont {Calandra}, \citenamefont
  {Car}, \citenamefont {Cavazzoni}, \citenamefont {Ceresoli}, \citenamefont
  {Cococcioni}, \citenamefont {Colonna}, \citenamefont {Carnimeo},
  \citenamefont {Corso}, \citenamefont {de~Gironcoli}, \citenamefont {Delugas},
  \citenamefont {Jr}, \citenamefont {Ferretti}, \citenamefont {Floris},
  \citenamefont {Fratesi}, \citenamefont {Fugallo}, \citenamefont {Gebauer},
  \citenamefont {Gerstmann}, \citenamefont {Giustino}, \citenamefont {Gorni},
  \citenamefont {Jia}, \citenamefont {Kawamura}, \citenamefont {Ko},
  \citenamefont {Kokalj}, \citenamefont {Küçükbenli}, \citenamefont
  {Lazzeri}, \citenamefont {Marsili}, \citenamefont {Marzari}, \citenamefont
  {Mauri}, \citenamefont {Nguyen}, \citenamefont {Nguyen}, \citenamefont {de-la
  Roza}, \citenamefont {Paulatto}, \citenamefont {Poncé}, \citenamefont
  {Rocca}, \citenamefont {Sabatini}, \citenamefont {Santra}, \citenamefont
  {Schlipf}, \citenamefont {Seitsonen}, \citenamefont {Smogunov}, \citenamefont
  {Timrov}, \citenamefont {Thonhauser}, \citenamefont {Umari}, \citenamefont
  {Vast}, \citenamefont {Wu},\ and\ \citenamefont {Baroni}}]{espresso}%
  \BibitemOpen
  \bibfield  {author} {\bibinfo {author} {\bibnamefont {Giannozzi},
  \bibfnamefont {P}}, \bibinfo {author} {\bibfnamefont {O}~\bibnamefont
  {Andreussi}}, \bibinfo {author} {\bibfnamefont {T}~\bibnamefont {Brumme}},
  \bibinfo {author} {\bibfnamefont {O}~\bibnamefont {Bunau}}, \bibinfo {author}
  {\bibfnamefont {M~Buongiorno}\ \bibnamefont {Nardelli}}, \bibinfo {author}
  {\bibfnamefont {M}~\bibnamefont {Calandra}}, \bibinfo {author} {\bibfnamefont
  {R}~\bibnamefont {Car}}, \bibinfo {author} {\bibfnamefont {C}~\bibnamefont
  {Cavazzoni}}, \bibinfo {author} {\bibfnamefont {D}~\bibnamefont {Ceresoli}},
  \bibinfo {author} {\bibfnamefont {M}~\bibnamefont {Cococcioni}}, \bibinfo
  {author} {\bibfnamefont {N}~\bibnamefont {Colonna}}, \bibinfo {author}
  {\bibfnamefont {I}~\bibnamefont {Carnimeo}}, \bibinfo {author} {\bibfnamefont
  {A~Dal}\ \bibnamefont {Corso}}, \bibinfo {author} {\bibfnamefont
  {S}~\bibnamefont {de~Gironcoli}}, \bibinfo {author} {\bibfnamefont
  {P}~\bibnamefont {Delugas}}, \bibinfo {author} {\bibfnamefont {R~A~DiStasio}\
  \bibnamefont {Jr}}, \bibinfo {author} {\bibfnamefont {A}~\bibnamefont
  {Ferretti}}, \bibinfo {author} {\bibfnamefont {A}~\bibnamefont {Floris}},
  \bibinfo {author} {\bibfnamefont {G}~\bibnamefont {Fratesi}}, \bibinfo
  {author} {\bibfnamefont {G}~\bibnamefont {Fugallo}}, \bibinfo {author}
  {\bibfnamefont {R}~\bibnamefont {Gebauer}}, \bibinfo {author} {\bibfnamefont
  {U}~\bibnamefont {Gerstmann}}, \bibinfo {author} {\bibfnamefont
  {F}~\bibnamefont {Giustino}}, \bibinfo {author} {\bibfnamefont
  {T}~\bibnamefont {Gorni}}, \bibinfo {author} {\bibfnamefont {J}~\bibnamefont
  {Jia}}, \bibinfo {author} {\bibfnamefont {M}~\bibnamefont {Kawamura}},
  \bibinfo {author} {\bibfnamefont {H-Y}\ \bibnamefont {Ko}}, \bibinfo {author}
  {\bibfnamefont {A}~\bibnamefont {Kokalj}}, \bibinfo {author} {\bibfnamefont
  {E}~\bibnamefont {Küçükbenli}}, \bibinfo {author} {\bibfnamefont
  {M}~\bibnamefont {Lazzeri}}, \bibinfo {author} {\bibfnamefont
  {M}~\bibnamefont {Marsili}}, \bibinfo {author} {\bibfnamefont
  {N}~\bibnamefont {Marzari}}, \bibinfo {author} {\bibfnamefont
  {F}~\bibnamefont {Mauri}}, \bibinfo {author} {\bibfnamefont {N~L}\
  \bibnamefont {Nguyen}}, \bibinfo {author} {\bibfnamefont {H-V}\ \bibnamefont
  {Nguyen}}, \bibinfo {author} {\bibfnamefont {A~Otero}\ \bibnamefont {de-la
  Roza}}, \bibinfo {author} {\bibfnamefont {L}~\bibnamefont {Paulatto}},
  \bibinfo {author} {\bibfnamefont {S}~\bibnamefont {Poncé}}, \bibinfo
  {author} {\bibfnamefont {D}~\bibnamefont {Rocca}}, \bibinfo {author}
  {\bibfnamefont {R}~\bibnamefont {Sabatini}}, \bibinfo {author} {\bibfnamefont
  {B}~\bibnamefont {Santra}}, \bibinfo {author} {\bibfnamefont {M}~\bibnamefont
  {Schlipf}}, \bibinfo {author} {\bibfnamefont {A~P}\ \bibnamefont
  {Seitsonen}}, \bibinfo {author} {\bibfnamefont {A}~\bibnamefont {Smogunov}},
  \bibinfo {author} {\bibfnamefont {I}~\bibnamefont {Timrov}}, \bibinfo
  {author} {\bibfnamefont {T}~\bibnamefont {Thonhauser}}, \bibinfo {author}
  {\bibfnamefont {P}~\bibnamefont {Umari}}, \bibinfo {author} {\bibfnamefont
  {N}~\bibnamefont {Vast}}, \bibinfo {author} {\bibfnamefont {X}~\bibnamefont
  {Wu}}, \ and\ \bibinfo {author} {\bibfnamefont {S}~\bibnamefont {Baroni}}}
  (\bibinfo {year} {2017}),\ \bibfield  {title} {\enquote {\bibinfo {title}
  {Advanced capabilities for materials modelling with {QUANTUM ESPRESSO}},}\
  }\href@noop {} {\bibfield  {journal} {\bibinfo  {journal} {J. Phys.: Condens.
  Matter}\ }\textbf {\bibinfo {volume} {29}}~(\bibinfo {number} {46}),\
  \bibinfo {pages} {465901}}\BibitemShut {NoStop}%
\bibitem [{\citenamefont {Giannozzi}\ \emph {et~al.}(1991)\citenamefont
  {Giannozzi}, \citenamefont {de~Gironcoli}, \citenamefont {Pavone},\ and\
  \citenamefont {Baroni}}]{giannozzi_1991}%
  \BibitemOpen
  \bibfield  {author} {\bibinfo {author} {\bibnamefont {Giannozzi},
  \bibfnamefont {P}}, \bibinfo {author} {\bibfnamefont {S.}~\bibnamefont
  {de~Gironcoli}}, \bibinfo {author} {\bibfnamefont {P.}~\bibnamefont
  {Pavone}}, \ and\ \bibinfo {author} {\bibfnamefont {S.}~\bibnamefont
  {Baroni}}} (\bibinfo {year} {1991}),\ \bibfield  {title} {\enquote {\bibinfo
  {title} {Ab initio calculation of phonon dispersions in semiconductors},}\
  }\href@noop {} {\bibfield  {journal} {\bibinfo  {journal} {Phys. Rev. B}\
  }\textbf {\bibinfo {volume} {43}},\ \bibinfo {pages}
  {7231--7242}}\BibitemShut {NoStop}%
\bibitem [{\citenamefont {Giesbertz}\ and\ \citenamefont {van
  Leeuwen}(2013)}]{natural_occupations}%
  \BibitemOpen
  \bibfield  {author} {\bibinfo {author} {\bibnamefont {Giesbertz},
  \bibfnamefont {K~J~H}}, \ and\ \bibinfo {author} {\bibfnamefont
  {R.}~\bibnamefont {van Leeuwen}}} (\bibinfo {year} {2013}),\ \bibfield
  {title} {\enquote {\bibinfo {title} {Natural occupation numbers: When do they
  vanish?}}\ }\href@noop {} {\bibfield  {journal} {\bibinfo  {journal} {J.
  Chem. Phys.}\ }\textbf {\bibinfo {volume} {139}}~(\bibinfo {number} {10}),\
  \bibinfo {pages} {104109}}\BibitemShut {NoStop}%
\bibitem [{\citenamefont {Gillespie}\ and\ \citenamefont
  {Robinson}(2007)}]{gillespie_robinson_2007}%
  \BibitemOpen
  \bibfield  {author} {\bibinfo {author} {\bibnamefont {Gillespie},
  \bibfnamefont {R~J}}, \ and\ \bibinfo {author} {\bibfnamefont {E.~A.}\
  \bibnamefont {Robinson}}} (\bibinfo {year} {2007}),\ \bibfield  {title}
  {\enquote {\bibinfo {title} {{Gilbert N. Lewis} and the chemical bond: The
  electron pair and the octet rule from 1916 to the present day},}\ }\href@noop
  {} {\bibfield  {journal} {\bibinfo  {journal} {J. Comput. Chem.}\ }\textbf
  {\bibinfo {volume} {28}}~(\bibinfo {number} {1}),\ \bibinfo {pages}
  {87--97}}\BibitemShut {NoStop}%
\bibitem [{\citenamefont {Goedecker}\ and\ \citenamefont
  {Umrigar}(1998)}]{umrigar_1998}%
  \BibitemOpen
  \bibfield  {author} {\bibinfo {author} {\bibnamefont {Goedecker},
  \bibfnamefont {S}}, \ and\ \bibinfo {author} {\bibfnamefont {C.~J.}\
  \bibnamefont {Umrigar}}} (\bibinfo {year} {1998}),\ \bibfield  {title}
  {\enquote {\bibinfo {title} {Natural orbital functional for the many-electron
  problem},}\ }\href {\doibase 10.1103/PhysRevLett.81.866} {\bibfield
  {journal} {\bibinfo  {journal} {Phys. Rev. Lett.}\ }\textbf {\bibinfo
  {volume} {81}},\ \bibinfo {pages} {866--869}}\BibitemShut {NoStop}%
\bibitem [{\citenamefont {Goniakowski}\ \emph {et~al.}(2008)\citenamefont
  {Goniakowski}, \citenamefont {Finocchi},\ and\ \citenamefont
  {Noguera}}]{goniakowski-rpp-2008}%
  \BibitemOpen
  \bibfield  {author} {\bibinfo {author} {\bibnamefont {Goniakowski},
  \bibfnamefont {J}}, \bibinfo {author} {\bibfnamefont {F.}~\bibnamefont
  {Finocchi}}, \ and\ \bibinfo {author} {\bibfnamefont {C.}~\bibnamefont
  {Noguera}}} (\bibinfo {year} {2008}),\ \bibfield  {title} {\enquote {\bibinfo
  {title} {Polarity of oxide surfaces and nanostructures},}\ }\href@noop {}
  {\bibfield  {journal} {\bibinfo  {journal} {Rep. Prog. Phys.}\ }\textbf
  {\bibinfo {volume} {71}}~(\bibinfo {number} {1}),\ \bibinfo {pages}
  {016501}}\BibitemShut {NoStop}%
\bibitem [{\citenamefont {Goniakowski}\ and\ \citenamefont
  {Noguera}(2011)}]{goniakowski-prb-2011}%
  \BibitemOpen
  \bibfield  {author} {\bibinfo {author} {\bibnamefont {Goniakowski},
  \bibfnamefont {J}}, \ and\ \bibinfo {author} {\bibfnamefont {C.}~\bibnamefont
  {Noguera}}} (\bibinfo {year} {2011}),\ \bibfield  {title} {\enquote {\bibinfo
  {title} {Polarity at the nanoscale},}\ }\href@noop {} {\bibfield  {journal}
  {\bibinfo  {journal} {Phys. Rev. B}\ }\textbf {\bibinfo {volume} {83}},\
  \bibinfo {pages} {115413}}\BibitemShut {NoStop}%
\bibitem [{\citenamefont {Goniakowski}\ and\ \citenamefont
  {Noguera}(2014)}]{goniakowski-2014}%
  \BibitemOpen
  \bibfield  {author} {\bibinfo {author} {\bibnamefont {Goniakowski},
  \bibfnamefont {J}}, \ and\ \bibinfo {author} {\bibfnamefont {C.}~\bibnamefont
  {Noguera}}} (\bibinfo {year} {2014}),\ \bibfield  {title} {\enquote {\bibinfo
  {title} {Conditions for electronic reconstruction at stoichiometric
  polar/polar interfaces},}\ }\href@noop {} {\bibfield  {journal} {\bibinfo
  {journal} {J. Phys.: Condens. Matter}\ }\textbf {\bibinfo {volume}
  {26}}~(\bibinfo {number} {48}),\ \bibinfo {pages} {485010}}\BibitemShut
  {NoStop}%
\bibitem [{\citenamefont {Goniakowski}\ and\ \citenamefont
  {Noguera}(2016)}]{goniakowski-2016}%
  \BibitemOpen
  \bibfield  {author} {\bibinfo {author} {\bibnamefont {Goniakowski},
  \bibfnamefont {J}}, \ and\ \bibinfo {author} {\bibfnamefont {C.}~\bibnamefont
  {Noguera}}} (\bibinfo {year} {2016}),\ \bibfield  {title} {\enquote {\bibinfo
  {title} {Insulating oxide surfaces and nanostructures},}\ }\href@noop {}
  {\bibfield  {journal} {\bibinfo  {journal} {C. R. Phys.}\ }\textbf {\bibinfo
  {volume} {17}}~(\bibinfo {number} {3–4}),\ \bibinfo {pages}
  {471--480}}\BibitemShut {NoStop}%
\bibitem [{\citenamefont {Gonze}\ \emph {et~al.}(2020)\citenamefont {Gonze},
  \citenamefont {Amadon}, \citenamefont {Antonius}, \citenamefont {Arnardi},
  \citenamefont {Baguet}, \citenamefont {Beuken}, \citenamefont {Bieder},
  \citenamefont {Bottin}, \citenamefont {Bouchet}, \citenamefont {Bousquet},
  \citenamefont {Brouwer}, \citenamefont {Bruneval}, \citenamefont {Brunin},
  \citenamefont {Cavignac}, \citenamefont {Charraud}, \citenamefont {Chen},
  \citenamefont {C\'ot\'e}, \citenamefont {Cottenier}, \citenamefont {Denier},
  \citenamefont {Geneste}, \citenamefont {Ghosez}, \citenamefont {Giantomassi},
  \citenamefont {Gillet}, \citenamefont {Gingras}, \citenamefont {Hamann},
  \citenamefont {Hautier}, \citenamefont {He}, \citenamefont {Helbig},
  \citenamefont {Holzwarth}, \citenamefont {Jia}, \citenamefont {Jollet},
  \citenamefont {Lafargue-Dit-Hauret}, \citenamefont {Lejaeghere},
  \citenamefont {Marques}, \citenamefont {Martin}, \citenamefont {Martins},
  \citenamefont {Miranda}, \citenamefont {Naccarato}, \citenamefont {Persson},
  \citenamefont {Petretto}, \citenamefont {Planes}, \citenamefont {Pouillon},
  \citenamefont {Prokhorenko}, \citenamefont {Ricci}, \citenamefont
  {Rignanese}, \citenamefont {Romero}, \citenamefont {Schmitt}, \citenamefont
  {Torrent}, \citenamefont {van Setten}, \citenamefont {Troeye}, \citenamefont
  {Verstraete}, \citenamefont {Z\'erah},\ and\ \citenamefont
  {Zwanziger}}]{abinit}%
  \BibitemOpen
  \bibfield  {author} {\bibinfo {author} {\bibnamefont {Gonze}, \bibfnamefont
  {X}}, \bibinfo {author} {\bibfnamefont {B.}~\bibnamefont {Amadon}}, \bibinfo
  {author} {\bibfnamefont {G.}~\bibnamefont {Antonius}}, \bibinfo {author}
  {\bibfnamefont {F.}~\bibnamefont {Arnardi}}, \bibinfo {author} {\bibfnamefont
  {L.}~\bibnamefont {Baguet}}, \bibinfo {author} {\bibfnamefont {J.-M.}\
  \bibnamefont {Beuken}}, \bibinfo {author} {\bibfnamefont {J.}~\bibnamefont
  {Bieder}}, \bibinfo {author} {\bibfnamefont {F.}~\bibnamefont {Bottin}},
  \bibinfo {author} {\bibfnamefont {J.}~\bibnamefont {Bouchet}}, \bibinfo
  {author} {\bibfnamefont {E.}~\bibnamefont {Bousquet}}, \bibinfo {author}
  {\bibfnamefont {N.}~\bibnamefont {Brouwer}}, \bibinfo {author} {\bibfnamefont
  {F.}~\bibnamefont {Bruneval}}, \bibinfo {author} {\bibfnamefont
  {G.}~\bibnamefont {Brunin}}, \bibinfo {author} {\bibfnamefont
  {T.}~\bibnamefont {Cavignac}}, \bibinfo {author} {\bibfnamefont {J.-B.}\
  \bibnamefont {Charraud}}, \bibinfo {author} {\bibfnamefont {W.}~\bibnamefont
  {Chen}}, \bibinfo {author} {\bibfnamefont {M.}~\bibnamefont {C\'ot\'e}},
  \bibinfo {author} {\bibfnamefont {S.}~\bibnamefont {Cottenier}}, \bibinfo
  {author} {\bibfnamefont {J.}~\bibnamefont {Denier}}, \bibinfo {author}
  {\bibfnamefont {G.}~\bibnamefont {Geneste}}, \bibinfo {author} {\bibfnamefont
  {P.}~\bibnamefont {Ghosez}}, \bibinfo {author} {\bibfnamefont
  {M.}~\bibnamefont {Giantomassi}}, \bibinfo {author} {\bibfnamefont
  {Y.}~\bibnamefont {Gillet}}, \bibinfo {author} {\bibfnamefont
  {O.}~\bibnamefont {Gingras}}, \bibinfo {author} {\bibfnamefont {D.~R.}\
  \bibnamefont {Hamann}}, \bibinfo {author} {\bibfnamefont {G.}~\bibnamefont
  {Hautier}}, \bibinfo {author} {\bibfnamefont {X.}~\bibnamefont {He}},
  \bibinfo {author} {\bibfnamefont {N.}~\bibnamefont {Helbig}}, \bibinfo
  {author} {\bibfnamefont {N.}~\bibnamefont {Holzwarth}}, \bibinfo {author}
  {\bibfnamefont {Y.}~\bibnamefont {Jia}}, \bibinfo {author} {\bibfnamefont
  {F.}~\bibnamefont {Jollet}}, \bibinfo {author} {\bibfnamefont
  {W.}~\bibnamefont {Lafargue-Dit-Hauret}}, \bibinfo {author} {\bibfnamefont
  {K.}~\bibnamefont {Lejaeghere}}, \bibinfo {author} {\bibfnamefont {M.~A.~L.}\
  \bibnamefont {Marques}}, \bibinfo {author} {\bibfnamefont {A.}~\bibnamefont
  {Martin}}, \bibinfo {author} {\bibfnamefont {C.}~\bibnamefont {Martins}},
  \bibinfo {author} {\bibfnamefont {H.~P.~C.}\ \bibnamefont {Miranda}},
  \bibinfo {author} {\bibfnamefont {F.}~\bibnamefont {Naccarato}}, \bibinfo
  {author} {\bibfnamefont {K.}~\bibnamefont {Persson}}, \bibinfo {author}
  {\bibfnamefont {G.}~\bibnamefont {Petretto}}, \bibinfo {author}
  {\bibfnamefont {V.}~\bibnamefont {Planes}}, \bibinfo {author} {\bibfnamefont
  {Y.}~\bibnamefont {Pouillon}}, \bibinfo {author} {\bibfnamefont
  {S.}~\bibnamefont {Prokhorenko}}, \bibinfo {author} {\bibfnamefont
  {F.}~\bibnamefont {Ricci}}, \bibinfo {author} {\bibfnamefont {G.-M.}\
  \bibnamefont {Rignanese}}, \bibinfo {author} {\bibfnamefont {A.~H.}\
  \bibnamefont {Romero}}, \bibinfo {author} {\bibfnamefont {M.~M.}\
  \bibnamefont {Schmitt}}, \bibinfo {author} {\bibfnamefont {M.}~\bibnamefont
  {Torrent}}, \bibinfo {author} {\bibfnamefont {M.~J.}\ \bibnamefont {van
  Setten}}, \bibinfo {author} {\bibfnamefont {B.~Van}\ \bibnamefont {Troeye}},
  \bibinfo {author} {\bibfnamefont {M.~J.}\ \bibnamefont {Verstraete}},
  \bibinfo {author} {\bibfnamefont {G.}~\bibnamefont {Z\'erah}}, \ and\
  \bibinfo {author} {\bibfnamefont {J.~W.}\ \bibnamefont {Zwanziger}}}
  (\bibinfo {year} {2020}),\ \bibfield  {title} {\enquote {\bibinfo {title}
  {The {ABINIT} project: Impact, environment and recent developments},}\
  }\href@noop {} {\bibfield  {journal} {\bibinfo  {journal} {Comput. Phys.
  Commun.}\ }\textbf {\bibinfo {volume} {248}},\ \bibinfo {pages}
  {107042}}\BibitemShut {NoStop}%
\bibitem [{\citenamefont {Gonze}\ and\ \citenamefont {Lee}(1997)}]{gonze_1997}%
  \BibitemOpen
  \bibfield  {author} {\bibinfo {author} {\bibnamefont {Gonze}, \bibfnamefont
  {X}}, \ and\ \bibinfo {author} {\bibfnamefont {C.}~\bibnamefont {Lee}}}
  (\bibinfo {year} {1997}),\ \bibfield  {title} {\enquote {\bibinfo {title}
  {Dynamical matrices, {Born} effective charges, dielectric permittivity
  tensors, and interatomic force constants from density-functional perturbation
  theory},}\ }\href@noop {} {\bibfield  {journal} {\bibinfo  {journal} {Phys.
  Rev. B}\ }\textbf {\bibinfo {volume} {55}},\ \bibinfo {pages}
  {10355--10368}}\BibitemShut {NoStop}%
\bibitem [{\citenamefont {Griffiths}(1999)}]{griffiths-book}%
  \BibitemOpen
  \bibfield  {author} {\bibinfo {author} {\bibnamefont {Griffiths},
  \bibfnamefont {D~J}}} (\bibinfo {year} {1999}),\ \href@noop {} {\emph
  {\bibinfo {title} {Introduction to Electrodynamics}}}\ (\bibinfo  {publisher}
  {Prentice Hall})\BibitemShut {NoStop}%
\bibitem [{\citenamefont {de~Groot}\ and\ \citenamefont
  {Vlieger}(1964)}]{degroot_1964}%
  \BibitemOpen
  \bibfield  {author} {\bibinfo {author} {\bibnamefont {de~Groot},
  \bibfnamefont {S~R}}, \ and\ \bibinfo {author} {\bibfnamefont
  {J.}~\bibnamefont {Vlieger}}} (\bibinfo {year} {1964}),\ \bibfield  {title}
  {\enquote {\bibinfo {title} {On the derivation of {Maxwell’}s equations},}\
  }\href@noop {} {\bibfield  {journal} {\bibinfo  {journal} {Il Nuovo Cimento}\
  }\textbf {\bibinfo {volume} {33}},\ \bibinfo {pages}
  {1225--1227}}\BibitemShut {NoStop}%
\bibitem [{\citenamefont {Grundmann}(2016)}]{grundmann_2016}%
  \BibitemOpen
  \bibfield  {author} {\bibinfo {author} {\bibnamefont {Grundmann},
  \bibfnamefont {M}}} (\bibinfo {year} {2016}),\ \href@noop {} {\emph {\bibinfo
  {title} {The Physics of Semiconductors - An Introduction Including
  Nanophysics and Applications}}},\ \bibinfo {edition} {3rd}\ ed.\ (\bibinfo
  {publisher} {Springer-Verlag Berlin Heidelberg})\BibitemShut {NoStop}%
\bibitem [{\citenamefont {Gu}\ \emph {et~al.}(2021)\citenamefont {Gu},
  \citenamefont {Murray},\ and\ \citenamefont {Tangney}}]{gu_murray_tangney}%
  \BibitemOpen
  \bibfield  {author} {\bibinfo {author} {\bibnamefont {Gu}, \bibfnamefont
  {F}}, \bibinfo {author} {\bibfnamefont {\'E.}\ \bibnamefont {Murray}}, \ and\
  \bibinfo {author} {\bibfnamefont {P.}~\bibnamefont {Tangney}}} (\bibinfo
  {year} {2021}),\ \bibfield  {title} {\enquote {\bibinfo {title}
  {Carrier-mediated control over the soft mode and ferroelectricity in
  {${\mathrm{BaTiO}_{3}}$}},}\ }\href@noop {} {\bibfield  {journal} {\bibinfo
  {journal} {Phys. Rev. Mater.}\ }\textbf {\bibinfo {volume} {5}},\ \bibinfo
  {pages} {034414}}\BibitemShut {NoStop}%
\bibitem [{\citenamefont {Heaviside}(1893)}]{heaviside-book}%
  \BibitemOpen
  \bibfield  {author} {\bibinfo {author} {\bibnamefont {Heaviside},
  \bibfnamefont {O}}} (\bibinfo {year} {1893}),\ \href@noop {} {\emph {\bibinfo
  {title} {Electromagnetic theory}}},\ Vol.~\bibinfo {volume} {I}\ (\bibinfo
  {publisher} {The Electrician Publishing})\BibitemShut {NoStop}%
\bibitem [{\citenamefont {Heil}(2018)}]{heil_2018}%
  \BibitemOpen
  \bibfield  {author} {\bibinfo {author} {\bibnamefont {Heil}, \bibfnamefont
  {C}}} (\bibinfo {year} {2018}),\ \href@noop {} {\emph {\bibinfo {title}
  {Metrics, Norms, Inner Products, and Operator Theory}}},\ Applied and
  numerical harmonic analysis\ (\bibinfo  {publisher} {Springer International
  Publishing})\BibitemShut {NoStop}%
\bibitem [{\citenamefont {Helbig}\ \emph {et~al.}(2010)\citenamefont {Helbig},
  \citenamefont {Tokatly},\ and\ \citenamefont {Rubio}}]{helbig_2010}%
  \BibitemOpen
  \bibfield  {author} {\bibinfo {author} {\bibnamefont {Helbig}, \bibfnamefont
  {N}}, \bibinfo {author} {\bibfnamefont {I.~V.}\ \bibnamefont {Tokatly}}, \
  and\ \bibinfo {author} {\bibfnamefont {A.}~\bibnamefont {Rubio}}} (\bibinfo
  {year} {2010}),\ \bibfield  {title} {\enquote {\bibinfo {title} {Physical
  meaning of the natural orbitals: Analysis of exactly solvable models},}\
  }\href {\doibase 10.1103/PhysRevA.81.022504} {\bibfield  {journal} {\bibinfo
  {journal} {Phys. Rev. A}\ }\textbf {\bibinfo {volume} {81}},\ \bibinfo
  {pages} {022504}}\BibitemShut {NoStop}%
\bibitem [{\citenamefont {Hohenberg}\ and\ \citenamefont
  {Kohn}(1964)}]{hohenberg_kohn}%
  \BibitemOpen
  \bibfield  {author} {\bibinfo {author} {\bibnamefont {Hohenberg},
  \bibfnamefont {P}}, \ and\ \bibinfo {author} {\bibfnamefont {W.}~\bibnamefont
  {Kohn}}} (\bibinfo {year} {1964}),\ \bibfield  {title} {\enquote {\bibinfo
  {title} {Inhomogeneous electron gas},}\ }\href@noop {} {\bibfield  {journal}
  {\bibinfo  {journal} {Phys. Rev.}\ }\textbf {\bibinfo {volume} {136}},\
  \bibinfo {pages} {B864--B871}}\BibitemShut {NoStop}%
\bibitem [{\citenamefont {H{\"o}rmander}(2015)}]{hormander}%
  \BibitemOpen
  \bibfield  {author} {\bibinfo {author} {\bibnamefont {H{\"o}rmander},
  \bibfnamefont {L}}} (\bibinfo {year} {2015}),\ \href@noop {} {\emph {\bibinfo
  {title} {The Analysis of Linear Partial Differential Operators {I:}
  Distribution Theory and {Fourier} Analysis}}},\ Classics in Mathematics\
  (\bibinfo  {publisher} {{Springer} {Berlin} {Heidelberg}})\BibitemShut
  {NoStop}%
\bibitem [{\citenamefont {H{\"o}rmann}\ \emph {et~al.}(2019)\citenamefont
  {H{\"o}rmann}, \citenamefont {Guo}, \citenamefont {Ambrosio}, \citenamefont
  {Andreussi}, \citenamefont {Pasquarello},\ and\ \citenamefont
  {Marzari}}]{mip_marzari}%
  \BibitemOpen
  \bibfield  {author} {\bibinfo {author} {\bibnamefont {H{\"o}rmann},
  \bibfnamefont {N~G}}, \bibinfo {author} {\bibfnamefont {Z.}~\bibnamefont
  {Guo}}, \bibinfo {author} {\bibfnamefont {F.}~\bibnamefont {Ambrosio}},
  \bibinfo {author} {\bibfnamefont {O.}~\bibnamefont {Andreussi}}, \bibinfo
  {author} {\bibfnamefont {A.}~\bibnamefont {Pasquarello}}, \ and\ \bibinfo
  {author} {\bibfnamefont {N.}~\bibnamefont {Marzari}}} (\bibinfo {year}
  {2019}),\ \bibfield  {title} {\enquote {\bibinfo {title} {Absolute band
  alignment at semiconductor-water interfaces using explicit and implicit
  descriptions for liquid water},}\ }\href@noop {} {\bibfield  {journal}
  {\bibinfo  {journal} {NPJ Comput. Mater.}\ }\textbf {\bibinfo {volume} {5}},\
  \bibinfo {pages} {100}}\BibitemShut {NoStop}%
\bibitem [{\citenamefont {Hund}(1926)}]{hund_1926}%
  \BibitemOpen
  \bibfield  {author} {\bibinfo {author} {\bibnamefont {Hund}, \bibfnamefont
  {F}}} (\bibinfo {year} {1926}),\ \bibfield  {title} {\enquote {\bibinfo
  {title} {Zur deutung einiger erscheinungen in den molekelspektren.}}\ }\href
  {\doibase 10.1007/BF01400155} {\bibfield  {journal} {\bibinfo  {journal} {Z.
  Physik}\ }\textbf {\bibinfo {volume} {36}},\ \bibinfo {pages}
  {657–674}}\BibitemShut {NoStop}%
\bibitem [{\citenamefont {Huse}\ \emph {et~al.}(2014)\citenamefont {Huse},
  \citenamefont {Nandkishore},\ and\ \citenamefont {Oganesyan}}]{mbloc_5}%
  \BibitemOpen
  \bibfield  {author} {\bibinfo {author} {\bibnamefont {Huse}, \bibfnamefont
  {D~A}}, \bibinfo {author} {\bibfnamefont {R.}~\bibnamefont {Nandkishore}}, \
  and\ \bibinfo {author} {\bibfnamefont {V.}~\bibnamefont {Oganesyan}}}
  (\bibinfo {year} {2014}),\ \bibfield  {title} {\enquote {\bibinfo {title}
  {Phenomenology of fully many-body-localized systems},}\ }\href {\doibase
  10.1103/PhysRevB.90.174202} {\bibfield  {journal} {\bibinfo  {journal} {Phys.
  Rev. B}\ }\textbf {\bibinfo {volume} {90}}~(\bibinfo {number} {17}),\
  10.1103/PhysRevB.90.174202}\BibitemShut {NoStop}%
\bibitem [{\citenamefont {Ibach}\ and\ \citenamefont
  {L{\"u}th}(2012)}]{ibach_and_luth}%
  \BibitemOpen
  \bibfield  {author} {\bibinfo {author} {\bibnamefont {Ibach}, \bibfnamefont
  {H}}, \ and\ \bibinfo {author} {\bibfnamefont {H.}~\bibnamefont {L{\"u}th}}}
  (\bibinfo {year} {2012}),\ \href@noop {} {\emph {\bibinfo {title}
  {{Solid-State Physics: An Introduction to Theory and Experiment}}}}\
  (\bibinfo  {publisher} {Springer Berlin Heidelberg})\BibitemShut {NoStop}%
\bibitem [{\citenamefont {Ibers}(1958)}]{Ibers_1958}%
  \BibitemOpen
  \bibfield  {author} {\bibinfo {author} {\bibnamefont {Ibers}, \bibfnamefont
  {J~A}}} (\bibinfo {year} {1958}),\ \bibfield  {title} {\enquote {\bibinfo
  {title} {{Atomic scattering amplitudes for electrons}},}\ }\href@noop {}
  {\bibfield  {journal} {\bibinfo  {journal} {Acta Crystallogr.}\ }\textbf
  {\bibinfo {volume} {11}}~(\bibinfo {number} {3}),\ \bibinfo {pages}
  {178--183}}\BibitemShut {NoStop}%
\bibitem [{\citenamefont {Iosevich}\ and\ \citenamefont
  {Liflyand}(2014)}]{decay_of_fourier}%
  \BibitemOpen
  \bibfield  {author} {\bibinfo {author} {\bibnamefont {Iosevich},
  \bibfnamefont {A}}, \ and\ \bibinfo {author} {\bibfnamefont {E.}~\bibnamefont
  {Liflyand}}} (\bibinfo {year} {2014}),\ \href@noop {} {\emph {\bibinfo
  {title} {Decay of the {Fourier} Transform: Analytic and Geometric Aspects}}}\
  (\bibinfo  {publisher} {Springer Basel})\BibitemShut {NoStop}%
\bibitem [{\citenamefont {Jackson}(1962)}]{jackson_firsted}%
  \BibitemOpen
  \bibfield  {author} {\bibinfo {author} {\bibnamefont {Jackson}, \bibfnamefont
  {J~D}}} (\bibinfo {year} {1962}),\ \href@noop {} {\emph {\bibinfo {title}
  {Classical electrodynamics}}},\ \bibinfo {edition} {1st}\ ed.\ (\bibinfo
  {publisher} {Wiley},\ \bibinfo {address} {New York, (NY)})\BibitemShut
  {NoStop}%
\bibitem [{\citenamefont {Jackson}(1975)}]{jackson-seconded}%
  \BibitemOpen
  \bibfield  {author} {\bibinfo {author} {\bibnamefont {Jackson}, \bibfnamefont
  {J~D}}} (\bibinfo {year} {1975}),\ \href@noop {} {\emph {\bibinfo {title}
  {Classical Electrodynamics}}},\ \bibinfo {edition} {2nd}\ ed.\ (\bibinfo
  {publisher} {Wiley})\BibitemShut {NoStop}%
\bibitem [{\citenamefont {Jackson}(1998)}]{jackson-book}%
  \BibitemOpen
  \bibfield  {author} {\bibinfo {author} {\bibnamefont {Jackson}, \bibfnamefont
  {J~D}}} (\bibinfo {year} {1998}),\ \href@noop {} {\emph {\bibinfo {title}
  {Classical Electrodynamics}}},\ \bibinfo {edition} {3rd}\ ed.\ (\bibinfo
  {publisher} {Wiley})\BibitemShut {NoStop}%
\bibitem [{\citenamefont {Jaynes}(1957{\natexlab{a}})}]{jaynes1}%
  \BibitemOpen
  \bibfield  {author} {\bibinfo {author} {\bibnamefont {Jaynes}, \bibfnamefont
  {E~T}}} (\bibinfo {year} {1957}{\natexlab{a}}),\ \bibfield  {title} {\enquote
  {\bibinfo {title} {Information theory and statistical mechanics},}\ }\href
  {\doibase 10.1103/PhysRev.106.620} {\bibfield  {journal} {\bibinfo  {journal}
  {Phys. Rev.}\ }\textbf {\bibinfo {volume} {106}},\ \bibinfo {pages}
  {620--630}}\BibitemShut {NoStop}%
\bibitem [{\citenamefont {Jaynes}(1957{\natexlab{b}})}]{jaynes2}%
  \BibitemOpen
  \bibfield  {author} {\bibinfo {author} {\bibnamefont {Jaynes}, \bibfnamefont
  {E~T}}} (\bibinfo {year} {1957}{\natexlab{b}}),\ \bibfield  {title} {\enquote
  {\bibinfo {title} {Information theory and statistical mechanics. {II}},}\
  }\href {\doibase 10.1103/PhysRev.108.171} {\bibfield  {journal} {\bibinfo
  {journal} {Phys. Rev.}\ }\textbf {\bibinfo {volume} {108}},\ \bibinfo {pages}
  {171--190}}\BibitemShut {NoStop}%
\bibitem [{\citenamefont {Jennison}\ and\ \citenamefont {Kunz}(1976)}]{nacl_1}%
  \BibitemOpen
  \bibfield  {author} {\bibinfo {author} {\bibnamefont {Jennison},
  \bibfnamefont {D~R}}, \ and\ \bibinfo {author} {\bibfnamefont {A.~B.}\
  \bibnamefont {Kunz}}} (\bibinfo {year} {1976}),\ \bibfield  {title} {\enquote
  {\bibinfo {title} {Electronic charge distribution and the degree of ionicity
  in crystalline {NaF}, {NaCl}, {Sr${\mathrm{F}}_{2}$}, and
  {Sr${\mathrm{Cl}}_{2}$} as found by the local-orbitals {Hartree-Fock}
  method},}\ }\href@noop {} {\bibfield  {journal} {\bibinfo  {journal} {Phys.
  Rev. B}\ }\textbf {\bibinfo {volume} {13}},\ \bibinfo {pages}
  {5597--5602}}\BibitemShut {NoStop}%
\bibitem [{\citenamefont {Jones}\ and\ \citenamefont
  {March}(1973)}]{Jones_and_March1}%
  \BibitemOpen
  \bibfield  {author} {\bibinfo {author} {\bibnamefont {Jones}, \bibfnamefont
  {W}}, \ and\ \bibinfo {author} {\bibfnamefont {N.~H.}\ \bibnamefont {March}}}
  (\bibinfo {year} {1973}),\ \href@noop {} {\emph {\bibinfo {title}
  {Theoretical solid state physics}}},\ Vol.~\bibinfo {volume} {{I}}\ (\bibinfo
   {publisher} {Dover Publications Inc.},\ \bibinfo {address} {New
  York})\BibitemShut {NoStop}%
\bibitem [{\citenamefont {Junquera}\ \emph {et~al.}(2007)\citenamefont
  {Junquera}, \citenamefont {Cohen},\ and\ \citenamefont
  {Rabe}}]{junquera_2007}%
  \BibitemOpen
  \bibfield  {author} {\bibinfo {author} {\bibnamefont {Junquera},
  \bibfnamefont {J}}, \bibinfo {author} {\bibfnamefont {M.~H.}\ \bibnamefont
  {Cohen}}, \ and\ \bibinfo {author} {\bibfnamefont {K.~M.}\ \bibnamefont
  {Rabe}}} (\bibinfo {year} {2007}),\ \bibfield  {title} {\enquote {\bibinfo
  {title} {Nanoscale smoothing and the analysis of interfacial charge and
  dipolar\ densities},}\ }\href {\doibase 10.1088/0953-8984/19/21/213203}
  {\bibfield  {journal} {\bibinfo  {journal} {J. Phys.: Condens. Matter}\
  }\textbf {\bibinfo {volume} {19}}~(\bibinfo {number} {21}),\ \bibinfo {pages}
  {213203}}\BibitemShut {NoStop}%
\bibitem [{\citenamefont {Junquera}\ and\ \citenamefont
  {Ghosez}(2003)}]{junquera_2003}%
  \BibitemOpen
  \bibfield  {author} {\bibinfo {author} {\bibnamefont {Junquera},
  \bibfnamefont {J}}, \ and\ \bibinfo {author} {\bibfnamefont {P}~\bibnamefont
  {Ghosez}}} (\bibinfo {year} {2003}),\ \bibfield  {title} {\enquote {\bibinfo
  {title} {Critical thickness for ferroelectricity in perovskite ultrathin
  films},}\ }\href {\doibase 10.1038/nature01501} {\bibfield  {journal}
  {\bibinfo  {journal} {Nature}\ }\textbf {\bibinfo {volume} {422}}~(\bibinfo
  {number} {6931}),\ \bibinfo {pages} {506--509}}\BibitemShut {NoStop}%
\bibitem [{\citenamefont {Kamenetskii}(1998)}]{kamenetskii-PRE-1998}%
  \BibitemOpen
  \bibfield  {author} {\bibinfo {author} {\bibnamefont {Kamenetskii},
  \bibfnamefont {E~O}}} (\bibinfo {year} {1998}),\ \bibfield  {title} {\enquote
  {\bibinfo {title} {Sampling theorem in macroscopic electrodynamics of crystal
  lattices},}\ }\href@noop {} {\bibfield  {journal} {\bibinfo  {journal} {Phys.
  Rev. E}\ }\textbf {\bibinfo {volume} {57}},\ \bibinfo {pages}
  {3556--3562}}\BibitemShut {NoStop}%
\bibitem [{\citenamefont {Kathmann}(2021)}]{mip_kathmann_2021}%
  \BibitemOpen
  \bibfield  {author} {\bibinfo {author} {\bibnamefont {Kathmann},
  \bibfnamefont {S~M}}} (\bibinfo {year} {2021}),\ \bibfield  {title} {\enquote
  {\bibinfo {title} {Electric fields and potentials in condensed phases},}\
  }\href@noop {} {\bibfield  {journal} {\bibinfo  {journal} {Phys. Chem. Chem.
  Phys.}\ }\textbf {\bibinfo {volume} {23}},\ \bibinfo {pages}
  {23836--23849}}\BibitemShut {NoStop}%
\bibitem [{\citenamefont {Kathmann}\ \emph {et~al.}(2011)\citenamefont
  {Kathmann}, \citenamefont {Kuo}, \citenamefont {Mundy},\ and\ \citenamefont
  {Schenter}}]{mip_mundy_2011}%
  \BibitemOpen
  \bibfield  {author} {\bibinfo {author} {\bibnamefont {Kathmann},
  \bibfnamefont {S~M}}, \bibinfo {author} {\bibfnamefont {I.-F.~W.}\
  \bibnamefont {Kuo}}, \bibinfo {author} {\bibfnamefont {C.~J.}\ \bibnamefont
  {Mundy}}, \ and\ \bibinfo {author} {\bibfnamefont {G.~K.}\ \bibnamefont
  {Schenter}}} (\bibinfo {year} {2011}),\ \bibfield  {title} {\enquote
  {\bibinfo {title} {Understanding the surface potential of water},}\
  }\href@noop {} {\bibfield  {journal} {\bibinfo  {journal} {J. Phys. Chem. B}\
  }\textbf {\bibinfo {volume} {115}}~(\bibinfo {number} {15}),\ \bibinfo
  {pages} {4369--4377}}\BibitemShut {NoStop}%
\bibitem [{\citenamefont {Kaufman}(1961)}]{kaufman_1961}%
  \BibitemOpen
  \bibfield  {author} {\bibinfo {author} {\bibnamefont {Kaufman}, \bibfnamefont
  {Allan~N}}} (\bibinfo {year} {1961}),\ \bibfield  {title} {\enquote {\bibinfo
  {title} {Definition of macroscopic electrostatic field},}\ }\href {\doibase
  10.1119/1.1937862} {\bibfield  {journal} {\bibinfo  {journal} {Am. J. Phys.}\
  }\textbf {\bibinfo {volume} {29}}~(\bibinfo {number} {9}),\ \bibinfo {pages}
  {626--630}}\BibitemShut {NoStop}%
\bibitem [{\citenamefont {Kaxiras}\ and\ \citenamefont
  {Joannopoulos}(2019)}]{kaxiras_2019}%
  \BibitemOpen
  \bibfield  {author} {\bibinfo {author} {\bibnamefont {Kaxiras}, \bibfnamefont
  {E}}, \ and\ \bibinfo {author} {\bibfnamefont {J.~D.}\ \bibnamefont
  {Joannopoulos}}} (\bibinfo {year} {2019}),\ \href@noop {} {\emph {\bibinfo
  {title} {Quantum Theory of Materials}}}\ (\bibinfo  {publisher} {Cambridge
  University Press})\BibitemShut {NoStop}%
\bibitem [{\citenamefont {Kent}\ and\ \citenamefont
  {Wittmann}(2021)}]{consciousness1}%
  \BibitemOpen
  \bibfield  {author} {\bibinfo {author} {\bibnamefont {Kent}, \bibfnamefont
  {L}}, \ and\ \bibinfo {author} {\bibfnamefont {M.}~\bibnamefont {Wittmann}}}
  (\bibinfo {year} {2021}),\ \bibfield  {title} {\enquote {\bibinfo {title}
  {{Time consciousness: the missing link in theories of consciousness}},}\
  }\href {\doibase 10.1093/nc/niab011} {\bibfield  {journal} {\bibinfo
  {journal} {Neuroscience of Consciousness}\ }\textbf {\bibinfo {volume}
  {2021}}~(\bibinfo {number} {2}),\ \bibinfo {pages} {niab011}}\BibitemShut
  {NoStop}%
\bibitem [{\citenamefont {King-Smith}\ and\ \citenamefont
  {Vanderbilt}(1993)}]{kingsmith-vanderbilt-prb-1993-1}%
  \BibitemOpen
  \bibfield  {author} {\bibinfo {author} {\bibnamefont {King-Smith},
  \bibfnamefont {R~D}}, \ and\ \bibinfo {author} {\bibfnamefont
  {D.}~\bibnamefont {Vanderbilt}}} (\bibinfo {year} {1993}),\ \bibfield
  {title} {\enquote {\bibinfo {title} {Theory of polarization of crystalline
  solids},}\ }\href@noop {} {\bibfield  {journal} {\bibinfo  {journal} {Phys.
  Rev. B}\ }\textbf {\bibinfo {volume} {47}}~(\bibinfo {number} {3}),\ \bibinfo
  {pages} {1651--1654}}\BibitemShut {NoStop}%
\bibitem [{\citenamefont {Kirkwood}(1936)}]{kirkwood_1936}%
  \BibitemOpen
  \bibfield  {author} {\bibinfo {author} {\bibnamefont {Kirkwood},
  \bibfnamefont {J~G}}} (\bibinfo {year} {1936}),\ \bibfield  {title} {\enquote
  {\bibinfo {title} {{On the Theory of Dielectric Polarization}},}\ }\href@noop
  {} {\bibfield  {journal} {\bibinfo  {journal} {J. Chem. Phys.}\ }\textbf
  {\bibinfo {volume} {4}}~(\bibinfo {number} {9}),\ \bibinfo {pages}
  {592--601}}\BibitemShut {NoStop}%
\bibitem [{\citenamefont {Kirkwood}(1940)}]{kirkwood-1940}%
  \BibitemOpen
  \bibfield  {author} {\bibinfo {author} {\bibnamefont {Kirkwood},
  \bibfnamefont {J~G}}} (\bibinfo {year} {1940}),\ \bibfield  {title} {\enquote
  {\bibinfo {title} {The local field in dielectrics},}\ }\href@noop {}
  {\bibfield  {journal} {\bibinfo  {journal} {Ann. N. Y. Acad. Sci.}\ }\textbf
  {\bibinfo {volume} {40}}~(\bibinfo {number} {5}),\ \bibinfo {pages}
  {315--320}}\BibitemShut {NoStop}%
\bibitem [{\citenamefont {Kittel}(2004)}]{kittel}%
  \BibitemOpen
  \bibfield  {author} {\bibinfo {author} {\bibnamefont {Kittel}, \bibfnamefont
  {C}}} (\bibinfo {year} {2004}),\ \href@noop {} {\emph {\bibinfo {title}
  {Introduction to Solid State Physics}}},\ \bibinfo {edition} {8th}\ ed.\
  (\bibinfo  {publisher} {Wiley})\BibitemShut {NoStop}%
\bibitem [{\citenamefont {Kohn}(1973)}]{kohn-prb-1973}%
  \BibitemOpen
  \bibfield  {author} {\bibinfo {author} {\bibnamefont {Kohn}, \bibfnamefont
  {W}}} (\bibinfo {year} {1973}),\ \bibfield  {title} {\enquote {\bibinfo
  {title} {Construction of {Wannier} functions and applications to energy
  bands},}\ }\href@noop {} {\bibfield  {journal} {\bibinfo  {journal} {Phys.
  Rev. B}\ }\textbf {\bibinfo {volume} {7}},\ \bibinfo {pages}
  {4388--4398}}\BibitemShut {NoStop}%
\bibitem [{\citenamefont {Kohn}\ and\ \citenamefont {Sham}(1965)}]{kohn_sham}%
  \BibitemOpen
  \bibfield  {author} {\bibinfo {author} {\bibnamefont {Kohn}, \bibfnamefont
  {W}}, \ and\ \bibinfo {author} {\bibfnamefont {L.~J.}\ \bibnamefont {Sham}}}
  (\bibinfo {year} {1965}),\ \bibfield  {title} {\enquote {\bibinfo {title}
  {Self-consistent equations including exchange and correlation effects},}\
  }\href@noop {} {\bibfield  {journal} {\bibinfo  {journal} {Phys. Rev.}\
  }\textbf {\bibinfo {volume} {140}},\ \bibinfo {pages}
  {A1133--A1138}}\BibitemShut {NoStop}%
\bibitem [{\citenamefont {Kvashnin}\ \emph {et~al.}(2019)\citenamefont
  {Kvashnin}, \citenamefont {Kvashnin},\ and\ \citenamefont {Oganov}}]{nacl_3}%
  \BibitemOpen
  \bibfield  {author} {\bibinfo {author} {\bibnamefont {Kvashnin},
  \bibfnamefont {A~G}}, \bibinfo {author} {\bibfnamefont {D.~G.}\ \bibnamefont
  {Kvashnin}}, \ and\ \bibinfo {author} {\bibfnamefont {A.~R.}\ \bibnamefont
  {Oganov}}} (\bibinfo {year} {2019}),\ \bibfield  {title} {\enquote {\bibinfo
  {title} {Novel unexpected reconstructions of {(100)} and {(111)} surfaces of
  {NaCl}: Theoretical prediction},}\ }\href {\doibase
  10.1038/s41598-019-50548-8} {\bibfield  {journal} {\bibinfo  {journal} {Sci.
  Rep.}\ }\textbf {\bibinfo {volume} {9}},\ \bibinfo {pages}
  {14267}}\BibitemShut {NoStop}%
\bibitem [{\citenamefont {Landauer}(1960)}]{landauer_1960}%
  \BibitemOpen
  \bibfield  {author} {\bibinfo {author} {\bibnamefont {Landauer},
  \bibfnamefont {R}}} (\bibinfo {year} {1960}),\ \bibfield  {title} {\enquote
  {\bibinfo {title} {Pyroelectric effect in the cubic {ZnS} structure},}\
  }\href@noop {} {\bibfield  {journal} {\bibinfo  {journal} {J. Chem. Phys.}\
  }\textbf {\bibinfo {volume} {32}}~(\bibinfo {number} {6}),\ \bibinfo {pages}
  {1784--1785}}\BibitemShut {NoStop}%
\bibitem [{\citenamefont {Landauer}(1981)}]{landauer_1981}%
  \BibitemOpen
  \bibfield  {author} {\bibinfo {author} {\bibnamefont {Landauer},
  \bibfnamefont {R}}} (\bibinfo {year} {1981}),\ \bibfield  {title} {\enquote
  {\bibinfo {title} {Pyroelectricity and piezoelectricity are not true volume
  effects},}\ }\href@noop {} {\bibfield  {journal} {\bibinfo  {journal} {Solid
  State Commun.}\ }\textbf {\bibinfo {volume} {40}}~(\bibinfo {number} {11}),\
  \bibinfo {pages} {971--974}}\BibitemShut {NoStop}%
\bibitem [{\citenamefont {de~Lange}\ and\ \citenamefont
  {Raab}(2006)}]{raab_2006}%
  \BibitemOpen
  \bibfield  {author} {\bibinfo {author} {\bibnamefont {de~Lange},
  \bibfnamefont {O~L}}, \ and\ \bibinfo {author} {\bibfnamefont {R.~E.}\
  \bibnamefont {Raab}}} (\bibinfo {year} {2006}),\ \bibfield  {title} {\enquote
  {\bibinfo {title} {Surprises in the multipole description of macroscopic
  electrodynamics},}\ }\href@noop {} {\bibfield  {journal} {\bibinfo  {journal}
  {Am. J. Phys.}\ }\textbf {\bibinfo {volume} {74}}~(\bibinfo {number} {4}),\
  \bibinfo {pages} {301--312}}\BibitemShut {NoStop}%
\bibitem [{\citenamefont {de~Lange}\ \emph {et~al.}(2012)\citenamefont
  {de~Lange}, \citenamefont {Raab},\ and\ \citenamefont {Welter}}]{raab_2012}%
  \BibitemOpen
  \bibfield  {author} {\bibinfo {author} {\bibnamefont {de~Lange},
  \bibfnamefont {O~L}}, \bibinfo {author} {\bibfnamefont {R.~E.}\ \bibnamefont
  {Raab}}, \ and\ \bibinfo {author} {\bibfnamefont {A.}~\bibnamefont {Welter}}}
  (\bibinfo {year} {2012}),\ \bibfield  {title} {\enquote {\bibinfo {title}
  {{On the transition from microscopic to macroscopic electrodynamics}},}\
  }\href@noop {} {\bibfield  {journal} {\bibinfo  {journal} {J. Math. Phys.}\
  }\textbf {\bibinfo {volume} {53}}~(\bibinfo {number} {1}),\ \bibinfo {pages}
  {013513}}\BibitemShut {NoStop}%
\bibitem [{\citenamefont {Langmuir}(1919)}]{langmuir_covalency}%
  \BibitemOpen
  \bibfield  {author} {\bibinfo {author} {\bibnamefont {Langmuir},
  \bibfnamefont {I}}} (\bibinfo {year} {1919}),\ \bibfield  {title} {\enquote
  {\bibinfo {title} {The arrangement of electrons in atoms and molecules.}}\
  }\href@noop {} {\bibfield  {journal} {\bibinfo  {journal} {J. Am. Chem.
  Soc.}\ }\textbf {\bibinfo {volume} {41}}~(\bibinfo {number} {6}),\ \bibinfo
  {pages} {868--934}}\BibitemShut {NoStop}%
\bibitem [{\citenamefont {Larmor}(1921)}]{larmor_1921}%
  \BibitemOpen
  \bibfield  {author} {\bibinfo {author} {\bibnamefont {Larmor}, \bibfnamefont
  {J}}} (\bibinfo {year} {1921}),\ \bibfield  {title} {\enquote {\bibinfo
  {title} {On electro-crystalline properties as conditioned by atomic
  lattices},}\ }\href@noop {} {\bibfield  {journal} {\bibinfo  {journal} {Proc.
  R. Soc. Lond. A}\ }\textbf {\bibinfo {volume} {99}},\ \bibinfo {pages}
  {1--10}}\BibitemShut {NoStop}%
\bibitem [{\citenamefont {van Leeuwen}(2003)}]{vanleeuwen_2003}%
  \BibitemOpen
  \bibfield  {author} {\bibinfo {author} {\bibnamefont {van Leeuwen},
  \bibfnamefont {R}}} (\bibinfo {year} {2003}),\ \enquote {\bibinfo {title}
  {Density functional approach to the many-body problem: Key concepts and exact
  functionals},}\ \ (\bibinfo  {publisher} {Academic Press})\ pp.\ \bibinfo
  {pages} {25--94}\BibitemShut {NoStop}%
\bibitem [{\citenamefont {Lennard-Jones}(1929)}]{lennard-jones_1929}%
  \BibitemOpen
  \bibfield  {author} {\bibinfo {author} {\bibnamefont {Lennard-Jones},
  \bibfnamefont {J~E}}} (\bibinfo {year} {1929}),\ \bibfield  {title} {\enquote
  {\bibinfo {title} {The electronic structure of some diatomic molecules},}\
  }\href {\doibase 10.1039/TF9292500668} {\bibfield  {journal} {\bibinfo
  {journal} {Trans. Faraday Soc.}\ }\textbf {\bibinfo {volume} {25}},\ \bibinfo
  {pages} {668--686}}\BibitemShut {NoStop}%
\bibitem [{\citenamefont {Leung}(2010)}]{mip_leung_2010}%
  \BibitemOpen
  \bibfield  {author} {\bibinfo {author} {\bibnamefont {Leung}, \bibfnamefont
  {K}}} (\bibinfo {year} {2010}),\ \bibfield  {title} {\enquote {\bibinfo
  {title} {Surface potential at the air-water interface computed using density
  functional theory},}\ }\href@noop {} {\bibfield  {journal} {\bibinfo
  {journal} {J. Phys. Chem. Lett.}\ }\textbf {\bibinfo {volume} {1}}~(\bibinfo
  {number} {2}),\ \bibinfo {pages} {496--499}}\BibitemShut {NoStop}%
\bibitem [{\citenamefont {Lewis}(1916)}]{lewis_1916}%
  \BibitemOpen
  \bibfield  {author} {\bibinfo {author} {\bibnamefont {Lewis}, \bibfnamefont
  {G~N}}} (\bibinfo {year} {1916}),\ \bibfield  {title} {\enquote {\bibinfo
  {title} {The atom and the molecule},}\ }\href@noop {} {\bibfield  {journal}
  {\bibinfo  {journal} {J. Am. Chem. Soc.}\ }\textbf {\bibinfo {volume}
  {38}}~(\bibinfo {number} {4}),\ \bibinfo {pages} {762--785}}\BibitemShut
  {NoStop}%
\bibitem [{\citenamefont {Li}\ \emph {et~al.}(2007)\citenamefont {Li},
  \citenamefont {Michaelides},\ and\ \citenamefont {Scheffler}}]{nacl_4}%
  \BibitemOpen
  \bibfield  {author} {\bibinfo {author} {\bibnamefont {Li}, \bibfnamefont
  {B}}, \bibinfo {author} {\bibfnamefont {A.}~\bibnamefont {Michaelides}}, \
  and\ \bibinfo {author} {\bibfnamefont {M.}~\bibnamefont {Scheffler}}}
  (\bibinfo {year} {2007}),\ \bibfield  {title} {\enquote {\bibinfo {title}
  {Density functional theory study of flat and stepped {NaCl(001)}},}\
  }\href@noop {} {\bibfield  {journal} {\bibinfo  {journal} {Phys. Rev. B}\
  }\textbf {\bibinfo {volume} {76}},\ \bibinfo {pages} {075401}}\BibitemShut
  {NoStop}%
\bibitem [{\citenamefont {Littlewood}(1980)}]{littlewood-1980}%
  \BibitemOpen
  \bibfield  {author} {\bibinfo {author} {\bibnamefont {Littlewood},
  \bibfnamefont {P~B}}} (\bibinfo {year} {1980}),\ \bibfield  {title} {\enquote
  {\bibinfo {title} {On the calculation of the macroscopic polarisation induced
  by an optic phonon},}\ }\href@noop {} {\bibfield  {journal} {\bibinfo
  {journal} {J. Phys. C: Solid State Phys.}\ }\textbf {\bibinfo {volume}
  {13}}~(\bibinfo {number} {26}),\ \bibinfo {pages} {4893}}\BibitemShut
  {NoStop}%
\bibitem [{\citenamefont {Littlewood}\ and\ \citenamefont
  {Heine}(1979)}]{littlewood_1979}%
  \BibitemOpen
  \bibfield  {author} {\bibinfo {author} {\bibnamefont {Littlewood},
  \bibfnamefont {P~B}}, \ and\ \bibinfo {author} {\bibfnamefont
  {V.}~\bibnamefont {Heine}}} (\bibinfo {year} {1979}),\ \bibfield  {title}
  {\enquote {\bibinfo {title} {The infrared effective charge in {IV-VI}
  compounds. {I.} a simple one-dimensional model},}\ }\href@noop {} {\bibfield
  {journal} {\bibinfo  {journal} {J. Phys. C: Solid State Phys.}\ }\textbf
  {\bibinfo {volume} {12}}~(\bibinfo {number} {21}),\ \bibinfo {pages}
  {4431}}\BibitemShut {NoStop}%
\bibitem [{\citenamefont {Lorentz}(1916)}]{lorentz}%
  \BibitemOpen
  \bibfield  {author} {\bibinfo {author} {\bibnamefont {Lorentz}, \bibfnamefont
  {H~A}}} (\bibinfo {year} {1916}),\ \href@noop {} {\emph {\bibinfo {title}
  {The Theory of electrons and its applications to the phenomena of light and
  radiant heat}}},\ \bibinfo {edition} {2nd}\ ed.\ (\bibinfo  {publisher}
  {Leipzig : B.G. Teubner ; New York : G.E. Stechert},\ \bibinfo {address} {New
  York})\BibitemShut {NoStop}%
\bibitem [{\citenamefont {Lounesto}(2001)}]{lounesto_2001}%
  \BibitemOpen
  \bibfield  {author} {\bibinfo {author} {\bibnamefont {Lounesto},
  \bibfnamefont {P}}} (\bibinfo {year} {2001}),\ \href@noop {} {\emph {\bibinfo
  {title} {Clifford Algebras and Spinors}}},\ \bibinfo {edition} {2nd}\ ed.,\
  London Mathematical Society Lecture Note Series\ (\bibinfo  {publisher}
  {Cambridge University Press})\BibitemShut {NoStop}%
\bibitem [{\citenamefont {L\"owdin}(1955)}]{lowdin_1955}%
  \BibitemOpen
  \bibfield  {author} {\bibinfo {author} {\bibnamefont {L\"owdin},
  \bibfnamefont {P-O}}} (\bibinfo {year} {1955}),\ \bibfield  {title} {\enquote
  {\bibinfo {title} {Quantum theory of many-particle systems. {I.} physical
  interpretations by means of density matrices, natural spin-orbitals, and
  convergence problems in the method of configurational interaction},}\
  }\href@noop {} {\bibfield  {journal} {\bibinfo  {journal} {Phys. Rev.}\
  }\textbf {\bibinfo {volume} {97}},\ \bibinfo {pages}
  {1474--1489}}\BibitemShut {NoStop}%
\bibitem [{\citenamefont {Lyddane}\ \emph {et~al.}(1941)\citenamefont
  {Lyddane}, \citenamefont {Sachs},\ and\ \citenamefont
  {Teller}}]{lyddane_sachs_teller}%
  \BibitemOpen
  \bibfield  {author} {\bibinfo {author} {\bibnamefont {Lyddane}, \bibfnamefont
  {R~H}}, \bibinfo {author} {\bibfnamefont {R.~G.}\ \bibnamefont {Sachs}}, \
  and\ \bibinfo {author} {\bibfnamefont {E.}~\bibnamefont {Teller}}} (\bibinfo
  {year} {1941}),\ \bibfield  {title} {\enquote {\bibinfo {title} {On the polar
  vibrations of alkali halides},}\ }\href@noop {} {\bibfield  {journal}
  {\bibinfo  {journal} {Phys. Rev.}\ }\textbf {\bibinfo {volume} {59}},\
  \bibinfo {pages} {673--676}}\BibitemShut {NoStop}%
\bibitem [{\citenamefont {Madsen}\ \emph {et~al.}(2021)\citenamefont {Madsen},
  \citenamefont {Pennycook},\ and\ \citenamefont {Susi}}]{mip_madsen_2021}%
  \BibitemOpen
  \bibfield  {author} {\bibinfo {author} {\bibnamefont {Madsen}, \bibfnamefont
  {J}}, \bibinfo {author} {\bibfnamefont {T.~J.}\ \bibnamefont {Pennycook}}, \
  and\ \bibinfo {author} {\bibfnamefont {T.}~\bibnamefont {Susi}}} (\bibinfo
  {year} {2021}),\ \bibfield  {title} {\enquote {\bibinfo {title} {{\em Ab
  initio} description of bonding for transmission electron microscopy},}\
  }\href@noop {} {\bibinfo  {journal} {Ultramicroscopy}\ ,\ \bibinfo {pages}
  {113253}}\BibitemShut {NoStop}%
\bibitem [{\citenamefont {Martin}(1974)}]{martin-prb-1974}%
  \BibitemOpen
\bibfield  {journal} {  }\bibfield  {author} {\bibinfo {author} {\bibnamefont
  {Martin}, \bibfnamefont {R~M}}} (\bibinfo {year} {1974}),\ \bibfield  {title}
  {\enquote {\bibinfo {title} {Comment on calculations of electric polarization
  in crystals},}\ }\href@noop {} {\bibfield  {journal} {\bibinfo  {journal}
  {Phys. Rev. B}\ }\textbf {\bibinfo {volume} {9}},\ \bibinfo {pages}
  {1998--1999}}\BibitemShut {NoStop}%
\bibitem [{\citenamefont {Marzari}\ \emph {et~al.}(2012)\citenamefont
  {Marzari}, \citenamefont {Mostofi}, \citenamefont {Yates}, \citenamefont
  {Souza},\ and\ \citenamefont {Vanderbilt}}]{mlwf_rmp}%
  \BibitemOpen
  \bibfield  {author} {\bibinfo {author} {\bibnamefont {Marzari}, \bibfnamefont
  {N}}, \bibinfo {author} {\bibfnamefont {A.~A.}\ \bibnamefont {Mostofi}},
  \bibinfo {author} {\bibfnamefont {J.~R.}\ \bibnamefont {Yates}}, \bibinfo
  {author} {\bibfnamefont {I.}~\bibnamefont {Souza}}, \ and\ \bibinfo {author}
  {\bibfnamefont {D.}~\bibnamefont {Vanderbilt}}} (\bibinfo {year} {2012}),\
  \bibfield  {title} {\enquote {\bibinfo {title} {Maximally localized {Wannier}
  functions: Theory and applications},}\ }\href@noop {} {\bibfield  {journal}
  {\bibinfo  {journal} {Rev. Mod. Phys.}\ }\textbf {\bibinfo {volume} {84}},\
  \bibinfo {pages} {1419--1475}}\BibitemShut {NoStop}%
\bibitem [{\citenamefont {Marzari}\ and\ \citenamefont
  {Vanderbilt}(1997)}]{marzari_mlwf}%
  \BibitemOpen
  \bibfield  {author} {\bibinfo {author} {\bibnamefont {Marzari}, \bibfnamefont
  {N}}, \ and\ \bibinfo {author} {\bibfnamefont {D.}~\bibnamefont
  {Vanderbilt}}} (\bibinfo {year} {1997}),\ \bibfield  {title} {\enquote
  {\bibinfo {title} {Maximally localized generalized {Wannier} functions for
  composite energy bands},}\ }\href@noop {} {\bibfield  {journal} {\bibinfo
  {journal} {Phys. Rev. B}\ }\textbf {\bibinfo {volume} {56}},\ \bibinfo
  {pages} {12847--12865}}\BibitemShut {NoStop}%
\bibitem [{\citenamefont {Maxwell}(1865)}]{maxwell-1865}%
  \BibitemOpen
  \bibfield  {author} {\bibinfo {author} {\bibnamefont {Maxwell}, \bibfnamefont
  {J~C}}} (\bibinfo {year} {1865}),\ \bibfield  {title} {\enquote {\bibinfo
  {title} {A dynamical theory of the electromagnetic field},}\ }\href@noop {}
  {\bibfield  {journal} {\bibinfo  {journal} {Phil. Trans. R. Soc. Lond.}\
  }\textbf {\bibinfo {volume} {155}},\ \bibinfo {pages} {459--512}}\BibitemShut
  {NoStop}%
\bibitem [{\citenamefont {Maxwell}(1873)}]{maxwell-book1}%
  \BibitemOpen
  \bibfield  {author} {\bibinfo {author} {\bibnamefont {Maxwell}, \bibfnamefont
  {J~C}}} (\bibinfo {year} {1873}),\ \href@noop {} {\emph {\bibinfo {title} {A
  treatise on electricity and magnetism}}},\ \bibinfo {edition} {3rd}\ ed.,\
  Vol.~\bibinfo {volume} {I}\ (\bibinfo  {publisher} {Oxford, Clarendon
  Press})\BibitemShut {NoStop}%
\bibitem [{\citenamefont {Maxwell}(1892)}]{maxwell-book2}%
  \BibitemOpen
  \bibfield  {author} {\bibinfo {author} {\bibnamefont {Maxwell}, \bibfnamefont
  {J~C}}} (\bibinfo {year} {1892}),\ \href@noop {} {\emph {\bibinfo {title} {A
  treatise on electricity and magnetism}}},\ \bibinfo {edition} {3rd}\ ed.,\
  Vol.~\bibinfo {volume} {II}\ (\bibinfo  {publisher} {Oxford, Clarendon
  Press})\BibitemShut {NoStop}%
\bibitem [{\citenamefont {Mazur}(1957)}]{mazur_1957}%
  \BibitemOpen
  \bibfield  {author} {\bibinfo {author} {\bibnamefont {Mazur}, \bibfnamefont
  {P}}} (\bibinfo {year} {1957}),\ \enquote {\bibinfo {title} {On statistical
  mechanics and electromagnetic properties of matter},}\ in\ \href@noop {}
  {\emph {\bibinfo {booktitle} {Adv. Chem. Phys.}}}\ (\bibinfo  {publisher}
  {John Wiley \& Sons, Ltd})\ pp.\ \bibinfo {pages} {309--360}\BibitemShut
  {NoStop}%
\bibitem [{\citenamefont {Mazur}\ and\ \citenamefont
  {Nijboer}(1953)}]{mazur_1953}%
  \BibitemOpen
  \bibfield  {author} {\bibinfo {author} {\bibnamefont {Mazur}, \bibfnamefont
  {P}}, \ and\ \bibinfo {author} {\bibfnamefont {B.~R.~A.}\ \bibnamefont
  {Nijboer}}} (\bibinfo {year} {1953}),\ \bibfield  {title} {\enquote {\bibinfo
  {title} {On the statistical mechanics of matter in an electromagnetic field.
  {I:} derivation of the {Maxwell} equations from electron theory},}\
  }\href@noop {} {\bibfield  {journal} {\bibinfo  {journal} {Physica}\ }\textbf
  {\bibinfo {volume} {19}}~(\bibinfo {number} {1}),\ \bibinfo {pages}
  {971--986}}\BibitemShut {NoStop}%
\bibitem [{\citenamefont {McQuarrie}\ \emph {et~al.}(2011)\citenamefont
  {McQuarrie}, \citenamefont {Rock},\ and\ \citenamefont
  {Gallogly}}]{chemistry1}%
  \BibitemOpen
  \bibfield  {author} {\bibinfo {author} {\bibnamefont {McQuarrie},
  \bibfnamefont {D~A}}, \bibinfo {author} {\bibfnamefont {P.~A.}\ \bibnamefont
  {Rock}}, \ and\ \bibinfo {author} {\bibfnamefont {E.~B.}\ \bibnamefont
  {Gallogly}}} (\bibinfo {year} {2011}),\ \href@noop {} {\emph {\bibinfo
  {title} {{General Chemistry}}}},\ \bibinfo {edition} {4th}\ ed.\ (\bibinfo
  {publisher} {Royal Society of Chemistry})\BibitemShut {NoStop}%
\bibitem [{\citenamefont {McWeeny}(1960)}]{mcweeny_1960}%
  \BibitemOpen
  \bibfield  {author} {\bibinfo {author} {\bibnamefont {McWeeny}, \bibfnamefont
  {R}}} (\bibinfo {year} {1960}),\ \bibfield  {title} {\enquote {\bibinfo
  {title} {Some recent advances in density matrix theory},}\ }\href {\doibase
  10.1103/RevModPhys.32.335} {\bibfield  {journal} {\bibinfo  {journal} {Rev.
  Mod. Phys.}\ }\textbf {\bibinfo {volume} {32}},\ \bibinfo {pages}
  {335--369}}\BibitemShut {NoStop}%
\bibitem [{\citenamefont {Messiah}(1961)}]{messiah_1961}%
  \BibitemOpen
  \bibfield  {author} {\bibinfo {author} {\bibnamefont {Messiah}, \bibfnamefont
  {A}}} (\bibinfo {year} {1961}),\ \href@noop {} {\emph {\bibinfo {title}
  {Quantum Mechanics Volume I}}}\ (\bibinfo  {publisher} {Elsevier Science
  B.V.})\BibitemShut {NoStop}%
\bibitem [{\citenamefont {Miller}(1981)}]{miller_1981}%
  \BibitemOpen
  \bibfield  {author} {\bibinfo {author} {\bibnamefont {Miller}, \bibfnamefont
  {A~I}}} (\bibinfo {year} {1981}),\ \href@noop {} {\emph {\bibinfo {title}
  {{Albert Einstein's Special Theory of Relativity: Emergence (1905) and Early
  Interpretation, 1905-1911}}}},\ Advanced book program\ (\bibinfo  {publisher}
  {{Addison-Wesley Publishing Company}})\BibitemShut {NoStop}%
\bibitem [{\citenamefont {Miyake}(1940)}]{miyake-1940}%
  \BibitemOpen
  \bibfield  {author} {\bibinfo {author} {\bibnamefont {Miyake}, \bibfnamefont
  {S}}} (\bibinfo {year} {1940}),\ \bibfield  {title} {\enquote {\bibinfo
  {title} {On the mean inner potential of crystals},}\ }in\ \href@noop {}
  {\emph {\bibinfo {booktitle} {Proceedings of the Physico-Mathematical Society
  of Japan}}},\ \bibinfo {series} {3rd}, Vol.~\bibinfo {volume} {22},\ pp.\
  \bibinfo {pages} {666--676}\BibitemShut {NoStop}%
\bibitem [{\citenamefont {Mulliken}(1928)}]{mulliken_1928}%
  \BibitemOpen
  \bibfield  {author} {\bibinfo {author} {\bibnamefont {Mulliken},
  \bibfnamefont {R~S}}} (\bibinfo {year} {1928}),\ \bibfield  {title} {\enquote
  {\bibinfo {title} {The assignment of quantum numbers for electrons in
  molecules. {I}},}\ }\href {\doibase 10.1103/PhysRev.32.186} {\bibfield
  {journal} {\bibinfo  {journal} {Phys. Rev.}\ }\textbf {\bibinfo {volume}
  {32}},\ \bibinfo {pages} {186--222}}\BibitemShut {NoStop}%
\bibitem [{\citenamefont {Nandkishore}\ and\ \citenamefont
  {Huse}(2015)}]{mbloc_2}%
  \BibitemOpen
  \bibfield  {author} {\bibinfo {author} {\bibnamefont {Nandkishore},
  \bibfnamefont {R}}, \ and\ \bibinfo {author} {\bibfnamefont {D.~A.}\
  \bibnamefont {Huse}}} (\bibinfo {year} {2015}),\ \bibfield  {title} {\enquote
  {\bibinfo {title} {Many-body localization and thermalization in quantum
  statistical mechanics},}\ }in\ \href {\doibase
  10.1146/annurev-conmatphys-031214-014726} {\emph {\bibinfo {booktitle} {Annu.
  Rev. Condens. Matter Phys.}}},\ \bibinfo {series} {Annu. Rev. Condens. Matter
  Phys.}, Vol.~\bibinfo {volume} {6},\ \bibinfo {editor} {edited by\ \bibinfo
  {editor} {\bibfnamefont {J.~S.}\ \bibnamefont {Langer}}},\ pp.\ \bibinfo
  {pages} {15--38}\BibitemShut {NoStop}%
\bibitem [{\citenamefont {von Neumann}(1955)}]{von_neumann_old}%
  \BibitemOpen
  \bibfield  {author} {\bibinfo {author} {\bibnamefont {von Neumann},
  \bibfnamefont {J}}} (\bibinfo {year} {1955}),\ \href@noop {} {\emph {\bibinfo
  {title} {Mathematical Foundations of Quantum Mechanics}}},\ Goldstine Printed
  Materials\ (\bibinfo  {publisher} {Princeton University Press})\BibitemShut
  {NoStop}%
\bibitem [{\citenamefont {von Neumann}\ \emph {et~al.}(2018)\citenamefont {von
  Neumann}, \citenamefont {Beyer},\ and\ \citenamefont
  {Wheeler}}]{von_neumann_new}%
  \BibitemOpen
  \bibfield  {author} {\bibinfo {author} {\bibnamefont {von Neumann},
  \bibfnamefont {J}}, \bibinfo {author} {\bibfnamefont {R.T.}\ \bibnamefont
  {Beyer}}, \ and\ \bibinfo {author} {\bibfnamefont {N.A.}\ \bibnamefont
  {Wheeler}}} (\bibinfo {year} {2018}),\ \href@noop {} {\emph {\bibinfo {title}
  {Mathematical Foundations of Quantum Mechanics: New Edition}}},\ Princeton
  Landmarks in Mathematics and Physics\ (\bibinfo  {publisher} {Princeton
  University Press})\BibitemShut {NoStop}%
\bibitem [{\citenamefont {Noguera}(2000)}]{noguera-2000}%
  \BibitemOpen
  \bibfield  {author} {\bibinfo {author} {\bibnamefont {Noguera}, \bibfnamefont
  {C}}} (\bibinfo {year} {2000}),\ \bibfield  {title} {\enquote {\bibinfo
  {title} {Polar oxide surfaces},}\ }\href@noop {} {\bibfield  {journal}
  {\bibinfo  {journal} {J. Phys.: Condens. Matter}\ }\textbf {\bibinfo {volume}
  {12}},\ \bibinfo {pages} {R367--R410}}\BibitemShut {NoStop}%
\bibitem [{\citenamefont {Noguera}\ and\ \citenamefont
  {Goniakowski}(2013)}]{noguera-chemrev-2013}%
  \BibitemOpen
  \bibfield  {author} {\bibinfo {author} {\bibnamefont {Noguera}, \bibfnamefont
  {C}}, \ and\ \bibinfo {author} {\bibfnamefont {J.}~\bibnamefont
  {Goniakowski}}} (\bibinfo {year} {2013}),\ \bibfield  {title} {\enquote
  {\bibinfo {title} {Polarity in oxide {nano-objects}},}\ }\href@noop {}
  {\bibfield  {journal} {\bibinfo  {journal} {Chem. Rev.}\ }\textbf {\bibinfo
  {volume} {113}}~(\bibinfo {number} {6}),\ \bibinfo {pages}
  {4073--4105}}\BibitemShut {NoStop}%
\bibitem [{\citenamefont {Northoff}\ and\ \citenamefont
  {Lamme}(2020)}]{consciousness3}%
  \BibitemOpen
  \bibfield  {author} {\bibinfo {author} {\bibnamefont {Northoff},
  \bibfnamefont {G}}, \ and\ \bibinfo {author} {\bibfnamefont {V.}~\bibnamefont
  {Lamme}}} (\bibinfo {year} {2020}),\ \bibfield  {title} {\enquote {\bibinfo
  {title} {Neural signs and mechanisms of consciousness: Is there a potential
  convergence of theories of consciousness in sight?}}\ }\href@noop {}
  {\bibfield  {journal} {\bibinfo  {journal} {Neuroscience \& Biobehavioral
  Reviews}\ }\textbf {\bibinfo {volume} {118}},\ \bibinfo {pages}
  {568--587}}\BibitemShut {NoStop}%
\bibitem [{\citenamefont {Oganesyan}\ and\ \citenamefont
  {Huse}(2007)}]{mbloc_3}%
  \BibitemOpen
  \bibfield  {author} {\bibinfo {author} {\bibnamefont {Oganesyan},
  \bibfnamefont {V}}, \ and\ \bibinfo {author} {\bibfnamefont {D.~A.}\
  \bibnamefont {Huse}}} (\bibinfo {year} {2007}),\ \bibfield  {title} {\enquote
  {\bibinfo {title} {Localization of interacting fermions at high
  temperature},}\ }\href {\doibase 10.1103/PhysRevB.75.155111} {\bibfield
  {journal} {\bibinfo  {journal} {Phys. Rev. B}\ }\textbf {\bibinfo {volume}
  {75}}~(\bibinfo {number} {15}),\ 10.1103/PhysRevB.75.155111}\BibitemShut
  {NoStop}%
\bibitem [{\citenamefont {Ott}(1979)}]{ott_1979}%
  \BibitemOpen
  \bibfield  {author} {\bibinfo {author} {\bibnamefont {Ott}, \bibfnamefont
  {E}}} (\bibinfo {year} {1979}),\ \bibfield  {title} {\enquote {\bibinfo
  {title} {Goodness of ergodic adiabatic invariants},}\ }\href {\doibase
  10.1103/PhysRevLett.42.1628} {\bibfield  {journal} {\bibinfo  {journal}
  {Phys. Rev. Lett.}\ }\textbf {\bibinfo {volume} {42}},\ \bibinfo {pages}
  {1628--1631}}\BibitemShut {NoStop}%
\bibitem [{\citenamefont {Owen}(2014)}]{chemistry_for_IB}%
  \BibitemOpen
  \bibfield  {author} {\bibinfo {author} {\bibnamefont {Owen}, \bibfnamefont
  {S}}} (\bibinfo {year} {2014}),\ \href@noop {} {\emph {\bibinfo {title}
  {{Chemistry for the {IB} diploma}}}},\ \bibinfo {edition} {2nd}\ ed.\
  (\bibinfo  {publisher} {Cambridge University Press})\BibitemShut {NoStop}%
\bibitem [{\citenamefont {Pal}\ and\ \citenamefont {Huse}(2010)}]{mbloc_4}%
  \BibitemOpen
  \bibfield  {author} {\bibinfo {author} {\bibnamefont {Pal}, \bibfnamefont
  {A}}, \ and\ \bibinfo {author} {\bibfnamefont {D.~A.}\ \bibnamefont {Huse}}}
  (\bibinfo {year} {2010}),\ \bibfield  {title} {\enquote {\bibinfo {title}
  {Many-body localization phase transition},}\ }\href {\doibase
  10.1103/PhysRevB.82.174411} {\bibfield  {journal} {\bibinfo  {journal} {Phys.
  Rev. B}\ }\textbf {\bibinfo {volume} {82}}~(\bibinfo {number} {17}),\
  10.1103/PhysRevB.82.174411}\BibitemShut {NoStop}%
\bibitem [{\citenamefont {Pasquarello}\ \emph {et~al.}(1992)\citenamefont
  {Pasquarello}, \citenamefont {Laasonen}, \citenamefont {Car}, \citenamefont
  {Lee},\ and\ \citenamefont {Vanderbilt}}]{pasquarello_1992}%
  \BibitemOpen
  \bibfield  {author} {\bibinfo {author} {\bibnamefont {Pasquarello},
  \bibfnamefont {A}}, \bibinfo {author} {\bibfnamefont {K.}~\bibnamefont
  {Laasonen}}, \bibinfo {author} {\bibfnamefont {R.}~\bibnamefont {Car}},
  \bibinfo {author} {\bibfnamefont {C.}~\bibnamefont {Lee}}, \ and\ \bibinfo
  {author} {\bibfnamefont {D.}~\bibnamefont {Vanderbilt}}} (\bibinfo {year}
  {1992}),\ \bibfield  {title} {\enquote {\bibinfo {title} {{\em Ab initio}
  molecular dynamics for d-electron systems: {Liquid} copper at $1500$ k},}\
  }\href {\doibase 10.1103/PhysRevLett.69.1982} {\bibfield  {journal} {\bibinfo
   {journal} {Phys. Rev. Lett.}\ }\textbf {\bibinfo {volume} {69}},\ \bibinfo
  {pages} {1982--1985}}\BibitemShut {NoStop}%
\bibitem [{\citenamefont {Pauling}(1926)}]{pauling_1926}%
  \BibitemOpen
  \bibfield  {author} {\bibinfo {author} {\bibnamefont {Pauling}, \bibfnamefont
  {L}}} (\bibinfo {year} {1926}),\ \bibfield  {title} {\enquote {\bibinfo
  {title} {The dynamic model of the chemical bond and its application to the
  structure of benzene},}\ }\href {\doibase 10.1021/ja01416a003} {\bibfield
  {journal} {\bibinfo  {journal} {J. Am. Chem. Soc.}\ }\textbf {\bibinfo
  {volume} {48}}~(\bibinfo {number} {5}),\ \bibinfo {pages}
  {1132--1143}}\BibitemShut {NoStop}%
\bibitem [{\citenamefont {Pauling}(1928)}]{pauling_1928}%
  \BibitemOpen
  \bibfield  {author} {\bibinfo {author} {\bibnamefont {Pauling}, \bibfnamefont
  {L}}} (\bibinfo {year} {1928}),\ \bibfield  {title} {\enquote {\bibinfo
  {title} {The shared-electron chemical bond},}\ }\href {\doibase
  10.1073/pnas.14.4.359} {\bibfield  {journal} {\bibinfo  {journal} {Proc. Nat.
  Acad. Sci.}\ }\textbf {\bibinfo {volume} {14}}~(\bibinfo {number} {4}),\
  \bibinfo {pages} {359--362}}\BibitemShut {NoStop}%
\bibitem [{\citenamefont {Pauling}(1931{\natexlab{a}})}]{pauling_1931_1}%
  \BibitemOpen
  \bibfield  {author} {\bibinfo {author} {\bibnamefont {Pauling}, \bibfnamefont
  {L}}} (\bibinfo {year} {1931}{\natexlab{a}}),\ \bibfield  {title} {\enquote
  {\bibinfo {title} {Quantum mechanics and the chemical bond},}\ }\href
  {\doibase 10.1103/PhysRev.37.1185} {\bibfield  {journal} {\bibinfo  {journal}
  {Phys. Rev.}\ }\textbf {\bibinfo {volume} {37}},\ \bibinfo {pages}
  {1185--1186}}\BibitemShut {NoStop}%
\bibitem [{\citenamefont {Pauling}(1931{\natexlab{b}})}]{pauling_1931_2}%
  \BibitemOpen
  \bibfield  {author} {\bibinfo {author} {\bibnamefont {Pauling}, \bibfnamefont
  {L}}} (\bibinfo {year} {1931}{\natexlab{b}}),\ \bibfield  {title} {\enquote
  {\bibinfo {title} {The nature of the chemical bond. application of results
  obtained from the quantum mechanics and from a theory of paramagnetic
  susceptibility to the structure of molecules},}\ }\href {\doibase
  10.1021/ja01355a027} {\bibfield  {journal} {\bibinfo  {journal} {J. Am. Chem.
  Soc.}\ }\textbf {\bibinfo {volume} {53}}~(\bibinfo {number} {4}),\ \bibinfo
  {pages} {1367--1400}}\BibitemShut {NoStop}%
\bibitem [{\citenamefont {Pauling}(1931{\natexlab{c}})}]{pauling_1931_3}%
  \BibitemOpen
  \bibfield  {author} {\bibinfo {author} {\bibnamefont {Pauling}, \bibfnamefont
  {L}}} (\bibinfo {year} {1931}{\natexlab{c}}),\ \bibfield  {title} {\enquote
  {\bibinfo {title} {The nature of the chemical bond. {II}. the one-electron
  bond and the three-electron bond},}\ }\href {\doibase 10.1021/ja01360a004}
  {\bibfield  {journal} {\bibinfo  {journal} {J. Am. Chem. Soc.}\ }\textbf
  {\bibinfo {volume} {53}}~(\bibinfo {number} {9}),\ \bibinfo {pages}
  {3225--3237}}\BibitemShut {NoStop}%
\bibitem [{\citenamefont {Pauling}(1932{\natexlab{a}})}]{pauling_1932_1}%
  \BibitemOpen
  \bibfield  {author} {\bibinfo {author} {\bibnamefont {Pauling}, \bibfnamefont
  {L}}} (\bibinfo {year} {1932}{\natexlab{a}}),\ \bibfield  {title} {\enquote
  {\bibinfo {title} {The nature of the chemical bond. {III}. the transition
  from one extreme bond type to another},}\ }\href {\doibase
  10.1021/ja01342a022} {\bibfield  {journal} {\bibinfo  {journal} {J. Am. Chem.
  Soc.}\ }\textbf {\bibinfo {volume} {54}}~(\bibinfo {number} {3}),\ \bibinfo
  {pages} {988--1003}}\BibitemShut {NoStop}%
\bibitem [{\citenamefont {Pauling}(1932{\natexlab{b}})}]{pauling_1932_2}%
  \BibitemOpen
  \bibfield  {author} {\bibinfo {author} {\bibnamefont {Pauling}, \bibfnamefont
  {L}}} (\bibinfo {year} {1932}{\natexlab{b}}),\ \bibfield  {title} {\enquote
  {\bibinfo {title} {The nature of the chemical bond. {IV}. the energy of
  single bonds and the relative electronegativity of atoms},}\ }\href {\doibase
  10.1021/ja01348a011} {\bibfield  {journal} {\bibinfo  {journal} {J. Am. Chem.
  Soc.}\ }\textbf {\bibinfo {volume} {54}}~(\bibinfo {number} {9}),\ \bibinfo
  {pages} {3570--3582}}\BibitemShut {NoStop}%
\bibitem [{\citenamefont {Pauling}(1960)}]{pauling_1960}%
  \BibitemOpen
  \bibfield  {author} {\bibinfo {author} {\bibnamefont {Pauling}, \bibfnamefont
  {L}}} (\bibinfo {year} {1960}),\ \href@noop {} {\emph {\bibinfo {title} {The
  Nature of the Chemical Bond and the Structure of Molecules and Crystals: An
  Introduction to Modern Structural Chemistry}}},\ George Fisher Baker
  Non-Resident Lecture Series\ (\bibinfo  {publisher} {Cornell University
  Press})\BibitemShut {NoStop}%
\bibitem [{\citenamefont {Peckham}(1967)}]{peckham_1967}%
  \BibitemOpen
  \bibfield  {author} {\bibinfo {author} {\bibnamefont {Peckham}, \bibfnamefont
  {G}}} (\bibinfo {year} {1967}),\ \bibfield  {title} {\enquote {\bibinfo
  {title} {The phonon dispersion relation for diamond},}\ }\href {\doibase
  10.1016/0038-1098(67)90280-3} {\bibfield  {journal} {\bibinfo  {journal}
  {Solid State Commun.}\ }\textbf {\bibinfo {volume} {5}}~(\bibinfo {number}
  {4}),\ \bibinfo {pages} {311--313}}\BibitemShut {NoStop}%
\bibitem [{\citenamefont {Peng}(1999)}]{peng_1999}%
  \BibitemOpen
  \bibfield  {author} {\bibinfo {author} {\bibnamefont {Peng}, \bibfnamefont
  {L-M}}} (\bibinfo {year} {1999}),\ \bibfield  {title} {\enquote {\bibinfo
  {title} {Electron atomic scattering factors and scattering potentials of
  crystals},}\ }\href@noop {} {\bibfield  {journal} {\bibinfo  {journal}
  {Micron}\ }\textbf {\bibinfo {volume} {30}}~(\bibinfo {number} {6}),\
  \bibinfo {pages} {625--648}}\BibitemShut {NoStop}%
\bibitem [{\citenamefont {Penz}\ \emph {et~al.}(2023)\citenamefont {Penz},
  \citenamefont {Tellgren}, \citenamefont {Csirik}, \citenamefont
  {Ruggenthaler},\ and\ \citenamefont {Laestadius}}]{v-rep-review_2023}%
  \BibitemOpen
  \bibfield  {author} {\bibinfo {author} {\bibnamefont {Penz}, \bibfnamefont
  {M}}, \bibinfo {author} {\bibfnamefont {E.~I.}\ \bibnamefont {Tellgren}},
  \bibinfo {author} {\bibfnamefont {M.~A.}\ \bibnamefont {Csirik}}, \bibinfo
  {author} {\bibfnamefont {M.}~\bibnamefont {Ruggenthaler}}, \ and\ \bibinfo
  {author} {\bibfnamefont {A.}~\bibnamefont {Laestadius}}} (\bibinfo {year}
  {2023}),\ \bibfield  {title} {\enquote {\bibinfo {title} {The structure of
  density-potential mapping. part {I}: Standard density-functional theory},}\
  }\href {\doibase 10.1021/acsphyschemau.2c00069} {\bibfield  {journal}
  {\bibinfo  {journal} {ACS Physical Chemistry Au}\ }\textbf {\bibinfo {volume}
  {3}}~(\bibinfo {number} {4}),\ \bibinfo {pages} {334--347}}\BibitemShut
  {NoStop}%
\bibitem [{\citenamefont {Pick}\ \emph {et~al.}(1970)\citenamefont {Pick},
  \citenamefont {Cohen},\ and\ \citenamefont {Martin}}]{pick_1970}%
  \BibitemOpen
  \bibfield  {author} {\bibinfo {author} {\bibnamefont {Pick}, \bibfnamefont
  {R~M}}, \bibinfo {author} {\bibfnamefont {M.~H.}\ \bibnamefont {Cohen}}, \
  and\ \bibinfo {author} {\bibfnamefont {R.~M.}\ \bibnamefont {Martin}}}
  (\bibinfo {year} {1970}),\ \bibfield  {title} {\enquote {\bibinfo {title}
  {Microscopic theory of force constants in the adiabatic approximation},}\
  }\href@noop {} {\bibfield  {journal} {\bibinfo  {journal} {Phys. Rev. B}\
  }\textbf {\bibinfo {volume} {1}},\ \bibinfo {pages} {910--920}}\BibitemShut
  {NoStop}%
\bibitem [{\citenamefont {Pratt}(1992)}]{mip_pratt_1992}%
  \BibitemOpen
  \bibfield  {author} {\bibinfo {author} {\bibnamefont {Pratt}, \bibfnamefont
  {L~R}}} (\bibinfo {year} {1992}),\ \bibfield  {title} {\enquote {\bibinfo
  {title} {Contact potentials of solution interfaces: phase equilibrium and
  interfacial electric fields},}\ }\href@noop {} {\bibfield  {journal}
  {\bibinfo  {journal} {J. Phys. Chem.}\ }\textbf {\bibinfo {volume}
  {96}}~(\bibinfo {number} {1}),\ \bibinfo {pages} {25--33}}\BibitemShut
  {NoStop}%
\bibitem [{\citenamefont {Raab}\ and\ \citenamefont
  {de~Lange}(2005)}]{raab_2005}%
  \BibitemOpen
  \bibfield  {author} {\bibinfo {author} {\bibnamefont {Raab}, \bibfnamefont
  {R~E}}, \ and\ \bibinfo {author} {\bibfnamefont {O.~L.}\ \bibnamefont
  {de~Lange}}} (\bibinfo {year} {2005}),\ \bibfield  {title} {\enquote
  {\bibinfo {title} {Transformed multipole theory of the response fields ${\D}$
  and ${\H}$ to electric octopole–magnetic quadrupole order},}\ }\href@noop
  {} {\bibfield  {journal} {\bibinfo  {journal} {Proc. R. Soc. A.}\ }\textbf
  {\bibinfo {volume} {461}},\ \bibinfo {pages} {595--608}}\BibitemShut
  {NoStop}%
\bibitem [{\citenamefont {Renteln}(2013)}]{renteln_2013}%
  \BibitemOpen
  \bibfield  {author} {\bibinfo {author} {\bibnamefont {Renteln}, \bibfnamefont
  {P}}} (\bibinfo {year} {2013}),\ \href@noop {} {\emph {\bibinfo {title}
  {Manifolds, Tensors, and Forms: An Introduction for Mathematicians and
  Physicists}}}\ (\bibinfo  {publisher} {Cambridge University
  Press})\BibitemShut {NoStop}%
\bibitem [{\citenamefont {Resta}(1992)}]{resta-1992}%
  \BibitemOpen
  \bibfield  {author} {\bibinfo {author} {\bibnamefont {Resta}, \bibfnamefont
  {R}}} (\bibinfo {year} {1992}),\ \bibfield  {title} {\enquote {\bibinfo
  {title} {Theory of the electric polarization in crystals},}\ }\href@noop {}
  {\bibfield  {journal} {\bibinfo  {journal} {Ferroelectrics}\ }\textbf
  {\bibinfo {volume} {136}}~(\bibinfo {number} {1}),\ \bibinfo {pages}
  {51--55}}\BibitemShut {NoStop}%
\bibitem [{\citenamefont {Resta}(1993)}]{resta-1993}%
  \BibitemOpen
  \bibfield  {author} {\bibinfo {author} {\bibnamefont {Resta}, \bibfnamefont
  {R}}} (\bibinfo {year} {1993}),\ \bibfield  {title} {\enquote {\bibinfo
  {title} {Macroscopic electric polarization as a geometric quantum phase},}\
  }\href@noop {} {\bibfield  {journal} {\bibinfo  {journal} {Europhys. lett.}\
  }\textbf {\bibinfo {volume} {22}}~(\bibinfo {number} {2}),\ \bibinfo {pages}
  {133--138}}\BibitemShut {NoStop}%
\bibitem [{\citenamefont {Resta}(1994)}]{resta-rmp-1994}%
  \BibitemOpen
  \bibfield  {author} {\bibinfo {author} {\bibnamefont {Resta}, \bibfnamefont
  {R}}} (\bibinfo {year} {1994}),\ \bibfield  {title} {\enquote {\bibinfo
  {title} {Macroscopic polarization in crystalline dielectrics: the geometric
  phase approach},}\ }\href@noop {} {\bibfield  {journal} {\bibinfo  {journal}
  {Rev. Mod. Phys.}\ }\textbf {\bibinfo {volume} {66}},\ \bibinfo {pages}
  {899--915}}\BibitemShut {NoStop}%
\bibitem [{\citenamefont {Resta}(2010)}]{Resta_2010}%
  \BibitemOpen
  \bibfield  {author} {\bibinfo {author} {\bibnamefont {Resta}, \bibfnamefont
  {R}}} (\bibinfo {year} {2010}),\ \bibfield  {title} {\enquote {\bibinfo
  {title} {Electrical polarization and orbital magnetization: the modern
  theories},}\ }\href {\doibase 10.1088/0953-8984/22/12/123201} {\bibfield
  {journal} {\bibinfo  {journal} {J. Phys.: Condens. Matter}\ }\textbf
  {\bibinfo {volume} {22}}~(\bibinfo {number} {12}),\ \bibinfo {pages}
  {123201}}\BibitemShut {NoStop}%
\bibitem [{\citenamefont {Resta}(2018)}]{Resta2018}%
  \BibitemOpen
  \bibfield  {author} {\bibinfo {author} {\bibnamefont {Resta}, \bibfnamefont
  {R}}} (\bibinfo {year} {2018}),\ \enquote {\bibinfo {title} {Electrical
  polarization and orbital magnetization: The position operator tamed},}\ in\
  \href {\doibase 10.1007/978-3-319-42913-7_12-1} {\emph {\bibinfo {booktitle}
  {Handbook of Materials Modeling : Methods: Theory and Modeling}}},\ \bibinfo
  {editor} {edited by\ \bibinfo {editor} {\bibfnamefont {Wanda}\ \bibnamefont
  {Andreoni}}\ and\ \bibinfo {editor} {\bibfnamefont {Sidney}\ \bibnamefont
  {Yip}}}\ (\bibinfo  {publisher} {Springer International Publishing},\
  \bibinfo {address} {Cham})\ pp.\ \bibinfo {pages} {1--31}\BibitemShut
  {NoStop}%
\bibitem [{\citenamefont {Resta}\ and\ \citenamefont
  {Vanderbilt}(2007)}]{resta-vanderbilt-2007}%
  \BibitemOpen
  \bibfield  {author} {\bibinfo {author} {\bibnamefont {Resta}, \bibfnamefont
  {R}}, \ and\ \bibinfo {author} {\bibfnamefont {D.}~\bibnamefont
  {Vanderbilt}}} (\bibinfo {year} {2007}),\ \enquote {\bibinfo {title} {Theory
  of polarization: A modern approach},}\ in\ \href@noop {} {\emph {\bibinfo
  {booktitle} {Physics of Ferroelectrics}}},\ \bibinfo {series} {Topics in
  Applied Physics}, Vol.\ \bibinfo {volume} {105}\ (\bibinfo  {publisher}
  {Springer Berlin, Heidelberg})\ pp.\ \bibinfo {pages} {31--68}\BibitemShut
  {NoStop}%
\bibitem [{\citenamefont {Rez}\ \emph {et~al.}(1994)\citenamefont {Rez},
  \citenamefont {Rez},\ and\ \citenamefont {Grant}}]{rez_1994}%
  \BibitemOpen
  \bibfield  {author} {\bibinfo {author} {\bibnamefont {Rez}, \bibfnamefont
  {D}}, \bibinfo {author} {\bibfnamefont {P.}~\bibnamefont {Rez}}, \ and\
  \bibinfo {author} {\bibfnamefont {I.}~\bibnamefont {Grant}}} (\bibinfo {year}
  {1994}),\ \bibfield  {title} {\enquote {\bibinfo {title} {{Dirac–Fock}
  calculations of {X‐ray} scattering factors and contributions to the mean
  inner potential for electron scattering},}\ }\href@noop {} {\bibfield
  {journal} {\bibinfo  {journal} {Acta Crystallogr. A}\ }\textbf {\bibinfo
  {volume} {50}}~(\bibinfo {number} {4}),\ \bibinfo {pages}
  {481--497}}\BibitemShut {NoStop}%
\bibitem [{\citenamefont {Riley}\ \emph {et~al.}(2006)\citenamefont {Riley},
  \citenamefont {Hobson},\ and\ \citenamefont
  {Bence}}]{Riley_Hobson_Bence_2006}%
  \BibitemOpen
  \bibfield  {author} {\bibinfo {author} {\bibnamefont {Riley}, \bibfnamefont
  {K~F}}, \bibinfo {author} {\bibfnamefont {M.~P.}\ \bibnamefont {Hobson}}, \
  and\ \bibinfo {author} {\bibfnamefont {S.~J.}\ \bibnamefont {Bence}}}
  (\bibinfo {year} {2006}),\ \href@noop {} {\emph {\bibinfo {title}
  {Mathematical Methods for Physics and Engineering: A Comprehensive Guide}}},\
  \bibinfo {edition} {3rd}\ ed.\ (\bibinfo  {publisher} {Cambridge University
  Press})\BibitemShut {NoStop}%
\bibitem [{\citenamefont {Robinson}(1971)}]{robinson_1971}%
  \BibitemOpen
  \bibfield  {author} {\bibinfo {author} {\bibnamefont {Robinson},
  \bibfnamefont {F~N~H}}} (\bibinfo {year} {1971}),\ \bibfield  {title}
  {\enquote {\bibinfo {title} {The microscopic and macroscopic equations of the
  electromagnetic field},}\ }\href@noop {} {\bibfield  {journal} {\bibinfo
  {journal} {Physica}\ }\textbf {\bibinfo {volume} {54}}~(\bibinfo {number}
  {3}),\ \bibinfo {pages} {329--341}}\BibitemShut {NoStop}%
\bibitem [{\citenamefont {Roche}(2000)}]{roche_2000}%
  \BibitemOpen
  \bibfield  {author} {\bibinfo {author} {\bibnamefont {Roche}, \bibfnamefont
  {J~J}}} (\bibinfo {year} {2000}),\ \bibfield  {title} {\enquote {\bibinfo
  {title} {{${\B}$} and {$\H$}, the intensity vectors of magnetism: A new
  approach to resolving a century-old controversy},}\ }\href@noop {} {\bibfield
   {journal} {\bibinfo  {journal} {Am. J. Phys.}\ }\textbf {\bibinfo {volume}
  {68}}~(\bibinfo {number} {5}),\ \bibinfo {pages} {438--449}}\BibitemShut
  {NoStop}%
\bibitem [{\citenamefont {Rosenfeld}(1965)}]{rosenfeld_1965}%
  \BibitemOpen
  \bibfield  {author} {\bibinfo {author} {\bibnamefont {Rosenfeld},
  \bibfnamefont {L}}} (\bibinfo {year} {1965}),\ \href@noop {} {\emph {\bibinfo
  {title} {Theory of Electrons}}},\ \bibinfo {edition} {dover ed.}\ ed.\
  (\bibinfo  {publisher} {Dover Publications},\ \bibinfo {address} {New
  York})\BibitemShut {NoStop}%
\bibitem [{\citenamefont {Royo}\ and\ \citenamefont
  {Stengel}(2021)}]{stengel_2021}%
  \BibitemOpen
  \bibfield  {author} {\bibinfo {author} {\bibnamefont {Royo}, \bibfnamefont
  {M}}, \ and\ \bibinfo {author} {\bibfnamefont {M.}~\bibnamefont {Stengel}}}
  (\bibinfo {year} {2021}),\ \bibfield  {title} {\enquote {\bibinfo {title}
  {Exact long-range dielectric screening and interatomic force constants in
  quasi-two-dimensional crystals},}\ }\href {\doibase
  10.1103/PhysRevX.11.041027} {\bibfield  {journal} {\bibinfo  {journal} {Phys.
  Rev. X}\ }\textbf {\bibinfo {volume} {11}},\ \bibinfo {pages}
  {041027}}\BibitemShut {NoStop}%
\bibitem [{\citenamefont {Russakoff}(1970)}]{russakoff-ajp-1970}%
  \BibitemOpen
  \bibfield  {author} {\bibinfo {author} {\bibnamefont {Russakoff},
  \bibfnamefont {G}}} (\bibinfo {year} {1970}),\ \bibfield  {title} {\enquote
  {\bibinfo {title} {A derivation of the macroscopic {Maxwell} equations},}\
  }\href@noop {} {\bibfield  {journal} {\bibinfo  {journal} {Am. J. Phys.}\
  }\textbf {\bibinfo {volume} {38}}~(\bibinfo {number} {10}),\ \bibinfo {pages}
  {1188--1195}}\BibitemShut {NoStop}%
\bibitem [{\citenamefont {Sagan}(1994)}]{sagan_1994}%
  \BibitemOpen
  \bibfield  {author} {\bibinfo {author} {\bibnamefont {Sagan}, \bibfnamefont
  {C}}} (\bibinfo {year} {1994}),\ \href@noop {} {\emph {\bibinfo {title} {Pale
  Blue Dot: A Vision of the Human Future in Space}}}\ (\bibinfo  {publisher}
  {Random House})\BibitemShut {NoStop}%
\bibitem [{\citenamefont {Saldin}\ and\ \citenamefont
  {Spence}(1994)}]{spence_1994}%
  \BibitemOpen
  \bibfield  {author} {\bibinfo {author} {\bibnamefont {Saldin}, \bibfnamefont
  {D~K}}, \ and\ \bibinfo {author} {\bibfnamefont {J.~C.~H.}\ \bibnamefont
  {Spence}}} (\bibinfo {year} {1994}),\ \bibfield  {title} {\enquote {\bibinfo
  {title} {On the mean inner potential in high- and low-energy electron
  diffraction},}\ }\href@noop {} {\bibfield  {journal} {\bibinfo  {journal}
  {Ultramicroscopy}\ }\textbf {\bibinfo {volume} {55}}~(\bibinfo {number}
  {4}),\ \bibinfo {pages} {397--406}}\BibitemShut {NoStop}%
\bibitem [{\citenamefont {Sanchez}\ and\ \citenamefont
  {Ochando}(1985)}]{mip_sanchez_1985}%
  \BibitemOpen
  \bibfield  {author} {\bibinfo {author} {\bibnamefont {Sanchez}, \bibfnamefont
  {A}}, \ and\ \bibinfo {author} {\bibfnamefont {M.~A.}\ \bibnamefont
  {Ochando}}} (\bibinfo {year} {1985}),\ \bibfield  {title} {\enquote {\bibinfo
  {title} {Calculation of the mean inner potential},}\ }\href@noop {}
  {\bibfield  {journal} {\bibinfo  {journal} {J. Phys. C: Solid State Phys.}\
  }\textbf {\bibinfo {volume} {18}}~(\bibinfo {number} {1}),\ \bibinfo {pages}
  {33--41}}\BibitemShut {NoStop}%
\bibitem [{\citenamefont {Saunders}\ \emph {et~al.}(1992)\citenamefont
  {Saunders}, \citenamefont {Freyria-Fava}, \citenamefont {Dovesi},
  \citenamefont {Salasco},\ and\ \citenamefont {Roetti}}]{saunders_1992}%
  \BibitemOpen
  \bibfield  {author} {\bibinfo {author} {\bibnamefont {Saunders},
  \bibfnamefont {V~R}}, \bibinfo {author} {\bibfnamefont {C.}~\bibnamefont
  {Freyria-Fava}}, \bibinfo {author} {\bibfnamefont {R.}~\bibnamefont
  {Dovesi}}, \bibinfo {author} {\bibfnamefont {L.}~\bibnamefont {Salasco}}, \
  and\ \bibinfo {author} {\bibfnamefont {C.}~\bibnamefont {Roetti}}} (\bibinfo
  {year} {1992}),\ \bibfield  {title} {\enquote {\bibinfo {title} {On the
  electrostatic potential in crystalline systems where the charge density is
  expanded in gaussian functions},}\ }\href@noop {} {\bibfield  {journal}
  {\bibinfo  {journal} {Mol. Phys.}\ }\textbf {\bibinfo {volume}
  {77}}~(\bibinfo {number} {4}),\ \bibinfo {pages} {629--665}}\BibitemShut
  {NoStop}%
\bibitem [{\citenamefont {Scharlau}(2012)}]{scharlau_2012}%
  \BibitemOpen
  \bibfield  {author} {\bibinfo {author} {\bibnamefont {Scharlau},
  \bibfnamefont {W}}} (\bibinfo {year} {2012}),\ \href@noop {} {\emph {\bibinfo
  {title} {Quadratic and {Hermitian} Forms}}},\ Grundlehren der mathematischen
  Wissenschaften\ (\bibinfo  {publisher} {Springer Berlin
  Heidelberg})\BibitemShut {NoStop}%
\bibitem [{\citenamefont {Schram}(1960)}]{schram_1960}%
  \BibitemOpen
  \bibfield  {author} {\bibinfo {author} {\bibnamefont {Schram}, \bibfnamefont
  {K}}} (\bibinfo {year} {1960}),\ \bibfield  {title} {\enquote {\bibinfo
  {title} {Quantum statistical derivation of the macroscopic {Maxwell}
  equations},}\ }\href@noop {} {\bibfield  {journal} {\bibinfo  {journal}
  {Physica}\ }\textbf {\bibinfo {volume} {26}}~(\bibinfo {number} {12}),\
  \bibinfo {pages} {1080--1090}}\BibitemShut {NoStop}%
\bibitem [{\citenamefont {Shannon}(1948)}]{shannon}%
  \BibitemOpen
  \bibfield  {author} {\bibinfo {author} {\bibnamefont {Shannon}, \bibfnamefont
  {C~E}}} (\bibinfo {year} {1948}),\ \bibfield  {title} {\enquote {\bibinfo
  {title} {A mathematical theory of communication},}\ }\href@noop {} {\bibfield
   {journal} {\bibinfo  {journal} {The {Bell} System Technical Journal}\
  }\textbf {\bibinfo {volume} {27}}~(\bibinfo {number} {3}),\ \bibinfo {pages}
  {379--423}}\BibitemShut {NoStop}%
\bibitem [{\citenamefont {Sokhan}\ and\ \citenamefont
  {Tildesley}(1997{\natexlab{a}})}]{mip_sokhan_1997}%
  \BibitemOpen
  \bibfield  {author} {\bibinfo {author} {\bibnamefont {Sokhan}, \bibfnamefont
  {V~P}}, \ and\ \bibinfo {author} {\bibfnamefont {D.~J.}\ \bibnamefont
  {Tildesley}}} (\bibinfo {year} {1997}{\natexlab{a}}),\ \bibfield  {title}
  {\enquote {\bibinfo {title} {The free surface of water: Molecular
  orientation, surface potential, and nonlinear susceptibility},}\ }\href@noop
  {} {\bibfield  {journal} {\bibinfo  {journal} {Mol. Phys.}\ }\textbf
  {\bibinfo {volume} {92}},\ \bibinfo {pages} {625}}\BibitemShut {NoStop}%
\bibitem [{\citenamefont {Sokhan}\ and\ \citenamefont
  {Tildesley}(1997{\natexlab{b}})}]{tildesley_1997}%
  \BibitemOpen
  \bibfield  {author} {\bibinfo {author} {\bibnamefont {Sokhan}, \bibfnamefont
  {V~P}}, \ and\ \bibinfo {author} {\bibfnamefont {D.~J.}\ \bibnamefont
  {Tildesley}}} (\bibinfo {year} {1997}{\natexlab{b}}),\ \bibfield  {title}
  {\enquote {\bibinfo {title} {The free surface of water: molecular
  orientation, surface potential and nonlinear susceptibility},}\ }\href@noop
  {} {\bibfield  {journal} {\bibinfo  {journal} {Mol. Phys.}\ }\textbf
  {\bibinfo {volume} {92}}~(\bibinfo {number} {4}),\ \bibinfo {pages}
  {625--640}}\BibitemShut {NoStop}%
\bibitem [{\citenamefont {Souza}\ \emph {et~al.}(2001)\citenamefont {Souza},
  \citenamefont {Marzari},\ and\ \citenamefont {Vanderbilt}}]{souza}%
  \BibitemOpen
  \bibfield  {author} {\bibinfo {author} {\bibnamefont {Souza}, \bibfnamefont
  {I}}, \bibinfo {author} {\bibfnamefont {N.}~\bibnamefont {Marzari}}, \ and\
  \bibinfo {author} {\bibfnamefont {D.}~\bibnamefont {Vanderbilt}}} (\bibinfo
  {year} {2001}),\ \bibfield  {title} {\enquote {\bibinfo {title} {Maximally
  localized {Wannier} functions for entangled energy bands},}\ }\href@noop {}
  {\bibfield  {journal} {\bibinfo  {journal} {Phys. Rev. B}\ }\textbf {\bibinfo
  {volume} {65}},\ \bibinfo {pages} {035109}}\BibitemShut {NoStop}%
\bibitem [{\citenamefont {Spence}(1993)}]{spence-1993}%
  \BibitemOpen
  \bibfield  {author} {\bibinfo {author} {\bibnamefont {Spence}, \bibfnamefont
  {J~C~H}}} (\bibinfo {year} {1993}),\ \bibfield  {title} {\enquote {\bibinfo
  {title} {On the accurate measurement of structure-factor amplitudes and
  phases by electron diffraction},}\ }\href@noop {} {\bibfield  {journal}
  {\bibinfo  {journal} {Acta Crystallogr., Sect. A: Found. Crystallogr.}\
  }\textbf {\bibinfo {volume} {49}}~(\bibinfo {number} {2}),\ \bibinfo {pages}
  {231--260}}\BibitemShut {NoStop}%
\bibitem [{\citenamefont {Spence}(1999)}]{spence-1999}%
  \BibitemOpen
  \bibfield  {author} {\bibinfo {author} {\bibnamefont {Spence}, \bibfnamefont
  {J~C~H}}} (\bibinfo {year} {1999}),\ \bibfield  {title} {\enquote {\bibinfo
  {title} {The future of atomic resolution electron microscopy for materials
  science},}\ }\href@noop {} {\bibfield  {journal} {\bibinfo  {journal} {Mater.
  Sci. Eng., R}\ }\textbf {\bibinfo {volume} {26}}~(\bibinfo {number} {1}),\
  \bibinfo {pages} {1 -- 49}}\BibitemShut {NoStop}%
\bibitem [{\citenamefont {Spohn}\ and\ \citenamefont
  {Teufel}(2001)}]{spohn_2001}%
  \BibitemOpen
  \bibfield  {author} {\bibinfo {author} {\bibnamefont {Spohn}, \bibfnamefont
  {H}}, \ and\ \bibinfo {author} {\bibfnamefont {S.}~\bibnamefont {Teufel}}}
  (\bibinfo {year} {2001}),\ \bibfield  {title} {\enquote {\bibinfo {title}
  {Adiabatic decoupling and time-dependent {Born–Oppenheimer} theory},}\
  }\href {\doibase 10.1007/s002200100535} {\bibfield  {journal} {\bibinfo
  {journal} {Commun. Math. Phys.}\ }\textbf {\bibinfo {volume} {224}},\
  \bibinfo {pages} {113--132}}\BibitemShut {NoStop}%
\bibitem [{\citenamefont {Stengel}(2011)}]{stengel-prb-2011}%
  \BibitemOpen
  \bibfield  {author} {\bibinfo {author} {\bibnamefont {Stengel}, \bibfnamefont
  {M}}} (\bibinfo {year} {2011}),\ \bibfield  {title} {\enquote {\bibinfo
  {title} {Electrostatic stability of insulating surfaces: Theory and
  applications},}\ }\href@noop {} {\bibfield  {journal} {\bibinfo  {journal}
  {Phys. Rev. B}\ }\textbf {\bibinfo {volume} {84}},\ \bibinfo {pages}
  {205432}}\BibitemShut {NoStop}%
\bibitem [{\citenamefont {Stengel}\ and\ \citenamefont
  {Vanderbilt}(2009)}]{stengel-vanderbilt-prb-2009}%
  \BibitemOpen
  \bibfield  {author} {\bibinfo {author} {\bibnamefont {Stengel}, \bibfnamefont
  {M}}, \ and\ \bibinfo {author} {\bibfnamefont {D.}~\bibnamefont
  {Vanderbilt}}} (\bibinfo {year} {2009}),\ \bibfield  {title} {\enquote
  {\bibinfo {title} {Berry-phase theory of polar discontinuities at oxide-oxide
  interfaces},}\ }\href@noop {} {\bibfield  {journal} {\bibinfo  {journal}
  {Phys. Rev. B}\ }\textbf {\bibinfo {volume} {80}},\ \bibinfo {pages}
  {241103}}\BibitemShut {NoStop}%
\bibitem [{\citenamefont {Stillinger}\ and\ \citenamefont
  {Ben‐Naim}(1967)}]{stillinger_1967}%
  \BibitemOpen
  \bibfield  {author} {\bibinfo {author} {\bibnamefont {Stillinger},
  \bibfnamefont {F~H, Jr}}, \ and\ \bibinfo {author} {\bibfnamefont
  {A.}~\bibnamefont {Ben‐Naim}}} (\bibinfo {year} {1967}),\ \bibfield
  {title} {\enquote {\bibinfo {title} {Liquid—vapor interface potential for
  water},}\ }\href@noop {} {\bibfield  {journal} {\bibinfo  {journal} {J. Chem.
  Phys.}\ }\textbf {\bibinfo {volume} {47}}~(\bibinfo {number} {11}),\ \bibinfo
  {pages} {4431--4437}}\BibitemShut {NoStop}%
\bibitem [{\citenamefont {Strichartz}(2003)}]{strichartz_distribution_theory}%
  \BibitemOpen
  \bibfield  {author} {\bibinfo {author} {\bibnamefont {Strichartz},
  \bibfnamefont {R~S}}} (\bibinfo {year} {2003}),\ \href@noop {} {\emph
  {\bibinfo {title} {A Guide To Distribution Theory And Fourier Transforms}}}\
  (\bibinfo  {publisher} {Scientific, World})\BibitemShut {NoStop}%
\bibitem [{\citenamefont {Sutter}\ \emph {et~al.}(2024)\citenamefont {Sutter},
  \citenamefont {Penz}, \citenamefont {Ruggenthaler}, \citenamefont {van
  Leeuwen},\ and\ \citenamefont {Giesbertz}}]{Sutter_2024}%
  \BibitemOpen
  \bibfield  {author} {\bibinfo {author} {\bibnamefont {Sutter}, \bibfnamefont
  {S~M}}, \bibinfo {author} {\bibfnamefont {M.}~\bibnamefont {Penz}}, \bibinfo
  {author} {\bibfnamefont {M.}~\bibnamefont {Ruggenthaler}}, \bibinfo {author}
  {\bibfnamefont {R.}~\bibnamefont {van Leeuwen}}, \ and\ \bibinfo {author}
  {\bibfnamefont {K.~J.~H.}\ \bibnamefont {Giesbertz}}} (\bibinfo {year}
  {2024}),\ \bibfield  {title} {\enquote {\bibinfo {title} {Solution of the
  v-representability problem on a one-dimensional torus},}\ }\href {\doibase
  10.1088/1751-8121/ad8a2a} {\bibfield  {journal} {\bibinfo  {journal} {J.
  Phys. A: Math. Theor.}\ }\textbf {\bibinfo {volume} {57}}~(\bibinfo {number}
  {47}),\ \bibinfo {pages} {475202}}\BibitemShut {NoStop}%
\bibitem [{\citenamefont {Sutton}(2024)}]{sutton_2024}%
  \BibitemOpen
  \bibfield  {author} {\bibinfo {author} {\bibnamefont {Sutton}, \bibfnamefont
  {A~P}}} (\bibinfo {year} {2024}),\ \bibfield  {title} {\enquote {\bibinfo
  {title} {Misconceptions about metals},}\ }\href {\doibase
  10.1021/acs.jchemed.4c00010} {\bibfield  {journal} {\bibinfo  {journal} {J.
  Chem. Educ.}\ }\textbf {\bibinfo {volume} {101}}~(\bibinfo {number} {7}),\
  \bibinfo {pages} {2710--2715}}\BibitemShut {NoStop}%
\bibitem [{\citenamefont {Tagantsev}(1991)}]{tagantsev_1991}%
  \BibitemOpen
  \bibfield  {author} {\bibinfo {author} {\bibnamefont {Tagantsev},
  \bibfnamefont {A~K}}} (\bibinfo {year} {1991}),\ \bibfield  {title} {\enquote
  {\bibinfo {title} {Electric polarization in crystals and its response to
  thermal and elastic perturbations},}\ }\href@noop {} {\bibfield  {journal}
  {\bibinfo  {journal} {Phase Transitions}\ }\textbf {\bibinfo {volume}
  {35}}~(\bibinfo {number} {3-4}),\ \bibinfo {pages} {119--203}}\BibitemShut
  {NoStop}%
\bibitem [{\citenamefont {Tang}(2006)}]{tang_2006}%
  \BibitemOpen
  \bibfield  {author} {\bibinfo {author} {\bibnamefont {Tang}, \bibfnamefont
  {K~T}}} (\bibinfo {year} {2006}),\ \href@noop {} {\emph {\bibinfo {title}
  {Mathematical Methods for Engineers and Scientists 3: Fourier Analysis,
  Partial Differential Equations and Variational Methods}}},\ Mathematical
  Methods for Engineers and Scientists\ (\bibinfo  {publisher} {Springer Berlin
  Heidelberg})\BibitemShut {NoStop}%
\bibitem [{\citenamefont {Tangney}(2006)}]{Tangney_JCP_2006}%
  \BibitemOpen
  \bibfield  {author} {\bibinfo {author} {\bibnamefont {Tangney}, \bibfnamefont
  {P}}} (\bibinfo {year} {2006}),\ \bibfield  {title} {\enquote {\bibinfo
  {title} {On the theory underlying the {Car-Parrinello} method and the role of
  the fictitious mass parameter},}\ }\href {\doibase 10.1063/1.2162893}
  {\bibfield  {journal} {\bibinfo  {journal} {J. Chem. Phys.}\ }\textbf
  {\bibinfo {volume} {124}}~(\bibinfo {number} {4}),\ \bibinfo {pages}
  {044111}}\BibitemShut {NoStop}%
\bibitem [{\citenamefont {Tangney}(2024)}]{tangney_bose_einstein}%
  \BibitemOpen
  \bibfield  {author} {\bibinfo {author} {\bibnamefont {Tangney}, \bibfnamefont
  {P}}} (\bibinfo {year} {2024}),\ \bibfield  {title} {\enquote {\bibinfo
  {title} {Derivation of {Bose-Einstein} statistics from the uncertainty
  principle},}\ }\href {\doibase https://doi.org/10.1088/1742-5468/ad74e9}
  {\bibfield  {journal} {\bibinfo  {journal} {J. Stat. Mech.: Theory Exp}\
  }https://doi.org/10.1088/1742-5468/ad74e9}\BibitemShut {NoStop}%
\bibitem [{\citenamefont {Tangney}\ and\ \citenamefont
  {Scandolo}(2002)}]{tangney_silica_2002}%
  \BibitemOpen
  \bibfield  {author} {\bibinfo {author} {\bibnamefont {Tangney}, \bibfnamefont
  {P}}, \ and\ \bibinfo {author} {\bibfnamefont {S.}~\bibnamefont {Scandolo}}}
  (\bibinfo {year} {2002}),\ \bibfield  {title} {\enquote {\bibinfo {title}
  {{An {\em ab initio} parametrized interatomic force field for silica}},}\
  }\href {\doibase 10.1063/1.1513312} {\bibfield  {journal} {\bibinfo
  {journal} {J. Chem. Phys.}\ }\textbf {\bibinfo {volume} {117}}~(\bibinfo
  {number} {19}),\ \bibinfo {pages} {8898--8904}}\BibitemShut {NoStop}%
\bibitem [{\citenamefont {Tasker}(1979)}]{tasker-1979}%
  \BibitemOpen
  \bibfield  {author} {\bibinfo {author} {\bibnamefont {Tasker}, \bibfnamefont
  {P~W}}} (\bibinfo {year} {1979}),\ \bibfield  {title} {\enquote {\bibinfo
  {title} {The stability of ionic crystal surfaces},}\ }\href@noop {}
  {\bibfield  {journal} {\bibinfo  {journal} {J. Phys. C: Solid State Phys.}\
  }\textbf {\bibinfo {volume} {12}}~(\bibinfo {number} {22}),\ \bibinfo {pages}
  {4977}}\BibitemShut {NoStop}%
\bibitem [{\citenamefont {Tognetti}\ and\ \citenamefont
  {Loos}(2016)}]{tognetti_2016}%
  \BibitemOpen
  \bibfield  {author} {\bibinfo {author} {\bibnamefont {Tognetti},
  \bibfnamefont {V}}, \ and\ \bibinfo {author} {\bibfnamefont {P.-F.}\
  \bibnamefont {Loos}}} (\bibinfo {year} {2016}),\ \bibfield  {title} {\enquote
  {\bibinfo {title} {Natural occupation numbers in two-electron quantum
  rings},}\ }\href {\doibase 10.1063/1.4940919} {\bibfield  {journal} {\bibinfo
   {journal} {J. Chem. Phys.}\ }\textbf {\bibinfo {volume} {144}}~(\bibinfo
  {number} {5}),\ \bibinfo {pages} {054108}}\BibitemShut {NoStop}%
\bibitem [{\citenamefont {Toumpanaki}\ \emph {et~al.}(2021)\citenamefont
  {Toumpanaki}, \citenamefont {Shah},\ and\ \citenamefont {Eichhorn}}]{wood}%
  \BibitemOpen
  \bibfield  {author} {\bibinfo {author} {\bibnamefont {Toumpanaki},
  \bibfnamefont {E}}, \bibinfo {author} {\bibfnamefont {D.~U.}\ \bibnamefont
  {Shah}}, \ and\ \bibinfo {author} {\bibfnamefont {S.~J.}\ \bibnamefont
  {Eichhorn}}} (\bibinfo {year} {2021}),\ \bibfield  {title} {\enquote
  {\bibinfo {title} {Beyond what meets the eye: Imaging and imagining wood
  mechanical–structural properties},}\ }\href@noop {} {\bibfield  {journal}
  {\bibinfo  {journal} {Adv. Mat.}\ }\textbf {\bibinfo {volume} {33}}~(\bibinfo
  {number} {28}),\ \bibinfo {pages} {2001613}}\BibitemShut {NoStop}%
\bibitem [{\citenamefont {Trave}\ \emph {et~al.}(2002)\citenamefont {Trave},
  \citenamefont {Tangney}, \citenamefont {Scandolo}, \citenamefont
  {Pasquarello},\ and\ \citenamefont {Car}}]{trave_2002}%
  \BibitemOpen
  \bibfield  {author} {\bibinfo {author} {\bibnamefont {Trave}, \bibfnamefont
  {A}}, \bibinfo {author} {\bibfnamefont {P.}~\bibnamefont {Tangney}}, \bibinfo
  {author} {\bibfnamefont {S.}~\bibnamefont {Scandolo}}, \bibinfo {author}
  {\bibfnamefont {A.}~\bibnamefont {Pasquarello}}, \ and\ \bibinfo {author}
  {\bibfnamefont {R.}~\bibnamefont {Car}}} (\bibinfo {year} {2002}),\ \bibfield
   {title} {\enquote {\bibinfo {title} {Pressure-induced structural changes in
  liquid {SiO$_2$} from $\mathrm{\text{ab initio}}$ simulations},}\ }\href
  {\doibase 10.1103/PhysRevLett.89.245504} {\bibfield  {journal} {\bibinfo
  {journal} {Phys. Rev. Lett.}\ }\textbf {\bibinfo {volume} {89}},\ \bibinfo
  {pages} {245504}}\BibitemShut {NoStop}%
\bibitem [{\citenamefont {Vanderbilt}(2018)}]{vanderbilt_2018}%
  \BibitemOpen
  \bibfield  {author} {\bibinfo {author} {\bibnamefont {Vanderbilt},
  \bibfnamefont {D}}} (\bibinfo {year} {2018}),\ \href@noop {} {\emph {\bibinfo
  {title} {Berry Phases in Electronic Structure Theory: Electric Polarization,
  Orbital Magnetization and Topological Insulators}}}\ (\bibinfo  {publisher}
  {Cambridge University Press})\BibitemShut {NoStop}%
\bibitem [{\citenamefont {Vanderbilt}\ and\ \citenamefont
  {King-Smith}(1993)}]{kingsmith-vanderbilt-prb-1993-2}%
  \BibitemOpen
  \bibfield  {author} {\bibinfo {author} {\bibnamefont {Vanderbilt},
  \bibfnamefont {D}}, \ and\ \bibinfo {author} {\bibfnamefont {R.~D.}\
  \bibnamefont {King-Smith}}} (\bibinfo {year} {1993}),\ \bibfield  {title}
  {\enquote {\bibinfo {title} {Electric polarization as a bulk quantity and its
  relation to surface charge},}\ }\href@noop {} {\bibfield  {journal} {\bibinfo
   {journal} {Phys. Rev. B}\ }\textbf {\bibinfo {volume} {48}},\ \bibinfo
  {pages} {4442--4455}}\BibitemShut {NoStop}%
\bibitem [{\citenamefont {Vaz}\ and\ \citenamefont
  {da~Rocha}(2016)}]{vaz_darocha_2016}%
  \BibitemOpen
  \bibfield  {author} {\bibinfo {author} {\bibnamefont {Vaz}, \bibfnamefont
  {J}}, \ and\ \bibinfo {author} {\bibfnamefont {R.}~\bibnamefont {da~Rocha}}}
  (\bibinfo {year} {2016}),\ \href@noop {} {\emph {\bibinfo {title} {An
  Introduction to {Clifford} Algebras and Spinors}}}\ (\bibinfo  {publisher}
  {Oxford University Press})\BibitemShut {NoStop}%
\bibitem [{\citenamefont {Vinogradov}\ and\ \citenamefont
  {Aivazyan}(1999)}]{vinogradov-PRE-1999}%
  \BibitemOpen
  \bibfield  {author} {\bibinfo {author} {\bibnamefont {Vinogradov},
  \bibfnamefont {A~P}}, \ and\ \bibinfo {author} {\bibfnamefont {A.~V.}\
  \bibnamefont {Aivazyan}}} (\bibinfo {year} {1999}),\ \bibfield  {title}
  {\enquote {\bibinfo {title} {Scaling theory for homogenization of the
  {Maxwell} equations},}\ }\href@noop {} {\bibfield  {journal} {\bibinfo
  {journal} {Phys. Rev. E}\ }\textbf {\bibinfo {volume} {60}},\ \bibinfo
  {pages} {987--993}}\BibitemShut {NoStop}%
\bibitem [{\citenamefont {van Vleck}(1937)}]{vanvleck-1937}%
  \BibitemOpen
  \bibfield  {author} {\bibinfo {author} {\bibnamefont {van Vleck},
  \bibfnamefont {J~H}}} (\bibinfo {year} {1937}),\ \bibfield  {title} {\enquote
  {\bibinfo {title} {On the role of dipole‐dipole coupling in dielectric
  media},}\ }\href@noop {} {\bibfield  {journal} {\bibinfo  {journal} {J. Chem.
  Phys.}\ }\textbf {\bibinfo {volume} {5}}~(\bibinfo {number} {7}),\ \bibinfo
  {pages} {556--568}}\BibitemShut {NoStop}%
\bibitem [{\citenamefont {Vogl}(1978)}]{vogl_1978}%
  \BibitemOpen
  \bibfield  {author} {\bibinfo {author} {\bibnamefont {Vogl}, \bibfnamefont
  {P}}} (\bibinfo {year} {1978}),\ \bibfield  {title} {\enquote {\bibinfo
  {title} {Dynamical effective charges in semiconductors: A pseudopotential
  approach},}\ }\href@noop {} {\bibfield  {journal} {\bibinfo  {journal} {J.
  Phys. C: Solid State Phys.}\ }\textbf {\bibinfo {volume} {11}}~(\bibinfo
  {number} {2}),\ \bibinfo {pages} {251}}\BibitemShut {NoStop}%
\bibitem [{\citenamefont {Walter}(2014)}]{walter_2014}%
  \BibitemOpen
  \bibfield  {author} {\bibinfo {author} {\bibnamefont {Walter}, \bibfnamefont
  {S}}} (\bibinfo {year} {2014}),\ \enquote {\bibinfo {title} {The historical
  origins of spacetime},}\ in\ \href {\doibase 10.1007/978-3-642-41992-8_2}
  {\emph {\bibinfo {booktitle} {Springer Handbook of Spacetime}}},\ \bibinfo
  {editor} {edited by\ \bibinfo {editor} {\bibfnamefont {A.}~\bibnamefont
  {Ashtekar}}\ and\ \bibinfo {editor} {\bibfnamefont {V.}~\bibnamefont
  {Petkov}}}\ (\bibinfo  {publisher} {Springer Berlin Heidelberg},\ \bibinfo
  {address} {Berlin, Heidelberg})\ pp.\ \bibinfo {pages} {27--38}\BibitemShut
  {NoStop}%
\bibitem [{\citenamefont {Wannier}(1937)}]{wannier}%
  \BibitemOpen
  \bibfield  {author} {\bibinfo {author} {\bibnamefont {Wannier}, \bibfnamefont
  {G~H}}} (\bibinfo {year} {1937}),\ \bibfield  {title} {\enquote {\bibinfo
  {title} {The structure of electronic excitation levels in insulating
  crystals},}\ }\href@noop {} {\bibfield  {journal} {\bibinfo  {journal} {Phys.
  Rev.}\ }\textbf {\bibinfo {volume} {52}},\ \bibinfo {pages}
  {191--197}}\BibitemShut {NoStop}%
\bibitem [{\citenamefont {{Wikipedia contributors}}(2025)}]{bloch_wiki}%
  \BibitemOpen
  \bibfield  {author} {\bibinfo {author} {\bibnamefont {{Wikipedia
  contributors}},}} (\bibinfo {year} {2025}),\ \href@noop {} {\enquote
  {\bibinfo {title} {Bloch's theorem --- {Wikipedia}{,} the free
  encyclopedia},}\ }\bibinfo {howpublished}
  {\url{https://en.wikipedia.org/w/index.php?title=Bloch\%27s_theorem\&oldid=1285946185}},\
  \bibinfo {note} {[Online; accessed 7-July-2025]}\BibitemShut {NoStop}%
\bibitem [{\citenamefont {Wilson}\ \emph {et~al.}(1987)\citenamefont {Wilson},
  \citenamefont {Pohorille},\ and\ \citenamefont {Pratt}}]{mip_pratt_1987}%
  \BibitemOpen
  \bibfield  {author} {\bibinfo {author} {\bibnamefont {Wilson}, \bibfnamefont
  {M~A}}, \bibinfo {author} {\bibfnamefont {A.}~\bibnamefont {Pohorille}}, \
  and\ \bibinfo {author} {\bibfnamefont {L.~R.}\ \bibnamefont {Pratt}}}
  (\bibinfo {year} {1987}),\ \bibfield  {title} {\enquote {\bibinfo {title}
  {Molecular dynamics of the water liquid-vapor interface},}\ }\href@noop {}
  {\bibfield  {journal} {\bibinfo  {journal} {J. Phys. Chem.}\ }\textbf
  {\bibinfo {volume} {91}}~(\bibinfo {number} {19}),\ \bibinfo {pages}
  {4873--4878}}\BibitemShut {NoStop}%
\bibitem [{\citenamefont {Wilson}\ \emph {et~al.}(1988)\citenamefont {Wilson},
  \citenamefont {Pohorille},\ and\ \citenamefont {Pratt}}]{mip_pratt_1988}%
  \BibitemOpen
  \bibfield  {author} {\bibinfo {author} {\bibnamefont {Wilson}, \bibfnamefont
  {M~A}}, \bibinfo {author} {\bibfnamefont {A.}~\bibnamefont {Pohorille}}, \
  and\ \bibinfo {author} {\bibfnamefont {L.~R.}\ \bibnamefont {Pratt}}}
  (\bibinfo {year} {1988}),\ \bibfield  {title} {\enquote {\bibinfo {title}
  {Surface potential of the water liquid-vapor interface},}\ }\href@noop {}
  {\bibfield  {journal} {\bibinfo  {journal} {J. Chem. Phys.}\ }\textbf
  {\bibinfo {volume} {88}}~(\bibinfo {number} {5}),\ \bibinfo {pages}
  {3281--3285}}\BibitemShut {NoStop}%
\bibitem [{\citenamefont {Wilson}\ \emph {et~al.}(1989)\citenamefont {Wilson},
  \citenamefont {Pohorille},\ and\ \citenamefont {Pratt}}]{mip_pratt_1989}%
  \BibitemOpen
  \bibfield  {author} {\bibinfo {author} {\bibnamefont {Wilson}, \bibfnamefont
  {M~A}}, \bibinfo {author} {\bibfnamefont {A.}~\bibnamefont {Pohorille}}, \
  and\ \bibinfo {author} {\bibfnamefont {L.~R.}\ \bibnamefont {Pratt}}}
  (\bibinfo {year} {1989}),\ \bibfield  {title} {\enquote {\bibinfo {title}
  {Comment on “{Study} on the liquid-vapor interface of water. 1.
  {Simulation} results of the thermodyanmic properties and orientational
  structure},}\ }\href@noop {} {\bibfield  {journal} {\bibinfo  {journal} {J.
  Chem. Phys.}\ }\textbf {\bibinfo {volume} {90}},\ \bibinfo {pages}
  {5211}}\BibitemShut {NoStop}%
\bibitem [{\citenamefont {Woo}(1971)}]{woo-prb-1971}%
  \BibitemOpen
  \bibfield  {author} {\bibinfo {author} {\bibnamefont {Woo}, \bibfnamefont
  {J~W~F}}} (\bibinfo {year} {1971}),\ \bibfield  {title} {\enquote {\bibinfo
  {title} {Piezoelectricity under hydrostatic pressure},}\ }\href@noop {}
  {\bibfield  {journal} {\bibinfo  {journal} {Phys. Rev. B}\ }\textbf {\bibinfo
  {volume} {4}},\ \bibinfo {pages} {1218--1220}}\BibitemShut {NoStop}%
\bibitem [{\citenamefont {Yesibolati}\ \emph {et~al.}(2020)\citenamefont
  {Yesibolati}, \citenamefont {Lagan\`a}, \citenamefont {Sun}, \citenamefont
  {Beleggia}, \citenamefont {Kathmann}, \citenamefont {Kasama},\ and\
  \citenamefont {M\o{}lhave}}]{mip_water_2020}%
  \BibitemOpen
  \bibfield  {author} {\bibinfo {author} {\bibnamefont {Yesibolati},
  \bibfnamefont {M~N}}, \bibinfo {author} {\bibfnamefont {S.}~\bibnamefont
  {Lagan\`a}}, \bibinfo {author} {\bibfnamefont {H.}~\bibnamefont {Sun}},
  \bibinfo {author} {\bibfnamefont {M.}~\bibnamefont {Beleggia}}, \bibinfo
  {author} {\bibfnamefont {S.~M.}\ \bibnamefont {Kathmann}}, \bibinfo {author}
  {\bibfnamefont {T.}~\bibnamefont {Kasama}}, \ and\ \bibinfo {author}
  {\bibfnamefont {K.}~\bibnamefont {M\o{}lhave}}} (\bibinfo {year} {2020}),\
  \bibfield  {title} {\enquote {\bibinfo {title} {Mean inner potential of
  liquid water},}\ }\href@noop {} {\bibfield  {journal} {\bibinfo  {journal}
  {Phys. Rev. Lett.}\ }\textbf {\bibinfo {volume} {124}},\ \bibinfo {pages}
  {065502}}\BibitemShut {NoStop}%
\bibitem [{\citenamefont {Z\"urcher}(2018)}]{zurcher_2018}%
  \BibitemOpen
  \bibfield  {author} {\bibinfo {author} {\bibnamefont {Z\"urcher},
  \bibfnamefont {U}}} (\bibinfo {year} {2018}),\ \href@noop {} {\emph {\bibinfo
  {title} {Electrostatics at the Molecular Level}}}\ (\bibinfo  {publisher}
  {{IOP} Concise Physics and Morgan \& Claypool Publishers})\BibitemShut
  {NoStop}%
\end{thebibliography}%

\end{document}